\documentclass[letterpaper, 12pt]{report}
\usepackage[margin=0.75in]{geometry}
\usepackage[T1]{fontenc}
\usepackage{graphicx, lipsum, textcomp, float} 
\usepackage{hyperref} 
\usepackage{chemformula}
\usepackage{amsmath}
\usepackage{amssymb}
\usepackage{multirow}
\usepackage{longtable}
\usepackage{xcolor}
\usepackage{listings}

\usepackage{adjustbox}
\usepackage{array, booktabs}
\usepackage{nameref}

\usepackage[frozencache,cachedir=.]{minted}

\usepackage[scaled]{beramono}
 
\definecolor{darkgreen}{rgb}{0.05, 0.3, 0.1}

\usepackage[htt]{hyphenat} 

\let\oldtexttt\texttt
\renewcommand{\texttt}[1]{\oldtexttt{\textcolor{darkgreen}{#1}}}

\usepackage{setspace}

\usepackage[
    style=ieee, 
    isbn=false, 
    url=true, 
    natbib=true, 
    backend=bibtex,
    maxcitenames=1,
    mincitenames=1
    ]{biblatex}
\usepackage{caption}
\usepackage{titlesec}

\titleformat{\chapter}[display]
{\Large\bfseries\centering}
{Chapter \thechapter}{0.5em}{}[\vspace{2ex}\titlerule]
\titlespacing*{\chapter}{0pt}{0pt}{30pt}

\titleformat{\section}[hang]
{\large\bfseries}
{\thesection}{0.5em}{}

\titleformat{\subsection}[hang]
{\large\bfseries}
{\thesubsection}{0.5em}{}

\usepackage{tocbasic}

\DeclareTOCStyleEntry[
  beforeskip=.2em plus 1pt,
  pagenumberformat=\textbf
]{tocline}{section}

\DeclareTOCStyleEntry[
  entrynumberformat=\entrywithprefix{\chaptername},
  dynnumwidth
]{tocline}{chapter}
\newcommand*\entrywithprefix[2]{#1~#2}

\usepackage[toc]{appendix}
\newcommand{\mypart}[1]{\thispagestyle{empty}\part*{#1}}

\definecolor{mypurple}{RGB}{140,54,140}
\definecolor{homered}{RGB}{127, 0, 10}
\definecolor{officeorange}{RGB}{204, 75, 0}
\definecolor{mauroblue}{RGB}{53, 48, 217}
\definecolor{citegreen}{RGB}{15, 133, 13}
\definecolor{hyperlinkpurple}{RGB}{42, 0, 163}
\definecolor{subtlegray}{gray}{0.98}
\definecolor{subduedgray}{gray}{0.75}
\hypersetup{
    colorlinks=true,
    linkcolor=hyperlinkpurple,
    filecolor=mypurple,      
    urlcolor=teal,
    citecolor=citegreen
}

\newcommand*{\fullref}[1]{\hyperref[{#1}]{\ref*{#1} on \nameref*{#1}}}
\newcommand{\acknowledge}[1]{\textit{
\small
Acknowledgment: #1
}}

\begin{document}

\pagenumbering{roman}
\thispagestyle{empty}
\setstretch{1}

{
\centering
The Pennsylvania State University\\
The Graduate School\\
\vfill
\setstretch{2}
{
\Large
\textbf{EFFICIENT MATERIALS INFORMATICS BETWEEN ROCKETS AND ELECTRONS}\\
}
\vfill
A Dissertation in\\
Materials Science and Engineering\\
by\\
Adam M. Krajewski\\
\vfill
© 2024 Adam M. Krajewski\\
\setstretch{1}
\vfill
Submitted in Partial Fulfillment\\
of the Requirements\\
for the Degree of\\

\vfill
Doctor of Philosophy\\
\vfill
August 2024\\
\vfill
}

\newpage
\setstretch{1.5}
\setlength\parindent{0pt} 

The dissertation of Adam M. Krajewski was reviewed and approved by the following:\\

\textbf{Zi-Kui Liu}\\
Dorothy Pate Enright Professor at the Department of Materials Science and Engineering\\
Director of the Phases Research Laboratory\\
Dissertation Advisor and Chair of the Committee\\

\textbf{Allison M. Beese}\\
Professor of Materials Science and Engineering\\
Professor of Mechanical Engineering\\
Director of Additive Manufacturing \& Design Graduate Program\\
Co-director of Center for Innovative Materials Processing through Direct Digital Deposition\\

\textbf{Ismaila Dabo}\\
Associate Professor of Materials Science and Engineering\\
Associate Professor of Physics\\

\textbf{Wenrui Hao}\\
Associate Professor of Mathematics\\

\textbf{John Mauro}\\
Program Head\\

\vfill

\newpage
\chapter*{Abstract}

The true power of computational research typically can lay in either what it accomplishes or what it enables others to accomplish. In this work, both avenues are simultaneously embraced across several distinct efforts existing at three general scales of abstractions of what a material is - atomistic, physical, and design. At each, an efficient materials informatics infrastructure is being built from the ground up based on (1) the fundamental understanding of the underlying prior knowledge, including the data, (2) deployment routes that take advantage of it, and (3) pathways to extend it in an autonomous or semi-autonomous fashion, while heavily relying on artificial intelligence (AI) to guide well-established DFT-based ab initio and CALPHAD-based thermodynamic methods.

The resulting multi-level discovery infrastructure is highly generalizable as it focuses on encoding problems to solve them easily rather than looking for an existing solution. To showcase it, this dissertation discusses the design of multi-alloy functionally graded materials (FGMs) incorporating ultra-high temperature refractory high entropy alloys (RHEAs) towards gas turbine and jet engine efficiency increase reducing \ch{CO_2} emissions, as well as hypersonic vehicles. It leverages a new graph representation of underlying mathematical space using a newly developed algorithm based on combinatorics, not subject to many problems troubling the community. Underneath, property models and phase relations are learned from optimized samplings of the largest and highest quality dataset of HEA in the world, called ULTERA. At the atomistic level, a data ecosystem optimized for machine learning (ML) from over 4.5 million relaxed structures, called MPDD, is used to inform experimental observations and improve thermodynamic models by providing stability data enabled by a new efficient featurization framework.

\setstretch{1}
\newpage
\tableofcontents

\newpage
\addcontentsline{toc}{chapter}{\listfigurename}
\renewcommand{\listfigurename}{List of Figures}
\listoffigures

\newpage
\addcontentsline{toc}{chapter}{\listtablename}
\renewcommand{\listtablename}{List of Tables}
\listoftables

\newpage
\chapter*{Acknowledgments}
\label{acknowledgments}
\addcontentsline{toc}{chapter}{\nameref{acknowledgments}}

I would like to thank all of my family, friends, and collaborators who supported me along the way, with the spotlight given to my parents, \textbf{Mikołaj Krajewski} and \textbf{Izabella Krajewska}, without whom I would not (statistically) become a scientist, let alone become a \emph{doctor}. However, the degree is just a classifier in a database somewhere without being backed by science created while completing it. Thus, I would like to thank my advisor, \textbf{Zi-Kui Liu}, for guiding me over the last five years of exceptionally productive research that pushed me to \emph{do better than my best}.

I want to thank my research group colleagues, whom I worked with over the years, including \textbf{Zi-Kui Liu}, who was a great colleague, in addition to being a great advisor, \textbf{ShunLi Shang, Yi Wang, Brandon Bocklund, Jorge Paz Soldan Palma, Hongyeun Kim, John Shimanek, Hui Sun, Rushi Gong, Shuang Lin, Alexander Richter, Luke Myers}, and \textbf{Ricardo Amaral}.

I would like to thank my colleagues who, to the best of their abilities, kept me from falling into an abyss of scientific insanity through hours spent on less technical conversations. In particular, but in no particular order, I would like to acknowledge several of them who regularly attended my weekly office hours over the years: \textbf{Stephen Holoviak, Alexander Richter, Luke Myers, Cooper Pan, Curtis Warner, Ian Cunningham, James Ricardo, Ellie Franklin, Hamdan Almarzooqi, Jorge Paz Soldan Palma, Brandon Bocklund}, and \textbf{Stephanie Castro Baldivieso.}

I would like to thank my Lawrence Livermore National Lab colleagues \textbf{Aurelien Perron, Brandon Bocklund, Kate Elder, Joseph McKeown}, and other amazing colleagues from the Materials Science Division at Lawrence Livermore National Lab (LLNL) for having the opportunity of working with them on solving challenging problems with great implementation flexibility which prompted me to deepen my understanding of highly dimensional design spaces, without which some of the work in this dissertation would never happen.

On the software side, I would like to thank (1) \textbf{Jinchao Xu} from PSU/KAUST for his contribution to the development of SIPFENN; (2) \textbf{Richard Otis} and \textbf{Brandon Bocklund} from \textbf{Materials Genome Foundation} for supporting my work since 2019 in many ways, including invaluable guidance in organizing community workshops.

In the Fall of 2023, I had an opportunity to be a Visiting PhD Student at the University of Cambridge, for which I am very grateful to \textbf{Gonville \& Caius College} which invited me, Caius Fellow \textbf{Gareth Conduit} for generously sponsoring said invitation, and \textbf{Peter and Carol Thrower} for sponsoring the fellowship enabling this travel.

This work was made possible by the financial support and training provided by US Department of Energy (DOE) via Awards DE-FE0031553 and DE-EE0008456, DOE Advanced Research Projects Agency-Energy (ARPA-E) via DE-AR0001435, the DOE BES (Theoretical Condensed Matter Physics) via DE-SC0023185, US Office of Naval Research (ONR) via N00014-17-1-2567 and N00014-23-2721, The Pennsylvania State University via ICDS Seed Grant, US National Science Foundation (NSF) via CMMI-1825538, and Pathways to Enable Open-Source Ecosystems (POSE) via FAIN-2229690. Any opinions, findings, conclusions, or recommendations expressed in this publication are those of the author and do not necessarily reflect the views of the funding agencies.


\newpage
\setlength\parindent{6pt} 
\setstretch{1.5}
\pagenumbering{arabic}

\chapter{Introduction} \label{sec:Introduction}

\section{Big Picture} \label{intro:sec:bigpicture}

The discovery of novel materials that solve societal challenges or otherwise improve human lives is arguably one of the most critical components of building a modern world. Starting from the bronze age, humans were able to reliably combine raw materials in a structured fashion to achieve desired results, even though, at the time, there was no mechanistic understanding of \emph{why} things happen. This has changed with gradual introduction of the scientific method, which standardized and systematized the discovery approach, with revolutionary advancements in materials happening every time a new technology for sharing and combining knowledge, such as propagation of the Greek language, printing press, or computer aided design (CAD), has been introduced and widely adopted.

In the current world, which went through the Internet revolution around 2000 and is currently going through the artificial intelligence (AI) revolution of the 2020s, one can point to the informatiztion of materials science as one such communication technology with a potential to revolutionize materials discovery by combining vast amounts of multidimensional data, intricate multidisciplinary domain knowledge, and ability to guide experiments beyond level achievable by a human being. In order to achieve this, one has to consider how to combine these \emph{efficiently}, mitigating problems such as inhomogenities between data sources, computational challenges related to vast design spaces, hidden uncertainties in the reported values, and many flavors of errors, unavoidably present in the complex datasets involved.

\begin{figure}[H]
    \centering
    \includegraphics[width=0.95\textwidth]{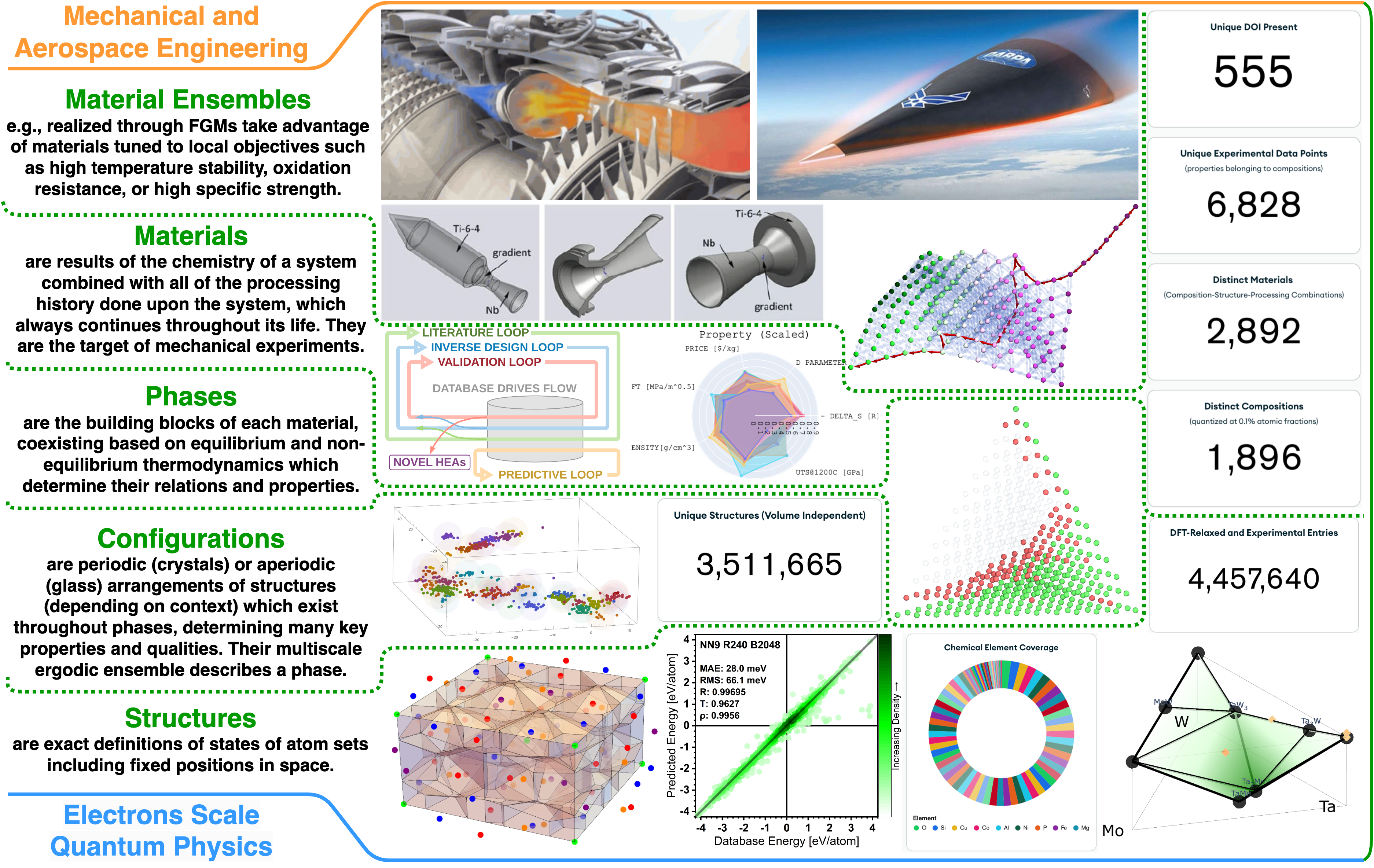}
    \caption{
    Intermediate material modeling scales bridging together quantum physics and aerospace engineering to enable high-technology solutions through excellence of underlying ensembles of materials. In this work, all of the scales are brought together to take advantage of data and knowledge from all relevant sources. Top render of hypersonic vehicle reproduced from DARPA under public domain and gray nozzle renders from Hofmann2014 under CC BY-NC-ND 4.0 License. Several images occur later in the manuscript in Figures \ref{pathplan:fig:lowgradientsquared}, \ref{ultera:fig:dashboard}, \ref{ultera:fig:dataloops}, \ref{inverse:fig:cgandemo}, \ref{infeasibilitygliding:fig:glide}, \ref{crystall:fig:ndbi2clusters}, \ref{pysipfenn:fig:ks2022}, \ref{sipfenn:fig:oqmdperformance}, and \ref{mpdd:fig:dataset}.
    }
    \label{intro:fig:bigpicture}
\end{figure}

While creating an efficient, high-performance, fully general ecosystem for materials informatics appears nearly impossible even for a large group of researchers, a case-specific approach can be constructed in a fashion prioritizing generalizability, which can then be adjusted to other problems. This Dissertation builds such a case-specific approach, embedding a more general blueprint through the development of methods that are rarely limited to the target application but are rather biased towards it through design choices, assumptions, and helpful simplifications. In the process, it introduces several novel individual pieces of software, including \texttt{pySIPFENN}, \texttt{MPDD}, \texttt{crystALL}, \texttt{ULTERA}, \texttt{PyQAlloy}, \texttt{nimCSO}, \texttt{nimplex}, and their derivatives to collectively bridge ab-initio methods, belonging to the domain of quantum physics, with engineering of devices placed in extreme environments, such as gas engine turbine blades or hypersonic vehicles, designed by aerospace engineers, through efficient materials informatics pipelines existing across scales, as summarized in the Figure~\ref{intro:fig:bigpicture} and described in detail in Section~\ref{intro:sec:flow}.

The motivation for this specific choice of application - \emph{metallic alloys targeting extreme environments}, has been twofold. First, several intrinsic challenges, including (1) competing property trends, (2) scarce experimental data relative to room temperature materials, and (3) compositional complexity of currently studied alloy families, make this problem very difficult. Thus, it is an excellent target for the design of advanced methods that can mitigate them while encountering and addressing otherwise hidden problems.

Secondly, such alloys are of great interest to the society as a whole. For instance, per the US Department of Energy's ARPA-E estimates, developing a standalone alloy that could continuously operate at $1300^oC$ has the potential to increase gas turbine efficiency up to $7\%$, significantly reducing wasted energy, and consequently carbon emissions, saving up to 20 quads of energy in electricity generation and civilian aviation between now and 2050 \cite{ULTIMATEArpa-e.energy.gov}. Such efficiency increase could prevent the release of approximately 1,000,000,000,000 kg of \ch{CO_2} from burning natural gas, or double that from coal; thus, becoming a critical effort in fighting global warming in applications, like airplanes, where green technologies cannot be directly adapted. 

Another extreme environment application, quite far from the first one, is the class of hypersonic vehicles that travel faster than 5 times the speed of sound \emph{through Earth's atmosphere for extended periods of time}, thus generating extreme sustained temperatures within structural components. This prompts the need for novel materials and engineering techniques, as evidenced by massive funding assigned to this research area by the US military, which increased its yearly budgets for hypersonic \emph{research} from \$3.8 billion in FY2022 to \$4.7 billion in FY2023, and to an undisclosed amount this year (FY2024) \cite{Sayler2024HypersonicCongress}, further demonstrating the criticality of such materials.

Lastly, in the near future, granted other related challenges are solved, extreme-environment alloys may be the missing key to constructing reliable fusion reactors where chamber walls must be capable of resisting extreme temperatures, constant plasma exposure, and irradiation, appreciably beyond our current materials \cite{CHADWICKArpa-e.energy.gov}. Thus, research into such materials may also one day enable an entirely green future, and has recently been endorsed in this direction by the United States White House \cite{FactHouse}.

\section{Flow of Material Discovery and This Work} \label{intro:sec:flow}

Throughout this work, all topics raised in Section \ref{intro:sec:bigpicture} will be discussed in a reversed order to progressively build from fundamentals to highly-specialized final applications, while retaining generality at every stage. This way, one will be able to build a holistic picture focused on how data flows within materials informatics research and converges together at consecutive length-scales to discover new materials in specific niches in a systematic, easy-to-automate approach, rather than build elaborate solutions that may \emph{happen} to work well but can also break the fundamentals - a common occurrence in our era of powerful computing and machine learning where tools \emph{always give an answer} but it may hold negative value.

\begin{figure}[H]
    \centering
    \includegraphics[width=0.7\textwidth]{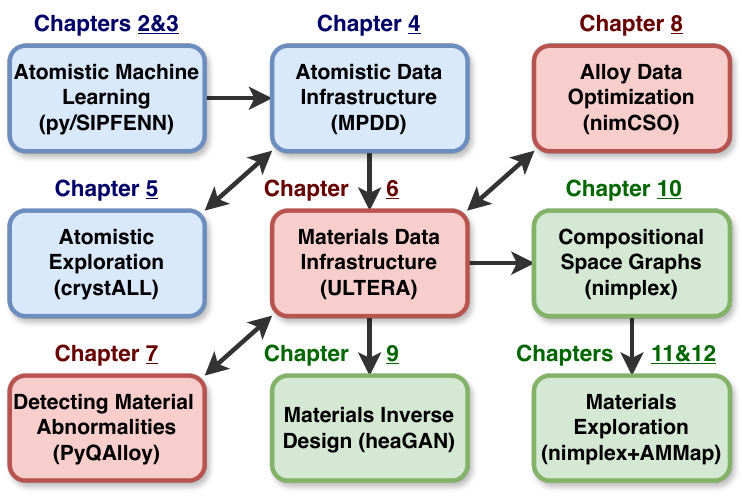}
    \caption{Schematic outline of this dissertation flowing through 3 overarching types of materials science research. It starts from atomistic treatment (blue) allowing modeling of physical materials (blue) and leading to design (green). For each category, three most significant advancements done in this work have been selected to showcase computational infrastructures and methods to extend our understanding or capabilities.}
    \label{intro:fig:outline}
\end{figure}

As shown in Figure \ref{intro:fig:outline}, the first 4 chapters (colored blue) cover atomistic treatment of materials, discussing how data at this level is collected, featurized, managed and expanded. \texttt{SIPFENN} approach and the latest \texttt{pySIPFENN} featurization package are first developed. Then \texttt{MPDD} database with 4.5 million DFT-relaxed or experimentally observed entries is set up to serve as a highly efficient deployment vehicle. Lastly, \texttt{crystALL} approach automatically extends it into chemical areas of interest.

All of \texttt{MPDD} is then harvested to model materials at the physical scale by (1) serving as inputs to thermodynamic model generation using \texttt{pycalphad} \cite{Otis2017Pycalphad:Python} and \texttt{ESPEI} \cite{Bocklund2019ESPEICuMg} or training of \texttt{pySIPFENN} ML models generating needed data, and (2) informing experimental observations by, for instance, automatically compiling a set of carbides stable in an alloy system at 0K. At the same time, the largest experimental HEA data infrastructure, called \texttt{ULTERA}, is compiled joining together over 6,800 property datapoints manually extracted from 555 literature publications. 

The experimental database is curated through novel \texttt{PyQAlloy} package created to detect abnormalities and dramatically reduce fraction of erroneous data relative to other similar ones in the literature. Once curated, the \texttt{nimCSO} package can guide ML efforts in terms of which components of the data (chemical elements) should be considered when modeling to optimize trade-off between applicability and data density available to the models. Lastly, compositional space representations generated through \texttt{nimplex} and inverse design workflows serve as deployment vehicles for the trained methods.

\section{Executive Summary} \label{intro:sec:summary}

First, Chapter \fullref{chap:sipfenn} introduces fundamental concepts critical to structure-informed modeling of atomic configurations from the perspective of machine learning (ML) modeling and presents design of such models employing artificial neural networks for the prediction of formation energies of atomic structures based on elemental and structural features of Voronoi-tessellated materials. It provide a concise overview of the connection between the machine learning and the true material-property relationship, how to improve the generalization accuracy by reducing overfitting, how new data can be incorporated into the model to tune it to a specific material system, and preliminary results on using models to preform local structure relaxations.

It results in three final models optimized for achieving (1) highest test accuracy on the Open Quantum Materials Database (OQMD), (2) high extrapolative performance in the discovery of new materials, and (3) high performance at a low computational cost. On a test set of 21,800 compounds randomly selected from OQMD, these models achieves a mean absolute error (MAE) of 28, 40, and 42 meV/atom, respectively. The first model represented the state-of-the-art performance on this problem when released in 2020 \cite{Krajewski2020SIPFENNModels} (see Table \ref{sipfenn:comparison-results}), the second model provides better predictions in a test case of interest not present in the OQMD, while the third one reduces the computational cost by a factor of 8 making it applicable to embedded and mobile applications. A transfer learning procedure was also demonstrated for the first time, showing dramatic improvements in extrapolation with just several datapoints (see Figure \ref{sipfenn:fig:transfersigmaVsDatapoints}).

The results were implemented into a new open-source tool called \texttt{SIPFENN} or \textit{Structure-Informed Prediction of Formation Energy using Neural Networks}, which not only improved the accuracy beyond existing models but also shipped in a ready-to-use form with pre-trained neural networks, which was first-of-a-kind at release, and a GUI interface allowing it be included in DFT calculations routines at nearly no cost.

Next, Chapter \fullref{chap:pysipfenn} expands upon \texttt{SIPFENN} and implements a fully-featured machine learning focused analysis framework called \texttt{pySIPFENN} or \textit{python toolset for Structure-Informed Property and Feature Engineering with Neural Networks} to fuel needs of structure-informed materials informatics - a rapidly evolving discipline of materials science relying on the featurization of atomic structures or configurations to construct vector, voxel, graph, graphlet, and other representations useful for machine learning prediction of properties, fingerprinting, and generative design. This chapter discusses how current featurizers typically perform redundant calculations and how their efficiency could be improved by considering (1) fundamentals of crystallographic (orbits) equivalency to optimize ordered cases and (2) representation-dependent equivalency to optimize cases of dilute, doped, and defect structures with broken symmetry. It also discusses and contrasts ways of (3) approximating random solid solutions occupying arbitrary lattices under such representations.

Efficiency improvements discussed in this work were implemented within \texttt{pySIPFENN} and shown to increase performance from 2 to 10 times for typical inputs just based on fundamentals of materials science. Throughout this work, the authors explicitly discuss how these advances can be applied to different kinds of similar tools in the community.

Chapter \fullref{mpdd:sec:mpdd} shifts focus from developing new techniques for data analysis to building an innovative atomistic data infrastructure to fuel machine learning model training and deployment process. This effort builds from the idea that, fundamentally, each atomistic ML study comprises of three elements: a dataset or database of \emph{materials}, a \emph{descriptor} or set of features known for each material, and an ML algorithm trained to predict a \emph{property} or a set of them. These three are combined in two steps. First, the data representation is calculated using the descriptor. Then the model is iteratively evaluated on this representation or adjusted to improve it. Both processes are nearly instantaneous compared to ab-initio based methods; however, with extensive databases or materials modeled with large super-cells (e.g., glasses), compute times can grow into days or years for elaborate analysis tools deployed over many millions of datapoints. 

\texttt{MPDD} is a tool that can speed up the total process for the end-user by orders of magnitude through removal of the most time-intensive step, i.e., the descriptor calculation. To accomplish that, it moves from the traditional practice of sharing only the material-properties data to sharing the descriptors-properties data corresponding to the material as well, employing a high-performance NoSQL \texttt{MongoDB} database with highly engineered indexing system and fully salable compute node deployment model. The latter is used to progressively extend it based on guidance received from tools described in later chapters of this dissertation.

\texttt{MPDD} deployment model is not only much faster but also serves as a tool for an automated and robust embodiment of prior knowledge about materials in a graph-like fashion. Lastly, since the descriptors are often reused for related properties, our database provides a tremendous speed-up in the design space exploration.

Since 2023, a stable, well-implemented endpoint of the \texttt{OPTIMADE} API available at \href{https://optimade.mpdd.org}{https://optimade.mpdd.org} allows \texttt{MPDD} to be seamlessly integrated with other community databases and serve atomistic feature data associated with their entries for synergistic merger of experimental observations, ab initio calculations, and machine learning predictions.

Chapter \fullref{chap:crystall}, introduces a simple yet powerful approach to discovery of new atomic structures and identification of experimentally observed ones that evade identification, which was already demonstrated in certain important chemical systems. It implements in a concise computational tool which leverages millions of previously calculated structures from \texttt{MPDD} Database and other databases accessed through the \texttt{OPTIMADE} API to identify all atomic structures that can accommodate target stoichiometry (or uncertain chemical composition form experiments) and perform all permutations of substitutions to arrive at a list of tens of thousands of candidate structures. 

It analyzes candidates in the \texttt{KS2022} feature space, introduced in Chapter \ref{chap:pysipfenn} as part of \texttt{pySIPFENN}, to detect unique set of structures underlying different atomic configurations through different clustering techniques. Lastly, it selects a member from each cluster based on minimum formation energy to arrive at an ensemble of candidate configurations then passed to ab initio calculations or experiments for validation. 

Such ensemble approach is shown to provide critical advantages when contrasted with convex-hull-structure searching approach which recently dominated the community. It can also be used to expand unexplored regions of \texttt{MPDD} database, providing valuable deployment target for machine learning models.

Chapter \fullref{chap:ultera} crosses from the atomistic-level materials science into "real" materials physically created and measured experimentally, while still focusing on similar aspects of efficient modeling, data handing, and model deployment. It discusses the \texttt{ULTERA} Database, developed under the ARPA-E's ULTIMATE program and aimed at collecting literature data on high entropy alloys (HEAs).

\texttt{ULTERA} facilitates rapid discovery of new alloys using forward and inverse design, with the primary focus on creep behavior, yield stress, ductility, and hardness. Its advanced architecture, composed of many intermediate databases, pipelines, and analysis tools, is designed to automatically integrate starting literature data in real-time with methods such as experiments, generative modeling, predictive modeling, and validations. Thanks to such automation, the experimental team can operate on the best candidates available, while generation of new ones based on incoming experiments can be delegated to the cloud.

As of April 2024, ULTERA contains over 6,800 property-datapoints, corresponding to 2,900 unique HEAs, manually collected from 550 source literature publications. All data is available through a high-performance API, following FAIR principles.

Chapter \fullref{chap:pyqalloy} takes the \texttt{ULTERA} Database and dramatically increases its value to the machine learning applications by removing or fixing approximately 5\% of the errors present in it and other state-of-the-art HEA datasets, arriving at \emph{several times less erroneous data}.

In the past, these challenge was not as severe nor impactfull because the effort placed in the data analysis was much greater than data extraction; however, as our community moves towards this large-scale multi-source data collection, it increasingly starts to surface impacting ML efforts. While present in datasets on all alloy classes, this problem is particularly visible in high entropy alloys (HEAs) due to their novelty and complexity. A number of factors causes errors, including a lack of standardized notations and general parsing mistakes, such as typos, but most of them can be caught through detection of abnormalities in relation to the correct data and its patterns.

In this chapter, we present an open-source Python tool called \texttt{PyQAlloy} that can be run on alloy datasets to screen them for a variety of such abnormalities. It puts each data point in a variety of scopes ranging from individual datapoint analysis, through single-study meta-data, to the database as a whole.

Chapter \fullref{chap:nimcso} builds on top of \texttt{ULTERA} datasets curated by \texttt{PyQAlloy} and recognizes that selecting dimensions to model, or in this case, the selection of chemical elements to restrict the modeling effort, is a combinatorically hard problem for complex compositions existing in highly dimensional spaces due to the interdependency of components being present. Consequentially, optimizing the data availability and density for applications such as machine learning becomes a challenge.

A novel tool, called \texttt{nimCSO}, has been introduced and implmenets several methods for performing this task over even millions of datapoints and extreme dimensionalities encountered, for instance, in materials science of Compositionally Complex Materials (CCMs) which often span 20-45 chemical elements, 5-10 processing types, and several temperature regimes, for up to 60 total data dimensions.

It achieves the extreme performance by leveraging the metaprogramming ability of the Nim language to optimize itself at the compile time, both in terms of speed and memory handling, to the specific problem statement and dataset at hand based on a human-readable configuration file. As demonstrated in the chapter, it reaches the physical limits of the hardware (L1 cache latency) and can outperform an efficient native Python implementation over 400 times in terms of speed and 50 times in terms of memory usage (\emph{not} counting interpreter), while also outperforming NumPy implementation 35 and 17 times, respectively, when checking a candidate solution. It was designed to be both a user-ready tool and a scaffold for building even more elaborate methods in the future, including heuristics going beyond data availability.

Chapter \fullref{chap:inversedesign} demonstrates how \texttt{ULTERA} dataset can be used to propose new alloys through inverse design. It leverages conditional Generative Adversarial Netowrks (cGAN) models to bias the predicted alloys in terms of (1) the property values by setting the conditioning and (2) the predicted compositions by recognizing \emph{concept vectors} in the latent space established at the training step. The former ability is implemented within a demonstrator Python package called \texttt{heaGAN} which can be used to design HEAs by setting design criteria. It also allows advanced users to re-train cGAN models based on their own data or a third-party property surrogate model.

Chapter \fullref{chap:nimplex} departs from both data and machine learning to focus on the mathematical abstractions of the alloy design problem. It begins by considering that many disciplines of science and engineering deal with problems related to compositions, ranging from chemical compositions in materials science to portfolio compositions in economics. They exist in non-Euclidean simplex spaces, causing many standard tools to be incorrect or inefficient, which is significant in combinatorically or structurally challenging spaces exemplified by Compositionally Complex Materials (CCMs) and Functionally Graded Materials (FGMs). In this chapter, we explore them conceptually in terms of problem spaces and quantitatively in terms of computational feasibility.

Several essential methods specific to the compositional (simplex) spaces are implemented, with most critical parts presented in chapter body as code listings, through a high-performance open-source library \texttt{nimplex}. Most significantly, we derive and implement an algorithm for constructing a novel n-dimensional simplex graph data structure, which contains all discretized compositions and all possible neighbor-to-neighbor transitions as pointer arrays. Critically, no distance or neighborhood calculations are performed, instead leveraging pure combinatorics and the ordering in procedurally generated simplex grids, keeping the algorithm $\mathcal{O}(N)$, so that graphs with billions of transitions take seconds to construct on a laptop. Furthermore, we demonstrate how such graph representations can be combined to express path-planning problem spaces and to incorporate prior knowledge while keeping the problem space homogeneous. This allows for efficient deployment of existing high-performance gradient descent, graph traversal search, and other path optimization algorithms.

Chapter \fullref{chap:infeasibilitygliding} leverages the novel graph-based representations of compositional spaces to dramatically improve our approach to screening and modifying compositionally complex alloys based on their structural constraint feasibility. It builds on the fact that the equilibrium existence of a particular phase in the phase-regions of chemical space (not necessarily corresponding to the present elements, as exploted in Chapter \ref{chap:nimplex}) is bound by a continuous surface of (generally) low curvature. Thus, performing the community-standard brute force screening type evaluation of all compositions will often calculate high numbers of points that cannot be reached.

Thus, \texttt{nimplex}-generated graphs are used to quickly re-implement the screening procedure into depth-first search traversals of the feasible regions that \emph{glide one the infeasibility boundaries}, reducing computation by half in semi-randomly picked "bad-case" which was known to produce relatively large feasible regions. In more elaborate screenings not taking advantage of prior knowledge, this improvement may very well reach between one and several orders of magnitude.

Lastly, Chapter \fullref{chap:pathplanning} takes additional advantage of the \texttt{nimplex}-generated graphs, which can be also extracted from feasible regions established in \ref{chap:infeasibilitygliding}, to deploy path planning algorithms and find sets of compositions that allow dissimilar alloys to be continuously combined in the least number of steps. Additional considerations are made in relation to an example property field of Root Mean Square Atomic Displacement (RMSAD) related to the yield stress in HEAs. In particular, its value, gradient, and magnitude of gradient are used to modify the graph weights encoding different design objectives, which are effortlessly solved with the same out-of-the box pathfinding approach.

Appendix \fullref{chap:supdiscussions} explores minor considerations tangential to the information flow or otherwise less critical, that may be of significant interest to some readers, while Appendix \fullref{chap:othersoft} discusses 8 pieces of software developed to during this work bur (1) were not directly used, (2) critical to discussion, or (3) were a contribution of a component to existing software.

Appendices \fullref{chap:nimplextutorial1} and \fullref{chap:nimplextutorial2} contain workshop material available under \texttt{nimplex} repository adapted to the formatting of this dissertation and go into details of both "Why?" and "How?" tools introduced between Chapters \ref{chap:nimplex} and \ref{chap:pathplanning} can be used by the end user. They can be run interactively with one-click cloud virtual environment by following current link at \href{https://nimplex.phaseslab.org}{nimplex.phaseslab.org} or by running them locally based on the supplementary materials.

Appendices \fullref{chap:pysipfenntutorial1} and \fullref{chap:pysipfenntutorial2} showcase how \texttt{pySIPFENN} integrates with other computational tools in the community and enables one to guide remote DFT calculations on a server cluster and adjust its models to a specific problem. Both were given by Adam Krajewski as guest lectures in the MatSE 580 course at Penn State in the Fall of 2023 and can be run interactively with one-click cloud virtual environment by following current link at \href{https://amkrajewski.github.io/MatSE580GuestLectures}{amkrajewski.github.io/MatSE580GuestLectures}, which also neatly presents the expected outcomes, or by running them locally based on the supplementary materials.


\chapter{Extensible Structure-Informed Prediction of Formation Energy with Improved Accuracy and Usability employing Neural Networks} \label{chap:sipfenn}

\acknowledge{
This chapter adapts verbatim the preprint version of \citet{Krajewski2022ExtensibleNetworks} published under \href{https://arxiv.org/abs/2008.13654v4}{arXiv:2008.13654v4} extended by additional discussions present in earlier versions published starting August 2020 under \href{https://arxiv.org/abs/2008.13654v1}{arXiv:2008.13654v1}. This work was co-authored with Jonathan Siegel, Jinhao-Xu, and Zi-Kui Liu. All text was written by Adam M. Krajewski with Jonathan Siegel co-writing Section \ref{sipfenn:ref:machinelearningoverview}. Jinhao Xu and Zi-Kui Liu provided edits and guidance.
}

\section{Introduction} \label{sipfenn:sec:Introduction}

\label{sipfenn:ssec:Motivation}
In recent years the field of material data informatics has been growing in importance thanks to the proliferation of open-access databases \cite{Saal2013MaterialsOQMD,Kirklin2015TheEnergies, vandeWalle2018TheDatabase,Jain2013Commentary:Innovation,Curtarolo2013AFLOW:Discovery,Toher2018TheDiscovery,Pizzi2016AiiDA:Science} and new methods being implemented to predict a wide variety of material properties \cite{Isayev2017UniversalCrystals, Legrain2017HowSolids, Pilania1987MachineSuperlattices, Jung2019BayesianSteels, Ouyang2020ComputationalConductors,Bucior2019Energy-based,Chandrasekaran2019SolvingLearning, Kim2018Machine-learning-acceleratedCompounds,Wen2019MachineProperty, Scime2019UsingProcess}. Within these methods, machine learning (ML) and, more broadly, artificial intelligence (AI) is becoming dominant, as noted in two recent reviews \cite{Schmidt2019RecentScience, Vasudevan2019MaterialsPhysics}, which listed a total of around 100 recent studies that attempted to solve material science problems using ML and AI techniques. These studies report benefits such as a 30-fold increase in material discovery rate when guided by an ML-model \cite{Kim2018Machine-learning-acceleratedCompounds}, or the ability to create new state-of-the-art materials in highly complex design spaces like 6-component alloys \cite{Wen2019MachineProperty}. They also dive into new paradigms of materials science by handling previously unthinkable amounts of data, allowing the creation and analysis of an energy convex-hull calculated for all elements \cite{Aykol2019NetworkDiscovery, I.Hegde2020TheMaterials}, or a concurrent analysis of all available literature texts to find paths for material synthesis \cite{Kononova2019Text-minedRecipes}. In addition, some studies promise to solve significant industrial challenges such as detection of additive manufacturing flaws with relatively simple and accessible data, but above-human pattern recognition quality and speed \cite{Scime2019UsingProcess}.

A common approach is to  focus on the discovery of candidate materials promising a new state-of-the-art performance, which must then be validated by experiment. The mismatch between the predictions and experiment measures the quality of the model, and reducing this gap is a major challenge due to the newly designed materials often being far from known materials, combined with attention placed on regions with extraordinary predictions.However, even if design models were perfectly accurate, many predicted materials cannot be physically made in the lab. An increasing number of studies attempt to solve this challenge by focusing not only on predicting how the material will perform but also on whether it can be manufactured \cite{Alberi2019TheRoadmap}. Generally, these include predicting materials' stability \cite{Balachandran2018PredictionsTheory, Li2019ThermodynamicLearning, I.Hegde2020TheMaterials, Im2022ThermodynamicModeling, Shang2021FormingJoints} and synthesizability \cite{Hattrick-Simpers2018AMaterials,Kononova2019Text-minedRecipes, Aykol2019NetworkDiscovery} with the stability being the more constraining parameter, as it determines whether the material could be stable or metastable in the use conditions, and therefore whether it can be synthesizable. Thus, predicting stability through prediction of fundamental thermodynamic properties such as formation energy is of special importance.

In the present work, new ML models and a tool to quickly use them are developed to improve the process of materials discovery by efficient prediction the formation energy and streamlined incorporation into materials discovery frameworks that aim to screen billions rather than hundreds of candidates available with cost-intensive calculations like first-principles calculations based on the density functional theory (DFT). 

\label{sipfenn:ssec:currentapproach}
In simple terms, every ML model is composed of three essential elements: a database, a descriptor, and an ML technique (also known as ML algorithm). The first element, databases, contain prior knowledge and are becoming increasingly shared between many studies, thanks to being open-access and often containing orders of magnitude more experimental or computational data than could be feasibly collected for a single study \cite{Saal2013MaterialsOQMD,Kirklin2015TheEnergies, vandeWalle2018TheDatabase,Jain2013Commentary:Innovation,Curtarolo2013AFLOW:Discovery,Toher2018TheDiscovery,Pizzi2016AiiDA:Science}. Databases used within the present paper are detailed in Section \ref{sipfenn:sssec:Data}.

The second element of an ML model is the descriptor (i.e., feature vector describing the material) which determines a representation of knowledge (data from the database) in a way relevant to the problem. It is typically built from many features, also known as attributes or vector components, which usually are determined through domain knowledge to be relevant or selected through correlation analysis. All combined, these features are a representation of some state whose meaning will be problem-specific.

When treating materials on the single atomic configuration level, descriptors can be generally divided into composition-based (also known as stoichiometric, structure-invariant, or elemental) \cite{Jha2018ElemNet:Composition, Ward2016AMaterials, , Legrain2017HowSolids} and structure-informed \cite{Ward2017IncludingTessellations, Seko2017RepresentationProperties,Schutt2014HowProperties}. The first type usually provides a more compact representation at a much lower computational cost, as calculating a composition-based descriptor often needs to involve only simple linear algebra operations such as matrix multiplication \cite{Ward2016AMaterials}, or prior-knowledge-incorporating attention-based analysis of a graph representation of the composition \cite{Goodall2020PredictingStoichiometry}. In cases where deep neural networks (DNNs) are employed, descriptor calculation can be skipped altogether by passing a composition vector directly \cite{Jha2018ElemNet:Composition}.

It is important to recognize that the descriptor choice impacts both the performance and applicability of the model. In the case of prediction of material properties, such as formation energy, selecting a composition-based descriptor, no matter how complex, limits the model to either a specific arrangement of atoms, such as BCC or amorphous, or some defined pattern of structures, such as the convex hull of lowest-energy structures. Such limitation of the problem domain, given a comparable amount of data, allows to quickly achieve much lower prediction error at a cost of fundamentally changing the problem, making a comparison between methods impossible. Furthermore, a composition-only representation is inherently unsuitable for the direct prediction of most material properties that depend on the atomic structure. The structure-informed descriptors can include much more information related to interatomic interactions, making them more robust and more physics-relevant. They also, implicitly or explicitly, include symmetries present in the material, which can be used to predict certain properties, such as zero piezoelectric response, with high confidence. Furthermore, such descriptors often include extensive composition-based arguments within them \cite{Ward2017IncludingTessellations}, making it possible to both recognize patterns in the property coming from different chemical species occupying the same structure and structural effects in the case of a single composition.

At the same time, it is important to consider that physically existing materials are rarely described by a single atomic configuration, usually requiring considerations for defects and coexisting configurations. Thus, like a traditional DFT-based modeling, in order to reproduce real material behavior, a structure-informed model will often require utilization of a method such as CALPHAD \cite{Kaufman1970ComputerMetals, Liu2018OceanLearning}. One of such methods, recently developed by authors and named "zentropy theory" shows the potential to connect individual configurations to predict macroscopic properties, such as colossal positive and negative thermal expansions \cite{Liu2022ZentropyExpansion}.

In some cases however, investigating all configurations can be a very challenging task (e.g., for high entropy alloys), necessitating the use of an elemental-only model trained to give predictions assuming future observations to be consistent with the past ones \cite{Debnath2021GenerativeAlloys}. 

\label{sipfenn:ssec:specificapproach}

The structure-informed representation which was the ground for the present work has been developed by Ward et al. based on information from the Voronoi tesselation of a crystal structure \cite{Ward2017IncludingTessellations}. Ward's descriptor contains 271 features that combine information from elemental properties of atoms, such as shell occurrences, with information about the their local environments, such as coordination number or bond lengths to neighbours. This approach was demonstrated to work excellently when comprehensively compared to two previous approaches based on the Coulomb matrix (CM) \cite{Schutt2014HowProperties} and on the partial radial distribution function (PRDF) \cite{Seko2017RepresentationProperties}, when trained on the same data from the Open Quantum Materials Database (OQMD) and with the same machine learning algorithm. A more detailed overview is given in \ref{sipfenn:ssec:descriptorused}.

Ward et al. used an automated Random Forest ML algorithm \cite{Ward2017IncludingTessellations} set to a fully automatic parameter selection. While fairly common, that approach without complexity limit for the model, and when trained on over 400,000 materials, resulted in a forest composed of 100 trees with approximately 700,000 nodes each. Such model requires over 27 GB of RAM memory to run, making it unusable on a typical personal or lab computer. Such size also results in a relatively low efficiency, requiring over 100 ms to run on a high-performance lab computer \cite{Ward2017IncludingTessellations}.

In the present work, aforementioned issues are resolved through a targeted design of the ML algorithm to fully utilize the data and its representation. This is done by consideration of the problem formulation and the deep neural network technique (see \ref{sipfenn:ref:machinelearningoverview}), combined with iterative model design (see \ref{sipfenn:sssec:NetDesign}), and by designing and testing over 50 neural networks belonging to around 30 designed architectures. Notably, in the time between Ward's work and the present paper, neural networks have been used in this application, e.g., \cite{Jha2019IRNet}, which uses residual neural networks. However, as we show in Section \ref{sipfenn:ssec:oqmdperformance}, the present paper provides more accurate predictions than both Ward's model and the state-of-the-art neural network model \cite{Jha2019IRNet}.

Additionally, the present work brings two further improvements. The first one is good transfer learning ability, described in \ref{sipfenn:ssec:transferlearningresults} allowing other researchers, at a relatively small cost, to adjust the model to small problem-specific databases, typically consisting of tens of DFT calculations or less. This method substantially improves predictions for similar materials while retaining the general knowledge learnt from the large data set and demonstrates that the model learns features related to underlying physics. The second  improvement is the end-user usability. While most of the materials-related ML model are reported in a reproducible way with an evaluation of the performance \cite{Ward2017IncludingTessellations, Schutt2014HowProperties, Schutt2018SchNetMaterials, Seko2017RepresentationProperties}, only a fraction goes beyond to make models accessible to the community. the present work has been focused on creating a Findable, Accessible, Interoperable, and Reusable tool, inspired by FAIR principles \cite{FAIRFAIR}, created open-source with common and convertible data formats as is described in more detail in \ref{sipfenn:sssec:SoftwareUsed}. This lead to many standalone components combined into an end-user tool, described in \ref{sipfenn:ssec:SIPFENN}, that is ready to use without any costly computation to create the model and can be run on any modern computer, as low-power as smartphones.

\section{Methodology} \label{sipfenn:sec:methodology}

\subsection{Descriptor Used} \label{sipfenn:ssec:descriptorused}
A descriptor of a material is a point in a well-defined multidimensional property space that can be used to represent knowledge associated with entries in a database in vector form. Within the present work, the property space has 271 dimensions (corresponding to 271 features) related to elemental properties and atomic structure of an arbitrary crystalline material, as designed by Ward et al. \cite{Ward2016AMaterials, Ward2017IncludingTessellations} utilizing the \texttt{voro++} code \cite{rycroft2009voro++}. These features can be categorized as:

\begin{itemize}
    \item \textbf{Elemental Attributes} (145 total): Attributes which only depend upon the elements present and their stoichiometry.
    \begin{itemize}
        \item \textbf{Stoichiometric Attributes} (6): Describe the components fractions.
        \item \textbf{Elemental Properties Attributes} (132): Contain statistics taken over the various elemental properties, weighted by the stoichiometry of the structure.
        \item \textbf{Attributes based on Valence Orbital Occupation} (4): Depend upon the distribution of valence electrons across different orbitals, i.e. on the total number of valence electrons in each orbital across the structure.
        \item \textbf{Ionic Character Attributes} (3): Attributes which encode whether the material is ionically bonded. 
    \end{itemize}
    \item \textbf{Structural Attributes} (126 total): Attributes which depend on the precise structural configuration, i.e. exactly how the atoms are arranged in space.
    \begin{itemize}
        \item \textbf{Geometry Attributes} (16): Attributes which depend upon the spatial configuration of atoms only.
        \item \textbf{Physical Property Differences Attributes} (110): Contain statistics taken over the differences between elemental properties of neighboring sites in the structure, weighted by the size of the Voronoi cell face between the neighbors.
    \end{itemize}
\end{itemize}

A complete table list of features is given in Table \ref{sipfenn:feature-table}. Further details can be found in \cite{Ward2016AMaterials, Ward2017IncludingTessellations}.

\begin{table}[H]
    \footnotesize
    \centering
    \caption{List of Features with Descriptions. Site Statistics refers the mean, range, mean absolute error, maximum, minimum, and mode unless otherwise stated in the description. Difference Statistics refers to the mean, mean absolute error, minimum, maximum and range of the differences between neighboring sites in a structure, weighted by the size of the face between them in the Voronoi tessellation.}
    \begin{tabular}{|p{2.25cm}|p{2cm}|c|c|}
        \hline
        \textbf{Site \hspace{0.3cm} Statistics} & \textbf{Difference \hspace{0.3cm} Statistics} & \textbf{Name} & \textbf{Description} \\
        \hline
        1-4 & - & Effective Coordination Number & mean, mean abs error, min, max\\
        \hline
        5-7 & - & Mean Bond Length & mean abs error, min, max\\
        \hline
        8-11 & - & Bond Length Variation & mean, mean abs error, min, max \\
        \hline
        12 & - & Cell Volume Variation & Variation in the voronoi cell volume\\
        & & & no statistics \\
        \hline
        13-15 & - & Mean WC Magnitude & shells 1-3, global non-backtracking \\
        \hline
        16 & - & Packing Efficiency & no statistics \\
        \hline
        133-138 & 17-21 & Atomic Number & \\
        \hline
        139-144 & 22-26 & Mendeleev Number & \\ 
        \hline
        145-150 & 27-31 & Atomic Weight & \\ 
        \hline
        151-156 & 32-36 & Melting Temperature & \\ 
        \hline
        157-162 & 37-41 & Column & Group in Periodic Table \\ 
        \hline
        163-168 & 42-46 & Row & Period in Periodic Table\\ 
        \hline
        169-174 & 47-51 & Covalent Radius & \\ 
        \hline
        175-180 & 52-56 & Electronegativity & \\ 
        \hline
        181-210 & 57-81 & Valence Electron Count & Listed for s,p,d,f orbitals and total \\ 
        \hline
        211-240 & 82-106 & Unfilled Count & Number of unfilled orbitals \\
         & & & Listed for s,p,d,f orbitals and total \\
         \hline
        241-246 & 107-111 & Ground State Volume & \\ 
        \hline
        247-252 & 112-116 & Ground State Band Gap & \\ 
        \hline
        253-258 & 117-121 & Ground State Magnetic Moment & \\ 
        \hline
        259-264 & 122-126 & Space Group Number & Index of Space group\\
        \hline
        127 & - & Number of Components & no statistics \\
        \hline
        128-132 & - & $\ell^p$-norms of Component Fractions & $p \in \{2,3,5,7,10\}$ \\
        \hline
        265-268 & - & Fraction of Valence Electrons & \\
        & & in s,p,d,f orbitals & no statistics\\
        \hline
        269 & - & Can Form Ionic Compound & boolean, no statistics\\
        \hline
        270-271 & - & Ionic Character & max, mean over pairs of species\\
        \hline
    \end{tabular}
    \label{sipfenn:feature-table}
\end{table}

\subsection{Machine Learning Techniques Overview} \label{sipfenn:ref:machinelearningoverview}
This section gives a brief overview of the employed machine learning techniques and terminology, described in more detail in the Appendix \ref{sipfenn:appedix1}. The interest is placed on the statistical problem of regression, whose goal is to learn a functional relationship $f:X\rightarrow Y$ which minimizes the risk (also known as loss or expected error) \cite{vapnik1999overview} given by

\begin{equation} \label{sipfenn:true_risk}
    R(f) = \mathbb{E}_{x,y\sim \mathcal{P}} l(y,f(x)).
\end{equation}

Here $X$ denotes a space of input features, $Y$ denotes an output space, the expectation above is taken over an unknown distribution $\mathcal{P}$ on $X\times Y$ (representing the true relationship between inputs and outputs), and $l$ is a given loss function. 

In the specific application considered here, the function $f$ which is to be learned, maps input material structures (arrangements of atoms) $x$ to the predicted formation energy $y$. The distribution $\mathcal{P}$ is unknown, but samples $(x_i,y_i)$ are given, consisting of structures $x_i$ and corresponding predictions $y_i$ which are used to learning $f$. In the present case, this data comes from the OQMD and other smaller materials databases.


In order to learn the relationship $f$ from the data, the empirical risk

\begin{equation} \label{sipfenn:empirical_risk}
    L(f) = \frac{1}{n}\displaystyle\sum_{i=1}^n l(y_i, f(x_i)),
\end{equation}

is minimized over a class of functions defined by a neural network architecture. A neural network architecture consists of a sequence of alternating linear functions and point-wise non-linear functions defined by an activation function (see \cite{goodfellow2016deep} for more information about neural networks). As the loss function $l$ in \eqref{sipfenn:empirical_risk} the $\ell^1$-loss function $l(y,x) = |x-y|$ is used. The neural networks are trained on this loss \eqref{sipfenn:empirical_risk} using the common ADAM optimizer \cite{kingma2014adam}. 

An important issue when training complex statistical models is the overfitting, which occurs when a model accurately fits the training data but fails to generalize well to new examples. In order to detect overfitting, the standard practice of dividing the data into training, validation, and test datasets \cite{hastie2009elements} is used. In order to mitigate overfitting, dropout \cite{srivastava2014dropout} and weight decay, two standard methods for regularizing neural networks, are used. In Section \ref{sipfenn:sssec:DesignedModels}, Figure \ref{sipfenn:fig:trainingvalidation-body} illustrates overfitting mitigation effects on the training process of neural networks designed in the present paper.

\subsection{Software Used} \label{sipfenn:sssec:SoftwareUsed}

The choice of software for the machine learning portion of \texttt{SIPFENN} was Apache \texttt{MXNet} \cite{ChenMXNet:Systems} due to it's open source nature, model portability, and state-of-the-art scalability, allowing the same code to run on a laptop with a low-power CPU/GPU and a supercomputer (e.g., ORNL Summit) with hundreds of powerful GPU's. It's portability allows trained networks to be converted and used with other popular frameworks such as Google \texttt{Tensorflow}, \texttt{PyTorch}, or even Apple's \texttt{Core ML}, making results of the present paper highly accessible.

\texttt{MXNet} framework was used through \texttt{Wolfram} and \texttt{Python} languages. \texttt{Wolfram} language was used primarily for the network architecture design, training, and testing, as it provides an excellent interface with detailed training results shown in real-time during the training process. It also provides good out-of-the-box performance due to its well-optimized memory handling when training on a single GPU setup. 

Python, on the other hand, was used when writing the end-user tool for running previously trained networks. This choice was made so that the software is completely open-source and can be easily reused for specific purposes or incorporated within other packages. Furthermore, Python allowed quick implementation of a Graphical User Interface (GUI) through the \texttt{wxpython package}.

As explored later in Chapter \ref{chap:pysipfenn}, over several years, the software evolved and developed into \texttt{pySIPFENN} framework which, as of April 2024, has moved to \texttt{PyTorch} for ML runtime and Open Neural Network Exchange (\texttt{ONNX}) format \cite{Bai2019ONNX:Exchange} for model storage and distribution.

\subsection{Data Acquisition and Curation} \label{sipfenn:sssec:Data}

Four sets of data were used within the present work. The largest by volume and significance was the Open Quantum Materials Database (OQMD) \cite{Kirklin2015TheEnergies, Saal2013MaterialsOQMD, Shen2022ReflectionsOQMD}, which contains the results of DFT calculations performed by the Vienna Ab Initio Simulation Package (VASP) \cite{Kresse1993AbMetals} for a broad spectrum of materials. The snapshot used here was extracted from the database by Ward et al. in 2017 and contained 435,792 unique compounds \cite{Ward2017IncludingTessellations}. The choice of 2017 snapshot rather than the current one was made to ensure direct performance comparison between new and previously reported methods. The second database was a part of the Inorganic Crystal Structure Database (ICSD), a subset of the OQMD with only experimentally obtained structures containing around 30,000 entries. ICSD was primarily used for the quick design of simple neural network architectures at the beginning, and OQMD used for more complex models designed later. 

Two smaller data sets were used, in addition to these large databases. The first small dataset contained DFT-calculated formation energies of Fe-Cr-Ni ternary $\sigma$-phase endmembers in the 5-sublattice model \cite{Feurer2019Cr-Fe-NiCalculations}. As this model contains 5 chemically distinct positions (Wyckoff positions), populated by one of 3 elements, in total it included 243 ($3^5$) structures with 30-atom basis each. This data served as an example of a relatively complex structure that was not included in the OQMD. Furthermore, it was a test case of a material that is highly industry-relevant, as it causes steel embrittelment \cite{Hsieh2012OverviewSteels} and is costly to investigate using traditional methods due to compositional and configurational complexity. The second small dataset included 13 Special Quasirandom Structures (SQS), which are the best periodic supercell approximations to the true disordered state of metal alloys \cite{Zunger1990SpecialStructures, Jiang2004First-principlesStructures, Shin2006ThermodynamicStructures}. SQS structures in this set were binary alloys containing Fe, Ni, Co, and V, laying on deformed FCC (A1), BCC (A2), or HCP (A3) lattices. The main purpose of these smaller datasets was to test the performance in extrapolation from OQMD, in a particular case of interest for the author's.

During the network design process described in \ref{sipfenn:sssec:NetDesign}, it was found that a small fraction of the OQMD dataset (under 0.03\%) contains anomalous values of formation energy above 10 eV/atom. In the extreme case of $CuO_2$ (OQMD ID: 647358) this value was 1123 eV/atom or 108350 kJ/mole. Since the source database contains hundreds of thousands of data points reported by many scientists, it can be expected that a small fraction of the data may contain some sort of errors and typos. In the present work, they were removed from all datasets used for training and evaluation.

\subsection{Neural Network Design Process} \label{sipfenn:sssec:NetDesign}

This section conceptually outlines the network design process leading to the final models. All essential details regarding the design and performance of intermediate models, useful for better understanding changes and for applying the similar approach in different problems, can be found in the Appendix \ref{sipfenn:appendix2}.

The design started with the simplest single-layer neural network (perceptron) with the Sigmoid activation function, trained on the ICSD and its smaller subset, to provide a baseline for the design. Then, the process was conducted in the following steps:

    \begin{figure}[H]
    \centering
    \includegraphics[width=0.56\textwidth]{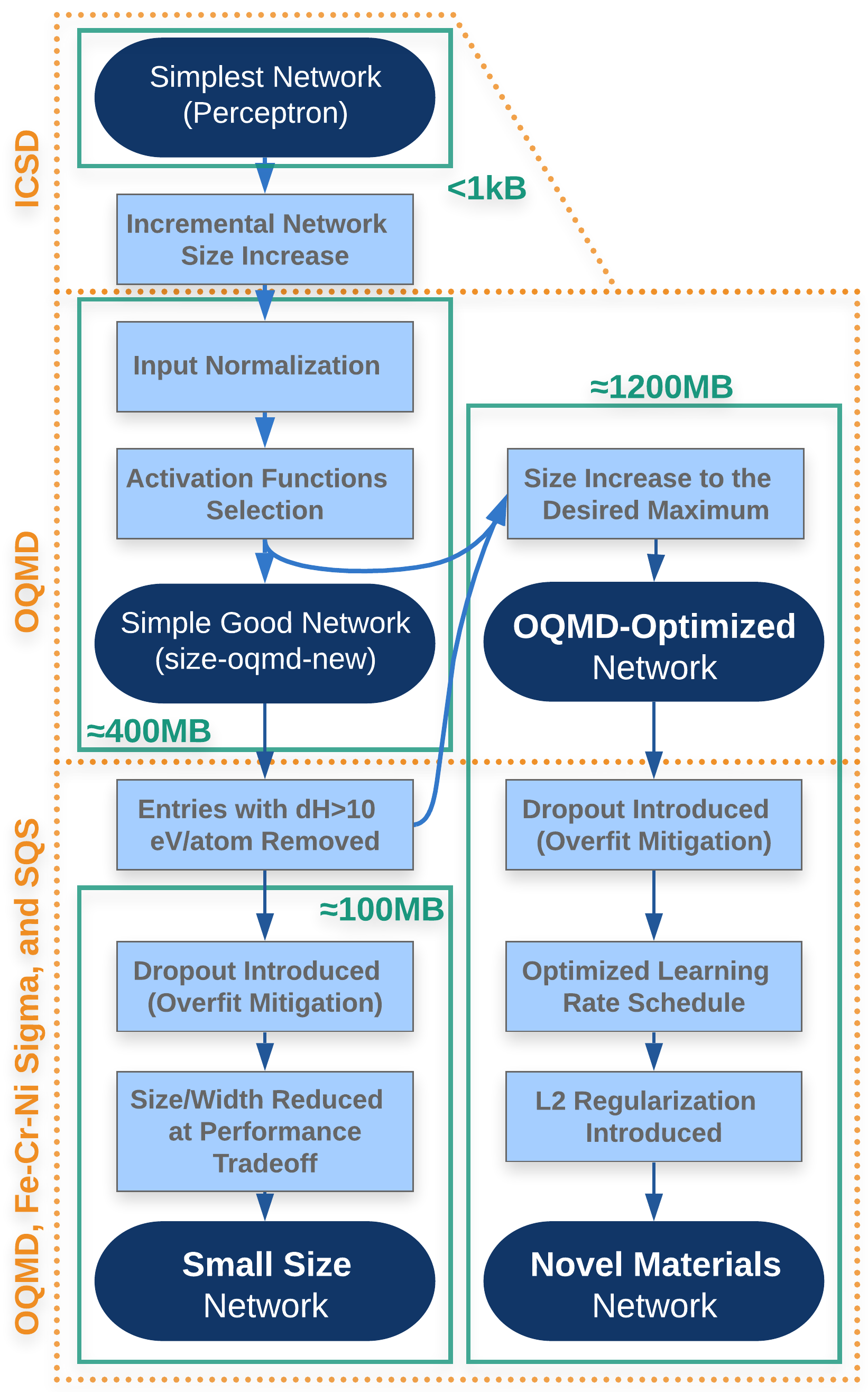}
    \caption{The model design process schematic.}
    \label{sipfenn:fig:designprocess}
    \end{figure}

\textbf{1. }The network size has been increased step-wise while training on the ICSD dataset (30k+ entries). Results were extrapolated to estimate network size suitable for larger OQMD (400k+) to be 4 hidden layers in a (10000, 10000, 1000, 100) configuration.

\textbf{2. }To improve convergence during the training, descriptor features values were normalized to their maximum values present in the OQMD dataset.

\textbf{3. }Performance and time to convergence were improved by moving from Sigmoid activation function to a mix of Soft Sign, Exponential Linear Unit, and Sigmoid. This relatively simple model has improved performance over the existing Random Forest model \cite{Ward2017IncludingTessellations}, achieving MAE of 42 meV/atom on the same dataset.

\textbf{4. }At this step, it was noticed that a small fraction (around 0.03\%) of data points exhibits extreme errors, as high as over 1,000,000 meV/atom causing some instability during the training process, despite the large batch size of 2048. They also caused a high deviation in test MAE values across repeated model training rounds. As describes in \ref{sipfenn:sssec:Data}, these were identified to be a few rare errors in the dataset and removed during later model design. 

\textbf{5. }The network size was increased to around 1GB limit (maximum size target) by the addition of two more 10,000-width layers. This \textbf{OQMD-optimized} network has achieved the best performance on the OQMD out of all designed in the present paper, with an MAE of 28 meV/atom. Performance analysis can be found in \ref{sipfenn:ssec:oqmdperformance} and in Figure \ref{sipfenn:fig:oqmdperformance}.

\textbf{6. }After the good performance on the OQMD was achieved, the design goals shifted to (1) reducing the training-set-to-validation-set error mismatch during the network training, while (2) keeping the test MAE on the OQMD on a suitable level (below 50 meV/atom), and (3) improving performance on datasets not presented to network before (see \ref{sipfenn:sssec:Data}). The first step was the introduction of Dropout layers\cite{srivastava2014dropout}, described in more detail in Appendix \ref{sipfenn:appedix1}, which allow for better distribution of knowledge across the network.

\textbf{7. }The introduction of strong Dropout\cite{srivastava2014dropout} made the network prone to falling in local minima, which was solved by the introduction of a changing learning rate schedule.

\textbf{8. }With optimized network architecture, lastly, the descriptor interpretation by the network has been modified through the introduction of L2 regularization \cite{L2Regularization}, a technique which assigns an error penalty for "attention" (input layer weights) to each of the descriptor features, effectively refining features in the descriptor to only the most significant ones. Figure \ref{sipfenn:fig:squaredweights} ranks them. The resulting \textbf{Novel Materials} model achieved a much lower training-set-to-validation-set error mismatch (1.15 vs 1.57 after 240 rounds), presented in Figure \ref{sipfenn:fig:trainingvalidation} as a function of training progress. On the OQMD test set, it achieved a higher, yet suitable 49 meV/atom.

\textbf{9. }To cater to applications requiring very high throughput or low memory consumption, an additional \textbf{Small Size} network was designed by adding Dropout to one of the earlier networks, designed before the size increase step, and then reducing its size to the desired level. It was found that after reduction of total size from around 400MB to around 100MB, the network retained MAE of 42 meV/atom on an OQMD test set and further reduction was possible if needed for the application.

\section{Results} \label{sipfenn:sec:Results}

\subsection{Final Predictive Models} \label{sipfenn:sssec:DesignedModels}

Throughout the architecture design process described in \ref{sipfenn:sssec:NetDesign}, detailed in Appendix \ref{sipfenn:appendix2}, and depicted in Figure \ref{sipfenn:fig:designprocess}, new networks were designed and tested in various ways, leading to about 50 predictive models (trained neural networks) with varying training parameters and training data. The majority of the intermediate networks were stored for the record, and are available upon request. Details regarding hyper-parameters and training routines used to obtain three resulting models can be found in the Appendix \ref{sipfenn:appedix1}.

\begin{figure}[H]
    \centering
    \frame{\includegraphics[width=0.30\textwidth]{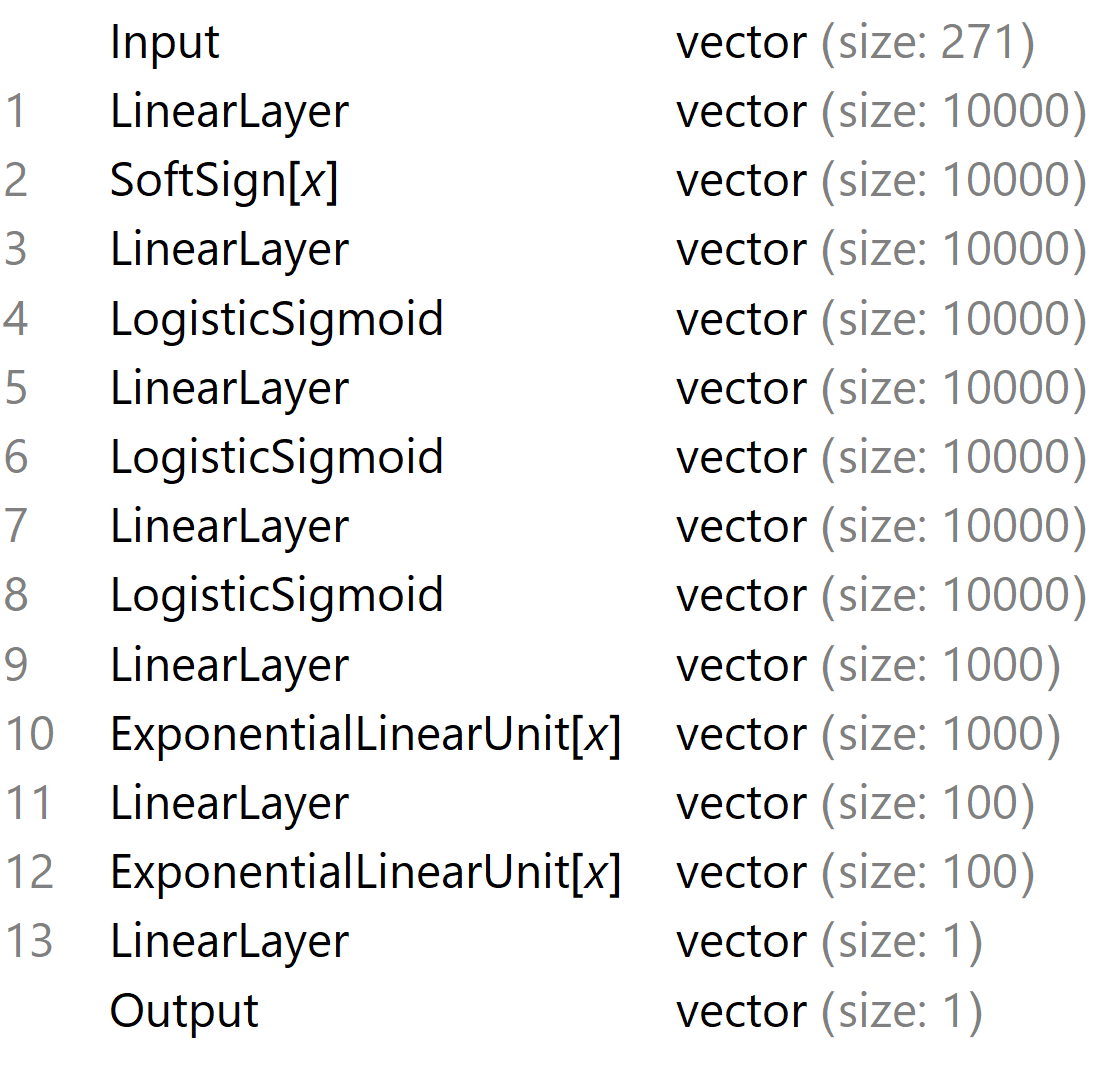}}
    \hspace{6pt}
    \frame{\includegraphics[width=0.30\textwidth]{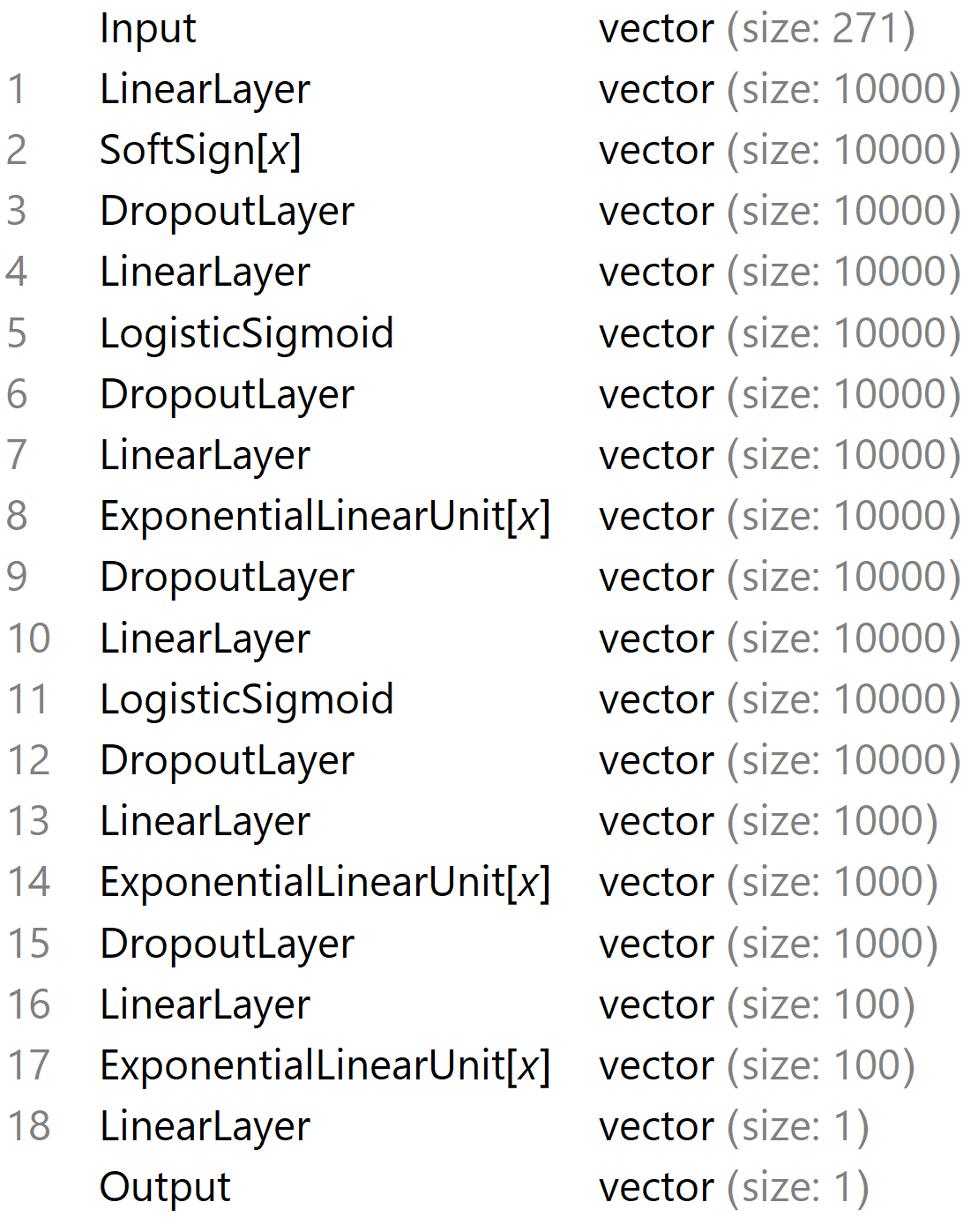}}
    \hspace{6pt}
    \frame{\includegraphics[width=0.30\textwidth]{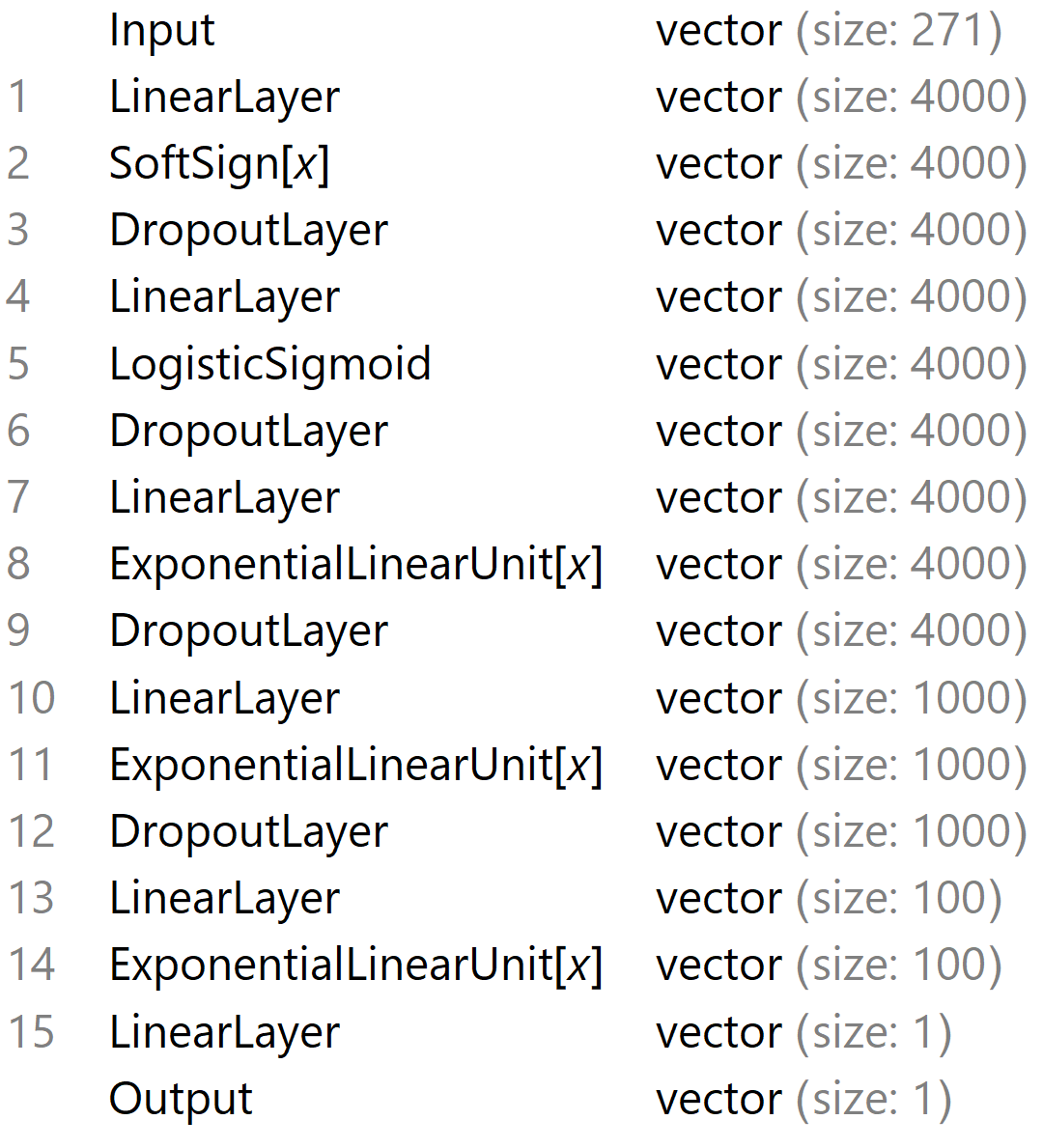}}
    \caption{Three selected architectures designed within the present work. Optimized for: (Left) OQMD performance, (Middle) predicting new materials, (Right) small size at good performance. Internally in the code, they are designated as NN9, NN20, and NN24.}
    \label{sipfenn:fig:architectures}
\end{figure}

Out of all trained neural networks, three were selected and can be considered final outcomes of the design process, optimized for different objectives. Their architectures are presented in Figure \ref{sipfenn:fig:architectures}. The first one, denoted NN9, was created specifically for the OQMD performance. This was the same objective as in the study by Ward et al. \cite{Ward2017IncludingTessellations} and its performance serves as a direct comparison to the Random Forest method employed in that paper \cite{Ward2017IncludingTessellations} and other works \cite{Schutt2014HowProperties, Seko2017RepresentationProperties}.

\begin{figure}[H]
    \centering
    \includegraphics[width=0.52\textwidth]{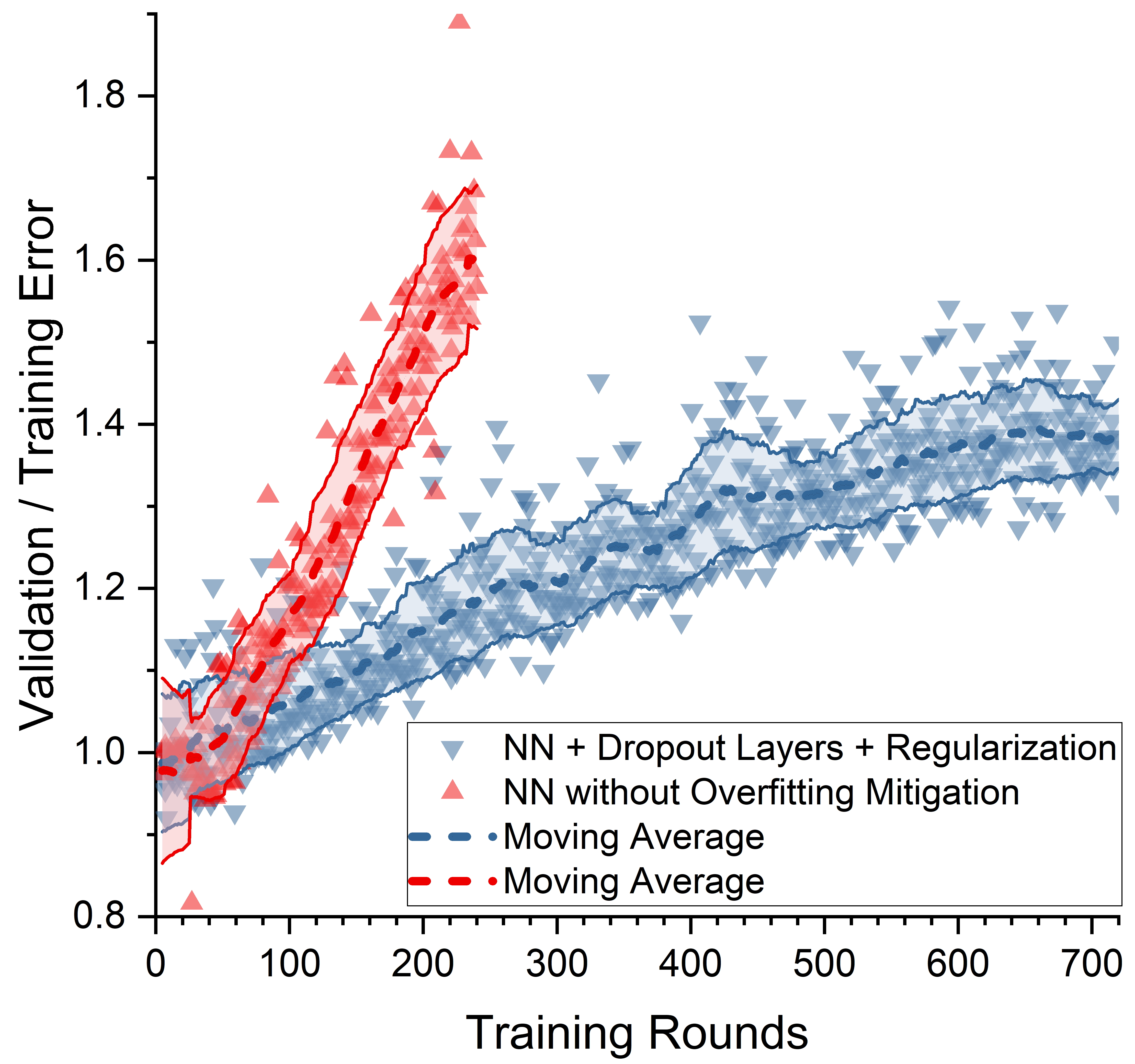}
    \caption{Training Loss to Validation Loss in a model that does without (NN9) and with overfitting mitigation (NN20), plotted versus training progress.}
    \label{sipfenn:fig:trainingvalidation-body}
\end{figure}

\begin{figure}[h]
    \centering
    \begin{minipage}[c]{0.65\textwidth}
    \includegraphics[width=\textwidth]{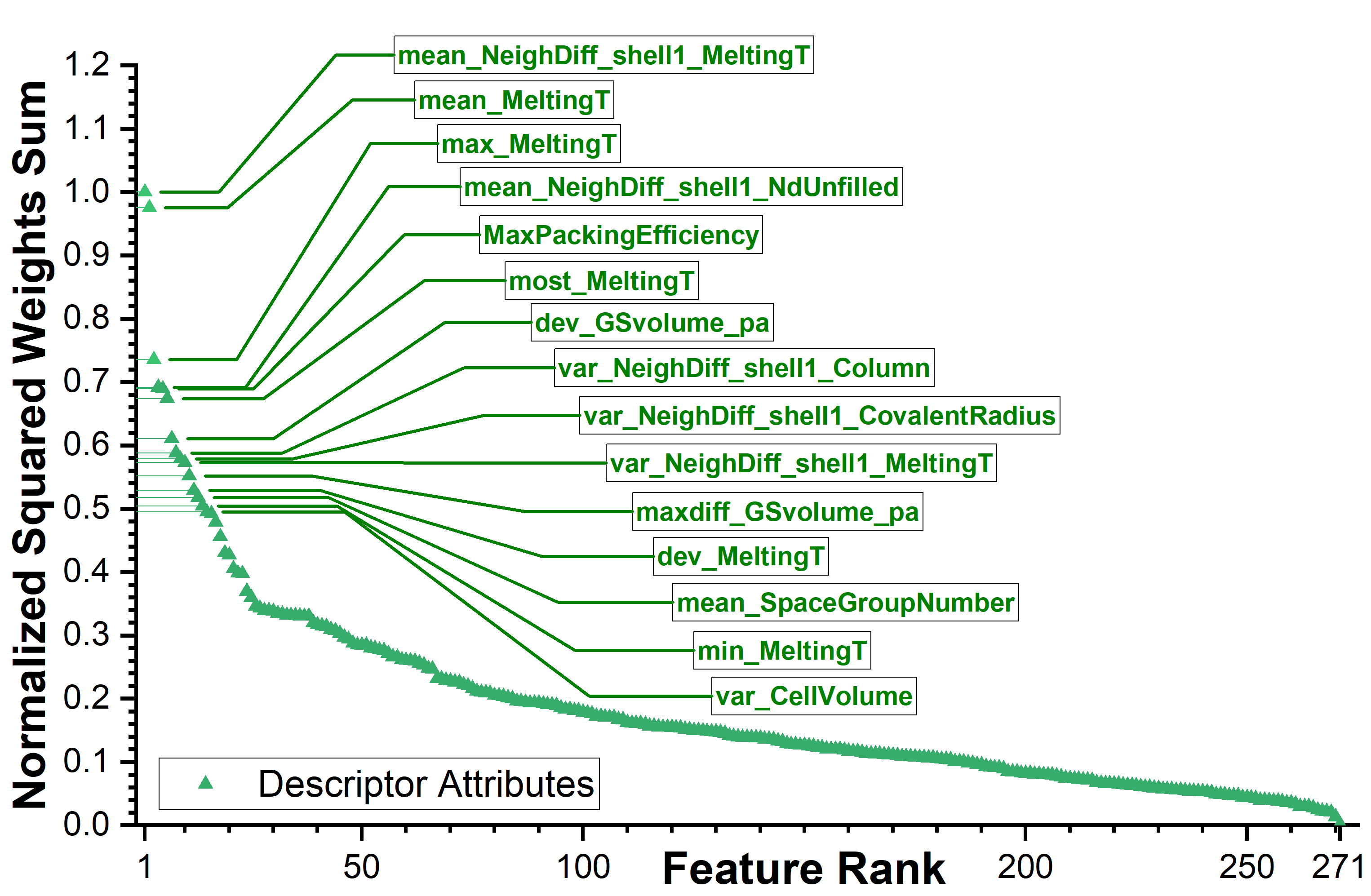}
    \end{minipage}\hfill
    \begin{minipage}[c]{0.33\textwidth}
    \caption{Distribution of sums of squared input weights. High values correspond to attributes that were not lowered due to their contribution to pattern recognition of the model. 15 attributes with the highest values are labeled. The labels are taken from the descriptor definition in  \cite{Ward2016AMaterials}.}
    \label{sipfenn:fig:squaredweights}
    \end{minipage}
\end{figure}

The second network was optimized for improved pattern recognition on OQMD and improved performance on non-OQMD datasets used in the present work (i.e. SQS/$\sigma$-phase datasets). This was achieved primarily through extensive overfitting mitigation, applied during design and training (see Figure \ref{sipfenn:fig:trainingvalidation-body}), which leads to a network with improved generalization/materials-discovery capability. Furthermore, one fo the overfitting mitigation methods, namely the regularization described in \ref{sipfenn:ref:machinelearningoverview}, have allowed identification of descriptor attributes that contributed the most to the predictive capability and the ones that were almost completely discarded once the penalty for considering them was assigned. Figure \ref{sipfenn:fig:squaredweights} presents the distribution of sums of squared weights between each neuron in the input layer (each of the 273 descriptor features) and all 10,000 neurons in the first hidden layer. 

Feature rankings, such as presented in Figure \ref{sipfenn:fig:squaredweights}, allow a more efficient selection of input features in future studies looking into the same problem; thus both reducing the number of features that need to be computed for each atomic configuration and the total number of weights in the network. Furthermore, it can be used to gain an insight into the model interpretability. Looking at the specific ranking for NN20, the high-impact features present a mix of elemental features, likely allowing the model to establish some formation energy baseline for a given composition, and structure-informed features allowing to distinguish between polymorphic configurations. High impact elemental features include different statistics on elemental melting temperatures and ground-state structure volume per atom. The structural features extend them by considering how they differ between neighboring atoms and also include purely structural features such as packing efficiency and variance in Wigner–Seitz cells volumes. A complete ranking of features is included in Appendix \ref{sipfenn:appendix3}.

The third network, denoted NN24, was created for memory/power-constrained applications requiring a balance between OQMD performance and memory intensity and processing power required. Model parameters contained in this architecture occupy only 145MB, over 8 times less than two other models and around 200 times less than the model reported by Ward et al. \cite{Ward2017IncludingTessellations}.

\subsection{OQMD Data Performance} \label{sipfenn:ssec:oqmdperformance}
\begin{figure}[H]
    \centering
    \includegraphics[width=0.31\textwidth]{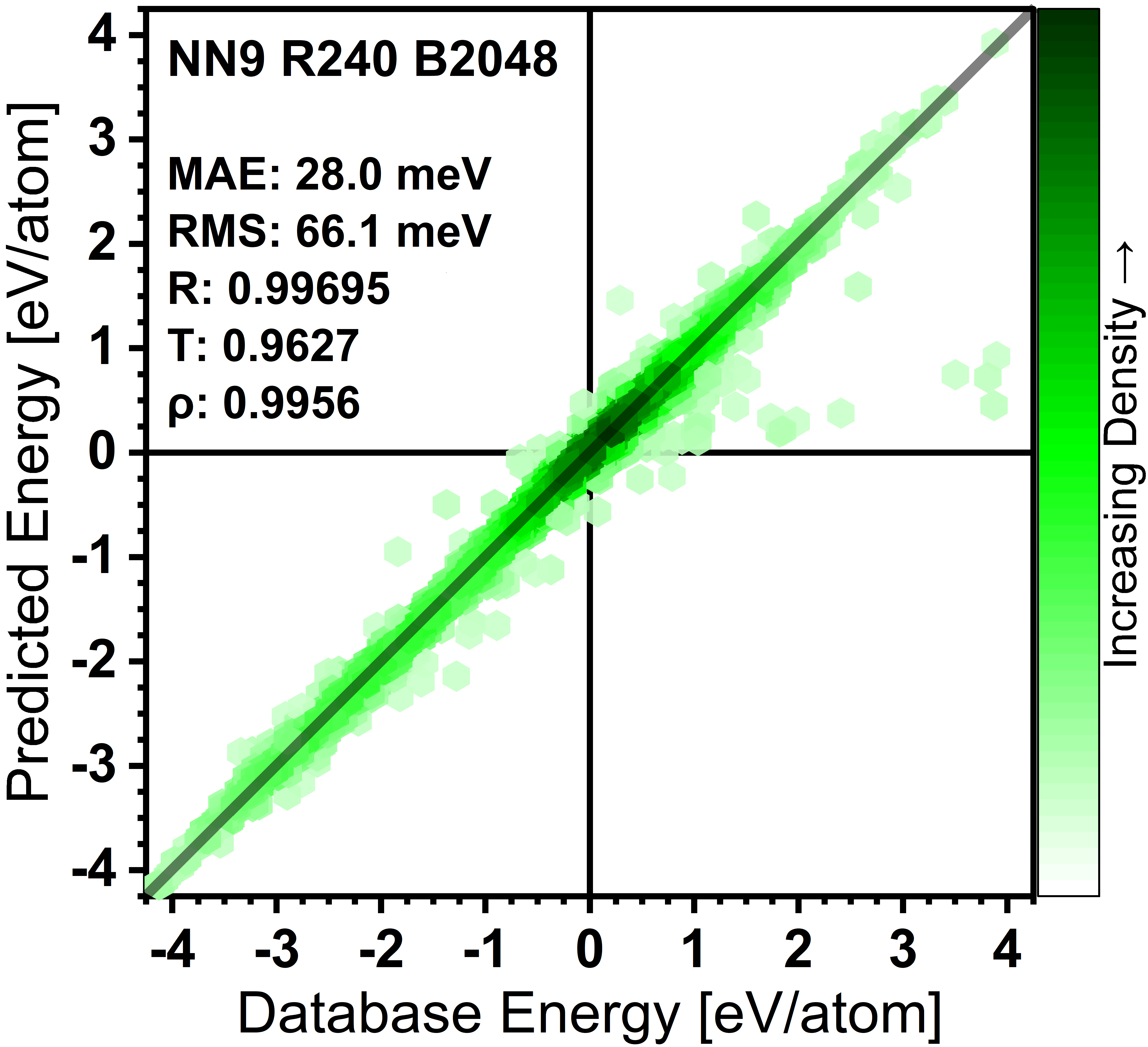}
    \hspace{0.01\textwidth}
    \includegraphics[width=0.31\textwidth]{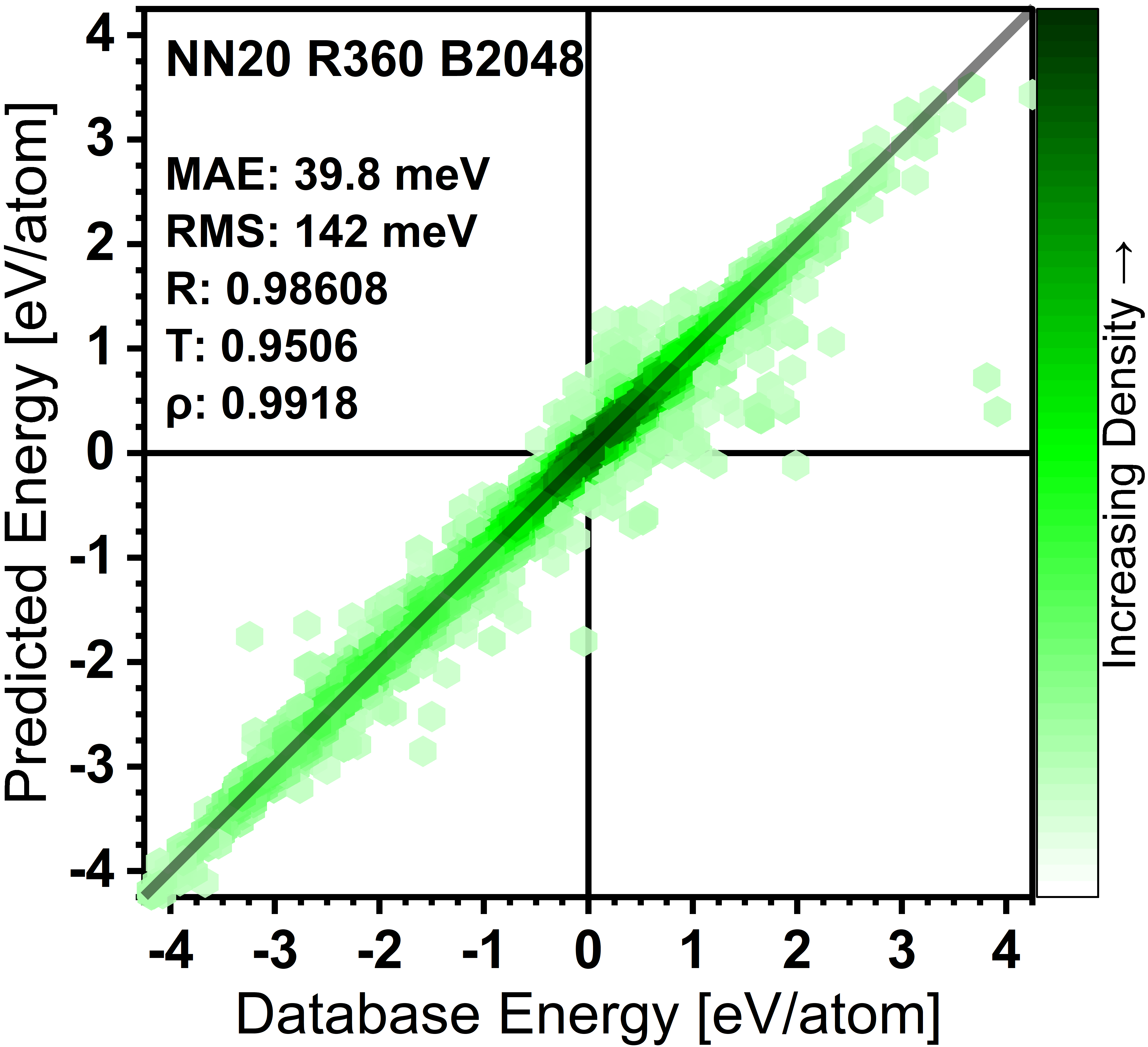}
    \hspace{0.01\textwidth}
    \includegraphics[width=0.31\textwidth]{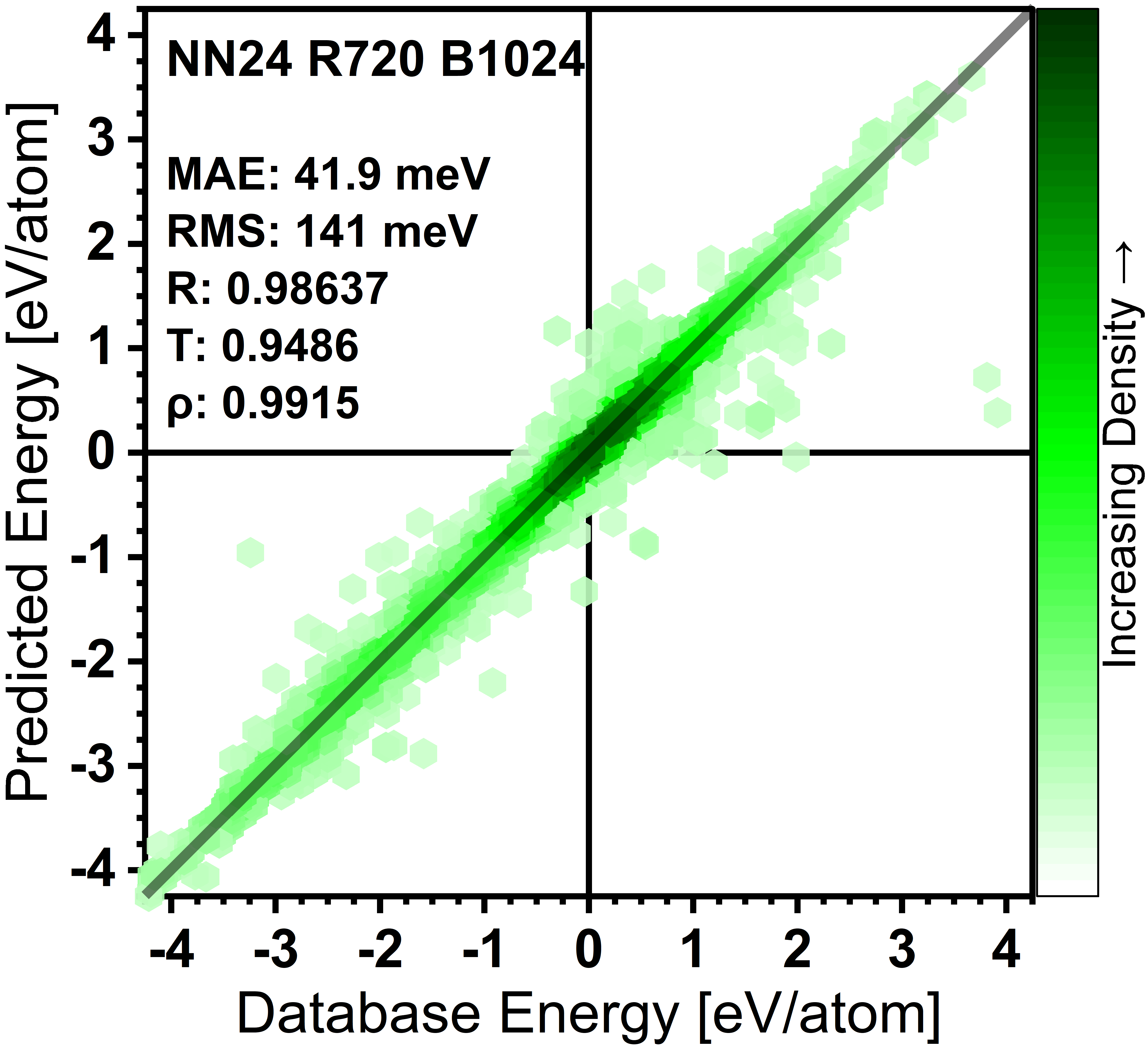}
    \includegraphics[width=0.31\textwidth]{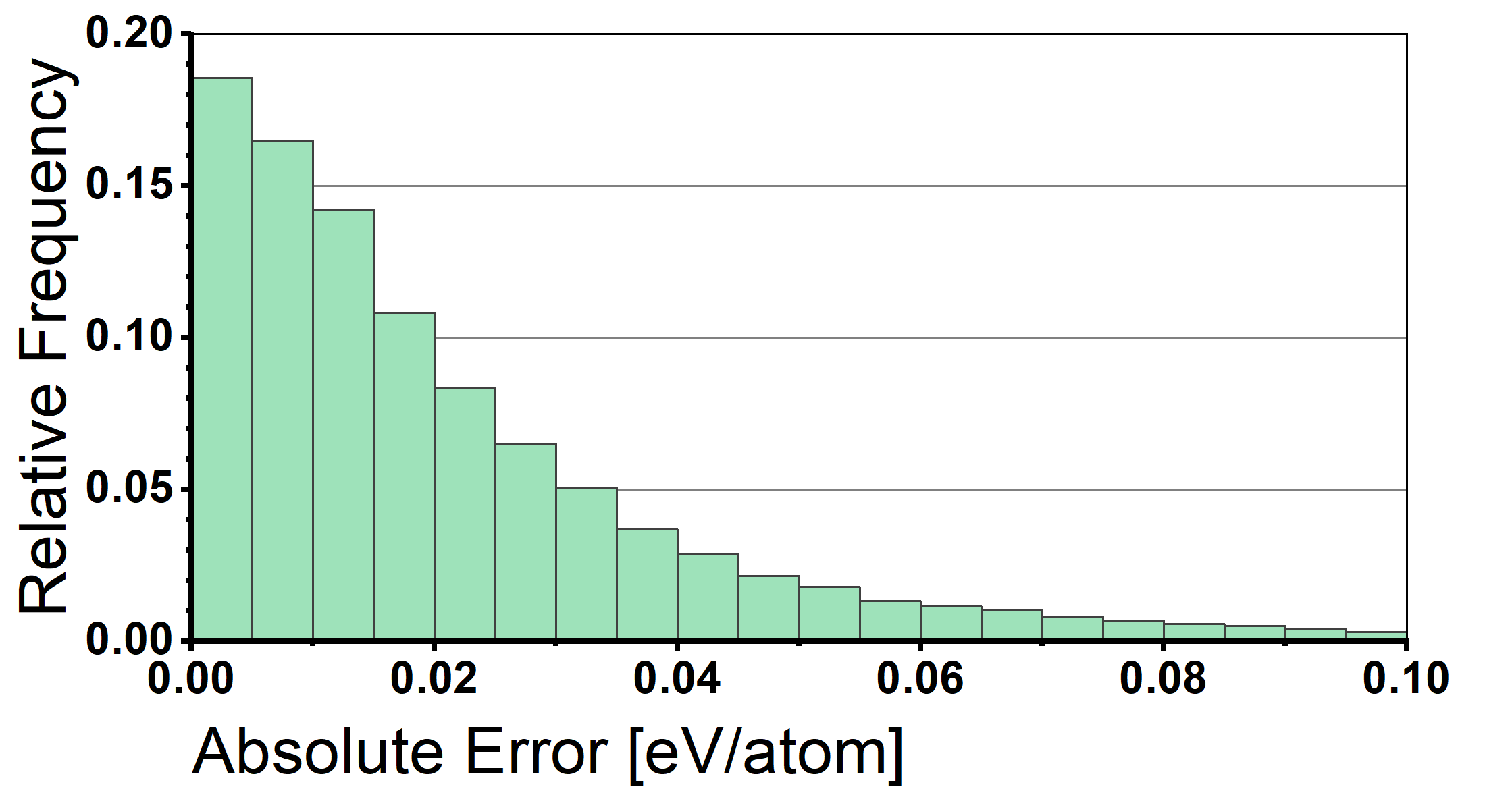}
    \hspace{0.01\textwidth}
    \includegraphics[width=0.31\textwidth]{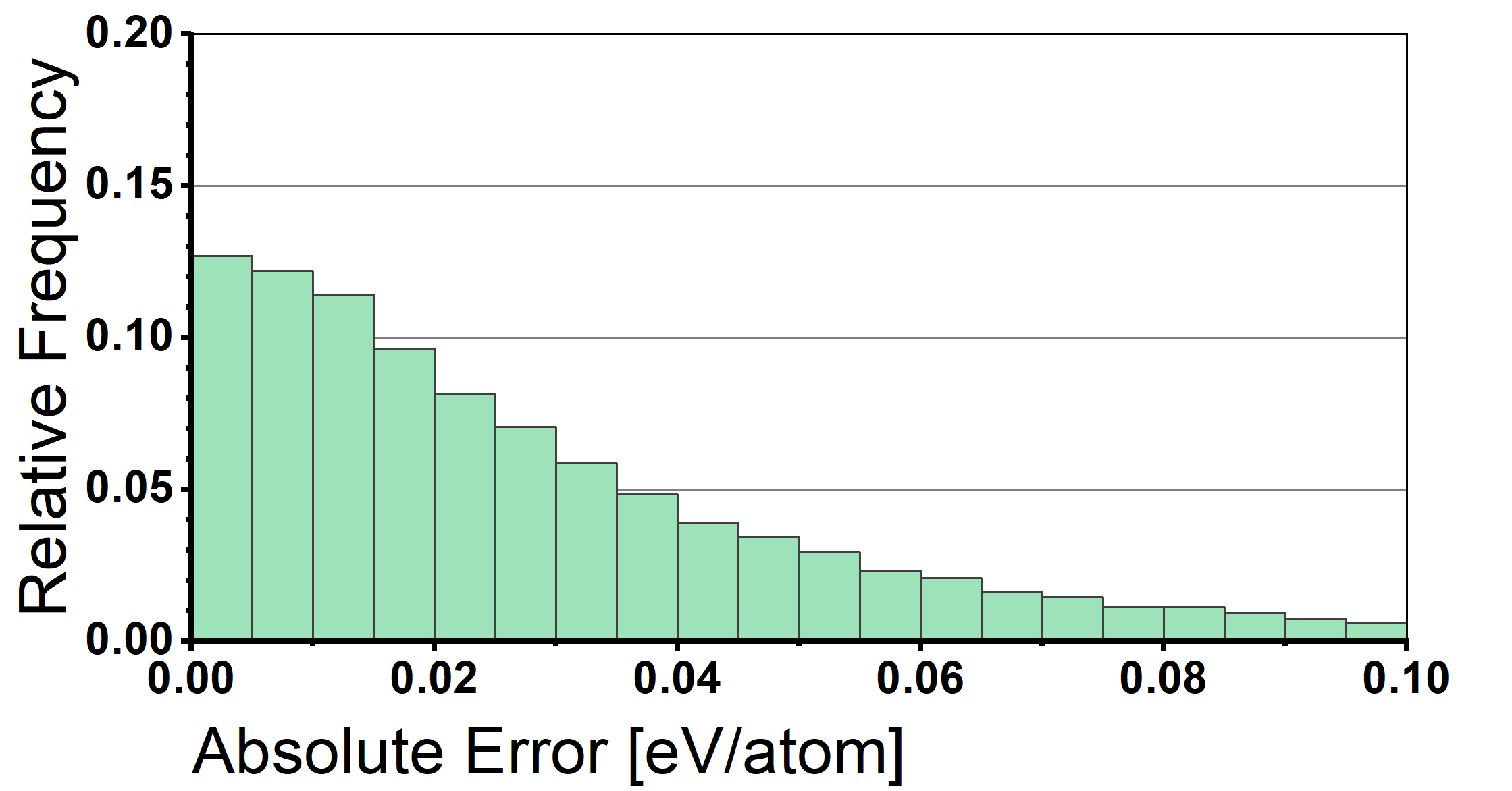}
    \hspace{0.01\textwidth}
    \includegraphics[width=0.31\textwidth]{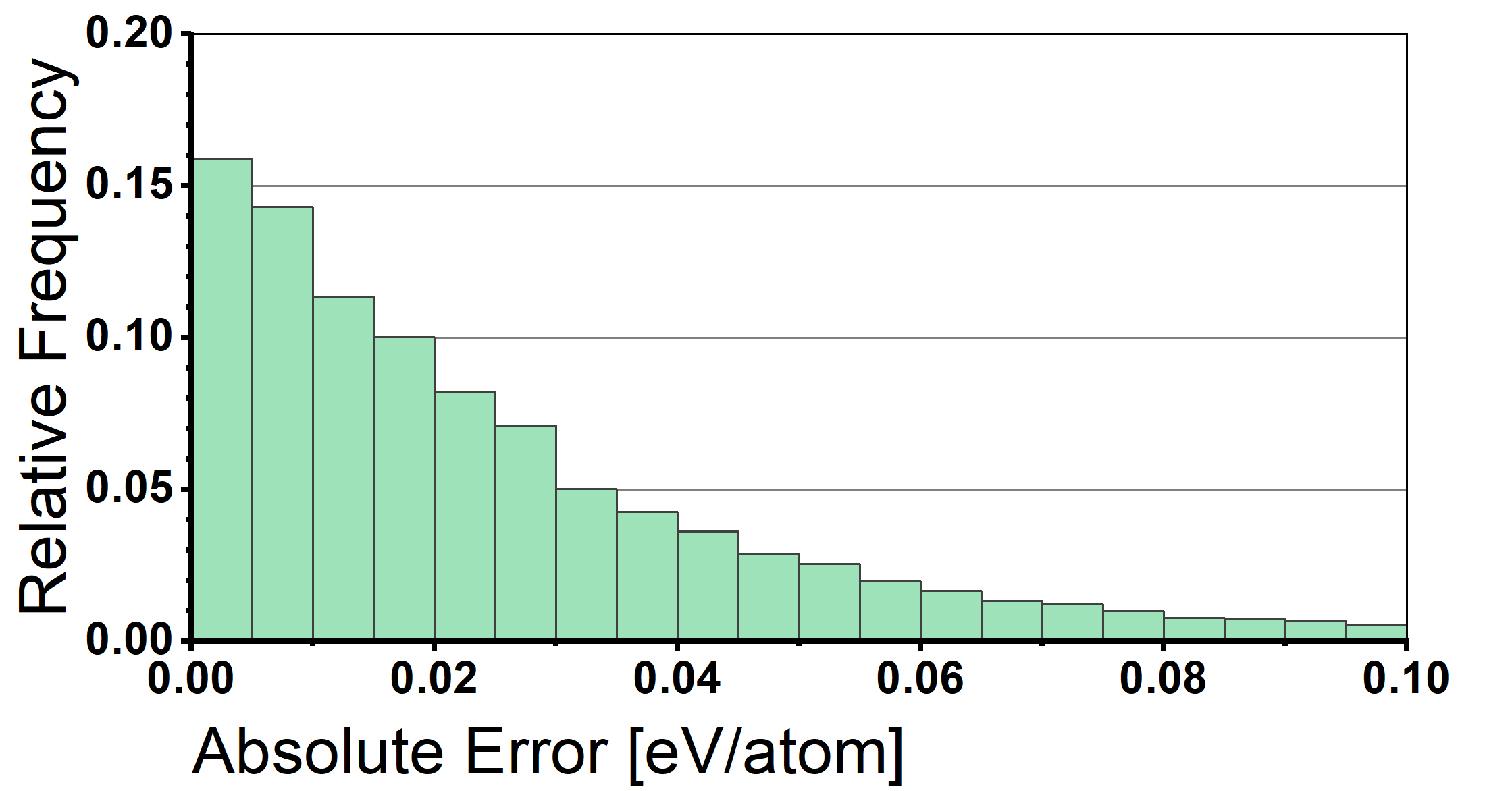}
    \caption{Performance of 3 selected neural networks on a random subset of 21,800 entries from OQMD. (Left) OQMD performance, (Middle) predicting new materials, (Right) small size at good performance. Internally in the code, they are designated as NN9, NN20, and NN24.}
    \label{sipfenn:fig:oqmdperformance}
\end{figure}

As described in \ref{sipfenn:sssec:NetDesign}, all three final networks were evaluated on a randomly selected subset of the OQMD to give a comparison between the state-of-the-art model presented by Ward et al. \cite{Ward2017IncludingTessellations} and the present ML method. This random subset consisted of 21,800 OQMD entries, constituting approximately $5\%$, which were not presented to the network, nor used for evaluation at any stage of the training process. This sample size was considered to be representative of the whole dataset once the small fraction ($0.026\%$) of likely incorrect entries were removed from the dataset as described in \ref{sipfenn:sssec:Data}. The random selection itself was initially performed separately for each training process and recorded after completion. Later, when networks were modified to mitigate overfitting, a single random subset was used for all of them to allow more careful design and more accurate comparative analysis of results. Figure \ref{sipfenn:fig:oqmdperformance} gives (1) prediction vs OQMD values of formation energy plot, (2) statistics related to the error in predictions relative to the OQMD values, and (3) a histogram of the absolute error in predictions relative to the OQMD values.

\subsection{Existing Methods Comparison} \label{sipfenn:ssec:existing}
In this section, the performance of the models is compared with a few similar existing approaches based on the OQMD dataset, when formation energy of a structure is predicted \cite{Ward2016AMaterials, Ward2017IncludingTessellations, Jha2019IRNet}, or its subset of the convex-hull structures, when formation energy of the most stable structure is predicted \cite{Jha2018ElemNet:Composition, Goodall2020PredictingStoichiometry}.  This division is made based on the reasoning presented in \ref{sipfenn:ssec:currentapproach}. While the latter type cannot be used to predict the formation energy of any arbitrary structure, the structure-informed models like \texttt{SIPFENN} (the present work) can be tested on the convex hull structures.

\begin{table}[H]
\begin{center}
\begin{tabular}{|c|c|c|}
\hline
 Method & Formation Energy MAE & Convex Hull MAE \\
 \hline
 \texttt{SIPFENN} (This Work) & \textbf{28.0 meV/atom} (OQMD Opt.) & 32meV/atom (Novel. Mat.) \\
 \texttt{Ward2017} \cite{Ward2016AMaterials, Ward2017IncludingTessellations} & 80 meV/at & N/M \\
 \texttt{ElemNet} \cite{Jha2018ElemNet:Composition} & N/A & 50 meV/at\\  
 \texttt{IRNet} \cite{Jha2019IRNet} & 38 meV & N/M \\
 \texttt{Roost} \cite{Goodall2020PredictingStoichiometry} & N/A & 29 meV/at | \textbf{24 meV/at}\\
 \hline
\end{tabular}
\caption{Comparison of our method with existing state-of-the-art methods, as of late 2020, with OQMD-optimized \texttt{SIPFENN} model exhibiting state-of-the-art performance. N/A and N/M respectively stand for not applicable (out of domain) and not measured.}
\label{sipfenn:comparison-results}
\end{center}
\vspace{-24pt}
\end{table}

The results are shown in Table \ref{sipfenn:comparison-results}. The \texttt{SIPFENN} convex hull MAE has been reported based on using the Novel Materials Model limiting the original test set to structures laying within 50meV/atom from the convex hull. From these results, we can see that the \texttt{SIPFENN} neural networks approach outperforms existing state-of-the-art methods for predicting the formation energy of any material. At the same time, while not being the best, it is capable of reaching performance levels of specialized models in predicting the formation energies of structures laying on the convex hull.

\subsection{Non-OQMD Data Performance} \label{sipfenn:ssec:sigmasqsperformance}


Models created in the present work, specifically the ones optimized for predicting the formation energy of new materials, were designed and implemented to serve as tools for materials discovery. Evaluating their performance on data from the same source as the training set done in \ref{sipfenn:ssec:oqmdperformance} is inherently biased towards favoring models that provide the best fit to the prior (training) knowledge. This is amplified by the fact that many entries in the database are reported in groups that come from common studies and span similar materials, causing high domain clustering, which in some cases effectively makes such evaluation more akin to interpolation than extrapolation of knowledge.

To partially mitigate the described issue, the performance of the models was also evaluated on two smaller non-OQMD databases, described in \ref{sipfenn:sssec:Data}, representing an example of chemistries and structures that were of interest to the authors project on Ni-based superalloys. At the same time, they were not directly presented to the network in any capacity during the training process.

\begin{figure}[H]
\centering
    \includegraphics[width=0.3\textwidth]{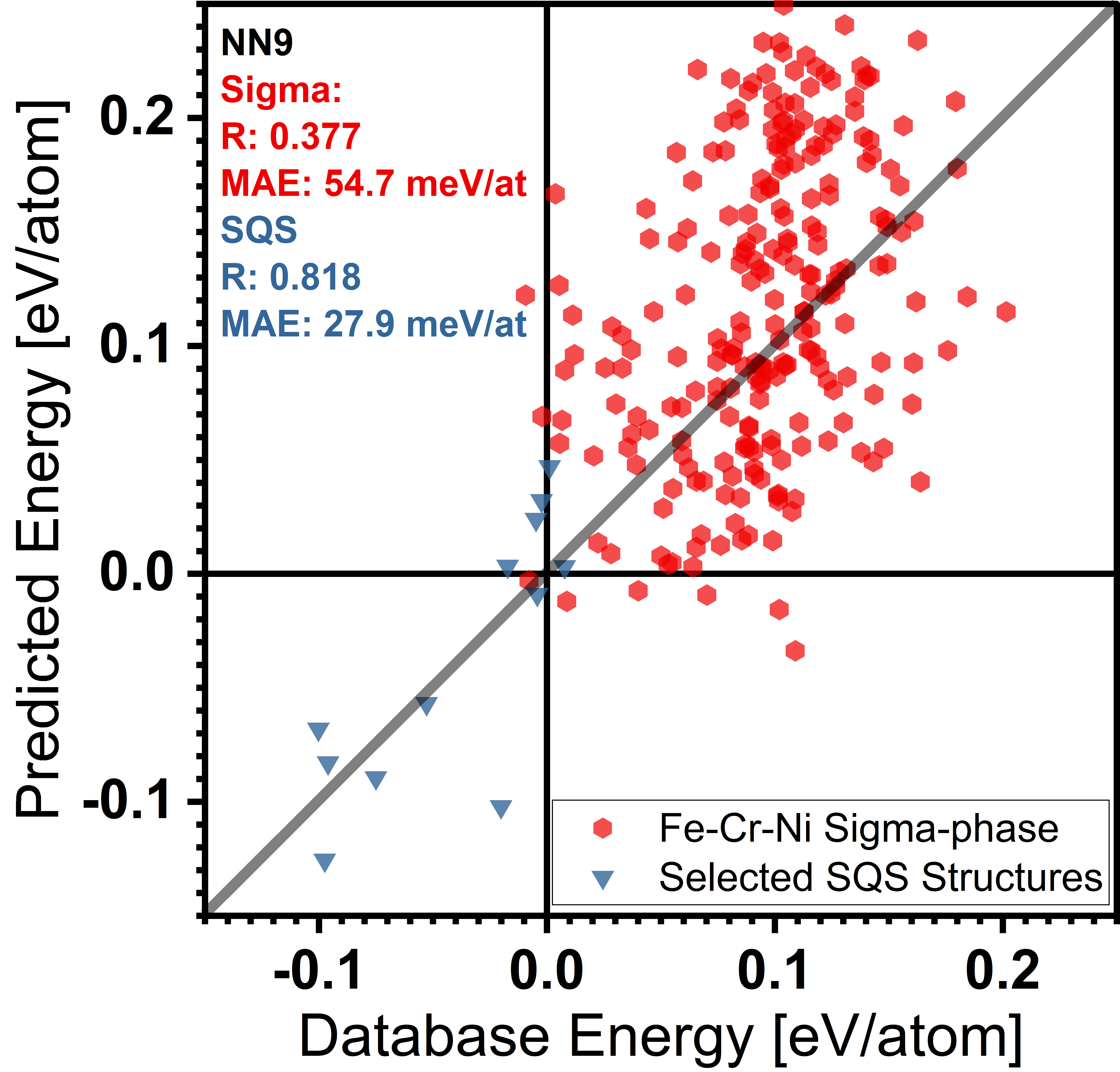}
    \hspace{0.03\textwidth}
    \includegraphics[width=0.3\textwidth]{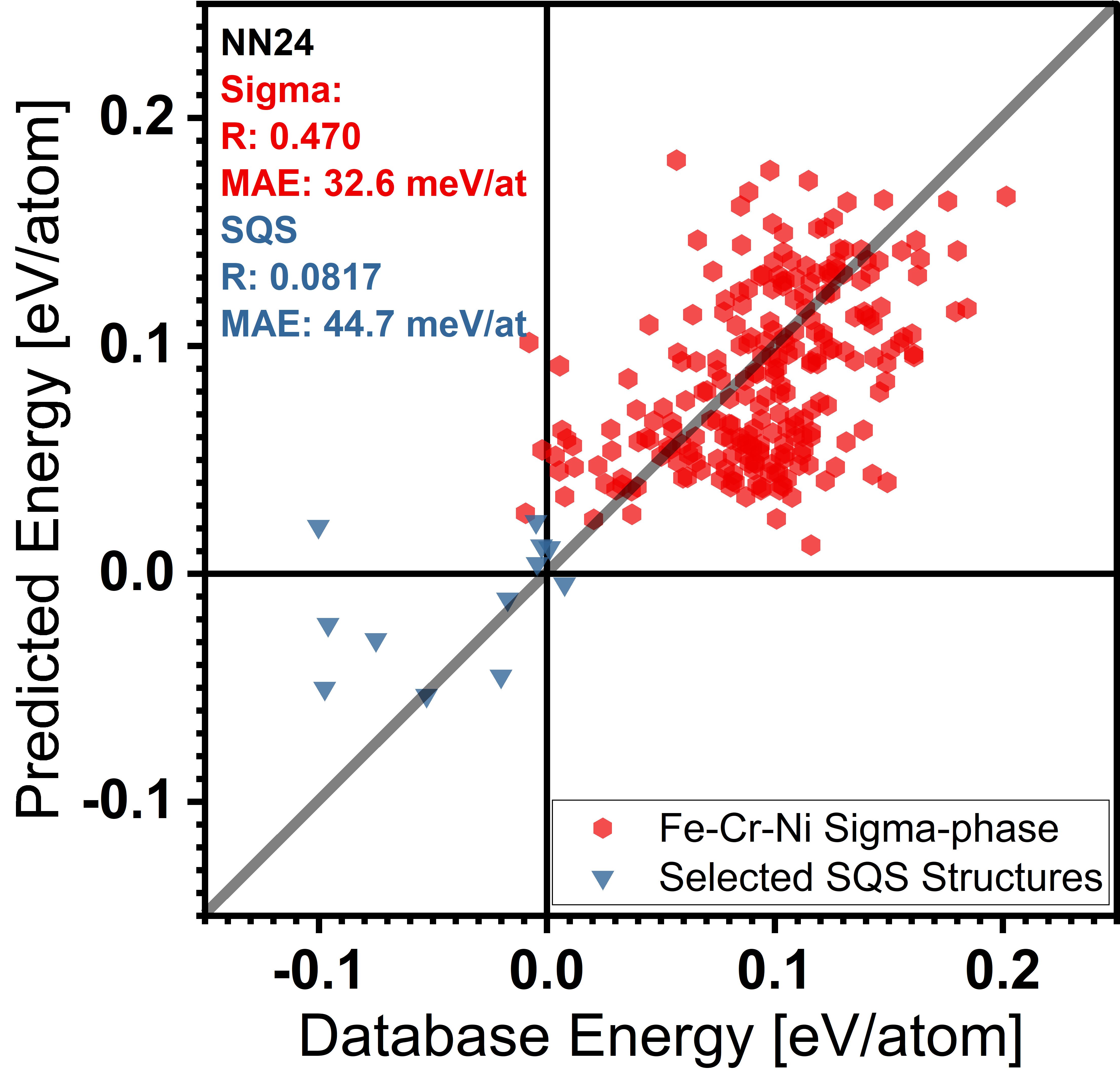}
    \hspace{0.03\textwidth}
    \includegraphics[width=0.3\textwidth]{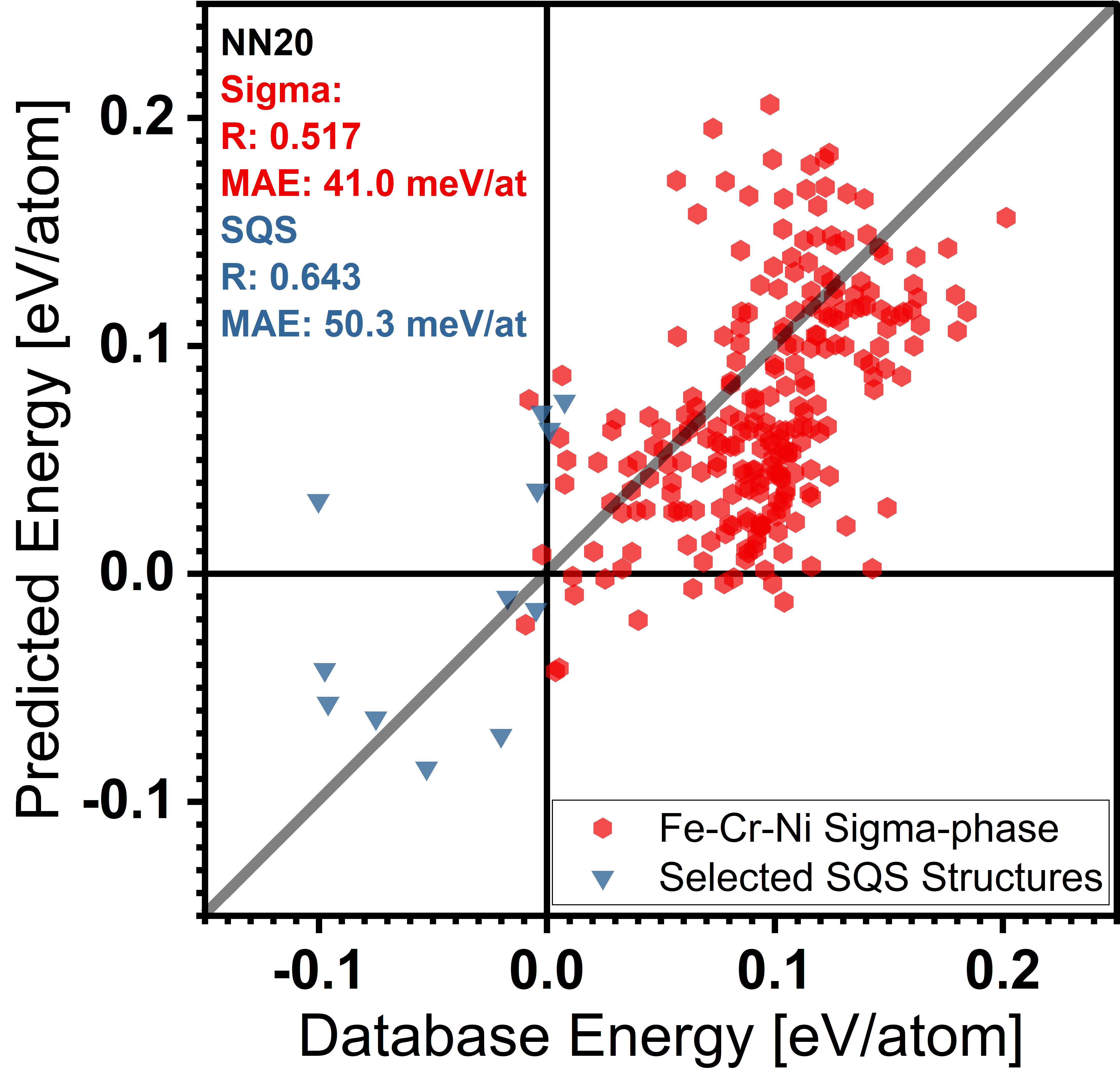}
    \caption{Performance of 3 selected neural networks on non-OQMD data described in \ref{sipfenn:sssec:Data}. Evaluated on (red) Fe-Cr-Ni $\sigma$-phase and (blue) SQS dataset. Networks organized by columns; optimized for (left) OQMD performance, (middle) predicting new materials, (right) size-constrained applications. Internally in the code, they are designated as NN9, NN20, and NN24 respectively.}
    \label{sipfenn:fig:sigmasqsperformance}
\end{figure}

In all cases, models created in the present paper were able to achieve approximately the same performance as on a random selection from the OQMD. To give a more in-depth analysis of the results, Figure \ref{sipfenn:fig:sigmasqsperformance} shows a magnified view of the predictions and basic statistics on the agreement between predictions and the database for the three models developed in the present work.

While all three models performed at around the same MAE level as for the OQMD, network optimized for new materials, the NN20 and NN24, performed better in the non-OQMD test cases of interest, providing major increases in correlations, significant for ranking of end-member configurations, except for 4 SQS configurations which were underestimated. The Pearson correlation slightly decreased in the first case and slightly increased in the second case. In both cases, the mean absolute error decreased by about 20\% compared to the OQMD-optimized model.

\subsection{Transfer Learning Capability} \label{sipfenn:ssec:transferlearningresults}
In this section, the technique of transfer learning is considered. It has been observed among deep learning models across a variety of domains \cite{tan2018survey,cirecsan2012transfer,chang2017unsupervised,george2018deep} and refers the to the ability of properly trained deep learning models to `transfer' their knowledge to related tasks. In the least complex approach, one does this by simply 'fine-tuning' the parameters of the model using new training data (from the new task). This methodology has shown in practice that deep neural networks are often able to transfer knowledge between different but related tasks. Such a problem is analogous to many others in materials science, where general knowledge is used to make meaningful statements without statistically significant patterns in locally available data. 

It is shown that a network trained on the OQMD database, which covers a broad yet limited spectrum of materials, can be quickly adjusted to materials outside of this spectrum with very little additional cost relative to the initial training. Specifically, the transfer learning capability of a network trained in this way on the set of all (243) Fe-Ni-Cr $\sigma$-phase 5-sublattice model endmembers, described in \ref{sipfenn:sssec:Data}, was tested. The ML model was first trained on a broad and general material dataset (OQMD) and then further trained (i.e., re-trained) for a given number of rounds on the new data (Fe-Ni-Cr $\sigma$-phase dataset) to adapt to the new system, while still conserving its broad knowledge, and can be thought of as fine-tuning a model to improve extrapolation outside of a prior knowledge space. 

\begin{figure}[H]
    \centering
    \includegraphics[width=0.6\textwidth]{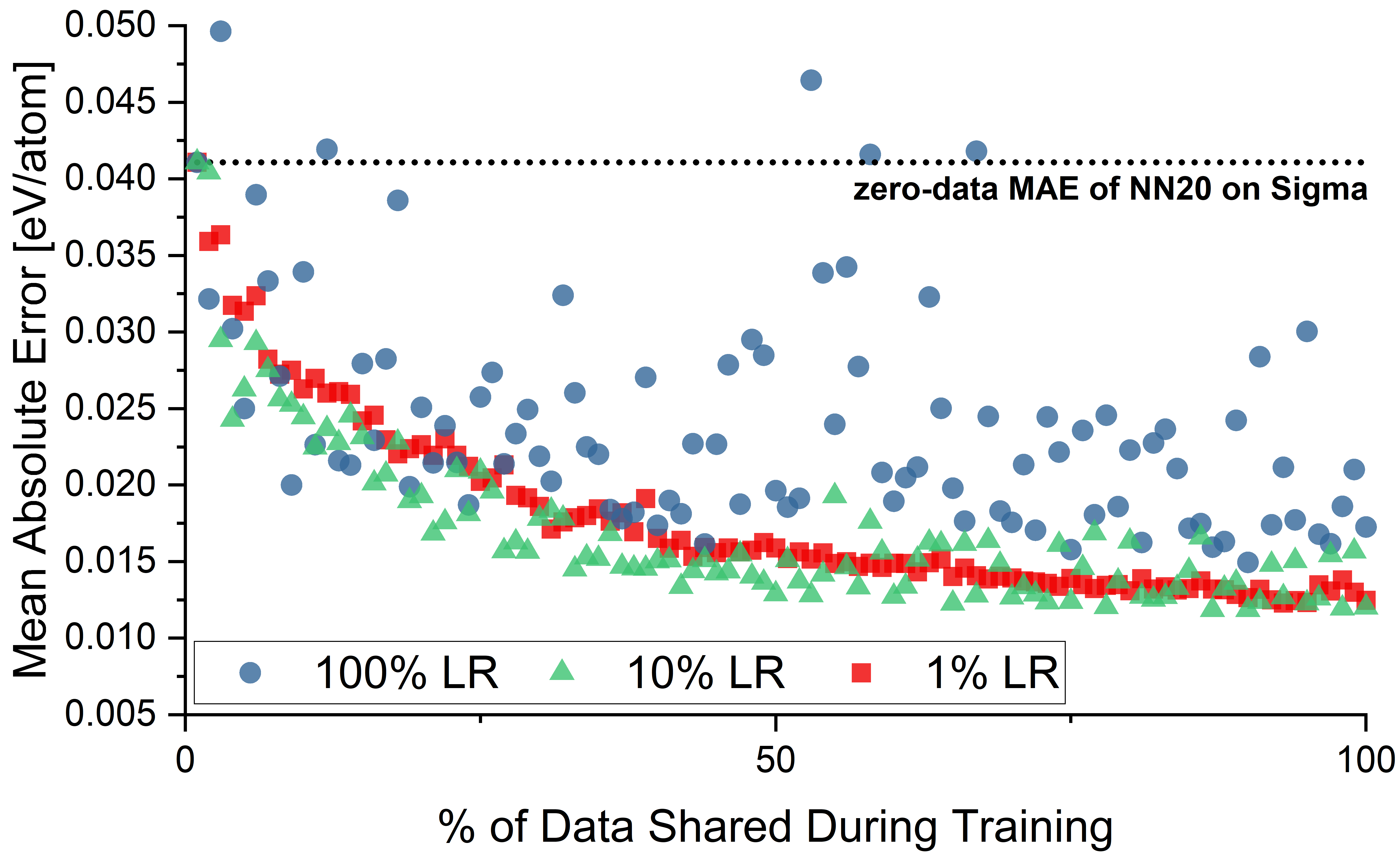}
    \caption{MAE evolution of NN20 model re-trained for 25 additional rounds on an increasing fraction of data from Fe-Cr-Ni $\sigma-$dataset. Presents the dependence of transfer learning from new data for different learning rates expressed as fractions of default ADAM learning rate (0.001).}
    \vspace{-12pt}
    \label{sipfenn:fig:transfersigmaLR}
\end{figure}

In order to achieve good performance, both the number of rounds and the learning rate have to be optimized. This can be accomplished by investigating the dependence of error on the fraction of available data while one of these parameters is fixed. Figure \ref{sipfenn:fig:transfersigmaLR} presents the dependence of transfer learning from new data for different learning rates expressed as fractions of default ADAM learning rate (0.001 shared across a vast majority of software). 

\begin{figure}[H]
    \centering
    \includegraphics[width=0.48\textwidth]{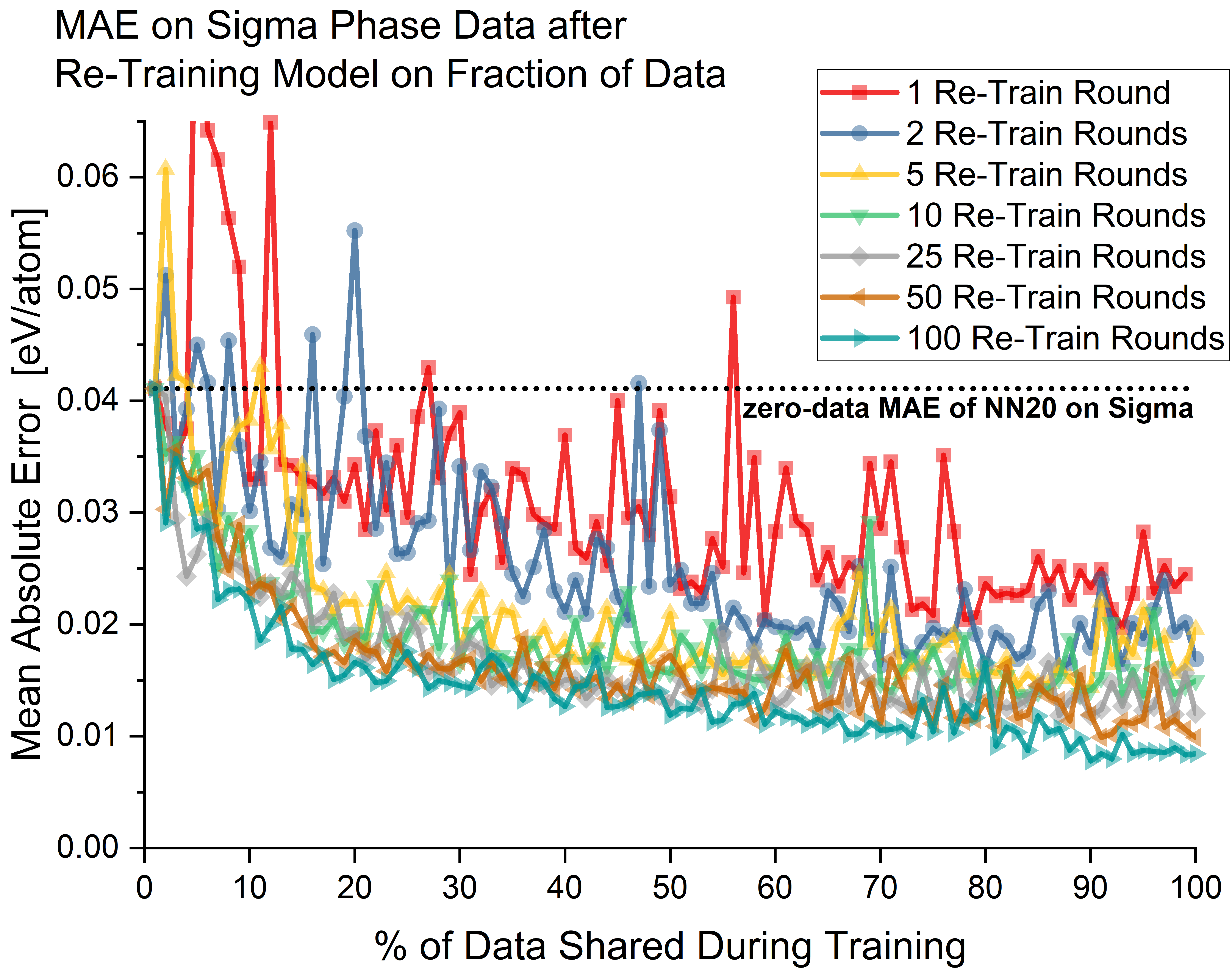}
    \includegraphics[width=0.48\textwidth]{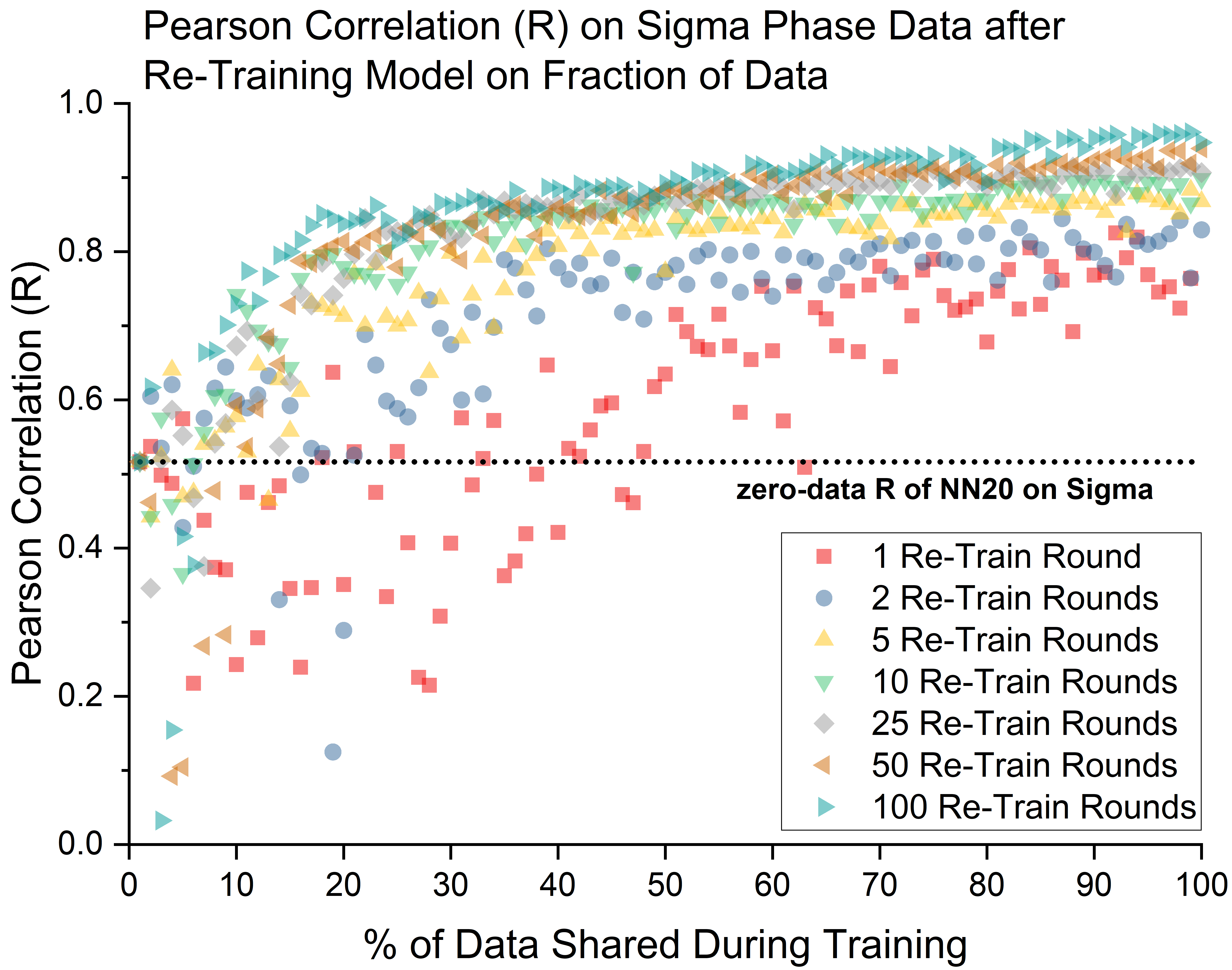}
    \caption{MAE and Person correlation (R) evolution of NN20 model re-trained at 10\% learning rate on an increasing fraction of data from Fe-Cr-Ni $\sigma-$dataset. Presents the dependence of transfer learning from new data for different re-training rounds numbers.}
    \vspace{-12pt}
    \label{sipfenn:fig:transfersigmaARR}
\end{figure}

As shown, in this case, the default learning rate (100\%)cannot be used for the transfer learning as it will adjust network parameters in both an unreliable and detrimental fashion, resulting in poor performance on the whole system of interest (both training and test sets) as shown in Figure \ref{sipfenn:fig:transfersigmaLR}. The same behavior would be observed if the process were conducted using an automated model design available in software such as \texttt{MATLAB} or \texttt{Mathematica}. The 10\% learning rate provided reliable enough outcomes and allowed a better performance improvement given little data, relative to using a 1\% learning rate (relative to the default). The second parameter to be optimized was the number of re-training rounds, as presented in Figure \ref{sipfenn:fig:transfersigmaARR}.

\begin{figure}[H]
    \centering
    \includegraphics[width=0.24\textwidth]{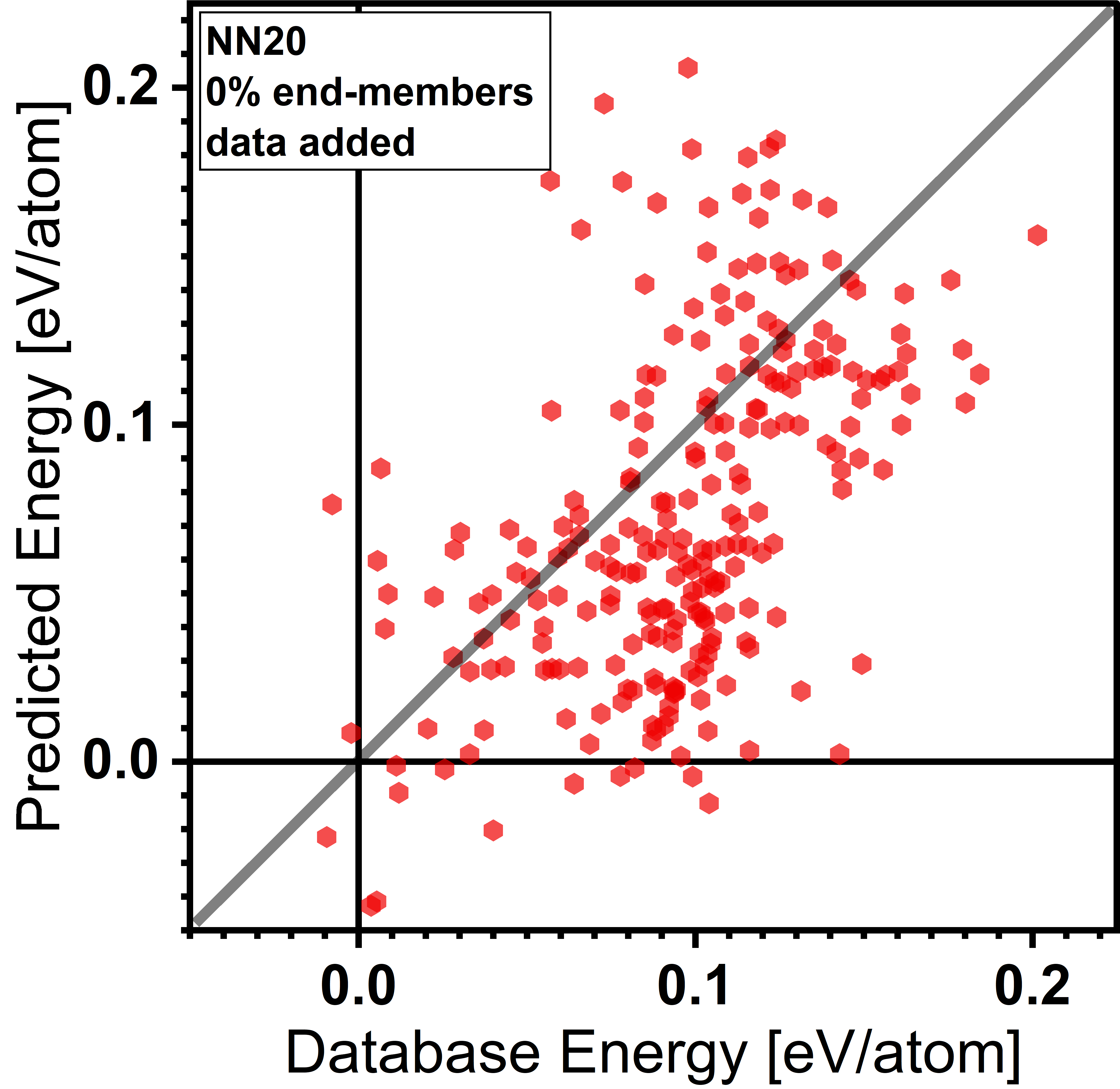}
    \includegraphics[width=0.24\textwidth]{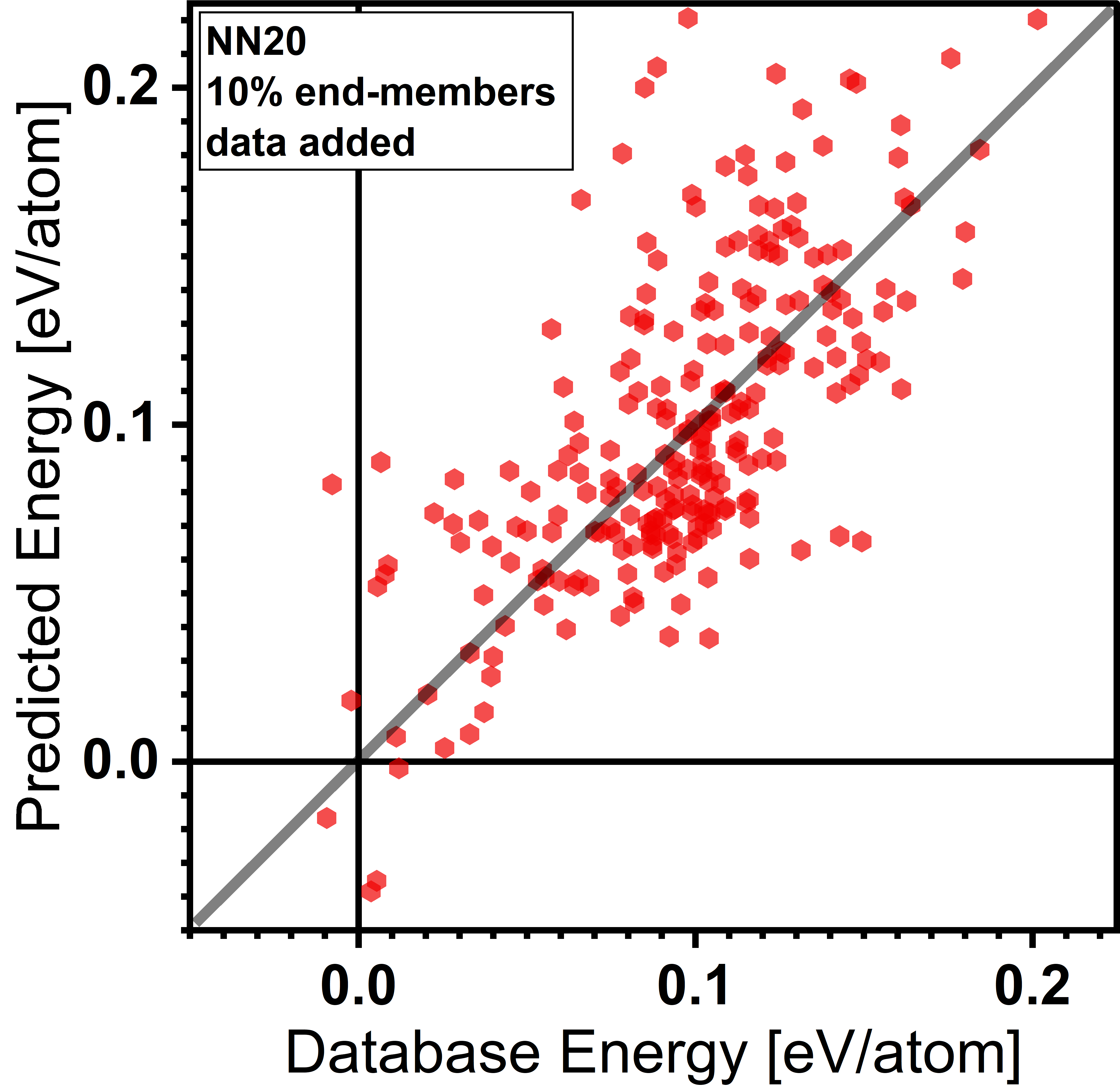}
    \includegraphics[width=0.24\textwidth]{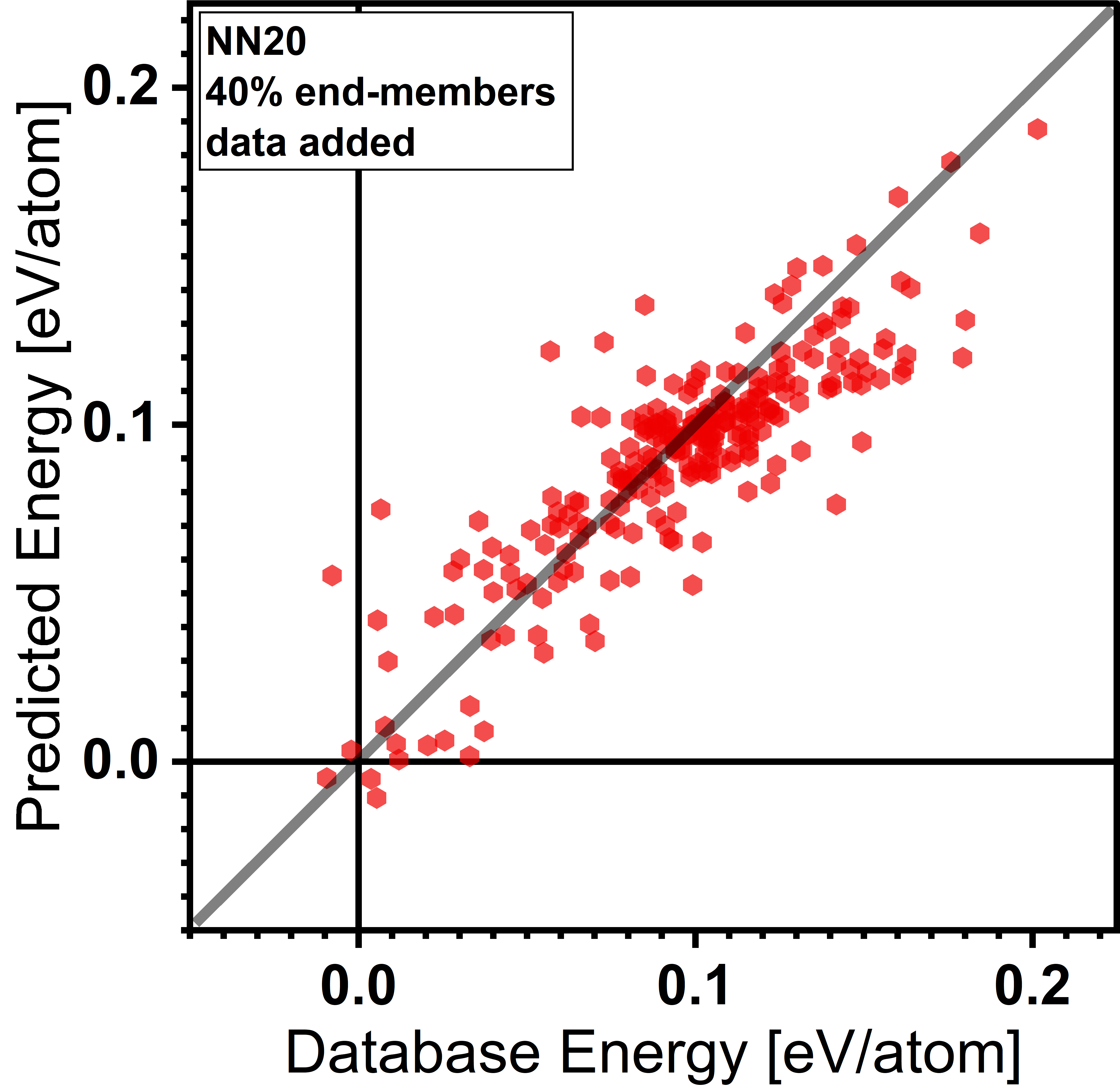}
    \includegraphics[width=0.24\textwidth]{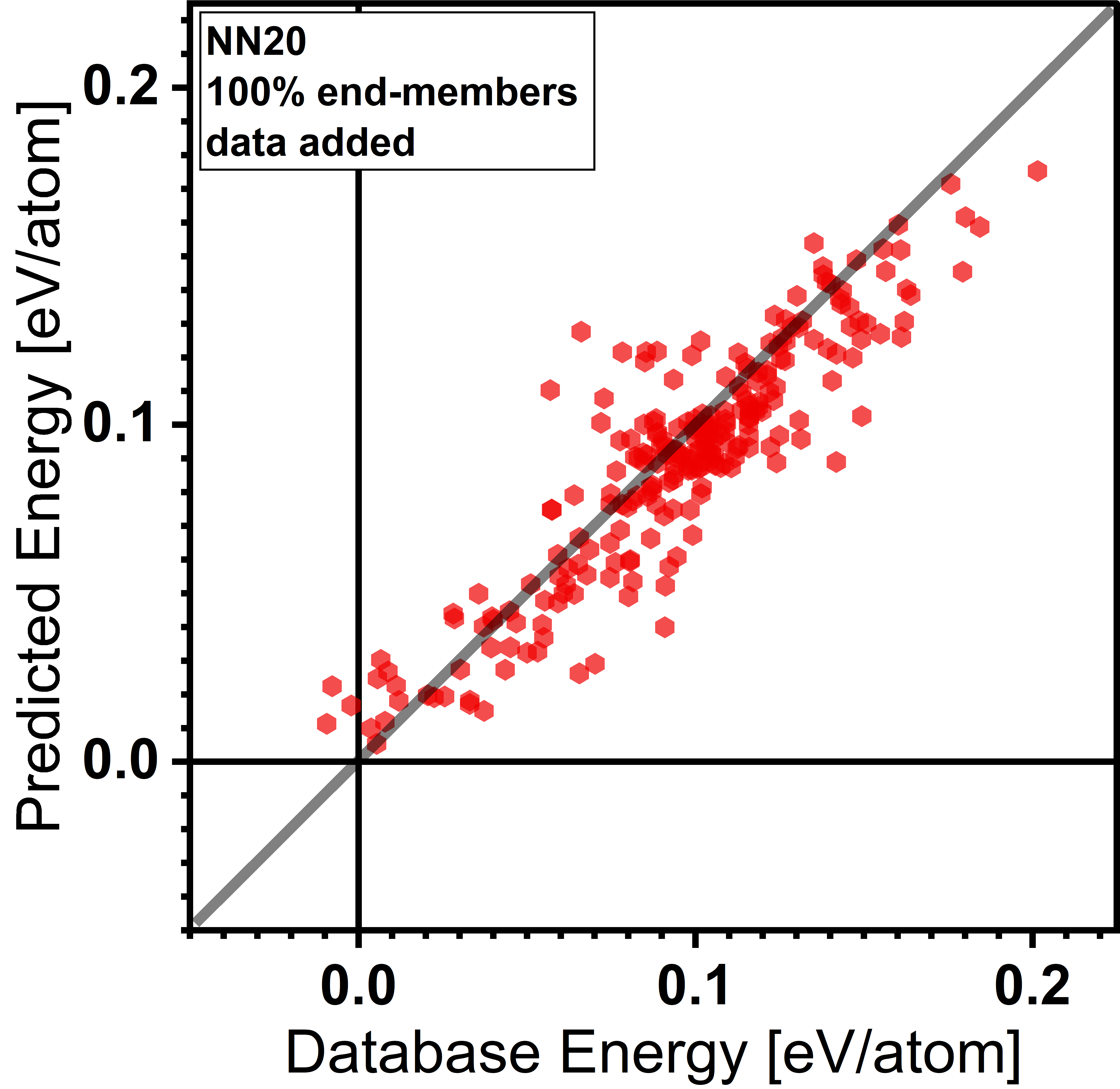}
    \caption{Performance of a new-materials-optimized network (NN20) on $\sigma$-phase data. Left-to-right: as trained on the OQMD, with additional training on 10\%, 40\%, and 100\% of the Fe-Cr-Ni $\sigma-$phase end-member data. The points on the figure correspond to all end-members (both training and testing data). Corresponding MAE and R are presented in Figure \ref{sipfenn:fig:transfersigmaARR} (gray rhombus points).}
    \vspace{-12pt}
    \label{sipfenn:fig:transfersigma}
\end{figure}

Figure \ref{sipfenn:fig:transfersigmaARR} shows that use of too few retraining rounds causes unreliable outcomes, while too many causes overfitting for low amounts of new data. In the case of Fe-Cr-Ni $\sigma-$dataset, retraining for 10 or 25 rounds provides balanced results across the whole dataset. With parameters for the process set to 10\% learning rate and 25 additional rounds, the performance can be evaluated graphically, as presented in Figure \ref{sipfenn:fig:transfersigma}.

As depicted, adding just 10\% of DFT-calculated data (24/243 endmembers) provided a significant improvement in the prediction quality over the system, including the other 90\%  was never shown to the model. This result indicates that the models in the present paper can be combined with partial data obtained through DFT calculations to create accurate predictive tools for a specific closed material system, such as sublattice endmembers, and potentially limit the number of calculations required within the study. This can then provide the ability to investigate broader material search spaces at a given computational cost.

\begin{figure}[H]
    \centering
    \includegraphics[width=0.48\textwidth]{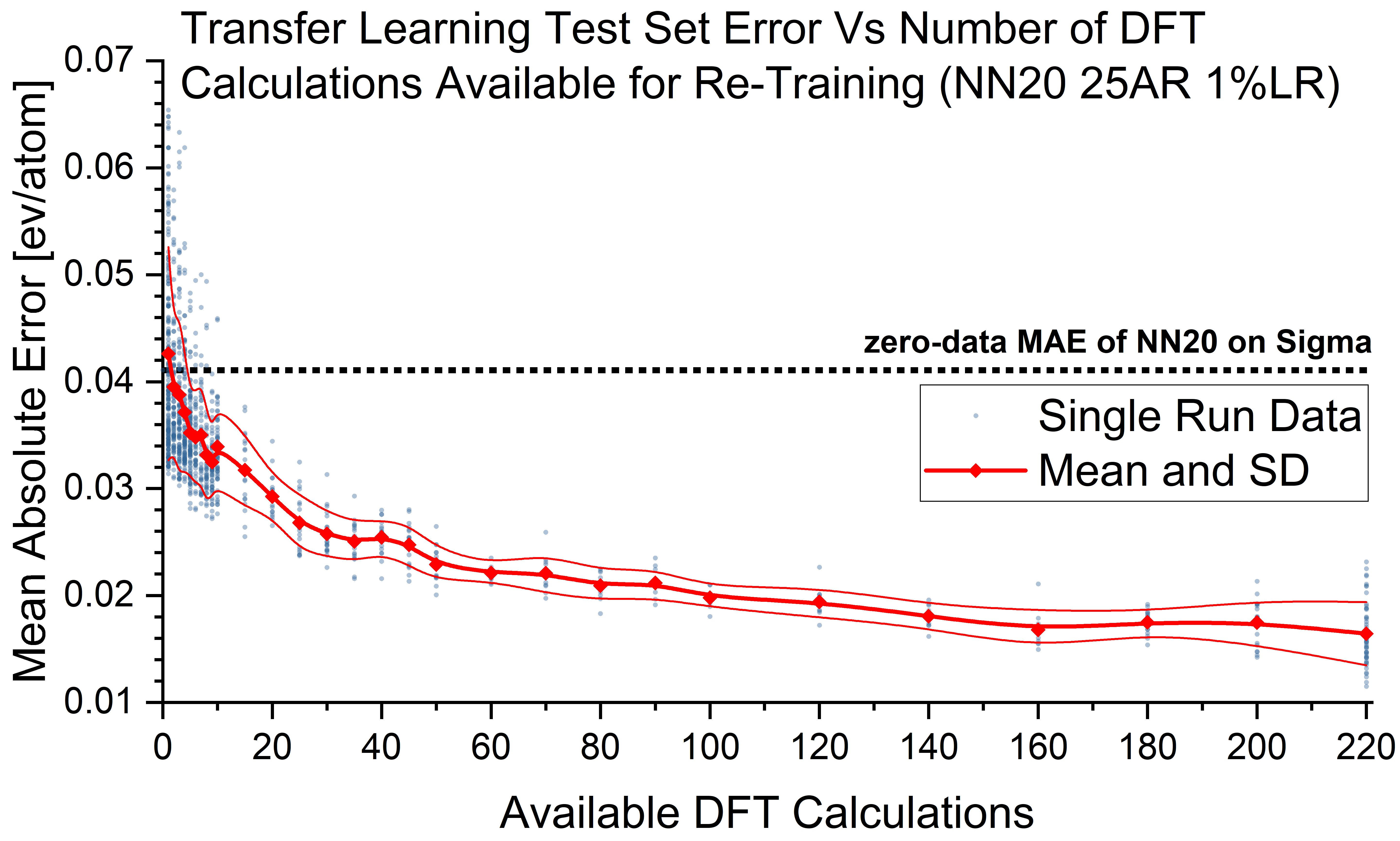}
    \includegraphics[width=0.48\textwidth]{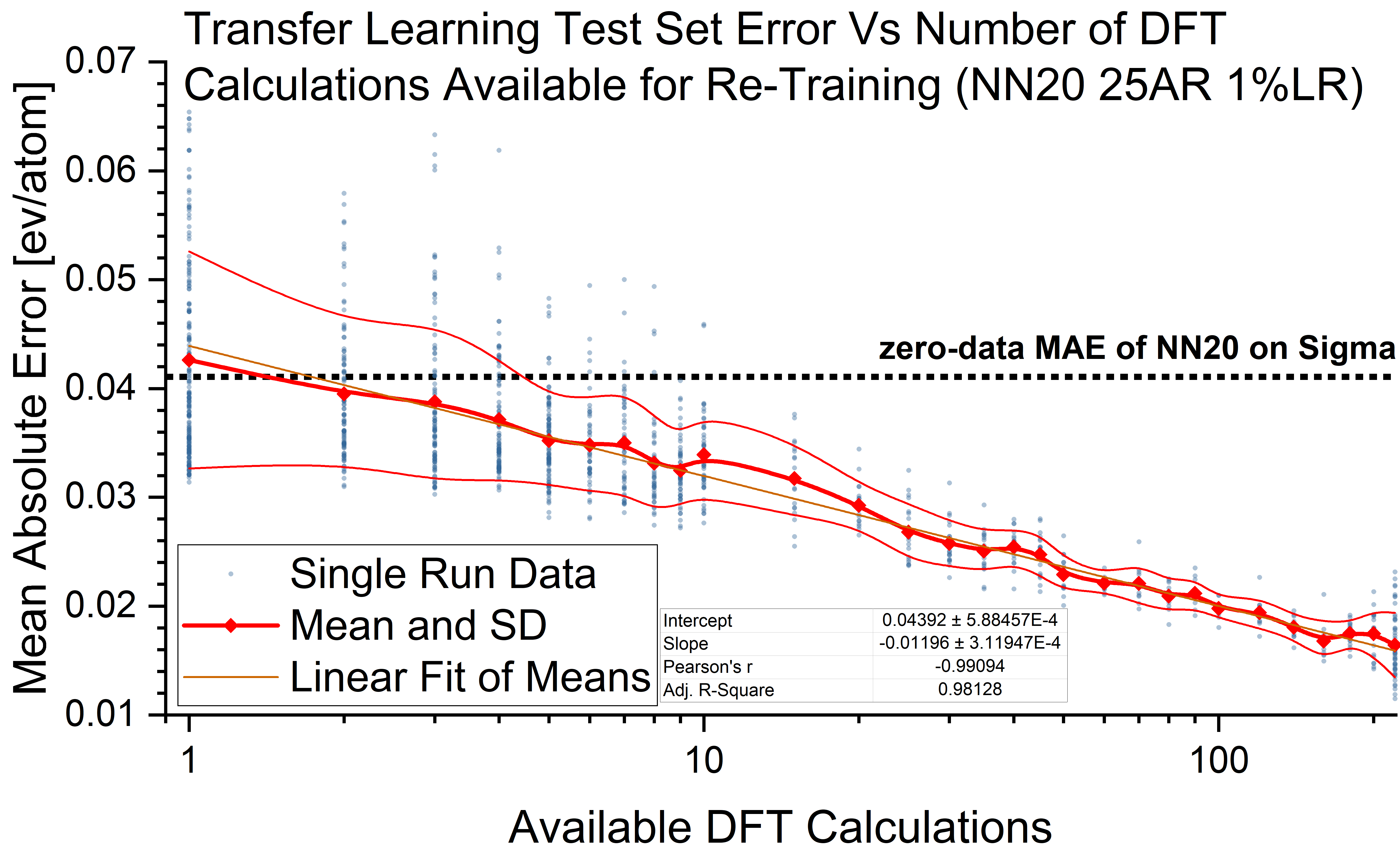}
    \caption{MAE of predictions evaluated on test set data vs number of newly available training datapoints. 1180 blue points correspond to single transfer learning processes. Red plot gives mean MAE and standard deviation. Both plots contain the same data.}
    \label{sipfenn:fig:transfersigmaVsDatapoints}
    \vspace{-12pt}
\end{figure}

Furthermore, the presented transfer learning capability could be used for a more broad materials exploration without a well-defined finite search space like the ternary Fe-Cr-Ni $\sigma-$phase. In such a case, it is better to evaluate and report the performance of the model on a test set that wasn't presented during the training and report, as a function of the number of added data points (new DFT calculations). With such a problem statement, the transfer learning process has been repeated 1180 for the statistical significance of the outcomes, which are presented in Figure \ref{sipfenn:fig:transfersigmaVsDatapoints}.

As presented in Figure \ref{sipfenn:fig:transfersigmaVsDatapoints}, adding just a small number of new data points allows to nearly half the MAE (around 20 datapoints). Furthermore, evident from the right plot, the mean performance increase is on average linear in log-lin scale and highly predictable ($R^2=0.98$).

\subsection{Model Limitations} \label{sipfenn:ssec:modellimitations}
As with any modeling tool, this modeling effort has some inherent limitations, coning from both data and methods used to create it. The most significant one comes from the type of data used for training of the model, where all data points correspond to DFT-relaxed structures, sitting in local minima in the configuration energy landscape. Thus, all energy predictions are given under an assumption that the input structure is fully relaxed with DFT settings inherited from the OQMD database \cite{Saal2013MaterialsOQMD}. At the same time, since the model was trained on many local energy minima configurations analyzed on the level of single-atom chemical environments, it should be able to approximate values for unrelaxed structures based on substitution from prototypes or similar compounds. Testing of this is performed by Ward 2017 \cite{Ward2017IncludingTessellations}, where it is shown that (a) in most of the test cases, the before-after relaxation energy difference is negligible in comparison to the DFT-ML difference for Ward 2017 model and usually much lower than the test MAE for models discussed in this work, and (b) in some particular cases ($Li_6CaCeO_6$) can be very high.

When faced with a new configuration, the model can thus either be used to (1) give an accurate prediction if the configuration is already relaxed or (2) give an approximate result that needs to be validated with DFT if confidence in the result is needed. This is inherent to all structure-informed ML models. One possible solution to partially mitigate this limitation is to perform relaxation using the model, which should work reasonably well for most materials. 

\begin{figure}[H]
    \centering
    \includegraphics[width=0.7\textwidth]{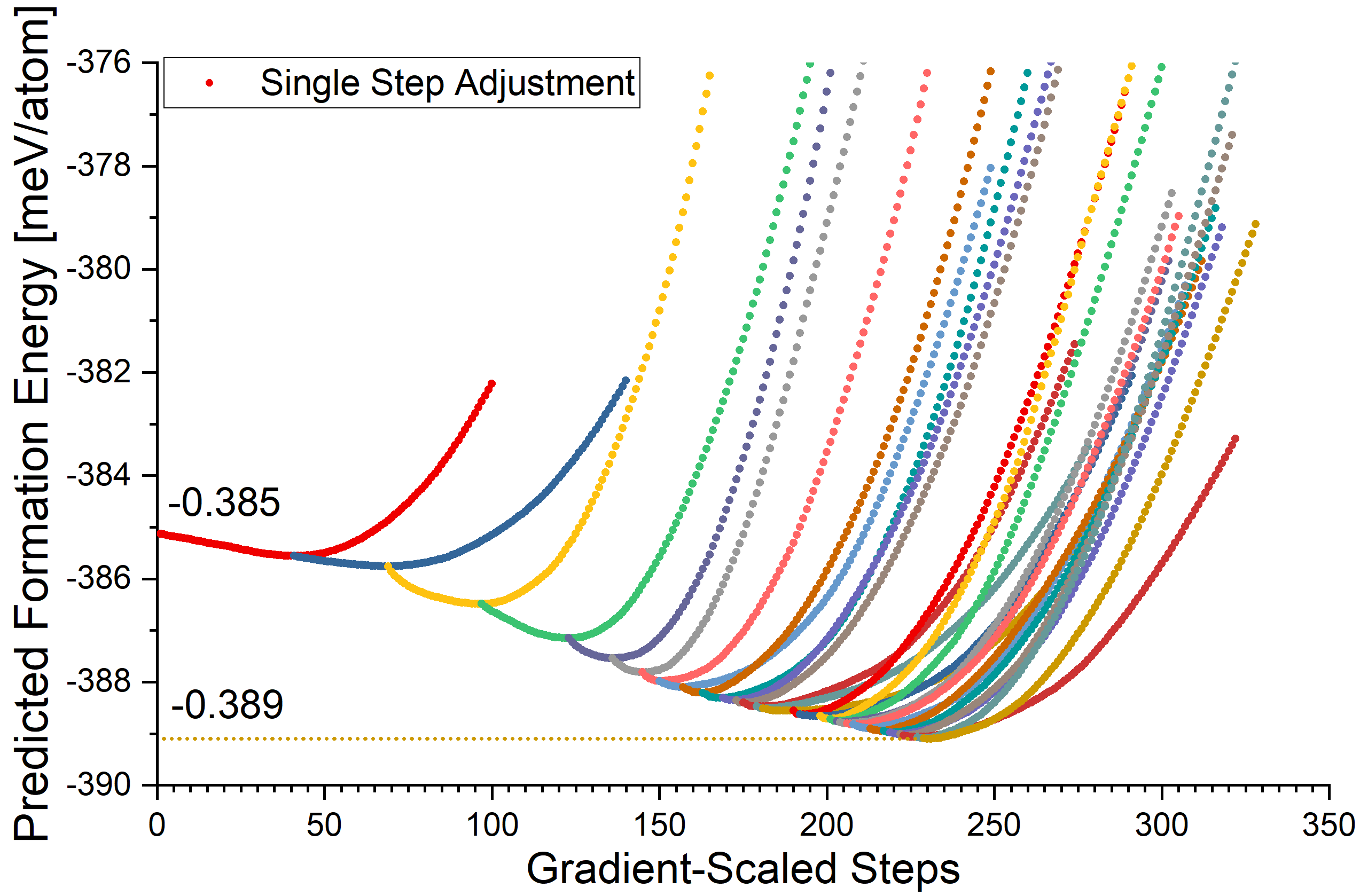}
    \caption{Local energy relaxation landscape of $Pd_8Zn_5$ $\gamma$-brass crystal structure from ideal positions guided iteratively by Novel Material Model (NN20).}
    \label{sipfenn:fig:localrelaxationpdzn}
\end{figure}

Discussion of such relaxation procedure in detail is extensive and beyond the scope of this work, yet a preliminary approach was constructed using the Novel Material Model (NN2) and deployed on all 16 end-members of Pd-Zn $\gamma$-brass crystal structure \cite{Dasgupta2022} in an iterative fashion. At each iteration, first, the local energy gradient for each atom was calculated by comparing the starting configuration with perturbations in x, y, z directions. Then, all atoms were displaced proportionally to the gradient in 100 discrete steps, reaching some local minimum, which acted as a starting point for the next iteration. An example for $Pd_8Zn_5$ is presented in Figure \ref{sipfenn:fig:localrelaxationpdzn}.

As shown in \ref{sipfenn:fig:localrelaxationpdzn}, the resulting relaxation reduced predicted formation energy by 4 meV/atom for this particular end-member. In the other 15 cases, results were similar, ranging between near 0 and 15 meV/atom, converging into fine local minima, expected to correspond with true local relaxations; however, extensive research into the problem is needed before conclusions can be drawn.

\subsection{End-User Implementation - SIPFENN} \label{sipfenn:ssec:SIPFENN}

One of the main objectives of the present paper was to create a tool that is transparent, easy to use by the research community, and easily modifiable. This lead to the creation of \texttt{SIPFENN} (Structure-Informed Prediction of Formation Energy using Neural Networks) software. \texttt{SIPFENN} provides the user with near-instant access to the models presented in \ref{sipfenn:sssec:DesignedModels}. In the future, this selection will likely be further expanded. On the user side, the use of the software is as easy as selecting one of the models, specifying a folder containing structure information files like \texttt{POSCAR}s \cite{POSCARFile} or \texttt{CIF}s \cite{Hall1991TheCrystallography}, running the predictions, and saving results.

\begin{figure}[H]
    \centering
    \includegraphics[width=0.67\textwidth]{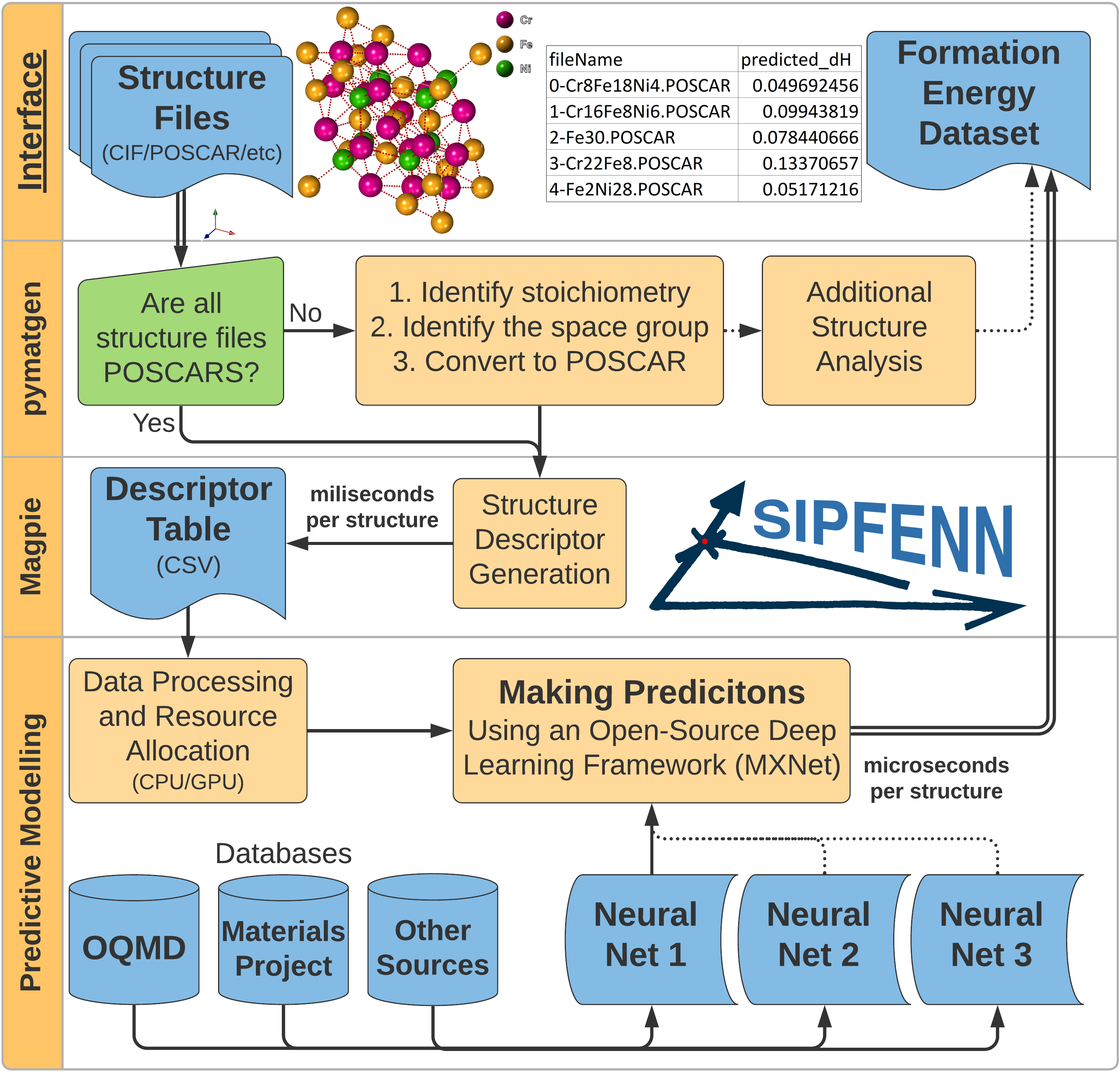}
    \caption{SIPFENN schematic description of operation.}
    \label{sipfenn:fig:sipfenn}
\end{figure}

SIPFENN was written entirely in Python to allow other researchers to easily modify it and adjust it to specific needs. Its schematic of operation is presented in Figure \ref{sipfenn:fig:sipfenn}. In broad scope, it first performs the structure analysis and modifications using the Python Materials Genomics library (\texttt{pymatgen}) \cite{Ong2013PythonAnalysis}. In the current implementation, it imports all structure files, analyzes the stoichiometry, creates unique names based on that, and exports them as POSCAR files. This is a rather simple task, however pymatgen is a powerful tool with a suit of more complex analytical tools that can be quickly implemented into \texttt{SIPFENN} by the user with even basic Python skills. Following the analysis, \texttt{SIPFENN} runs java-based Magpie \cite{Ward2016AMaterials} which calculates a descriptor for every imported structure and exports the result as a CSV file. This file is a descriptor table, where each row corresponds to a single material, and which can be stored and re-used later to run multiple predictive models at a fraction of the original computation time. It can also be used to create datasets for training procedures by replacing the last column with calculated or experimental values of formation energy.

Finally, the descriptor table is imported into the \texttt{MXNet} library framework, allocated into the CPU or GPU memory based on user selection, and evaluated using the selected predictive model. Once results are obtained, they are exported in CSV format and can be analyzed by any spreadsheet software such as Microsoft Excel.

\begin{figure}[H]
    \centering
    \frame{\includegraphics[width=0.55\textwidth]{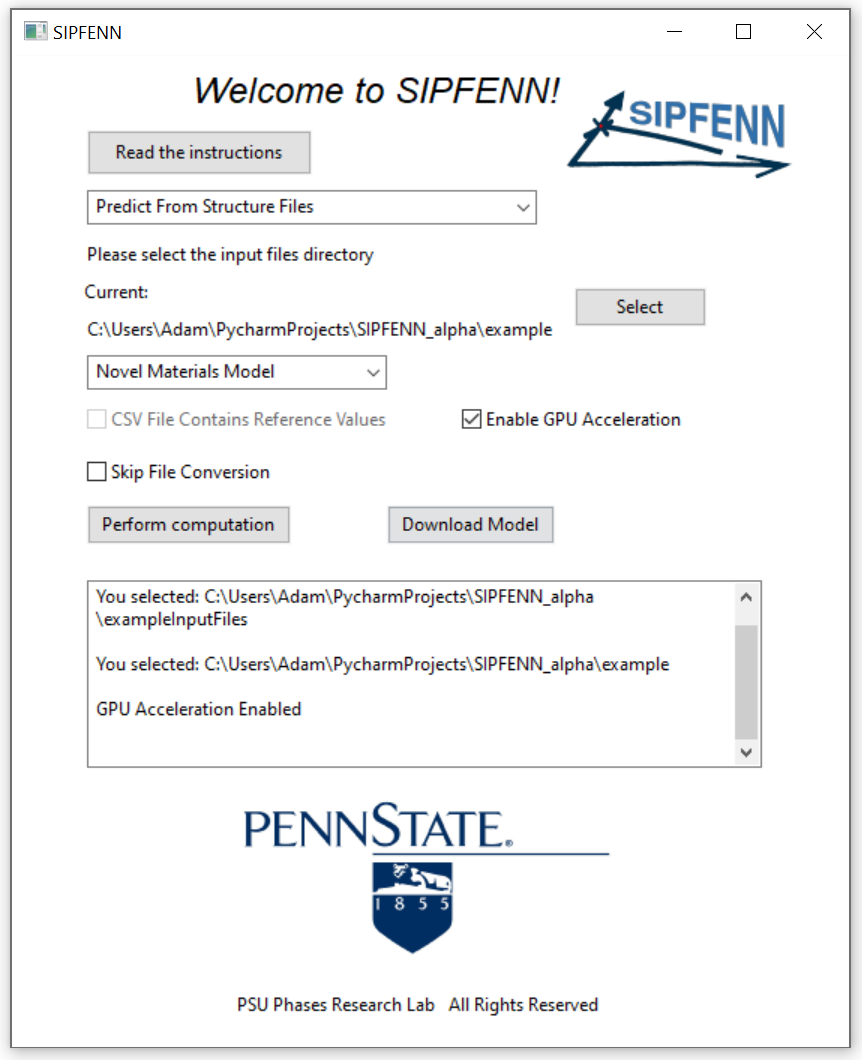}}
    \caption{A snapshot of the graphical user interface of SIPFENN.}
    \label{sipfenn:fig:sipfennGUI}
\end{figure}

\texttt{SIPFENN} was planned as a command-line tool, however, it was recognized that some users, especially those with little computational background, may find that difficult. Therefore, a simple graphical user interface (GUI) was created using \texttt{wxPython} library. It incorporates all the capabilities of the command line version. Furthermore, it lets the user download the predictive models from a repository in a single click. A sample snapshot of the GUI before performing calculations is presented in Figure \ref{sipfenn:fig:sipfennGUI}.

\section{Conclusions} \label{sipfenn:ssec:Conclusions}
In the present paper new machine learning models and a ready-to-use tool were created, based on the dataset and descriptor design by Ward et al. \cite{Ward2017IncludingTessellations}. Models reported in this work significantly improve upon existing methods, both in terms of performance and accessibility. For the most direct comparison, one of the designed models has been optimized for performing well on a random subset of the OQMD database and achieved an MAE of 28 meV/atom, compared to 80 meV/atom in the original  Ward et al. paper \cite{Ward2017IncludingTessellations}, and to 38 meV/atom in the most recent model called \texttt{IRNet} \cite{Jha2019IRNet}. Furthermore, it was shown that the error of the model is lowered when applied to the problem of finding the convex hull energy, achieving levels comparable with the current state-of-the-art approaches \cite{Jha2018ElemNet:Composition, Goodall2020PredictingStoichiometry}.

In addition, using appropriate overfitting mitigation efforts, such as Dropout and L2 regularization, models tuned for generalization to other types of materials datasets were developed. To test this, the models were evaluated on two datasets not contained within the OQMD, namely all end-members (243) of 5-sublattice topologically-close-packed Fe-Cr-Ni Sigma-phase \cite{Feurer2019Cr-Fe-NiCalculations, Hsieh2012OverviewSteels} and a few selected random-solution-approximating SQS \cite{Zunger1990SpecialStructures, Shin2006ThermodynamicStructures, Jiang2004First-principlesStructures}. The MAE values for these two test sets were found to be close to the values obtained on a test set from the OQMD. This exemplifies that the models are able to generalize to new datasets.

Furthermore, it was shown that models created within the present paper can be used for transfer learning, where vast knowledge of a broad spectrum of materials is combined with as little as a few DFT-datapoints from a specific materials system to provide excellent results within that specific system. Such at least partially process mitigates the issue of low data availability, present in numerous materials science problems, and consequently allows users to investigate a broader scope of materials at the same computational cost.

Finally, the three neural network models designed within the present paper were used, in conjunction with additional software, to create an end-user tool called \texttt{SIPFENN}. \texttt{SIPFENN}'s capabilities extend far beyond allowing validation of the presented results. It is implemented to work without any intensive computations on the user side, using models accessible from a repository, requiring only a quick one-click model download  to run. It is very fast thanks to using one of the industry's leading ML frameworks capable of well-optimized computations on GPUs. Furthermore, it is an open-source tool written in \texttt{Python}, which can be easily modified to specific needs in a straightforward way without extensive changes in the code.

\section{Software and Data Availability}

As of 2022, the \texttt{SIPFENN} code has been deprecated in favor of highly improved \texttt{pySIPFENN} which was re-written with great stability, reliability, and deeper integration with other community tools. Chapter \ref{chap:pysipfenn} describes it in detail and Section \ref{pysipfenn:sec:softwareavaialbility} describes its availability along related workshops and tutorial materials.

For archival purposes, the last version of \texttt{SIPFENN} code is available through Penn State's Phases Research Lab website at 
\href{https://phaseslab.org/sipfenn}{https://phaseslab.org/sipfenn} in (1) a minimal version that can be run on pre-computed descriptors in CSV format as well as (2) ready-to-use version with pre-compiled Magpie \cite{Ward2016AMaterials}. \texttt{SIPFENN} contains hard-coded links to neural networks stored in the cloud that can be downloaded at a single-click (see Figure \ref{sipfenn:fig:sipfennGUI}). 

All neural networks were made available in (1) open-source \texttt{MXNet} format maintained by Apache Foundation, used within \texttt{SIPFENN}, (2) closed-source \texttt{WLNet} format maintained by Wolfram Research and having the advantage of even easier deployment, as well as guaranteed forward compatibility with future versions of Wolfram Language, and in (3) Open Neural Network Exchange (\texttt{ONNX}) format \cite{Bai2019ONNX:Exchange} distributed through \texttt{pySIPFENN}, as of April 2024.

For ensured longevity of results, original \texttt{SIPFENN} neural networks are stored through the courtesy of Zenodo.org service under DOI:~\href{https://doi.org/10.5281/zenodo.4006803}{10.5281/zenodo.4006803} at the CERN’s Data Centre.

\pagebreak


\chapter{Efficient Structure-Informed Featurization and Property Prediction of Ordered, Dilute, and Random Atomic Structures} \label{chap:pysipfenn}

\acknowledge{
This chapter adapts verbatim a preprint by \citet{Krajewski2024EfficientStructures} submitted for publication to the journal of Computational Materials Science. This work was co-authored with Jonathan Siegel and Zi-Kui Liu. All text was written by Adam M. Krajewski with Jonathan Siegel co-writing Section \ref{pysipfenn:ssec:Ward2017Translation} and associated software. Zi-Kui Liu provided edits and guidance.
}

\section{Introduction} \label{pysipfenn:sec:Introduction}

\texttt{SIPFENN} or \textit{Structure-Informed Prediction of Formation Energy using Neural Networks} software, first introduced by the authors in 2020 \cite{Krajewski2022ExtensibleNetworks, Krajewski2020SIPFENNModels} and described extensively in Chapter \ref{chap:sipfenn}, is one of several open-source tools available in the literature \cite{Ward2017, Jha2019IRNet, Chen2019GraphCrystals, Choudhary2021AtomisticPredictions, Deng2023CHGNetModelling, Davariashtiyani2023FormationRepresentation, Davariashtiyani2023FormationRepresentation, Schmidt2023Machine-Learning-AssistedMaterials} which train machine learning (ML) models on the data from large Density Functional Theory (DFT) based datasets like \texttt{OQMD} \cite{Saal2013MaterialsOQMD, Kirklin2015TheEnergies, Shen2022ReflectionsOQMD}, \texttt{AFLOW} \cite{Curtarolo2013AFLOW:Discovery, Toher2018TheDiscovery}, Materials Project \cite{Jain2013Commentary:Innovation}, NIST-\texttt{JARVIS}\cite{Choudhary2020TheDesign}, \texttt{Alexandria} \cite{Schmidt2022AFunctionals}, or \texttt{GNoME} \cite{Merchant2023ScalingDiscovery} to predict formation energies of arbitrary atomic structures, with accuracy high enough to act as a low-cost surrogate in the prediction of thermodynamic stability of ground and non-ground state configurations at 0K temperature. The low runtime cost allows such models to efficiently screen through millions of different atomic structures of interest on a personal machine in a reasonable time. 

In addition to high-accuracy neural network models trained on \texttt{OQMD} \cite{Saal2013MaterialsOQMD, Kirklin2015TheEnergies, Shen2022ReflectionsOQMD}, \texttt{SIPFENN} included a number of features not found in competing tools available at the time, such as the ability to quickly readjust models to a new chemical system based on just a few DFT data points through transfer learning and a selection of models optimized for different objectives like extrapolation to new materials instead of overfitting to high-data-density regions or low memory footprint \cite{Krajewski2022ExtensibleNetworks}.

\texttt{SIPFENN}'s usefulness has been demonstrated, for instance, in the cases where the structure of an experimentally observed compound could not be identified in industry-relevant Nd-Bi \cite{Im2022ThermodynamicModeling} and Al-Fe \cite{Shang2021FormingJoints} systems and had to be predicted. This was accomplished by (1) high-throughput generation of hundreds of thousands of possible candidates with the exact stoichiometry based on elemental substitutions into structures from both open DFT-based databases \cite{Saal2013MaterialsOQMD, Kirklin2015TheEnergies, Shen2022ReflectionsOQMD, Curtarolo2013AFLOW:Discovery, Toher2018TheDiscovery, Jain2013Commentary:Innovation, Choudhary2020TheDesign, Schmidt2022AFunctionals, Merchant2023ScalingDiscovery} and experimentally observed ones from Crystallography Open Database (COD) \cite{Grazulis2009CrystallographyStructures, Grazulis2012CrystallographyCollaboration, Grazulis2019CrystallographyPerspectives}, followed by (2) selection of thousands of low-energy candidates, (3) down-selection of tens of unique candidates based on clustering in the \texttt{SIPFENN}'s feature space, and (4) final validation with DFT and experiments. It has also been deployed in several thermodynamic modeling studies, e.g. of Nb-Ni system \cite{Sun2023ThermodynamicESPEI}, in conjunction with DFT and experimental data processed through \texttt{ESPEI} \cite{Bocklund2019ESPEICuMg} to automatically fit parameters of CALPHAD \cite{Olson2023GenomicDynamics} models deployed in \texttt{pycalphad} \cite{Otis2017Pycalphad:Python}.

\section{General Structure Featurization Improvements} \label{pysipfenn:sec:featurization}

\subsection{pySIPFENN Overview and Core Advantages} \label{pysipfenn:ssec:coreimprovements}

Being able to predict the thermodynamic stability of arbitrary atomic structures and their modifications is one of the most critical steps in establishing whether hypothetical candidates can be made in real life \cite{Zunger2019BewareMaterials}; however, it is certainly not the only task of interest to the community \cite{Jha2023MachineChallenges, Tao2021MachineDiscovery}. These diverse needs, combined with increasing interest in multi-property modeling, have shifted the focus of \texttt{SIPFENN} tool from model training \cite{Krajewski2022ExtensibleNetworks} toward the development of reliable, easy-to-use, and efficient general-purpose featurizers existing in a framework, which can be used by researchers and companies to quickly develop and deploy property-specific models, or use features directly in exploring similarity and trends in materials.

Thus, while the acronym has been retained, the name of the software has been changed to \textit{python toolset for Structure-Informed Property and Feature Engineering with Neural Networks} or \texttt{pySIPFENN}, and the software component has been carefully re-implemented in its entirety to make it as general as possible and enable the following core advantages:

\begin{enumerate}
    
    \item Reliable featurization, which can be immediately transferred to other tools thanks to standalone submodule implementations based only on two common libraries (\texttt{NumPy} \cite{Harris2020ArrayNumPy} and \texttt{pymatgen} \cite{Ong2013PythonAnalysis}). These include completely re-written \texttt{Ward2017} Java-based featurizer \cite{Ward2017} (see Section \ref{pysipfenn:ssec:Ward2017Translation}) and 3 new ones, described in Sections \ref{pysipfenn:sec:ordered}, \ref{pysipfenn:sec:dilute}, and \ref{pysipfenn:sec:randomsolutions}.

    \item Effortless plug-and-play deployment of neural network (and other) ML models (for any property) utilizing any of the defined feature vectors, enabled by the use of Open Neural Network Exchange (\texttt{ONNX}) open-source format \cite{Bai2019ONNX:Exchange} which can be exported from nearly every modern ML framework and is then loaded into \texttt{pySIPFENN}'s \texttt{PyTorch} backend \cite{Paszke2019PyTorch:Library} through \texttt{onnx2torch} \cite{Kalgin2021Onnx2torch:PyTorch}. Furthermore, implementing custom predictors, beyond those supported by \texttt{PyTorch}, is made easy by design.

    \item Dedicated \texttt{ModelExporters} submodule makes it easy to export trained models for publication or deployment on a different device while also enabling weight quantization and model graph optimizations to reduce memory requirements.

    \item The ability to acquire data and adjust or completely retrain model weights through automated \texttt{ModelAdjusters} submodule. Its applications include:
    \begin{enumerate}
        \item Fine-tuning models based on additional local data to facilitate transfer learning ML schemes of the domain adaptation kind \cite{Ben-David2010ADomains}, where a model can be adjusted to new chemistry and specific calculation settings, introduced by \texttt{SIPFENN} back in 2020 \cite{Krajewski2022ExtensibleNetworks}, which is also being adopted by other models like \texttt{ALIGNN} \cite{Gupta2024Structure-awareDatasets}. Such an approach can also be used iteratively in active learning schemes where new data is obtained and added.
        
        \item Tuning or retraining of the models based on other atomistic databases, or their subsets, accessed through \texttt{OPTIMADE} \cite{Andersen2021OPTIMADEData, Evans2024DevelopmentsExchange} to adjust the model to a different domain, which in the context of DFT datasets could mean adjusting the model to predict properties with DFT settings used by that database or focusing its attention to specific chemistry like, for instance, all compounds of Sn and all perovskites.
        
        \item Knowledge transfer learning \cite{Torrey2010HandbookLearning} to adjust models to entirely new, often less available properties while harvesting the existing pattern recognition.
    \end{enumerate}    
\end{enumerate}

\begin{figure}[h!]
    \centering
    \includegraphics[width=0.85\textwidth]{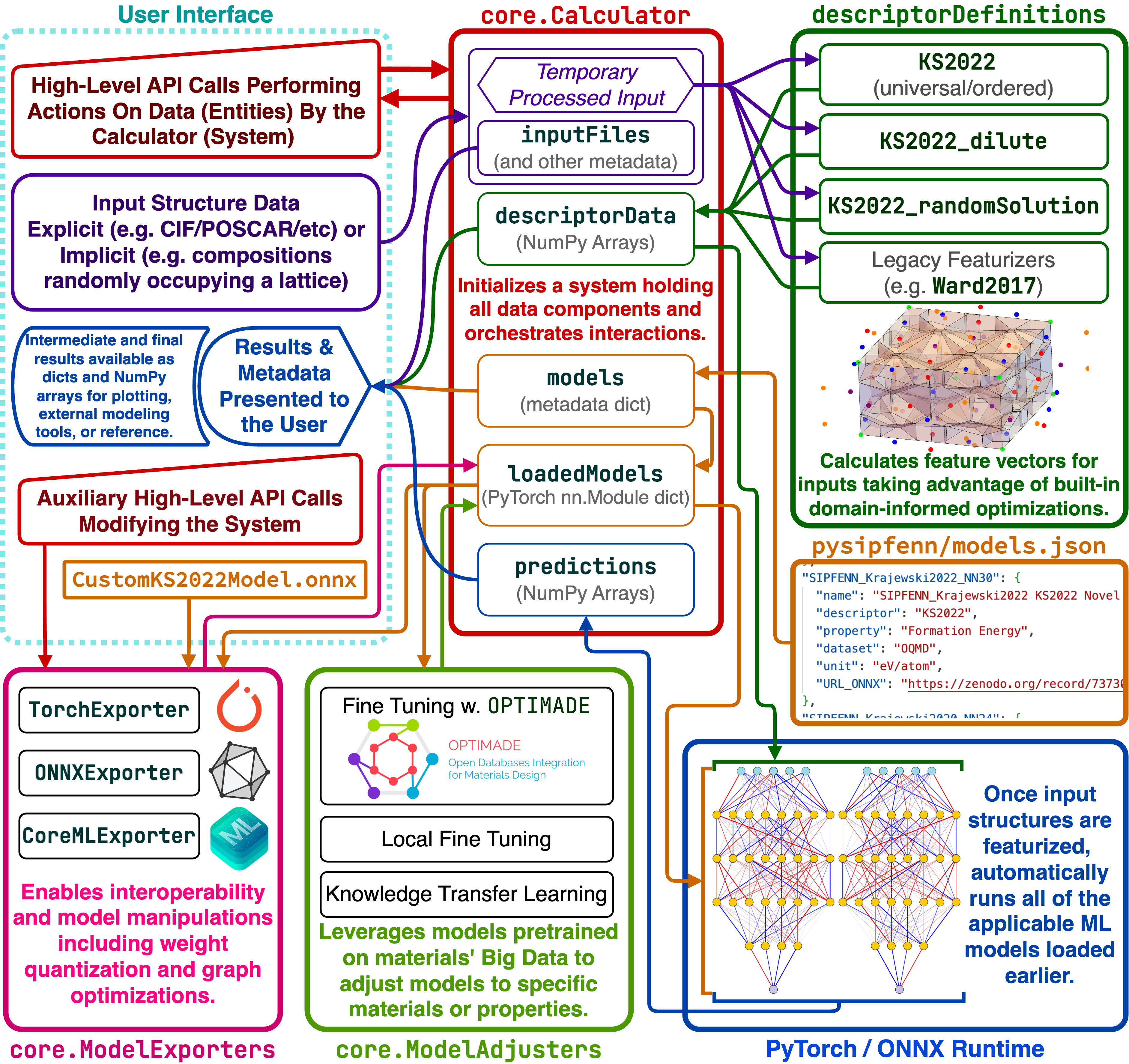}
    \caption{Main schematic of \texttt{pySIPFENN} framework detailing the interplay of internal components described in Section \ref{pysipfenn:ssec:coreimprovements}. The user interface provides a high-level API to process structural data within \texttt{core.Calculator}, pass it to featurization submodules in \texttt{descriptorDefinitions} to obtain vector representation, then passed to models defined in \texttt{models.json} and (typically) run automatically through all available models. All internal data of \texttt{core.Calculator} is accessible directly, enabling rapid customization. An auxiliary high-level API enables advanced users to operate and retrain the models.}
    \label{pysipfenn:fig:pySIPFENNMainSchematic}
\end{figure}

The resulting \texttt{pySIPFENN} computational framework is composed of several components, as depicted in Figure \ref{pysipfenn:fig:pySIPFENNMainSchematic}, and is available through several means described in Section \ref{pysipfenn:sec:softwareavaialbility}, alongside high-quality documentation and examples.

\subsection{Ward2017 Reimplementation} \label{pysipfenn:ssec:Ward2017Translation}

In their 2017 work \citet{Ward2017} introduced a novel atomic structure featurization concept based on establishing and weighting neighbor interactions by faces from 3D Voronoi tesselation to describe local chemical environments (LCEs) of atomic sites and then performing statistics over them to obtain a global feature vector. The original \texttt{SIPFENN} models \cite{Krajewski2020SIPFENNModels} built on top of this while utilizing an improved, carefully designed deep neural network models to obtain up to 2.5 times lower prediction error on the same dataset \cite{Krajewski2022ExtensibleNetworks}. A detailed description of the descriptor can be found in Section 2.1 of \citet{Krajewski2022ExtensibleNetworks}. In general, the calculation of the \texttt{Ward2017} descriptor consists of three parts:

\begin{itemize}
    \item Calculation of attributes based upon global averages over the components of the structure.
    \item Calculation of attributes based upon local neighborhood averages for each site in the structure.
    \item Calculation of more complex attributes based upon averages over paths in the structure.
\end{itemize}

\citet{Ward2017} implemented the above calculations in Java, which was popular at the time; while most of the current machine-learning packages use almost exclusively Python (e.g., \texttt{scikit-learn} \cite{PedregosaFABIANPEDREGOSA2011Scikit-learn:Python} and \texttt{PyTorch} \cite{Paszke2019PyTorch:Library}), making it cumbersome to use Java. Even more critically, the original Java implementation was not computationally efficient (as explored in Sections \ref{pysipfenn:sec:ordered}, \ref{pysipfenn:sec:dilute}, and \ref{pysipfenn:sec:randomsolutions}), and enabling tools were not supported in Java.

In the present work, authors have reimplemented \citet{Ward2017} from scratch in Python as a standalone submodule for \texttt{pySIPFENN}, which calculates all 271 features within numerical precision, except for three performing a random walk on the structure, which is stochastic in nature and results in slightly different final values due to a different seed. The Voronoi tessellation has been implemented with \texttt{Voro++} \cite{Rycroft2007MultiscaleFlow, Rycroft2009Voro++:C++, Lu2023AnCells} and all numerical operations were written using \texttt{NumPy} \cite{Harris2020ArrayNumPy} arrays to greatly speed up the calculations and make the efficient utilization of different computing resources, such as GPUs, easy to implement.

\subsection{KS2022 Feature Optimization} \label{pysipfenn:ssec:ks2022features}

Typically, during feature engineering, researchers first attempt to collect all features expected to enable good inference and then remove some based on the interplay of several factors:
\begin{enumerate}
    \item \textbf{Low impact} on the performance, which increases the representation memory requirements and possibly increases the risk of overfitting to both systematic and random trends. 
    \label{pysipfenn:item:featureoptimize1}
    \item \textbf{High computational cost}, which limits the throughput of the method deployment.
    \label{pysipfenn:item:featureoptimize2}
    \item \textbf{Unphysical features or feature representations} which can improve model performance against well-behaving benchmarks covering a small subset of the problem domain but compromise model interpretability and extrapolation ability in unpredictable ways.
    \label{pysipfenn:item:featureoptimize3}
\end{enumerate}

The \texttt{KS2022} feature set, added in \texttt{pySIPFENN v0.10} in November 2022, is a significant modification of the \texttt{Ward2017} \cite{Ward2017}, which focuses on points \ref{pysipfenn:item:featureoptimize2} and \ref{pysipfenn:item:featureoptimize3} above, while enabling optimizations described in Sections \ref{pysipfenn:sec:ordered} through \ref{pysipfenn:sec:randomsolutions} and delegating the removal of low-impact features to modeling efforts and keeping featurization as problem-independent as possible.

First, all 11 features relying on representation of crystal symmetry space groups with space group number \texttt{float}s rather than classes (e.g. using one-hot vectors) have been removed due to the unphysical nature of such representation leading to, for instance, BCC ($229$) being much closer to FCC ($225$) than to just slightly uniaxially distorted BCC ($139$), which itself would be very close to trigonal structures. 

Next, featurization code has been thoroughly profiled in regard to time spent on the execution of feature-specific subroutines and analyzed in the context of feature importance identified in the past work \cite{Krajewski2022ExtensibleNetworks}. This led to the removal of the 1 \textit{CanFormIonic} feature, which relied on combinatorically expensive guessing of oxidation states, and 3 features based on Warren-Cowley (WC) parameters \cite{Cowley1950AnAlloys}, which were relatively very expensive without significantly contributing to the performance due to scarcity of disordered structures in most atomistic datasets. However, the authors intend to add them back in future problem-specific feature sets using a recently released high-performance library by \citet{Gehringer2023ModelsSimple}. 

Together, 15 features were removed, bringing the total number of the \texttt{KS2022} features to $256$ while disproportionately improving the featurization speed. For instance, in the case of featurization of 30 sites in a disordered (no symmetry) structure, \texttt{KS2022} is $2.3$ times faster than \texttt{Ward2017} ($430$ms vs $990$ms single-threaded on Apple M2 Max).

\section{Optimizations for Ordered Structures} \label{pysipfenn:sec:ordered}

Modeling of disordered materials is a critical area of research \cite{Zaki2023Glassomics:Intelligence}; however, the vast majority of atomistic ab initio datasets used for ML studies focuses on highly symmetric ordered structures because of their high availability and ability to model complex phenomena in a holistic fashion if local ergodicity can be established \cite{Liu2022TheoryTheorem, Liu2023ThermodynamicsPerspectives}. One evidence of the focus on such data is the fact that out of $4.4$ million atomic structures in \texttt{MPDD} \cite{Krajewski2021MPDD:Database}, which includes both DFT-based \cite{Saal2013MaterialsOQMD, Kirklin2015TheEnergies, Shen2022ReflectionsOQMD, Curtarolo2013AFLOW:Discovery, Toher2018TheDiscovery, Jain2013Commentary:Innovation, Choudhary2020TheDesign, Merchant2023ScalingDiscovery} and experimental \cite{Grazulis2009CrystallographyStructures, Grazulis2012CrystallographyCollaboration, Grazulis2019CrystallographyPerspectives} data, only $54$ thousand or $1.25\%$ lack any symmetry. It is also worth noting that this number used to be much lower before the recent publication of the \texttt{GNoME} dataset by Google DeepMind \cite{Merchant2023ScalingDiscovery}, which accounts for around $\frac{3}{4}$ of them. 

In the case of remaining $98.75\%$ structures, a 3-dimensional crystallographic spacegroup is defined for each of them along with corresponding \emph{Wyckoff positions} (designated by letters) which are populated with either zero (empty), one (when symmetry-fixed), or up to infinitely many (typically up to a few) atoms forming a set of symmetry-equivalent sites called \emph{crystallographic orbits} \cite{Muller2006RemarksPositions}. When these crystallographic orbits are collapsed into atoms occupying a unit cell, each is repeated 
based on the \emph{multiplicity} associated with the Wyckoff position it occupies, which can range from 1 up to 192 (e.g., position l in Fm-3m/225), with values 1, 2, 3, 4, 6, 8, 16, 24, 32, 48, and 96 being typical \cite{Mehl2016ThePrototypes} even in compositionally simple materials like one of the experimentally observed allotropes of pure silicon with atoms at the 8a, 32e, and 96g positions \cite{Gryko2000Low-densityGap}. For certain crystal lattice types, the multiplicity can be somewhat reduced by redefining their spatial periodicity with so-called \textit{primitive} unit cells, like in the case of the aforementioned Si allotrope, in which primitive unit cell has 4 times fewer (34) sites but still over 10 times more than the 3 unique crystallographic orbits.

\begin{figure}[h]
    \centering
    \includegraphics[width=0.98\textwidth]{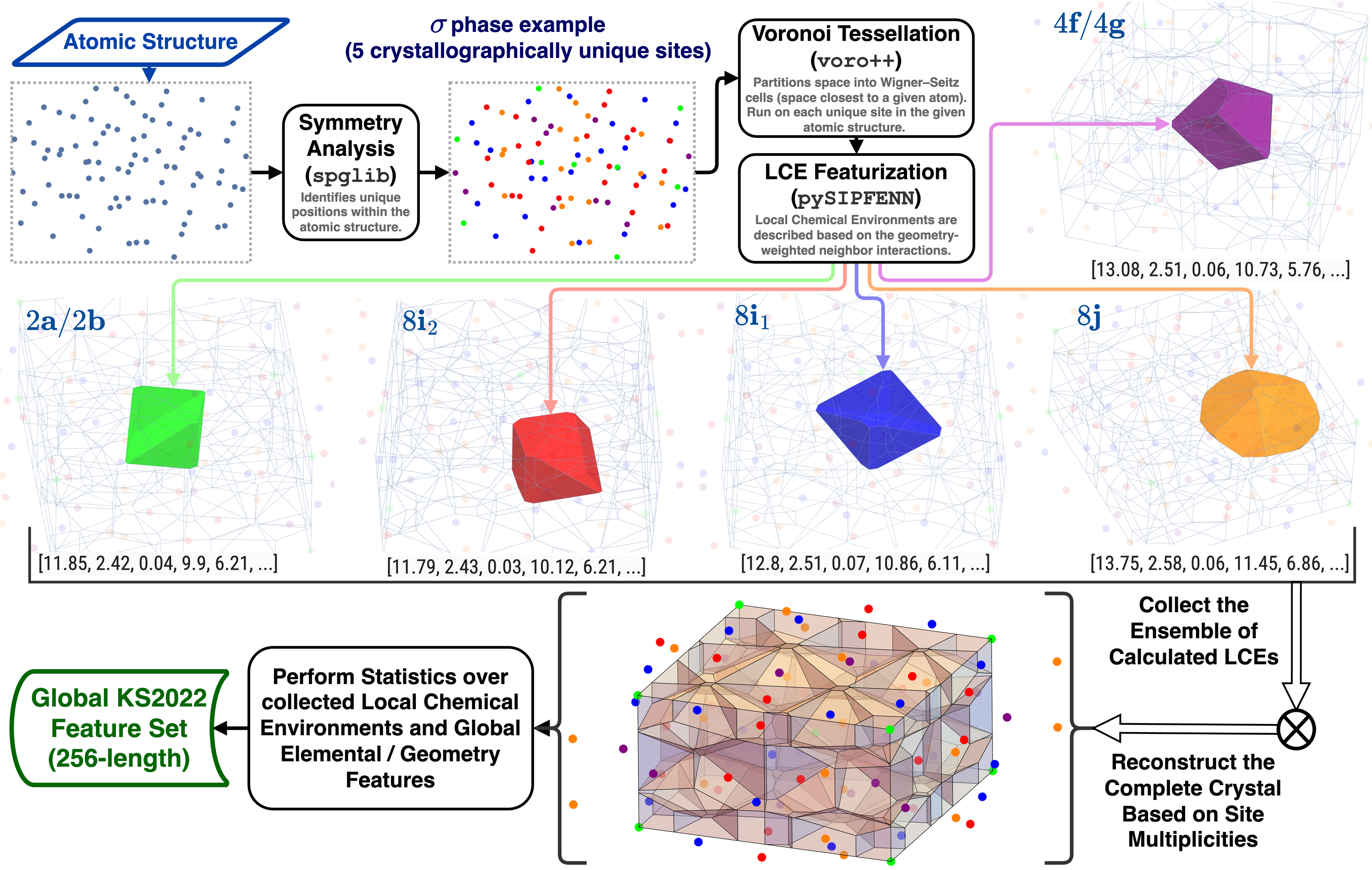}
    \caption{
    Schematic of the general-purpose \texttt{KS2022} featurization routine with built-in optimization for ordered structures. First, the atomic structure (in \texttt{pymatgen Structure} object format \cite{Ong2013PythonAnalysis}) is loaded, and sites in it are annotated with their crystallographic orbits using \texttt{spglib} \cite{Togo2018Spglib:Search}. Then, one site is selected from each orbit to form a set of unique sites, for which Wigner-Seitz cells (depicted as colored polyhedra) are calculated with \texttt{Voro++} \cite{Rycroft2007MultiscaleFlow, Rycroft2009Voro++:C++, Lu2023AnCells} and featurized to get site-specific local chemical environment (LCE) descriptors. The complete site ensemble is then reconstructed based on multiplicities of Wyckoff positions corresponding to the sites. A non-trivial example of $\sigma$-phase with 30 atoms belonging to 5 crystallographic orbits with interesting Wigner-Seitz cells (relative to usually shown FCC/BCC ones \cite{Bohm1996VoronoiLattices}) has been depicted.
    }
    \label{pysipfenn:fig:ks2022}
\end{figure}

This presents an immediate and previously untapped opportunity for multiplying the computational performance of most atomistic featurizers (e.g., \texttt{Matminer} \cite{Ward2018Matminer:Mining}) and ML models \cite{Ward2017, Chen2019GraphCrystals, Jha2019IRNet, Krajewski2022ExtensibleNetworks, Choudhary2021AtomisticPredictions, Deng2023CHGNetModelling, Davariashtiyani2023FormationRepresentation, Schmidt2023Machine-Learning-AssistedMaterials, Banik2024EvaluatingMaterials, Hu2021Atomtransmachine:Learning}, which nearly always process all atoms given in the input structure occasionally converting to primitive unit cell in certain routines (\texttt{CHGNet} \cite{Deng2023CHGNetModelling}), unless they operate on different occupancies of the same structure \cite{Crivello2022SupervisedExample}. This allows for a proportional decrease in both CPU/GPU time and memory footprint. The general-purpose \texttt{KS2022} in \texttt{pySIPFENN} uses high-performance symmetry analysis library \texttt{spglib} \cite{Togo2018Spglib:Search} to automatically take advantage of this whenever possible, as depicted in the schematic in Figure \ref{pysipfenn:fig:ks2022}. It shows an interesting example of a topologically close-packed $\sigma$ phase, which is critical to model in a wide range of metallic alloys \cite{Joubert2008CrystalPhase} but challenging in terms of combinatorics because of 5 unique sites that can be occupied by many elements \cite{Choi2019ADesign, Ostrowska2020ThermodynamicW} making it a very active area of ML modeling efforts \cite{Crivello2022SupervisedExample, Zha2024ApplyingEnergy} in the thermodynamics community.

In the case of \texttt{KS2022} featurizer, running the same 30-atom test as in Section \ref{pysipfenn:ssec:ks2022features} but on $\sigma$ phase takes on average $84$ms or is 5.1 times faster thanks to processing 6 times less sites. Similar results should be (a) quickly achievable with any other featurizer processing individual sites, including most graph representations embedding local environments (e.g., \texttt{MEGNet} \cite{Chen2019GraphCrystals}) or deconstructing graphs into graphlets (e.g., \texttt{minervachem} molecule featurizer \cite{Tynes2024LinearCheminformatics}), and (b) possible with convolution-based models operating on graphs (e.g., \texttt{ALIGNN} \cite{Choudhary2021AtomisticPredictions}) or voxels \cite{Davariashtiyani2023FormationRepresentation} through custom adjustments to the specific convolution implementation. In the case of voxel representations and any other memory-intense ones, it may also be beneficial to utilize this approach to compress them when transferring between devices like CPU and GPU or across an HPC network.

\section{Optimizations for Dilute, Defect, and Doped Structures} \label{pysipfenn:sec:dilute}

The optimization strategy in Section \ref{pysipfenn:sec:ordered} ensures that only the sites that are \emph{guaranteed} to be \emph{crystallographically unique} are processed through featurization or graph convolution and is directly applicable to the vast majority of both data points and literature methods. However, in the case of methods relying on describing the immediate neighbors, whether through Wigner-Seitz cell (see Fig. \ref{pysipfenn:fig:ks2022}) or subgraph (see, e.g., \cite{Chen2019GraphCrystals}), one can achieve further efficiency improvements by considering which sites are \emph{guaranteed} to be \emph{unique under the representation}.

There are several classes of atomic structures where the distinction above makes a difference, but the room to improve is exceptionally high when one site in an otherwise highly symmetric structure is modified, leading to a structure that, depending on the context, will be typically called \emph{dilute} when discussing alloys \cite{Chong2021CorrelationAlloys}, \emph{doped} when discussing electronic materials \cite{Chen2022InteractionStudy}, or said to have \emph{defect} in a more general sense \cite{Castleton2009DensitySupercells}. Throughout \texttt{pySIPFENN}'s codebase and the rest of this work, the single term \emph{dilute} is used to refer to all of such structures due to authors' research focus on Ni-based superalloys at the time when optimizations below were made public in February 2023.

\begin{figure}[h]
    \centering
    \includegraphics[width=0.85\textwidth]{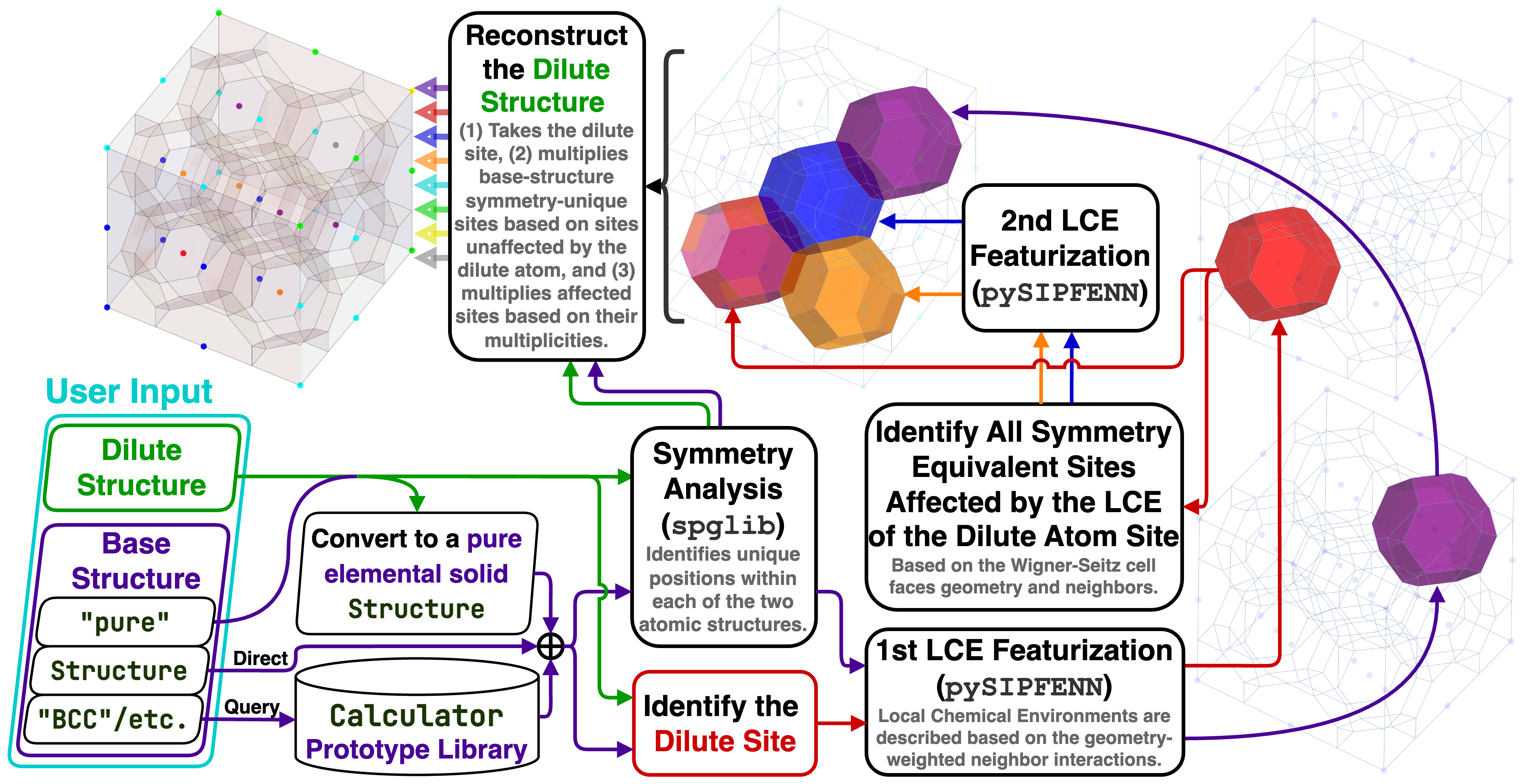}
    \caption{Core schematic of the \texttt{KS2022\_dilute} featurizer. The dilute structure is compared to either the explicit or implicit base structure to identify the dilute site, which is then featurized alongside all crystallographically unique sites in the base structure. Information extracted from dilute structure featurization is then used to identify previously-equivalent sites affected by it, which go through the second round of featurization. Lastly, the complete ensemble is reconstructed, and \texttt{KS2022} are obtained. BCC supercell is used as an example.}
    \label{pysipfenn:fig:KS2022dilute}
\end{figure}

To visualize the concept, one can consider, for instance, a 3x3x3 body-centered cubic (BCC) conventional supercell (54 sites) and call it \textit{base structure}. If it only contains a single specie, then \texttt{KS2022} from Section \ref{pysipfenn:sec:ordered} will recognize that there is only one crystallographic orbit and only process that one. However, if a substitution is made at any of the 54 equivalent sites, the space group will change from Im-3m (229) to Pm-3m (221), with 8 crystallographic orbits on 7 Wyckoff positions; thus, the default \texttt{KS2022} featurizer will process 8 sites. 
At the same time, several of these crystallographic orbits will be differentiated \emph{only} by the orientation and distance to the dilute (substitution) site, which \emph{does} affect ab initio calculation results (e.g., vacancy formation energy vs supercell size \cite{Hargather2022ANi}), but is \emph{guaranteed} to \emph{have no effect on the model's representation} because of the exact same neighborhood configuration (including angles and bond lengths) if conditions given earlier are met. Thus, it only requires adjustments to the site multiplicities or convolution implementation (simplified through, e.g., a Markov chain). In the particular dilute-BCC example at hand, depicted in Figure \ref{pysipfenn:fig:KS2022dilute}, there are 4 such \emph{representation-unique} crystallographic orbits, i.e., 1 with the dilute atom, 2 neighboring the dilute atom sharing either large hexagonal (1st nearest neighbor shell) or small square face (2nd nearest neighbor shell), and 1 non affected by the dilute atom which is equivalent to the remaining 4 orbits; thus reducing number of sites that need to be considered by a factor of 2.

The \texttt{KS2022\_dilute} featurization routine, schematically depicted in Figure \ref{pysipfenn:fig:KS2022dilute}, conveniently automates the above process for both simple cases like aforementioned substitution in pure element and complex cases like introducing a dilute atom at the 2a/2b orbit in $\sigma$-phase (green cell in Fig. \ref{pysipfenn:fig:ks2022}), by performing independent identification of crystallographic orbits in the dilute structure and base structure, followed by identification of the dilute site and its configuration to establish orbit equivalency under \texttt{pySIPFENN}'s \texttt{KS2022} representation, to finally reconstruct complete site ensemble of the dilute structure.

In the case of \texttt{KS2022\_dilute} implementation run on the dilute BCC supercell shown in Figure \ref{pysipfenn:fig:KS2022dilute}, the efficiency is improved nearly proportionally to the reduction in the number of considered sites, averaging $51$ms vs $98$ms \texttt{KS2022}, signifying 1.9 computational cost reduction relative to calculating all crystallographically unique sites. Or around 10-fold computational cost reduction relative to the standard \cite{Ward2017, Chen2019GraphCrystals, Jha2019IRNet, Krajewski2022ExtensibleNetworks, Choudhary2021AtomisticPredictions, Deng2023CHGNetModelling, Davariashtiyani2023FormationRepresentation, Schmidt2023Machine-Learning-AssistedMaterials} approach of processing all sites ($494$ms), while producing precisely the same results (within the numerical precision).

\section{Optimizations for Random Solid Solutions} \label{pysipfenn:sec:randomsolutions}

Sections \ref{pysipfenn:sec:ordered} and \ref{pysipfenn:sec:dilute} have demonstrated how recognition of symmetry in ordered structures can guarantee equivalency of sites and how understanding the character of featurization can further extend that notion of equivalency so that the ML representations of all sites can be obtained efficiently up to an order of magnitude faster. Random solid solutions are the conceptually opposite class of atomic structures, where the \emph{lack of} symmetry or site equivalency is \emph{guaranteed}, yet featurizing them requires one to solve the same problem of efficiently obtaining the ML representations of all sites present, which also happen to be infinite.

Typically, in the ab initio community, random solid solutions are represented using Special Quasirandom Structures (SQS) introduced in landmark 1990 work by \citet{Zunger1990SpecialStructures}, which are \emph{the} best structures to match neighborhood correlations in a purely random state given component ratios and number of atoms to use, hence the name \emph{special}. For many years, finding SQS structures required exponentially complex enumeration of all choices and was limited to simple cases until another critical work by \citet{VanDeWalle2013EfficientStructures}, which used simulated annealing Monte Carlo implemented through \texttt{ATAT} software to find these special cases much faster, exemplified through the relatively complex $\sigma$-phase and enabling the creation of SQS libraries used in thermodynamic modeling \cite{vandeWalle2017SoftwareData}.

However, the direct use of an SQS may not be the optimal choice for structure-informed random solid solution featurization due to several reasons. Firstly, as discussed by \citet{VanDeWalle2013EfficientStructures}, SQS can be expected to perform well on purely fundamental grounds for certain properties like total energy calculations, but one has to treat them with caution because different properties will depend differently on the correlation and selecting the SQS may be suboptimal. Building up on that, one could, for instance, imagine a property that depends strongly on the existence of low-frequency, high-correlation regions catalyzing a surface reaction or enabling nucleation of a dislocation. In terms of ML modeling, this notion is taken to the extreme with calculated features being both very diverse and numerous while being expected to be universal surrogates for such mechanistically complex properties.

Secondly, SQSs that can be generated in a reasonable time are limited in terms of the number of atoms considered, causing quantization of the composition. This is not an issue if a common grid of results is needed, e.g., to fit CALPHAD model parameters \cite{vandeWalle2017SoftwareData} or to train a single-purpose ML model \cite{Tandoc2023MiningAlloys}, but it becomes a critical issue if one needs to accept an arbitrary composition as the ML model and SQS would have to be obtained every time. This issue is further amplified by the rapidly growing field of compositionally complex materials (CCMs), which exist in vast many-component compositional spaces prohibiting SQS reuse even at coarse quantizations \cite{Krajewski2024Nimplex} while being a popular deployment target for both forward and inverse artificial intelligence methods \cite{Catal2023MachineProperties, Rao2022MachineDiscovery, Debnath2023ComparingAlloys} due to their inherent complexity.

Based on the above, it becomes clear that costly computing of an SQS structure would have to be done for every ML model, and it would not be consistent between chemistries and complexities. At the same time, the primary motivation for limiting the number of sites for ab initio calculations is gone since \texttt{KS2022} can featurize over 1,000 sites per second on a laptop (Apple M2 Max run in parallel). 

\begin{figure}[h]
    \centering
    \includegraphics[width=0.95\textwidth]{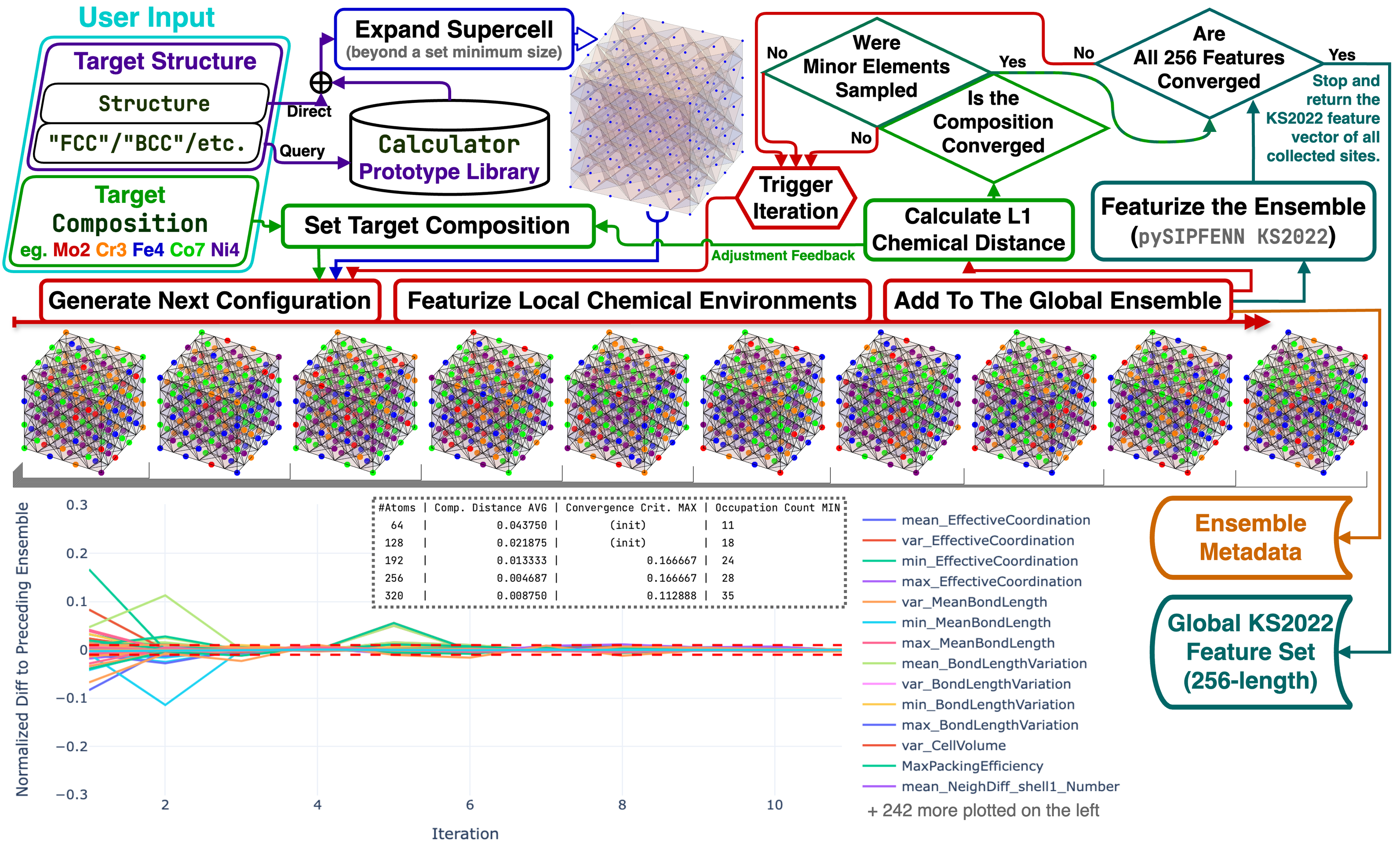}
    \caption{
    Core schematic of the \texttt{KS2022\_randomSolutions} featurizer. The target structure given explicitly or implicitly is expanded to form a (lattice) (i.e. template) supercell. It is then iteratively populated with target composition (slightly adjusted each time) and divided into individual sites, which are featurized (like in \texttt{KS2022}) and added to the global ensemble. The process repeats until the composition is converged, all species have had a chance to occur, and \emph{every individual feature} has converged. Lastly, the global \texttt{KS2022} feature vector and metadata are returned. FCC supercell is used as an example.
    }
    \label{pysipfenn:fig:KS2022randomSolution}
\end{figure}

Thus, the objective of optimization is shifted towards consistency in convergence to feature vector values at infinity. To accomplish that, \texttt{pySIPFENN} goes back to random sampling but at a large scale and \emph{individually monitoring the convergence of every feature} during the expansion procedure, implemented through \texttt{KS2022\_randomSolutions} and depicted in Figure \ref{pysipfenn:fig:KS2022randomSolution}, to ensure individual convergence.

Such a representation-centered approach can also efficiently account for (1) the dissimilarity of any set of chemical elements and (2) the neighbor weight during featurization, where some may be much more important than others (see highly-anisotropic $\sigma$-phase Wigner-Seitz cells in Figure \ref{pysipfenn:fig:ks2022}). It is also flexible in accepting any target structure, even a distorted one, since no assumptions are made about the neighborhood geometry.

At the same time, it is important to note that such an approach is not a replacement for SQS in a general sense. It is, instead, a complementary method, as it does not result in a defined approximation of random structure but its representation for machine learning.

\section{Summary and Conclusions} \label{pysipfenn:sec:summaryconclusions}

\begin{itemize}
    \item \texttt{pySIPFENN} or \textit{python toolset for Structure-Informed Property and Feature Engineering with Neural Networks} is a free open-source software (FOSS) modular framework extending authors' past work \cite{Krajewski2022ExtensibleNetworks} described in Section \ref{pysipfenn:sec:Introduction} by including many key improvements in the structure-informed featurization, machine learning model deployment, different types of transfer learning (connected to OPTIMADE API \cite{Evans2024DevelopmentsExchange}), rewrite of key literature tools (e.g., \texttt{Ward2017} Java-based featurizer \cite{Ward2017}) into Python+\texttt{NumPy} \cite{Harris2020ArrayNumPy}, and optimizations of past feature set as described in Sections \ref{pysipfenn:ssec:coreimprovements}, \ref{pysipfenn:ssec:Ward2017Translation}, and \ref{pysipfenn:ssec:ks2022features}.
    
    \item \texttt{pySIPFENN} framework is uniquely built from tightly integrated yet highly independent modules to allow easy use of essential functions without limiting advanced researchers from taking specific components they need, like a specific featurizer, and simply copying it into their software, reducing dependencies to the minimum (including \texttt{pySIPFENN} itself).
    
    \item Section \ref{pysipfenn:sec:ordered} discusses how featurization of atomic structures (or configurations) to construct vector, voxel, graph, graphlet, and other representations is typically performed inefficiently because of redundant calculations and how their efficiency could be improved by considering fundamentals of crystallographic (orbits) equivalency to increase throughout of literature machine learning model, typically between 2 to 10 times. Critically, this optimization applies to $98.75\%$ of 4.4 million stored in \texttt{MPDD} \cite{Krajewski2021MPDD:Database}, which includes both DFT-based \cite{Saal2013MaterialsOQMD, Kirklin2015TheEnergies, Shen2022ReflectionsOQMD, Curtarolo2013AFLOW:Discovery, Toher2018TheDiscovery, Jain2013Commentary:Innovation, Choudhary2020TheDesign, Merchant2023ScalingDiscovery} and experimental \cite{Grazulis2009CrystallographyStructures, Grazulis2012CrystallographyCollaboration, Grazulis2019CrystallographyPerspectives} data, showing massive impact if deployed. \texttt{KS2022} featurizer implements these advances in \texttt{pySIPFENN} using \texttt{spglib} \cite{Togo2018Spglib:Search} and \texttt{Voro++} \cite{Rycroft2007MultiscaleFlow, Rycroft2009Voro++:C++, Lu2023AnCells}, while retaining ability to process arbitrary structures.

    \item Section \ref{pysipfenn:sec:dilute} explores how symmetry is broken in dilute, doped, and defect structures, to then discuss site equivalency under different representations and how this notion can be used to improve efficiency by skipping redundant calculations of sites which are not guaranteed to be equivalent based on crystallographic symmetry alone but need to be contrasted with defect-free representation. \texttt{KS2022\_dilute} featurizer implements these advances in \texttt{pySIPFENN}.

    \item Section \ref{pysipfenn:sec:randomsolutions} discusses featurization of perfectly random configuration of atoms occupying an arbitrary atomic structure and, for the first time, considers fundamental challenges with using SQS approach in the context of forward and inverse machine learning model deployment by extending past discussion on SQS limitations given by \citet{VanDeWalle2013EfficientStructures}, which do not typically appear in ab initio and thermodynamic studies. \texttt{KS2022\_randomSolutions} featurizer 
    has been developed to efficiently featurize solid solutions of any compositional complexity by expanding the local chemical environments (LCEs) ensemble until standardized convergence criteria are met.

    \item As described in Section \ref{pysipfenn:sec:softwareavaialbility}, software introduced in this work is continuously tested, well documented, regularly maintained, and 

    \item Throughout this work, the authors explicitly discuss how advances in featurization efficiency described in this work can be applied to different kinds of similar tools in the community, including those using voxel, graph, or graphlet representations.
    
\end{itemize}

\section{Software Availability and Accessibility} \label{pysipfenn:sec:softwareavaialbility}

\texttt{pySIPFENN} or \textit{python toolset for Structure-Informed Property and Feature Engineering with Neural Networks} is an easily extensible free, open-source software (FOSS) under \href{https://opensource.org/license/lgpl-3-0}{OSI-approved LGPL-3.0 license}, available as (1) source code hosted in a \texttt{GitHub} repository (\href{https://git.pysipfenn.org/}{git.pysipfenn.org}), (2) a python package through \href{https://pypi.org/project/pysipfenn/}{\texttt{PyPI} index}, and (3) a conda package hosted through \href{https://anaconda.org/conda-forge/pysipfenn}{\texttt{conda-forge} channel}.

It is very well-documented through (1) API reference, (2) detailed changelog, (3) install instructions, (4) tutorials and task-specific notes, and (5) FAQ, compiled for development (\href{https://pysipfenn.org/en/latest/}{pysipfenn.org/en/latest}), stable (\href{https://pysipfenn.org/en/stable/}{pysipfenn.org/en/stable}), and past (e.g., \href{https://pysipfenn.org/en/v0.12.0/}{pysipfenn.org/en/v0.12.0}) versions.

\texttt{pySIPFENN} has been built from the ground up to be a reliable user tool. It is automatically tested across a range of platforms (Linux / Windows / Mac (Intel) / Mac (M1)) and Python versions on every change, as well as on a weekly schedule.

It has been actively disseminated to its target audience through two large workshops organized with support from the Materials Genome Foundation (MGF / \href{https://materialsgenomefoundation.org}{materialsgenomefoundation.org}). The first one, covering \texttt{v0.10.3} and held online on March 2nd 2023, had over 300 users registered and over 100 following all exercises. It has been recorded and published on MGF's YouTube channel \cite{Krajewski20232023YouTube}. The second one, using \texttt{v0.12.1}, was held in-person on June 25th 2023 at the \href{https://calphad.org/calphad-2023}{CALPHAD 2023 conference} in Boston, as a part of Materials Genome Toolkit Workshops, covering its integration with \texttt{ESPEI} \cite{Bocklund2019ESPEICuMg} and \texttt{pycalphad} \cite{Otis2017Pycalphad:Python}. In November 2023, it was also employed in a pair of workshop-style graduate-level guest lectures introducing materials informatics (\href{https://amkrajewski.github.io/MatSE580GuestLectures/}{amkrajewski.github.io/MatSE580GuestLectures}), which can be used as an advanced tutorial.


\chapter{Handling Millions of Atomic Structures} \label{chap:mpdd}

\acknowledge{
This chapter adapts parts of a manuscript draft planned for publication before dissertation submission, co-authored with Ricardo Amaral, Jonathan W. Siegel, and Zi-Kui Liu. All of included text was written by Adam M. Krajewski. Described software has been developed by Adam M. Krajewski since 2020 with assistance from Jonathan W. Siegel and Ricardo Amaral. Zi-Kui Liu provided edits and guidance. It also adapts excerpt written by Adam M. Krajewski for \citet{Evans2024DevelopmentsExchange} reproduced from Digital Discovery journal under CC BY 3.0 license. 
}

\section{Introduction} \label{mpdd:sec:background}

Traditionally, the field of materials science deals with highly complex data, in terms of both input and output descriptions, acquired through laborious experiments in the laboratories, such as mechanical or electrochemical tests performed according to professional standards or expensive computations on supercomputer clusters, including ones based on the density functional theory (DFT)~\cite{Kohn1996DensityStructure} or finite element method (FEM)~\cite{Liu2022EightyFuture}. Because of this high-cost aspect, the number of data points being generated is usually very limited, especially after raw data is fitted into models and disseminated through reports and scientific publications as typeset tables, text, or figures, which are then carefully interpreted by other researchers.

In the last few decades, however, the community has become increasingly engaged in large-scale efforts to construct extensive collections of data points, especially after endorsement and special funding was provided by the Materials Genome Initiative in 2011 \cite{SubcommitteeontheMaterialsGenomeInitiative2021MaterialsPlan, Agren2023CALPHADAnniversary, Olson2023GenomicDynamics}. It significantly accelerated many efforts to combine materials-specific scientific literature sources into a homogeneous structure were attempted, with perhaps the most successful being the \texttt{Pauling File} developed since 1995 with as of 2024, was built with 1000 years of full-time academic effort \cite{Blokhin2018TheGenome} and underlies several databases, including ASM Phase Diagram Database \cite{ASMInternational} and Materials Platform for Data Science (MPDS) \cite{Blokhin2018TheGenome}. Furthermore, several computational databases listed later in Section~\ref{mpdd:ssec:dataset} were also created and, in some cases, grew beyond one million entries.

With the rise of such large-scale databases, it becomes critical to be able to efficiently operate on them and utilize them, as even seemingly fast calculations of $1s$ grow to over 8 weeks when performed 5 million times. Furthermore, this scale is likely to sharply increase in the near future due to a large volume of data coming from machine learning (ML) studies, which, even if filtered out, will likely accelerate search efforts using more traditional computation by constantly providing guidance.

\section{The Material-Property-Descriptor Database} \label{mpdd:sec:mpdd}

\subsection{Motivation} \label{mpdd:ssec:motivation}

The Material-Property-Descriptor Database (MPDD) is an extensive (4.5M+) database of \emph{ab initio} relaxations of 3D crystal structures,  combined with an infrastructure of tools allowing efficient descriptor calculation (featurization), as well as the deployment of ML models like \texttt{SIPFENN} \cite{Krajewski2024EfficientStructures} described in Chapter \ref{chap:sipfenn}, and other developed by the community including \texttt{ALIGNN} \cite{Choudhary2021AtomisticPredictions} and \texttt{CHGNet} \cite{Deng2023CHGNetModelling}.

The most critical motivation behind MPDD is the retention of intermediate modeling data (atomistic features), including structure-informed descriptors described in Chapter \ref{chap:pysipfenn}, which typically cost orders of magnitude more computational time than any of the other steps performed during ML model deployment~\cite{Krajewski2022ExtensibleNetworks}. Thus, many ML models can be run at a small fraction of the original cost if the same descriptor (or, more commonly, a subset chosen through feature selection) is used. This benefit applies regardless of whether a model is just another iteration, e.g., fine-tuned to a specific class of materials like perovskites, or an entirely new model for a different property. Thanks to this, machine learning researchers can effortlessly take advantage of MPDD to deploy numerous ML models directly to the community without needing to construct individual deployment targets, which are typically both much smaller and redundant relative to existing datasets.

Furthermore, MPDD's access to stored atomic structures and associated metadata has been shown to be useful, for instance, in the fully data-driven prediction of atomic structures (validated with DFT and experiments). It, for instance allowed quick identification of unknown structures in Nd-Bi~\cite{Im2022ThermodynamicModeling} and Al-Fe~\cite{Shang2021FormingJoints} systems, what is discussed in more detail in Chapter \ref{chap:crystall}.

\subsection{Dataset} \label{mpdd:ssec:dataset}

\begin{figure}[H]
    \centering
    \includegraphics[width=0.4\textwidth]{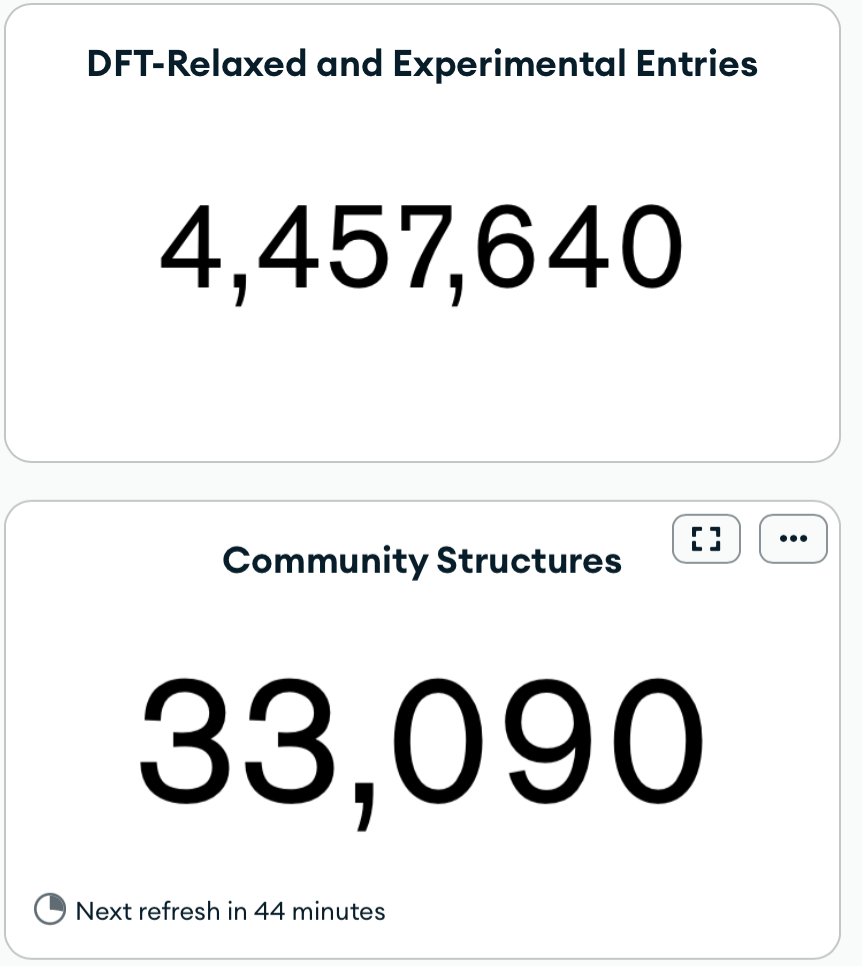}
    \includegraphics[width=0.4\textwidth]{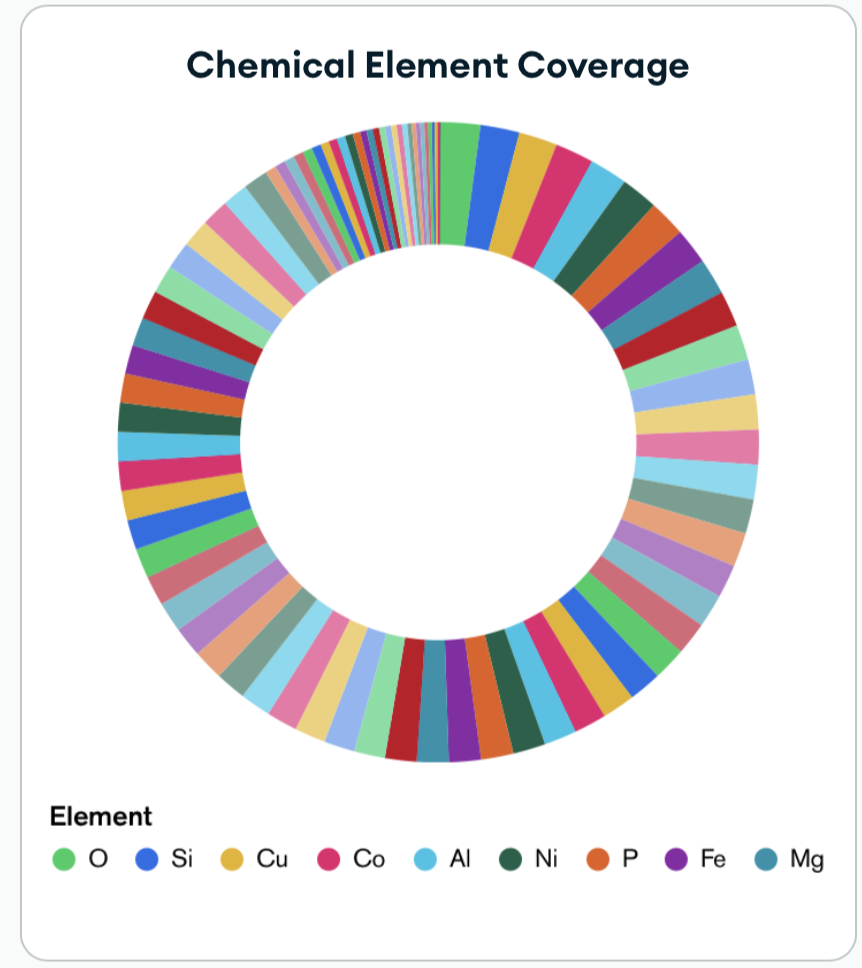}
    \caption{Key statistics over MPDD dataset as of April 2024 demonstrating (1) extent of the dataset and (2) high diversity of chemical space coverage over all elements. Figure~\ref{sup:mpdd:dashboard} presents a snapshot of the full dashboard.}
    \label{mpdd:fig:dataset}
\end{figure}

To act as the starting point for both (a) the generative design of new data points and (b) the deployment target for forward methods, an extensive collection of atomic structures has been collected and homogenized into the \texttt{MPDD}. It contained \textit{DFT-based datasets} including OQMD \cite{Saal2013MaterialsOQMD, Kirklin2015TheEnergies, Shen2022ReflectionsOQMD}, AFLOW \cite{Curtarolo2013AFLOW:Discovery, Toher2018TheDiscovery}, Materials Project \cite{Jain2013Commentary:Innovation}, NIST-JARVIS \cite{Choudhary2020TheDesign}, Alexandria \cite{Schmidt2022AFunctionals}, CAMD \cite{Ye2022NovelAgents}, GNoME \cite{Merchant2023ScalingDiscovery}, \textit{in-house DFT-based datasets} including phonon calculations created with DFTTK \cite{Wang2021DFTTK:Calculations}, as well as hundreds of thousands of experimentally observed entries from Crystallography Open Database (COD) \cite{Grazulis2009CrystallographyStructures, Grazulis2012CrystallographyCollaboration, Grazulis2019CrystallographyPerspectives}. As shown in Figure~\ref{mpdd:fig:dataset}, the collected dataset does not exhibit a bias towards any particular chemical element, but rather, it covers tens of different elements almost uniformly.

Figure~\ref{mpdd:fig:dataset} is a small subset of the larger dashboard available at \href{https://phaseslab.org/mpdd}{phaseslab.org/mpdd} web page, which is presented in Figure~\ref{sup:mpdd:dashboard} in Appendix Section~\ref{mpdd:app1}, which can be consulted for analysis of the dataset in terms of different levels of "coarse-graining" of the uniqueness criteria, as well as coverage of chemical systems of different order.

Notably, motivated by striving for completeness, a substantial effort was made to cover \emph{all 118 elements} by specifically collecting a unique collection of recent experimental observations and DFT-based lattice stability calculations of many uncommon elements from several sources. These included Einsteinium (Es), Californium (Cf), Curium (Cm), Berkelium (Bk) based on \citet{2006TheElements}, Lawrencium (Lr), Rutherfordium (Rf), Dubnium (Db), Seaborgium (Sg), Bohrium (Bh), Hassium (Hs), Meitnerium (Mt), Darmstadtium (Ds), Roentgenium (Rg) based on \citet{Gyanchandani2011PhysicalMetals}, Copernicium (Cn) based on \citet{Atta-Fynn2015DensityElements}, Flerovium (Fl) based on \citet{MaizHadjAhmed2017RevisitingFlerovium}, Nihonium (Nh) based on \citet{Atarah2020FirstNihonium}, Copernicium (Cn), Flerovium (Fl), Livermorium (Lv), Moscovium (Mc), Nihonium (Nh), Oganesson (Og), Tennessine (Ts) based on \citet{Trombach2019ExploringTheory}, Francium (Fr) based on \citet{Koufos2013ElectronicFrancium}, and Astatine (At) based on \citet{Hermann2013CondensedMetallic}, to cover 115 elements. The last 3 missing elements -  Fermium (Fm), Mendelevium (Md), and Nobelium (No) were not yet experimentally measured nor calculated with DFT-based methods, best to the author's knowledge; however, they were approximated based on the rationale in \cite{2006TheElements} combining assumption of crystal structure based on electronic structure with predicted atomic radius.

\subsection{Infrastructure} \label{mpdd:ssec:infrastructure}

\begin{figure}[H]
    \centering
    \includegraphics[width=0.4\textwidth]{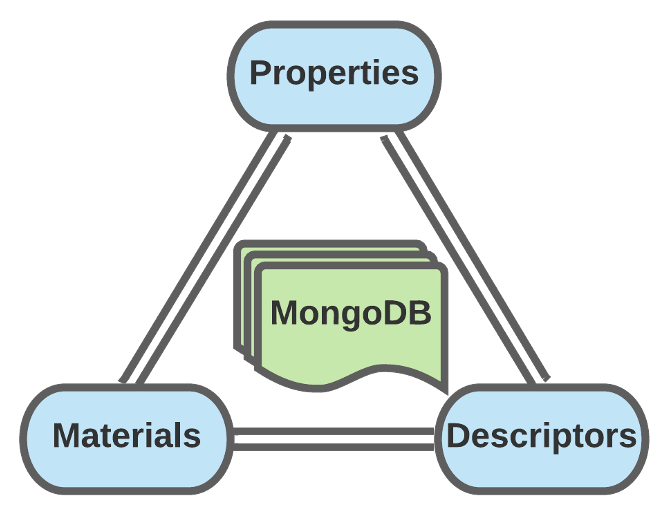}
    \caption{Three entities at the core of MPDD are treated as "first-class citizens" interacting with each other. Going counter-clockwise \emph{Materials} cover our past sampling of the problem domain, \emph{Descriptors} cover our understanding of it, and \emph{Properties} determine utility. Going clockwise desired \emph{Properties} guide analysis leading to understanding encoded in \emph{Descriptors}, which inform us of unexplored regions of problem domain in their individual contexts.}
    \label{mpdd:fig:core}
\end{figure}

The designed infrastructure of \texttt{MPDD}, built on top of a \texttt{MongoDB} database framework, is focused on independent methods operating on each datapoint composed of (1) a definition of material, (2) a descriptor set, and (3) a property set, in a way such that one part of the data is augmented by another, as depicted in Figure~\ref{mpdd:fig:core}. This is typically accomplished by using special \texttt{MongoDB} indexes, which enable to effectively present the data to a given method as a collection of either of the three, e.g., to select all materials fulfilling requirements) or lack thereof, e.g., to operate on a small subset of incomplete newly inserted data for which a task needs to be run.

\begin{figure}[H]
    \centering
    \includegraphics[width=0.9\textwidth]{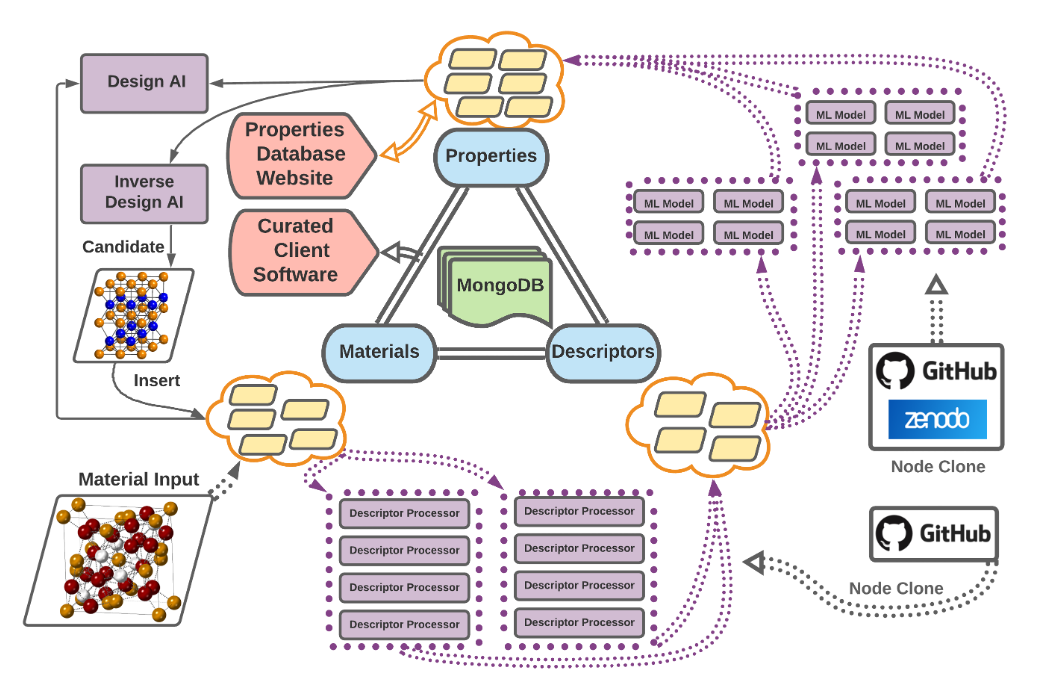}
    \caption{Main schematic of the MPDD Database infrastructure.}
    \label{mpdd:fig:schematic}
\end{figure}

When filled with a large dataset described in Section~\ref{mpdd:ssec:dataset} and combined with generative models extending it (see Chapter~\ref{chap:crystall}), such ecosystem becomes cyclic in nature and thanks to built-in automations can continuously grow, as shown schematically in Figure~\ref{mpdd:fig:schematic}. At the same time, thanks to its decentralized nature, it can be highly parallelized across many devices, ranging from a slow single-board computer to scalable high performance computing (HPC) cluster allocations, enabling handling of variable workloads.

\section{Open Databases Integration for Materials Design (OPTIMADE) API of MPDD} \label{mpdd:sec:optimade}

\texttt{OPTIMADE} or the Open Databases Integration for Materials Design consortium has been established to make materials databases interoperable by developing a specification for a ubiquitous REST API \cite{Evans2024DevelopmentsExchange}. That way, databases can remain independently maintained and tuned to specific needs, like ab initio data or ML data, while at the same time reporting on the contained knowledge so that redundant calculations are not performed, and the efficiency of the materials informatics community at large is dramatically improved.

MPDD has a stable \texttt{OPTIMADE} API that serves the entire core MPDD dataset, fully implementing \texttt{v1.1.0} of the OPTIMADE standard, as of April 2024, through a cloud server based on \texttt{optimade-python-tools} \cite{Evans2021}.
Making the MPDD available via OPTIMADE was initially challenging, as MPDD stores and exchanges data in a way that prioritises high throughput and low storage requirements, including binary data, making it difficult or slow to make MPDD queryable as an OPTIMADE API on-the-fly.
However, issues have been resolved by establishing a self-updating mirror of the dataset where structures are made OPTIMADE-compliant during transfer, which can occur within the same virtual machine or other integrated computing environment. Most of the MPDD-specific data is available under the \texttt{mpdd} namespace of \texttt{OPTIMADE}, including dictionaries of metadata (e.g., \texttt{\_mpdd\_atomicvolume}), properties (e.g., \texttt{\_mpdd\_formationenergy\_sipfenn\_krajewski2020\_lightmodel}), and descriptors (e.g., \texttt{\_mpdd\_descriptors.KS2022}), described earlier in this chapter.

The base URL is available at:

\hspace{24pt} \href{https://optimade.mpdd.org}{https://optimade.mpdd.org}

and one can see a sample dataset response by following the \texttt{structures} endpoint at:

\hspace{24pt} \href{https://optimade.mpdd.org/v1/structures}{https://optimade.mpdd.org/v1/structures}

Further discussion on MPDD's \texttt{OPTIMADE} API is given in Appendix Section~\ref{mpdd:app1}, with Figures~\ref{sup:mpdd:optimade} and \ref{sup:mpdd:optimadeout} showing outputs of, respectively, base URL and \texttt{structures} endpoint being queried.

\section{MPDD-eXchange} \label{mpdd:sec:mpddx}

In addition to the \texttt{OPTIMADE} API serving as the primary endpoint for the end-users, \texttt{MPDD} has recently been extended through an experimental platform available at \href{https://contrib.mpdd.org}{contrib.mpdd.org} based on GitHub cloud-based automations to facilitate data exchange; thus, called \texttt{MPDD-eXchange} or \texttt{MPDD-X}. The central exchange happens when the user uploads their data, which gets validated and ingested into the data ecosystem, enriching the \texttt{MPDD}, while the user gets presented with machine learning predictions associated with it and gets persistent credit associated with their account. 

\begin{figure}[H]
    \centering
    \includegraphics[width=0.85\textwidth]{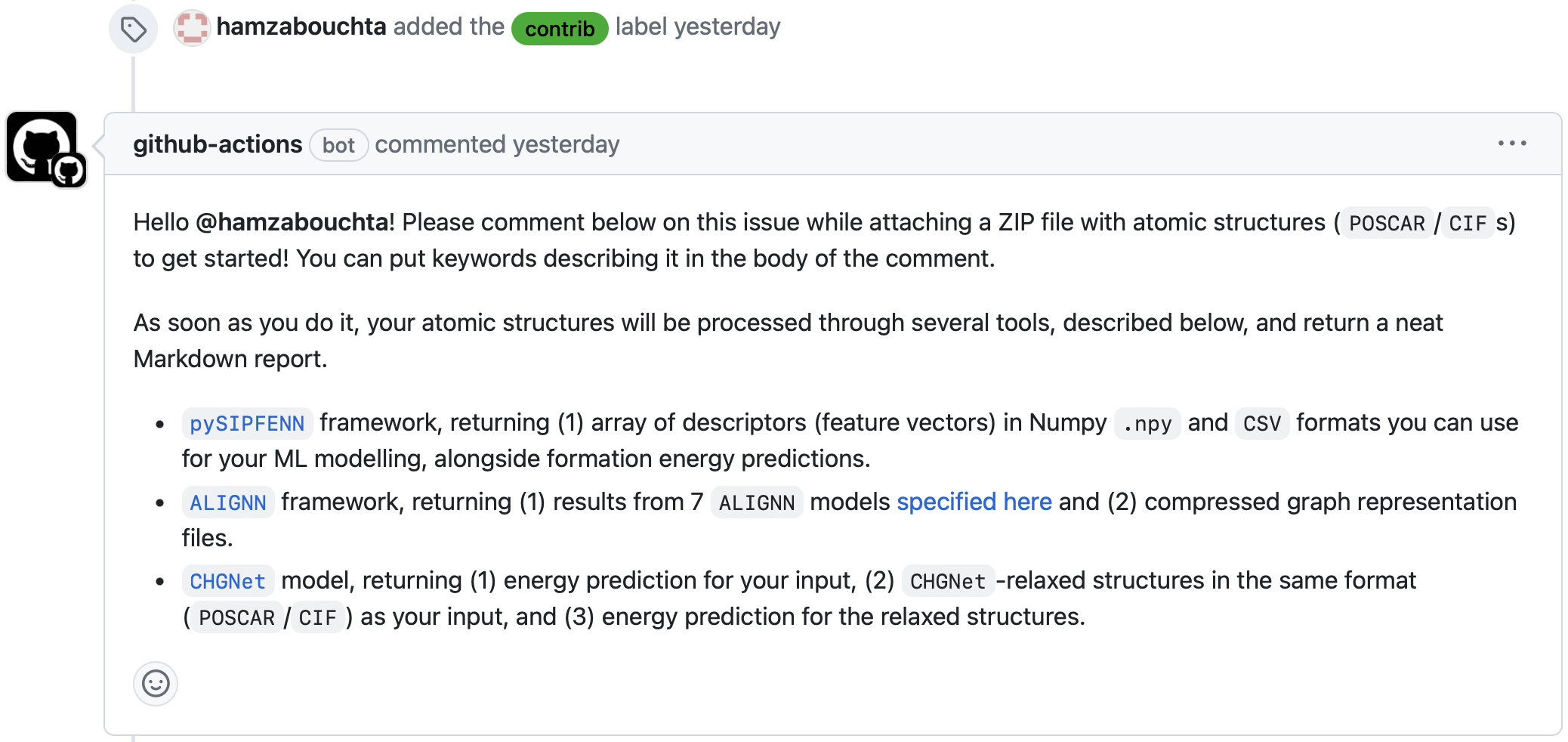}
    \caption{Printout of the greeting message on issues opened with auto-assigned \texttt{contrib} label instructing user how to send a contribution and what models will be run.}
    \label{mpdd:fig:mpddx1}
\end{figure}

From the user perspective, all work is being performed entirely within GitHub Issues of the \texttt{MPDD-X} repository, enabling anyone with a free account to start it, as shown in Figure~\ref{mpdd:fig:mpddx1}. 

\begin{figure}[H]
    \centering
    \includegraphics[width=0.85\textwidth]{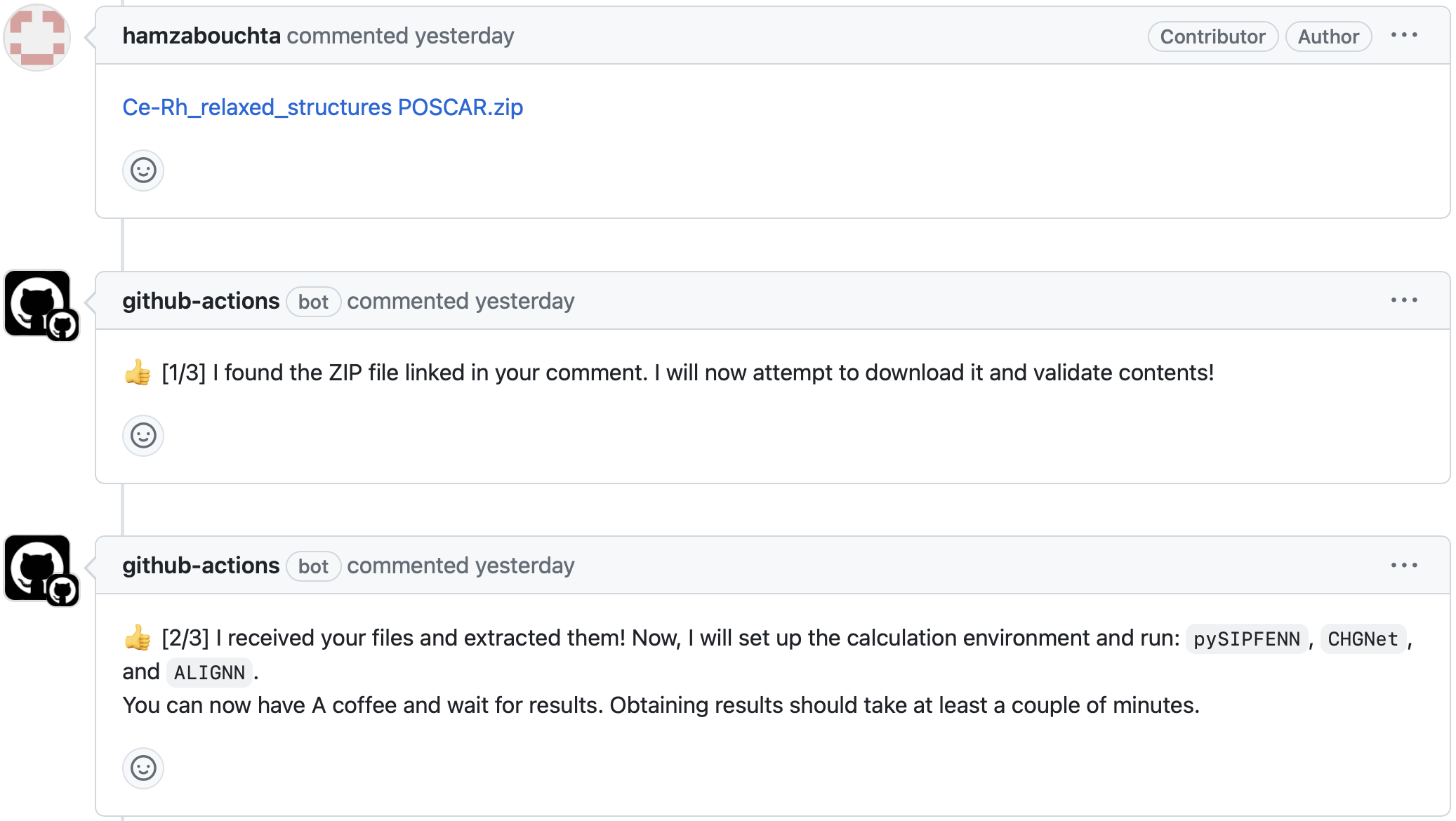}
    \caption{Printout of the intermediate messages informing that validation checks are passing, if the user sent a properly formatted CIF or POSCAR files in a ZIP file. Otherwise (not depicted) messages would provide feedback on errors.}
    \label{mpdd:fig:mpddx2}
\end{figure}

Once the user opens a GitHub \texttt{Issue}, they only need to provide a \texttt{ZIP} file containing either \texttt{CIF} \cite{Hall1991TheCrystallography} or \texttt{POSCAR} \cite{VASPPOSCAR} files, commonly used in the community, what can be done by simply dragging a file into the comment. Momentarily after that, (1) the ZIP file is posted, and (2) an automated action processes it, informing the user of the progress, as shown in the example in Figure~\ref{mpdd:fig:mpddx2}.

\begin{figure}[H]
    \centering
    \includegraphics[width=0.85\textwidth]{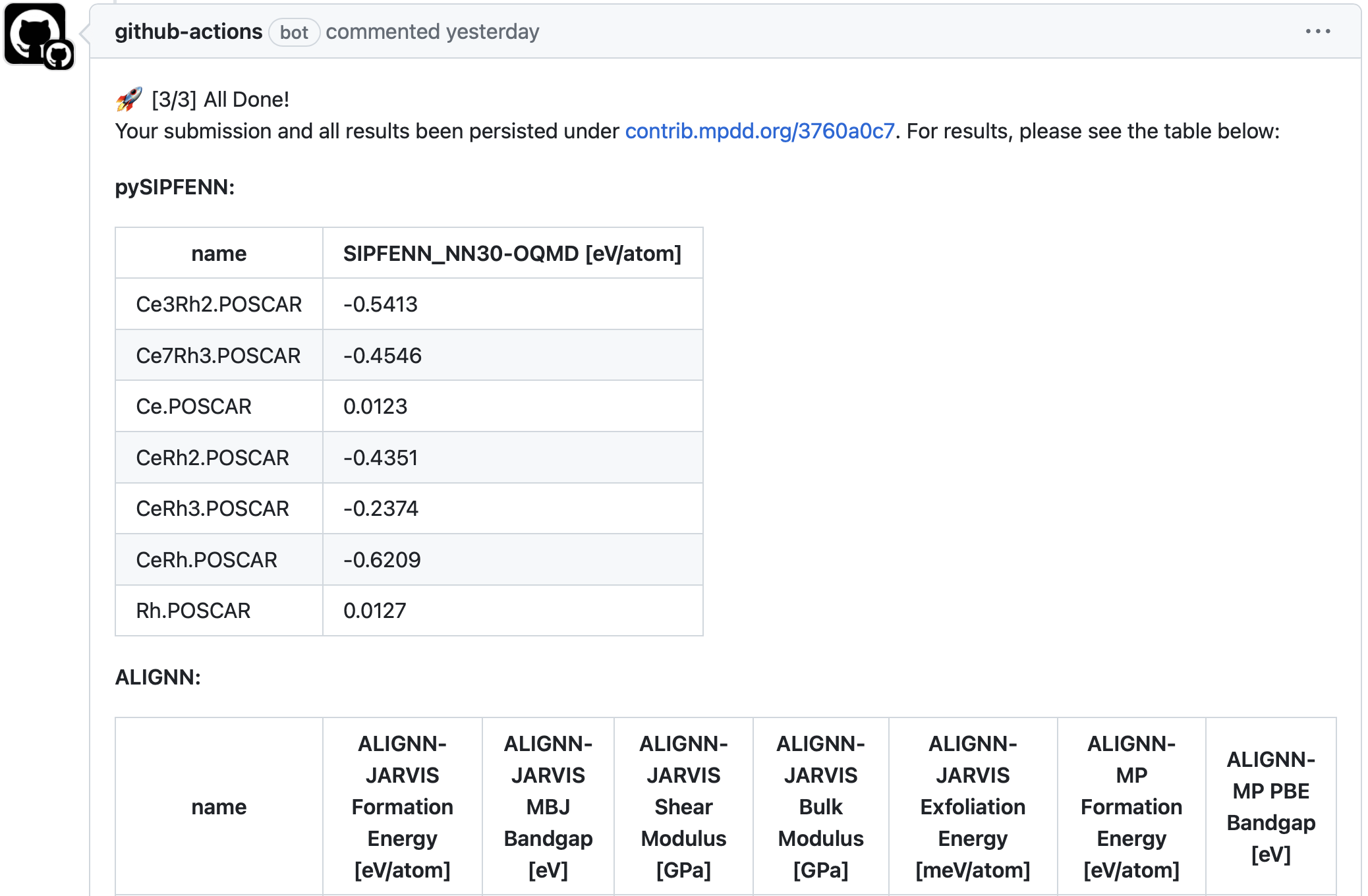}
    \caption{Printout of final message after all computation is successfully completed. User is presented with (1) outputs of the ML models deployed on the data and (2) a unique contribution ID based on the commit hash, which can be cited as \texttt{contrib.mpdd.org/3760a0c7} or \texttt{mat-x.org/mpdd-3760a0c7} and points to persisted data record appended with ML results and calculation metadata.}
    \label{mpdd:fig:mpddx3}
\end{figure}

Next, typically after 5 to 20 minutes, depending on the task complexity and server load, the user gets presented with, as shown in Figure~\ref{mpdd:fig:mpddx3}, (1) a result table with outputs of several literature models, as well as, (2) a persistent identifier which can be used to reference both the results and details of how they were obtained.

\begin{figure}[H]
    \centering
    \includegraphics[width=0.85\textwidth]{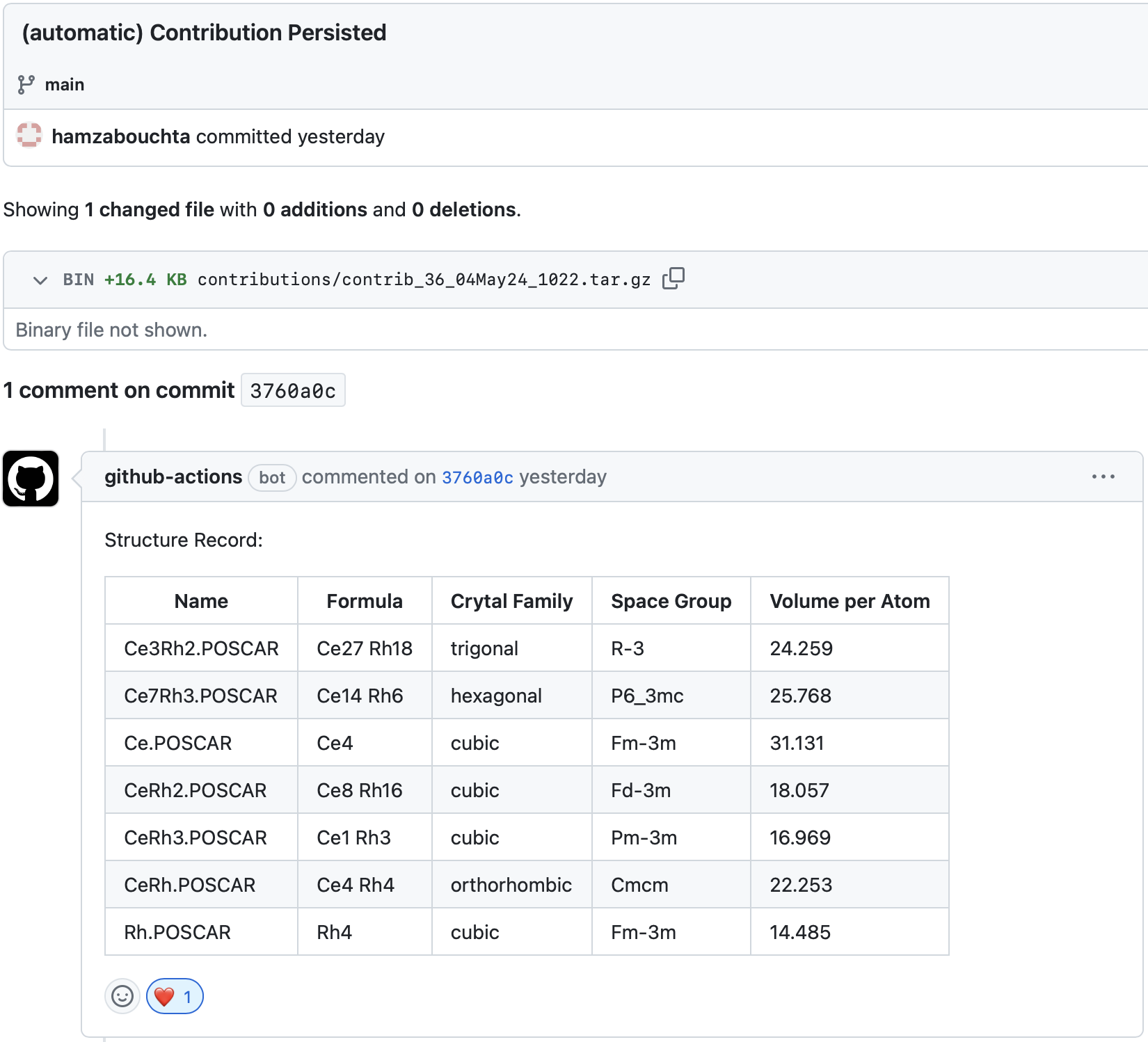}
    \caption{Printout of contribution record stored within git repository as auto-generated commit. Right after a successful commit, the system automatically generates a comment on it describing the structures included.}
    \label{mpdd:fig:mpddx4}
\end{figure}

Internally, the contribution is persisted as a commit credited to the submitting user rather than the computation bot, as shown in Figure~\ref{mpdd:fig:mpddx4}, which consists of a highly-compressed record of the task and its results.

Finally, at the end of a successful run, the bot comments on the contribution commit with some additional metadata, shown in Figure~\ref{mpdd:fig:mpddx4}, which makes it easier to recognize a particular contribution without its identifier.

\section{Software Availability} \label{mpdd:sec:softwareavaialbility}

\texttt{MPDD} is built as a free community platform which gets progressively open-sourced under MIT and LGPL-3.0 licenses, as new components are being finalized and published. As described in Section~\ref{mpdd:sec:optimade}, all data is available to fetch though the state-of-the-art domain-specific API (\texttt{OPTIMADE} \cite{Evans2024DevelopmentsExchange}) at \href{https://optimade.mpdd.org}{optimade.mpdd.org}, and an experimental input pipeline can be used by anyone with a GitHub account through \href{https://contrib.mpdd.org}{contrib.mpdd.org}.


\chapter{Ergodic Ensemble Approach to the Material Discovery} \label{chap:crystall}

\acknowledge{
This chapter adapts parts of a manuscript draft planned for publication around the dissertation submission, co-authored with Ricardo Amaral, Jonathan W. Siegel, Rushi Gong, ShunLi Shang, and Zi-Kui Liu. All of included text was written by Adam M. Krajewski. Described software has been developed by Adam M. Krajewski since 2020 and by Ricardo Amaral since 2024, with Rushi Gong and ShunLi Shang developing it through testing-feedback loop. ShunLi Shang and Zi-Kui Liu provided edits and guidance. It also adapts excerpt written by Adam M. Krajewski for \citet{Im2022ThermodynamicModeling}. 
}

\section{New Stable Structure Prediction?} \label{crystall:sec:newstable}

The ability to determine atomic structure of existing or hypothetical solids formed by chemical elements is one of the fundamental challanges in theoritical crystallography and in materials science as a whole. For the last several decades \cite{Hawthorne1994}, the two remaining challenges are predictions of (a) compositions forming compounds of interest and (b) candidate topologies or approximate atomic arrangements they can form. Given these, the exact structure can be solved experimentally, by matching X-ray diffraction profile to candidates using the method by \citet{LeBail1988} and interpreting the profile, or using deep learning \cite{Yue2024}. Similar analysis can be done, e.g., on the Raman spectrum \cite{Ferrari2013}. For hypothetical compounds in their pure states, the exact structure can also be solved using ab initio quantum mechanical methods, such as Kohn-Sham density functional theory (DFT) \cite{Kohn1965}, which are computationally expensive but can provide accurate results and continue to improve \cite{Kothakonda2023}.

The latter problem of proposing the topologies has been traditionally approached by finding geometries satisfying constraints, like the principles systematized by [@Pauling1929], which included then-novel concepts like coördination. Over decades, this has evolved into automated computational tools like the \texttt{GRINSP} by \citet{LeBail2005} which can propose such structures. More recently, increasing computational power shifted focus towards testing many options, with some of the methods starting from random arrangements and collapsing them into local minima, like \texttt{AIRSS} by \citet{Pickard2011}, or by evolutionary techniques, like \texttt{USPEX} \cite{Glass2006}. 

Several efforts have been made to tackle both topological and compositional challenges simultaneously, such as high-throughput construction of the \texttt{AFLOW} \cite{Curtarolo2012} DFT Database \cite{Toher2018} (and others, described in Chapter \ref{chap:mpdd}) built by systematically populating the earlier, yet continuously growing, library of prototype structures \cite{Mehl2016} with new chemistries based on the expert knowledge. However, even for a limited set of elements, this problem becomes combinatorically challenging \cite{Krajewski2024Nimplex}. This prompted recent efforts into (1) systematic similarity-driven prediction of substitutions which would result in finding new low-energy structures \cite{Wang2021}; (2) into brute-force substitutions filtered by machine learning (ML) models \cite{Schmidt2023}; and (3) the combination thereof in \citet{Ye2022} and \citet{Merchant2023}, leading to rapid discovery of new structures stable at 0K temperature and 0Pa pressure against competing arrangements.

While finding new candidates for zero-temperature zero-pressure hypothetical structures certainly opens the world for new discoveries, it warrants some key questions. Firstly, *"To what degree have we biased the discovery of new structures?"* through many iterations of biasing the search to our expectations, which is philosophically analogous to how retraining Large Language Models (LLMs) and other ML models on generated data makes them underestimate improbable events causing, so called, *Model Collapse* \cite{Shumailov2023}, and drift from the distribution of the real data \cite{Alemohammad2023}. In the context of materials, the latter is depicted well by Fig. 5 in \citet{Ye2022} showing relatively small overlap between experimentally known structures and new ones predicted to be stable.

In case the data becomes biased towards low-energy solutions, one needs to ask \textit{"Do we need the diverse energy landscapes to discover new materials?"} to which the answer is a multifaceted yes. First, the metastable are critical to fitting thermodynamic models as protodata, as highlighted in study by \citet{Olson2023}, which also stresses that at the same time, computational thermodynamics models \emph{phases} which, at finite temperatures, represent ensembles of underlying configurations and structures. 

Furthermore, one has to consider that going beyond highly-stable materials is needed to create realistic model, as concisely explained by \citet{Zunger2019}, because of the several complexities involved in the synthesis of new materials, such as defect modeling. 

Once such realistic models are obtained, coexistance of many configurations can be used to accurately (1) take properties to extremes, e.g. infinitely negative thermal expansion modeled by \citet{Liu2022} to match the experiments without arbitrary fitting parameters, or (2) model superconductors \cite{Ferreira2023}. Thus, efforts to generate ensembles of possible structure candidates need to propose diverse polymorph populations, such as one in Fig. 6 in \citet{Pickard2011} generateed through \texttt{AIRSS}.

\section{\texttt{crystALL} - Purely Data-Driven Structure Prediction for Unidentified Chemical Compounds} \label{crystall:sec:crystall}

\begin{figure}[H]
    \centering
    \includegraphics[width=0.75\textwidth]{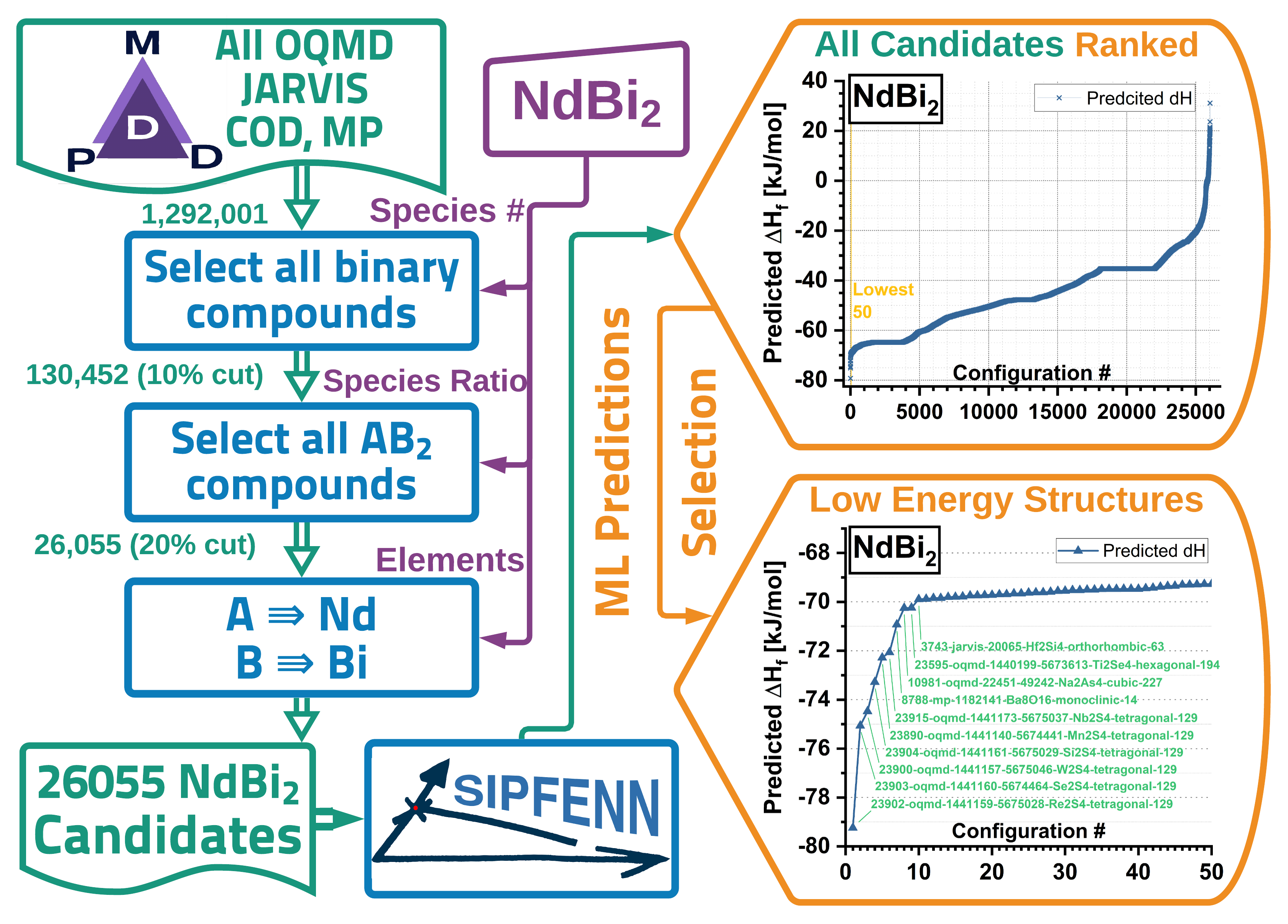}
    \caption{Simplified \texttt{crystALL} core schematic of operation based around performing all permutations of elemental substitutions and energy predictions, exemplified in the case of $NdBi_2$ intermetallic.}
    \label{crystall:fig:crystallcompound}
\end{figure}

At its core, \texttt{crystALL} leverages large local and remote collections of atomic structures, such as \texttt{MPDD} described in Chapter~\ref{chap:mpdd}, to propose new candidates based on matching stoichiometry, i.e., the anonymized formula, as shown in Figure~\ref{crystall:fig:crystallcompound}, while generating all possible permutations of such substitutions (e.g., 6 for $ABCD_4$ compounds). These candidates are then passed into the \texttt{pySIPFENN} \citet{Krajewski2024EfficientStructures}, featurized using \texttt{KS2022} described in Section~\ref{pysipfenn:ssec:ks2022features}, analyzed for uniqueness, and then passed to next stages of filtering.

During the filtering step, selection of candidates is primarily based on the clustering analysis inside the \texttt{KS2022} feature space, with lowest-predicted-energy member of each cluster being selected and \emph{assumed} to be a metastable local minimum of structures belonging to a configuration. Such set of configurations is then passed into DFT-based methods for validation, in the order of predicted stability, until it appears \emph{ergodic} or a structure of interest has been identified. Section~\ref{crystall:ndbi2} discusses all of the above steps on a specific example of \ch{NdBi_2}.

\section{Successfully Identifying \ch{NdBi_2} Structure} \label{crystall:ndbi2}

The first deployment of the \texttt{crystALL} method happened within \citet{Im2022ThermodynamicModeling} work, which re-assessed thermodynamic models of Nd-Bi chemical system, which was an essential step towards the rational design of rare-earth alloys for clean energy technologies through, e.g., electrochemical recovery of rare-earth elements and prospects of novel energy storage devices like liquid metal batteries. As noted in the publication, the \ch{NdBi_2} compound has long been known to be thermodynamically stable \cite{Yoshihara1975Rare-earthBismuthides}; however, its crystal structure has remained unknown, presenting a significant obstacle in using ab initio methods to study it. Using \texttt{crystALL}'s data-mining approach deployed on \texttt{MPDD}, the possible configurations of \ch{NdBi_2} were determined and later validated by DFT-based methods, as depicted earlier in Figure \ref{crystall:fig:crystallcompound}.

\begin{figure}[H]
    \centering
    \includegraphics[width=0.8\textwidth]{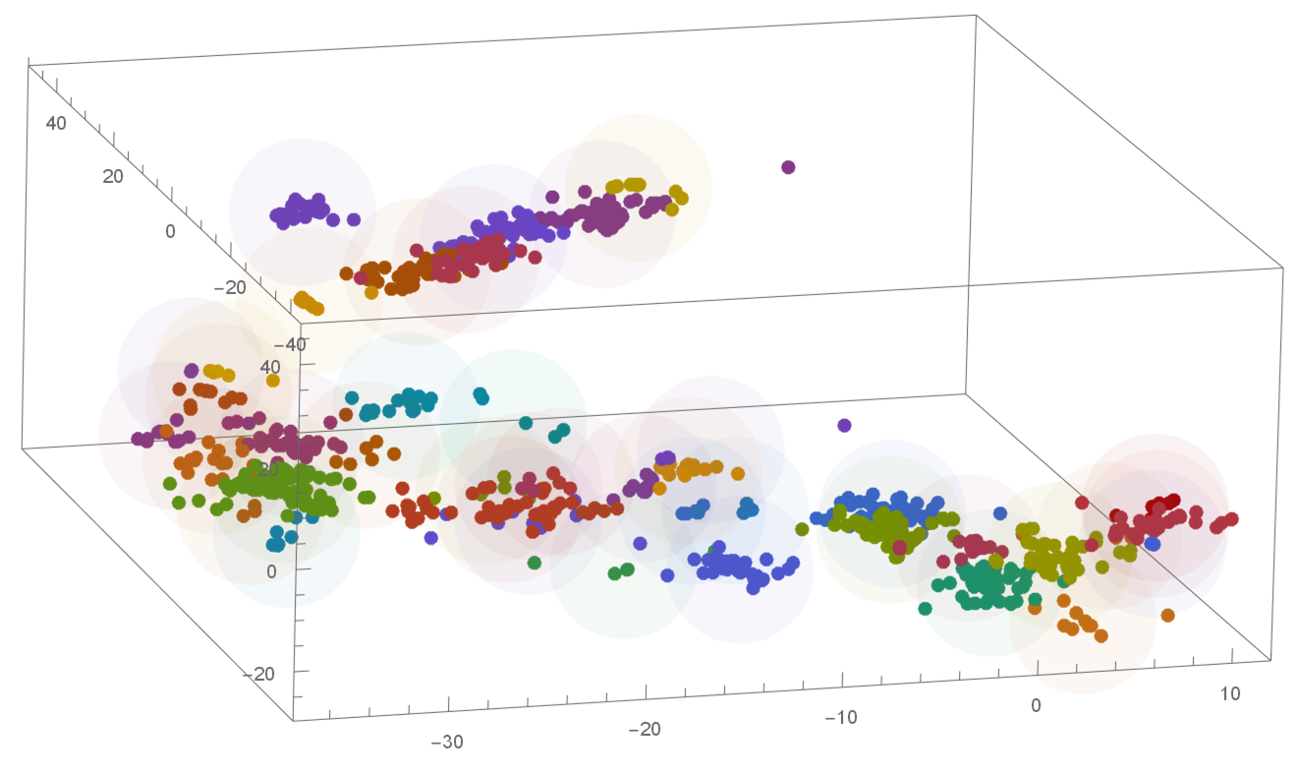}
    \caption{\texttt{crystALL}'s t-SNE low-dimensional (3D) embedding of \texttt{Ward2017} feature vectors of 1,000 lowest energy \ch{NdBi_2} candidates, as predicted by \texttt{SIPFENN}, clustered with k-means approach to determine the ensemble of unique structures.}
    \label{crystall:fig:ndbi2clusters}
\end{figure}

First, all of the 26,055 \ch{AB_2}-type configurations were extracted from the contemporary mid-2020 snapshot of \texttt{MPDD}, which at the time had approximately 1.3 million total configurations dataset of DFT-relaxed or experimental structures covering all materials contained in the Open Quantum Materials Database (OQMD), the Materials Project (MP), the Joint Automated Repository for Various Integrated Simulations (JARVIS), and the Crystallography Open Database (COD), described in Chapter \ref{chap:mpdd}. Following the extraction and substitution, all generated candidates have been featurized using \texttt{Ward2017} \cite{Ward2017IncludingTessellations}, and their energies were predicted through \texttt{SIPFENN} Novel Materials Model (NN20) described in Chapter \ref{chap:sipfenn}. The 1,000 lowest energy candidates were selected, and their feature-space representations were embedded into lower dimensional space (3D) using popular t-distributed stochastic neighbor embedding (t-SNE) \cite{HintonStochasticEmbedding} and clustered using k-means approach, as depicted in Figure \ref{crystall:fig:ndbi2clusters}.

\begin{figure}[H]
    \centering
    \includegraphics[width=0.8\textwidth]{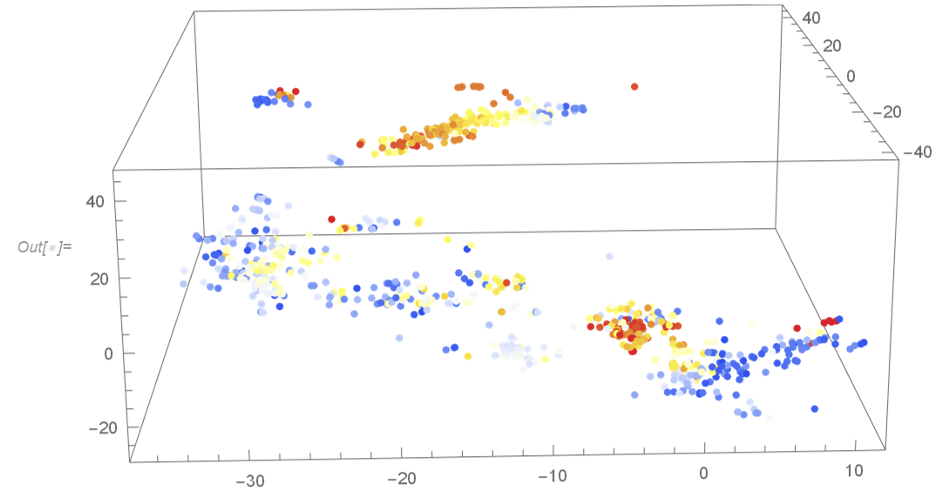}
    \caption{\texttt{crystALL}'s t-SNE low-dimensional (3D) embedding of \texttt{Ward2017} feature vectors of 1,000 lowest energy \ch{NdBi_2} candidates overlaid with \texttt{SIPFENN} formation energies depicting low energy ("hot") regions.}
    \label{crystall:fig:ndbi2energies}
\end{figure}

Next, the \texttt{SIPFENN}-predicted formation energies were overlaid on the candidate space, as shown in Figure \ref{crystall:fig:ndbi2energies}, and the most stable candidate from each cluster has been selected for further evaluation under 3 different DFT settings in the order of predicted stability.

\begin{figure}[H]
    \centering
    \includegraphics[width=0.6\textwidth]{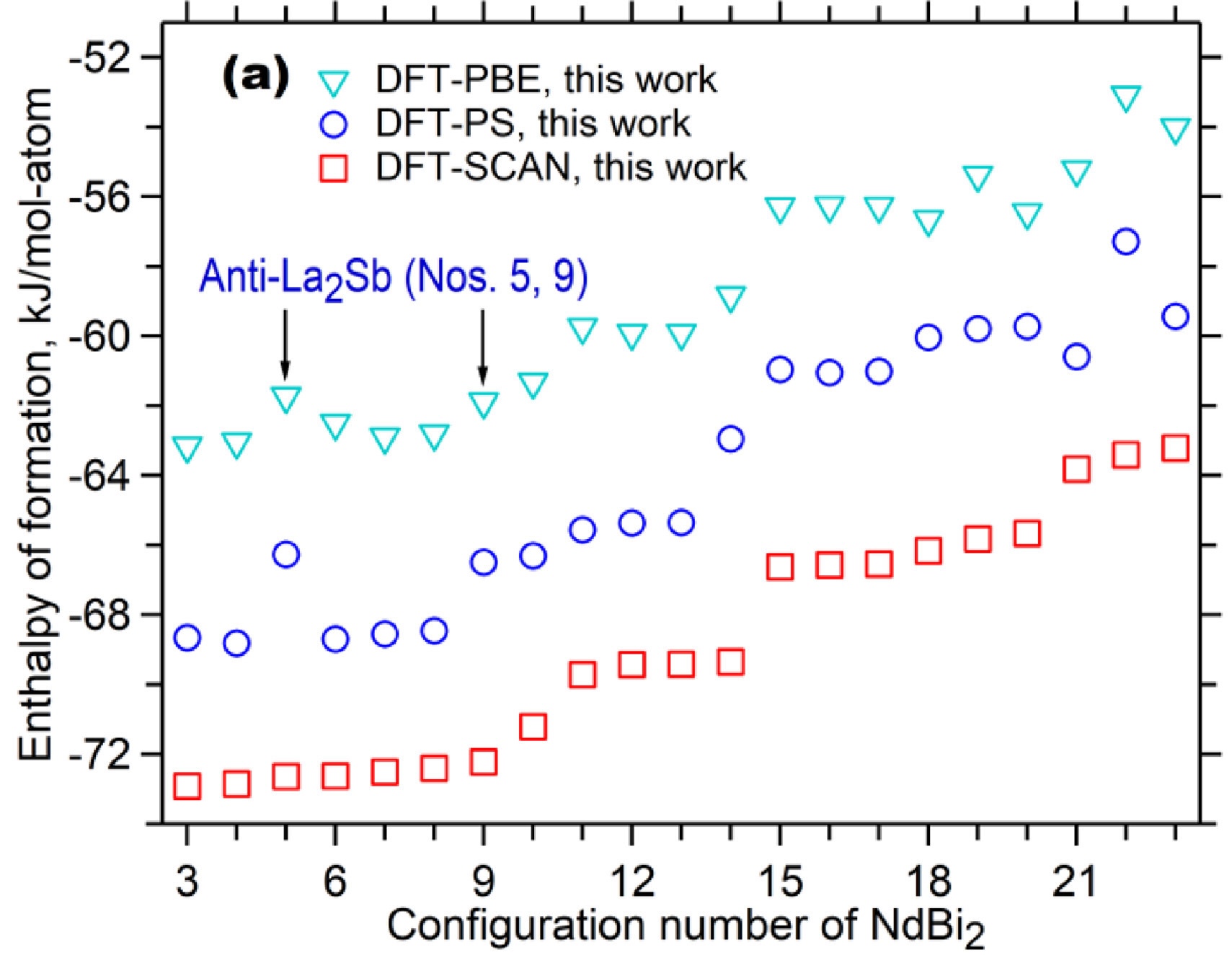}
    \caption{0K formation energies of structure candidates predicted by USPEX (\#3) and \texttt{crystALL} (\#4-23). The mixture of 2nd and 6th lowest-energy candidates has matched the experimental observations of \ch{NdBi_2} demonstrating both \texttt{crystALL}'s predictive ability and criticality of 0K metastable structures.}
    \label{crystall:fig:ndbi2dft}
\end{figure}

Simultaneously to \texttt{crystALL} method, an alternative, \texttt{USPEX} evolutionary structure search \cite{Lyakhov2013NewUSPEX}, mentioned earlier in Section \ref{crystall:sec:newstable}, has also been deployed and, as shown in Figure \ref{crystall:fig:ndbi2dft}, and found a lower energy structure than any of the 20 generated by \texttt{crystALL}, as shown in Figure \ref{crystall:fig:ndbi2dft}. However, as explained in Section \ref{crystall:sec:newstable}, the 0K lowest energy structure is not necessarily the one being searched for. This has been precisely the case in \citet{Im2022ThermodynamicModeling}, as the XRD patterns have been matched to a mixture of two 0K metastable configurations, denoted by arrows in \ref{crystall:fig:ndbi2dft}, which were present in the \texttt{crystALL}-generated ensemble of states.

\section{Predicting Compounds of Uncertain Compositions} \label{sec:crystallcompositional}

Compound-based searches, discussed in Section~\ref{crystall:sec:crystall} and exemplified in Section~\ref{crystall:ndbi2}, operate based on anonymous chemical formulas which are commonly used for such purpose, either based on expert knowledge \cite{Mehl2016ThePrototypes} or through screenings \cite{Curtarolo2013AFLOW:Discovery, Schmidt2023, Ye2022NovelAgents}, and can be accessed from nearly any atomistic database thanks to \texttt{OPTIMADE} API \citet{Evans2024DevelopmentsExchange}. While powerful, they suffer from a major drawback when used for structure predictions, i.e., they require researcher to specify exact composition of a compound, while the experimental observations can often match many stoichiometries.

\begin{figure}[h]
    \centering
    \includegraphics[width=0.95\textwidth]{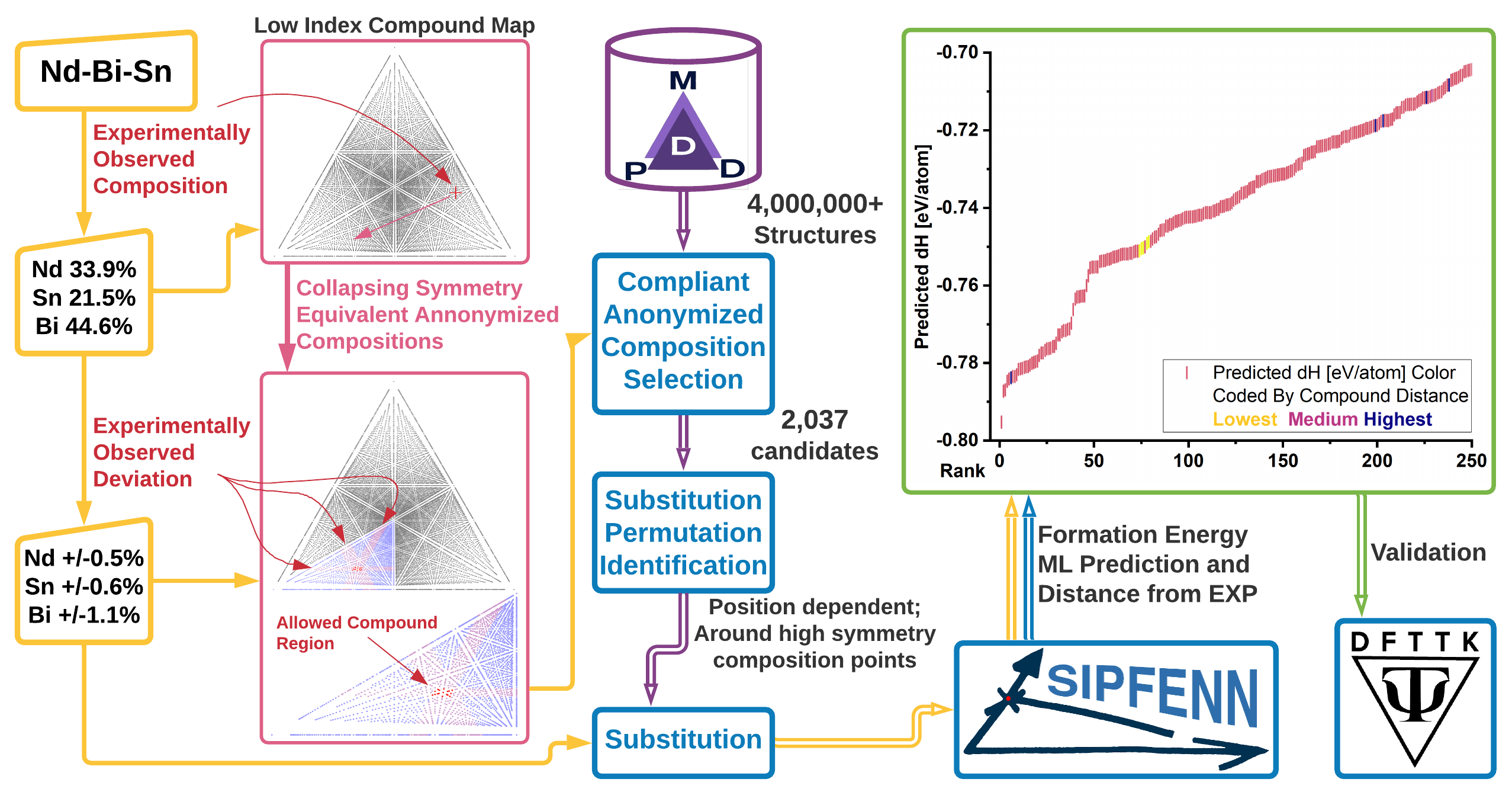}
    \caption{\texttt{crystALL} schematic of operation in cases of "compositional" searches where measured composition can be given alongside uncertainty bounds. Efficient handling of such queries is a unique feature of MPDD.}
    \label{fig:crystallcomposition}
\end{figure}

Thus, to address this need, \texttt{MPDD} has been extended with a new \texttt{sortedFractionsDict} field holding an ordered list of anonymized and quantized ($0.1\%$) atomic fractions, with associated \texttt{MongoDB} indexing. This enables \texttt{crystALL} to rapidly query for candidate structures falling within a certain range, which can be set accounting for experimentally observed deviations in the measurement, as shown in Figure~\ref{fig:crystallcomposition}.

Such approach generates output with additional information of the compositional distance to the \textit{reported} composition value, enabling researchers to make a better-informed decision on what to pass into the validation steps based on belief whether the hypothetical compound is the one being observed.

\section{Software Availability} \label{crystall:sec:softwareavaialbility}

The \texttt{crystALL} source code is currently developed closed-sourced; however, it is planned to be released as a free open-source software (FOSS) in the Fall of 2024 through outlets including GitHub repository at \href{https://github.com/PhasesResearchLab/crystALL}{github.com/PhasesResearchLab/crystALL}, alongside a scientific publication describing it, high-quality documentation, and a workshop-style tutorial.

%


\chapter{Creating an Efficient Database Infrastructure for Discovery of Real Materials Exemplified with High Entropy Alloys} \label{chap:ultera}

\section{Introduction} \label{ultera:sec:intro}

Compared to atomic structures discussed in Chapters \ref{chap:sipfenn} through \ref{chap:crystall}, which are precisely defined, "real" materials which were physically made in a lab tends to be (a) much less homogeneous in terms of \textit{how} they are reported and (b) the data belonging to them is much less defined in terms of the \textit{completeness} of description, i.e., many critical parameters like phases present can be missing or misreported due to a plethora of reasons including lack of equipment, limited precision, or human errors. Because of these, handling them generally requires much more flexible and elaborate ecosystem capable of:
\begin{enumerate}
    \item Handling voids in the knowledge with a structured, well-defined, reproducible procedures. 
    \item Filling in the gaps using the prior knowledge and available domain coverage.
    \item Detect unavoidable errors coming from humans, machines, and miscommunications along a long path from when the data is conceived (e.g., \textit{"This alloy should be strong and ductile"}), through initial report (e.g., \textit{"Sample 17 (A17) was X [...] Tensile tests were performed. [...] A17 fractured at 0.172"}), to the database (e.g., \textit{"X has tensile ductility of 17.2\%"}).
\end{enumerate}

One of the most challenging examples of handling "real" materials are compositionally complex materials (CCMs), and their sub-class of high entropy alloys (HEAs) \cite{Yeh2004NanostructuredOutcomes, Cantor2004MicrostructuralAlloys}, also known as multi principle element alloys (MPEAs) \cite{Borg2020ExpandedAlloys} which cover a broad spectrum of chemical elements with several of them present simultaneously, making their exploration both conceptually and computationally challenging, as explored in detail in later chapters (see Section \ref{nimplex:ssec:compositionallycomplex}). Further complexity arises from relatively high experimental and computational cost of studying them, causing researchers to typically focus on 1 to 4 property measurements (2.8 on average), while leaving at least several properties of interest undefined. Thus, it becomes critical to be able to combine multiple studies into a homogeneous structure that can be used collectively in as complete as possible form.

\begin{figure}[H]
    \centering
    \includegraphics[width=0.2\textwidth]{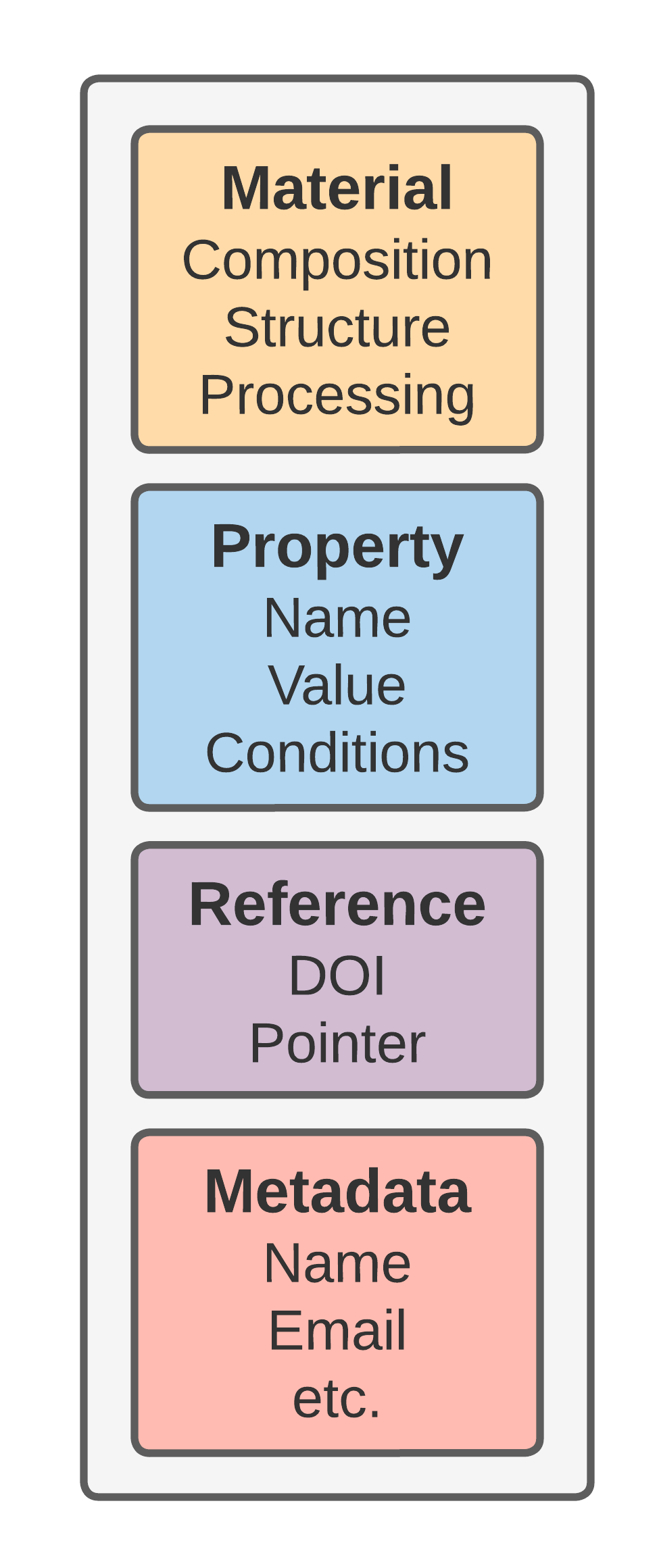}
    \caption{Schematic overview of the underlying core data structure in \texttt{ULTERA}. See Figure~\ref{ultera:fig:datapipeline} for details.}
    \label{ultera:fig:material}
\end{figure}

This has been largely accomplished for HEAs within the ULTERA, or ULtrahigh TEmperature Refractory Alloys project performed under the ARPA-E ULTIMATE program \cite{ULTIMATEArpa-e.energy.gov}, described in this Chapter, that has been continuously built and expanded since 2021 \cite{Debnath2021GenerativeAlloys} alongside associated machine learning efforts \cite{Debnath2023ComparingAlloys} and experimental validations \cite{Li2024DesignExperiments}. Within ULTERA, every property data serving as the core of the database is structured into Material - Property - Reference - Metadata quadruplet, as depicted along with sub-fields in Figure~\ref{ultera:fig:material}.

This core structure is then populated with dataset in Section~\ref{ultera:sec:datadescription} using infrastructure described in Sections~\ref{ultera:sec:infrastructure} to \ref{ultera:sec:contributions}, while being expanded and filled using several automated modeling methods described in Section \ref{ultera:sec:automodel}.

\section{Dataset} \label{ultera:sec:datadescription}
\newcommand{\statisticstime}{April 2024}

The dataset of the ULTERA Database has been manually collected through efforts of several ULTERA team members specializing in different properties, augmented with a few manually collected literature datasets processed through curation and aggregation data pipelines described in Section \ref{ultera:sec:pipeline}. All data was then further curated through \texttt{PyQAlloy} software described in Chapter \ref{chap:pyqalloy}, typically resulting in $5-10\%$ of the data points being modified to match original publications or removed.

\begin{figure}[H]
    \centering
    \includegraphics[width=0.95\textwidth]{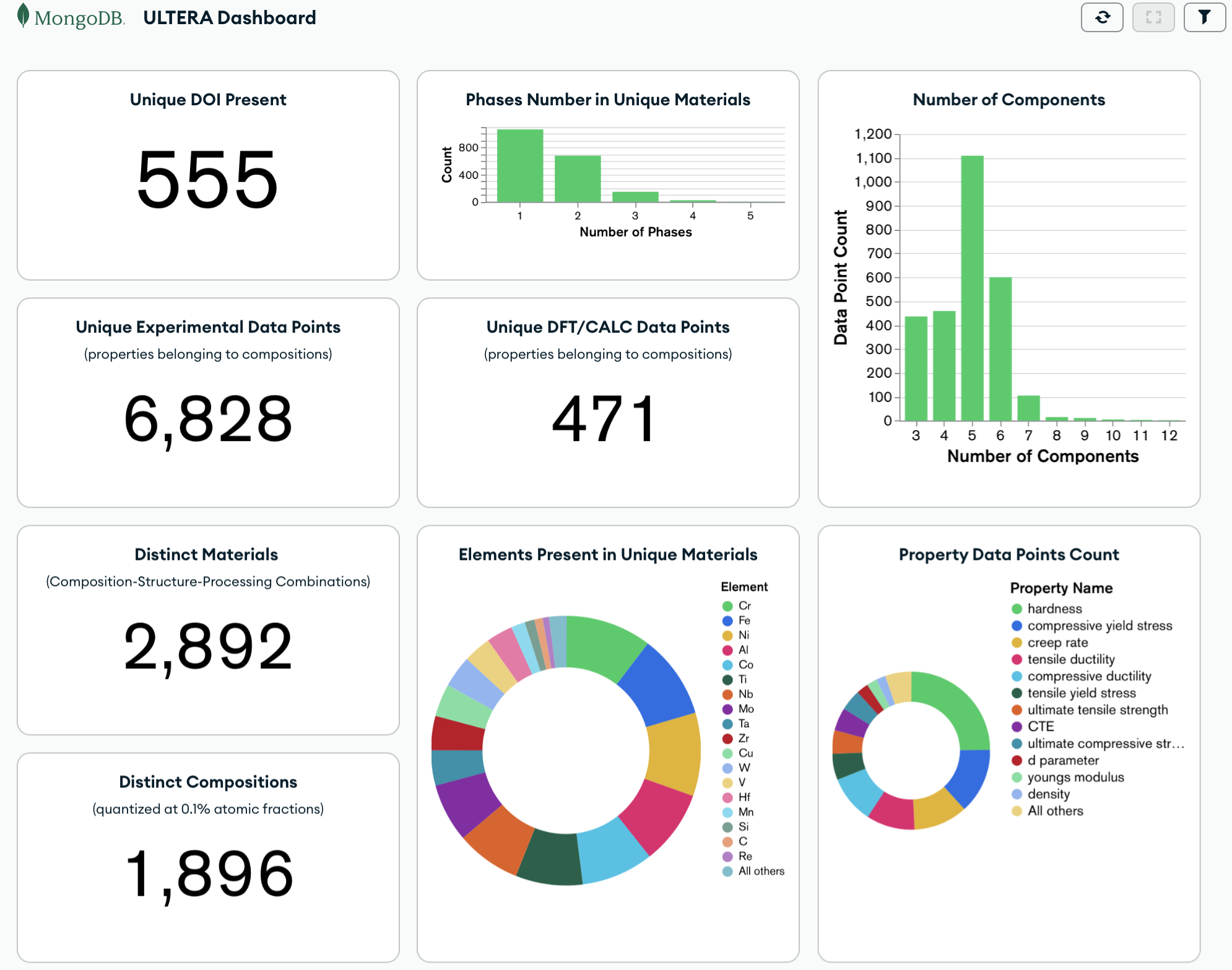}
    \caption{The main section of \texttt{ULTERA} Database dashboard at \href{https://ultera.org}{ultera.org}; presents statistics as of \statisticstime. All included figures are live and automatically recalculated every 1h. They are interactive allowing users to, e.g., select, highlight, or export the plot data in machine-readable format.}
    \label{ultera:fig:dashboard}
\end{figure}

After all the data processing steps, which generally reduced the number of datapoints and prioritized original studies over larger collections, as of \statisticstime, $\approx54\%$ of the property datapoints has been collected internally by ULTERA team, $\approx39\%$ from internally-improved version of dataset by \citet{Borg2020ExpandedAlloys}, $\approx5\%$ from \citet{Yang2022AHardness}, and $\approx3\%$ from \citet{Wang2023SearchingExperiments}. Collectively, over 550 scientific publications have been parsed to arrive at nearly 7,000 individual \textit{experimental} property datapoints, with further 500 coming from computer-based methods, belonging to nearly 3,000 unique materials spanning nearly 2,000 distinct chemical compositions, as presented in Figure~\ref{ultera:fig:dashboard}, which can be accessed in its most current form under \href{https://ultera.org}{ultera.org} project website.

As depicted in the bottom part of Figure~\ref{ultera:fig:dashboard}, the dataset covers a diverse set of (a) properties and (b) chemical elements. The property set spans 12 with at least 130 datapoints present, with 10 coming from experiments and 2 from computation, with the most common being hardness ($24.6\%$), tensile or compressive ductility ($19.7\%$), tensile or compressive yield stress ($18.9\%$), creep rate ($11.3\%$), and compressive or tensile ultimate strength ($9.1\%$).

\begin{figure}[H]
    \centering
    \includegraphics[width=0.95\textwidth]{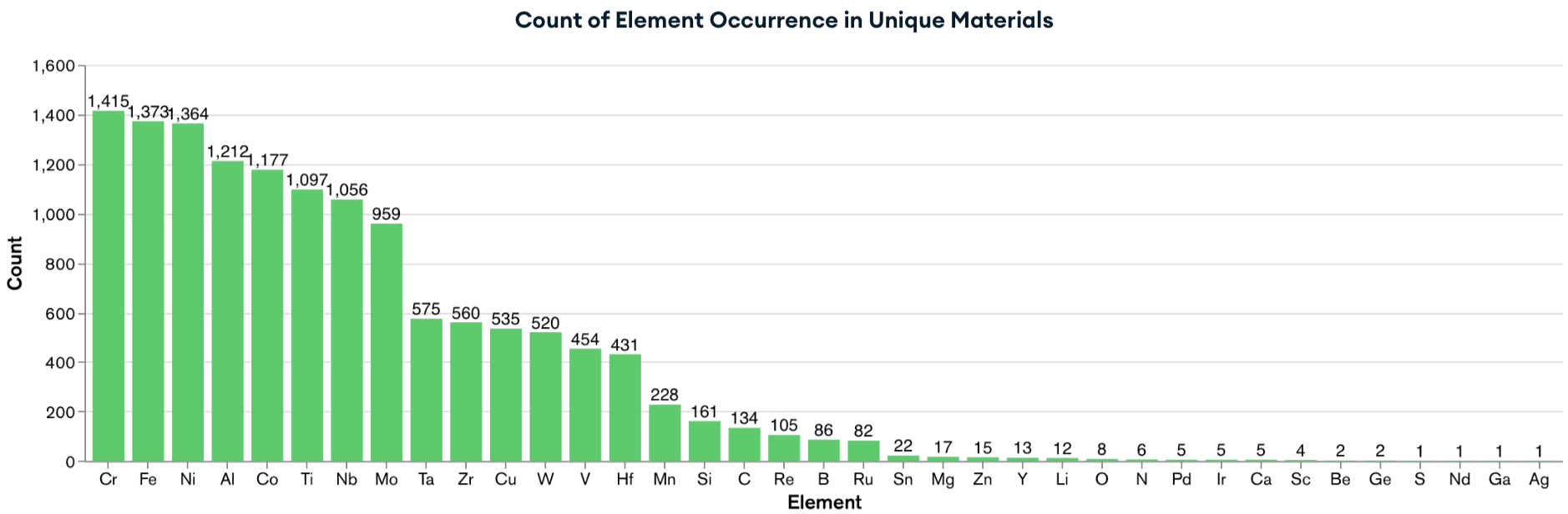}
    \caption{Chemical elements in the unique materials collection of the ULTERA Database as of \statisticstime. Please note that the same formula-processing-structure triplet can often be reported by many groups and is counted here as 1 point.}
    \label{ultera:fig:chemistries}
\end{figure}

The chemical space coverage spans 37 elements, with no bias towards a small subset of them, that most commonly co-occur in 5-component systems, as shown in Figure~\ref{ultera:fig:dashboard}. Out of the chemical elements, 20 are present in at least 80 unique alloys, as listed in Figure~\ref{ultera:fig:chemistries}, making them suitable for ML studies.

\begin{figure}[H]
    \centering
    \includegraphics[width=0.7\textwidth]{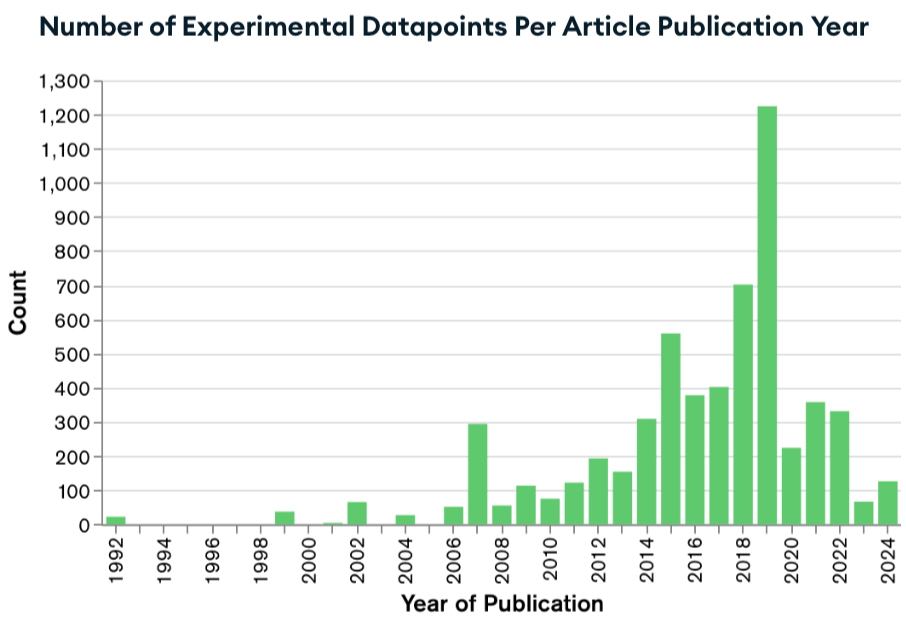}
    \caption{Number of experimental datapoints collected in ULTERA as of \statisticstime vs the year they were published, showing rapid growth. The lower numbers in the last 5 years reported can be attributed to significant portion of the data coming from compilations delayed by 1-3 years and height of COVID pandemic in 2020 delaying experiments.}
    \label{ultera:fig:publicationyears}
\end{figure}

For each experimental datapoint, the Crossref (\href{https://www.crossref.org}{crossref.org}) service is automatically queried based on the associated DOI (once per unique DOI) to retrieve a set of metadata associated with the study. This allows ULTERA to also be analyzed in terms of time-distribution of the experimental data which can reveal trends in the community regarding its research output, as shown in Figure~\ref{ultera:fig:publicationyears}, or to analyze trends in the explored alloys chemical spaces over the years. The significance of the latter will likely grow in the future, as it can prevent implicit biases into exploration of what "should work", which can limit innovation and design space exploration.

\begin{figure}[H]
    \centering
    \includegraphics[width=0.95\textwidth]{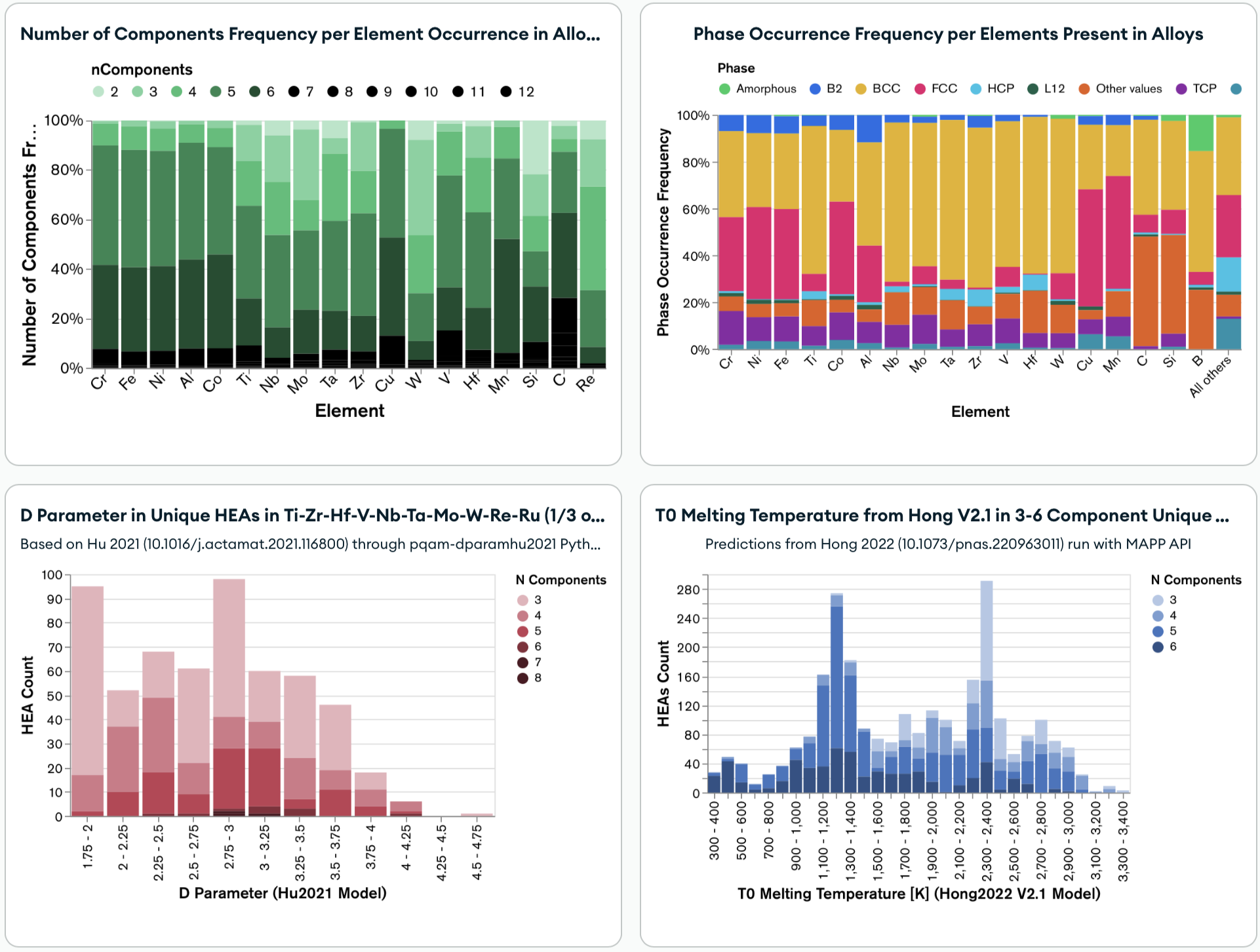}
    \caption{A large compiled dataset allows insights into prior expert knowledge driving the discovery and possible biases models generating new alloys will be subject to. The automated data infrastructure, described in Section \ref{ultera:sec:infrastructure}, enables efficient deployment of many tools, such as community models described in Subsection \ref{ultera:ssec:communitymodels}.}
    \label{ultera:fig:insights}
\end{figure}

Lastly, it is critically beneficial that every datapoint inserted into the ULTERA's data ecosystem is automatically extended through a number of data homogenization tools, machine learning models, and empirical models, described in later sections, which become immediately available to both modeling researchers (through an API) and data end-users. While much less accurate compared to proper investigations, their complete (or near-complete) coverage is a unique asset, as it puts all materials in the same context, visualizing trends over the entire database, as shown in Figure~\ref{ultera:fig:insights}.

\section{Alloy Discovery Infrastructure} \label{ultera:sec:infrastructure}

\begin{figure}[H]
    \centering
    \includegraphics[width=0.5\textwidth]{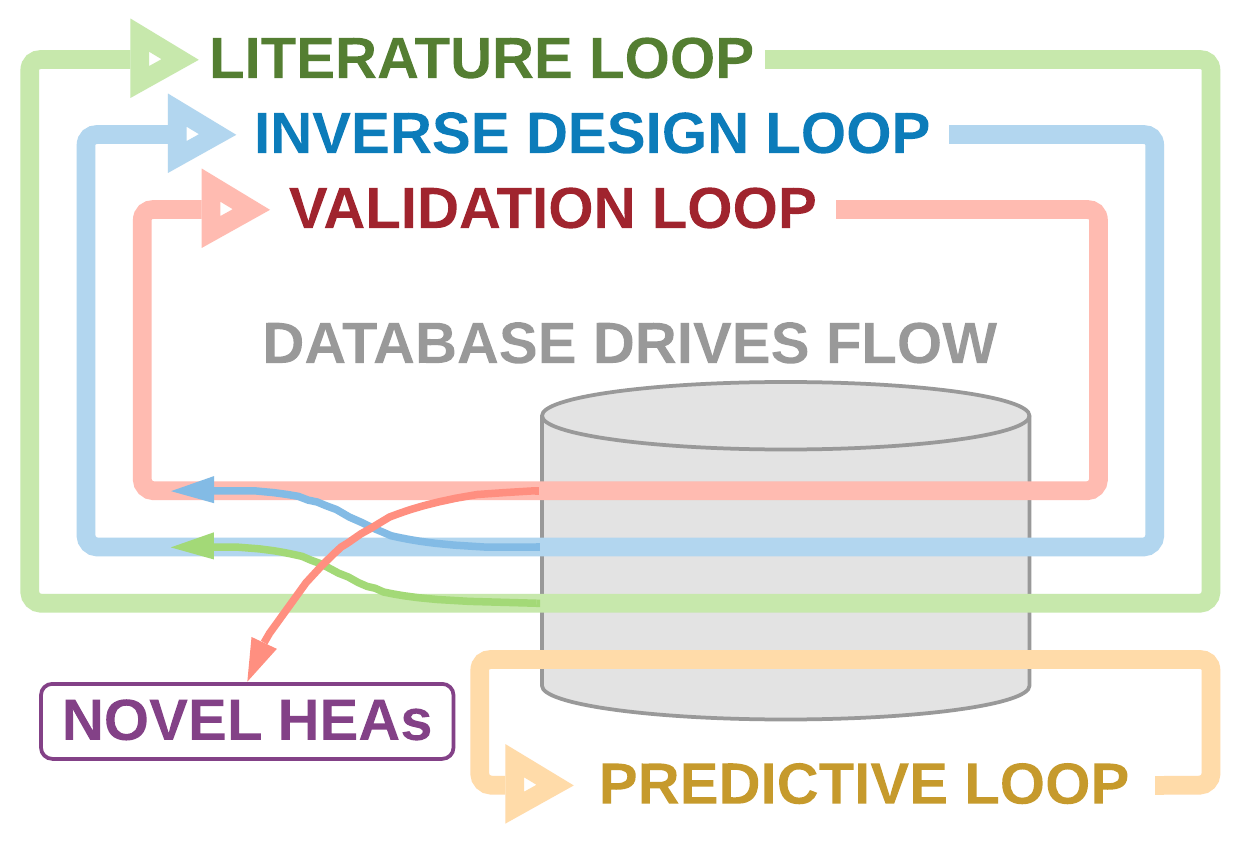}
    \caption{Four \emph{data loops} associated with different parts of the alloy discovery efforts and the database driving information flow between them to arrive at novel high entropy alloys.}
    \label{ultera:fig:dataloops}
\end{figure}

The ULTERA Alloy Discovery Infrastructure, first published conceptually in 2021 \cite{Debnath2021GenerativeAlloys}, is based on a multi-loop paradigm, where the \emph{literature} loop, which collects and extracts as much past knowledge as possible, feeds into the \emph{inverse design} loop, which proposes new alloys based on machine learning (ML) and articifical intelligence (AI) techniques, which are lastly verified experimentally and computationally through \emph{validation loop}, while forward models in the \emph{predictive loop} are developed to fill in the gaps in the knowledge, as depicted in Figure~\ref{ultera:fig:dataloops}. All of the loops are handled by largely independent sub-teams with specific expertise, while the communication of the data between them is facilitated through the ULTERA Database, from which novel alloys are "harvested" after verification.

\begin{figure}[H]
    \centering
    \includegraphics[width=0.9\textwidth]{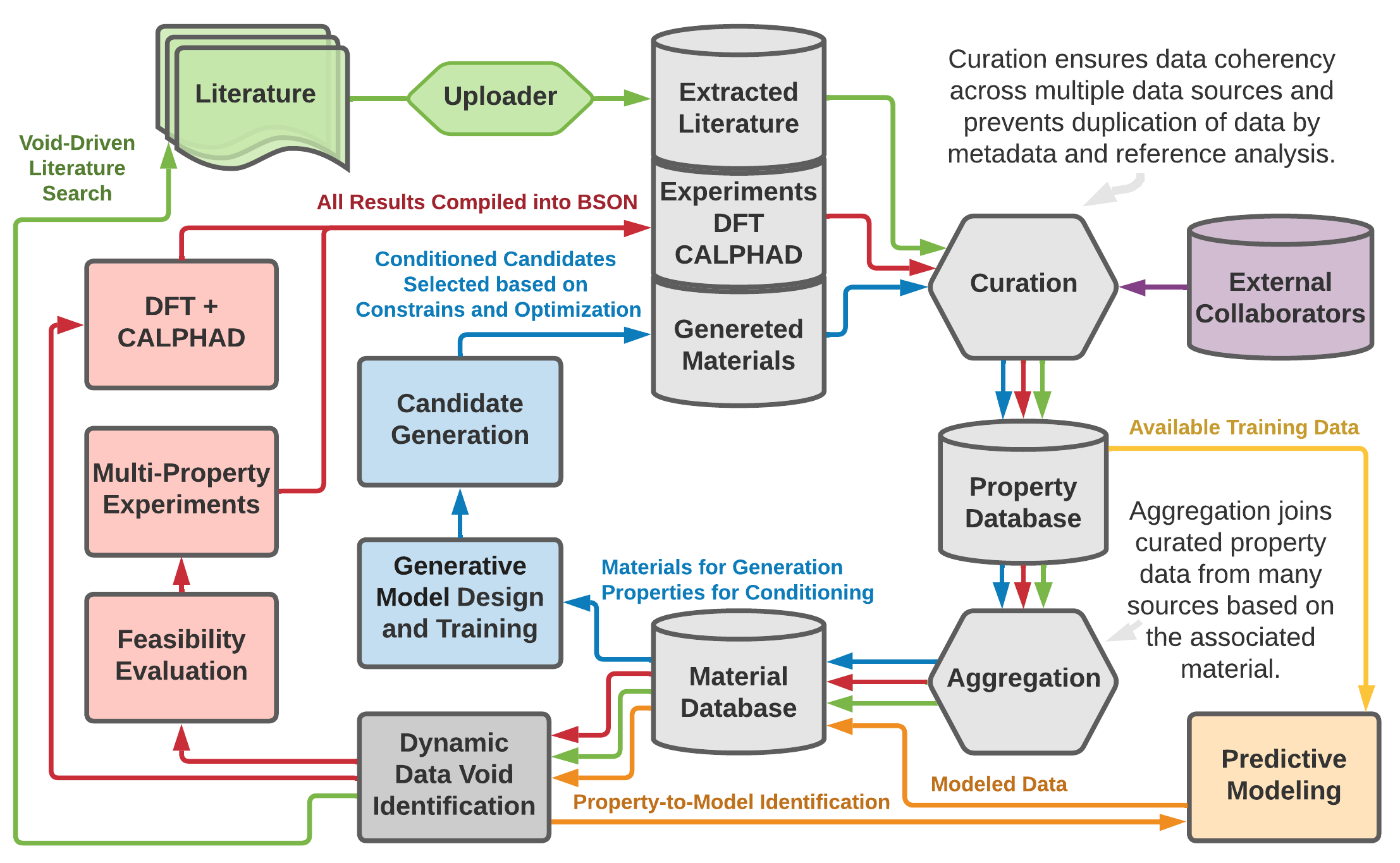}
    \caption{Big picture schematic of the ULTERA Data Infrastructure composed of the literature loop (green) collecting available external knowledge from many sources, the predictive loop (orange) filling in the gaps in current state of knowledge with modeling data, the generative loop (blue) proposing new candidate alloys to evaluate, validation loop (red) performing calculations and experiments to validate candidates. In the process databases are created, containing \texttt{CURATED} subset of materials property data, then \texttt{AGGREGATED} around unique materials for multi-property learning. The underlying infrastructure includes many more data collections hidden from users to enable efficient pipelines.}
    \label{ultera:fig:dataschematic}
\end{figure}

The flow presented in Figure~\ref{ultera:fig:dataloops} can be expanded in terms of the most critical tasks in each loop to arrive at the more complete picture presented in Figure~\ref{ultera:fig:dataschematic}. As shown, the data infrastructure component of the ULTERA ecosystem focuses on efficiently combining multiple internal and external data sources and then reorganizes it, as explored in detail later in Section~\ref{ultera:sec:pipeline}. In the process, two individual "databases" are created from the end-user perspective. 

The first one, called \texttt{CURATED} collection, is a "database" centered around individual property datapoints, which allows quick investigation in a format akin to literature publication tables and can be used to efficiently query for property data following certain criteria, such as property name being "tensile yield strength", containing chemical element "Mo", and being measured at temperatures between $270K$ and $300K$, to establish datasets for single-property machine learning models.

The second one, called \texttt{AGGREGATED} collection, is a "database" centered around unique materials (composition-structure-processing triplets), which allow efficient merging of properties from multiple studies and models to construct as complete picture as possible around the defined alloys. It serves as the query point for (1) selection of next experiments and other valudations, (2) identification of what critical data may be missing and should be looked for in the literature, and (3) more elaborate machine learning efforts that leverage correlations between properties to improve individual performance.

Together, all of the loops in ULTERA Infrastructure form a cyclic ecosystem with majority of the steps being automated (except for experiments and literature parsings), thus creating nearly self-driving driving infrastructure capable of automatically extending itself.

\section{Data Pipeline} \label{ultera:sec:pipeline}

\begin{figure}[H]
    \centering
    \includegraphics[width=0.95\textwidth]{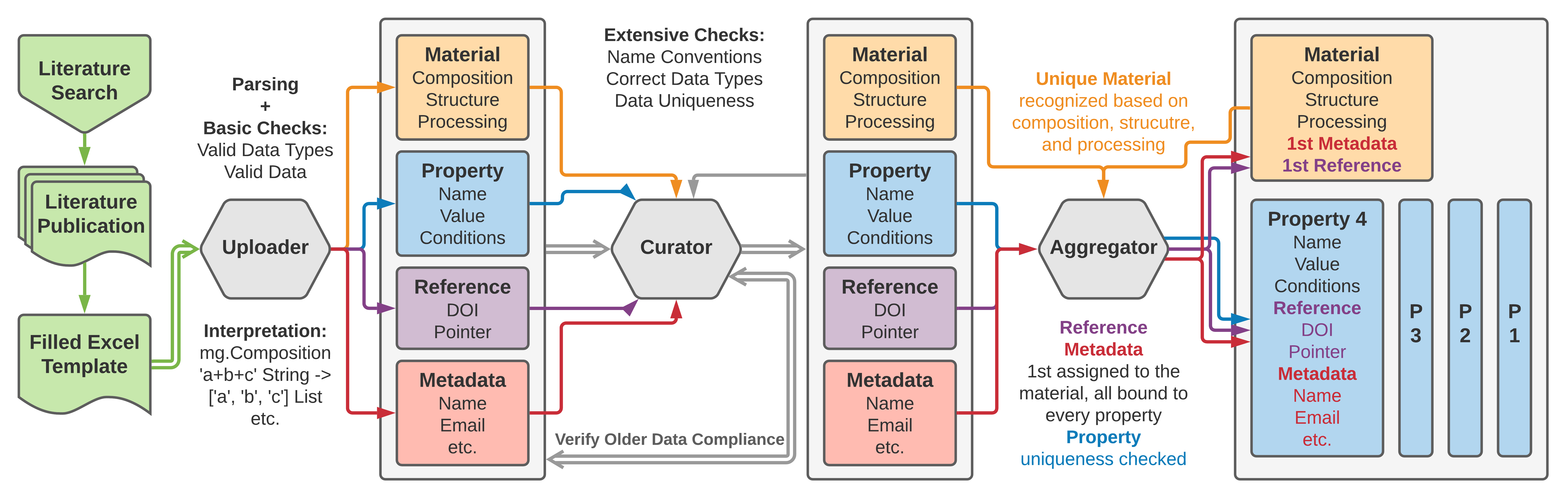}
    \caption{Schematic of the forward pipeline applied to data ingested into the system. For conciseness, intermediate steps of curation process and associated intermediate datasets are not depicted. Critically, a number of data points can converge at each step, so backward trace would be highly branched.}
    \label{ultera:fig:datapipeline}
\end{figure}

Between the data being inserted into the ULTERA ecosystem and presented to the end-user, several key operations are performed, as schematically depicted in Figure~\ref{ultera:fig:datapipeline}. First, the raw data point is passed through an \texttt{uploader} script, which attempts to interpret its the material definition and associated property data into a standardized structure while assigning metadata related to its provenance. At the same time, it performs a number of rule-based and pattern-based checks, related to the data schema, which verify data \emph{can} be valid based on, e.g., presence of required fields and correct data types.

Next, a set of curating scripts perform a set of checks both forward (as data gets uploaded) and backwards (on past uploaded data) to ensure that all of it is compliant to the current specification. While there is some overlap with the uploading step, the curation is much more extensive and puts data points in the context of the entire database to enable, e.g., ensuring the uniqueness of the data present. This approach enables ULTERA to avoid both obvious duplicates (accidental second upload) and hidden duplicates (one study reporting data from another without acknowledging the source). It also features homogenization mechnisms, such as converting names based on conventions, e.g., mapping "tensile ultimate strength" and "UTS" to "ultimate tensile strength", while providing reports of compliance to the data administrator. After the curation step, the data is cast into a schema exemplified with the code below and cast into the \texttt{CURATED} collection.

\begin{minted}[xleftmargin=3\parindent, linenos=true, fontsize=\small]{json}
{
  "_id": {
    "$oid": "638a980130b7ffbd9fa64947"
  },
  "meta": {
    "source": "LIT",
    "name": "Hui Sun",
    "email": "suh960@psu.edu",
    "directFetch": "T",
    "handFetch": "T",
    "comment": 
      "Data originally fetched by Hui, but heavily edited and re-uploaded by Adam",
    "timeStamp": {
      "$date": "2022-12-03T00:27:40.005Z"
    },
    "dataSheetName": "HuiAdam_FractureToughnessHardnessDataUpdate_Jan2022_AdamFix.xlsx",
    "sourcePriority": 2,
    "duplicates": [],
    "dataPointsCount": 1
  },
  "material": {
    "rawFormula": "ZrNbMoHfV",
    "formula": "Hf1 Zr1 Nb1 V1 Mo1",
    "compositionDictionary": {
      "Zr": 0.2,
      "Nb": 0.2,
      "Mo": 0.2,
      "Hf": 0.2,
      "V": 0.2
    },
    "percentileFormula": "Hf20 Zr20 Nb20 V20 Mo20",
    "relationalFormula": "Hf1 Zr1 Nb1 V1 Mo1",
    "compositionVector": 
        [ 0, 0, 0, 0, 0, 0, 0.2, 0.2, 0.2, 0.2, 0, 0, 0.2, 0, 0, 0, 0, 0, 0, 0
    ],
    "anonymizedFormula": "ABCDE",
    "reducedFormula": "HfZrNbVMo",
    "system": "Hf-Mo-Nb-V-Zr",
    "elements": [ "Hf", "Mo", "Nb", "V", "Zr" ],
    "nComponents": 5,
    "structure": [
      "BCC",
      "laves"
    ],
    "nPhases": 2,
    "observationTemperature": 298,
    "singleSolidSolution": false,
    "solidSolution": false
  },
  "property": {
    "name": "hardness",
    "value": 5285973000,
    "source": "EXP",
    "temperature": 298
  },
  "reference": {
    "doi": "10.1179/1432891714Z.000000000785",
    "pointer": "P769",
    "doiCount": 1,
    "citation": "Guo, N.N., Luo, L.S., Su, Y.Q. and Guo, J.J. 2014. Microstructure ...",
    "bibentry": " @article{Guo_2014, title={Microstructure and mechanical properties ...",
    "authors": [
      "N. N. Guo",
      "L. S. Luo",
      "Y. Q. Su",
      "J. J. Guo"
    ],
    "dateCreated": {
      "$date": "2014-08-05T00:00:00.000Z"
    },
    "title": "Microstructure and mechanical properties of ZrNbMoHfV high entropy alloy"
  },
  ...
}
\end{minted}

Finally, the aggregation step extracts the material definition from each datapoint present in the \texttt{CURATED} collection and groups them based on their uniqueness, while applying priority rules to establish precedence of original studies over reviews and reviews over larger datasets. During the grouping, most of the auxiliary data (metadata and reference information) are added to a set, while actual data is organized as ordered lists, to arrive at a schema exemplified with the code below and cast into the \texttt{AGGREGATED} collection.

\begin{minted}[xleftmargin=3\parindent, linenos=true, fontsize=\small]{json}
{
  "_id": {
    "$oid": "666956c247f65c310dae9e94"
  },
  "material": {
    "formula": "Zr1 Ti1 Ta1 Nb1 Mo1",
    "compositionDictionary": {
      "Ti": 0.2,
      "Ta": 0.2,
      "Nb": 0.2,
      "Mo": 0.2,
      "Zr": 0.2
    },
    "percentileFormula": "Zr20 Ti20 Ta20 Nb20 Mo20",
    "relationalFormula": "Zr1 Ti1 Ta1 Nb1 Mo1",
    "compositionVector": 
      [ 0, 0, 0, 0, 0, 0.2, 0.2, 0.2, 0.2, 0, 0, 0.2, 0, 0, 0, 0, 0, 0, 0, 0
    ],
    "anonymizedFormula": "ABCDE",
    "reducedFormula": "ZrTaTiNbMo",
    "system": "Mo-Nb-Ta-Ti-Zr",
    "elements": [
      "Mo",
      "Nb",
      "Ta",
      "Ti",
      "Zr"
    ],
    "nComponents": 5,
    "structure": [
      "BCC",
      "BCC"
    ],
    "nPhases": 2,
    "processes": [
      "AC"
    ],
    "nProcessSteps": 1,
    "singleSolidSolution": false,
    "solidSolution": true,
    "rawFormulas": [
      "TiTaNbMoZr",
      "MoNbTaTiZr"
    ],
    "observationTemperatures": [ 1273, 800, 323, 298, 298, 298 ]
  },
  "properties": [
    {
      "name": "CTE",
      "value": 0.0000075,
      "source": "EXP",
      "temperature": 323,
      "unitName": "K^-1",
      "sourcePriority": 3,
      "meta": {
        "source": "LIT",
        "name": "Adam Krajewski",
        "email": "ak@psu.edu",
        "directFetch": "T",
        "handFetch": "T",
        "comment": "HEA CTE search effort",
        "timeStamp": {
          "$date": "2023-07-08T00:02:37.218Z"
        },
        "dataSheetName": "ULTERA-contribute-amkrajewski/ThermalExpansionSearch_Oct22.xlsx",
        "sourcePriority": 2,
        "duplicates": [],
        "dataPointsCount": 1
      },
      "reference": {
        "doi": "10.1016/j.jallcom.2021.162154",
        "pointer": "F6",
        "doiCount": 3
      }
    },
    {
      "name": "ultimate compressive strength",
      "value": 1450000000,
      "source": "EXP",
      "temperature": 298,
      "sourcePriority": 3,
      "meta": {
        "source": "LIT",
        "parentDatabase": "MPEA",
        "name": "Adam Krajewski",
        "email": "ak@psu.edu",
        "directFetch": "F",
        "handFetch": "F",
        "comment": "Alloys from Citrine / UCSB HEA Database as of August 2021",
        "timeStamp": {
          "$date": "2022-12-02T23:53:44.918Z"
        },
        "dataSheetName": "ULTERA_MEPA_combineddata.xlsx",
        "sourcePriority": 2,
        "duplicates": [],
        "dataPointsCount": 1
      },
      "reference": {
        "doi": "10.1016/j.scriptamat.2016.10.028",
        "doiCount": 4
      }
    },
    ...
  ],
  "parents": [
    {
      "$oid": "64a8a7ab9e7f829350cce73e"
    },
    {
      "$oid": "64a8a7ab9e7f829350cce73f"
    },
    ...
  ],
  "uploaders": [
    "Adam Krajewski"
  ],
  "metaSet": [
    {
      "source": "LIT",
      "name": "Adam Krajewski",
      "email": "ak@psu.edu",
      "directFetch": "T",
      "handFetch": "T",
      "comment": "HEA CTE search effort",
      "timeStamp": {
        "$date": "2023-07-08T00:02:37.218Z"
      },
      "dataSheetName": 
        "ULTERA-contribute-amkrajewski/ThermalExpansionSearch_Oct22.xlsx",
      "sourcePriority": 2,
      "duplicates": [],
      "dataPointsCount": 1
    },
    ...
  ],
  "referenceSet": [
    {
      "doi": "10.1016/j.msec.2016.12.057",
      "doiCount": 1
    },
    {
      "doi": "10.1016/j.scriptamat.2016.10.028",
      "doiCount": 4
    },
    {
      "doi": "10.1016/j.jallcom.2021.162154",
      "pointer": "F6",
      "doiCount": 3
    }
  ],
}
\end{minted}

In addition to the \texttt{CURATED} and \texttt{AGGREGATED} collections depicted in Figure~\ref{ultera:fig:dataschematic}, there are several other that are hidden from the end-user but perform critical functions as intermediate stages, deployment targets, and provenance metadata sources. These are:

\begin{itemize}
    \item Many (20+) individual user-specific or source-specific data collections with fine-grained permissions assigned based on current stewardship distribution.

    \item \texttt{ALL} - combines (nearly) all source collections into one for further processing using MongoDB aggregation pipelines, after trimming fields that were empty (due to, e.g., single whitespace in spreadsheet) and removing test data entries marked with comments like \texttt{'test'}.

    \item \texttt{STRUCTURAL} - which groups the \texttt{CURATED} collection around unique materials \emph{except for processing history, comments, and observation temperature} to act as a deployment target for models taking into account alloy composition and structure.

    \item \texttt{COMPOSITIONAL} - which groups the \texttt{CURATED} collection around unique materials \emph{except for structure-related fields, processing history, comments, and observation temperature} to act as a deployment target for models taking into account only the alloy composition.

    \item \texttt{DOI} - used to store references based on the DOIs and link back to the \texttt{CURATED} collection. The reference fetching is performed only when new DOI is inserted, but linking is refreshed every time \texttt{CURATED} data changes.

    \item \texttt{CURATED\_Jul2023} and other versioned-backup collections - used to produce results presented in scientific publications.
\end{itemize}



\section{Community Contributions} \label{ultera:sec:contributions}

\begin{figure}[H]
    \centering
    \includegraphics[width=0.97\textwidth]{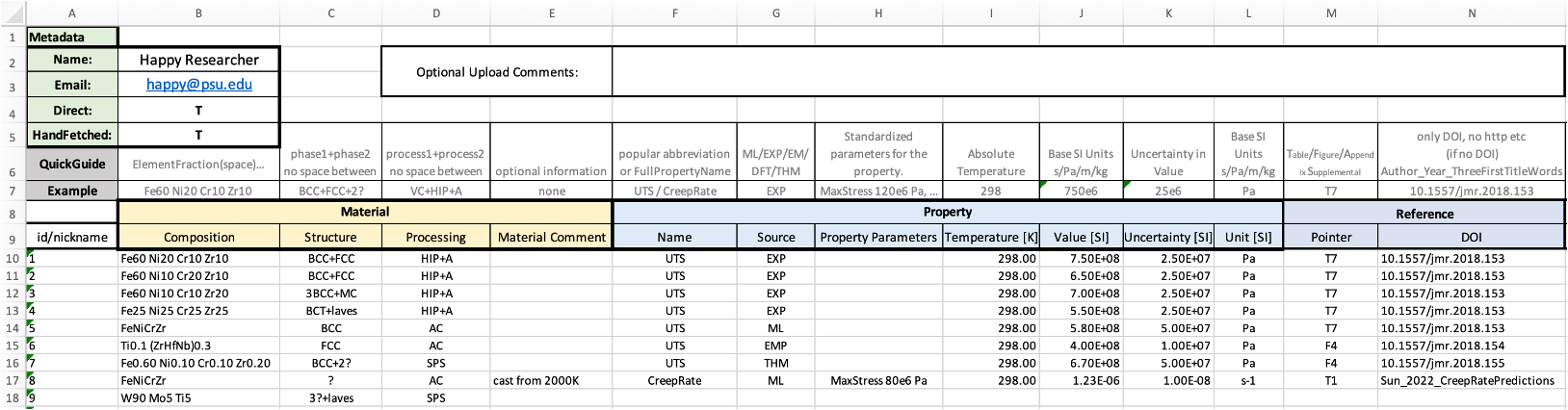}
    \caption{Header and example data rows of a formatted Excel spreadsheet template used to streamline and increase accessibility to the contribution system for non-technical members of the community. An automated system processes them on the cloud into git-tracked plain-text CSV records then passed to uploading system.}
    \label{ultera:fig:contributiontemplate}
\end{figure}

While ULTERA is internally organized based on a cloud non-relational database with elaborate data processing, explored in Sections \ref{ultera:sec:infrastructure} and \ref{ultera:sec:pipeline}, the majority of data ingested into the system has to pass through researchers not familiar with the ecosystem specifics or programming in general. Thus, a contribution template, shown in Figure~\ref{ultera:fig:contributiontemplate}, has been developed to contain \emph{unprocessed} source data mapping fields shown in Figure~\ref{ultera:fig:material}, needed to construct ULTERA entries.

Such spreadsheet templates can be either (a) stored locally and passed manually by users who want to make a small singular contribution, e.g., to accompany a publication, or (b) stored in a \texttt{git} repository on GitHub, which can be linked and automatically ingested into ULTERA based on the internal meta-repository. To simplify the latter, an origin repository has been made available at \href{http://contribute.ultera.org/}{contribute.ultera.org} for forking.

Furthermore, the provided template repository also includes several automations described in its \texttt{README.md}, which are built around GitHub Actions and enable functionalities, such as git-tracking of the Excel's \texttt{XLSX} spreadsheet contents by casting it into plain-text \texttt{CSV} on every push to a branch. In the future when ULTERA will be made fully open, data quality analysis methods, described in Chapter \ref{chap:pyqalloy}, will also be included at this step.

\section{Automatic Modeling} \label{ultera:sec:automodel}

\subsection{Multi-Structure Linear Combinations} \label{ultera:ssec:autolc}

Calculating a linear combination (LC) of "elemental" (or otherwise component) properties is the most straightforward approach to estimating property values, especially in cases where phase stability differences play a secondary role, as it presents near-zero computational expense while requiring only readily available data that has been likely tabulated in some computer-readable format. In most studies, these elemental values are assumed to have one-to-one correspondence with the chemical element. However, while this may hold true for, e.g., price or atomic weight, many elements will behave very differently when in different crystal structures (e.g., Fe in BCC vs FCC).

\begin{figure}[H]
    \centering
    \includegraphics[width=0.6\textwidth]{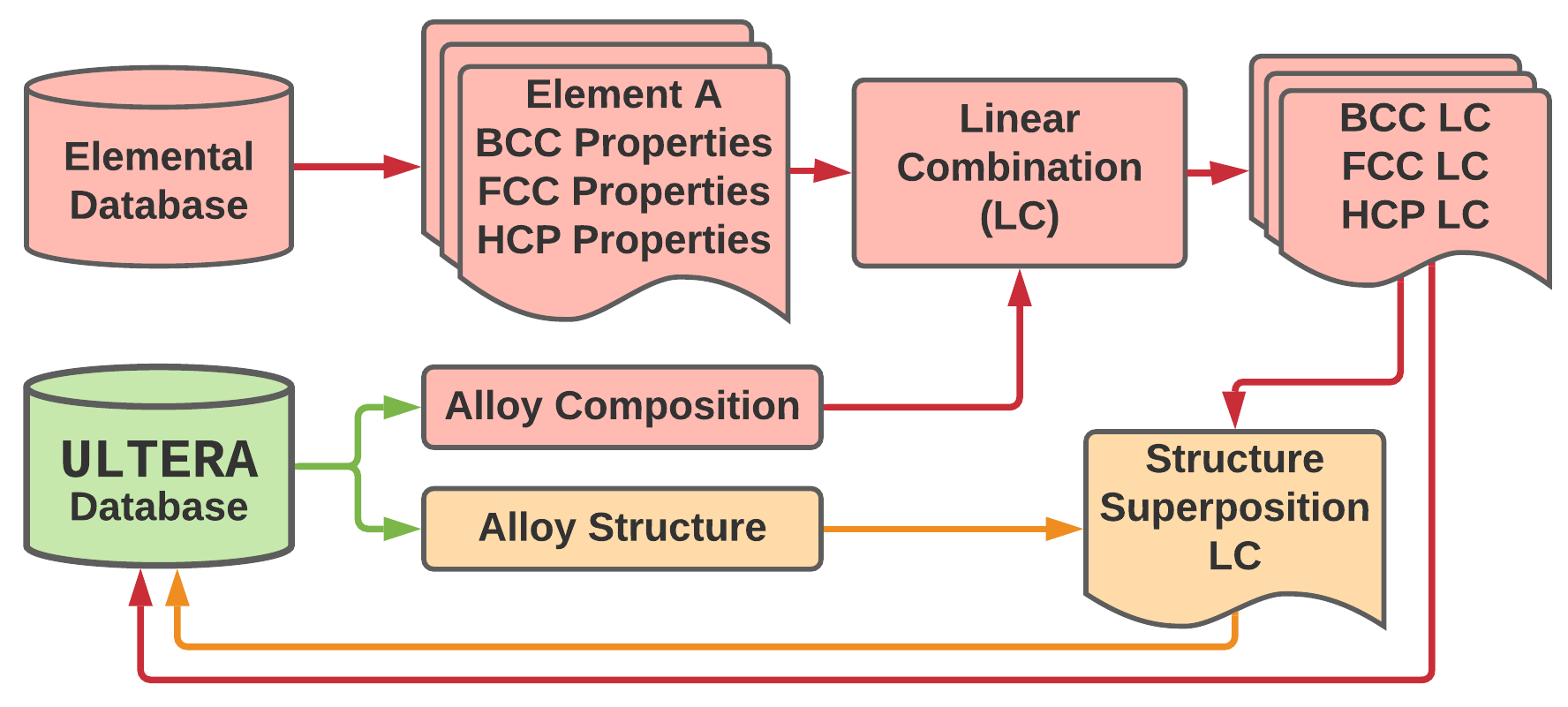}
    \caption{Simplified schematic of automatic linear combination modeling of properties. For every chemical composition, a linear combination of elemental properties is calculated for BCC, FCC, and HCP structures based on best-matched elemental polymorph data coming from experiments and DFT-based pure element calculations. If an applicable structure set (e.g., FCC+FCC+HCP) has been reported for a given input datapoint, an average of the respective linear combinations is reported as structure-informed LC.}
    \label{ultera:fig:autolc}
\end{figure}

Thus, in ULTERA, whenever possible, LCs for each alloy are calculated on structure-specific data coming from experiments or multi-lattice ab initio calculations \cite{Chong2021CorrelationAlloys} stored in a dedicated \texttt{ELEMENTAL} database of ULTERA, to establish different LC values under BCC, FCC, and HCP assumptions, as shown in Figure~\ref{ultera:fig:autolc}. During LC calculations, some element-structure pairs are unavoidably missing (e.g., BCC Cu \cite{Chong2021CorrelationAlloys}), but in such cases, the missing data with most similar one while tracking the level of it under \texttt{replacementLevel} field. 

If a datapoint has a structure (either single- or multi-phase) reported for it, then an additional LC is calculated to reflect an average of the structure-specific LCs weighted by the number of given phases being reported or when possible, by the exact phase fractions coming from CALPHAD equilibrium or solidification (e.g., Scheil \cite{Bocklund2020ExperimentalMaterials}) calculations run through \texttt{pycalphad} \cite{Otis2017Pycalphad:Python}. All LCs are then collectively reported under the \texttt{structuralProperties}, arriving at a schema exemplified with the code below.

\begin{minted}[xleftmargin=3\parindent, linenos=true, fontsize=\small]{json}
{
  "_id": {
    "$oid": "638a980130b7ffbd9fa64947"
  },
  ...
  "structuralProperties": {
    "LC_BCC": {
      "replacementLevel": 0.4,
      "SQusf": 0.68966,
      "G111C": 63.489999999999995,
      "DFTC12": 117.67000000000002,
      "DFTGvb": 18.1292,
      "NfUnfill": 0,
      "Heat_Sublimation": 623414.4,
      "Ion_Pot_2": 14.328000000000001,
      "Surf": 2.2020000000000004,
      ...
    },
    "LC_FCC": {
      "replacementLevel": 0.6,
      "SQusf": 0.6138399999999999,
      "G111C": 57.824,
      "DFTC12": 123.278,
      "DFTGvb": 21.0182,
      "NfUnfill": 0,
      "Heat_Sublimation": 623414.4,
      "Ion_Pot_2": 14.328000000000001,
      "Surf": 2.116,
      ...
    },
    "LC_HCP": {
      "replacementLevel": 0.6,
      "SQusf": 0.68966,
      "G111C": 63.489999999999995,
      "DFTC12": 117.67000000000002,
      "DFTGvb": 18.1292,
      "NfUnfill": 0,
      "Heat_Sublimation": 623414.4,
      "Ion_Pot_2": 14.328000000000001,
      "Surf": 2.2020000000000004,
      ...
    }
  }
}
\end{minted}

\subsection{Community Model Deployment} \label{ultera:ssec:communitymodels}

As mentioned in Section \ref{ultera:sec:intro}, the ULTERA Infrastructure automatically deploys several literature models onto all applicable data points. These include melting temperature ($T_0$) predictions by \citet{Hong2022MeltingMaterials}, intrinsic ductility prediction by \citet{Hu2021ScreeningAlloys}, root mean square atomic displacement in BCC HEA by \citet{Tandoc2023MiningAlloys}, or formation energy by \citet{Krajewski2022ExtensibleNetworks}. The utilized pipeline has been designed to be relatively straightforward and can be defined to deploy a model on both a complete description of a material and a partial one (e.g., in \texttt{COMPOSITIONAL} collection) to optimize many modes of model evaluations.

\begin{minted}[xleftmargin=3\parindent, linenos=true, fontsize=\small]{json}
{
  "_id": {
    "$oid": "638a980130b7ffbd9fa64947"
  },
  ...
  "compositionalProperties": {
    "TM[K]_HongV2_1": {
      "mean": 2322,
      "standard error": 294
    },
    "rmsad_Tandoc2023": 0.18933359511558134,
    "dparam_Hu2021": 2.884705648963465,
    "gfse_Hu2021": 0.6816671231315389,
    "surf_Hu2021": 1.966409000810224,
    "CH_SIPFENN_NN30": -0.03186363223940135,
    ...
  },
  "structuralProperties": {
    "DPARAM_AllData": 2.6416306495666504,
    "DPARAM_Validated": 2.774862051010132,
    "EF_SIPFENN_NN30_BCC": -0.009082297794520855,
    "EF_SIPFENN_NN30_FCC": -0.0027202172204852104,
    "EF_SIPFENN_NN30_HCP": 0.03427823260426521,
    ...
  }
}
\end{minted}

\subsection{Automated CALPHAD Modeling} \label{ultera:ssec:autocalphad}

A more robust pipeline has been created to automatically deploy results of CALPHAD \cite{Olson2023GenomicDynamics} based calculations into the \emph{modeled} \texttt{compositionalProperties} fields of end-user collections, whenever such calculation can be performed given chemical compositions scope of a given thermodynamic database (TDB), such as one created by Shuang Lin and shared through a GitHub repository \cite{LinShuangLin212/refractory-elements-database:ZR}. Lin's TDB was used to generate the example of data structure below, where one can see that the reported data contains several phase equilibrium calculation results, including metadata, phase fractions under \texttt{phaseDict} (assuming no miscibility gaps, i.e., two BCC phases are counted as one) and \texttt{phaseFullDict}, presented in two formats, as well as \texttt{zpfPositions} which give zero-phase-fraction (ZPF) points critical to interpretation of experimental results \cite{Li2024DesignExperiments} and modeling phase stability with AI approaches \cite{Wu2023EstimatingApproach}.

\begin{minted}[xleftmargin=3\parindent, linenos=true, fontsize=\small]{json}
{
  "_id": {
    "$oid": "638a980130b7ffbd9fa64947"
  },
  ...
  "compositionalProperties": {
    ...
    "calphad_Feb2023": {
      "T": 1473,
      "TDB": "CrHfMoNbTaTiVWZr_9element_Feb2023",
      "calphadStructure": "BCC+C15",
      "phaseDict": {
        "BCC_A2": 0.817,
        "LAVES_C15": 0.183
      },
      "phaseFullDict": {
        "BCC_A2": 0.817,
        "LAVES_C15": 0.183
      },
      "nPhases": 2,
      "phaseDictString": "{'BCC_A2': 0.817, 'LAVES_C15': 0.183}",
      "phaseFullDictString": "{'BCC_A2': 0.817, 'LAVES_C15': 0.183}",
      "zpfPositions": {
        "BCC_A2": {
          "HF": 0.201232305,
          "MO": 0.184393688,
          "NB": 0.244705403,
          "V": 0.15600952,
          "ZR": 0.213659083
        },
        "LAVES_C15": {
          "HF": 0.194433862,
          "MO": 0.269743594,
          "NB": 0.000002969,
          "V": 0.396690129,
          "ZR": 0.139129446
        }
      },
      "BCC_SS": 0,
      "BCC_SSS": 0
    }
  }
}
\end{minted}

In the example shown, the persisted output is not semantically versioned but rather marked with a unique name \texttt{calphad\_Feb2023} corresponding to a well-defined effort of the team. This design choice has been selected based on the character of CALPHAD modeling, where iterations of the TDBs are not numerous, while end-users traditionally favor descriptive names, such as \texttt{Shuang\_May2024\_RefractoryBCO}, which are then fixed and backed by a literature publication.

\subsection{MPDD Atomic Configuration Data Fetching} \label{ultera:ssec:mpdd}

\begin{figure}[H]
    \centering
    \includegraphics[width=0.8\textwidth]{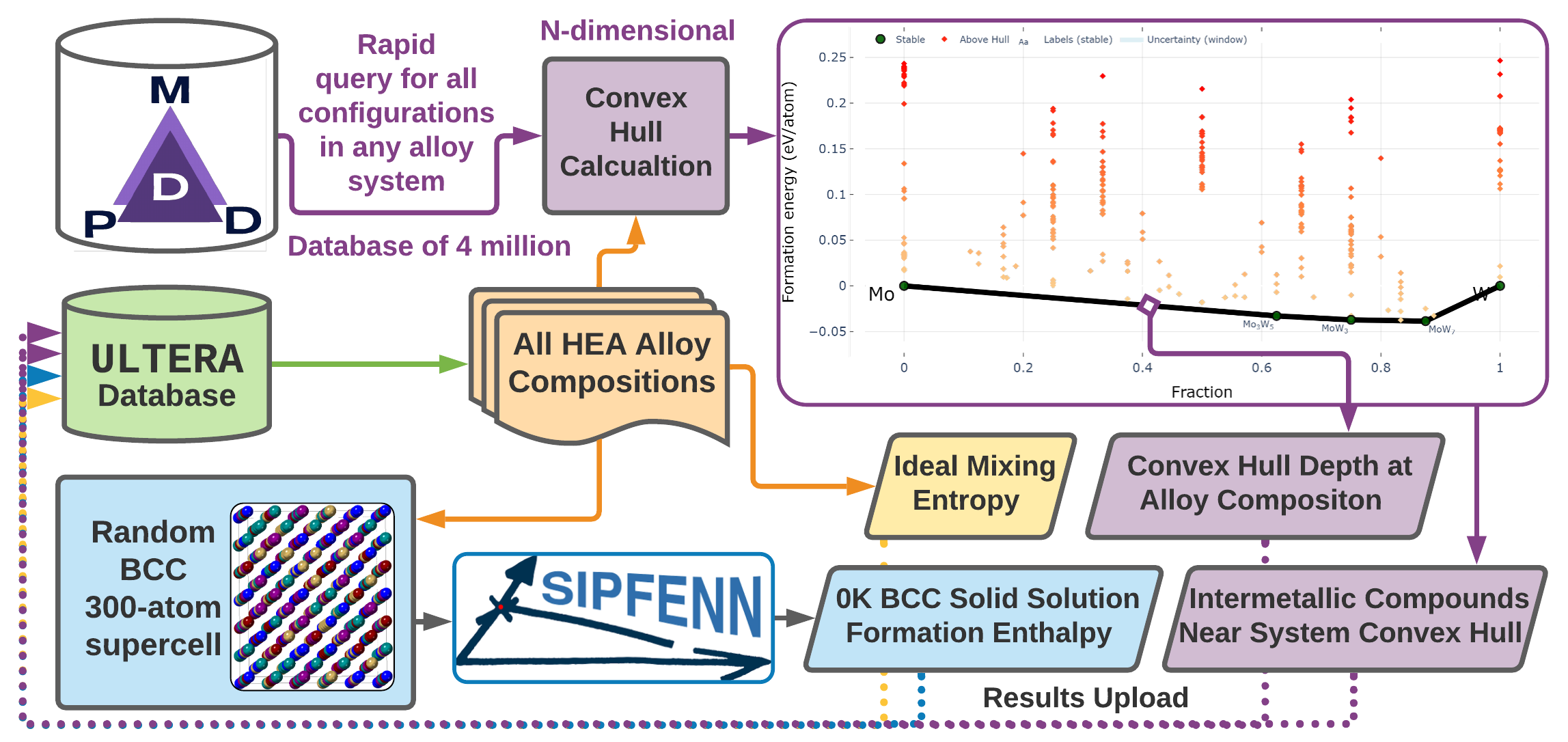}
    \caption{Conceptual schematic of how MPDD is directly utilized within ULTERA to include basic thermodynamic information in addition to indirect interaction through CALPHAD models created by \texttt{ESPEI}. For each composition, a convex hull of compounds present in the corresponding chemical system is calculated based on MPDD data and can be used (a) to immediately identify candidates for experimentally observed compounds based on 0K low-energy configurations or (b) convex hull depth can be used as an input to ML model indicating strength of interatomic interactions.}
    \label{ultera:fig:mpdd}
\end{figure}

Lastly, the ULTERA ecosystem leverages the 4.5 million atomistic data points discussed in Chapter~\ref{chap:mpdd} to extract some information about the 0K stability of structures in the given alloy's chemical system or its derivatives (including carbides, oxides, and nitrides), as schematically shown in Figure~\ref{ultera:fig:mpdd}. This data can be used to augment both (a) generative modeling efforts \cite{Debnath2021GenerativeAlloys}, as convex hull depth will likely be correlated with the formation of stable intermetallics, and (b) interpretation of experiments producing compositional maps \cite{Li2024DesignExperiments}, enabling easy identification of observed compounds of alloy's base elements or carbides formed by impurities. 

As shown in the example code below, the acquired data is structured as lists of names constructed from reduced chemical formulas, MPDD unique ID, and, if applicable, the identifier of the parent source. This allows one to quickly trace back where data came from and fetch additional information. Such schema also allows one to make queries over experimental data based on a specific compound to, for instance, find all literature publications where it may have occurred.

\begin{minted}[xleftmargin=3\parindent, linenos=true, fontsize=\small]{json}
{
  "_id": {
    "$oid": "638a980130b7ffbd9fa64947"
  },
  ...
  "compositionalProperties": {
    "carbides_EF_SIPFENN_NN30": [
      0,
      -1.143834464251995,
      -0.2136915735900402,
      -0.10025567188858986,
      -0.1811651661992073,
      -0.4918593931943178,
      ...
    ],
    "carbides_names": [
      "C-60762ac1b003796270f88dcb-cod-1512497",
      "HfC-60da0964462eba27079ee464-aflow:092f750896295d94",
      "Mo4C-60da968d462eba2707adb339-aflow:5c392f4f5bcb399b",
      "MoC-60dbd9c0462eba2707bd846c-aflow:b505b7a2c655c77c",
      "Mo2C-60dc26bd462eba2707c97ba4-aflow:f88a1b395cd63584",
      "NbC-60db3075462eba2707b7bfe9-aflow:949b754227969d87",
      ...
    ],
    "hydrides_EF_SIPFENN_NN30": [
      0,
      -0.4664096124470234,
      -0.30396447144448757,
      -0.44926807284355164,
      -0.26328401267528534,
      -0.25981644354760647,
      ...
    ],
    "hydrides_names": [
      "H2-6075af3cb003796270f4772d-mp-1181189",
      "HfH2-6074fcdbb003796270e70e2d-JVASP-18459",
      "HfH-60752d54b003796270e9d665-OQMD-1234971",
      "HfV2H-60752ef9b003796270ea2c41-OQMD-1287044",
      "MoH2-60752d76b003796270e9e11f-OQMD-1238065",
      "Mo2H-60752d76b003796270e9e11e-OQMD-1238064",
      ...
    ],
    "nitrides_EF_SIPFENN_NN30": [
      0,
      -1.6111129950731993,
      -1.6122573120519519,
      -1.8664569286629558,
      0,
      -0.7383877914398909,
      ...
    ],
    "nitrides_names": [
      "Hf-60752ce4b003796270e9ac2e-OQMD-1214798",
      "HfN-60752ad4b003796270e90b2f-OQMD-1108170",
      "HfVN-60da6fd8462eba2707ab2b4f-aflow:4e079c56a1634e26",
      "HfZrN-60da0169462eba27079d79af-aflow:0144b323e66b8b75",
      "Mo-60da6b40462eba2707aaa2a2-aflow:4b0a71282920e55b",
      "Mo2N-60752d88b003796270e9e5d4-OQMD-1239597",
      ...
    ],
    "oxides_EF_SIPFENN_NN30": [
      0,
      -3.924865636974573,
      -2.6255424302071333,
      0,
      -1.3926446568220854,
      -2.30870558321476,
      ...
    ],
    "oxides_names": [
      "Hf-60752ce4b003796270e9ac2e-OQMD-1214798",
      "HfO2-60dbe621462eba2707bfde2d-aflow:c24b2b7185250bc5",
      "HfO-60dbe7d6462eba2707c0306b-aflow:c41bd454b8f0284b",
      "Mo-60da6b40462eba2707aaa2a2-aflow:4b0a71282920e55b",
      "Mo2O-60752da2b003796270e9eb94-OQMD-1241669",
      "MoO2-6075bbdeb003796270f60c46-mvc-5806",
      ...
    ],
    "stable_EF_SIPFENN_NN30": [
      0,
      -0.16737285256385803,
      -0.14167926087975502,
      -1.0520951068028808,
      -0.046786464750766754,
      -0.07561314664781094,
      ...
    ],
    "stable_names": [
      "Hf-60752ce4b003796270e9ac2e-OQMD-1214798",
      "HfMo2-6074fcb1b003796270e70146-JVASP-14517",
      "Hf2Mo-60752d80b003796270e9e382-OQMD-1238828",
      "HfMo22-60dc048f462eba2707c46fee-aflow:dc1b0e16d62c30a5",
      "HfNb2-60dc12e9462eba2707c6529d-aflow:e6be499b110376f2",
      "HfNb5-60dc29b6462eba2707ca10ea-aflow:fbd6fc861701f121",
      ...
    ],
    ...
  }
}
\end{minted}


\chapter{Detecting Abnormalities in Materials Data}  \label{chap:pyqalloy}

\section{Introduction} \label{pyqalloy:sec:intro}

As explored in Chapter~\ref{chap:ultera}, handling of the "real" materials data requires many complex steps that introduce many potential points of failure, prompting investigations into the quality of the resulting datasets and the provenance of errors once they are identified. The first component of such efforts includes \emph{validity} of the data, which can generally be accomplished based on conformance to some defined specification and implemented in the data pipelines, like the ones discussed earlier in Section~\ref{ultera:sec:pipeline} or solutions built around frameworks like \texttt{pydantic} \cite{WelcomePydantic}.

Validating the data, however, does not, per se, make it correct, and most of the errors known to be common in alloy datasets, in fact, cannot be caught. Furthermore, even secondary hand-checking of the data, often taken as ground truth in parsing efforts, can often be very flawed since data may appear perfectly valid at a glance or be incorrectly interpreted.

\begin{figure}[H]
    \centering
    \includegraphics[width=0.6\textwidth]{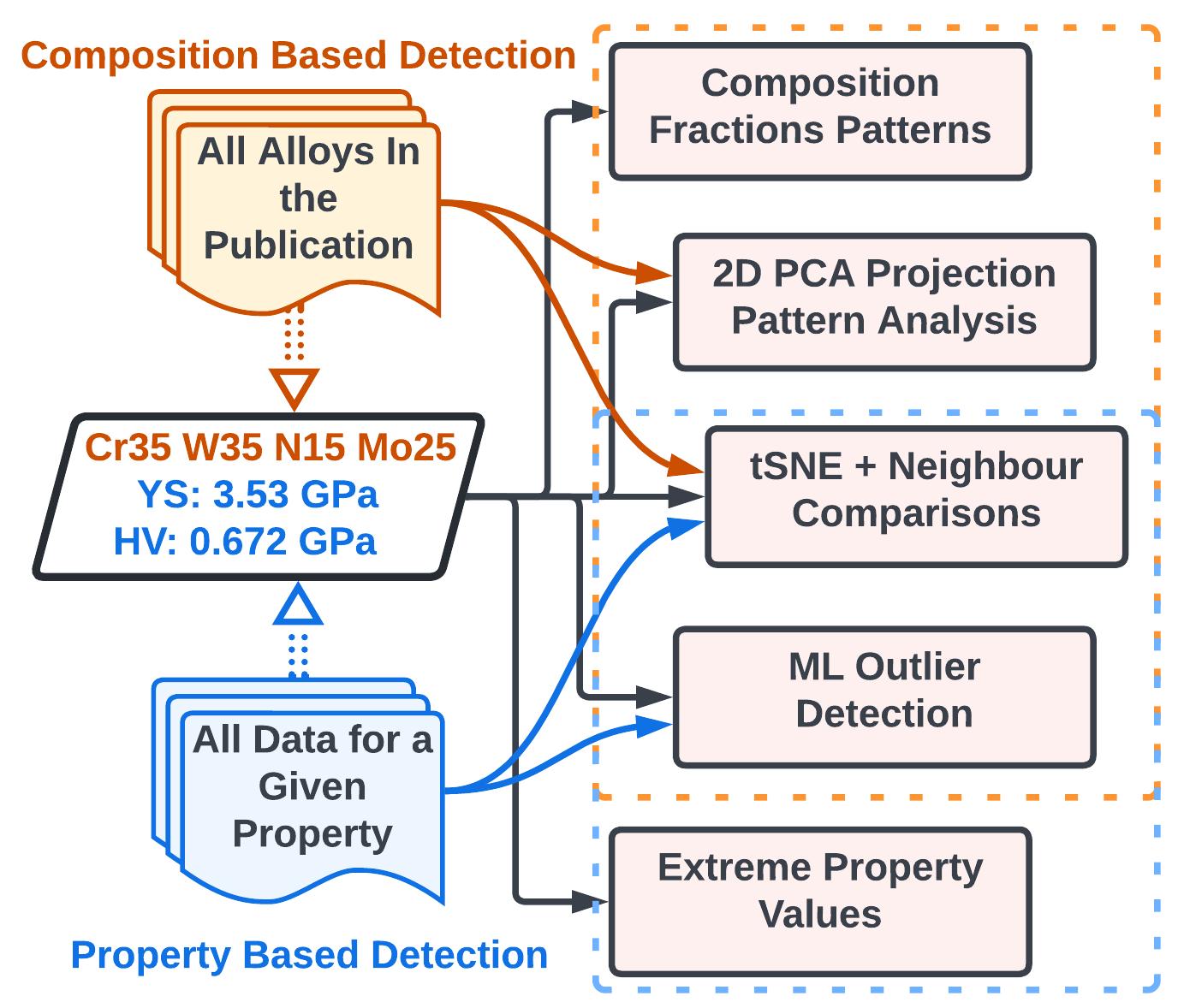}
    \caption{Schematic of \texttt{PyQAlloy} software operating in several contexts to detect and investigate common abnormalities discussed in Subsections \ref{pyqalloy:ssec:extreme} through \ref{pyqalloy:ssec:global}.}
    \label{pyqalloy:fig:schematic}
\end{figure}

This necessitates an alternative approach to ensuring high quality based around detection of abnormalities in the validated datasets, followed by re-verification of the suspicious data. In ULTERA Database, this has been implemented through \texttt{PyQAlloy} software, or \textit{Python toolset for Quality of Alloy data}, which connects to it and deploys several methods custom-designed for compositionally complex materials (CCMs) or high entropy alloys (HEAs) data, as shown schematically in Figure~\ref{pyqalloy:fig:schematic} and discussed throughout Section~\ref{pyqalloy:sec:abnormalities}.

Internally, \texttt{PyQAlloy} leverages several contexts each data exists in, ranging from a single aspect of it, through individual composition-structure-property relations, to the entire ULTERA database. Thanks to such a multi-context approach, a wide variety of errors can be caught.

\section{Common Abnormalities and Detection of Errors} \label{pyqalloy:sec:abnormalities}

Most commonly, the errors present in properly validated alloy datasets have to do with misreported compositions, numerical values, or both, which can add noise to the dataset or systematically skew it. The latter is particularly problematic to machine learning (ML) studies, as it can very significantly alter the extrapolation ability of the model. Throughout this Section, several common abnormalities are discussed alongside examples of errors that can cause them, starting from the least complex ones.

\subsection{Extreme Values} \label{pyqalloy:ssec:extreme}

\begin{figure}[H]
    \centering
    \includegraphics[width=0.95\textwidth]{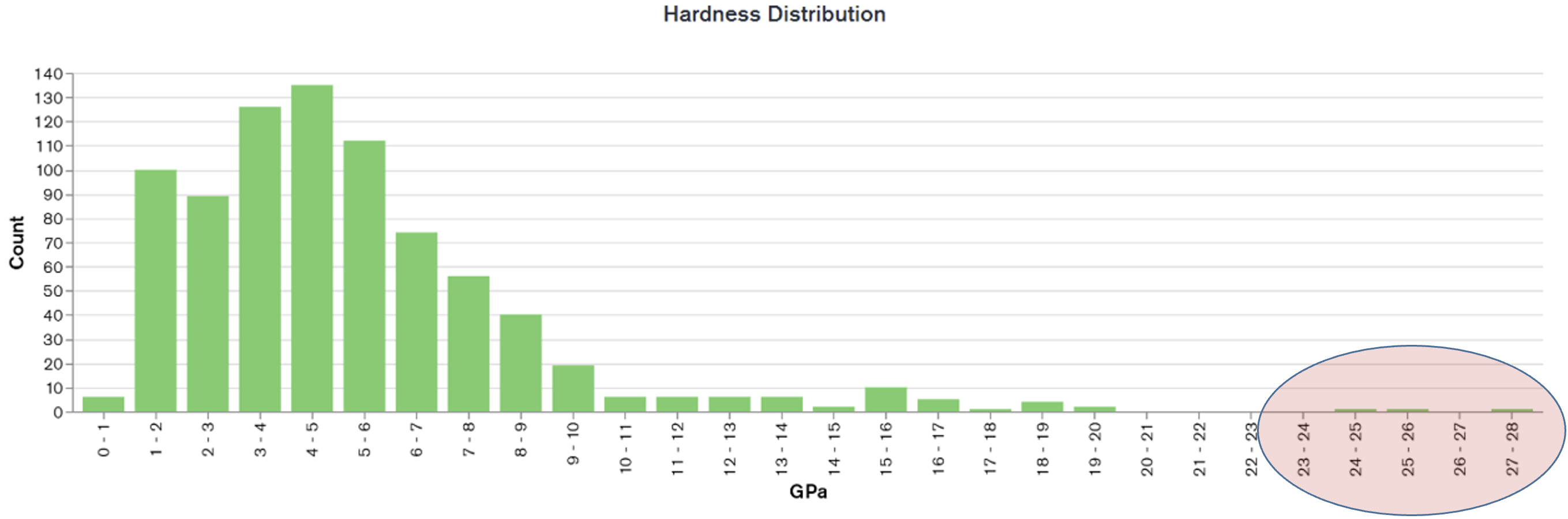}
    \caption{A legacy, pre-curation ULTERA hardness data histogram. The extreme, out-of-distribution values (highlighted) indicate possible misinterpretation. Two out of three were misreported due to the "extra 0" typo, while the highest one (27.5 GPa) was properly reported \ch{Mo_{40.5}Ni_{40.5}B_{10}Si_9} extremely hard metallic glass \cite{Kim2016DevelopmentRatios}. A similar analysis applies to extremely low values.}
    \label{pyqalloy:fig:extreme}
\end{figure}

The simplest abnormality one can detect, both in terms of concept and implementation, has to do with abnormally high or low values being present, which can be readily monitored through histograms, such as the one shown for old ULTERA hardness data in Figure~\ref{pyqalloy:fig:extreme}, constructed for each individual numerical value encountered in the database. Within ULTERA, such histograms are maintained alongside more descriptive extreme data tables and updated daily to inform responsible team members of possible errors.

\begin{figure}[H]
    \centering
    \includegraphics[width=0.95\textwidth]{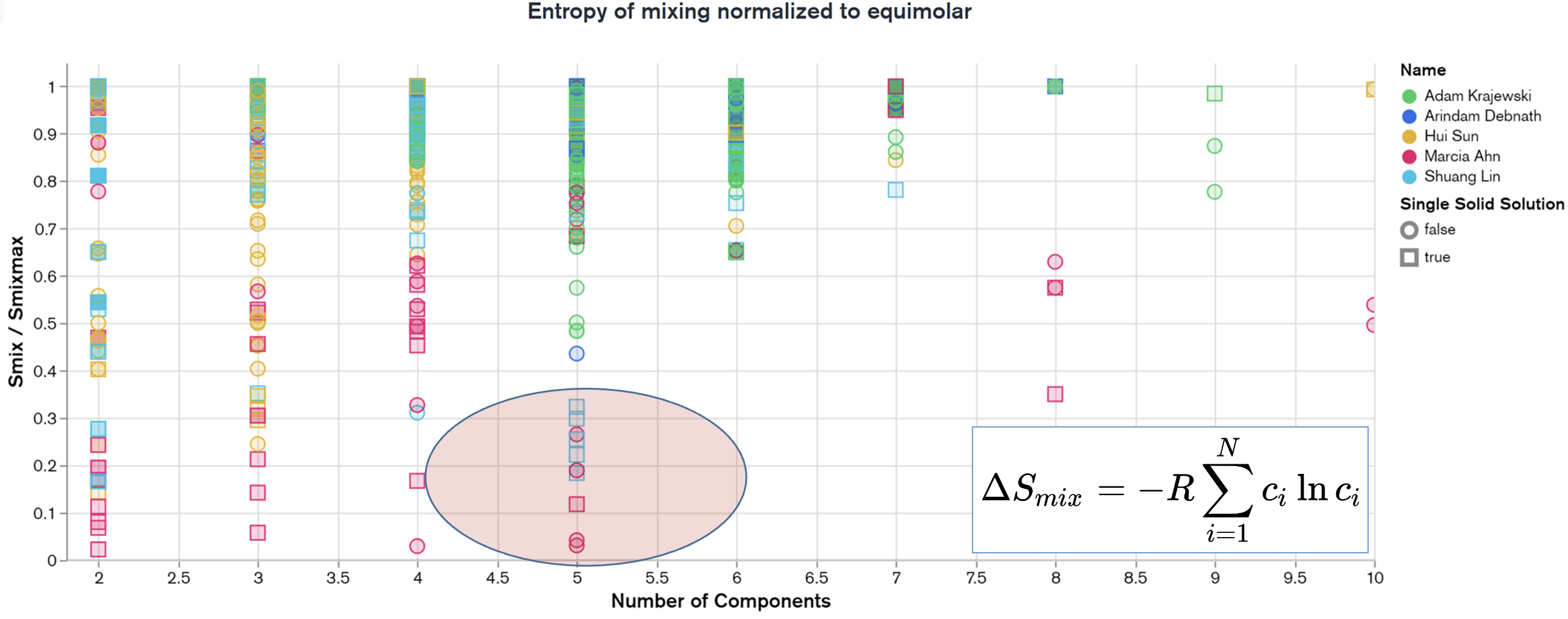}
    \caption{An example of extreme value detection in secondary data characteristics relative to expectations. The legacy, pre-curation ULTERA data plot of the ideal mixing entropy of a given alloy divided by the maximum ideal mixing entropy corresponding to the number of components present (value of $1$ indicates equimolar alloy). Extremely low values in this metric indicate a high likelihood of "double click" or "missing comma" typos at data parsing, which resulted in one element becoming highly dominant.}
    \label{pyqalloy:fig:lowentropy}
\end{figure}

More elaborate extreme value detection tools can also analyze the values in the context of expectations of what should be contained in the database. For instance, since the ULTERA database is focused on \emph{high} entropy alloys, \emph{low} entropy multi-component alloys, like the ones shown in Figure~\ref{pyqalloy:fig:lowentropy}, may be considered extreme in this context. This exact approach has been successfully used in ULTERA to fix several data points for which a fraction of one of the components has been accidentally inflated, causing the alloy to have a single principal element.

\subsection{Single Composition Patterns} \label{pyqalloy:ssec:singlecomp}

Similar to detecting abnormalities in the individual numerical values, one can also, relatively simply, detect specific abnormalities in the individual reported compositions. The most obvious yet the most common abnormality pattern deals with sums of atomic fractions being both (1) around $100$ or $1$ (within, e.g., $10$ or $20\%$ of them), and (2) not precisely $100$ or $1$, with minor deviations (e.g., $0.2\%$) allowed to account for numerical precision.

For instance, in the published (unprocessed) version of one of the literature datasets mentioned in Section~\ref{ultera:sec:datadescription} one can find a number of alloys reported from study by \citet{Stepanov2019EffectContent}, including \texttt{Cr20 Mn25 Fe40 Ni15 Al14}, which look fine at first but sum to $114$ what gets immediately detected by \texttt{PyQAlloy}'s \texttt{SingleCompositionAnalyzer} class. In this case, the researcher parsing the publication consistently forgot to subtract a fraction of added Al from the base alloy, causing the error to appear in 12 data points systematically and causing models to overestimate the strengthening effect of Al reported for these alloys \cite{Stepanov2019EffectContent}.

Another example, with a different root cause, has to do with 9 data points extracted from a well-known study by \citet{Lu2017DirectlyRange}, which included compositions like \texttt{Cr16 Fe16 Co16 Ni34.4 Al16} summing to $98.4$. In this case, the underlying cause for the error was likely incorrect normalization of the formula to sum to $100$, which in most circumstances could be considered noise, but in this study, it introduced significant error as the study investigated a narrow compositional range around a eutectic point; thus the errors became on the order of the study's domain and skewed the interpretation dramatically.

\subsection{Single Study Patterns}   \label{pyqalloy:ssec:singlestudy}

Going beyond individual values, one can leverage the fact that individual studies typically follow a limited number of approaches when it comes to setting up a hypothesis to be verified. The three that cover the vast majority of high entropy alloy studies are (1) gradually modifying an alloy in one or more ways, like adding a new element or mixing it with another alloy, (2) combinations or permutations of elements following particular pattern, like equimolar (equal fractions) or quantized (e.g., all fractions are $20\%$ multiples) alloy screenings \cite{Elder2023ComputationalDown-selection, Elder2023ComputationalValidation}, or (3) are actively driven and follow no pattern at all \cite{Rao2022MachineDiscovery}. 

\begin{figure}[H]
    \centering
    \includegraphics[width=0.95\textwidth]{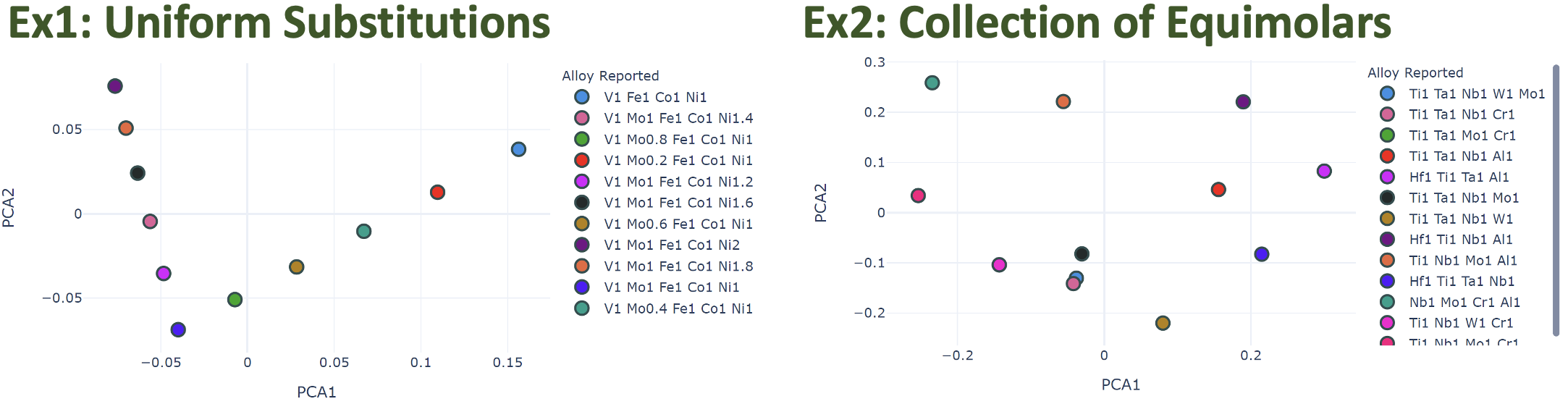}
    \caption{Expected patterns in the PCA projections of high entropy alloy composition vectors onto a 2D plane common for the vast majority of alloy design studies, which either (left) take an alloy and progressively modify it through elemental substitutions or mixing with another alloy in one or more ways, resulting in one or more linear patterns, or (right) test many different elemental combinations that are thought to possibly work well in the application in an anti-systematic fashion (in chemical space) does not follow any lower dimensional pattern and results in a point cloud. Breaks in these patterns, like out-of-line points or anisotropic point clouds, indicate possible errors and should be screened.}
    \label{pyqalloy:fig:expectedpatterns}
\end{figure}

When linearly projected to some lower-dimensional space of choice, using, for instance, principle component analysis (PCA), the resulting patterns fall into two categories. The first case typically follows one or more linear patterns with uniform distance between points (not required, per se), as linear trends in the high dimensional space can be preserved. Meanwhile, the second and third cases will typically result in a largely disordered pattern (given enough components) as there is no underlying low-dimensional pattern based on which the high-dimensional elemental space is being populated. Figure~\ref{pyqalloy:fig:expectedpatterns} depicts examples of both of these expected patterns.

While the second case of disordered embedding patterns cannot be easily exploited for abnormality detection, the breaks in single or multiple linear patterns can be taken advantage of in a semi-automated way through \texttt{PyQAlloy}'s \texttt{SingleDOIAnalyzer} class.

\begin{figure}[H]
    \centering
    \includegraphics[width=0.95\textwidth]{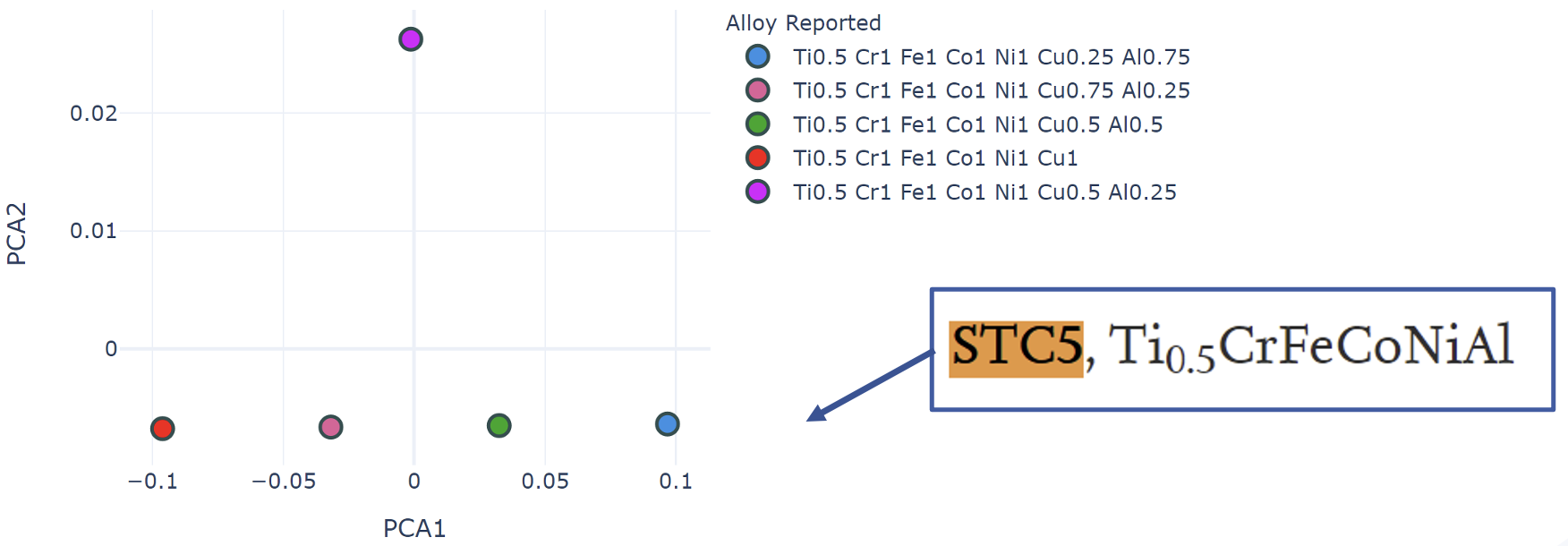}
    \caption{An example of an out-of-line pattern was detected in a literature review study. It was caused by the researcher parsing a publication incorrectly, noting the composition relative to the source \cite{Wang2009AtomicAlloy}. Composition similar to other points makes it look right; thus, such errors are nearly impossible to catch using other methods.}
    \label{pyqalloy:fig:patternbreak1}
\end{figure}

For instance, in an older version of the ULTERA dataset, a set of alloys from the study by \citet{Wang2009AtomicAlloy}, included \texttt{Ti0.5 Cr1 Fe1 Co1 Ni1 Cu0.5 Al0.25}, which appeared correct on its own, but when put in the context of other alloys from the study it broke the expected linear pattern. It was caught by \texttt{PyQAlloy} and then found to be \texttt{Ti0.5 Cr1 Fe1 Co1 Ni1 Al1} in the publication; as depicted in Figure~\ref{pyqalloy:fig:patternbreak1}.

\begin{figure}[H]
    \centering
    \includegraphics[width=0.95\textwidth]{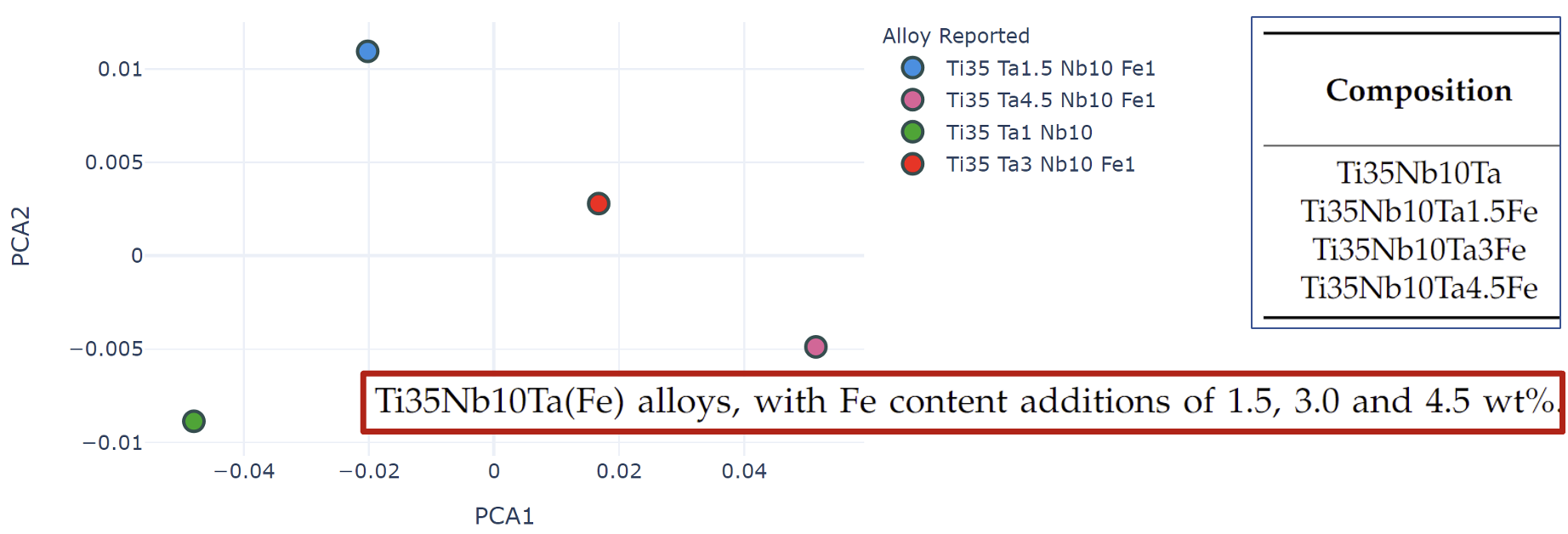}
    \caption{An example of an out-of-line pattern detected in a literature review study, in which chemical formulas present in the source publication \cite{Amigo2019MechanicalApplications} are correctly parsed. However, they are shorthand composition notations rather than actual chemical formulas of the studied material and must be interpreted; thus, they are incorrectly reported. Such misinterpretations typically follow some incorrect patterns locally but fail to do so if any component is removed or added to the mix, as depicted here.}
    \label{pyqalloy:fig:patternbreak2}
\end{figure}

In addition to typing or copying errors, such single study pattern analysis can also highlight interpretation errors, as such cases tend to break every time a fraction of some component goes to $0$. For instance, in an older version of the ULTERA dataset, a set of 4 alloys from a study by \citet{Amigo2019MechanicalApplications} was reported after passing validations. However, when processed through \texttt{PyQAlloy}, it was found that one of the alloys, with no \texttt{Fe} reported, broke the pattern as the fraction of \texttt{Ta} was expected to drop to $0$ instead, as illustrated in Figure~\ref{pyqalloy:fig:patternbreak2}. Further verification revealed that all 4 have been misreported, as the original study used composition notation which (a) reported atomic fractions as prefixes rather than suffixes of the element symbol, (b) assumed Ti to have the "balance" fraction of $100$ less the sum of other elements, and (c) used weight fractions, which collectively broke the pattern.

\subsection{Global Patterns} \label{pyqalloy:ssec:global}

\begin{figure}[H]
    \centering
    \includegraphics[width=0.95\textwidth]{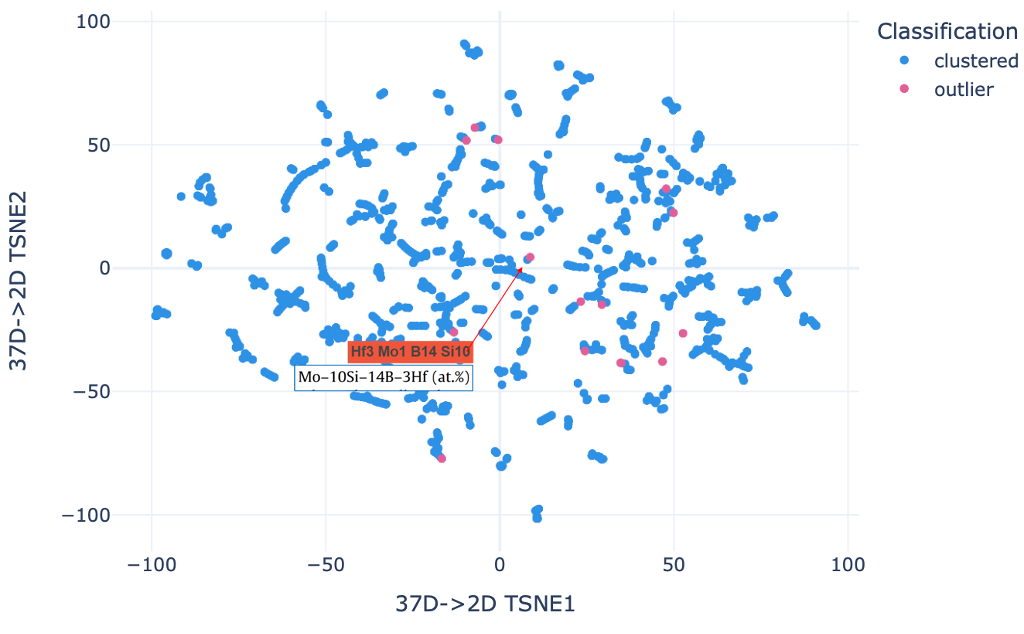}
    \caption{2D tSNE embedding of all chemical compositions present in ULTERA based on alloy neighborhoods overlaid with outliers detected through the DBSCAN method operating in the high-dimensional real composition space. Singular outliers between tSNE clusters are expected and indicate novel compositions, while outliers within clusters indicate far-removed members of an alloy family, which are likely incorrect. The highlighted \ch{B-Hf-Mo-Si} alloy is close to other alloys in that system, but a fraction of \ch{Mo} was omitted by the parsing researcher. In the depicted case, lower-level methods did not detect abnormality because only one alloy was reported.}
    \label{pyqalloy:fig:patternglobal}
\end{figure}

Lastly, one can also leverage the collected ULTERA Database to detect abnormalities between different studies, which can be a powerful technique, especially in cases where they report few data points individually. For instance, a single alloy of \texttt{Hf3 Mo1 B14 Si10} has been previously reported from a study by \citet{Yu2012TensileTemperatures} and did not appear suspicious, except to a person possessing expert knowledge. However, when \emph{entire database} was passed through \texttt{PyQAlloy}'s \texttt{AllDataAnalyzer} class, it was found to be in the neighborhood of other Hf-Mo-B-Si alloys reported in other works based on its t-distributed stochastic neighbor embedding (t-SNE) \cite{HintonStochasticEmbedding}, but at the same time be very far from any other member of this neighborhood based on the DBSCAN clustering \cite{Ester1996ANoise}, as depicted in Figure~\ref{pyqalloy:fig:patternglobal}. After verification, it was found that the fraction of Mo ($73\%$) has been missing.

\section{Software Implementation} \label{pyqalloy:sec:software}

The majority of the abnormality detection methods developed for ULTERA have been implemented as a user-tool \texttt{PyQAlloy}, which was released as a free, open-source software (FOSS) under the MIT license. It is available (1) as source code through a GitHub repository at \href{https://github.com/PhasesResearchLab/PyQAlloy}{github.com/PhasesResearchLab/PyQAlloy}, (2) through the PyPI, and (3) through the \texttt{conda-forge} channel. It is automatically tested across several platforms on a periodic basis. Furthermore, its documentation and use examples were made available through \href{https://pyqalloy.ultera.org/}{pyqalloy.ultera.org} web page.


\chapter{Optimization of Compositional Dataset Domain towards Reliable Machine Learning Training and Deployment} \label{chap:nimcso}

\acknowledge{
This chapter adapts verbatim a preprint by \citet{Krajewski2024NimCSO:Optimization} submitted for publication to the Journal of Open Source Software. This work was co-authored with Arindam Debnath, Wesley F. Reinhart, Allison M. Beese, and Zi-Kui Liu. All text and associated software was written by Adam M. Krajewski, while co-authors provided edits and guidance.
}

\section{Background and Motivation} \label{nimcso:sec:background}

\texttt{nimCSO} is an interdisciplinary tool applicable to any field
where data is composed of a large number of independent components and
their interaction is of interest in a modeling effort, ranging from
market economics, through medicine where drug interactions can have a
significant impact on the treatment, to materials science, where the
composition and processing history are critical to resulting properties.
The latter has been the root motivation for the development of
\texttt{nimCSO} within the \href{https://ultera.org}{ULTERA Project
(ultera.org)} carried under the
\href{https://arpa-e.energy.gov/?q=arpa-e-programs/ultimate}{US DOE
ARPA-E ULTIMATE} program, which aims to develop a new generation of
ultra-high temperature materials for aerospace applications, through
generative machine learning models \cite{Debnath2021} driving
thermodynamic modeling, alloy design, and manufacturing \cite{Li2024}.

One of the most promising materials for such applications are the
aforementioned CCMs and their metal-focused subset of Refractory High
Entropy Alloys (RHEAs) \cite{Senkov2018}, which have rapidly grown since first proposed by
\citet{Cantor2004} and
\citet{Yeh2004}. Contrary to most of
the traditional alloys, they contain many chemical elements (typically
4-9) in similar proportions in the hope of thermodynamically stabilizing
the material by increasing its configurational entropy
(\(\Delta S_{conf} = \Sigma_i^N x_i \ln{x_i}\) for ideal mixing of \(N\)
elements with fractions \(x_i\)), which encourages sampling from a large
palette of chemical elements. At the time of writing, the ULTERA
Database is the largest collection of HEA data, containing over 6,300
points manually extracted from almost 550 publications. It covers 37
chemical elements resulting in extremely large compositional spaces (see Section \ref{nimplex:ssec:combinatorialcomplexities});
thus, it becomes critical to answer questions like \emph{``Which
combination of how many elements will unlock the most expansive and
simultaneously dense dataset?''} which has \(2^{37}-1\) or 137 billion
possible solutions.

Another significant example of intended use is to perform similar
optimizations over large (many millions) datasets of quantum mechanics
calculations spanning 93 chemical elements and accessible through
OPTIMADE API \cite{Evans2024} described in Section \ref{mpdd:sec:optimade}.

\section{nimCSO - A Nim Package for Compositional Space Optimization} \label{nimcso:sec:nimcsosoftware}

\texttt{nimCSO} is a high-performance tool implementing several methods
for selecting components (data dimensions) in compositional datasets,
which optimize the data availability and density for applications such
as machine learning. Making said choice is a combinatorically hard
problem for complex compositions existing in highly dimensional spaces
due to the interdependency of components being present. Such spaces are
encountered, for instance, in materials science, where datasets on
Compositionally Complex Materials (CCMs) often span 20-45 chemical
elements, 5-10 processing types, and several temperature regimes, for up
to 60 total data dimensions.

At its core, \texttt{nimCSO} leverages the metaprogramming ability of
the Nim language \cite{Rumpf2006Nim:github.com/nim-lang/Nim} to
optimize itself at the compile time, both in terms of speed and memory
handling, to the specific problem statement and dataset at hand based on
a human-readable configuration file. As demonstrated in Section \ref{nimcso:sec:methods}, \texttt{nimCSO} reaches the physical limits of the hardware (L1
cache latency) and can outperform an efficient native Python
implementation over 400 times in terms of speed and 50 times in terms of
memory usage (\emph{not} counting interpreter), while also outperforming
NumPy implementation 35 and 17 times, respectively, when checking a
candidate solution.

\texttt{nimCSO} is designed to be both (1) a user-ready tool,
implementing two efficient brute-force approaches (for handling up to 25
dimensions), a custom search algorithm (for up to 40 dimensions), and a
genetic algorithm (for any dimensionality), and (2) a scaffold for
building even more elaborate methods in the future, including heuristics
going beyond data availability. All configuration is done with a simple
human-readable \texttt{YAML} config file and plain text data files,
making it easy to modify the search method and its parameters with no
knowledge of programming and only basic command line skills.

\section{Novel Methods and their Performance} \label{nimcso:sec:methods}

\subsection{Overview} \label{nimcso:sec:methodsoverview}

\begin{figure}[H]
\centering
\includegraphics[width=0.95\textwidth]{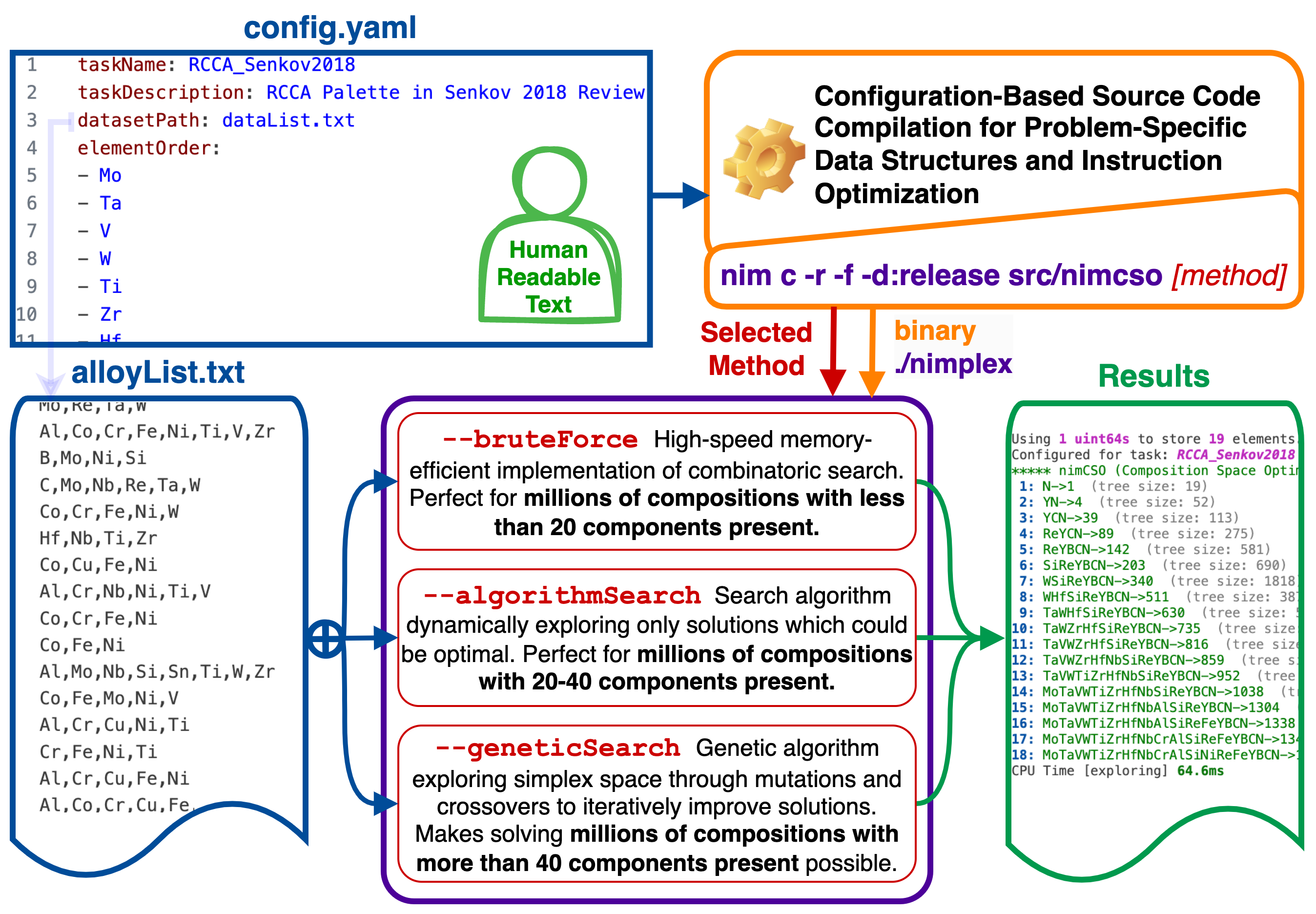}
\caption{Schematic of core nimCSO data flow with a description of key
methods. Metaprogramming techniques are used to compile the software optimized to
the human-readable data and configuration files at hand.}
\label{nimcso:fig:main}
\end{figure}

As shown in Figure \ref{nimcso:fig:main}, \texttt{nimCSO} can be used as a
user-tool based on human-readable configuration and a data file
containing data ``elements'' which can be any strings representing
problem-specific names of, e.g., market stocks, drug names, or chemical
formulas. A single command is then used to recompile
(\texttt{nim\ c\ -f}) and run (\texttt{-r}) problem
(\texttt{-d:configPath=config.yaml}) with \texttt{nimCSO}
(\texttt{src/nimcso}) using one of several methods. Advanced users can
also quickly customize the provided methods with brief scripts using the
\texttt{nimCSO} as a data-centric library.

Internally, \texttt{nimCSO} is built around storing the data and
solutions in one of two ways. The first is as bits inside an integer
(\texttt{uint64}), which allows for the highest speed and lowest memory
consumption possible but is limited to 64 dimensions and does not allow
for easy extension to other use cases; thus, as of publication, it is
used only in a particular \texttt{bruteForceInt} routine. The second
one, used in \texttt{bruteForce}, \texttt{algorithmSearch}, and
\texttt{geneticSearch}, implements a custom easily extensible
\texttt{ElSolution} type containing heuristic value and
\texttt{BitArray} payload, which is defined at compile time based on the
configuration file to minimize necessary overheads. Both encodings
outperform typical native Python and NumPy implementations, as shown in
Table \ref{nimcso:tab:methodbenchmarks}.

\begin{longtable}[]{@{}
  >{\raggedright\arraybackslash}p{(\columnwidth - 8\tabcolsep) * \real{0.1478}}
  >{\raggedright\arraybackslash}p{(\columnwidth - 8\tabcolsep) * \real{0.1478}}
  >{\centering\arraybackslash}p{(\columnwidth - 8\tabcolsep) * \real{0.1652}}
  >{\centering\arraybackslash}p{(\columnwidth - 8\tabcolsep) * \real{0.2609}}
  >{\centering\arraybackslash}p{(\columnwidth - 8\tabcolsep) * \real{0.2522}}@{}}
\tabularnewline
\toprule\noalign{}
\begin{minipage}[b]{\linewidth}\raggedright
Tool
\end{minipage} & \begin{minipage}[b]{\linewidth}\raggedright
Object
\end{minipage} & \begin{minipage}[b]{\linewidth}\centering
Time per Dataset
\end{minipage} & \begin{minipage}[b]{\linewidth}\centering
Time per Entry \emph{(Relative)}
\end{minipage} & \begin{minipage}[b]{\linewidth}\centering
Database Size \emph{(Relative)}
\end{minipage} \\
\midrule\noalign{}
\endfirsthead
\toprule\noalign{}
\begin{minipage}[b]{\linewidth}\raggedright
Tool
\end{minipage} & \begin{minipage}[b]{\linewidth}\raggedright
Object
\end{minipage} & \begin{minipage}[b]{\linewidth}\centering
Time per Dataset
\end{minipage} & \begin{minipage}[b]{\linewidth}\centering
Time per Entry \emph{(Relative)}
\end{minipage} & \begin{minipage}[b]{\linewidth}\centering
Database Size \emph{(Relative)}
\end{minipage} \\
\midrule\noalign{}
\endhead
\midrule\noalign{}
\texttt{Python}\textsuperscript{3.11} & \texttt{set} & 327.4 µs & 152.3
ns \emph{(x1)} & 871.5 kB \emph{(x1)} \\
\texttt{NumPy}\textsuperscript{1.26} & \texttt{array} & 40.1 µs & 18.6
ns \emph{(x8.3)} & 79.7 kB \emph{(x10.9)} \\
\texttt{nimCSO}\textsuperscript{0.6} & \texttt{BitArray} & 9.2 µs & 4.4
ns \emph{(x34.6)} & 50.4 kB \emph{(x17.3)} \\
\texttt{nimCSO}\textsuperscript{0.6} & \texttt{uint64} & 0.79 µs & 0.37
ns \emph{(x413)} & 16.8 kB \emph{(x52)} \\
\bottomrule\noalign{}
\caption{Benchmarks of average time to evaluate how many datapoints
would be lost if 5 selected components were removed from a dataset with
2,150 data points spanning 37 components (10,000 run average), and
the size of the data structure representing the dataset. Values were
obtained by running scripts in \texttt{benchmarks} on Apple M2
Max CPU.}
\label{nimcso:tab:methodbenchmarks}
\endlastfoot
\end{longtable}

\subsection{High-Performance Brute-Force through Compile Time Metaprogramming} \label{nimcso:ssec:bruteforce}

The brute-force search is a naïve method of evaluating all
possibilities; however, its near-zero overhead can make it the most
efficient for small problems. In this implementation, all entries in the
\emph{power set} of \(N\) considered elements are represented as a range
of integers from \(0\) to \(2^{N} - 1\), and used to initialize
\texttt{uint64}/\texttt{BitArray}s on the fly. To minimize the memory
footprint of solutions, the algorithm only keeps track of the best
solution for a given number of elements present in the solution. Current
implementations are limited to 64 elements, as it is not feasible beyond
approximately 30 elements; however, the one based on \texttt{BitArray}
could be easily extended if needed.

\subsection{Algorithmic Searches} \label{nimcso:ssec:algorithmicsearches}

The algorithm implemented in the \texttt{algorithmSearch} routine,
targeting high dimensional problems (20-50), iteratively expands and
evaluates candidates from a priority queue (implemented through an
efficient binary heap \cite{Williams1964AlgorithmHeapsort} while leveraging the fact that \emph{the number of data points lost when removing elements \texttt{A} and \texttt{B} from the dataset
has to be at least as large as when removing either \texttt{A} or
\texttt{B} alone} to delay exploration of candidates until they can
contribute to the solution. Furthermore, to (1) avoid revisiting the
same candidate without keeping track of visited states and (2) further
inhibit the exploration of unlikely candidates, the algorithm
\emph{assumes} that while searching for a given order of solution,
elements present in already expanded solutions will not improve those
not yet expanded. This effectively prunes candidate branches requiring
two or more levels of backtracking. In the authors' tests, this method
has generated the same results as \texttt{bruteForce}, except for
occasional differences in the last explored solution.

\subsection{Searches Based on Genetic Algorithms} \label{nimcso:ssec:geneticsearches}

\begin{figure}[H]
  \centering
  \includegraphics[width=0.3\textwidth]{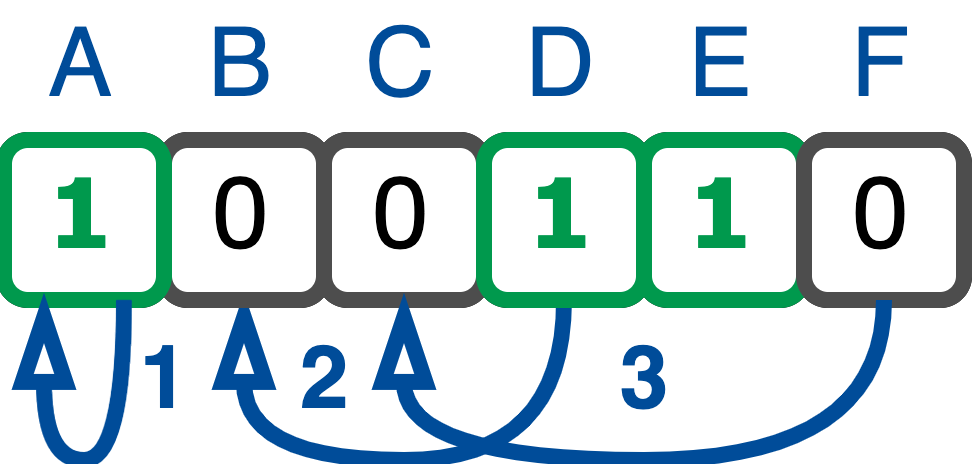}
  \caption{Schematic of \texttt{mutate} procedure where bits are swapping
  randomly, so that (1) bit can swap itself, (2) bits can swap causing a
  flip, or (3) bits can swap with no effect.}
  \label{nimcso:fig:mutate}
\end{figure}

Beyond 50 components, the
\protect\hyperlink{algorithm-based-search}{algorithm-based} method will
likely run out of memory on most personal systems. The
\texttt{geneticSearch} routine resolves this issue through an evolution
strategy to iteratively improve solutions based on custom
\texttt{mutate} and \texttt{crossover} procedures. Both are of uniform
type \cite{Goldberg1989} with
additional constraint of Hamming weight
\cite{Knuth} preservation in order to
preserve number of considered elements in parents and offspring. In
\texttt{mutate} this is achieved by using purely random bit swapping,
rather than more common flipping, as demonstrated in the Figure
\ref{nimcso:fig:mutate}.

\begin{figure}[H]
\centering
\includegraphics[width=0.85\textwidth]{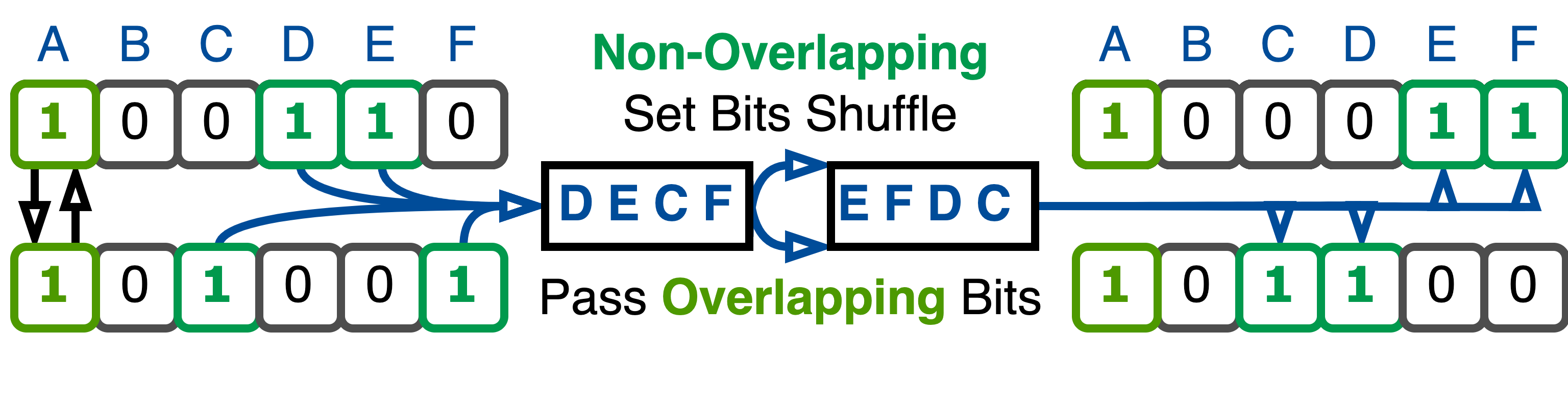}
\caption{Schematic of uniform \texttt{crossover} procedure preserving
Hamming weight implemented in \texttt{nimCSO}.}
\label{nimcso:fig:crossover}
\end{figure}

Meanwhile, in \texttt{crossover}, this constraint is satisfied by
passing overlapping bits directly, while non-overlapping bits are
shuffled and distributed at positions present in one of the parents, as
shown in Figure \ref{nimcso:fig:crossover}.

The above are applied iteratively, with best solutions carried to next
generation, until the solution converges or the maximum number of
iterations is reached. Unlike the other methods, the present method is
not limited by the number of components and lets user control both time
and memory requirements, either to make big problems feasible or to get
a good-enough solution quickly in small problems. However, it comes with
no optimality guarantees.

\subsection{Use Examples and Benchmarks}

The tool comes with two pre-defined example problems to demonstrate its
use. The first one is defined in the default \texttt{config.yaml} file
and goes through the complete dataset of 2,150 data points spanning 37
components in \texttt{dataList.txt} based on the ULTERA Dataset, described in Chapter \ref{chap:ultera}, as of January 2024. It is
intended to showcase \texttt{algorithmSearch}/\texttt{-as} and
\texttt{geneticSearch}/\texttt{-gs} methods, as brute-forcing would take
around one day. The second one is defined in \texttt{config\_rhea.yaml}
and uses the same dataset but a limited scope of components critical to
RHEAs \cite{Senkov2018} and is
intended to showcase \texttt{bruteForce}/\texttt{-bf} and
\texttt{bruteForceInt}/\texttt{-bfi} methods. With four simple commands
(see Table \ref{nimcso:tab:benchmarksmethods} below), the user can compare the methods' performance and the solutions' quality.

\begin{longtable}[]{@{}
  >{\raggedright\arraybackslash}p{(\columnwidth - 4\tabcolsep) * \real{0.6623}}
  >{\raggedright\arraybackslash}p{(\columnwidth - 4\tabcolsep) * \real{0.1429}}
  >{\centering\arraybackslash}p{(\columnwidth - 4\tabcolsep) * \real{0.1818}}@{}}
\caption{Four example tasks alongside typical CPU time and memory usage
on Apple M2 Max.}
\label{nimcso:tab:benchmarksmethods}
\tabularnewline
\toprule\noalign{}
\begin{minipage}[b]{\linewidth}\raggedright
Task Definition (\texttt{nim\ c\ -r\ -f\ -d:release\ ...})
\end{minipage} & \begin{minipage}[b]{\linewidth}\raggedright
Time (s)
\end{minipage} & \begin{minipage}[b]{\linewidth}\centering
Memory (MB)
\end{minipage} \\
\midrule\noalign{}
\endfirsthead
\toprule\noalign{}
\begin{minipage}[b]{\linewidth}\raggedright
Task Definition (\texttt{nim\ c\ -r\ -f\ -d:release\ ...})
\end{minipage} & \begin{minipage}[b]{\linewidth}\raggedright
Time (s)
\end{minipage} & \begin{minipage}[b]{\linewidth}\centering
Memory (MB)
\end{minipage} \\
\midrule\noalign{}
\endhead
\bottomrule\noalign{}
\endlastfoot
\texttt{-d:configPath=config.yaml\ src/nimcso\ -as} & 302s & 488 MB \\
\texttt{-d:configPath=config.yaml\ src/nimcso\ -gs} & 5.8s & 3.2 MB \\
\texttt{-d:configPath=config\_rhea.yaml\ src/nimcso\ -as} & 0.076s & 2.2
MB \\
\texttt{-d:configPath=config\_rhea.yaml\ src/nimcso\ -gs} & 0.429s & 2.1
MB \\
\texttt{-d:configPath=config\_rhea.yaml\ src/nimcso\ -bf} & 4.171s & 2.0
MB \\
\texttt{-d:configPath=config\_rhea.yaml\ src/nimcso\ -bfi} & 0.459s &
2.0 MB \\
\end{longtable}


\chapter{Inverse Design of Compositionally Complex Alloys} \label{chap:inversedesign}

\acknowledge{
This chapter is primarily a new piece of writing which intended to put ULTERA Database described in Chapter \ref{chap:ultera} in the target application context. It adapts verbatim selected critical excerpts and figures of past co-authored publications led by Arindam Debnath and Wenjie Li published under \citet{Debnath2021GenerativeAlloys}, \citet{Debnath2023ComparingAlloys}, and \citet{Li2024DesignExperiments}. Permission to reproduce was obtained from respective journals or was admissible under license. Other authors on these works were Lavanya Raman, Marcia Ahn, Shuang Lin, Shunli Shang, Hui Sun, Shashank Priya, Jogender Singh, Wesley F Reinhart, Zi-Kui Liu and Allison M Beese. Majority of content adopted in this chapter is has specifically been written by Adam M. Krajewski or co-written with Arindam Debnath. The majority of described ML efforts has been led by Arindam Debnath, while Adam M. Krajewski led training data acquisition and processing, as well as model deployment pipelines.
}

\section{Principles of Machine Learning Driven Design Patterns} \label{inverse:sec:principles}

Traditionally, over thousands of years, humanity performed various design tasks either with no approach (i.e., by chance) or utilizing scientific-method-based \emph{rational approach} where results of experiments were used to update the design and achieve desired results. With the advance of science, engineering, and communication methods, this approach has evolved into the de-facto standard for most studies, with experts guiding them.

Thanks to the improvements in a number of key computational methods developed over the last several decades, including machine learning (ML) modeling, nowadays design efforts are often guided through \emph{screening patterns or optimizations algorithms} which dictate what to perform next and can leverage surrogate models, which often run orders of magnitude faster relative to experiments in a laboratory or expensive computational methods.

\begin{figure}[H]
    \centering
    \includegraphics[width=0.96\textwidth]{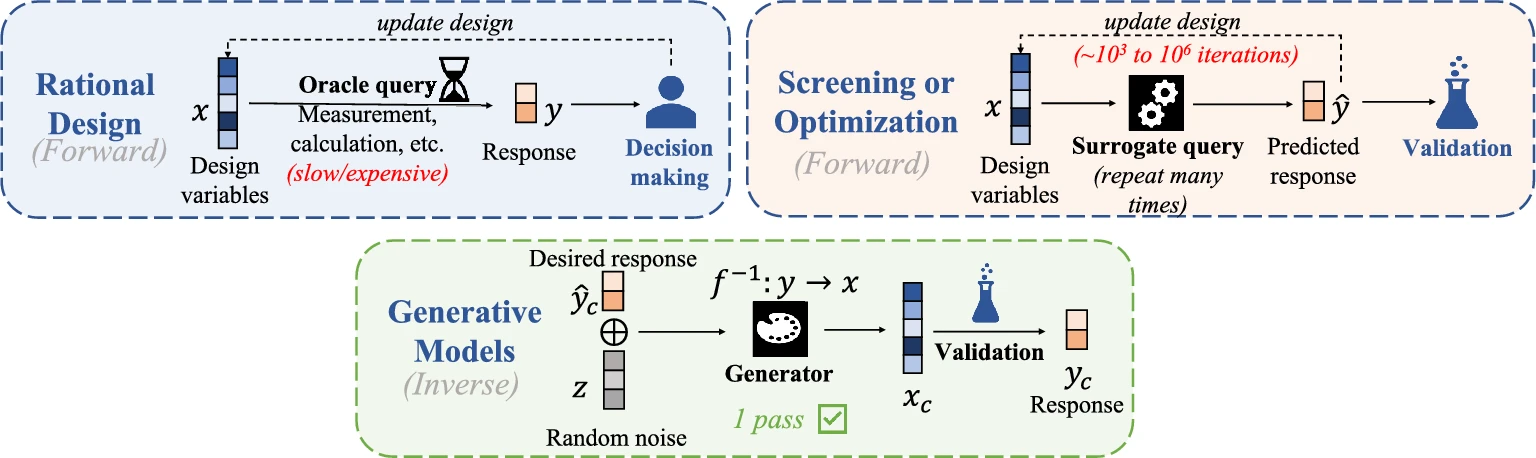}
    \caption{A schematic contrasting the design approaches based on (blue) traditional, \emph{rational} forward design, (orange) \emph{screening} forward design, and (blue) \emph{generative} inverse design. Reproduced from \cite{Debnath2023ComparingAlloys} under permission.}
    \label{inverse:fig:designs}
\end{figure}

Most recently, the inverse design paradigm, based around \emph{generative models}, is becoming increasingly popular thanks to further increases in computational power and novel machine learning models, such as generative adversarial networks (GANs) \cite{Goodfellow2020GenerativeNetworks}, which can generate distributions of outputs mimicking prior data, while finding underlying patterns that may be hard to find for even an expert due to high data complexity or quantity. Figure~\ref{inverse:fig:designs} summarizes and contrasts these three design paradigms.

\section{Conditional Generative Design} \label{inverse:sec:cgan}

\begin{figure}[H]
    \centering
    \includegraphics[width=0.55\textwidth]{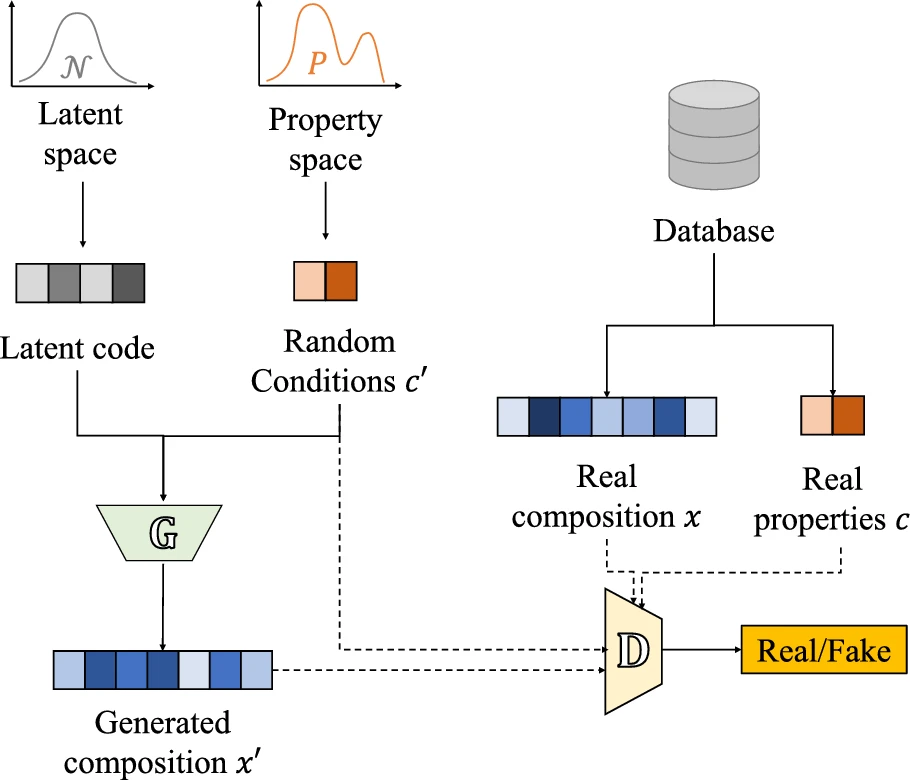}
    \caption{The conditional GAN (cGAN) architecture. It is a variation of the GAN architecture, with the modification that an additional conditioning vector is provided as an input to both $\mathbb{G}$ and $\mathbb{D}$, allowing biasing the model to certain property values. Reproduced from \cite{Debnath2023ComparingAlloys} under permission.}
    \label{inverse:fig:cgan}
\end{figure}

Materials design tends to focus on the ability to obtain specific key properties, unlike some other ML-related studies like cat or dog image generation, that may be primarily concerned with realistic sampling of the underlying distribution \cite{Ching-YuanBaiJuliaElliott2019GenerativeImages}. Thus, the creation of generator models for materials is usually accomplished with a slightly more elaborate architecture, shown in Figure~\ref{inverse:fig:cgan} and called conditional GAN or \emph{cGAN}, which enables biasing the generated outputs (e.g., compositions) towards specific property values \cite{Mirza2014ConditionalNets, Arjovsky2017WassersteinNetworks}.

\section{Building and Evaluating Generative Models in ULTERA} \label{inverse:sec:buildinggan}

\subsection{Data and Training} \label{inverse:ssec:datatrain}

Any material design effort requires close integration with existing literature data and scientific techniques to validate generated samples beyond the known set. This is especially true for generative design studies, as the models are specifically tasked with reproducing the underlying distribution. 

Thus, if the underlying data possesses systematically incorrect artifacts, like the ones explored in Chapter \ref{chap:pyqalloy}, one can expect a generative model to learn them. In the case of errors that bring data outside of the typical range, like the ones demonstrated in Figures \ref{pyqalloy:fig:extreme}, \ref{pyqalloy:fig:patternbreak2}, or \ref{pyqalloy:fig:patternbreak2}, the out-of-range values may lead to phenomena like (1) substantial overestimation of errors if they are present in the test set or (2) strongly bias the model to repeatedly generate them for a wide range of latent space samplings because of locally exceptional values. Thus, it becomes critical to use tools like \texttt{PyQAlloy} to ensure that data has few systematic errors present in it.

Furthermore, the data sparsity in terms of compositional (generator output) coverage may strongly contribute to the model overly exploiting some regions under strong conditioning due to insufficient data to establish underlying distribution. Thus, tools like \texttt{nimCSO} become critical in establishing which dimensions of the dataset to model in order to preserve the most knowledge while rejecting data points from which learning may be impossible.

As explored in \citet{Debnath2021GenerativeAlloys}, once a sufficient dataset has been collected from ULTERA, it was passed to the inverse design component of the data ecosystem. To demonstrate novel refractory HEAs with some desired property sets, a cGAN model based on a simple feed-forward NN architecture with four fully connected layers was trained using 529 HEA literature-derived compositions from an early ULTERA Database snapshot. To generate new compositions, the cGAN was conditioned on the shear modulus and fracture toughness values obtained from the empirical model, taking as an input linear combinations of pure element ab initio values from \citet{Chong2021CorrelationAlloys} passed through the \texttt{ULTERA} 's pipeline described in Section \ref{ultera:ssec:autolc}.

The values of these properties were normalized to ensure that the importance of each feature is equivalently reflected in the model. The conditioning values were sampled using the probability distribution of the property values. Batches of normally distributed sixteen-dimensional latent vectors and the sampled conditioning vectors were then provided as input to the generator. One advantage of the adversarial loss of GANs over other competing methods like reconstructive loss of VAEs is the simplicity of the objective function - here, the generator receives the negative critic score as its loss, such that it maximizes the "realism," or the underlying distribution match, of the generated samples. Because the critic is trained in tandem with the generator, there is no need to define a metric for this "realism", which is learned directly from the observed distribution. We used the Wasserstein GAN \cite{Arjovsky2017WassersteinNetworks} loss to avoid vanishing gradients and the unrolled GAN \cite{Metz2016UnrolledNetworks} strategy to avoid mode collapse.

\subsection{Biasing the Predictions to Property Values} \label{inverse:ssec:propbias}

\begin{figure}[H]
    \centering
    \includegraphics[width=0.85\textwidth]{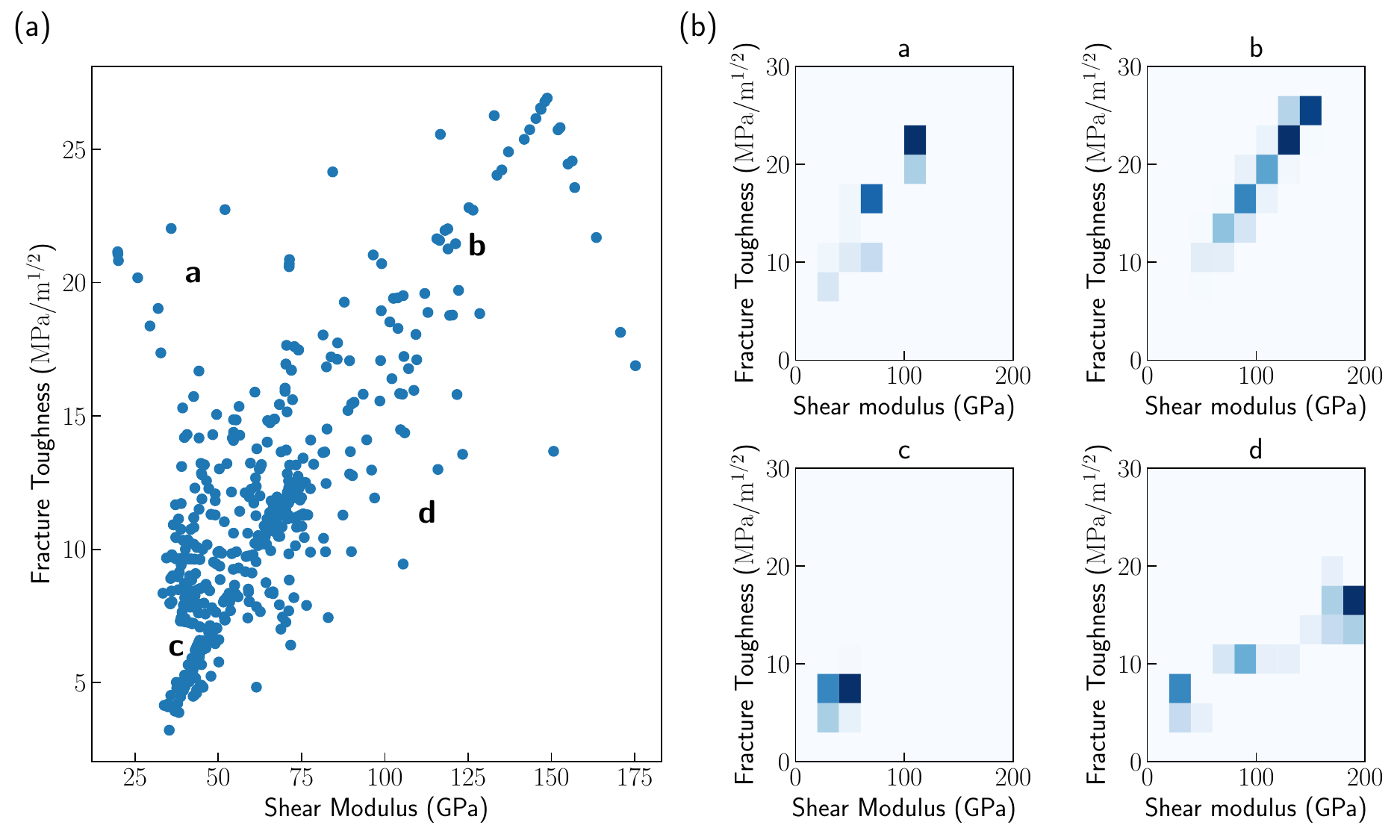}
    \caption{(A) Correlation between training shear modulus and fracture toughness values of the compositions from an old version of ULTERA. Points a, b, c, and d represent four conditioning cases of interest. (B) Histograms of shear modulus and fracture toughness for generated compositions. The intensity of blue in the histograms indicates a greater number of compositions with the corresponding values of shear modulus and fracture toughness. In addition to targeted bias, the cross-property correlations are visible. Reproduced from \cite{Debnath2021GenerativeAlloys} under permission.}
    \label{inverse:fig:propbias}
\end{figure}

With a trained conditional generative model $\mathbb{G}$, one can begin to assess the effects of the conditioning by, for instance, biasing it to the prediction of certain property values. As demonstrated in \citet{Debnath2021GenerativeAlloys}, this process is not exact; however, it does result in the sampling of regions of the latent space that are better aligned with the desired outcome, thus making it a valuable design tool. As shown in Figure~\ref{inverse:fig:propbias}, one can obtain very different distributions of values in the space of fracture toughness and shear modulus values when different targets are selected. Notably, given targets for which the cGAN had a lot of starting data (point c), it can generally generate narrow distributions, while it struggles in the sparse regions (points a, b, d), highlighting the need for robust training datasets, like one described in Section~\ref{ultera:sec:datadescription}.

This effect was observed to usually follow generally expected trends, e.g., in increasing value of shear modulus target causing the frequency of elements like W, Re, and Ru with high elemental shear modulus (173, 150, and 149 GPa, respectively) to increase, while elements like Hf, Mo, and Zr with low elemental shear modulus (30.4, 19.7, and 32.7 GPa) decreased.

\subsection{Biasing the Predictions to Compositional Ranges} \label{inverse:ssec:compbias}

As depicted in Figures \ref{inverse:fig:designs} and \ref{inverse:fig:cgan}, the two spaces sampled by the conditional generative model $\mathbb{G}$ are \emph{latent space} and \emph{property space}. While biasing of the latter, discussed in Subsection \ref{inverse:ssec:propbias}, is the most common approach in material design studies, the biasing of property space has some critical advantages.

For instance, one can utilize it to control the fraction of desired or undesired chemical elements, as shown in \citet{Debnath2023ComparingAlloys} for the case of \ch{W} content control based on recognizing a \emph{concept vector} corresponding to the higher-dimensional direction of high-variance in \ch{W} in the latent space embedding present after training. 

When biasing generative models in such a way, it is critical to consider that, similar to cross-property correlations discussed in Subsection \ref{inverse:ssec:propbias}, biasing the latent space alongside \ch{W} concept vector will also bias all other chemical element contents in the generator outputs.

\section{Demonstrating Generative Design with \texttt{heaGAN}} \label{inverse:sec:heagan}

To create a demonstrator platform for inverse design of high entropy alloys, generative models described in \citet{Debnath2023ComparingAlloys} were extended to several additional properties and deployed as a demonstrator packaged under the name \texttt{heaGAN} and available through interactive cloud notebooks deployed through Binder and GitHub Codespace services, reachable under:

\hspace{36pt}\href{https://cgandemo.ultera.org}{https://cgandemo.ultera.org}

\begin{figure}[H]
    \centering
    \includegraphics[width=0.7\textwidth]{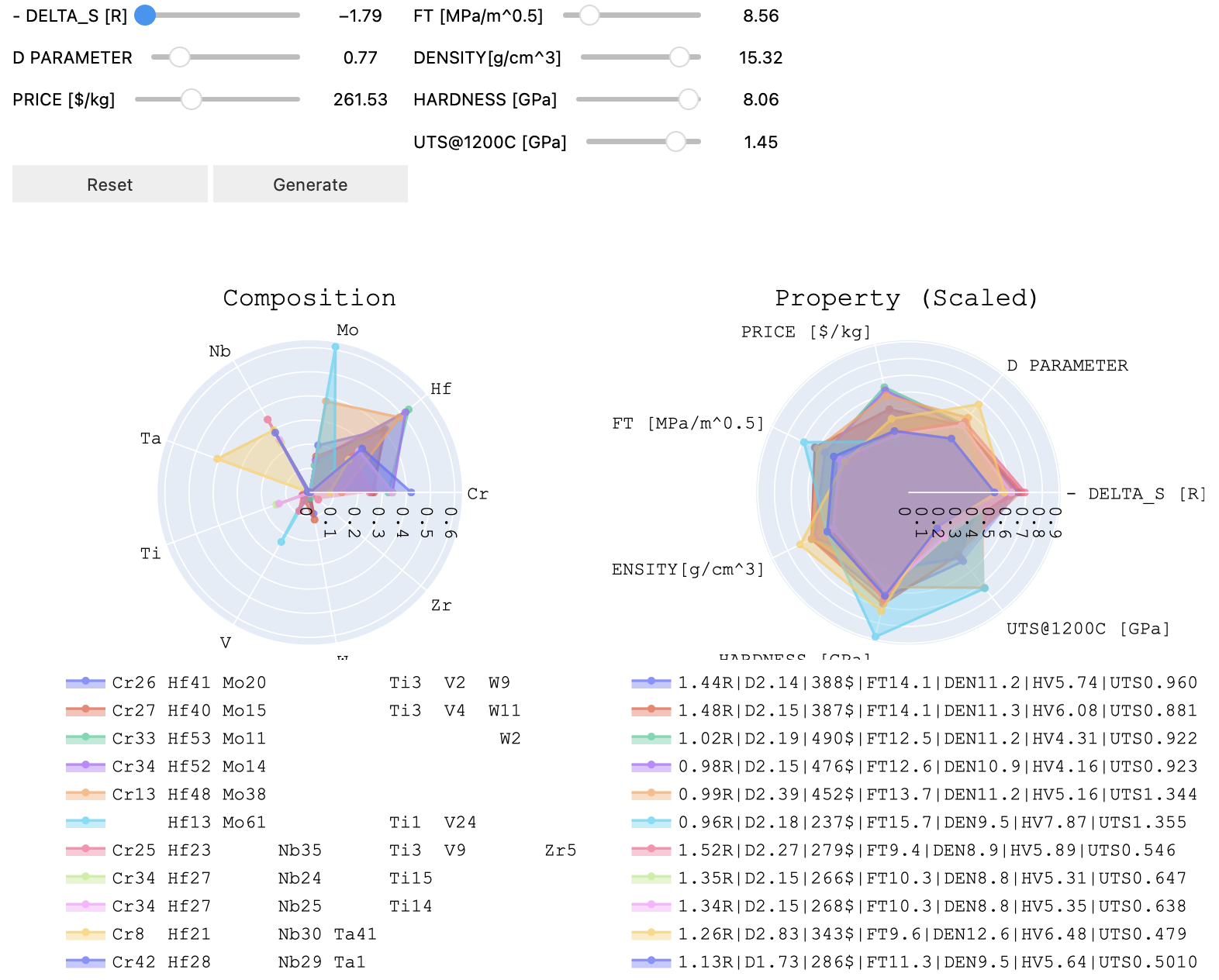}
    \caption{Printout of the cGAN demonstrator showing several controls user can move to bias the generator to different property values. The depicted plots and lists correspond to a series of biasing values selected semi-randomly to show different values.}
    \label{inverse:fig:cgandemo}
\end{figure}

As depicted in Figure \ref{inverse:fig:cgandemo}, the user of the demonstrator is given several parameters to control, including alloy configurational entropy, price, and five physical properties. Each press of the "Generate" button samples a new candidate in a split second.

\begin{figure}[H]
    \centering
    \includegraphics[width=0.4\textwidth]{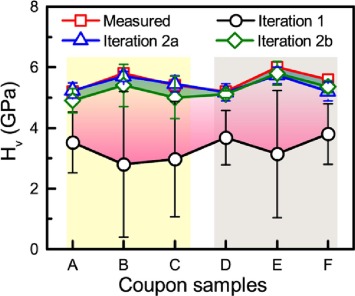}
    \caption{Vickers hardness predicted for 6 \ch{Mo-Nb-W} alloys in \cite{Li2024DesignExperiments} generated by a cGAN model both before the first measurement and after just 3 measurements (first or second iteration) become available. The original dataset contained no ternary of interest nor its binaries nor many related systems. Reproduced from \cite{Li2024DesignExperiments}.}
    \label{inverse:fig:activelearn}
\end{figure}

The forward surrogate models, used to generate inputs for GAN retraining, are stored under \texttt{saved\_surrogates} directory in a free open-source \texttt{ONNX} format to enable both (1) easy swapping for other community models, mostly out-of-the-box and (2) retraining by the end-user on additional datasets. Critically, this can be used to easily implement active learning approaches, dramatically improving the predictive ability of the underlying surrogate models and, by extension, generative models trained on them. As shown in Figure \ref{inverse:fig:activelearn} depicting hardness surrogate refitting, just three data points can be used to fit such a model to a new ternary chemical system.

\chapter{Efficient generation of grids and traversal graphs in compositional spaces towards exploration and path planning exemplified in materials} \label{chap:nimplex}

\acknowledge{
This chapter adapts verbatim a preprint by \citet{Krajewski2024Nimplex} submitted for publication to the npj Unconventional Computing. This work was co-authored with Wesley F. Reinhart, Allison M. Beese, and Zi-Kui Liu. All text and associated software was written by Adam M. Krajewski, while co-authors provided edits and guidance.
}

\section{Introduction} \label{nimplex:sec:introduction}

\subsection{Compositional Spaces} \label{nimplex:ssec:compositionalspaces}

The term \emph{composition} refers to \emph{a} way an entity can be split into a set of distinct components, and it plays a critical role in many disciplines of science, engineering, and mathematics. For instance, in combinatorics, the composition will refer to \emph{a} way a positive integer is split into a sequence of other positive integers. In materials science, chemical composition refers to how a material (or, more generally, matter) is split into distinct components, such as chemical elements, based on considerations such as fraction of atoms, occupied volume, or contributed mass. In economics, portfolio composition may refer to how finite capital is split across assets, such as cash, equity instruments, real estate, and commodities, based on their monetary value.

The definition of a composition will typically allow for the definition of a finite space in which such a composition exists. In the typical case of the composition defined in terms of a sequence of $d$ fractions, such space will be a standard simplex - a $(d-1)$-dimensional polytope of unit length edges defined for points \textbf{x} which satisfy $x_i>0$ and $\sum_{i=0}^d x_i = 1$. Or, in simple terms, the space where all fractions are positive, treated equally, and add up to 1. Some special cases of $d$=2,3,4, corresponding to 1-simplex, 2-simplex, and 3-simplex, are also known as line segment, triangle, and tetrahedron, respectively.

Working within compositional (simplex) spaces requires several additional considerations relative to the more common Euclidean spaces for which most tools were designed. Otherwise, numerous problems can be introduced, ranging from sampling points outside the space, through incorrect density estimates, to incorrect gradient calculations caused by modifying every $x_{j\neq i}$ when changing $x_i$ assumed to be independent.

This work introduces a new high-performance library called \texttt{nimplex} or \textit{NIM library for simPLEX spaces}, created exclusively for working with such spaces. It was written in low-level Nim language, allowing for careful optimizations, and then compiled with a native Python interface for general use. It provides an efficient implementation of (a) existing methods from literature (see Sec.~\ref{nimplex:ssec:mc}~and~\ref{nimplex:ssec:fullgrid}), (b) modifications of existing methods (see Sec.~\ref{nimplex:ssec:internalgrid}), and (c) entirely new capabilities developed in this paper (see Sec.~\ref{nimplex:sec:simplexgraph}).

Neither compositional space nor \texttt{nimplex} is exclusive to any discipline; however, to better showcase its capabilities, two complex, highly-dimensional materials-related problems of high impact are highlighted. At the same time, translating them and their solutions to other areas, such as economics, can be done directly and is discussed.

\subsection{Compositionally Complex Materials} \label{nimplex:ssec:compositionallycomplex}

An exemplar of how tackling highly-dimensional problems allows researchers to unlock novel solutions is the class of Compositionally Complex Materials (CCMs), which includes several sub-classes, such as Multi Principle Element Alloys (MPEAs), High Entropy Alloys (HEAs), High Entropy Ceramics (HECs), and High Entropy Metallic Glasses (HEMGs). CCMs are materials with at least several elements in significant fractions and was initiated by two pioneering 2004 works on HEAs by Yeh et al. \cite{Yeh2004NanostructuredOutcomes} and by Cantor et al. \cite{Cantor2004MicrostructuralAlloys}, who independently proposed that equimolar (equal fractions) alloys with more than 5 (Yeh) or between 6 and 9 (Cantor) elements, could form single solid solutions (SSS) thanks to the configurational entropy stabilizing them. Other notable definitions include all materials with idealized configurational entropy $\Delta S_{conf}  \geq R \ln{5} = 1.61R$ \cite{Li2019MechanicalAlloys} ($\approx2.32$ bits of information in the composition \textbf{x}) or $\Delta S_{conf}  \geq 1R$  \cite{Senkov2019HighAlloys} ($\approx1.44$ bits).

Regardless of the exact definition, while individual CCMs contain a few components, they always occupy very high dimensional problem spaces relative to other materials because they are not as restricted in terms of which elements are present. This results in homogeneous datasets occupying over 30-dimensional spaces (or 10-20 for specific problems, like refractory HEA \cite{Senkov2019HighAlloys}), which are orders of magnitude larger compared to traditional alloys with one or two primary elements. This introduces opportunities for finding exceptional alloys in little-explored chemical spaces, as demonstrated by some cases of excellent hardness \cite{Senkov2010RefractoryAlloys}, ductility \cite{Zhang2019PrecipitationAlloy}, room temperature strength \cite{Long2019AProperties}, and refractory strength \cite{Senkov2016DevelopmentSuperalloy, Kang2021SuperiorProcess}. 

In recent years, high-throughput thermodynamics-driven combinatorial studies on CCMs have been successfully performed to generate high-performance materials \cite{Elder2023ComputationalValidation, Elder2023ComputationalDown-selection}, utilizing  CALPHAD thermodynamic databases for CCMs/HEAs (e.g., \cite{Ostrowska2020ThermodynamicW, Ostrowska2022ThermodynamicExperiments, GambaroCombinedAlloys}). However, they are often limited to coarse sampling (e.g., spaced at $5/10$at.\%) due to the combinatorial complexity in number of alloys and low-dimensional points (e.g., $d=4$) due to the combinatorial complexity in component interactions tracked in CALPHAD calculations increasing single evaluation cost \cite{Elder2023ComputationalValidation, Elder2023ComputationalDown-selection}, sometimes limited further to particular points such as equimolar alloys \cite{Yan2021AcceleratedLearning}.

To somewhat alleviate these computational cost challenges, ML models started to be used as surrogates for thermodynamic calculations and experiments \cite{Debnath2023ComparingAlloys, Tandoc2023MiningAlloys} or in the role of candidate selection from ML latent space \cite{Rao2022MachineDiscovery}. They are extremely fast relative to traditional methods, usually taking microseconds per prediction, and they may seem to work near-instantly when used as a drop-in replacement. However, when one tries to deploy ML models on more complex systems, the combinatorial complexities involved (discussed in Section~\ref{nimplex:ssec:combinatorialcomplexities}) may quickly make ML deployment very expensive, prompting optimization of the approach. 

While the ML inference is typically optimized to the computational limits in state-of-the-art tools like \texttt{PyTorch} \cite{Paszke2019PyTorch:Library}, the rest of the customized composition space infrastructure, traditionally designed for thousands of evaluations taking seconds, may become a severe bottleneck when moving to billions of evaluations taking microseconds, as explored throughout this paper. In particular, being able to do the following tasks in nanosecond to microsecond region typically becomes critical and needs to be considered:

\begin{enumerate}

    \item Efficient random sampling from the uniform grids and continuous distributions (Monte Carlo in Section~\ref{nimplex:ssec:mc}) to facilitate approaches including active learning \cite{Rao2022MachineDiscovery} and generative design \cite{Debnath2021GenerativeAlloys}.
    
    \item Efficient generation of the uniform grids in simplex spaces to facilitate complete screenings, quantitatively explored in Sections~\ref{nimplex:ssec:fullgrid} and \ref{nimplex:ssec:internalgrid}.
    
    \item Efficient generation of high-dimensional graph representations with complete connectivity to all adjacent CCM compositions, explored in detail throughout Section~\ref{nimplex:sec:simplexgraph}, to deterministically allocate problem space structure and facilitate neighborhood-based exploration. This is particularly beneficial for the gradient calculations between neighboring grid points, where one typically has to either (a) na\"ively compute all possible compositional changes despite redundancy (e.g., if at point 1 gradient $+1\%B \atop -1\%A$ from point 1 to 2 and gradient $+1\%C \atop -1\%A$ from point 1 to 3, then at point 2 the gradient $+1\%C \atop -1\%B$ to point 3 can be known) at least doubling the number of required evaluations, or (b) keep track of all visited states through an associative array (dictionary). The latter can, in principle, scale well with the number of visited points ($\mathcal{O}(1)$ avg. time for hash map) but is many times more computationally intensive compared to directly accessing known memory location through a pointer as one can do with a graph data structure.
    
\end{enumerate}

\subsection{Path Planning in Functionally Graded Materials} \label{nimplex:ssec:functionallygraded}

Another class of materials where complex compositional spaces have to be considered, even if intermediate compositions may not be complex themselves, is the class of Functionally Graded Materials (FGMs), sometimes narrowed to Compositionally Graded Materials (CGMs). In them, a set of compositions is traversed to form a compositional path inside a single physical part in order to spatially leverage combinations of properties that may not be possible or feasible with a homogeneous material \cite{Saleh202030Challenges}. In the simplest binary example, this could mean increasing the porosity fraction as a function of depth from the part surface to achieve a higher performance-to-weight ratio. 

This paper focuses on the computational design of advanced FGMs, which enable solutions to otherwise impossible challenges. An example of such is the design of compositional pathways between stainless steel and titanium alloys to allow for additive manufacturing (AM) of aerospace and nuclear components, combining these alloys within a single print \cite{Bobbio2022DesignCompositions}. Such a task is highly non-trivial as the simple linear mixing causes several brittle or otherwise detrimental Fe-Ti and Cr-Ti phases to form, regardless of how gradual the compositional gradient is \cite{Reichardt2016DevelopmentManufacturing}. Formation of such phases in significant quantities is not specific to this alloy pair; thus, all possible ternary systems in Cr-Fe-Ni-Ti-V space had to be considered and manually arranged together by experts to obtain a pathway navigating through feasible regions \cite{Bobbio2022DesignCompositions}.

\begin{figure}[H]
    \centering
    \includegraphics[width=0.38\textwidth]{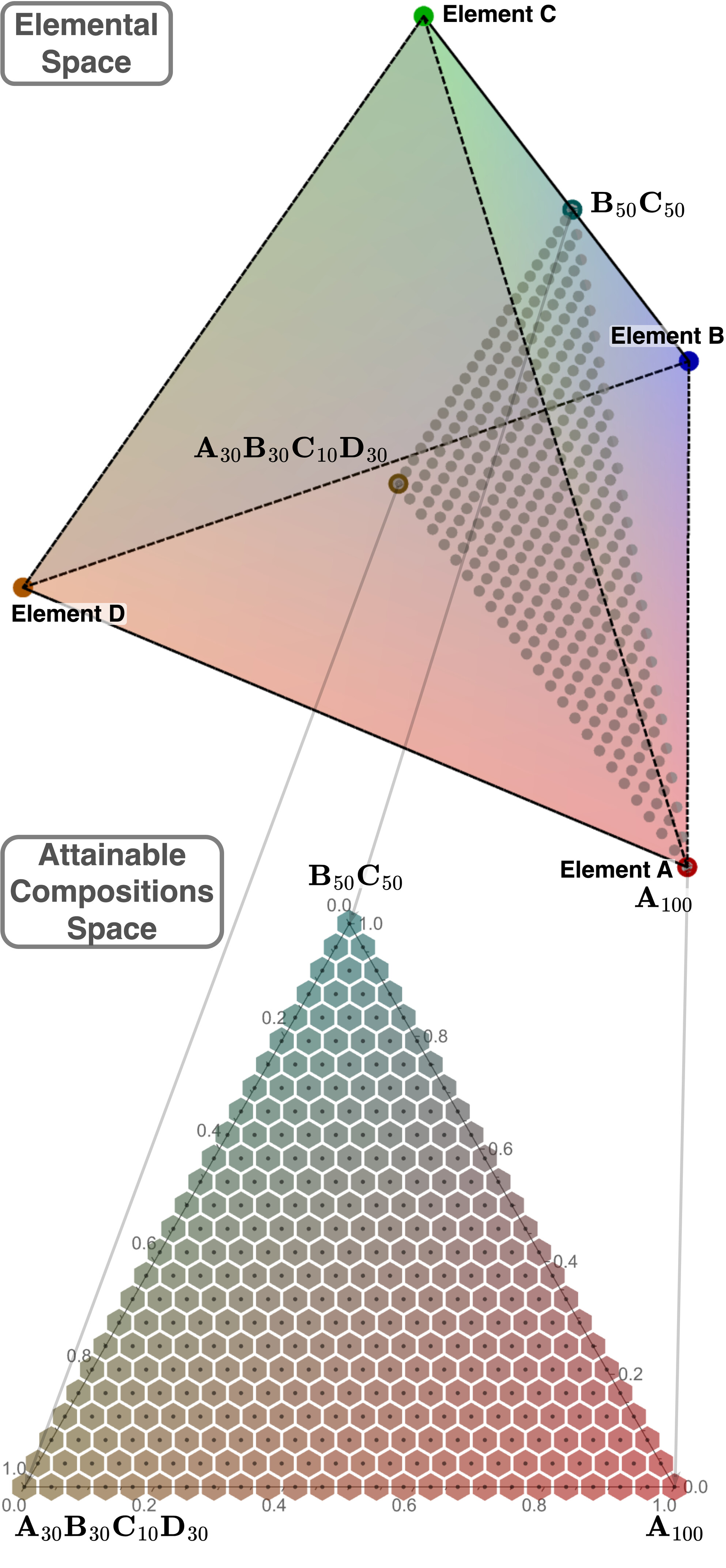}
    \caption{Three available compositions existing in a quaternary (d=4) compositional space forming a ternary (d=3) compositional space which can be attained with them; sampled with a uniform grid with 24 divisions. The hexagonal tiling emerges based on the distance metric in 2-simplex and would become rhombic dodecahedral in 3-simplex.} 
    \label{nimplex:fig:fgmspaces}
\end{figure}

While in recent years, the fabrication of FGMs has become dominated by Directed Energy Deposition AM for various geometries (e.g., radial deposition \cite{Hofmann2014DevelopingManufacturing}), several other notable manufacturing techniques allow the deployment of such pathways. These include deposition-based methods for high-precision applications, casting-based methods for high-volume fabrication \cite{Saleh202030Challenges}, and recently, brazing consecutive metallic foils \cite{Wu2023ATechnology} to create relatively thin compositionally graded interfaces on mass.

In a typical FGM manufacturing scenario, a discrete set of compositions (individual available materials) exists in a compositional (simplex) space formed by a union of all components (usually chemical elements or compounds - not affecting later steps), as depicted in the top of Figure~\ref{nimplex:fig:fgmspaces}, which one could call the \emph{elemental space}. The position in this elemental space is fundamental and is usually the one considered in both mechanistic (e.g., thermodynamic CALPHAD-type models \cite{Olson2023GenomicDynamics}) and predictive (ML/empirical-rule) modeling. However, during the FGM design, it is more convenient to consider another compositional space formed by treating the original available compositions as components, as depicted on the bottom of Figure~\ref{nimplex:fig:fgmspaces}, which one could call \emph{attainable compositions space} or more generally the \emph{design space}.

Within an FGM manufacturing apparatus, it is common for each of the available compositions to be treated equally, e.g., powder hoppers \cite{Reichardt2021AdvancesMaterials}, sputtering targets \cite{Wu2023ATechnology}, or other flow sources are symmetrically arranged and offer the same degree of control. Thus, as depicted in Figure~\ref{nimplex:fig:fgmspaces}, the attainable compositional space can be treated as a standard simplex for design purposes and partitioned equally across dimensions, reflecting the nature of the problem even though equidistant points in it may not be equidistant in the original (elemental) space.

The attainable spaces used in the final design tend to be lower-dimensional relative to the corresponding elemental spaces, especially when the available compositions are CCMs/HEAs or the number of flow sources is limited. However, this trend is not fundamentally required, and going against it may be beneficial in many contexts. For instance, one may conceptualize a ternary ($d=3$) elemental compositional space where 4 compositions are available, arranged as vertices of some tetragon; thus, forming a quaternary ($d=4$) attainable compositions space tetrahedron. In such a case, some regions have to overlap in the elemental space, while new regions are guaranteed to be unlocked relative to 3 available compositions if the formed tetragon is strictly convex. This seemingly oversamples; however, it is critical to consider that there is no oversampling in the design space because the available materials can possess properties that are not a function of the composition alone, such as the $CO_2$ footprint or price. 

A clear and industry-significant example of the above happens during FGM design in elemental spaces containing Hf and Zr. The two are very difficult to separate, causing both price and greenhouse emissions to rise sharply as a function of the separation purity requirements. Furthermore, physical form factors available from suppliers tend to be limited or at lower demand for pure-Zr and pure-Hf, furthering the cost. In 
the case of AM using wires as feedstock (WAAM) \cite{Shen2016FabricationProcess}, as explored in detail in Section \ref{nimplex:app1}, using pure Zr in place of the more industry-common alloy with $4.5\%$Hf can be somewhere from a few times to over 100 times more expensive. In a typical, manual FGM design, a researcher selects one of the two grades based on their expertise. However, by considering the two grades as independent components of the higher-dimensional design space, one can avoid forcing a decision before exploring the space, thus limiting human bias and allowing exploration of both options simultaneously, allowing their combination where regions of space insensitive to the Hf content utilize the cheaper grade while the pure Zr is used when necessary or favorable based on some path heuristic.

\begin{figure}[H]
    \centering
    \includegraphics[width=0.475\textwidth]{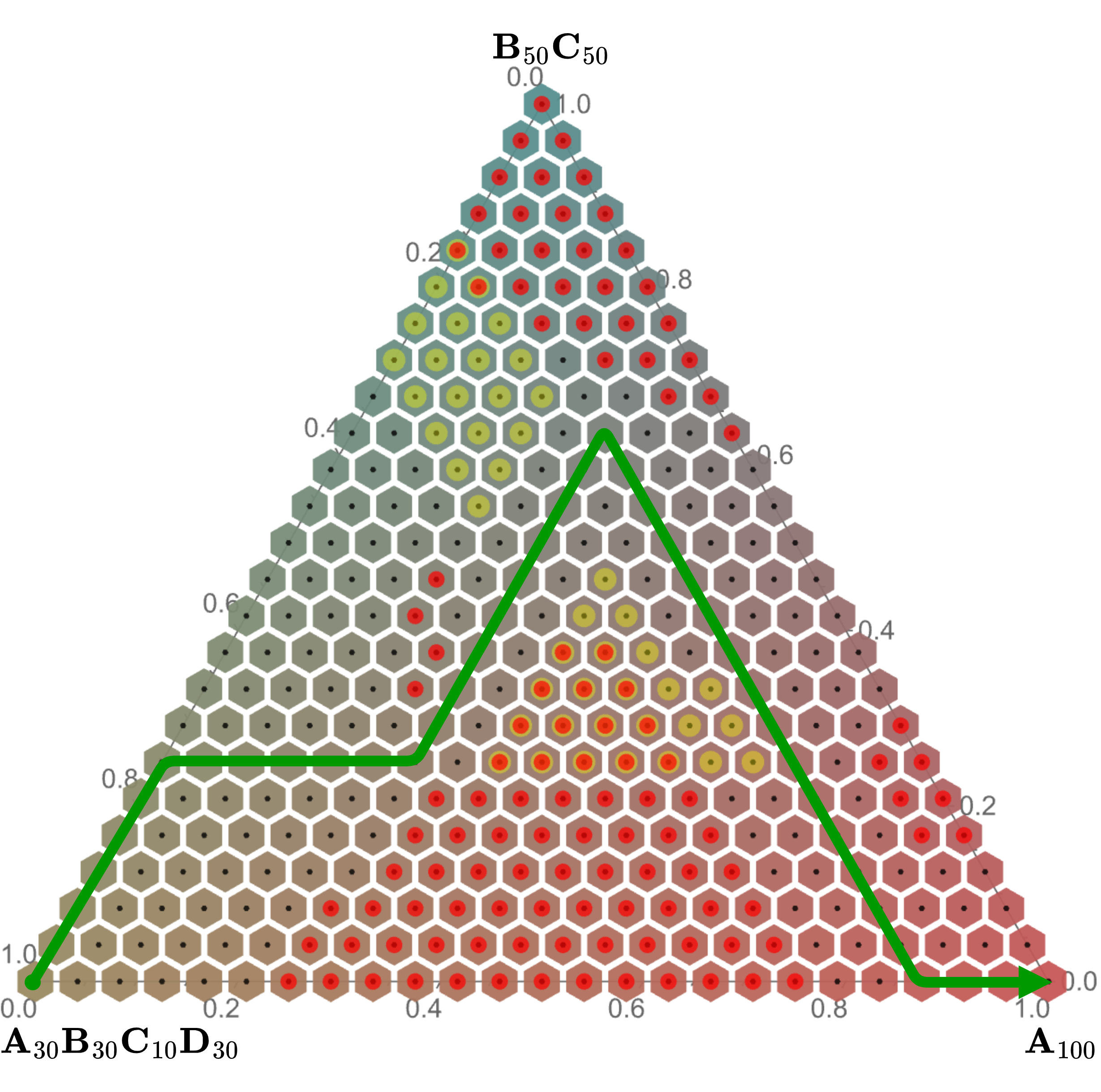}
    \vspace{-6pt}
    \caption{A path example which avoids infeasible (red) and undesirable (yellow) regions, or their combination (orange).} 
    \label{nimplex:fig:pathplanning1}
\end{figure}

With the design space carefully set up, one can start to evaluate different paths across it. Typically, the core considerations deal with meeting specific feasibility (hard) constraints. In the case of metal alloy FGMs, these can be (1) formation of detrimental phases based on thermodynamic equilibrium \cite{Reichardt2021AdvancesMaterials}, (2) formation of detrimental phases based on non-equilibrium predictions of solidification results based on Scheil–Gulliver method, which better describes the as-made material \cite{Bocklund2020ExperimentalMaterials}, or (3) a combination of the two \cite{Bobbio2022DesignCompositions}. In the future, these will likely be extended through (4) precipitation modeling improving the metastable material design, thanks to the recent release of open-source high-performance software Kawin \cite{Ury2023Kawin:Model}, and (5) automated modeling of manufacturing constraints, such as printability in AM \cite{SheikhAnAlloys}. Furthermore, one can also try to meet desirability (soft) constraints, such as the physical appearance of a composite material, which can be broken if needed. These two types of constraints are depicted in Figure~\ref{nimplex:fig:pathplanning1}, alongside an example path navigating through them.

In Figure~\ref{nimplex:fig:pathplanning1}, all infeasible points violating the constraints are annotated for visualization. However, doing so may be unnecessary when path-planning, especially iteratively based on neighbor connectivity, as the insides of the infeasible space could not be reached, thus reducing the total number of evaluations.

\begin{figure}[H]
    \centering
    \includegraphics[width=0.475\textwidth]{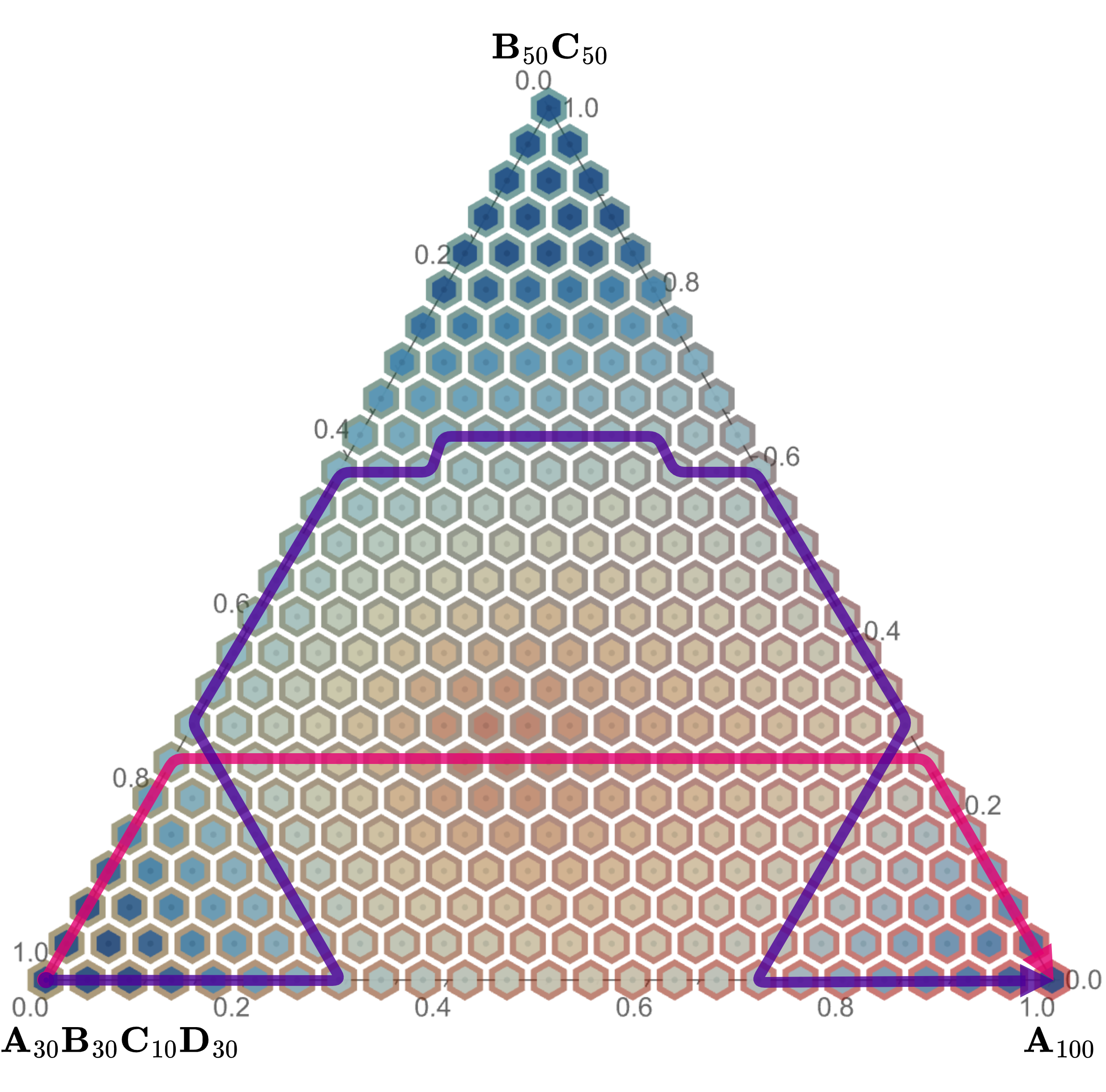}
    \caption{Two path examples in the attainable compositional space annotated with some virtual property. One (pink/inner) minimizes/maximizes the average property value given a number of fixed path lengths, and another (purple/outer) minimizes the gradient in the property along the path.} 
    \label{nimplex:fig:pathplanning2}
\end{figure}

In addition to the feasibility and desirability constraints, further considerations are often made to how the path optimizes values of a set of properties of interest, either individually or through some heuristics combining them. Usually, this optimization constitutes finding the path that minimizes or maximizes either average or extreme values over the set of visited states, exemplified by the pink path in Figure~\ref{nimplex:fig:pathplanning2}. In the case of metal alloy FGMs, this can mean, for instance, minimizing the average evaporation rate of the molten metal \cite{Mukherjee2016PrintabilityManufacturing}, minimizing the maximum susceptibility to different cracking mechanisms \cite{Yang2023DesignCracking}, or maximizing the ductility \cite{Hu2021ScreeningAlloys}.

The last, fundamentally different, property optimization task has to do with the gradient, or more generally, the character, of transitions between intermediate states, which will be critical later in the context of graphs in Section~\ref{nimplex:sec:simplexgraph}. Most commonly, one optimizes the path to minimize value function gradients, exemplified by the purple path in Figure~\ref{nimplex:fig:pathplanning2}, in order to, for instance, minimize the thermal expansion coefficient mismatch and by extension stresses induced by temperature changes \cite{Kirk2021ComputationalMonotonicity}.

\subsection{Combinatorial Complexities} \label{nimplex:ssec:combinatorialcomplexities}

As eluded to in Sections~\ref{nimplex:ssec:compositionallycomplex} and \ref{nimplex:ssec:functionallygraded}, when sampling compositions or partitioning corresponding spaces, the resulting combinatorial complexities have to be considered to determine whether a method will be computationally feasible. There are two key equations governing these complexities based on (1) the dimensionality of the space (number of components) $d$ and (2) the number of equal divisions made in each dimension $n_d$, which can be found for every feasible fractional step size (such that it can add to $100\%$). 

The first, very intuitive equation gives the number of samples $N_C$ on a Cartesian grid in $d-1$ dimensions, with $-1$ term due to one of the components being considered dependent.
\begin{equation}
    N_C(d, n_d) = (n_d+1)^{d-1}
    \label{nimplex:eq:nc}
\end{equation}
The second equation gives the number of ways $n_d$ balls can be arranged in $d$ bins, which is well known to be equivalent to much simpler problems of choosing $d-1$ moves or $n_d$ placements from $d-1+n_d$ possible options (see \cite{Nijenhuis1978CombinatorialCalculators} or \cite{Chasalow1995AlgorithmPoints}). While these may seem unrelated to compositions, the former problem is precisely equivalent to finding a composition of an integer or distributing $n_d$ compositional fractions $\frac{1}{n_d}$ across components or chemical elements, giving the number $N_S$ of unique discrete compositions in the simplex space.
\begin{equation}
    N_S(d, n_d) = \binom{d-1+n_d}{d-1} = \binom{d-1+n_d}{n_d}
    \label{nimplex:eq:ns1}
\end{equation}
In terms of factorials, both expressions can be simplified to the same
\[N_S(d, n_d) = \frac{(d - 1 + n_d)!}{(d-1)!n_d!}\]
Throughout Sections~\ref{nimplex:sec:simplexgrid}~and~\ref{nimplex:sec:simplexgraph}, the interplay between these equations will be utilized to contrast different computational methods, and their direct results will allow computational feasibility evaluation.

\section{Simplex Uniform Random Sampling} \label{nimplex:sec:randomuniformsampling}

\subsection{Monte Carlo} \label{nimplex:ssec:mc}

Performing a uniform random sampling, also known as the Monte Carlo method, over a simplex space is a prevalent task; however, it is also one of the most common sources of inefficiency, bias, or errors when implemented incorrectly.

\begin{figure}[h]
    \centering
    \includegraphics[width=0.270\textwidth]{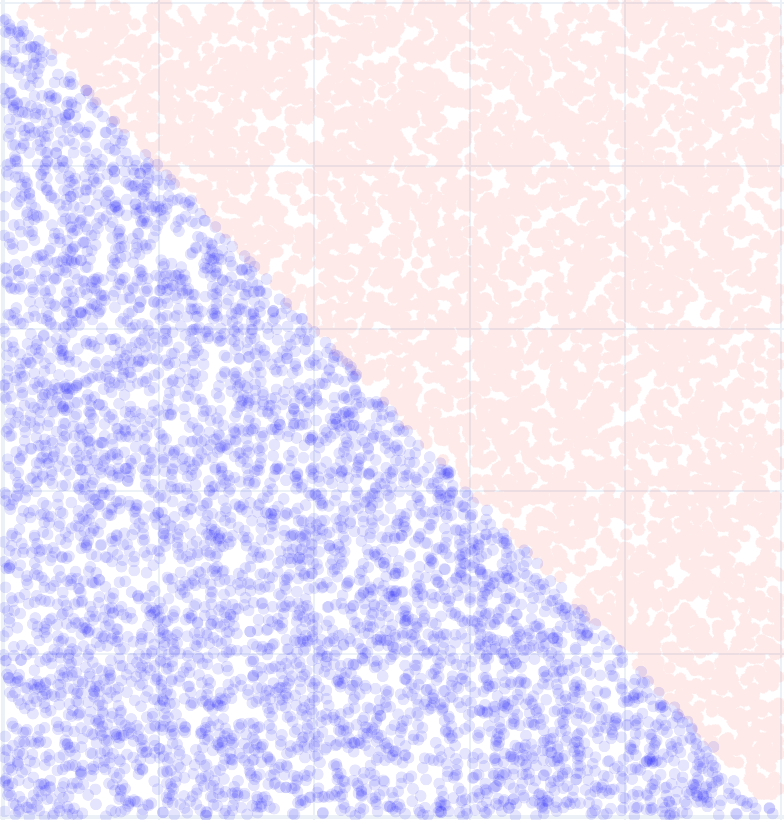}
    \hfill
    \includegraphics[width=0.320\textwidth]{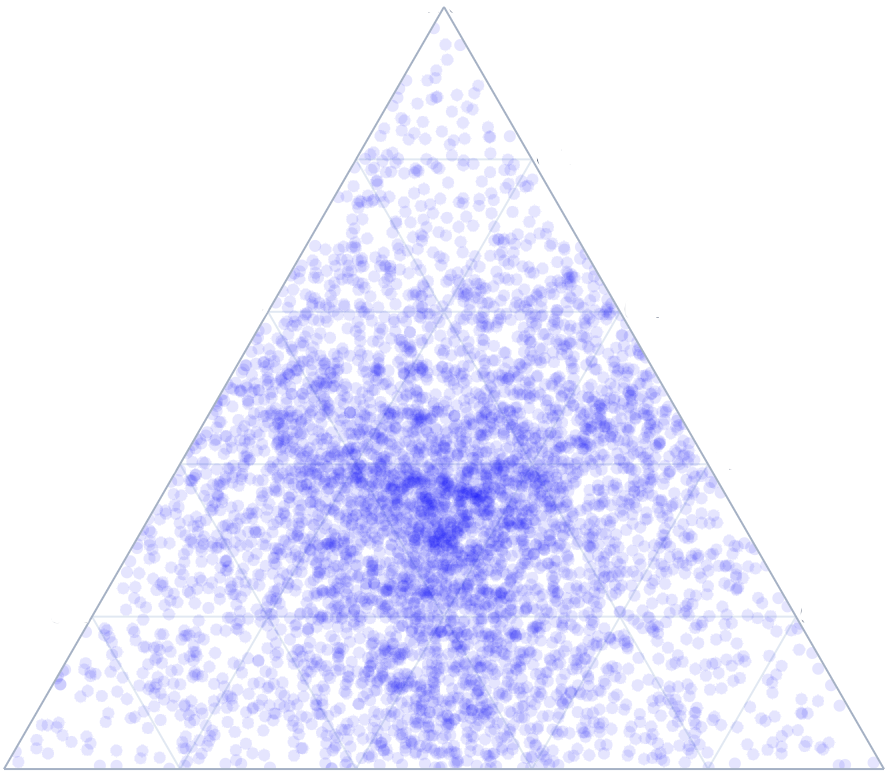}
    \hfill
    \includegraphics[width=0.320\textwidth]{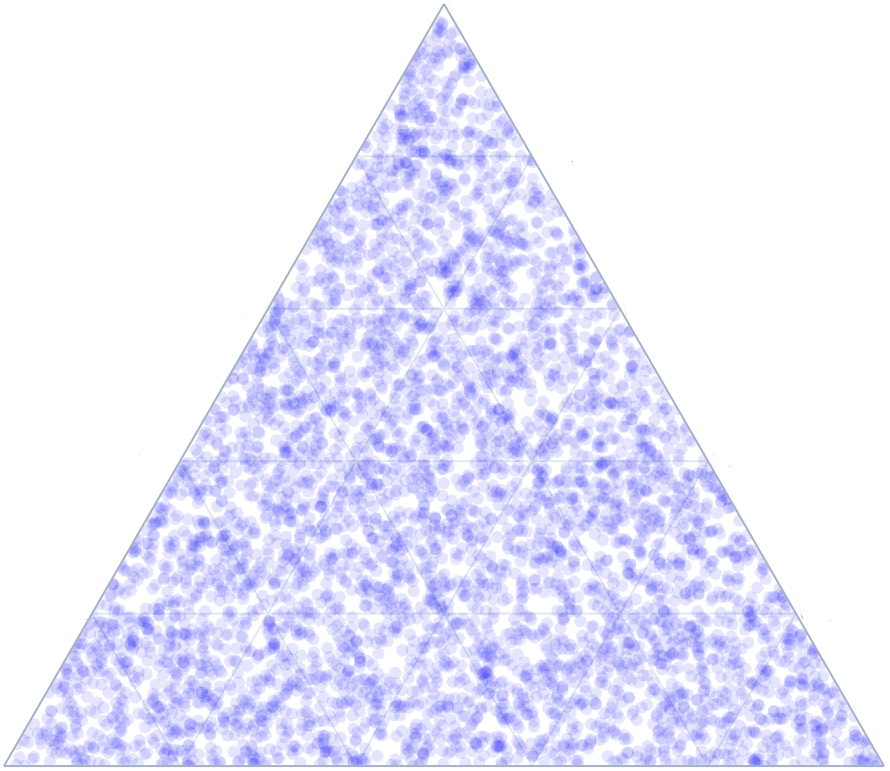}
    \caption{(left) Uniform random sampling in 2-cube (square) filtered to fall onto a 2-simplex (ternary composition), showing 50\% rejection rate, (middle) random sampling in 3-cube projected onto 2-simplex by normalizing coordinates, showing oversampling in the center of each dimension, and (right) ideal uniform random sampling of a simplex.} 
    \label{nimplex:fig:samplinginternary}
\end{figure}

Software (e.g., \texttt{alchemyst/ternplot} in Matlab \cite{Sandrock2017Alchemyst/ternplothttps://github.com/alchemyst/ternplot}) and methods dealing with low-dimensional or otherwise small compositional spaces, often utilize a na\"ive approach of sampling uniformly distributed points from a Cartesian space/grid in $d-1$ dimensions and then rejecting some infeasible points ($\sum^d_i x_i > 1$), as depicted in the left part of Figure~\ref{nimplex:fig:samplinginternary}, which for small ($d \leq 4$) can be both easiest and computationally fastest. 

However, this method becomes inefficient for large $d$ because the fraction of rejected points increases with the dimensionality. While this problem is widely noted in the literature \cite{Allen2022AAlloys}, best to the authors' knowledge, it has yet to be discussed quantitatively despite being critical to estimating the sampling's computational complexity. Thus, it is derived herein.

One can consider that a grid of $N_S$ simplex-feasible points is a subset of a grid of $N_C$ points distributed uniformly in the Cartesian space so that random selection from this grid should have a $\frac{N_S}{N_C}$ probability of falling within the simplex. Thus, as shown below, one can find the acceptance rate by considering an infinitely fine grid ($n_d\rightarrow\inf$). Section \ref{nimplex:app2} gives an alternative, intuitive method for finding $f(4)$ using geometry, which agrees with this result.

\begin{equation}
    \begin{aligned}
        f(d) &= \lim_{n_d\rightarrow\inf} \frac{N_S}{N_C} = \lim_{n_d\rightarrow\inf} \frac{\binom{d-1+n_d}{d-1}}{(n_d+1)^{d-1}}\\
        &= \Gamma(d)^{-1} = \frac{1}{(d-1)!} = \frac{d}{d!}
    \end{aligned}
    \label{nimplex:eq:fd}
\end{equation}
As one can see in Equation~\ref{nimplex:eq:fd}, the rejection rate exhibits factorial growth, and while it is not a significant obstacle for low-dimensional cases like ternary $f(3)=\frac{1}{2}$ or a quaternary $f(4) = \frac{1}{6}$, it will relatively quickly become a problem when compositionally complex materials are considered. For instance, in the case of nonary chemical space $f(9) = \frac{1}{40320}$ or only $\approx0.0025\%$ of points will fall into the feasible space. Such a rejection rate could have a particularly severe effect on ML-driven methods, such as generative CCM design.

To circumvent the rejection problem, one may randomly sample from N-cube and normalize to 1; however, as shown in the center of Figure~\ref{nimplex:fig:samplinginternary} and commonly known in the literature \cite{Otis2017AnSystems}, this leads to oversampling in the center of each underlying dimension.

Thus, to achieve a uniform random sampling, \texttt{nimplex} and other carefully designed methods (e.g., \cite{Allen2022AAlloys} and \cite{Otis2017AnSystems}) tend to take Dirichlet distribution, where one samples points $\textbf{y}$ from Gamma distributions with density $\frac{y_i^{\alpha-1} e^{-y_i}}{\Gamma(\alpha)}$ and consider its special "flat" case, where $\alpha=1$ simplifies the density equation to just $\frac{1 e^{-y_i}}{1} = e^{-y_i}$. This is equivalent to sampling $\textbf{z}$ from linear distributions and calculating $y_i=-\log(z_i)$, which then can be normalized to obtain $\textbf{x}$ as $x_i = y_i / \sum \textbf{y}$. The following snippet shows \texttt{nimplex}'s implementation of this, which samples \textbf{z} with the high-performance \texttt{xoroshiro128+} random number generator \cite{Blackman2018ScrambledGenerators} underlying \texttt{randomTensor} function from the \texttt{Arraymancer} tensor library \cite{RatsimbazafyMratsim/Arraymancer:Backends}.

\begin{minted}[xleftmargin=3\parindent, linenos=true, mathescape]{nim}
proc simplex_sampling_mc(dim: int, samples: int): Tensor[float] =
  let neglograndom = 
    randomTensor[float]([samples, dim], 1.0
    ).map(x => -ln(x))
  let sums = neglograndom.sum(axis=1)
  return neglograndom /. sums
\end{minted}

An alternative approach worth mentioning, sometimes found in this context, is based on (1) generating a $(d+1)$-length list composed of $0$, $d-1$ random numbers, and $1$, (2) sorting it, and (3) obtaining $d$-length list of differences between consecutive elements, which is guaranteed to be uniformly distributed over a simplex as shown in \cite{Rubin1981TheBootstrap}. While this approach may be easier to conceptualize, it is much more computationally expensive due to the sorting step. On the author's laptop, for $d=9$, the method implemented in \texttt{nimplex} (involving calculation of 9 logarithms and normalizing them) takes $3.6$ns while the above (implemented with merge sort) takes $74.5$ns per iteration, i.e., over 20 times longer while not providing any clear benefit. Furthermore, their complexities are $\mathcal{O}(N)$ and $\mathcal{O}(N \ln N)$, respectively, so the computational cost difference will also slowly widen with increasing $d$.

\subsection{Quasi Monte Carlo} \label{nimplex:ssec:qmc}
While beyond the current implementation scope of \texttt{nimplex}, it is beneficial to consider quasi-Monte Carlo (QMC) sampling methods, where quasi-random sequences of low discrepancy (having highly uniform coverage of all regions) are used to sample the space deterministically. Such an approach is guaranteed to be very beneficial in low-dimensional ($d\leq3$) problems and has been implemented in thermodynamic tools, including \texttt{pycalphad} \cite{Otis2017Pycalphad:Python, Otis2017AnSystems} improving sampling of ternary systems. However, the QMC can become problematic as one moves to higher dimensional problems. 

Firstly, the upper discrepancy bounds for QMC quickly increase with increasing $N$, unlike MC, which depends only on the number of samples; thus, MC \textit{can} outperform it (thanks to better guarantees) unless a quickly (often exponentially) growing number of samples is taken (see discussion on p.271 in \cite{Asmussen2007StochasticAnalysis}). Because of this, even for quaternary ($d=4$) spaces, MC may be preferred for a low number of samples, even though QMC, especially with additional scrambling, \textit{can} outperform it, as shown in \cite{Otis2017AnSystems}. 

Another significant problem in QMC is the unequal sampling of different dimensions, which can be very severe in high dimensions (see p.154 in \cite{Lemieux2009MonteSampling}). In addition to causing under-performance in space-filling, such bias, combined with the standard alphabetical ordering of chemical components, can cause systematically worse exploration of, e.g., titanium compared to aluminum in CCMs, just based on their names.

\section{Simplex Grid} \label{nimplex:sec:simplexgrid}

\subsection{Full} \label{nimplex:ssec:fullgrid}

\begin{figure}[h]
    \centering
    \includegraphics[width=0.270\textwidth]{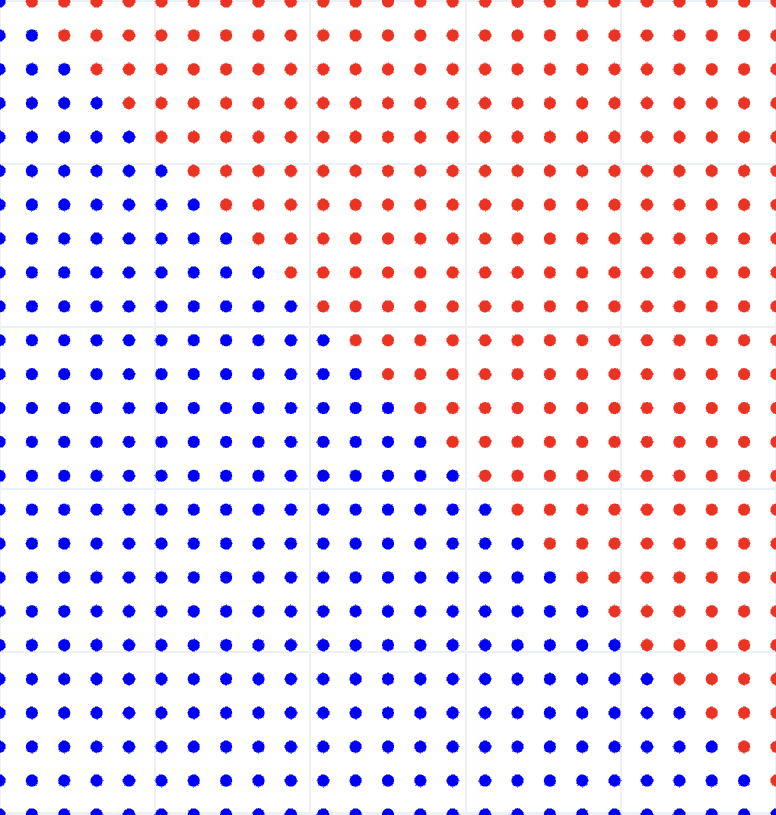}
    \hspace{48pt}
    \includegraphics[width=0.320\textwidth]{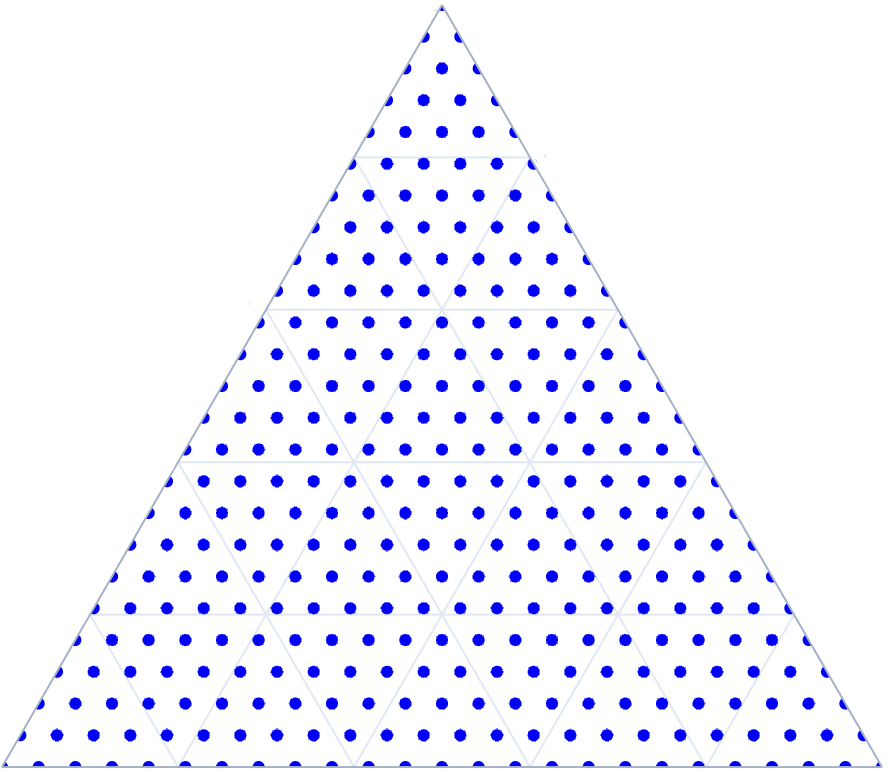}
    \caption{(left) Uniform grid ($n_d=24$) in 2-cube (square) filtered to fall onto a 2-simplex (ternary composition), showing $\frac{12}{25}=48\%$ rejection rate, (right) uniform grid in the corresponding simplex.} 
    \label{nimplex:fig:simplexgrid}
\end{figure}

Next, one can consider the creation of a grid of uniformly distributed points, which is known to contain $\binom{d-1+n_d}{d-1}$ points, as discussed in Section~\ref{nimplex:ssec:combinatorialcomplexities}. Similar to the random sampling discussed in Section~\ref{nimplex:sec:randomuniformsampling}, such a compositional grid cannot be constructed by simply projecting a Cartesian grid in $(N-1)$-cube as patterns will emerge (explored in detail in \cite{Otis2017AnSystems}), but it can be quickly constructed through rejecting infeasible points, as shown in Figure~\ref{nimplex:fig:simplexgrid}. However, it will suffer from a nearly as bad rejection rate, quantitatively dependent on both $d$ and $n_d$. For instance, if we consider $5\%$ spaced compositions in 9-components, the fraction of points summing to $100\%$ is $f_{M=20}(9) \approx \frac{1}{12,169}$ or $0.0082\%$. 

Fortunately, in their 1978 textbook, \citet{Nijenhuis1978CombinatorialCalculators} explored the problem and gave an efficient algorithm/routine called NEXCOM to procedurally generate these simplex lattice points for arbitrary $d$ and $n_d$, resulting in the grid shown in Figure~\ref{nimplex:fig:simplexgrid} on the right.

In the following years, several authors made various modifications to the algorithm, and the most recent one by \citet{Chasalow1995AlgorithmPoints} improves performance without sacrificing simplicity. Over the years, it has been implemented in relatively modern languages such as FORTRAN90, C, MATLAB, and Python. Now, it has been implemented in Nim language as well, with the Nim code snippet shown below.

\begin{minted}[xleftmargin=3\parindent, linenos=true, mathescape]{nim}
proc simplex_grid(dim: int, ndiv: int): Tensor[int] =
  let N: int = binom(ndiv+dim-1, dim-1)
  result = newTensor[int]([N, dim])
  var x = zeros[int](dim)
  x[dim-1] = ndiv
  for j in 0..dim-1:
    result[0, j] = x[j]
  var h = dim
  for i in 1..N-1:
    h -= 1
    let val = x[h]
    x[h] = 0
    x[dim-1] = val - 1
    x[h-1] += 1
    for j in 0..dim-1:
      result[i, j] = x[j]
    if val != 1:
      h = dim
  return result
\end{minted}

As one can deduce from above, the algorithm proceeds through the simplex space starting from $[0, 0, ..., n_d]$ and redistributes one $\frac{1}{n_d}$ fraction $N_S-1$ times across dimensions, forming a zig-zag path to $[n_d, 0, ..., 0]$.

\subsection{Internal} \label{nimplex:ssec:internalgrid}

To the best of the authors' knowledge, something that has not been implemented before, but that is significant to exploration of CCMs  (see Sec \ref{nimplex:ssec:compositionallycomplex}) is an algorithm to obtain only internal points of the simplex grid, i.e., points with non-zero values in all dimensions, to allow, e.g., generating all 7-component HEAs rather than all alloys in 7-component space. In principle, one can filter the output of the algorithm presented in Section~\ref{nimplex:ssec:fullgrid}; however, this may quickly become inefficient, especially for $n_d$ low enough as to approach $d$. 

The number of points can be found by, again, considering the surrogate problem of ball compositions mentioned in Section~\ref{nimplex:ssec:combinatorialcomplexities} and noting that if the last ball cannot be removed from any position, there will be $d$ fewer possible options to perform $d-1$ moves, thus resulting in $N_I$ samples:
\begin{equation}
    N_I(d, n_d) = \binom{n_d-1}{d-1}
\end{equation}
This can be quickly double-checked through summation of internal points of all lower $\delta$ dimensional spaces enclosed in $d$ space:
\[\sum_{\delta=1}^d \Biggl[ \binom{n_d-1}{\delta-1} \times \binom{d}{\delta} \Biggr] =  \frac{(d - 1 + n_d)!}{(d-1)!n_d!} = N_S(d, n_d)\]
We can now look at $N_I(d, n_d)$ to $N_S(d, n_d)$ ratio for the aforementioned case of generating all 7-component alloys. For $5\%$ grid ($n_d=20$) we get $\approx \frac{1}{8.5}$, and for $10\%$ grid ($n_d=10$) we get $\approx \frac{1}{95}$, showing a clear benefit of implementing the new method. This can be done by taking the modified-NEXCOM algorithm \cite{Chasalow1995AlgorithmPoints} from Section~\ref{nimplex:ssec:fullgrid} and:
\begin{enumerate}
    \item Adjusting procedure length from $N_S$ to $N_I$.
    \item Initializing first states in \textbf{x} to 1.
    \item Adjusting the starting point from [1, 1, ..., $n_{div}$] to [1, 1, ..., $n_{div}-d_{im}+1$].
    \item Jumping to the next dimension one step earlier ($val \neq 2$).
\end{enumerate}

To implement the following \texttt{nimplex} snippet.

\begin{minted}[xleftmargin=3\parindent, linenos=true, mathescape]{nim}
proc simplex_internal_grid(dim: int, ndiv: int): Tensor[int] =
  let N: int = binom(ndiv-1, dim-1)
  result = newTensor[int]([N, dim])
  var x = ones[int](dim)
  x[dim-1] = ndiv+1-dim
  for j in 0..dim-1:
    result[0, j] = x[j]
  var h = dim
  for i in 1..N-1:
    h -= 1
    let val = x[h]
    x[h] = 1
    x[dim-1] = val - 1
    x[h-1] += 1
    for j in 0..dim-1:
      result[i, j] = x[j]
    if val != 2:
      h = dim
  return result
\end{minted}

\section{Simplex Graph} \label{nimplex:sec:simplexgraph}

The simplex grid algorithm presented in Section~\ref{nimplex:ssec:fullgrid} is used commonly; however, it has an important feature that has not been utilized yet and was only briefly noted by its authors \cite{Chasalow1995AlgorithmPoints}. Namely, the fact that generated points are sorted in a lexicographic order (forward or reverse, depending on convention) which opens the door for using pure combinatorics for finding certain interesting relations between points at near-zero costs compared to other popular methods.

\subsection{Binary} \label{nimplex:ssec:binarygraph}

\begin{figure}[H]
    \centering
    \includegraphics[width=0.55\textwidth]{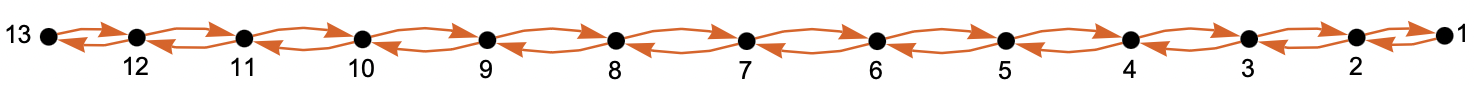}
    \caption{1-simplex graph corresponding to a binary system ($n_d=12$) with 13 nodes/compositions and 24 edges/transitions.} 
    \label{nimplex:fig:binarysimplexgraph}
\end{figure}

In the simplest possible case, which will be expanded upon later, one can look at a binary ($d=2$ / 1-simplex) compositional grid and write a straightforward function that will find all neighboring points (\emph{transitions} to them) to create a graph representation of the binary system like one presented in Figure~\ref{nimplex:fig:binarysimplexgraph}, without any notion of distance calculations.

Such a function, shown below, can be implemented by setting up a \texttt{neighbors} list of lists ($N_S$ of $\leq2$ length) of integer positions and then, at the end of every $i$-th iteration, populating it with forward ($i+1$) and backward ($i-1$) transitions unless start ($[0,1]$) or end ($[1,0]$) points $\textbf{x}$ respectively, corresponding to lack of some component, have been reached.

\begin{minted}[xleftmargin=3\parindent, linenos=true, mathescape=true]{nim}
proc neighborsLink2C(
    i:int, x:Tensor, neighbors: var seq[seq[int]]): void =
  if x[0] != 0:
    neighbors[i].add(i+1)
  if x[1] != 0:
    neighbors[i].add(i-1)
\end{minted}

While the above is trivial, it clearly demonstrates that the graph can be constructed within the original $\mathcal{O}(N)$ computational complexity of the simplex grid algorithm, unlike a similarly trivial distance matrix calculation, which would be $\mathcal{O}(N^2)$; thus, unlocking efficient generation of even massive graphs of this kind.

\subsection{Ternary} \label{nimplex:ssec:ternarygraph}

With the core of the approach set up in Section~\ref{nimplex:ssec:binarygraph}, one can move to the more complex ternary ($d=3$  / 2-simplex) case, which can be conceptualized as a series of $13$ binary systems (already solved individually in Sec. \ref{nimplex:ssec:binarygraph}) of lengths from $13$ to $1$ and with simple modification of positional coordinates shifted forward by 1 to accommodate for the new dimension. 

The newly allowed neighbor transitions across these binaries can be quickly noticed to be dependent on which of these binaries is considered; however, they can be intuitively found by considering that each transition in the 3rd dimension (increasing $x_0$) limits the size of the binary simplex by 1 from the original size of $\binom{d-1+n_d}{d-1} = \binom{2-1+n_d}{2-1} = n_d+1$. Thus, one can define two convenient jump lengths:
\[
\begin{aligned}
    J_0^{d=3} &= 1\\
    J_1^{d=3}(x_0) &= 1 + n_d - x_0
\end{aligned}
\]
Then, one can quickly visualize that (1) unless $x_2=0$, a transition by jump $J_1$ should be possible, (2) unless $x_1=0$, a transition by jump $J_1$ combined with backward jump $J_0$ in the target binary should be possible, and (3) unless $x_0=0$ (the first traversed binary is considered), transitions by both backward jump $J_1$ and backward jump $J_1+J_0$ (extra step within the earlier binary) should be possible. Thus, one arrives at the following algorithm, which requires additional $n_d$ ("ndiv") input on top of the one from Section~\ref{nimplex:ssec:binarygraph} but retains its structure.

\begin{minted}[xleftmargin=3\parindent, linenos=true, mathescape=true]{nim}
proc neighborsLink3C(
    i:int, x:Tensor, neighbors: var seq[seq[int]], ndiv: int): void =
  let jump0 = 1
  let jump1 = 1+ndiv-x[0]
  if x[0] != 0:
    neighbors[i].add(i-jump1)
    neighbors[i].add(i-jump1-jump0)
  if x[1] != 0:
    neighbors[i].add(i-jump0)
    neighbors[i].add(i+jump1-jump0)
  if x[2] != 0:
    neighbors[i].add(i+jump0)
    neighbors[i].add(i+jump1)
\end{minted}

\begin{figure}[h]
    \centering
    \includegraphics[width=0.55\textwidth]{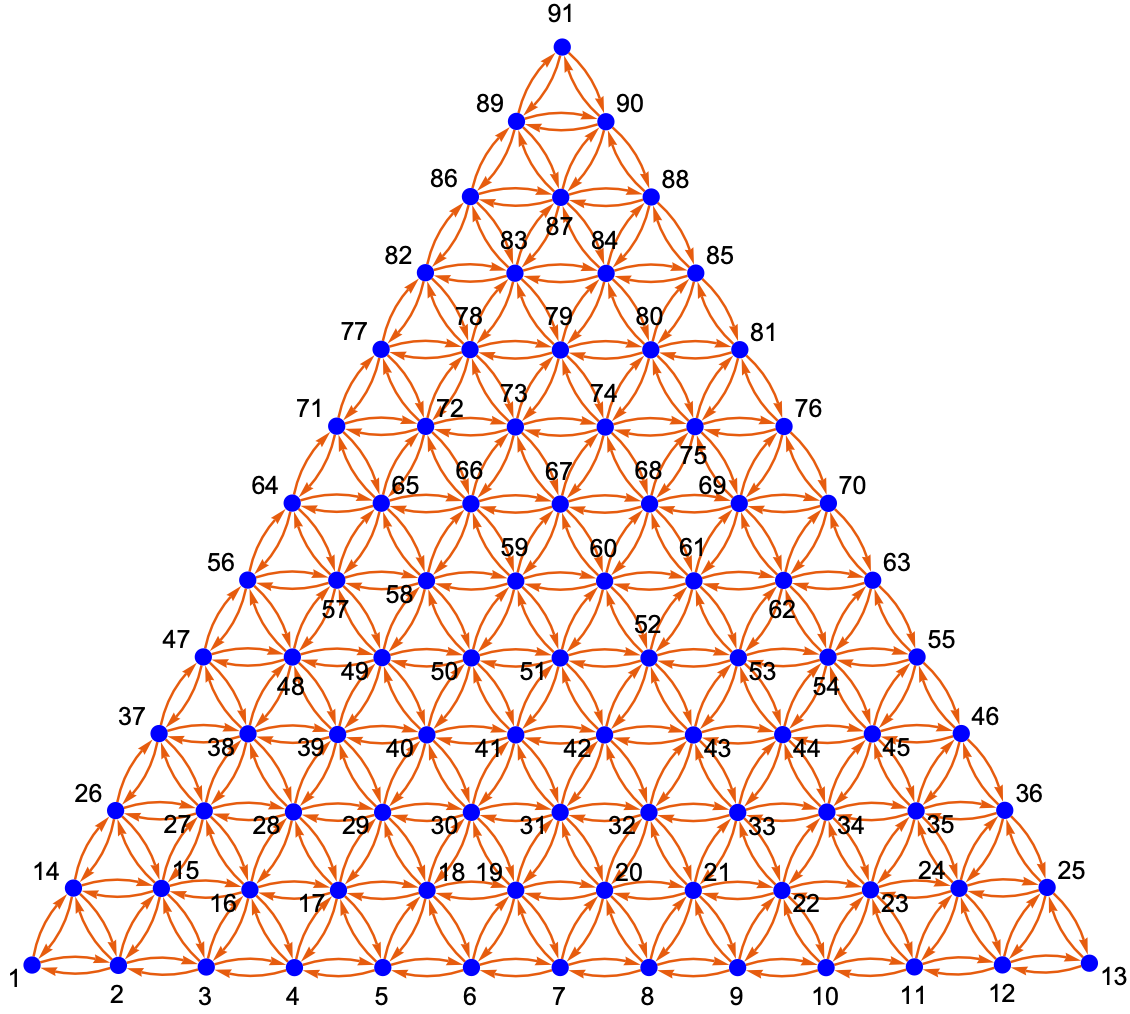}
    \caption{2-simplex graph corresponding to a ternary system ($n_d=12$) with 91 nodes/compositions and 468 edges/transitions.} 
    \label{nimplex:fig:ternarysimplexgraph}
\end{figure}

Utilizing the above, the result presented in Figure~\ref{nimplex:fig:ternarysimplexgraph} can be quickly obtained for any number of divisions. The numbering of points can help to visualize how the transitions were obtained.

\subsection{N-Dimensional} \label{nimplex:ssec:ndimgraph}

Moving beyond ternary systems, one has to increase the number of tracked transitions to higher dimensions, which can be counted for every jump length $J_j$ with $\sum_0^{(d-j-2)}x_i$, and then utilized to obtain a general equation for all $d-1$ elements of jump length array $\textbf{J}$ as a function of current point $\textbf{x}$.
\begin{equation}
    J_{j}(\textbf{x})= \binom{j+n_d- \sum_{i=0}^{(d-j-2)} x_i}{j}
    \label{nimplex:eq:jj}
\end{equation}
As expected, for the special cases of $d=3$, the above agrees with $J_0$ and $J_1$ found for the ternary case in Section~\ref{nimplex:ssec:ternarygraph}. One can also note that $J_0$ always equals to $1$ as $\binom{a}{0} = 1$ for any $a$.

With $\textbf{J}$ defined, one can take a quaternary system (d=4 / 3-simplex) and perform a similar visualization thought exercise in the head as in Section~\ref{nimplex:ssec:ternarygraph}, but in 3D, considering the new transitions to 3 neighbors above and 3 neighbors below, in order to set up \texttt{\textbf{neighborsLink4C}} procedure which is presented in Section \ref{nimplex:app3}. 

Such an approach of visualizing and counting the possible jumps in the head becomes (a) challenging for quinary systems (d=5 / 4-simplex) case where one has to visualize 4 forward and 4 backward jumps to and from points inscribed in every tetrahedron formed by the 3-simplex tetrahedral grids, and (b) near impossible for higher orders, both because of the visualization dimensionality and the growing number of neighbors to track, given by $\sum_\delta^d 2(\delta-1)= d(d+1)$ or for $d=$ 6, 7, 8, and 9 corresponding to 30, 42, 56, and 72 neighbors respectively; thus prompting for an alternative.

Fortunately, while performing the above thought exercises for increasing $d$, with transition lengths \textbf{T} expressed as compositions of jump lengths described by \textbf{J}, a careful observer can quickly note that for any dimensionality of the simplex grid, the main challenge in finding the higher-dimensional \textbf{T} lies in distributing the $d-1$ new forward ($x_0$ increment) transitions across all previous $x_i=0$ constraints, while the $d-1$ new backward ($x_0$ decrease) transitions are always possible for $x_0>0$ and follow a relatively simple trend of transition lengths $J_d$, $\sum_{j=d-1}^{d}J_j$, ..., $\sum_{j=0}^{d}J_j$. This allows a relatively simple construction of all backward transitions by stacking them together across all $d-2$ considered dimensions. 

Finally, a simple notion that every backward transition $b \rightarrow a$ of grid point $b$ is associated with a forward transition $a \rightarrow b$ of point $a$ allows for the complete construction of the simplex graph representation of the compositional space.

This is implemented very concisely in the \texttt{nimplex} snippet below, where for every considered dimension $\delta$ from $d$ (highest at $0$th index of $\textbf{x}$) down to $2$ ($(d-2)$th index), the $\delta$ of backward and $\delta$ of forward transitions of lengths $t_k$ are found by iteratively summing jump lengths $J_{\delta}$, $\sum_{j=\delta-1}^{\delta}J_j$, ..., $\sum_{j=0}^{\delta}J_j$, and then used to assign neighborhood.

\begin{minted}[xleftmargin=3\parindent, linenos=true, mathescape=true]{nim}
proc neighborsLink(
    i:int, x:Tensor, neighbors: var seq[seq[int]], ndiv: int): void =
  var jumps = newSeq[int](dim-1)
  jumps[0] = 1     #binom(a,0)=1
  for j in 1..<(dim-1):
    jumps[j] = binom(
    j+ndiv-sum(x[0..(dim-2-j)]), j)
  var trans: int
  for order in 0..(dim-2): 
    trans = 0
    if x[order] != 0:
      for dir in 0..(dim-2-order): 
        temp += jumps[dim-2-order-dir]
        neighbors[i].add(i - trans)
        neighbors[i - trans].add(i)           
\end{minted}

\begin{figure}[h]
    \centering
    \includegraphics[width=0.6\textwidth]{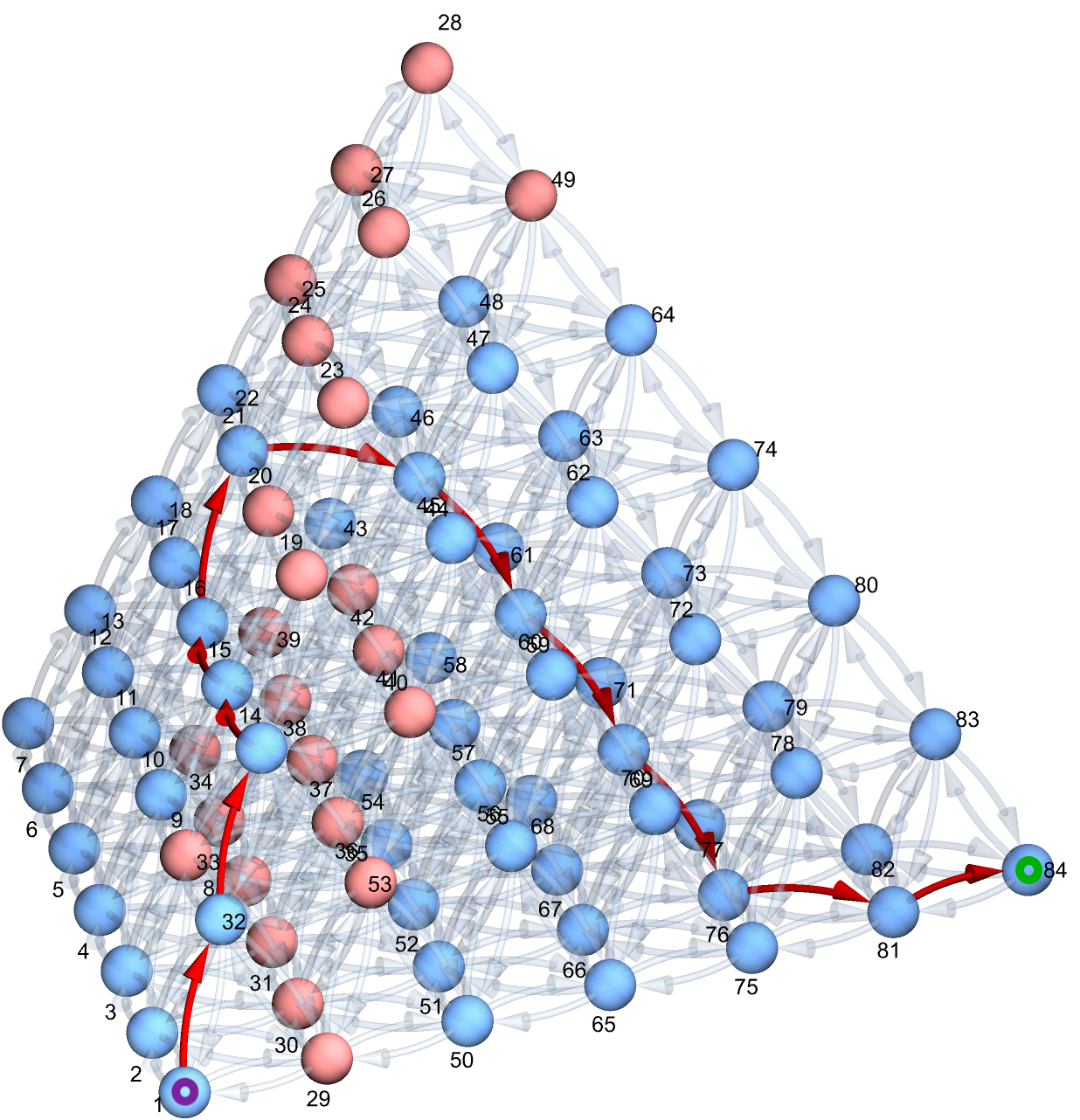}
    \caption{A quaternary (d=4 / 3-simplex) simplex graph ($n_d=6$) with 84 nodes (compositions) and 672 edges (possible moves). A set of nodes has been manually selected (highlighted in pink) to depict a toy example of infeasible points (similarly to Figure~\ref{nimplex:fig:pathplanning1}), which forces a non-trivial path (highlighted in red) to traverse from the bottom-left corner at 1 to the bottom-right corner at 84.} 
    \label{nimplex:fig:quaternarysimplexgraph}
\end{figure}

The result of running the above algorithm with $d=4$ and relatively low $n_d$ is shown in Figure~\ref{nimplex:fig:quaternarysimplexgraph} to help visualize neighbor-neighbor transitions despite the overlap when printed in 2D.

It is critical to note that the above algorithm is still within the $\mathcal{O}(N)$ computational complexity for $N$ grid points, just like the forward/backward jumps discussed in Section~\ref{nimplex:ssec:binarygraph}. Thus, for instance, the task of constructing 1\% resolution graph for a 6-component chemical space containing $N_S(d=6, n_d=100)$ or nearly \textit{100 million unique vertices} requiring \textit{2.76 billion edges} (possible chemistry changes) takes as little as \textit{23s} tested on author's laptop computer. This stands in stark contrast with $\mathcal{O}(N^2)$ distance-based graph construction, which, even when well implemented to take around $3ns$ per comparison, would take approximately 1 year on the same machine.

Furthermore, the method scales excellently with the increasing problem dimensionality. For a 12-component chemical space with $n_d=12$ divisions per dimension, even though up to $132$ neighbors have to be considered for all $N_S=1.35$ million vertices, the 93 million edges are constructed in 950 milliseconds.

\subsection{Simplex Graph Complexes} \label{nimplex:ssec:complexes}

Once able to rapidly set up simplex graphs in arbitrary dimensions, one can also efficiently combine them to construct more complex graphs representing non-trivial problem statements where many different paths are possible to explore, and prior knowledge can be incorporated as assumptions in the problem solution space if needed. At the same time, it allows the dimensionality of the intermediate compositional spaces to be kept within manufacturing feasibility, i.e., the number of material flow sources.

Suppose one tries to connect elemental compositions A and F, but assumes prior knowledge that they cannot be combined directly in any quantity, and also knows that (1) A is compatible with B and C, (2) F is compatible with D and E, but (3) B and E are incompatible in any quantity, (4) C and D are incompatible in any quantity. Furthermore, (5) G and H are individually compatible with B and D, and (6) I and J are individually compatible with C and E. 

\begin{figure}[H]
    \centering
    \begin{adjustbox}{valign=c}
        \includegraphics[width=0.45\textwidth]{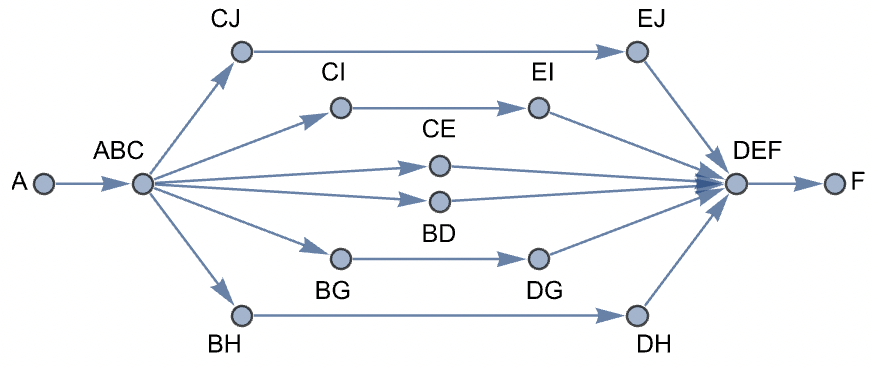}
    \end{adjustbox}
    \hspace{6pt}
    \begin{adjustbox}{valign=c}
        \includegraphics[width=0.5\textwidth]{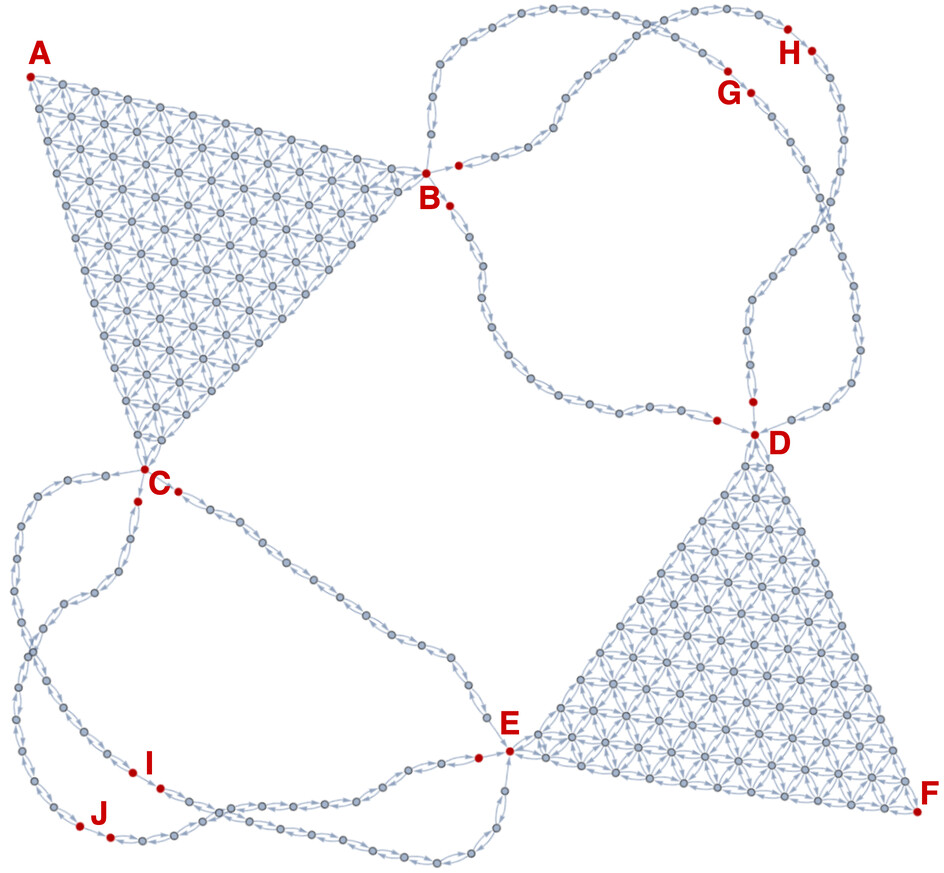}
    \end{adjustbox}
    \caption{Graph Complex Example \#1 depicting a problem space where 2 ternary systems can be connected through 6 different binary paths.} 
    \label{nimplex:fig:graphcomplex1}
\end{figure}

These rules can be used to set up a problem graph like in the top of Figure~\ref{nimplex:fig:graphcomplex1}, encoding everything that is known about the system \textit{a priori} and limiting the solution space from full $\binom{10-1+12}{12} \approx 300,000$ to $2\times\binom{3-1+12}{12} + 10\times\binom{2-1+12}{12} = 312$, or three orders of magnitude.

The space constructed in Figure~\ref{nimplex:fig:graphcomplex1} is kept very minimal in terms of going beyond known assumptions and dimensionality to illustrate the concept in a plane. However, real examples of this technique can be highly non-trivial and essential in bringing the number of considered points into a computationally feasible regime when tens of available compositions can be considered. 

Furthermore, unlike in Figure~\ref{nimplex:fig:graphcomplex1} where spaces are simply connected through single-components, the interfaces between the individual compositional spaces can be along any subspace (e.g., the ternary face of quaternary tetrahedron), allowing one to quickly set up search problems where one or more components are unknown, but their relations to others are fixed. 

\begin{figure}[h]
    \centering
    \begin{adjustbox}{valign=c}
        \includegraphics[width=0.37\textwidth]{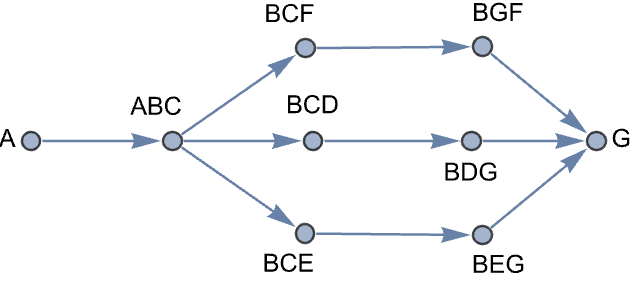}
    \end{adjustbox}
    \hspace{6pt}
    \begin{adjustbox}{valign=c}
        \includegraphics[width=0.57\textwidth]{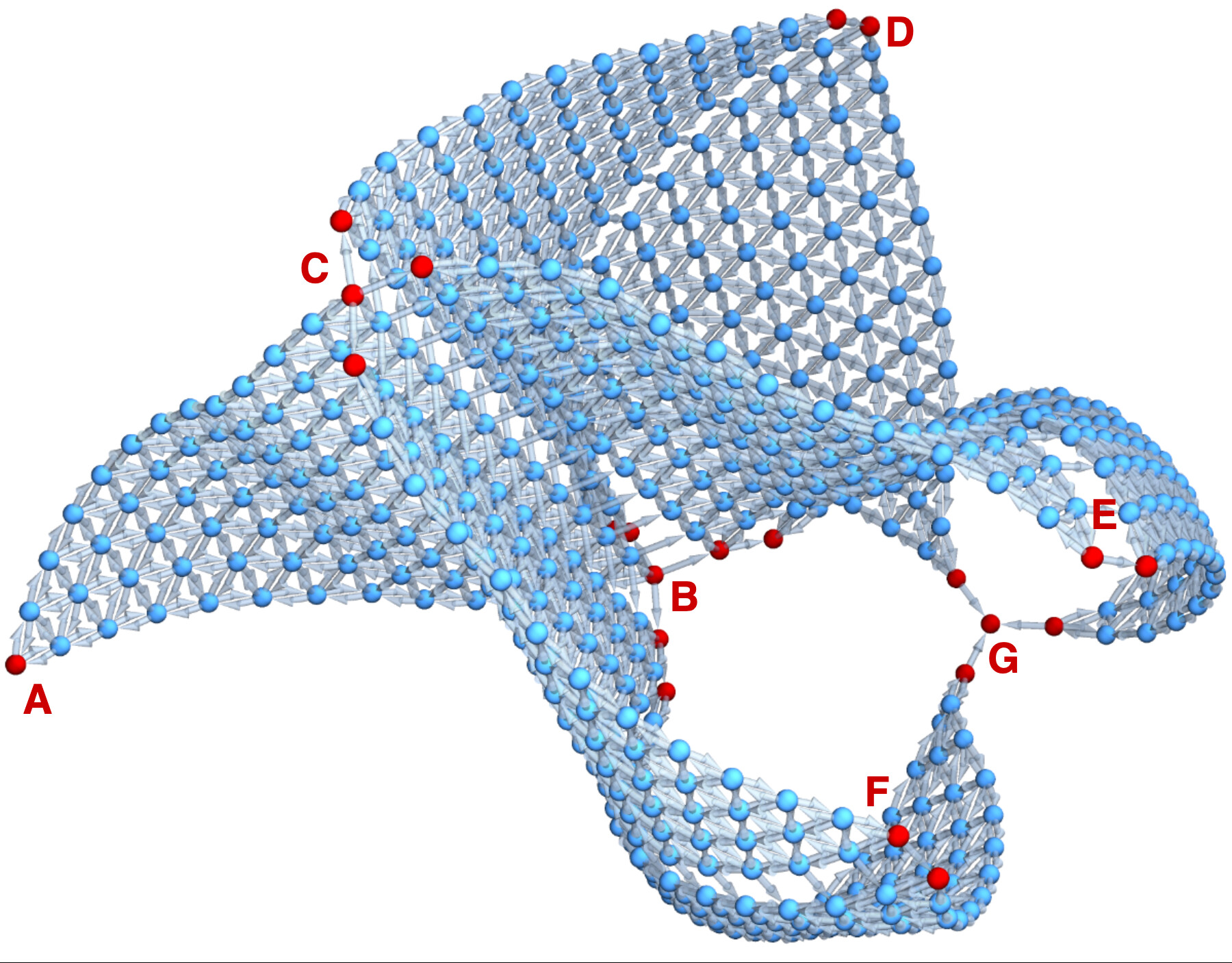}
    \end{adjustbox}
    \caption{Graph Complex Example \#2 depicting a problem where 3 choices (D/E/F) can be made to traverse from ABC to G through dual ternary systems containing B. Vertices were spread in 3D to depict three possible ABC to G paths, which would exactly overlap in a plane.} 
    \label{nimplex:fig:graphcomplex2}
\end{figure}

One can quickly demonstrate the benefits of such ability by looking at the SS316 to Ti-6Al-4V problem studied by \citet{Bobbio2022DesignCompositions}. After idealizing and anonymizing the components, it becomes a problem where one tries to combine compositions A with G, which cannot be combined directly in almost any quantity, and also knows that (1) system ABC is highly feasible across it, but (2) C cannot be combined directly with G in any quantity, and (3) a complete path from pure B to G is not possible. In this case, a simple problem setup is to look at several BC\texttt{?} and BG\texttt{?} pairs, forming parallel pathways from ABC to G. This is depicted in Figure~\ref{nimplex:fig:graphcomplex2} for 3 candidates D, E, F, forming 6 ternary spaces to consider, but nothing limits the method to be extended to an arbitrary number of candidates while still retaining its linear complexity.

\begin{figure}[h]
    \centering
    \begin{adjustbox}{valign=c}
        \includegraphics[width=0.35\textwidth]{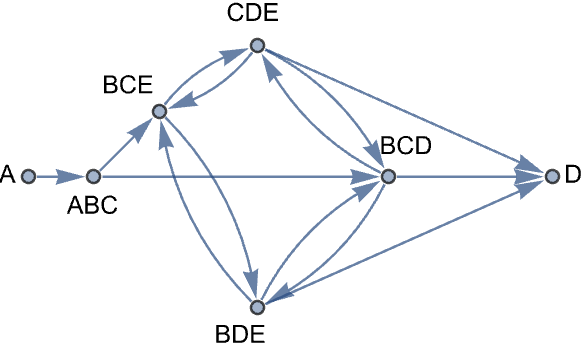}
    \end{adjustbox}
    \hspace{6pt}
    \begin{adjustbox}{valign=c}
        \includegraphics[width=0.62\textwidth]{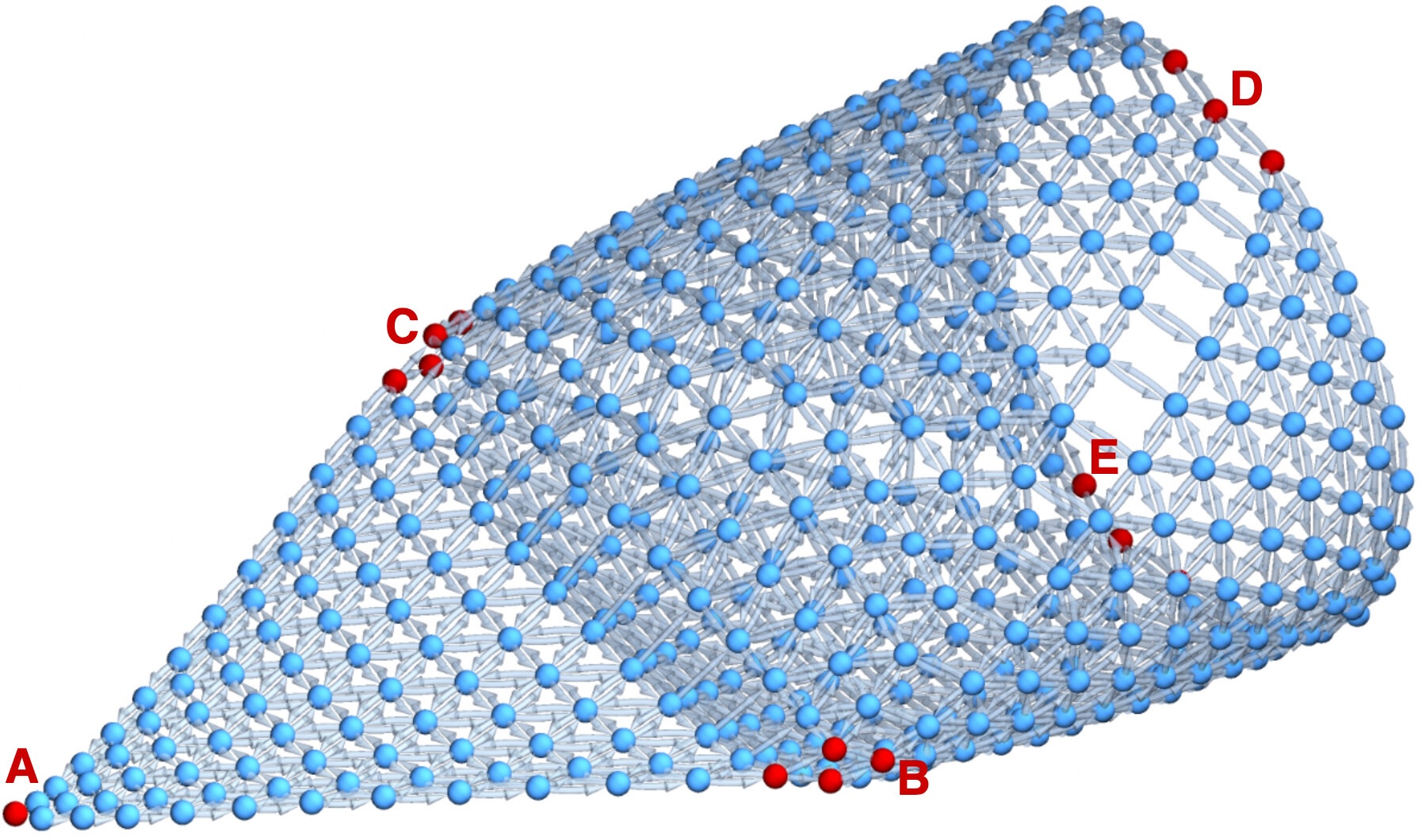}
    \end{adjustbox}
    \caption{Graph Complex Example \#3 depicting the possibility of competing paths, including cycles.} 
    \label{nimplex:fig:graphcomplex3}
\end{figure}

In the above examples in Figures~\ref{nimplex:fig:graphcomplex1}~and~\ref{nimplex:fig:graphcomplex2}, all connections between compositional spaces were directional; however, that is not necessary, and in some problems it may be beneficial to allow bidirectional movement. Suppose one tries to combine compositions A with D, which cannot be combined directly in any quantity, and also knows that (1) system ABC is highly feasible across it, but (2) system BCD is not traversable on its own. Thus, E can be introduced to set up intermediate spaces BDE and CDE, allowing obstacles in BCD to be avoided. Furthermore, BCE can also be set up as an alternative, possibly shortening the total path. Figure~\ref{nimplex:fig:graphcomplex3} depicts such a problem setup.

Notably, while the above example in Figure~\ref{nimplex:fig:graphcomplex3} depicts a single 5th component E to help visualize cycling between spaces, these concepts can be extended to many possible intermediate components. At the same time, the maximum dimensionality of individual compositional spaces is kept constant ($d=3$). Thus, it provides a powerful method to keep the problem solvable, even experimentally, while considering many possible pathways formally defined prior to path planning to fit within the feasibility of evaluation and manufacturing.

\subsection{Discussion of Exploration} \label{nimplex:sec:discussion}

Critically, creating such a homogeneous problem structure through graph representation allows one to deploy the same exploration strategies across many dimensionalities and even combinations of individual spaces shown in Section~\ref{nimplex:ssec:complexes}. Furthermore, in the described graphs, points are on an equidistant grid; thus, it is easy to set up a heuristic function that can be both consistent and admissible. 

This, in turn, enables one to harvest many general-purpose graph traversal algorithms, which are actively researched and available through high-performance libraries. For instance, to navigate against constraints, the $A^*$ algorithm \cite{Hart1968APaths} can be used with such a heuristic and is mathematically guaranteed to find \textit{the} shortest feasible compositional path while exploring the least number of nodes \cite{Dechter1985GeneralizedA}, what can be critical if the shortest path is necessary while each evaluation is relatively expensive. Then, if one tries to find \emph{a} feasible solution first and then improve on it, modifications of the $A^*$ algorithms such as the $RWA^*$ \cite{Bhatia2021OnA} can be used to first make it more greedy and then gradually move towards $A^*$ to obtain the optimal solution if sufficient resources are available. Alternatively, for highly complex problems where exploration needs to proceed quickly towards the goal but the optimality guarantees are not needed, one can use a search algorithm based on the Monte Carlo tree search (MCTS), which has been famously used in conjunction with an ML model to master the game of Go \cite{Silver2016MasteringSearch}.

\section{Summary and Conclusion} \label{nimplex:sec:summary}

This work starts by providing an abstract description of compositional spaces applicable to a wide range of disciplines while formalizing several vital concepts. Then, Section~\ref{nimplex:ssec:compositionallycomplex} discusses complex compositional spaces, using Compositionally Complex Materials (CCMs) as a real-world application and considers the challenges of exploring such spaces using different methods. Section~\ref{nimplex:ssec:functionallygraded} uses another real-world application of Functionally Graded Materials (FGMs) to expand on that by discussing compositional spaces formed from compositions in other spaces and when these spaces are preferred for design. It also discusses key concepts related to path planning in relation to types of constraint and property optimizations. Last in the Introduction, Section \ref{nimplex:ssec:combinatorialcomplexities} discusses some equations critical to investigating the combinatorial complexities in these problems.

Next, discussions and implementations are given for several methods for efficiently solving compositional problems through random sampling in Section~\ref{nimplex:sec:randomuniformsampling}, grid-based methods in Section~\ref{nimplex:sec:simplexgrid}, graph-based methods, including graphs combining multiple compositional spaces, in Section~\ref{nimplex:sec:simplexgraph}. The three most critical contributions introduced in this process are:

\begin{enumerate}
    \item Novel algorithm for rapid procedural generation of \emph{N-dimensional graph representations} of compositional spaces where uniformly distributed simplex grid points in $d$ dimensions are completely connected to up to $d(d-1)$ neighbors representing all possible component-pair changes. For instance, in economics, this could represent all possible compositions of a financial portfolio of 12 assets and, for each one of them, all 132 transfer choices that can be made to modify it. Critically, this method scales linearly with the number of points and generates graphs with billions of connections between millions of points in just seconds. Furthermore, this algorithm allows deterministic memory allocation during the graph construction, where arrays of pointers to neighboring compositions represent allowed transitions, resulting in a very high-performance data structure.
    
    \item The new free, open-source software (FOSS) package \texttt{nimplex} (\href{https://nimplex.phaseslab.org}{nimplex.phaseslab.org}), which gives high-performance implementations of both essential existing methods and all new methods introduced within this work including the simplex graphs.
    
    \item The novel concept of combining many compositional spaces using graph representations to create homogeneous problem spaces, both simplifying the path planning and allowing for efficient incorporation of constraints and assumptions about problem spaces as demonstrated in Section~\ref{nimplex:ssec:complexes}. 
    
\end{enumerate}

In addition to the above, three other new contributions are given in this work:

\begin{enumerate}
    
    \item Sections~\ref{nimplex:sec:randomuniformsampling}~and~\ref{nimplex:sec:simplexgrid} discuss random sampling and grid construction in simplex spaces in the context of the composition of chemical spaces. In the process, several theoretical results critical to the problem, which have not been discussed previously in this context, are presented. For instance, the commonly found random sampling of a $d-1$ hypercube and rejection of compositions $>100\%$ to sample a $d$-component space, commonly found in software, has a rejection rate exhibiting factorial growth and can severely impact when deploying ML models.
    
    \item In Section~\ref{nimplex:ssec:internalgrid}, a new algorithm was developed to efficiently create internal (subspace-exclusive) grids in simplex spaces based on an algorithm from the literature (modified-NEXCOM \cite{Chasalow1995AlgorithmPoints}). It is beneficial to performance in cases of, for instance, sampling only $d$-component materials in $d$-component chemical space without considering lower-order points.

    \item In a few areas, Section~\ref{nimplex:ssec:functionallygraded} leverages its general character to go beyond the usual FGM literature introduction. For instance, it contrasts elemental spaces with attainable design spaces and discusses the use of similar compositions (alloy grades) in the design process to reduce cost and greenhouse emissions without making prior assumptions.
    
\end{enumerate}

\section{Code Availability} \label{nimplex:sec:code}

The \texttt{nimplex} software described in this work has been published as free open-source software (FOSS) under the MIT license. It can be effortlessly used as a \emph{native} Nim library, \emph{native} Python library, or Command Line Interface (CLI) tool interfacing with nearly any language through binary data or plain text.

All distributions of the source contain (1) the core library, (2) additional utilities, (3) testing procedures, (4) use examples, (5) quick-start guide using Python/CLI in the form of a Jupyter notebook, (6) \texttt{devcontainer.json} specification, and (7) documentation. They are available through:

\begin{itemize}
    \item The documentation page at \href{https://nimplex.phaseslab.org}{nimplex.phaseslab.org}, which contains (1) installation instructions, (2) usage instructions in Python, Nim, and CLI, and (3) Application Programming Interface (API) reference. It also links to a public GitHub repository hosting the latest code (\href{https://github.com/amkrajewski/nimplex}{github.com/amkrajewski/nimplex}) at the time of writing.


    \item (Selected Major Versions) A public repository archive on Zenodo under DOI: \href{https://doi.org/10.5281/zenodo.10611931}{10.5281/zenodo.10611931}.
\end{itemize}


\chapter{Infeasibility Gliding in Compositional Spaces} \label{chap:infeasibilitygliding}

\acknowledge{
This chapter adapts parts of a manuscript draft planned for future publication, co-authored with Arindam Debnath, Shuang Lin, Alexander Richter, Ricardo Amaral, Wesley F. Reinhart, Allison M. Beese, and Zi-Kui Liu. All of included text was written by Adam M. Krajewski. Described software has been developed by Adam M. Krajewski extending \texttt{nimplex} described in Chapter \ref{chap:nimplex} and used to generate results employing thermodynamic models developed by Shuang Lin. All other authors provided edits and guidance.
}

\section{Introduction} \label{infglide:sec:intro}

As explored in Chapter~\ref{chap:nimplex}, exploration of high-dimensional compositional spaces, needed for many materials discovery tasks, is a challenging task, both conceptually and computationally, due to several inherent complexities. Typically, this forces efforts like screening and path planning to include as much prior knowledge (i.e., assumptions) as possible to bring these complexities down as much as possible, which has been explored in detail in Section~\ref{nimplex:ssec:complexes} on three individual examples, including real-world one based on \citet{Bobbio2022DesignCompositions}.

It is essential, however, to note that the assumptions imposed on the design space to reject spaces unlikely to work, like "\textit{Boron cannot be added because it will precipitate borides}", by the same assumptions do not significantly increase the volume of feasible (or desired) space. Thus, an approach that would explore only such regions while skipping the rest could, in principle, consider such design space at a low additional cost, reducing the number of assumptions and possibly identifying high-performing materials that would otherwise be skipped.

\section{Exploiting Compositional Graph Representation} \label{infglide:sec:exploitgraph}

To set up an approach exploring only feasible or otherwise desirable spaces, one can begin by leveraging \texttt{nimplex}'s compositional graph representation, described in Section~\ref{nimplex:sec:simplexgraph}, which enables one to easily traverse all compositions based on their adjacency, starting from one or more points, akin to typical high-throughput screenings that exhaust the design space population \cite{Feng2021High-throughputAlloys, Wang2023SearchingExperiments, Yang2022AHardness, Maresca2020Mechanistic1900K}.

\begin{figure}[H]
    \centering
    \includegraphics[width=0.7\textwidth]{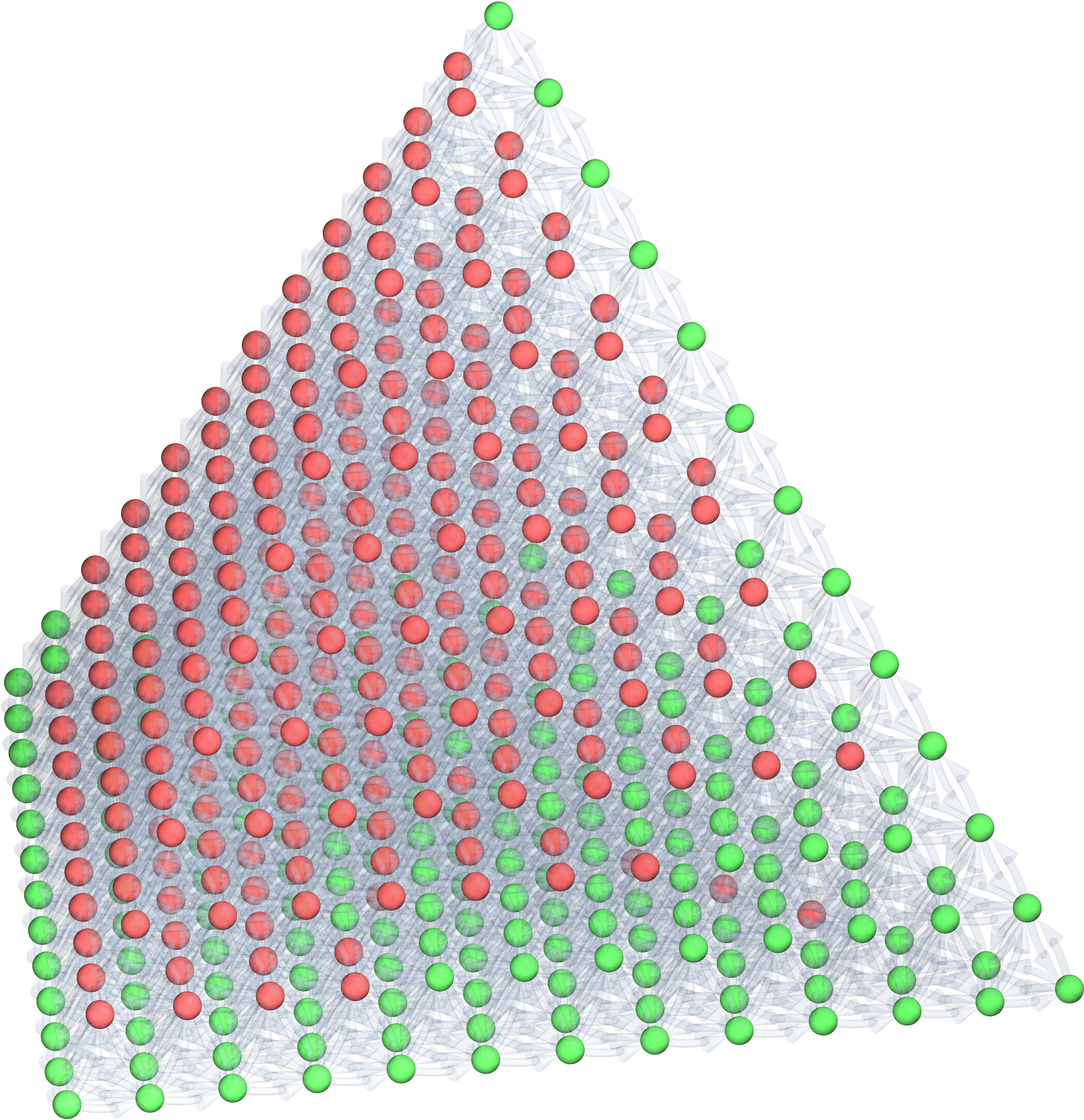}
    \caption{Feasibility map over compositional tetrahedron (3-simplex) formed by all combinations of Ti50 Zr50, Hf95 Ti5, Mo33 Nb33 Ta33, Mo80 Nb10 W10 discretized at 12 divisions per dimension. The positions in the 7-component elemental space obtained from \texttt{nimplex}, described in Chapter \ref{chap:nimplex}, were used to run \texttt{pycalphad} \cite{Otis2017Pycalphad:Python} evaluations and constrained by limiting phases present at equilibrium at 1000K to single or many solid solution phases. Roughly half of the compositions are infeasible, with most of them forming a single large region.}
    \label{infeasibilitygliding:fig:fullcomputation}
\end{figure}

Figure~\ref{infeasibilitygliding:fig:fullcomputation} depicts an example result of such an approach for a specific 4-component design space in a 7-component elemental space, with roughly half of the points being feasible (green), forming a complex concave shape, and half being infeasible (red).

\section{Gliding on the Boundaries of Infeasibility} \label{infglide:sec:glide}

As shown in Figure~\ref{infeasibilitygliding:fig:fullcomputation}, the infeasible region of space is generally continuously bounded by a single smooth surface, with only a few other small infeasible points. Thus, in principle, the infeasible space could be efficiently navigated around through only surface point calculations, without considering the bulk of internal points that cannot be accessed, accomplishing the goals set in Section~\ref{infglide:sec:intro}. This work coins the term \emph{Infeasibility Gliding} to describe such an approach.

\subsection{Underlying Assumptions} \label{infglide:ssec:assumptions}

\begin{figure}[H]
    \centering
    \includegraphics[width=0.5\textwidth]{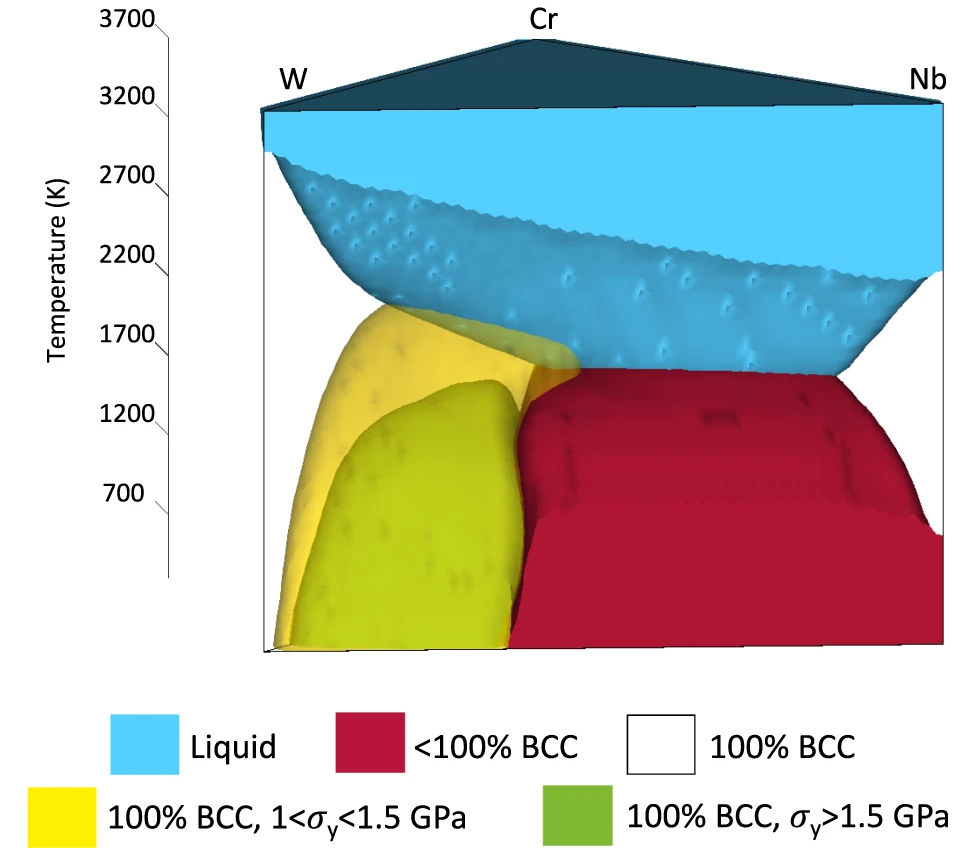}
    \caption{A view of ternary Cr-Nb-W phase diagram projected across a temperature range of phase classes, further augmented by imposing predicted property value constraints. It depicts the smoothness of the infeasible region boundary (red) and the increasingly smooth boundary of the property-constrained region. Taken from Figure 2b in \citet{Elder2023ComputationalValidation} under CC BY 4.0 license.}
    \label{infeasibilitygliding:fig:katesphasemap}
\end{figure}

One core assumption that needs to be considered in exploration based on the infeasibility gliding is that the high-dimensional surface bounding the infeasible space is highly smooth, or in terms of phase stability, that the region where a given infeasible phase exists (often spanning multiple phase regions) is smoothly bound. While not possible to be proven to be valid in every system, it can be shown to be reasonable for exploration problems, as it (1) is \emph{not required} for the method to glide around the boundary, but only to argue for low computational cost, and (2) it is generally true for metallic systems of interest, as depicted in an example in Figure~\ref{infeasibilitygliding:fig:katesphasemap} from \citet{Elder2023ComputationalValidation}, as well as many other studies, like ones by \citet{Bobbio2022DesignCompositions}, \citet{Sun2024MaterialsMap:Ag-Al-Cu}, \citet{Gao2016SenaryHfNbTaTiVZr}, or \citet{Zhao2014ExperimentalSystem}.

\subsection{Unbiased Exploration Searches} \label{infglide:ssec:unbiasedexplore}

With the infeasibility gliding approach, one can now perform the same traversal over a graph as done in Section~\ref{infglide:sec:exploitgraph}; however, it is limited to exploring only the neighborhood of the feasible points and, thus, not going into the inside of the infeasible region. This highly desirable behavior, reducing computation by a factor of roughly 2, is shown in Figure~\ref{infglide:sec:glide}, in contrast to the earlier Figure~\ref{infeasibilitygliding:fig:fullcomputation}. The presented results are taken from the second \texttt{nimplex} workshop, which has been adapted as Appendix~\ref{chap:nimplextutorial2} and can be consulted for step-by-step details of an example implementation.

\begin{figure}[H]
    \centering
    \includegraphics[width=0.7\textwidth]{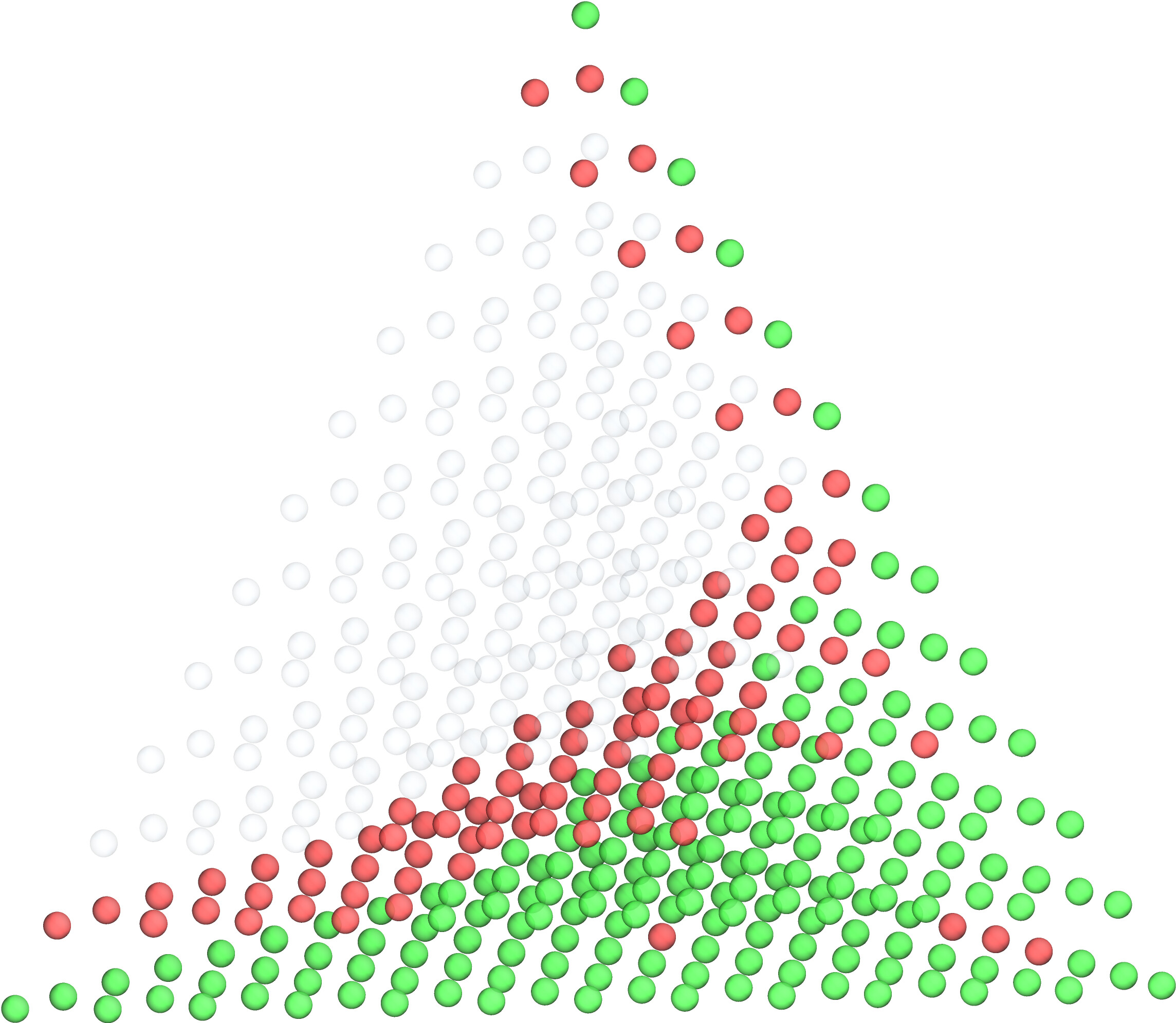}
    \caption{The same problem as in Figure \ref{infeasibilitygliding:fig:fullcomputation} solved by iteratively exploring all feasible paths in the compositional graph in a depth-first approach, which can be started from one or multiple points, and terminated once the goal is reached or once all of the feasible space is explored.}
    \label{infeasibilitygliding:fig:glide}
\end{figure}


\chapter{Path Planing In Compositional Spaces through Graph Traversals} \label{chap:pathplanning}

\acknowledge{
This chapter is part of work described in Chapter \ref{chap:infeasibilitygliding} and carries the same acknowledgments.
}

\section{Introduction} \label{pathplan:sec:intro}

Once (1) models developed in Chapters \ref{chap:sipfenn} through \ref{chap:nimcso} are available, and (2) the approach to traversing the compositional space established in Chapters \ref{chap:nimplex} and \ref{chap:infeasibilitygliding} is deployed, one can begin to design multi-composition entities, such as coexisting compositional regions on paths joining multiple compositions in Functionally Graded Alloys (FGMs) discussed in Section~\ref{nimplex:ssec:functionallygraded}. This task will generally entail constructing continuous composition sub-graphs based on certain value rules, allowing one to optimize the design.

All results and figures presented throughout this Chapter have been obtained and can be reproduced based on the second \texttt{nimplex} workshop, which has been adapted as Appendix~\ref{chap:nimplextutorial2} and can be consulted for step-by-step details, including an example implementation, which can be easily modified to work in a variety of problems with minimal changes.

\section{Shortest Path Planning} \label{pathplan:sec:shortest}

The most common and most simply defined task in the FGM path planning is to identify a short \cite{Bobbio2022DesignCompositions} or the shortest path. Figure~\ref{pathplan:sec:shortest} presents an example of such a path, constructed using Dijkstra's algorithm \cite{Dijkstra1959AGraphs} between two points of choice, over the feasible space graph extracted from results underlying Figure~\ref{infeasibilitygliding:fig:glide} discussed in the last chapter. 

\begin{figure}[H]
    \centering
    \includegraphics[width=0.9\textwidth]{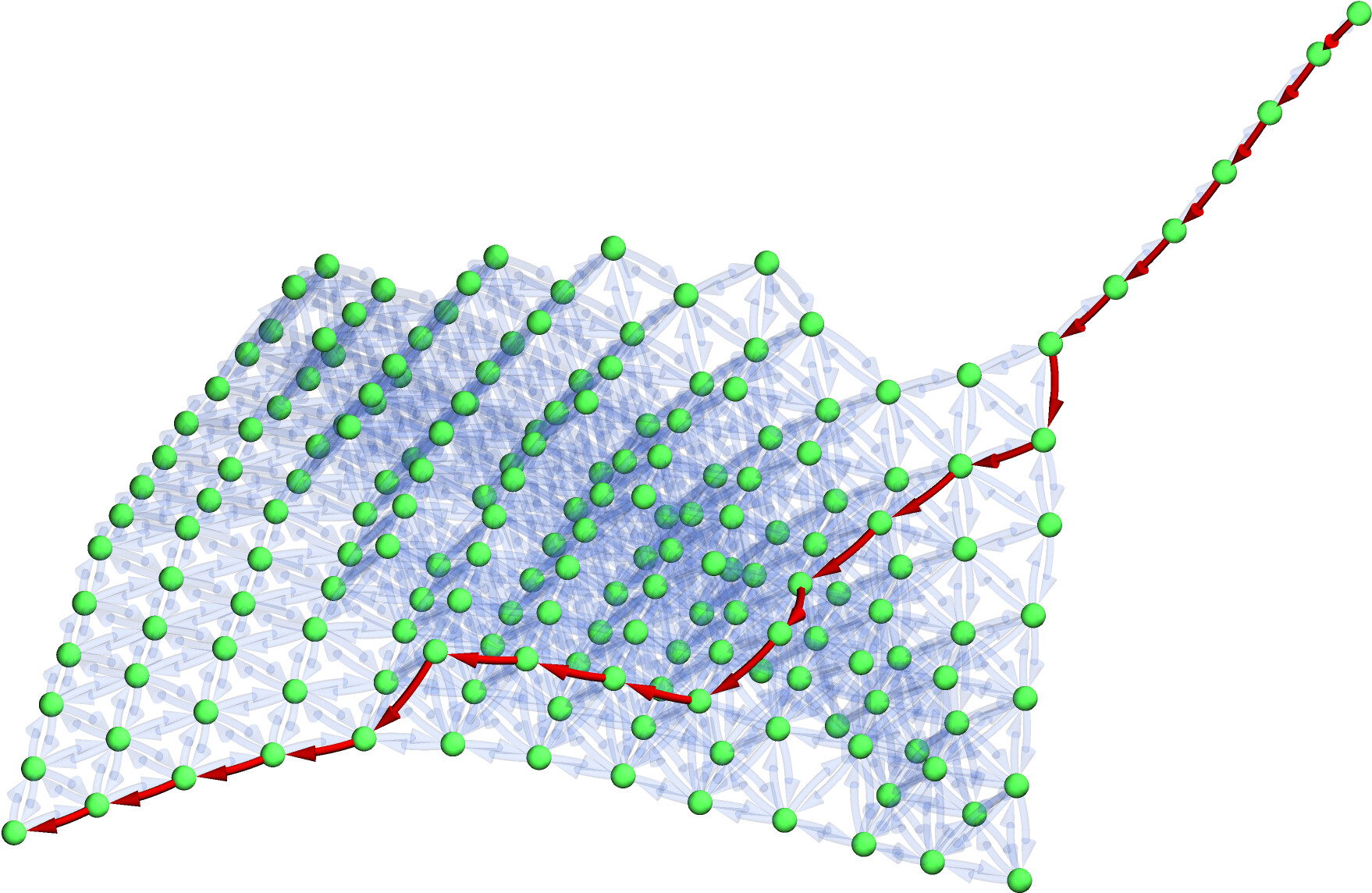}
    \caption{The subgraph of feasible space extracted from the full 3-simplex (tetrahedral) graph in Figure \ref{infeasibilitygliding:fig:glide} constructed by gliding around an infeasible region of compositional space. The red path overlaid over directional edges indicates the optimal (least number of transitions) path was identified by the common Dijkstra's algorithm \cite{Dijkstra1959AGraphs}.}
    \label{pathplan:fig:shortestpath}
\end{figure}

Notably, the depicted path belongs to a \emph{set of several equivalent shortest paths} between the two considered points. This situation is similar to more common path plannings on rectangular grids; however, as explored in Chapter~\ref{chap:nimplex}, the number of possible transitions from each point in increasingly dimensional simplex space grows much faster ($d(d-1)$) relative to corresponding Cartesian space ($2d$), what can have a significant effect on memory requirements of certain planning algorithms.

\section{Property Gradient Minimization} \label{pathplan:sec:gradientmin}

As demonstrated in detail in Appendix~\ref{chap:nimplextutorial2}, \texttt{nimplex} graph representations offer a very flexible framework for encoding additional characteristics of the problem at hand, like the common FGM-related task of minimization of thermal expansion coefficient (TCE) gradients \cite{Kirk2021ComputationalMonotonicity} to avoid cracking at elevated temperatures, or elastic modulus gradients to avoid cracking in certain stress conditions.

\subsection{Stretching the Space} \label{pathplan:ssec:gradientstretch}

\begin{figure}[H]
    \centering
    \includegraphics[width=0.9\textwidth]{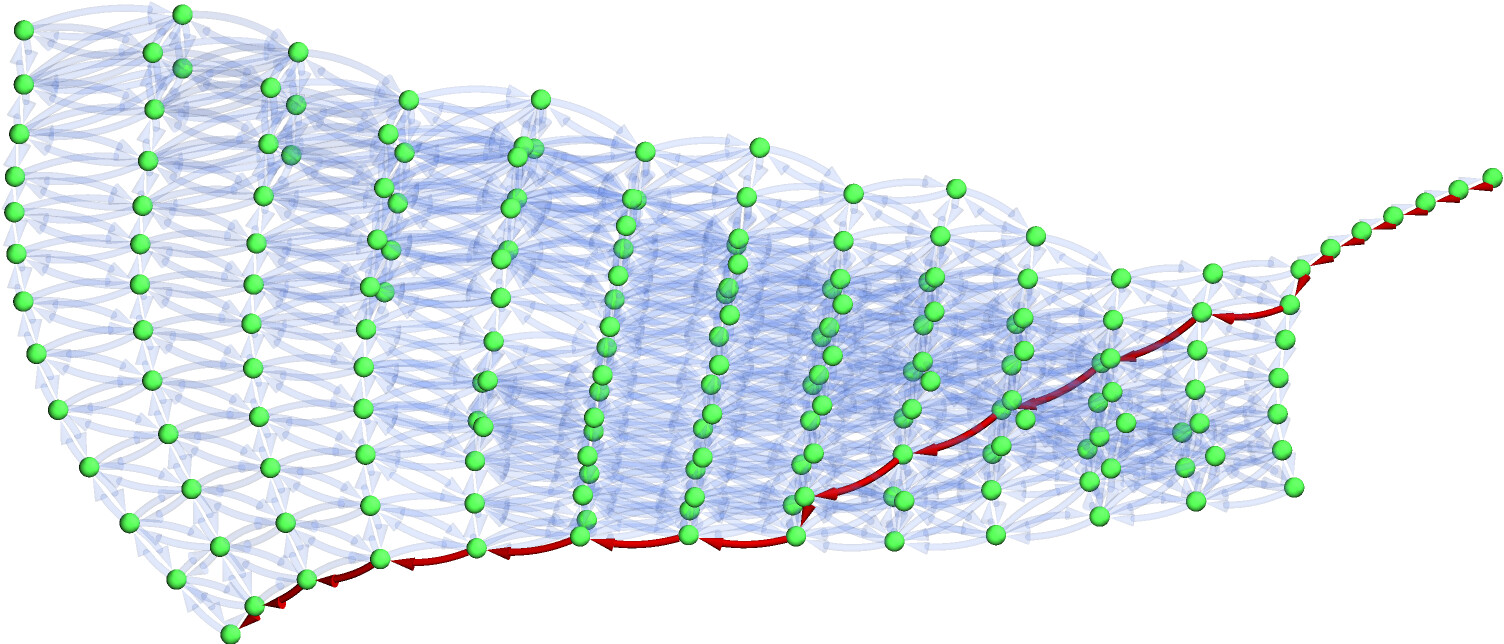}
    \caption{The graph from Figure \ref{pathplan:fig:shortestpath} stretched through distance increases from penalizing high magnitude of gradient in a property value. The graph is approximately relaxed through spring-type energy minimization performed in \texttt{Wolfram} Language. The shortest path is still equally optimal in terms of the number of steps, but the selection has been biased towards the low-gradient region.}
    \label{pathplan:fig:lowgradient}
\end{figure}

One can, for instance, stretch the edges between points shown earlier in Figure~\ref{pathplan:fig:shortestpath} by adding a penalty linearly proportional to the magnitude of the gradient of a property value. In the Appendix~\ref{chap:nimplextutorial2}, an example property of choice has been the Root Mean Square Average Deviation (RMSAD) model for BCC solid solutions by \citet{Tandoc2023MiningAlloys}, modified to fit ULTERA model pattern. The result of such stretching is shown in Figure~\ref{pathplan:fig:lowgradient}, where the path has been biased towards low-gradient regions, achieving the singular shortest design path under such conditions. 

Notably, while the resulting path is different compared to one in Figure~\ref{pathplan:fig:shortestpath}, it still belongs to the set of shortest paths in terms of un-stretched distance (number of transitions) for any scale of gradient penalty, as neighboring derivative paths are similarly affected by the penalty.

\subsection{Non-Linear Penalties Escaping Embedding} \label{pathplan:ssec:gradientsquare}

\begin{figure}[H]
    \centering
    \includegraphics[width=0.9\textwidth]{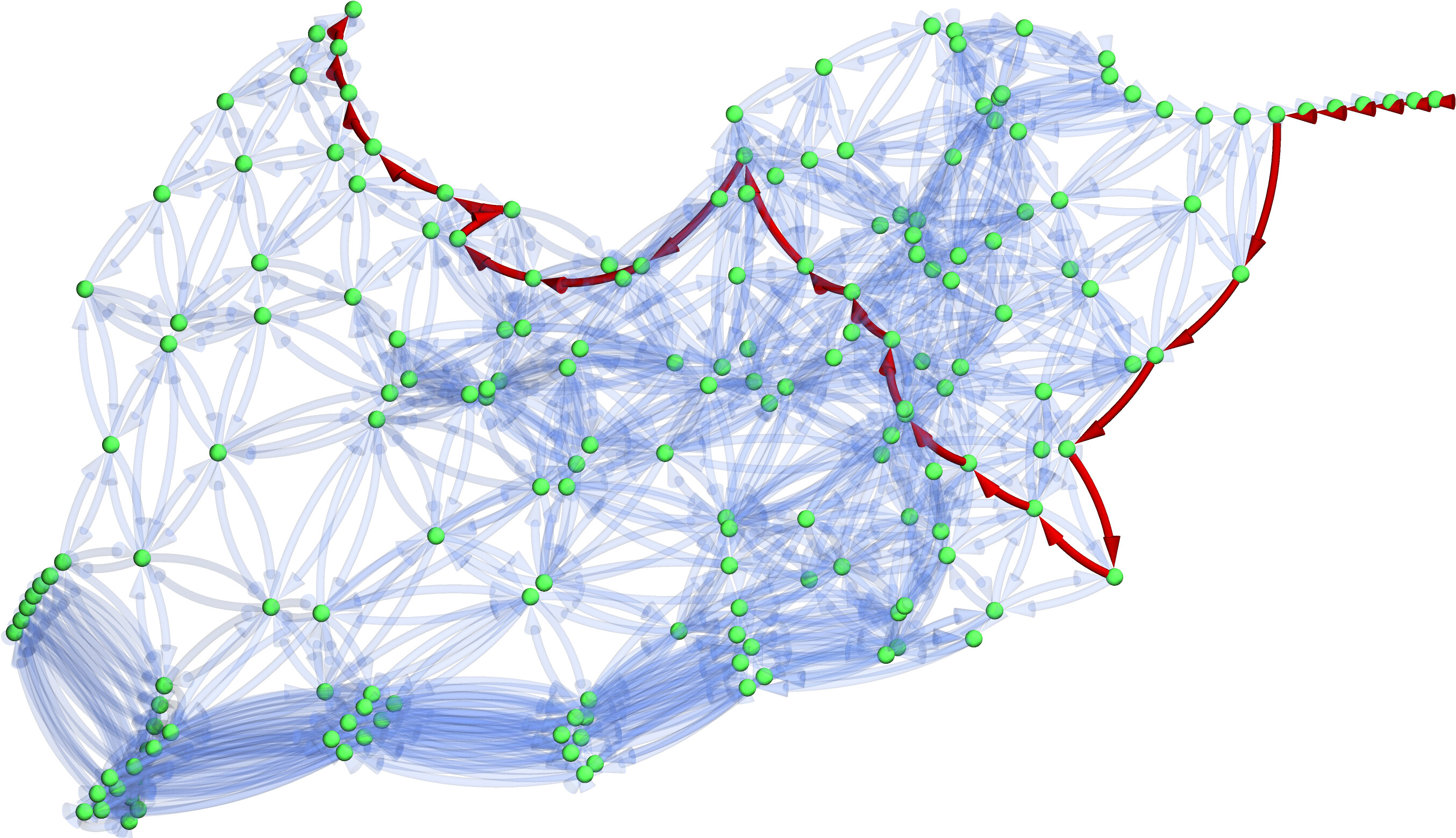}
    \includegraphics[width=0.9\textwidth]{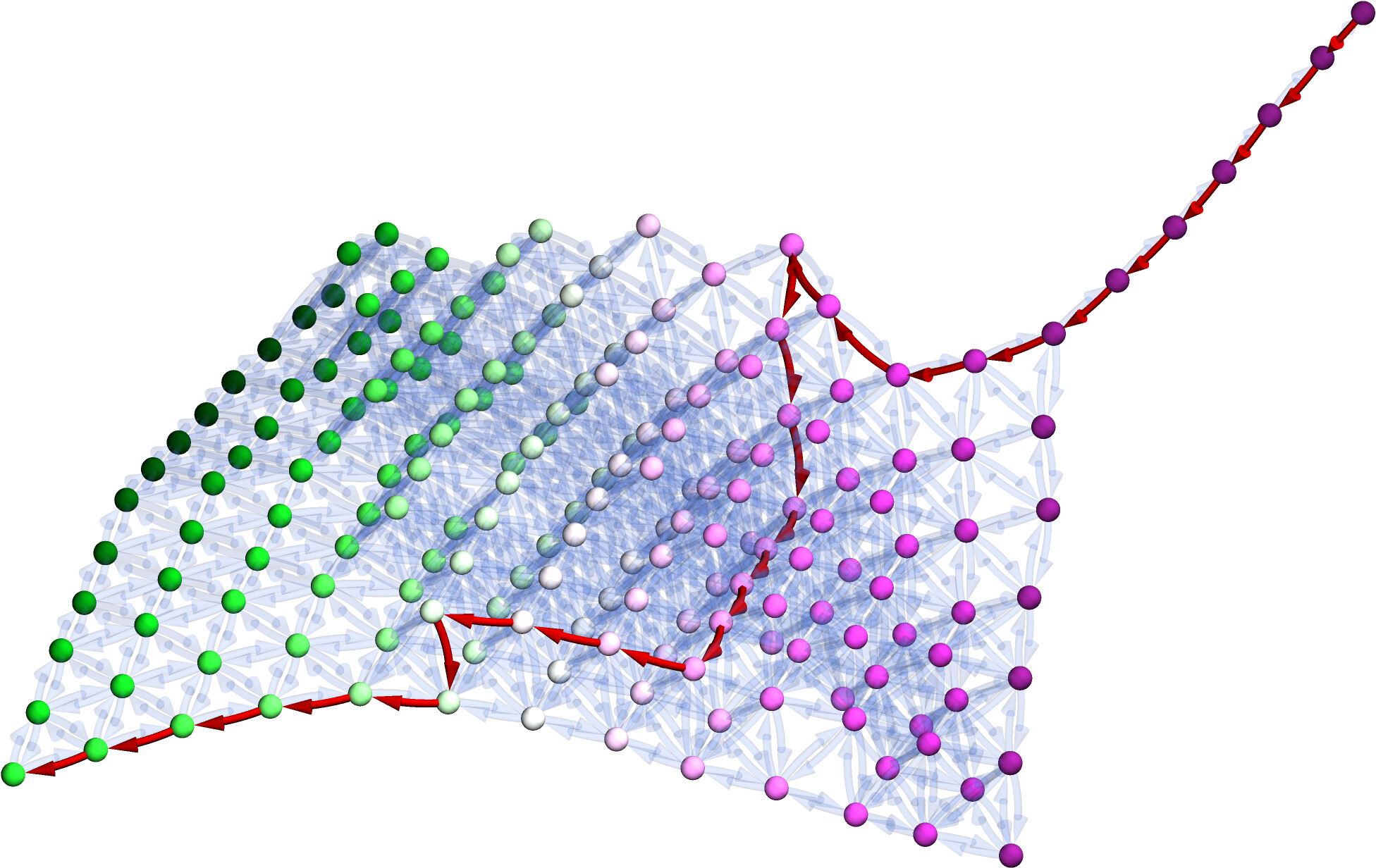}
    \caption{Application of a highly non-linear (squared) property gradient magnitude penalty to the inter-node distances causing (top) unrelaxable (figure shows a local minimum) spatial arrangement of graph nodes, which can be visualized (bottom) through the color encoding of property field with path forming "switchbacks" akin to mountain roads that minimize sharp gradients at the cost of 3 additional steps.}
    \label{pathplan:fig:lowgradientsquared}
\end{figure}

The introduction of a non-linear gradient results in distances that cannot be easily visualized as stretching of the space, as depicted in Figure~\ref{pathplan:fig:lowgradientsquared}, but can encode avoidance of singular gradient changes, which can be much more critical in the design of FGMs deposited layer-by-layer where resistance to cracking may be primarily dependent on a few most venerable transitions.

\section{Property Value Min/Maximization} \label{pathplan:sec:minmax}

\begin{figure}[H]
    \centering
    \includegraphics[width=0.9\textwidth]{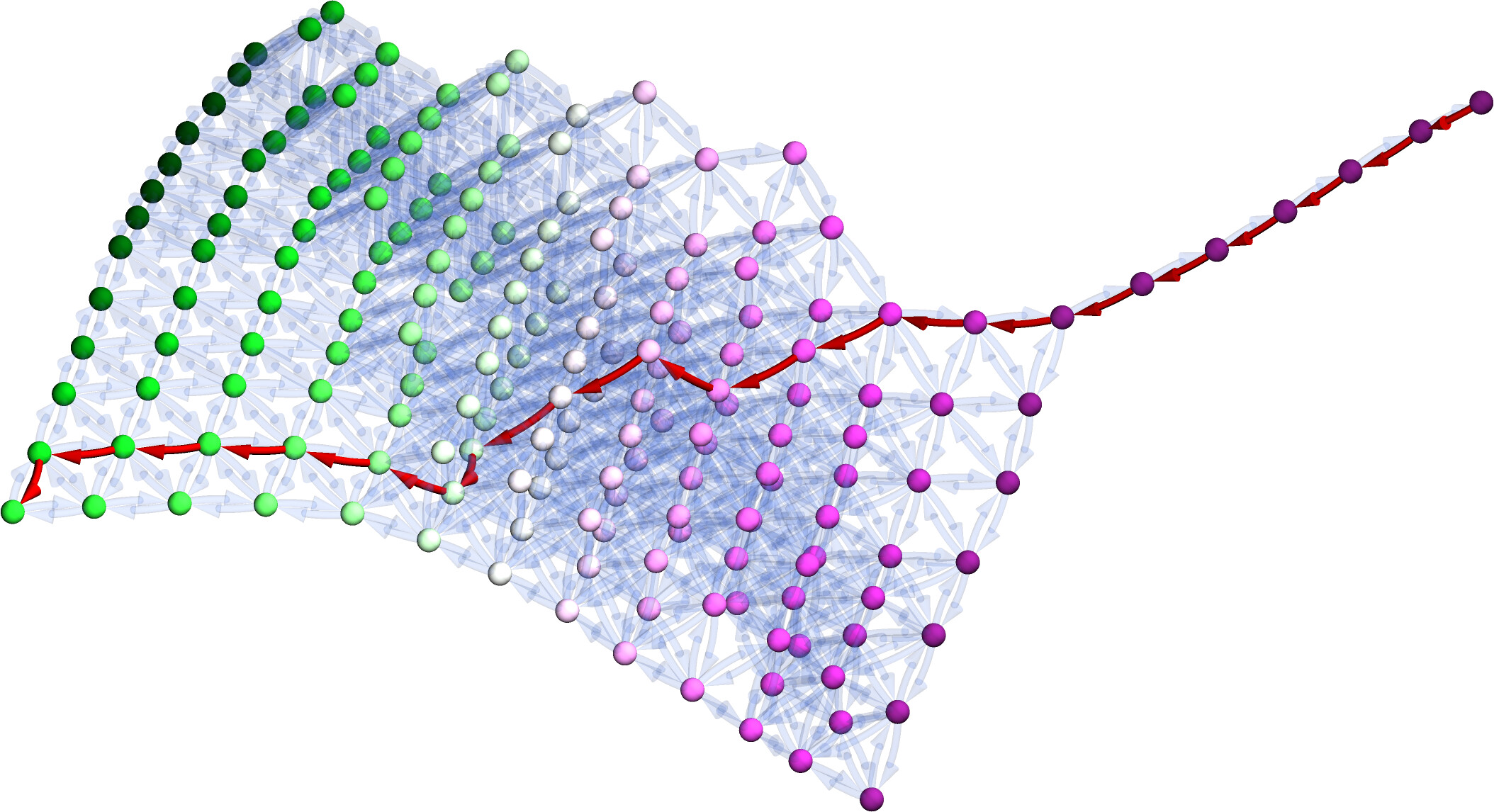}
    \caption{A selection of an optimal path, similar to one in Figure \ref{pathplan:fig:shortestpath} but biased towards high property value regions (green) by penalizing going to lower property regions.}
    \label{pathplan:fig:highrmsad}
\end{figure}

Lastly, since the transitions between neighboring compositions generated by \texttt{nimplex} are described through two unidirectional edges, the distance bias can also be used to encode directional properties, such as the property gradient itself. With such an approach, one can bias the path into crossing through regions of high property value, and similar to those previously described, can be tuned to a desired trade-off with path length. Figure~\ref{pathplan:fig:highrmsad} presents an example result of such bias in the high-strength region (green). A similar approach could be used to, e.g., minimize the cost along a path or other cost function encoding many different aspects of the problem.

\section{Software Availability} \label{pathplan:sec:softwareavaialbility}

As mentioned in Section~\ref{pathplan:sec:intro}, all results presented in this Chapter are based on the \texttt{nimplex} tutorial, which has been adapted as Appendix~\ref{chap:nimplextutorial2}. It can be obtained and run on the cloud (GitHub Codespaces) with a single click by following \texttt{nimplex} distribution channels (see Section~\ref{nimplex:sec:code}).




\begin{appendices}

\mypart{Appendices}

\chapter{Supplementary Discussions} \label{chap:supdiscussions}

\section{Machine Learning Overview towards Atomistic Materials Science} \label{sipfenn:appedix1}
The class of deep learning methods has been remarkably successful in recent years in applications ranging from computer vision to natural language processing and simulations of quantum systems \cite{lecun2015deep,silver2016mastering,devlin2018bert,carleo2017solving}. Although deep neural networks have existed for a long time \cite{rosenblatt1958perceptron}, and had been successfully applied to computer vision tasks \cite{lecun1995comparison,lecun1990handwritten,lecun1998gradient}, a major breakthrough was the \texttt{AlexNet} network \cite{krizhevsky2012imagenet}, which dramatically improved the accuracy achievable on large-scale image classification. Following this success, deep neural networks have been very intensively studied and applied to a variety of problems \cite{lecun2015deep,silver2016mastering,devlin2018bert}. Deep neural networks are particularly effective when applied to regression problems, where one is learning a functional relationship between a feature and a prediction. For many problems, deep neural networks are able to achieve significantly better performance than competing machine learning methods, due to their ability to learn more complex relationships. With materials science being a field where many complex dependencies are intertwined, it is to be expected that this superior pattern recognition can carry over to the improvement in the prediction of material properties.

\subsection{Regression Problem Formulation and Artificial Neural Networks}
\label{sipfenn:ssec:regressionformulation}

The general formulation of a regression problem in statistical machine learning is to find a function $f:X\rightarrow Y$ which minimizes the risk \cite{vapnik1999overview}, also known as loss or expected error.
\begin{equation}\label{sipfenn:true_risk_app}
    R(f) = \mathbb{E}_{x,y\sim \mathcal{P}} l(y,f(x)).
\end{equation}
Here $X$ denotes a space of input features, $Y$ denotes an output space, the expectation above is taken over an unknown distribution $\mathcal{P}$ on $X\times Y$ (representing the true relationship between inputs and outputs), and $l$ is a given loss function. The goal is to find a function $f$ which accurately predicts the (potentially random) output $y$ given an input $x$.

In the present work, $x\in X$ represents the input features (descriptor) characteristic of the material, and $y\in Y$ represents the formation energy. The distribution $\mathcal{P}$ represents the true material-property relationship between given descriptor $x$ and corresponding formation energy. This relation may not be as simple as mapping a given structure to an energy since different DFT methodologies may give different results, based on many variables, such as employed functionals. \cite{CharlesW.BauschlicherJr.1995AFunctionals, Alturk2017ComparisonMaterial} Consequently it is useful to describe this relationship via a probability distribution. Furthermore, the loss function considered in the present paper is the commonly used $\ell^1$ or absolute error (AE) loss function $l(y_1,y_2) = |y_1-y_2|$. 

In practice, the distribution $\mathcal{P}$ is not known. Indeed it is this relationship that one is trying to learn in the first place. Instead, what is available is data $\{(y_i,x_i)\}_{i=1}^n$, which is sampled from $\mathcal{P}$. From this one forms the empirical risk \cite{hastie2009elements,vapnik2013nature}
\begin{equation}\label{sipfenn:empirical_risk_app}
    L(f) = \frac{1}{n}\displaystyle\sum_{i=1}^n l(y_i, f(x_i)),
\end{equation}
and seeks a function $f$ which minimizes the empirical risk, also known as the training error.

In addition, one must specify the type of relationship that is expected to be found between the inputs $x_i\in X$ and the predictions $y_i\in Y$. This is done by restricting the function $f$ to a specific class. For instance, by restricting $f$ to be linear, which corresponds to looking for a linear relationship between $x_i$ and $y_i$, one obtains a linear regression. On the other hand, choosing $\mathcal{F}$ to be a reproducing kernel Hilbert space of functions on $X$ with the same loss $l$ one obtains the kernel ridge regression method. Thus in order to fit the model, the training error is minimized over a specific class of function $\mathcal{F}$, i.e. one solves the optimization problem
\begin{equation}\label{sipfenn:empirical_risk_min_eq}
    f^* = \arg\min_{f\in \mathcal{F}} L(f) = \arg\min_{f\in \mathcal{F}} \frac{1}{n}\displaystyle\sum_{i=1}^n l(y_i, f(x_i)).
\end{equation}

\begin{figure}[H]
    \centering
    \includegraphics[width=0.85\textwidth]{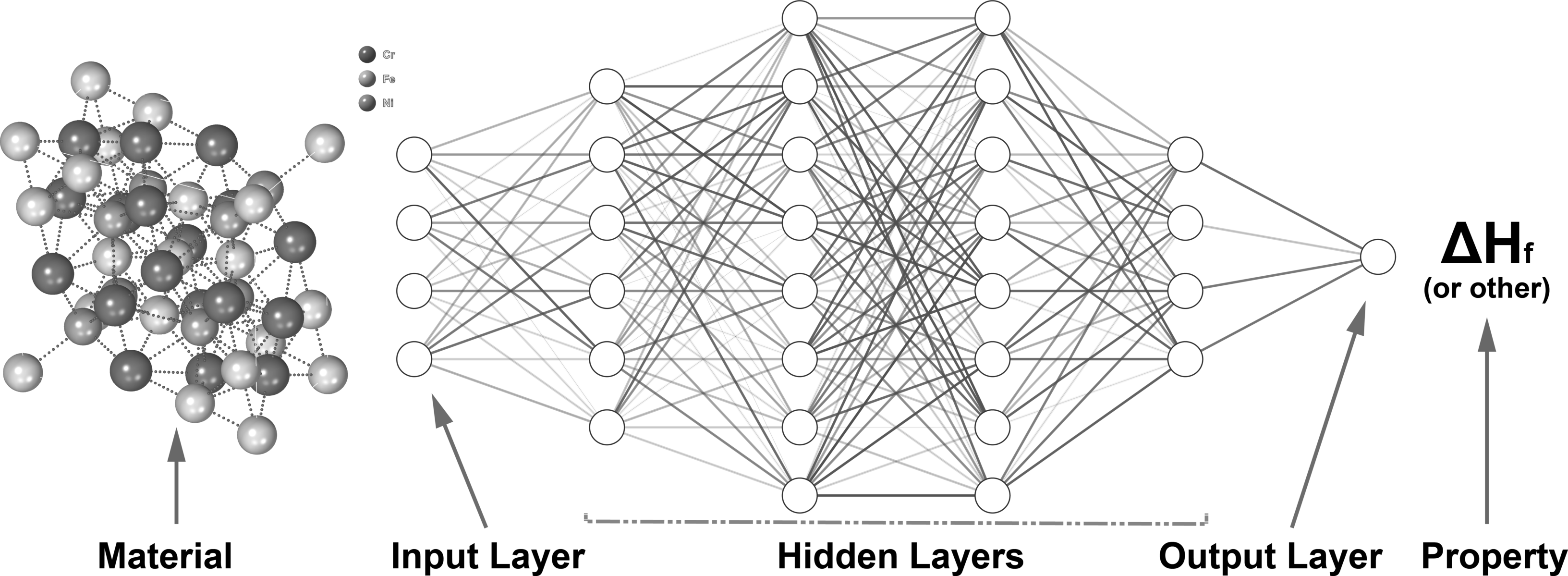}
    \caption{Simplified artificial neural network schematic}
    \label{sipfenn:fig:nnschematic}
\end{figure}

In this the class of functions $\mathcal{F}$ is chosen as the set of functions defined by a neural network architecture (schematic in Figure \ref{sipfenn:fig:nnschematic}), which leads to a deep learning method. A neural network architecture consists of a sequence of alternating linear functions and point-wise non-linear functions \cite{goodfellow2016deep}. In the figure \ref{sipfenn:fig:nnschematic} the nodes, or neurons, represent applications of a point-wise non-linear function, called an activation function, and the connections between nodes represent linear functions from the output of the nodes in one layer to the input of the next layer.

The class of functions represented by the neural network consists of the functions obtained by substituting different linear maps between each layer. Specifically, given weight matrices $W_1,...,W_n$ and biases $b_1,...,b_n$, which are parameters of the network, the corresponding neural network function is given by the composition

\begin{equation}
    f_{W_1,...,W_n,b_1,...,b_n}(x) = W_n\cdots\sigma(W_3\sigma(W_2\sigma(W_1x+b_1)+b_2)+b_3)\cdots + b_n
\end{equation}

where $\sigma$, called the activation function, is applied pointwise to each entry of the vector input (previous layer output). The neural network architecture is determined by the type, dimensionality, activation function $\sigma$, and arrangement of intermediate layers. This can potentially introduce some additional restrictions on the linear maps $W_i$, see for instance convolutional neural networks, where the linear maps $W_i$ are restricted to be convolutions with small kernels \cite{krizhevsky2012imagenet,lecun1995comparison,lecun1998gradient}.

Once the neural network architecture has been set, one must fit the values of the parameters $W_1,...,W_n$ and $b_1,...,b_n$ by optimizing the training loss $L$,

\begin{equation}
    \arg\min_{W_1,...,W_n,b_1,...,b_n} L(f_{W_1,...,W_n,b_1,...,b_n}).
\end{equation}

This optimization problem is typically solved using stochastic gradient descent \cite{lecun1998gradient}, or a more robust method such as ADAM \cite{kingma2014adam}, which was used in the present work. To solve the problem faster and to mitigate overfitting, which is discussed in the next sections, these methods form an estimate of the loss function gradient by considering a small subset of the data, called a batch. Each training step is done over all of the data in the batch, so parameters ($w$ and $b$) are updated based on many data points, rather than a single one. Most of the models created in the present work used a batch size of 2,048 data points.

This methodology has been successfully applied to a variety of practical machine learning problems \cite{krizhevsky2012imagenet,goodfellow2013multi,dahl2011context}. Specifically relevant to the present work, neural networks have been applied to problems in computational materials science \cite{Huang2019Machine-learningAlloys,Feng2019UsingDefects}. For example, in \cite{Huang2019Machine-learningAlloys} neural networks are used to classify the phases of high-entropy alloys. For this application, their neural network models compare favorably to other machine learning algorithms such as $k$-nearest neighbor (KNN) and support vector machines (SVM). Furthermore, in \cite{Feng2019UsingDefects} it is shown that even when training on small datasets which are typical of certain materials science problems, specifically in the prediction of solidification defects from optical microscopy data, deep neural networks can achieve better performance than other machine learning models. This is enabled by using a stacked auto-encoder (shallow neural network) to pre-train the deep neural network, whose weights are then fine-tuned on the small dataset. the present work complements these studies by applying deep neural networks to the prediction of thermodynamic quantities from atomic structure descriptors.

\subsection{Overfitting and its Mitigation}
\label{sipfenn:ssec:overfitting}

\begin{figure}[h]
    \centering
    \includegraphics[width=0.5\textwidth]{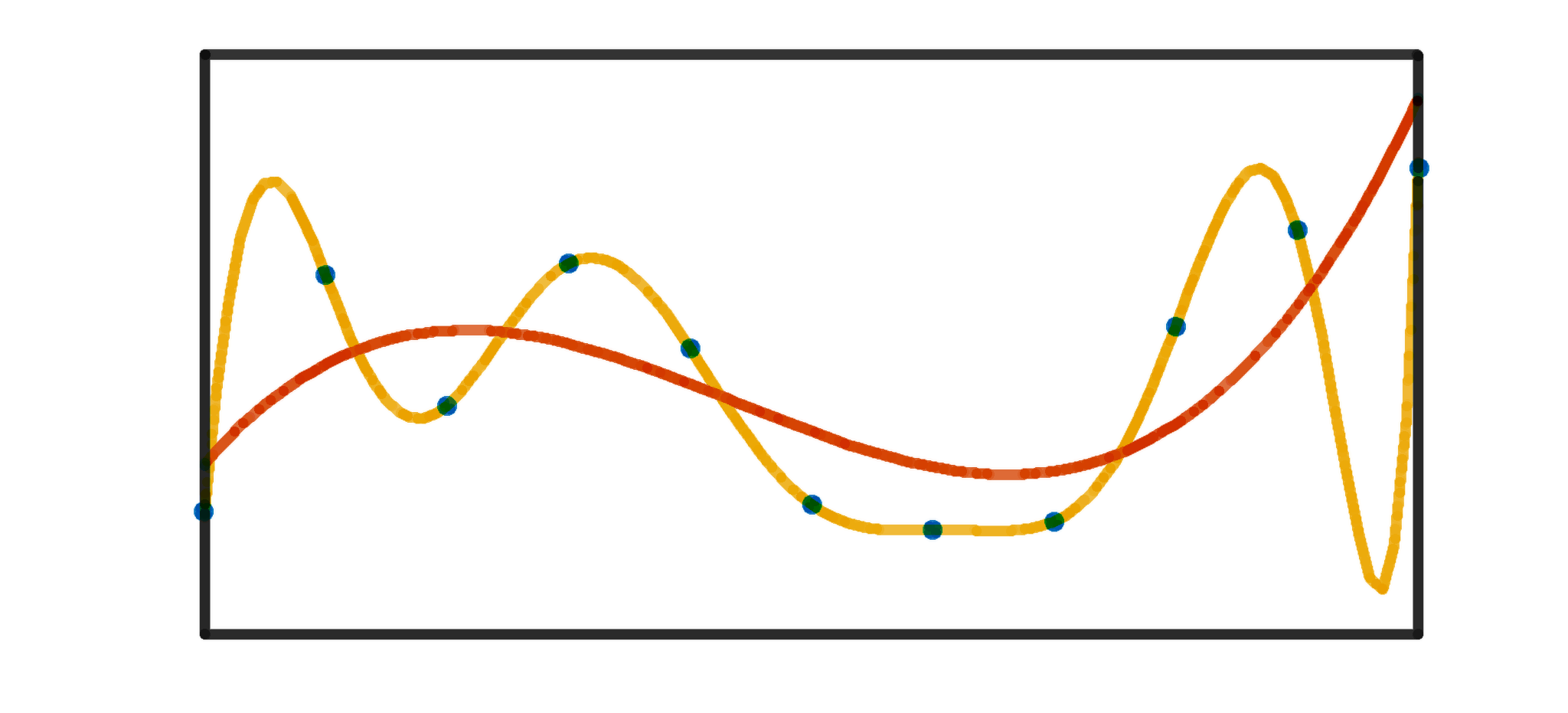}
    \caption{A schematic of overfitting. The overfit model (yellow) is too complex and memorizes the training data. This results in very low training error, but also very poor performance when predicting hidden data (test error) that follows the underlying phenomena (red).}
    \label{sipfenn:fig:overfitting}
\end{figure}

A major problem in statistical learning is avoiding overfitting \cite{hastie2009elements}, which, in simple terms, signifies that the model memorizes the training data instead of learning the true  relationship between descriptors $x$ and predictions $y$, as depicted in Figure~\ref{sipfenn:fig:overfitting}. This occurs when the class of functions $\mathcal{F}$ is too large, and at the optimal function $f^*$ in \eqref{sipfenn:empirical_risk_min_eq} the empirical \eqref{sipfenn:empirical_risk_app} and true risk \eqref{sipfenn:true_risk} diverge sharply. This results in very low training error, but poor performance on data that was not presented to the network.

Overfitting is typically detected by separating the training data into two sets, the data used in \eqref{sipfenn:empirical_risk_min_eq} to learn the function $f^*$, called the training data, and a separate set of data used to evaluate the performance of $f^*$, called the validation set. Consequently, in addition to the training loss in \eqref{sipfenn:empirical_risk_min_eq}, the validation error

\begin{equation}\label{sipfenn:validation_loss}
    L_{val} = \frac{1}{m}\displaystyle\sum_{i=1}^m l(\tilde{y}_i, f(\tilde{x}_i)),
\end{equation}

where $(\tilde{y}_i,\tilde{x}_i)$ for $i=1,...,m$ is the validation set, which was not presented to the network when adjusting its parameters, is used to detect overfitting. 
The fraction of the data set aside for validation set should be large enough to be representative of the whole dataset to provide statistically significant conclusions, yet small enough so that knowledge loss in the process is minimized. In the present work, a randomly selected 15\% of every dataset has been used as validation sets for all training. This corresponded to 65,300 data points in the case of the OQDM dataset described in \ref{sipfenn:sssec:Data}.

\begin{figure}{H}
    \centering
    \includegraphics[width=0.65\textwidth]{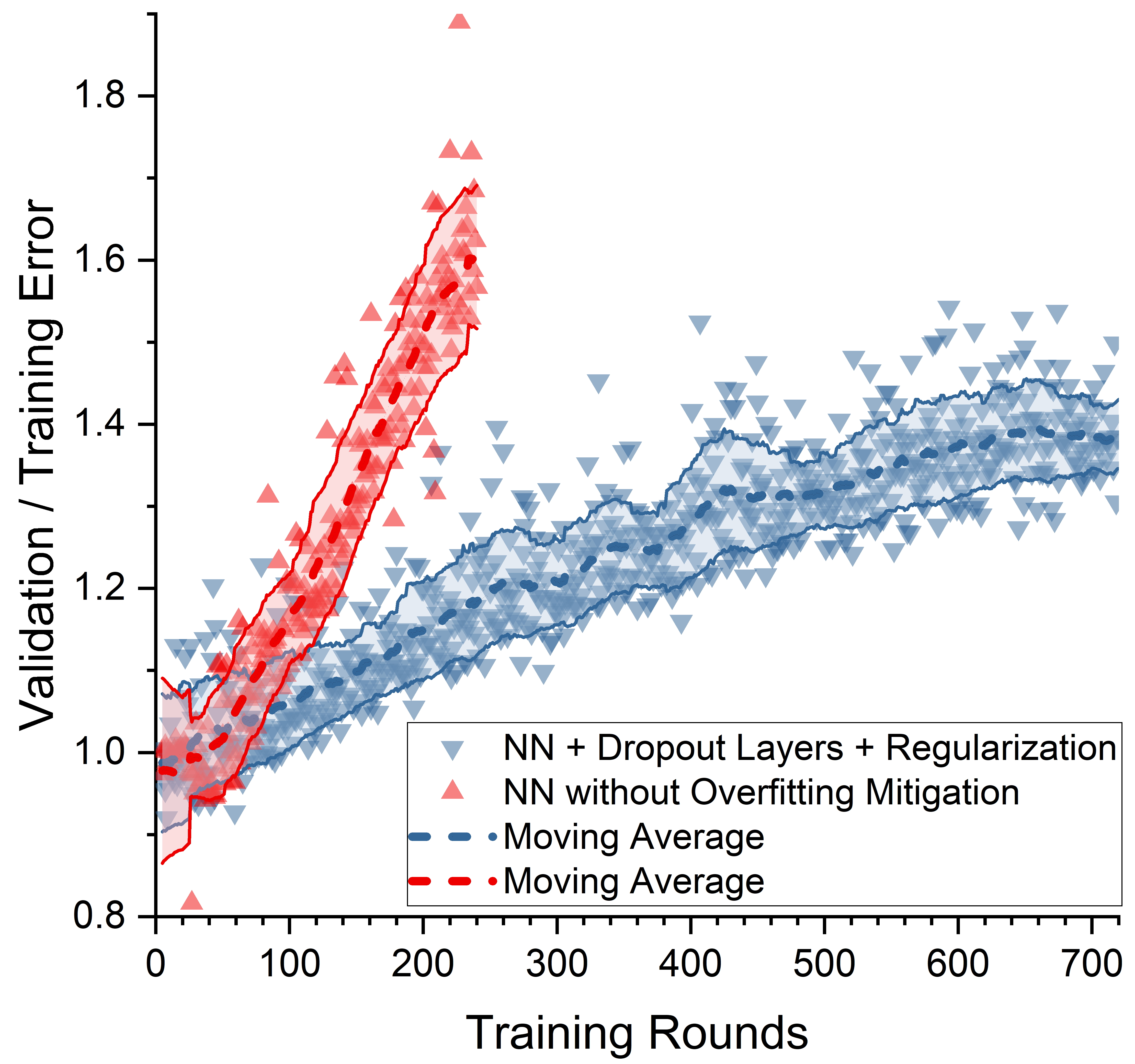}
    \caption{Training Loss to Validation Loss in a model that does without (NN9) and with overfitting mitigation (NN20), plotted versus training progress.}
    \label{sipfenn:fig:trainingvalidation}
\end{figure}

Typically, the validation loss will be greater than the training loss, as the validation set is not available for training. This is illustrated in Figure \ref{sipfenn:fig:trainingvalidation}, where the ratio between the validation loss \eqref{sipfenn:validation_loss} and test loss \eqref{sipfenn:empirical_risk_min_eq} during the course of two trainings of similar NN architectures on the same data with the same learning rate schedule has been plotted. This figure indicates that as the training proceeds, the gap between the training and validation errors widens and then increases. The size of this gap is an estimated measure of how much the model has overfitted to the data. In one of the models in this figure, extensive techniques to mitigate overfitting have been used, and for this model, the figure shows that the rate at which the model overfits to the data is much lower. At the same time both models exhibit similar performance on the test set.

There are numerous techniques used to prevent the issue of overfitting \cite{hastie2009elements,everitt2002cambridge}. These include utilization of a regularization term $\lambda R(\theta)$ added to the training error \eqref{sipfenn:empirical_risk_min_eq} to give the regularized empirical loss function

\begin{equation}
    f^* = \arg\min_{f\in \mathcal{F}} R_{emp}(f) + \lambda R(\theta).
\end{equation}

A standard regularizer typically added to the linear regression is the $\ell^2$-norm $R(\theta) = \|\theta\|_2^2$, which is often called Tikhonov regularization \cite{tikhonov1963solution} or ridge regression \cite{hoerl1970ridge}. The $\ell^2$-norm is also a popular regularizer in deep learning problems, where it is referred to as weight decay \cite{goodfellow2016deep}. In the context of the present work, it is implemented as a part of the training process, rather than network architecture, and causes rejection of some features in the descriptor that are not contributing to pattern recognition. Results of its implementation are shown throughout Section \ref{sipfenn:sssec:NetDesign}.

Another important method used to prevent overfitting in machine learning is the Dropout technique \cite{srivastava2014dropout}. The concept behind Dropout is to prevent neurons in the network from becoming overly dependent on the output from a specific neuron in the previous layer, often referred to as hard-wiring neuron paths. A Dropout layer, placed within a neural network, is implemented as a function operating during the training process and randomly discarding a specified fraction $p$ of previous layer outputs and multiplying the remaining values by $1/(1-p)$. This forces the pattern recognition ability to be dispersed across the network, as during evaluation of every training step, a random part of the network is acting as if it was not gone. Once the training is completed, all Dropout layers are deactivated and simply pass all information forward, so that the model returns to its deterministic character.

In the experiments performed in the present work, as later discussed in \ref{sipfenn:sssec:NetDesign}, both Dropout and weight decay were used to mitigate overfitting, with good effects shown in particular in Figure \ref{sipfenn:fig:trainingvalidation}.

Methods for avoiding overfitting typically come with one or more "hyperparameters" (i.e. parameters which control the training process) that can represent how much confidence is given to the training data versus prior knowledge. For instance, if a regularizer is used, the strength of the regularizer, $\lambda$, would be a hyperparameter. In the terms of the present work, it generally corresponds to how many features in the material descriptor can be considered non-essential to making predictions and therefore discarded systematically throughout the training. Furthermore, when using Dropout, the probability $p$ is also a hyperparameter. 

One typically trains the model on the training dataset using a number of different hyperparameters and then subsequently chooses the best set of them using the validation error. This allows the determination of hyperparameter values that are appropriate to the problem at hand. However, in order to ensure that the determined hyperparameter values are not overly specific to the validation set, the final accuracy of the model is evaluated on a test set that was not used at all during training \cite{hastie2009elements}.

\begin{figure}[H]
    \centering
    \includegraphics[width=0.7\textwidth]{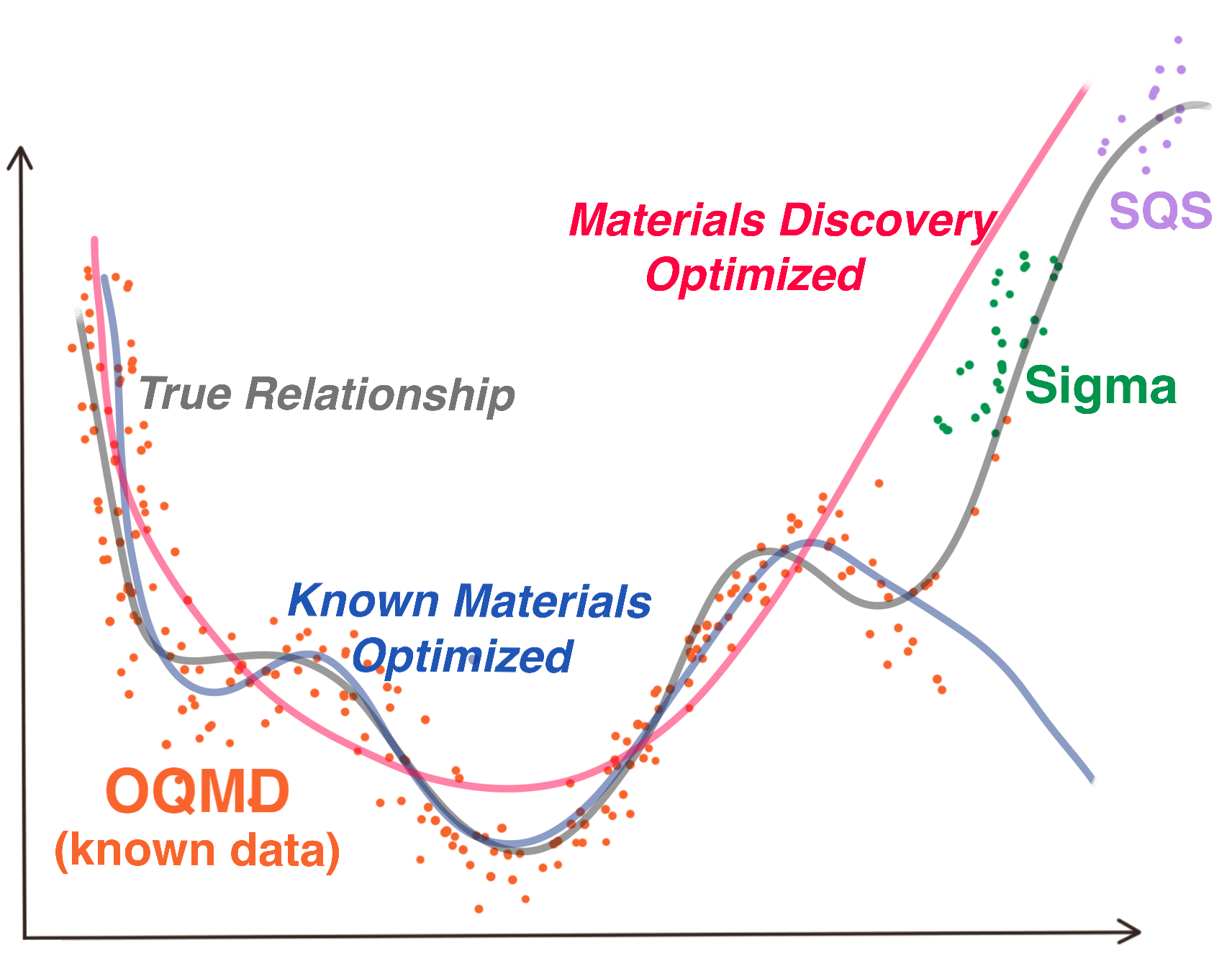}
    \caption{A conceptual drawing depicting how overfitting mitigation effort can improve performance beyond regions with high known data density.}
    \label{sipfenn:fig:overfitting_newregions}
\end{figure}

An additional advantage of mitigating overfitting to known data can be increased performance during extrapolation, as depicted conceptually in Figure \ref{sipfenn:fig:overfitting_newregions}. This is thanks to reduced model complexity, which forces recognition of stronger and more broadly exhibited patterns rather than small deviations present in the training data, whether real or due to noise, that can significantly degrade the extrapolation capability of the ML model. It is important to recognize that cost of such model simplification is often reduced performance on previously unseen data that lays within the known region.

\subsection{Transfer Learning} \label{sipfenn:ssec:transferlearning}
Finally, one should consider the technique of transfer learning, which has been observed among deep learning models across a variety of domains \cite{tan2018survey,cirecsan2012transfer,chang2017unsupervised,george2018deep}. Transfer learning refers the to the ability of properly trained deep learning models to `transfer' their knowledge to related tasks. In the least complex approach, one does this by simply `fine-tuning' the parameters of the model using new training data (from the new task). This has to be done using a small learning rate and a small number of iterations on a loss function defined by the new training data. It has been observed that this often produces accurate results on the new task for a relatively small amount of additional data. 

As an illustrative example, in \cite{cirecsan2012transfer}, a network is first trained to recognize lower case handwritten characters. It is then shown that with minimal `fine-tuning,' such a network can be made to accurately recognize upper case characters. The same phenomenon was also observed with a network that was first trained to recognize Chinese characters. Considering that this behavior has been widely observed \cite{tan2018survey,chang2017unsupervised,george2018deep}, this shows that deep neural networks are often able to transfer knowledge between different but related tasks. 

the present work adds to this evidence by showing that a network trained on the knowledge from the OQMD database covering a broad yet limited spectrum of material, can be easily adjusted to materials outside this spectrum with very little cost relative to the initial training. Specifically, the set of all (243) Fe-Ni-Cr $\sigma$-phase endmembers, described in \ref{sipfenn:sssec:Data}, is shown in \ref{sipfenn:ssec:transferlearningresults} to require transfer of only a few examples from that set to dramatically improve model performance on the rest.

\section{Intermediate \texttt{SIPFENN} Models} \label{sipfenn:appendix2}

The neural network design process was conducted in incremental fashion, starting from a perceptron, which is the simplest type of neural network proposed by Frank Rosenblatt in 1957 \cite{Rosenblatt1957TheAutomaton}. It effectively operates as a linear function $f(\vec{d}) = A(w_1 d_1 + w_2 d_2 + ... + w_n d_n)$ where $d_i$ is i-th element of the descriptor $\vec{d}$, $w_i$ is the weight associated with it, and $A$ is an activation function that can introduce non-linearity or turn it into a classifier. Here, the popular Sigmoid activation function was used. 

\begin{figure}[H]
    \centering
    \includegraphics[width=0.35\textwidth]{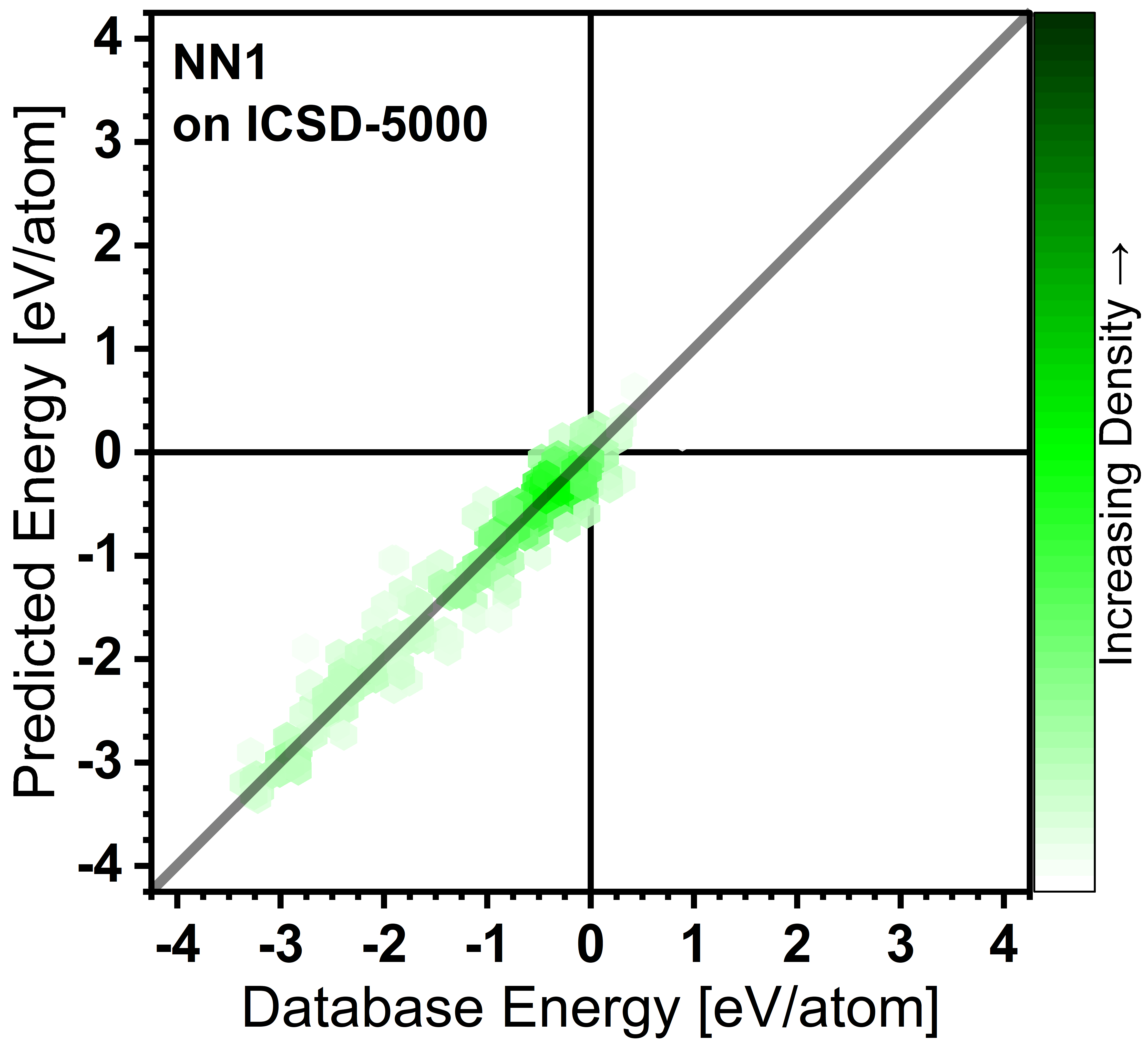}
    \caption{Test of perceptron trained on the data from the first 5000 entries in the ICSD dataset and evaluated on the test set of 230 randomly selected entries ($\approx5\%$)}
    \label{sipfenn:fig:nn1performance}
\end{figure}

The perceptron was first trained on the data from the first 5000 entries in the ICSD, to check whether the training was set up correctly. It achieved a MAE of 195 meV/atom on the test set of 230 randomly selected entries ($\approx 5\% \text{ from } 5000$). Results are shown in Figure \ref{sipfenn:fig:nn1performance}. 

When trained on the data from all entries in the ICSD, it achieved an MAE of 364 meV/atom on the test set ($\approx5\% \text{ from } 32116$). This error is comparable to the performance of a random forest model based on PRDF (370 meV/atom), is slightly worse than a CM (250 meV/atom), and is significantly worse than a random-forest model trained on the same descriptor (90 meV/atom), as reported by Ward et al. \cite{Ward2017IncludingTessellations}. Part of the significance of these results is the evident quality of the descriptor, as the model achieved performance that would be considered excellent just a few years prior to the present work while being much less complex and computationally costly. Furthermore, it is important to note the time- and space-complexity of the perceptron model. Training the final network took less than 8 seconds compared to around 10,000 seconds reported for the aforementioned random-forest methods, and the resulting model occupied less than 1kb of memory. Following the testing of a perceptron, which allowed rough estimation of the a good size of the network (i.e. number of weights), the design of the actual architecture began. All of these steps are schematically depicted in Figure \ref{sipfenn:fig:designprocess}.

\begin{figure}[H]
    \centering
    \includegraphics[width=0.5\textwidth]{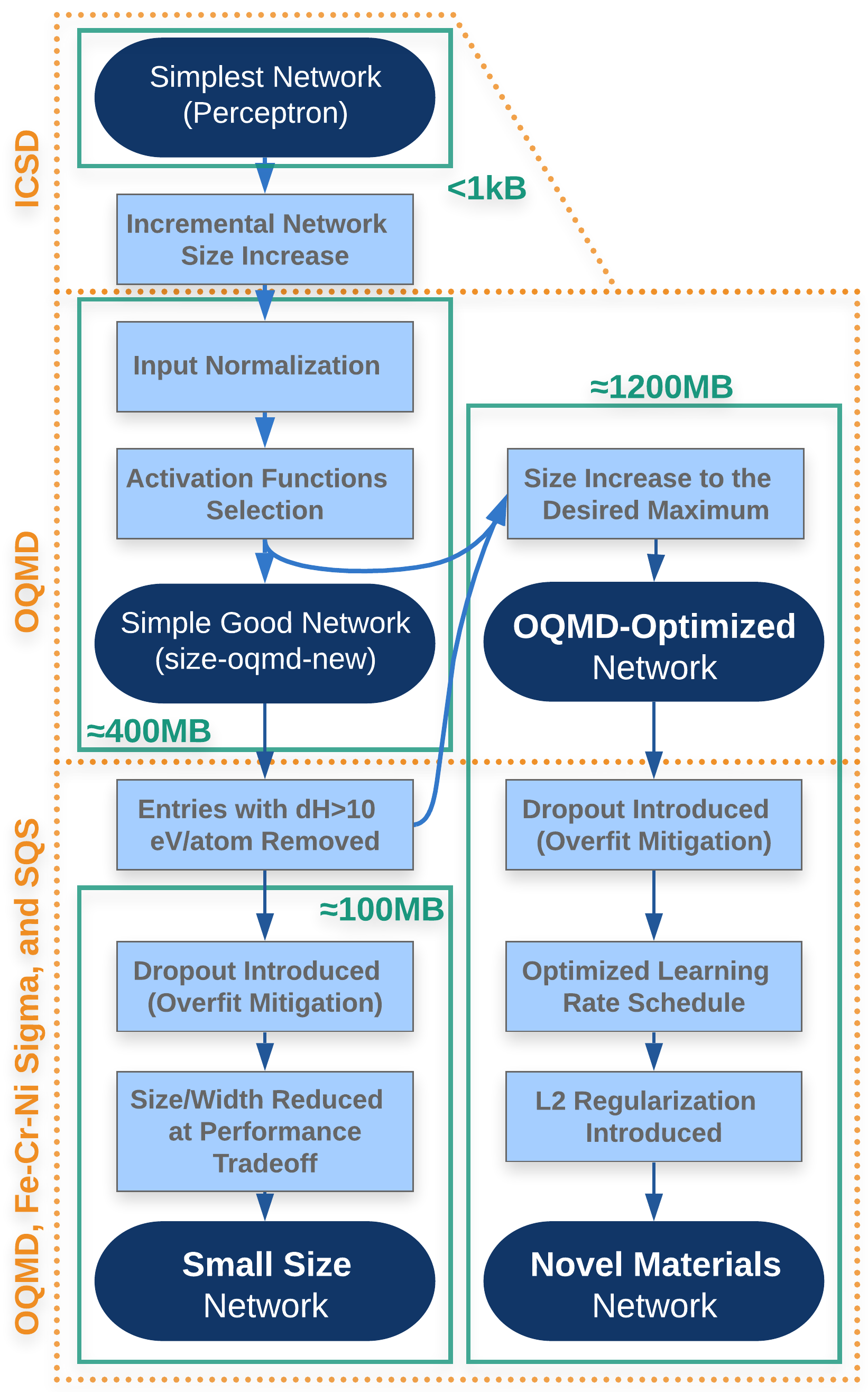}
    \caption{The network design process schematic leading to the three final models. This figure is Figure \ref{sipfenn:fig:designprocess} cloned to the Appendix for conveninece.}
\end{figure}

Next, in a few steps, the size of the network was incrementally increased. First, a layer of 1000 neurons was introduced. This reduced the performance on the first 5000 entries in the ICSD, likely due to overfitting issues, as the data was very limited. Performance on the ICSD was improved, reducing the test MAE to 305 meV/atom on the test set, however. The introduction of the next two 1000-width layers further reduced the MAE to 215 meV/atom. Based on these results, it was estimated that introducing 4 hidden layers with Sigmoid activation function and widths of 10000, 10000, 1000, and 100 would provide good results when trained on the much larger OQMD.

After switching to OQMD, the network exhibited issues with convergence, often predicting a single value for all of the entries. To mitigate this, the descriptor (i.e. network input) was normalized by dividing every element by its maximum value across the whole dataset. This solved the issue. Next, to improve the training behavior, the activation functions were changed from only the Sigmoid function to a mix of Soft Sign, Exponential Linear Unit, and Sigmoid, which was found to work well. These steps improved both the predictive performance and reduced the time required to converge. The network architecture resulting from these steps (internally designated NN8 / Simple Good Network in Figure \ref{sipfenn:fig:designprocess}) was the first to improve performance compared to the Ward et. al approach \cite{Ward2017IncludingTessellations}, achieving an MAE of 42 meV/atom on the test set of random subset 5\% of the OQMD dataset. When testing this network, a small fraction of around 0.03\% of likely incorrect entries in the OQMD was found, as described in \ref{sipfenn:sssec:Data}, and was removed from the dataset used later in the design process.

Once a network with desired performance was obtained, the network size was increased until it either exceeded 1GB or showed signs of extensive overfitting. At the first step of this process, two layers of width 10,000 were added, resulting in a network size of 1.2GB and reduced overfitting, as indicated by the ratio of validation-to-training error lowered from 2.2 to 1.6, relative to NN8. The resulting network (internally designated NN9 / OQMD-Optimized Network in Figure \ref{sipfenn:fig:designprocess}), achieved an MAE of 28 meV/atom on the test set of random subset 5\% of OQMD, which was the best performance on OQMD out of all the networks created in this project. If the 0.03\% of abnormal data wasn't removed as described in \ref{sipfenn:sssec:Data}, it would correspond to, on average, 6 data points which in one tested instance increased the MAE to 35 meV/atom. Important to point out, the training of this network was prone to staying in local minima at the beginning. The reported instance of the trained network exhibited no training progress between around rounds 5 and 25, after which it's performance quickly increased.  Detailed analysis of the performance is given in \ref{sipfenn:ssec:oqmdperformance}.

Once the main objective of the design process was obtained, i.e. the performance on the OQMD has improved appreciably beyond existing methods, the design process was focused on creating a tool for modeling materials that were not reported in the OQMD. Therefore, the objective changed from achieving the lowest MAE on a random subset 5\% of OQMD to (1) reducing the mismatch between training and validation sets errors (i.e. difference between training accuracy and validation accuracy) during the training process, (2) keeping the test MAE on the OQMD below 50 meV/atom, and (3) improving performance on two material groups significantly different from the OQMD data, namely Special Quasirandom Structures (SQS) and Fe-Cr-Ni $\sigma$-phase (see \ref{sipfenn:sssec:Data}).

\begin{figure}[H]
\centering
    \includegraphics[width=0.35\textwidth]{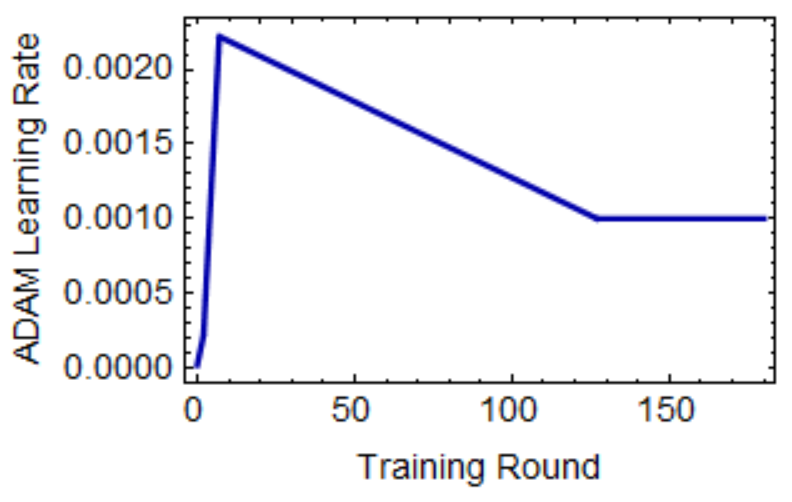}
    \caption{The learning rate schedule used for training of more complex networks in the later stage of the design process (e.g., NN18).}
    \label{sipfenn:fig:learningrate}
\end{figure}

With these new objectives, two Dropout layers in the middle part of the network were introduced to promote the distribution of pattern recognition abilities across the network. \cite{Srivastava2014Dropout:Overfitting} This introduced a problem with convergence as the network became more likely to fall into local minima at the initial stages of the training, which was solved by introducing custom learning rate schedules. Specifically, the learning rate was initially set to a value orders of magnitude lower than during the default initial training and then ramped up to the previous (ADAM default setting in the majority of frameworks) learning rate of 0.001 (or above) after around 2 rounds of training. This type of learning rate schedule is known as a warm-up in the deep learning literature \cite{gotmare2018closer}. The schedule found to perform the best is presented in Figure \ref{sipfenn:fig:learningrate}.

The next step was the introduction of $\ell^2$ regularization, which is a technique that favors simplification of the descriptor and effectively rejects features of the descriptor that do not contribute to prediction performance \cite{L2Regularization}. An overview on it is given in Section \ref{sipfenn:ref:machinelearningoverview}. In the models reported in the present work an $\ell^2$ value of $10^{-6}$ was used. Higher values were found to stop the training at early stages, impairing the pattern recognition, or in extreme cases (above $10^{-3}$) force the network to discard the input completely, resulting in constant or near-constant output (i.e. mean value from the training dataset predicted for any structure).

The final step was small curation of the training data based on the OQMD-reported structure stability, i.e. the energy difference between the formation energy and the energy convex hull. The motivation for that was the notion that DFT results are inherently less accurate for unstable phases. In this step, all entries with energies of more than 2000 meV/atom above the convex hull were removed from the training set. Importantly, the validation and testing sets were not modified for consistent performance reporting.

All of these changes resulted in a neural network that has been optimized for predicting new materials. In the code and Supplementary materials, it is designated as NN20 (Novel Materials Network in Figure \ref{sipfenn:fig:designprocess}). Compared to the OQMD-optimized network it was derived from, the test MAE on the OQMD increased from 28 to 49 meV/atom. However, at the same time, the mismatch between the training and validation set was reduced from 1.57 to 1.38. Or, as presented earlier in Figure \ref{sipfenn:fig:trainingvalidation}, reduced to about 1.15 for the same training duration. Furthermore, a relatively large portion of this error can be attributed to some unstable structures that were removed from the training set, but not from the test set. Once entries with formation energies of more than 1000 meV/atom above the convex hull were removed, the test MAE decreased to only 38 meV/atom. Restricting the test set further to only somewhat stable structures (stability below 250 meV/atom) resulted in an MAE of 30 meV/atom.

While the new-material-optimized network presented an increased MAE across a random subset of the OQMD, performance has significantly improved on the Fe-Cr-Ni $\sigma-$phase described in \ref{sipfenn:sssec:Data}. The MAE has decreased from 55 to 41 meV/atom, indicating that the model based on this neural network is more capable of making predictions for new materials.

Once two performance-oriented models were developed, increasing the performance-to-cost ratio has been explored with the motivation that some studies would benefit from many times higher throughput at minor accuracy decrease. Architecture design started from the selection of a network with a balanced size-to-performance ratio (NN8) and the introduction of an overfitting mitigation technique (Dropout \cite{srivastava2014dropout}) used for the network optimized for new materials, as depicted in Figure \ref{sipfenn:sssec:NetDesign}. Next, the network was gradually narrowed (fewer neurons in layers) until the performance started to noticeably deteriorate (41.9 meV/atom for 5000- and 4000-width vs 42.1 for 3000-width). This approach allowed a significant reduction of the network size (and the computational intensity to run it) from around 1,200MB of the two other models to around 145MB. If an application demands even more of a reduction in model size and computational cost, the same procedure could be continued until some minimum required performance is retained.

\section{Feature Ranking Learned During Formation Energy Modeling} 
\label{sipfenn:appendix3}

\begin{longtable}{|l|l|}
\caption{\texttt{SIPFENN}'s \texttt{NN20} Model Input Feature Ranking Learned During Formation Energy Modeling}
\label{sipfenn:appendix3:featureranking}\\
\hline
\multicolumn{1}{|c|}{\textbf{Descriptor Feature}} & \multicolumn{1}{c|}{\textbf{Normalized Squared Weights Sum}} \\ \hline
\endfirsthead
\endhead
mean\_NeighDiff\_shell1\_MeltingT & 1 \\ \hline
mean\_MeltingT & 0.97502 \\ \hline
max\_MeltingT & 0.73512 \\ \hline
mean\_NeighDiff\_shell1\_NdUnfilled & 0.69157 \\ \hline
MaxPackingEfficiency & 0.68889 \\ \hline
most\_MeltingT & 0.67373 \\ \hline
dev\_GSvolume\_pa & 0.61042 \\ \hline
var\_NeighDiff\_shell1\_Column & 0.58782 \\ \hline
var\_NeighDiff\_shell1\_CovalentRadius & 0.57826 \\ \hline
var\_NeighDiff\_shell1\_MeltingT & 0.57259 \\ \hline
maxdiff\_GSvolume\_pa & 0.55156 \\ \hline
dev\_MeltingT & 0.5286 \\ \hline
mean\_SpaceGroupNumber & 0.51761 \\ \hline
min\_MeltingT & 0.50437 \\ \hline
var\_CellVolume & 0.49467 \\ \hline
var\_NeighDiff\_shell1\_MendeleevNumber & 0.492 \\ \hline
min\_NeighDiff\_shell1\_MeltingT & 0.47853 \\ \hline
mean\_NeighDiff\_shell1\_Column & 0.45566 \\ \hline
maxdiff\_CovalentRadius & 0.42998 \\ \hline
var\_NeighDiff\_shell1\_Electronegativity & 0.42642 \\ \hline
var\_EffectiveCoordination & 0.40506 \\ \hline
min\_NeighDiff\_shell1\_Column & 0.39822 \\ \hline
dev\_NdUnfilled & 0.39739 \\ \hline
dev\_CovalentRadius & 0.36935 \\ \hline
range\_NeighDiff\_shell1\_Column & 0.35956 \\ \hline
range\_NeighDiff\_shell1\_CovalentRadius & 0.34585 \\ \hline
mean\_WCMagnitude\_Shell1 & 0.34275 \\ \hline
mean\_NeighDiff\_shell1\_MendeleevNumber & 0.33911 \\ \hline
mean\_EffectiveCoordination & 0.33899 \\ \hline
mean\_Number & 0.33769 \\ \hline
mean\_NdUnfilled & 0.33408 \\ \hline
maxdiff\_MeltingT & 0.33348 \\ \hline
mean\_AtomicWeight & 0.33149 \\ \hline
mean\_NeighDiff\_shell1\_NdValence & 0.33142 \\ \hline
range\_NeighDiff\_shell1\_MeltingT & 0.33107 \\ \hline
max\_NfUnfilled & 0.33041 \\ \hline
dev\_Electronegativity & 0.33001 \\ \hline
mean\_NeighDiff\_shell1\_CovalentRadius & 0.32999 \\ \hline
var\_NeighDiff\_shell1\_NdUnfilled & 0.31973 \\ \hline
dev\_Column & 0.31662 \\ \hline
var\_NeighDiff\_shell1\_NdValence & 0.31481 \\ \hline
mean\_WCMagnitude\_Shell2 & 0.31359 \\ \hline
most\_NfUnfilled & 0.30916 \\ \hline
MeanIonicChar & 0.30732 \\ \hline
mean\_NeighDiff\_shell1\_Electronegativity & 0.30277 \\ \hline
min\_EffectiveCoordination & 0.29705 \\ \hline
min\_NeighDiff\_shell1\_CovalentRadius & 0.29392 \\ \hline
max\_NeighDiff\_shell1\_GSvolume\_pa & 0.2875 \\ \hline
most\_SpaceGroupNumber & 0.28472 \\ \hline
max\_NdUnfilled & 0.28424 \\ \hline
maxdiff\_NdUnfilled & 0.28405 \\ \hline
var\_NeighDiff\_shell1\_GSvolume\_pa & 0.28008 \\ \hline
min\_BondLengthVariation & 0.27922 \\ \hline
var\_MeanBondLength & 0.2768 \\ \hline
dev\_NdValence & 0.27566 \\ \hline
max\_NeighDiff\_shell1\_MeltingT & 0.27097 \\ \hline
max\_BondLengthVariation & 0.26565 \\ \hline
mean\_NfValence & 0.26558 \\ \hline
mean\_NsUnfilled & 0.2612 \\ \hline
max\_NeighDiff\_shell1\_CovalentRadius & 0.26026 \\ \hline
max\_GSvolume\_pa & 0.25985 \\ \hline
min\_GSvolume\_pa & 0.25895 \\ \hline
mean\_NdValence & 0.25573 \\ \hline
mean\_NeighDiff\_shell1\_GSvolume\_pa & 0.25299 \\ \hline
max\_NValance & 0.24749 \\ \hline
range\_NeighDiff\_shell1\_NdUnfilled & 0.24643 \\ \hline
max\_CovalentRadius & 0.23136 \\ \hline
CanFormIonic & 0.23135 \\ \hline
min\_NeighDiff\_shell1\_Electronegativity & 0.22873 \\ \hline
min\_SpaceGroupNumber & 0.22766 \\ \hline
max\_Electronegativity & 0.22609 \\ \hline
max\_NdValence & 0.22576 \\ \hline
most\_NdUnfilled & 0.22198 \\ \hline
min\_NeighDiff\_shell1\_MendeleevNumber & 0.21991 \\ \hline
var\_NeighDiff\_shell1\_NpValence & 0.21609 \\ \hline
min\_NeighDiff\_shell1\_NdUnfilled & 0.2114 \\ \hline
dev\_SpaceGroupNumber & 0.2099 \\ \hline
most\_NfValence & 0.20888 \\ \hline
min\_MeanBondLength & 0.2086 \\ \hline
mean\_BondLengthVariation & 0.20507 \\ \hline
var\_NeighDiff\_shell1\_Row & 0.20454 \\ \hline
max\_NeighDiff\_shell1\_NdUnfilled & 0.20318 \\ \hline
min\_NeighDiff\_shell1\_NdValence & 0.20123 \\ \hline
min\_CovalentRadius & 0.19974 \\ \hline
range\_NeighDiff\_shell1\_MendeleevNumber & 0.19591 \\ \hline
min\_NeighDiff\_shell1\_GSvolume\_pa & 0.19565 \\ \hline
most\_NpUnfilled & 0.19457 \\ \hline
maxdiff\_NUnfilled & 0.19316 \\ \hline
max\_NeighDiff\_shell1\_NdValence & 0.19307 \\ \hline
max\_NpValence & 0.1929 \\ \hline
range\_NeighDiff\_shell1\_GSvolume\_pa & 0.19166 \\ \hline
most\_NdValence & 0.1904 \\ \hline
max\_MeanBondLength & 0.19021 \\ \hline
maxdiff\_NfUnfilled & 0.18897 \\ \hline
max\_NeighDiff\_shell1\_Column & 0.18518 \\ \hline
range\_NeighDiff\_shell1\_Electronegativity & 0.18322 \\ \hline
var\_NeighDiff\_shell1\_SpaceGroupNumber & 0.18313 \\ \hline
dev\_NpValence & 0.18099 \\ \hline
mean\_NpUnfilled & 0.18091 \\ \hline
range\_NeighDiff\_shell1\_SpaceGroupNumber & 0.17858 \\ \hline
dev\_MendeleevNumber & 0.17753 \\ \hline
MaxIonicChar & 0.176 \\ \hline
mean\_Column & 0.17206 \\ \hline
min\_Electronegativity & 0.17164 \\ \hline
mean\_WCMagnitude\_Shell3 & 0.17077 \\ \hline
mean\_Row & 0.17035 \\ \hline
min\_NeighDiff\_shell1\_SpaceGroupNumber & 0.17031 \\ \hline
most\_NsUnfilled & 0.16714 \\ \hline
var\_BondLengthVariation & 0.16653 \\ \hline
var\_NeighDiff\_shell1\_NfUnfilled & 0.16223 \\ \hline
range\_NeighDiff\_shell1\_NdValence & 0.16094 \\ \hline
frac\_fValence & 0.1609 \\ \hline
maxdiff\_Column & 0.16083 \\ \hline
max\_NUnfilled & 0.15916 \\ \hline
mean\_NpValence & 0.15639 \\ \hline
maxdiff\_NpValence & 0.15637 \\ \hline
mean\_MendeleevNumber & 0.15491 \\ \hline
most\_Electronegativity & 0.15469 \\ \hline
mean\_Electronegativity & 0.15458 \\ \hline
max\_SpaceGroupNumber & 0.15429 \\ \hline
dev\_Row & 0.15382 \\ \hline
maxdiff\_MendeleevNumber & 0.15373 \\ \hline
var\_NeighDiff\_shell1\_NpUnfilled & 0.15135 \\ \hline
max\_NeighDiff\_shell1\_Electronegativity & 0.15115 \\ \hline
most\_NUnfilled & 0.14955 \\ \hline
max\_GSbandgap & 0.14945 \\ \hline
mean\_NeighDiff\_shell1\_NUnfilled & 0.14891 \\ \hline
maxdiff\_NValance & 0.14819 \\ \hline
mean\_NeighDiff\_shell1\_NpValence & 0.14768 \\ \hline
maxdiff\_NdValence & 0.14735 \\ \hline
max\_NpUnfilled & 0.14647 \\ \hline
maxdiff\_Electronegativity & 0.14523 \\ \hline
min\_MendeleevNumber & 0.14119 \\ \hline
mean\_CovalentRadius & 0.14049 \\ \hline
mean\_NeighDiff\_shell1\_Row & 0.13945 \\ \hline
maxdiff\_GSbandgap & 0.13891 \\ \hline
max\_NeighDiff\_shell1\_MendeleevNumber & 0.13858 \\ \hline
most\_Number & 0.13823 \\ \hline
most\_AtomicWeight & 0.13798 \\ \hline
max\_NeighDiff\_shell1\_NpValence & 0.13757 \\ \hline
Comp\_L10Norm & 0.13598 \\ \hline
min\_Row & 0.13596 \\ \hline
range\_NeighDiff\_shell1\_NpValence & 0.13524 \\ \hline
mean\_GSvolume\_pa & 0.1331 \\ \hline
max\_NeighDiff\_shell1\_NUnfilled & 0.13205 \\ \hline
mean\_NeighDiff\_shell1\_NfValence & 0.12888 \\ \hline
min\_NeighDiff\_shell1\_NpUnfilled & 0.12778 \\ \hline
mean\_NeighDiff\_shell1\_SpaceGroupNumber & 0.12722 \\ \hline
mean\_NsValence & 0.12642 \\ \hline
most\_CovalentRadius & 0.12616 \\ \hline
var\_NeighDiff\_shell1\_NUnfilled & 0.12525 \\ \hline
mean\_NeighDiff\_shell1\_Number & 0.12466 \\ \hline
Comp\_L7Norm & 0.12293 \\ \hline
mean\_NeighDiff\_shell1\_AtomicWeight & 0.12229 \\ \hline
min\_NeighDiff\_shell1\_NpValence & 0.12026 \\ \hline
max\_EffectiveCoordination & 0.11995 \\ \hline
min\_NdValence & 0.11984 \\ \hline
maxdiff\_NpUnfilled & 0.11976 \\ \hline
mean\_NeighDiff\_shell1\_NsUnfilled & 0.11836 \\ \hline
max\_NeighDiff\_shell1\_GSbandgap & 0.11657 \\ \hline
min\_NUnfilled & 0.11648 \\ \hline
most\_Column & 0.1164 \\ \hline
var\_NeighDiff\_shell1\_Number & 0.11483 \\ \hline
most\_MendeleevNumber & 0.11312 \\ \hline
max\_NeighDiff\_shell1\_SpaceGroupNumber & 0.11292 \\ \hline
var\_NeighDiff\_shell1\_AtomicWeight & 0.11234 \\ \hline
most\_NpValence & 0.11231 \\ \hline
frac\_dValence & 0.11126 \\ \hline
NComp & 0.11097 \\ \hline
min\_Number & 0.11062 \\ \hline
range\_NeighDiff\_shell1\_NpUnfilled & 0.11002 \\ \hline
dev\_NValance & 0.10868 \\ \hline
min\_Column & 0.10846 \\ \hline
max\_NeighDiff\_shell1\_NpUnfilled & 0.10837 \\ \hline
maxdiff\_Row & 0.10735 \\ \hline
Comp\_L5Norm & 0.10726 \\ \hline
mean\_NeighDiff\_shell1\_NpUnfilled & 0.10682 \\ \hline
maxdiff\_SpaceGroupNumber & 0.10604 \\ \hline
dev\_GSbandgap & 0.10604 \\ \hline
max\_AtomicWeight & 0.10495 \\ \hline
max\_GSmagmom & 0.10416 \\ \hline
maxdiff\_GSmagmom & 0.1039 \\ \hline
dev\_NUnfilled & 0.10336 \\ \hline
var\_NeighDiff\_shell1\_NfValence & 0.10059 \\ \hline
dev\_GSmagmom & 0.10046 \\ \hline
most\_GSbandgap & 0.09997 \\ \hline
var\_NeighDiff\_shell1\_NValance & 0.09842 \\ \hline
min\_NeighDiff\_shell1\_Row & 0.09798 \\ \hline
min\_NeighDiff\_shell1\_NUnfilled & 0.09563 \\ \hline
most\_Row & 0.09538 \\ \hline
max\_Number & 0.0925 \\ \hline
most\_GSvolume\_pa & 0.09166 \\ \hline
mean\_GSbandgap & 0.09097 \\ \hline
range\_NeighDiff\_shell1\_Row & 0.09081 \\ \hline
mean\_NValance & 0.0889 \\ \hline
mean\_NeighDiff\_shell1\_NsValence & 0.08449 \\ \hline
min\_NsValence & 0.08408 \\ \hline
frac\_pValence & 0.08403 \\ \hline
mean\_NUnfilled & 0.08244 \\ \hline
mean\_NfUnfilled & 0.08194 \\ \hline
dev\_NpUnfilled & 0.0818 \\ \hline
dev\_Number & 0.08065 \\ \hline
max\_NeighDiff\_shell1\_GSmagmom & 0.08049 \\ \hline
max\_Column & 0.07989 \\ \hline
min\_AtomicWeight & 0.07959 \\ \hline
Comp\_L3Norm & 0.07913 \\ \hline
max\_NeighDiff\_shell1\_Row & 0.0776 \\ \hline
mean\_NeighDiff\_shell1\_NValance & 0.07619 \\ \hline
mean\_NeighDiff\_shell1\_NfUnfilled & 0.07413 \\ \hline
range\_NeighDiff\_shell1\_NfUnfilled & 0.07381 \\ \hline
min\_NValance & 0.07297 \\ \hline
max\_NeighDiff\_shell1\_NValance & 0.0726 \\ \hline
range\_NeighDiff\_shell1\_NfValence & 0.07163 \\ \hline
min\_NdUnfilled & 0.07145 \\ \hline
most\_NsValence & 0.07114 \\ \hline
mean\_NeighDiff\_shell1\_GSbandgap & 0.06709 \\ \hline
max\_NfValence & 0.06661 \\ \hline
dev\_AtomicWeight & 0.06581 \\ \hline
maxdiff\_Number & 0.06576 \\ \hline
max\_NeighDiff\_shell1\_NfUnfilled & 0.06523 \\ \hline
dev\_NfUnfilled & 0.06477 \\ \hline
dev\_NfValence & 0.06373 \\ \hline
range\_NeighDiff\_shell1\_GSmagmom & 0.06305 \\ \hline
var\_NeighDiff\_shell1\_NsUnfilled & 0.06288 \\ \hline
min\_NeighDiff\_shell1\_Number & 0.0623 \\ \hline
frac\_sValence & 0.06099 \\ \hline
min\_NeighDiff\_shell1\_NfValence & 0.06033 \\ \hline
max\_Row & 0.05998 \\ \hline
min\_NeighDiff\_shell1\_NValance & 0.05844 \\ \hline
range\_NeighDiff\_shell1\_NUnfilled & 0.05819 \\ \hline
var\_NeighDiff\_shell1\_GSbandgap & 0.05683 \\ \hline
range\_NeighDiff\_shell1\_AtomicWeight & 0.0568 \\ \hline
Comp\_L2Norm & 0.05638 \\ \hline
min\_NeighDiff\_shell1\_NsUnfilled & 0.05541 \\ \hline
most\_NValance & 0.0553 \\ \hline
maxdiff\_NsValence & 0.05459 \\ \hline
range\_NeighDiff\_shell1\_NValance & 0.0537 \\ \hline
min\_NeighDiff\_shell1\_AtomicWeight & 0.05369 \\ \hline
max\_NsValence & 0.05329 \\ \hline
range\_NeighDiff\_shell1\_GSbandgap & 0.05299 \\ \hline
min\_NeighDiff\_shell1\_NfUnfilled & 0.05266 \\ \hline
maxdiff\_NfValence & 0.05147 \\ \hline
dev\_NsUnfilled & 0.04884 \\ \hline
max\_MendeleevNumber & 0.04844 \\ \hline
maxdiff\_AtomicWeight & 0.04814 \\ \hline
max\_NeighDiff\_shell1\_NsUnfilled & 0.04675 \\ \hline
max\_NeighDiff\_shell1\_NsValence & 0.04663 \\ \hline
var\_NeighDiff\_shell1\_GSmagmom & 0.04635 \\ \hline
range\_NeighDiff\_shell1\_Number & 0.04416 \\ \hline
max\_NeighDiff\_shell1\_NfValence & 0.04376 \\ \hline
mean\_NeighDiff\_shell1\_GSmagmom & 0.0433 \\ \hline
most\_GSmagmom & 0.04239 \\ \hline
range\_NeighDiff\_shell1\_NsUnfilled & 0.03954 \\ \hline
min\_NeighDiff\_shell1\_NsValence & 0.03932 \\ \hline
max\_NeighDiff\_shell1\_AtomicWeight & 0.03905 \\ \hline
max\_NeighDiff\_shell1\_Number & 0.03815 \\ \hline
min\_NfValence & 0.03794 \\ \hline
dev\_NsValence & 0.0373 \\ \hline
maxdiff\_NsUnfilled & 0.03558 \\ \hline
min\_NfUnfilled & 0.03537 \\ \hline
min\_NeighDiff\_shell1\_GSmagmom & 0.03353 \\ \hline
var\_NeighDiff\_shell1\_NsValence & 0.02948 \\ \hline
min\_NpValence & 0.02946 \\ \hline
max\_NsUnfilled & 0.02933 \\ \hline
min\_NeighDiff\_shell1\_GSbandgap & 0.02735 \\ \hline
mean\_GSmagmom & 0.02402 \\ \hline
min\_NpUnfilled & 0.02233 \\ \hline
range\_NeighDiff\_shell1\_NsValence & 0.02171 \\ \hline
min\_NsUnfilled & 0.02051 \\ \hline
min\_GSbandgap & 0.01299 \\ \hline
min\_GSmagmom & 0.00132 \\ \hline
\end{longtable}

\section{Extended Statistics and Visualizations of Materials-Property-Descriptor Database} \label{mpdd:app1}

\begin{figure}[H]
    \centering
    \includegraphics[width=0.95\textwidth]{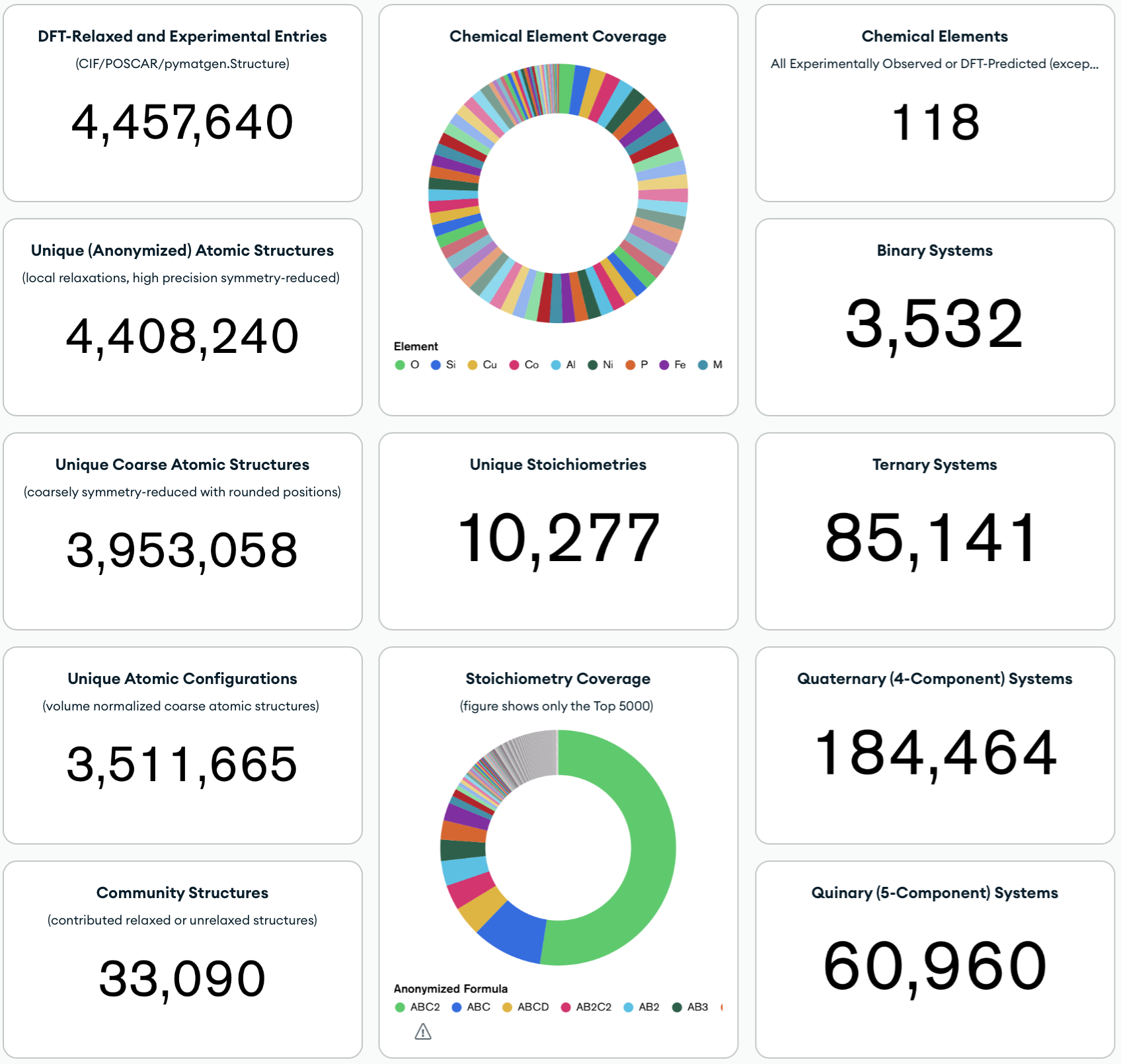}
    \caption{The main MPDD dashboard with statistics over MPDD dataset, as of April 2024, demonstrating the dataset in terms of different levels of "corese-graining" the uniqueness criteria, and coverage of chemical systems of different order.}
    \label{sup:mpdd:dashboard}
\end{figure}

The three key statistics presented in Figure~\ref{mpdd:fig:dataset} are a small subset of the larger dashboard available at \href{https://phaseslab.org/mpdd}{phaseslab.org/mpdd} web page, which is presented in Figure~\ref{sup:mpdd:dashboard}.

\begin{figure}[H]
    \centering
    \includegraphics[width=0.95\textwidth]{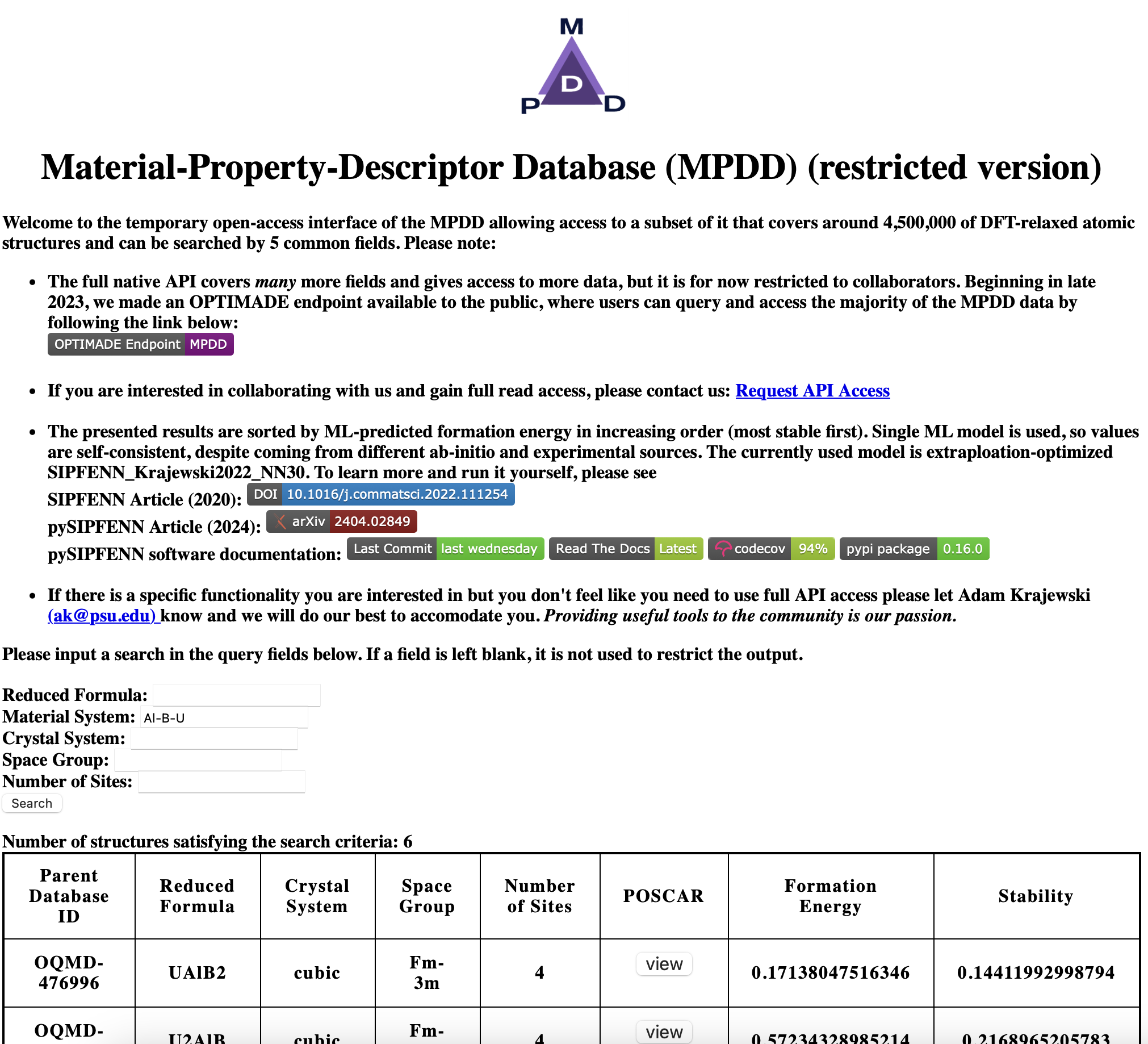}
    \caption{A basic MPDD graphical user interface (GUI) set up for easy access to the data based on a couple of common query fields.}
    \label{sup:mpdd:gui}
\end{figure}

Partial data on all of the MPDD data points can be accessed through a graphical user interface (GUI), available at \href{https://mpdd.org}{mpdd.org}, presented in Figure~\ref{sup:mpdd:gui}. It enables users to query the database based on fields including chemical formula, chemical system, or space group number, to facilitate easy access for users not familiar with programming and those who only need to access a small subset of it.

\begin{figure}[H]
    \centering
    \includegraphics[width=0.9\textwidth]{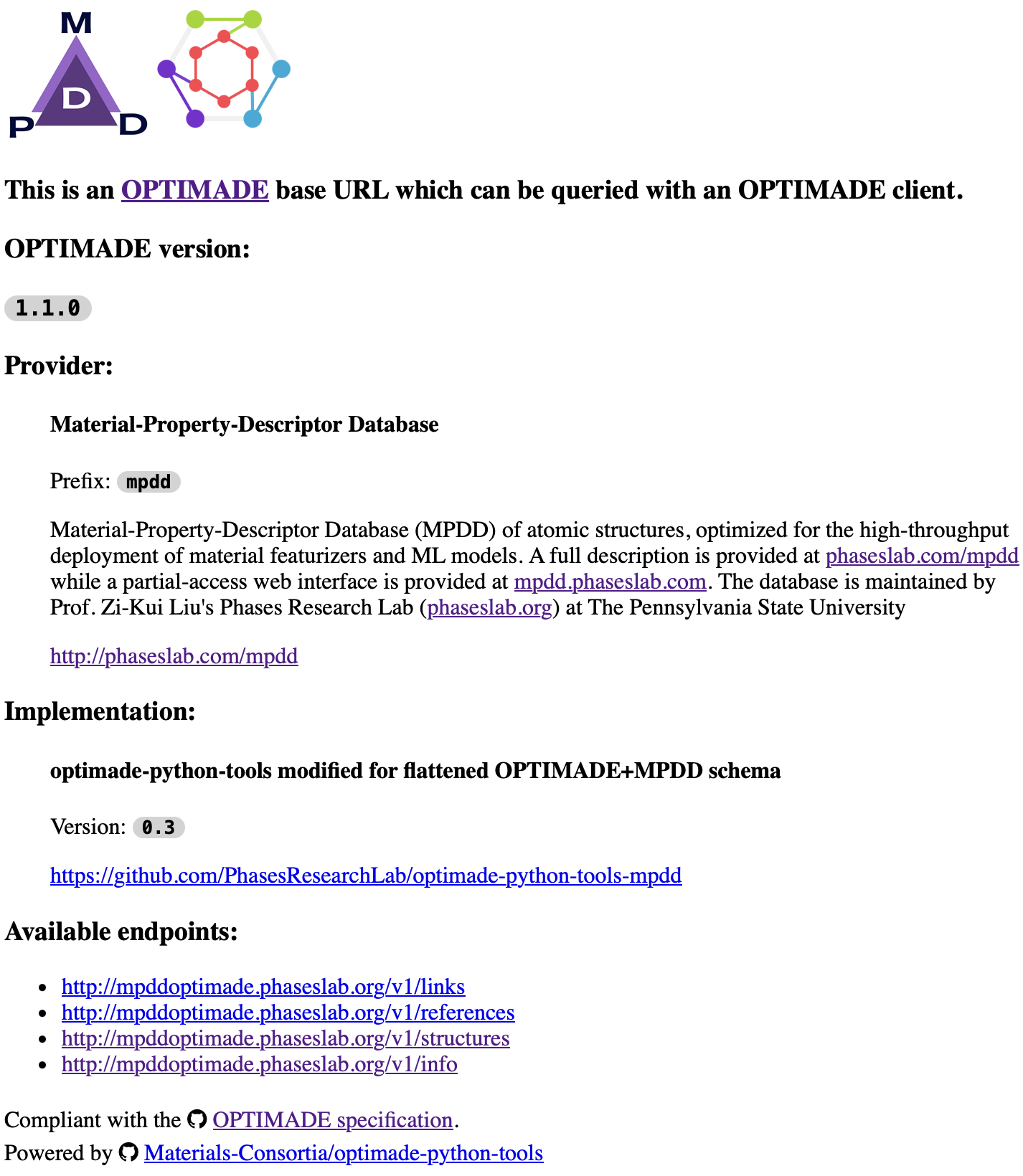}
    \caption{Printout of the MPDD OPTIMADE API endpoint page.}
    \label{sup:mpdd:optimade}
\end{figure}

The OPTIMADE API \cite{Evans2024DevelopmentsExchange}, as discussed in Section~\ref{mpdd:sec:optimade}, can be accessed using software like \texttt{optimade-python-tools} \cite{Evans2021} or by going to the \href{http://optimade.mpdd.org}{optimade.mpdd.org} endpoint. Figure~\ref{sup:mpdd:optimade} depicts the expected web page, as of May 2024.

\begin{figure}[H]
    \centering
    \includegraphics[width=0.95\textwidth]{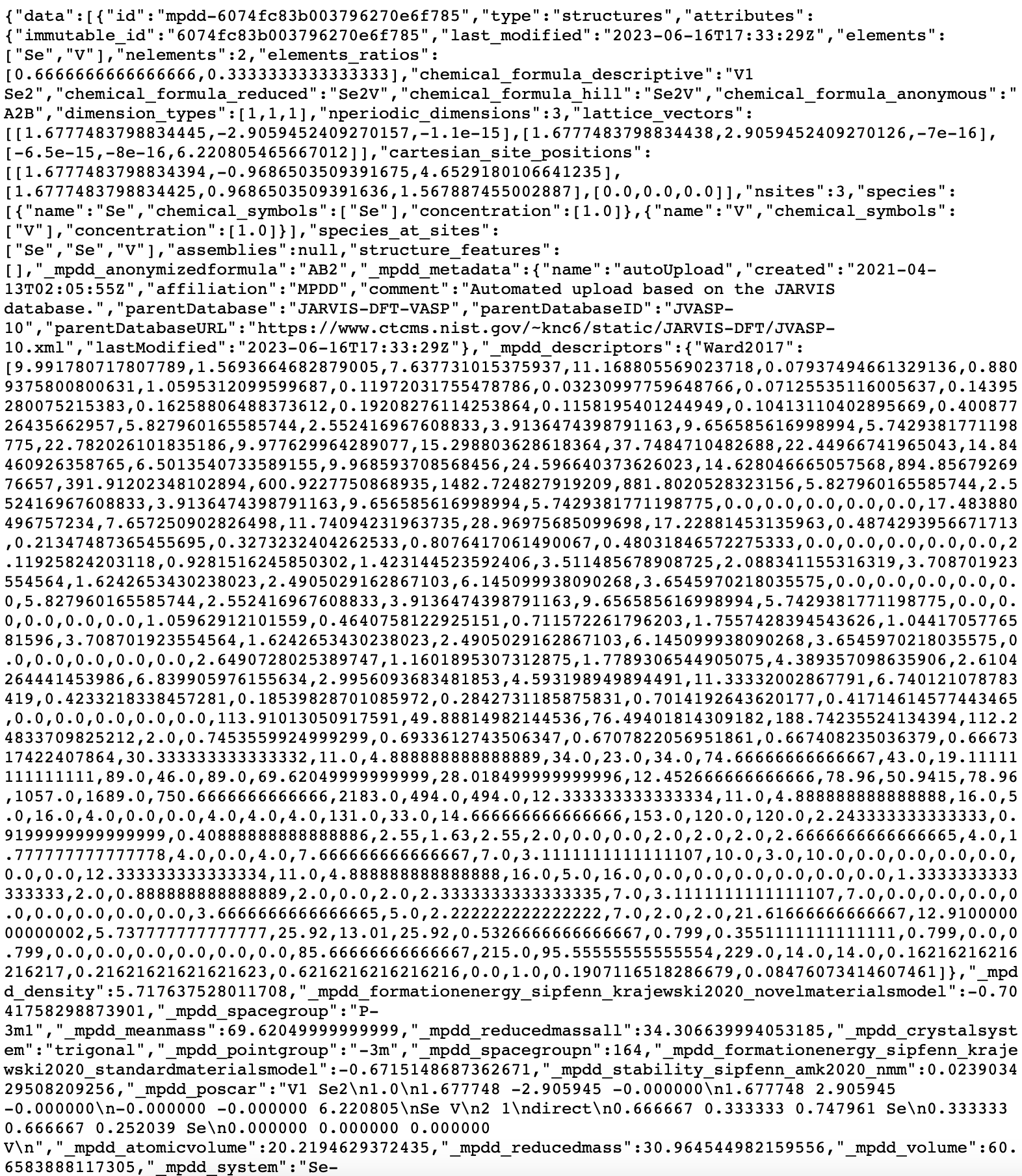}
    \caption{An example printout of (partial) MPDD's OPTIMADE response upon a query to the \texttt{http://optimade.mpdd.org/v1/structures} endpoint, depicting human-unreadable but machine-readable output.}
    \label{sup:mpdd:optimadeout}
\end{figure}

One can use the endpoint page shown in Figure~\ref{sup:mpdd:optimade} to quickly check (1) accessibility of the database, and (2) investigate its schema, by following endpoint like \href{http://optimade.mpdd.org/v1/structures}{optimade.mpdd.org/v1/structures} to see example data response, like the one shown in Figure~\ref{sup:mpdd:optimadeout}, or \href{http://optimade.mpdd.org/v1/info}{optimade.mpdd.org/v1/info} to see MPDD provider information page.

\section{Motivation for Multi-Grade Compositional Design of Materials Exemplified with Hf-Zr Powders} \label{nimplex:app1}

As mentioned in Section~\ref{nimplex:ssec:functionallygraded}, different grades of base metals may have very different costs associated with them. For instance, as of December 2023, at \href{https://www.fishersci.com}{Fisher Scientific online store (fishersci.com)}, one can purchase:
\begin{itemize}
    \item High-purity Zr wire: $250cm$ of $0.25mm$-diameter (AA00416CB) for $\$317$ or $\approx 390\frac{\$}{g}$
    \item $99.2\%$ (Zr+$4.5\%$Hf) wire: $200cm$ of $0.25mm$-diameter (AA43334G2) for $\$63$ or $\approx 100\frac{\$}{g}$. 
    \item $99.97\%$ (Hf+$3\%$Zr) wire: $200cm$ of $0.25mm$-diameter (AA10200G2) for $\$200$ or $\approx 156\frac{\$}{g}$. 
\end{itemize} 

Now, if one tries to create FGMs which navigates Zr-rich regions in Hf-containing space, there are two possible choices for Zr source, namely, pure Zr or the (Zr+$4.5\%Hf$) alloy. The first one enables all possible Zr fractions, unlike the latter which establishes the minimum Hf fraction at 4.5\% at the "Zr" corner of the attainable space tetrahedron (anonymous example of this is in Figure~\ref{nimplex:fig:fgmspaces}). Such an ability may be necessary, e.g., to avoid infeasible regions of space, but if not, it represents an unnecessary cost. 

For instance, to obtain (Zr+$10\%wt$Hf) alloy, one can combine high-purity Zr and (Hf+$3\%Zr$) for $\approx 360\frac{\$}{g}$ or equivalently from (Zr+$4.5\%$Hf) and (Hf+$3\%$Zr) for $\approx 103\frac{\$}{g}$, representing 3.5 times cost reduction.

At Fisher Scientific, as of writing this, the pure-Zr wire is only available in $0.25mm$ diameter, thus, the above considerations were restricted to it to keep the comparisons fair. However, for the (Zr+$4.5\%$Hf) grade, many less-expensive form factors are available as it is much more industry relevant. Furthermore, larger package sizes ($\geq50g$) are available driving the cost down further. For instance, the following $1mm$ wires can be purchased:

\begin{itemize}
    \item $99.2\%$ (Zr+$4.5\%$Hf) wire: $10m$ of $1mm$-diameter  (AA14627H2) for $\$130$ or $\approx 2.5\frac{\$}{g}$. 
    \item $99.97\%$ (Hf+$3\%$Zr) wire: $5m$ of $1mm$-diameter (AA10205CC) for $\$580$ or $\approx 11.3\frac{\$}{g}$. 
\end{itemize}

If the above are used, one can now obtain the same (Zr+$10\%wt$Hf) alloy for $\approx 3\frac{\$}{g}$ or 120 times cheaper relative to using high purity Zr in the only available physical form factor.

\section{Geometric Cross-Evidence for Factorial Decay of Simplex Space in Equally Dimensional Cartesian Space}  \label{nimplex:app2}

The equation for the fraction of a cube bound by [111] plane, equivalent to the result for $f(4)$ in Section~\ref{nimplex:sec:randomuniformsampling}, can be quickly obtained by considering that the volume of a pyramid is given by $V = \frac{A_B h}{3}$, where $A_B$ is the base area of equilateral triangle $\frac{\sqrt{3}}{4}\times\sqrt{2}^2 = \frac{\sqrt{3}}{2}$ and $h$ is $\frac{1}{\sqrt{3}}$. Thus we get 
$$V = \frac{\frac{\sqrt{3}}{2} \frac{1}{\sqrt{3}}}{3} = \frac{1}{6}$$ 
agreeing with the aforementioned result in Section~\ref{nimplex:sec:randomuniformsampling}.

\section{Bidirectional \texttt{neighborsLink4C} Algorithm found Conceptually}  \label{nimplex:app3}

The equation for "forward" and "backward" jumps in 3-simplex graph corresponding to a quaternary chemical system. 

\begin{minted}[xleftmargin=3\parindent, fontsize=\small]{nim}
proc neighborsLink4C(
    i:int, x:Tensor, neighbors: var seq[seq[int]], ndiv: int): void =
  let jump0 = 1  #binom(x, 0)=1
  let jump1 = binom(1+ndiv-x[0]-x[1], 1)
  let jump2 = binom(2+ndiv-x[0], 2)
  
  if x[0] != 0:
    # quaternary
    neighbors[i].add(i - jump2)
    # quaternary
    neighbors[i].add(i - jump2 - jump1)
    # quaternary
    neighbors[i].add(i - jump2 - jump1 - jump0) 
  
  if x[1] != 0:
    # ternary
    neighbors[i].add(i - jump1)    
    # ternary
    neighbors[i].add(i - jump1 - jump0)         
    # quaternary
    neighbors[i].add(i + jump2 - jump1 - x[1])  
  
  if x[2] != 0:
    # binary
    neighbors[i].add(i - jump0)     
    # ternary
    neighbors[i].add(i + jump1 - jump0)    
    # quaternary
    neighbors[i].add(i + jump2 - jump0 - x[1])  
  
  if x[3] != 0:
    # binary
    neighbors[i].add(i + jump0)        
    # ternary
    neighbors[i].add(i + jump1)          
    # quaternary     
    neighbors[i].add(i + jump2 - x[1])              
\end{minted}


\chapter{Additional Developed Software} \label{chap:othersoft}

\section{\texttt{TDB-Highlighter}} \label{osoft:sec:tdbhighlighter}

\hspace{24pt} Motto: \textit{You didn't know you can't do CALPHAD without it.}

This VS Code Language Extension, developed from scratch by Adam M. Krajewski, provides syntax highlighting for the Thermodynamic DataBase (\texttt{TDB}) files used in the CALPHAD community to describe thermodynamic properties of materials. 

As of June 2024, it was installed by 173 users of VS Code and holds 5/5 stars rating in the VS Code Marketplace under \href{https://marketplace.visualstudio.com/items?itemName=amkrajewski.tdb-highlighter#review-details}{AMKrajewski.tdb-highlighter} software identifier.

\textbf{Most critically, if something is highlighted differently than you expect, you know something may be wrong with it}. E.g., in \texttt{+1.95033E-07*T**3-.01798O876*T**2+78.5244752*T*LN(T)-722.59722*T+229382.886-93813648*T**(-1); 3290 Y} one of the numbers is not a number and will be red. One can guess which one... Likewise, that \texttt{LAVE\_C36} phase one just typed will stand out.

Features include:
\begin{itemize}
    \item Phases, species, elements and other critical elements of the database, are formatted with highly visible fonts and colors, speeding up the process of navigating, reading, and editing the files tremendously.
    \item References, notes, comments, optionals, and otherwise less critical parts of the database, are formatted with subtle formatting to reduce visual clutter.
    \item Temperature (T) and its powers/logs/exps in equations are formatted with bright red colors to immediately pinpoint where a given part of the polynomial is located.
    \item Function names are exposed within the equations to make them stand out from numbers and be easier to read.
    \item A number of TDB syntax keywords are specifically highlighted in neutral colors matching their functions to make them stand out in case of typos.
\end{itemize}

\begin{figure}[H]
    \centering
    \includegraphics[width=0.9\textwidth]{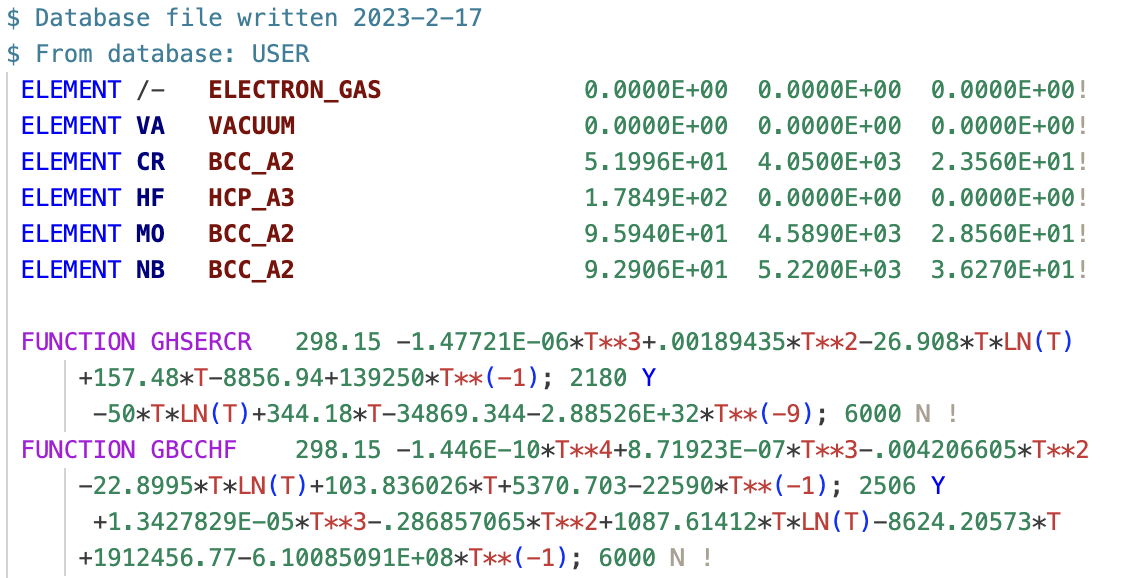}
    \caption{Header of am example TDB file (including some function definitions) syntax-highlighted using \texttt{TDB-Highlighter}.}
    \label{osoft:fig:tdbheader}
\end{figure}

Together, they yield the following neat-looking TDB header shown in Figure~\ref{osoft:fig:tdbheader}.

\begin{figure}[H]
    \centering
    \includegraphics[width=0.9\textwidth]{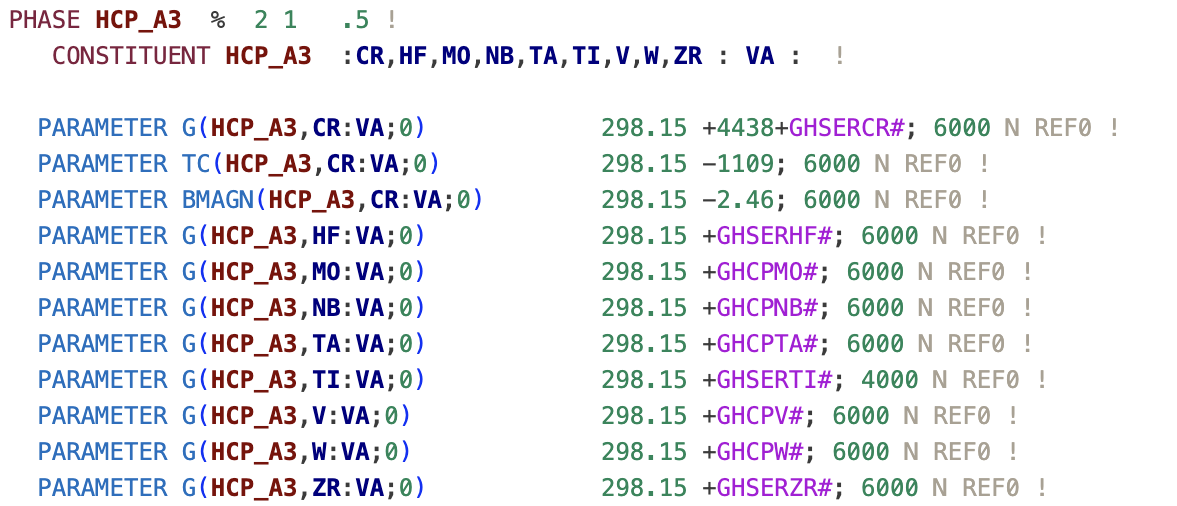}
    \caption{Thermodynamic parameter definitions in an example TDB file syntax-highlighted using \texttt{TDB-Highlighter}.}
    \label{osoft:fig:tdbfunctions}
\end{figure}

As well as easy-to-read function definitions shown in Figure~\ref{osoft:fig:tdbfunctions}.

\begin{figure}[H]
    \centering
    \includegraphics[width=0.9\textwidth]{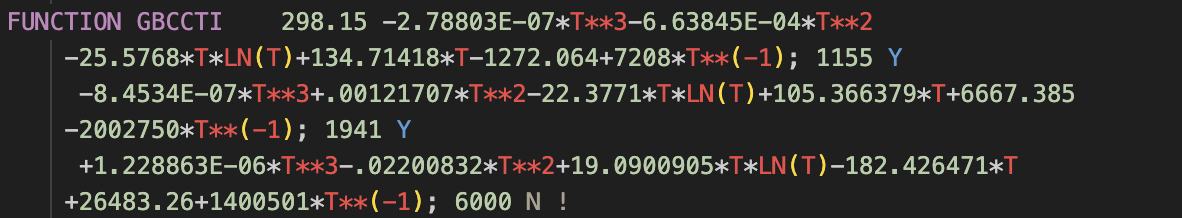}
    \caption{An example of \texttt{TDB-Highlighter} output under a "dark-mode" color scheme example.}
    \label{osoft:fig:tdbdarkmode}
\end{figure}

Furthermore, since the syntax-highlighting is not encoding colors, but rather tokens classifying TDB based on patterns, \texttt{TDB-Highlighter} works with any color scheme giving it improved customizability and accessibility. These include high-contrast schemes for people with poor vision, different types of color-blind schemes, or their dark versions, such as the one shown in Figure~\ref{osoft:fig:tdbdarkmode}.

\hspace{24pt} 
\textit{Source available at:} 
\href{https://github.com/amkrajewski/TDB-Highlighter}{https://github.com/amkrajewski/TDB-Highlighter}

\section{\texttt{pymatgen.core} Contributions} \label{osoft:sec:pymatgen}

In addition to several discussion and bug reporting meta-contributions, two critical contributions to the \texttt{core} of \texttt{pymatgen} were made. One was an \emph{enhancement} to the \texttt{core.Composition} class enabling construction of these objects from weight fraction dictionaries, which was a long-missing feature needed by metallurgists and other scientists working with real/physical materials. The second one was a \emph{fix} of a bug in the \texttt{core.Composition} causing incompatibility with \texttt{OPTIMADE} due to incorrect calculations of the Hill formula.

\hspace{24pt} 
\textit{Source available at:} 
\href{https://github.com/materialsproject/pymatgen}{https://github.com/materialsproject/pymatgen}

\section{LLNL \texttt{MAP} Platform} \label{osoft:sec:map}

Exploration of the MPEA or HEA design space is a challenging task, partly due to the combinatorial intensity in composition. For non-equiatomic alloys with more than a few components, methods investigating a grid of compositions fail sooner or later depending on how fine the grid is.

\texttt{MAP} is an unreleased internal Lawrence Livermore National Lab high entropy alloy discovery platform, core of which has been designed from scratch by Adam M. Krajewski during MaCI and CCMS programs in Summer 2022 and Summer 2023 under Contract DE-AC52-07NA27344.

It implements a method of efficient searching for novel alloys that combines (1) highly optimized Black-Box Optimization (BBO) multi-stage investigation of individual alloy systems using CALPHAD modeling and surrogate models for targeting properties with (2) on-the-fly decisions on systems to investigate and resources to allocate to each. 

The developed toolset runs parallel across systems, allowing rapid calculations on high-performance computers (HPCs), and is agnostic of the surrogate models, thus can be quickly re-used to target any properties. Here, we demonstrate targeting high yield strength at elevated temperatures in 10-component refractory systems.

\section{\texttt{pqam-dparamhu2021}} \label{osoft:sec:dparam}

A Python binding library has been added to a community model for D Parameter prediction by Hu et al. 2021, originally written in R. In the process, the underlying model has been significantly re-written, without affecting its output values, to enable large scale predictions needed by \texttt{ULTERA} ecosystem. The resulting version is 10-1000x times faster depending on the task volume and machine.

\hspace{24pt} 
\textit{Source available at:} 
\href{https://github.com/amkrajewski/pqam-dparamhu2021}{https://github.com/amkrajewski/pqam-dparamhu2021}

\section{\texttt{pqam-rmsadtandoc2023}} \label{osoft:sec:rmsad}

A community model by Tandoc et al. 2023 designed for prediction of Root Mean Square Atomic Displacement (RMSAD) has been restructured as a PyQAlloyModel (\texttt{PQAM}) and trimmed down to provide minimal interface needed to run inference within the \texttt{ULTERA} ecosystem. In this dissertation, it is used later in Appendices \ref{chap:nimplextutorial1} and \ref{chap:nimplextutorial2} as a demonstrator property model.

\hspace{24pt} 
\textit{Source available at:} 
\href{https://github.com/amkrajewski/pqam-dparamhu2021}{https://github.com/amkrajewski/pqam-dparamhu2021}

\section{\texttt{MaterialsMap} Contributions} \label{osoft:sec:materialsmap}

\texttt{MaterialsMap} is an effort parallel to the \texttt{nimplex} which performs traditional community approach to similar problems but with much lower generality that enables application of more in-depth crafted methods.

\hspace{24pt} 
\textit{Source available at:} 
\href{https://github.com/PhasesResearchLab/MaterialsMap}{https://github.com/PhasesResearchLab/MaterialsMap}

\section{\texttt{MPDD-ALIGNN}} \label{osoft:sec:mpddalignn}

This fork of the \href{https://github.com/usnistgov/alignn}{JARVIS-ALIGNN} tool has been modified in several key ways to improve it towards \texttt{MPDD} and similar applications. 

\begin{itemize}
    \item A set of models of interest has been selected and defined in a \texttt{config.yaml} for consistency, readability, and easy tracking. These are the models which will be populating MPDD.
    \item \textbf{Dependency optimizations for running models}, skipping by default installation of several packages needed only for training and auxiliary tasks. Full set can still be installed by \texttt{pip install "mpdd-alignn[full]"}.
    \item The process of model fetching was reimplemented, making it similar to \texttt{pySIPFENN} by multi-threading connection to Figshare. Model \textbf{download is now 7 times faster}, fetching all 7 default models in 6.1 vs 41.4 seconds. Optimized what is included in the built package. Now, its \textbf{package size is reduced 33.5 times}, from 21.7MB to 0.65MB.
    \item Streamlined operation, where we can get results for a directory of POSCARS for all default models in just 3 quick lines.
    \begin{minted}[xleftmargin=3\parindent, linenos=true, fontsize=\small]{python}
    from alignn import pretrained
    pretrained.download_default_models()
    result = pretrained.run_models_from_directory(
      'example.SigmaPhase', mode='serial')
    \end{minted}
    which give a neat:
    \begin{minted}[xleftmargin=3\parindent, fontsize=\small, bgcolor=subtlegray]{output}
    [{
        'name': '9-Pb8O12.POSCAR',
        'ALIGNN-JARVIS Bulk Modulus [GPa]': 98.06883239746094,
        'ALIGNN-JARVIS Exfoliation Energy [meV/atom]': 101.71208190917969,
        'ALIGNN-JARVIS Formation Energy [eV/atom]': -1.1146986484527588,
        'ALIGNN-JARVIS MBJ Bandgap [eV]': 0.5845542550086975,
        'ALIGNN-JARVIS Shear Modulus [GPa]': 39.18968963623047,
        'ALIGNN-MP Formation Energy [eV/atom]': -1.4002774953842163,
        'ALIGNN-MP PBE Bandgap [eV]': 1.074204921722412
    },
    ...
    \end{minted}
\end{itemize}

\hspace{24pt} 
\textit{Source available at:} 
\href{https://github.com/amkrajewski/mpdd-alignn}{https://github.com/amkrajewski/mpdd-alignn}

\section{\texttt{spgnim}} \label{osoft:sec:spgnim}

\texttt{spgnim} is a \texttt{nim} library built around the state-of-the-art symmetry analysis \texttt{spglib} C library (\href{https://github.com/spglib/spglib}{https://github.com/spglib/spglib}) with the goals of (1) streamlining and optimizing internal calls for improved performance, (2) easy-to-use interface for non-C programmers, and (3) future interfacing with low-level \texttt{nim}-based atomistic featurization methods for \texttt{pySIPFENN}.

\hspace{24pt} 
\textit{Source available at:} 
\href{https://github.com/amkrajewski/spgnim}{https://github.com/amkrajewski/spgnim}


\chapter{Nimplex Workshop No.1 - Quick Start Guide to using Nimplex through Python and Command Line Interface} \label{chap:nimplextutorial1}

This quick start guide will walk you through the of using Nimplex
through Python (basic) and CLI. It is not meant to be a comprehensive guide to Nimplex, but rather a quick way to get started.

\section{Basic Functions (in Python) - Grids and Random
Samples} 
\label{nimplextutorial1:basic-functions-in-python---grids-and-random-samples}

If you are running this notebook through pre-configured Codespaces, you
are ready to go as several steps have already been completed for you. If
you picked it up from GitHub on your own, make sure you (1) installed it
correctly (per the
\href{../README.md\#reproducible-installation-recommended}{README's
Reproducible Installation (recommended)} section), (2) compiled (for
Python) and moved \texttt{nimplex} into the
\texttt{examples/nimplex.so} and (3) compiled (for
Python) and moved \texttt{utils/plotting} into
\texttt{examples/utils/plotting.so}, and (4) installed
\texttt{pqam-rmsadtandoc2023} needed for some examples.
To be precise, the code below sets everyhting up for you requiring only
that you have some \texttt{conda} installed.

\begin{minted}[breaklines, xleftmargin=3\parindent, fontsize=\small, bgcolor=subtlegray]{output}
conda install -y -c conda-forge nim
conda install -y python=3.11 liblapack jupyter numpy pandas plotly
pip install pqam-rmsadtandoc2023
nimble install -y arraymancer nimpy
nim c --d:release --out:examples/nimplex nimplex
nim c --d:release --threads:on --app:lib --out:examples/nimplex.so nimplex
nim c --d:release --threads:on --app:lib --out:examples/utils/plotting.so utils/plotting
\end{minted}

And now, you can simply import \texttt{nimplex} just
like any other Python module:

\begin{minted}[xleftmargin=3\parindent, linenos=true, fontsize=\small]{python}
import nimplex
\end{minted}

You should be able to just type \texttt{nimplex.} in
the code cell below (without running it) and a dropdown menu of all the
functions available in the module should pop up. If it doesn't,
something went wrong with the installation.

You can start with the most basic functionalities of Nimplex, such as
creating a \textbf{simplex grid} with \textbf{fractional} positions in
\texttt{4}-component space quantized at 20\% or
\texttt{5} divisions per dimension:

\begin{minted}[xleftmargin=3\parindent, linenos=true, fontsize=\small]{python}
grid1 = nimplex.simplex_grid_fractional_py(4,5)
\end{minted}

Lets look at the first 10 points of the grid:

\begin{minted}[xleftmargin=3\parindent, linenos=true, fontsize=\small]{python}
grid1[0:10]
\end{minted}

\begin{minted}[xleftmargin=3\parindent, fontsize=\small, bgcolor=subtlegray]{output}
[[0.0, 0.0, 0.0, 1.0],
 [0.0, 0.0, 0.2, 0.8],
 [0.0, 0.0, 0.4, 0.6],
 [0.0, 0.0, 0.6, 0.4],
 [0.0, 0.0, 0.8, 0.2],
 [0.0, 0.0, 1.0, 0.0],
 [0.0, 0.2, 0.0, 0.8],
 [0.0, 0.2, 0.2, 0.6],
 [0.0, 0.2, 0.4, 0.4],
 [0.0, 0.2, 0.6, 0.2]]
\end{minted}

As you can see, all points are in the range
\texttt{[0, 1]} and the sum of all components is
\texttt{1} (within numerical precision), as the grid
exists in the simplex, not Cartesian/Euclidean (aka hypercube) space.
You can also note that the grid does include the corners of the simplex
or pure components, such as \texttt{[1, 0, 0, 0]} or
\texttt{[0, 1, 0, 0]}. If you want to exclude them, you
can generate \textbf{internal} grid points:

\begin{minted}[xleftmargin=3\parindent, linenos=true, fontsize=\small]{python}
grid2 = nimplex.simplex_internal_grid_fractional_py(4,5)
grid2
\end{minted}

\begin{minted}[xleftmargin=3\parindent, fontsize=\small, bgcolor=subtlegray]{output}
[[0.2, 0.2, 0.2, 0.4],
 [0.2, 0.2, 0.4, 0.2],
 [0.2, 0.4, 0.2, 0.2],
 [0.4, 0.2, 0.2, 0.2]]
\end{minted}

which in this case happens to be fairly small because the grid is so
coarse. Changing the number of divisions to \texttt{10}
per dimension gives us a much denser grid:

\begin{minted}[xleftmargin=3\parindent, linenos=true, fontsize=\small]{python}
grid3 = nimplex.simplex_internal_grid_fractional_py(4,10)
grid3[:10]
\end{minted}

\begin{minted}[xleftmargin=3\parindent, fontsize=\small, bgcolor=subtlegray]{output}
[[0.1, 0.1, 0.1, 0.7],
 [0.1, 0.1, 0.2, 0.6],
 [0.1, 0.1, 0.3, 0.5],
 [0.1, 0.1, 0.4, 0.4],
 [0.1, 0.1, 0.5, 0.3],
 [0.1, 0.1, 0.6, 0.2],
 [0.1, 0.1, 0.7, 0.1],
 [0.1, 0.2, 0.1, 0.6],
 [0.1, 0.2, 0.2, 0.5],
 [0.1, 0.2, 0.3, 0.4]]
\end{minted}

And, if we want to express it in terms of integer coordinates (number of
quantization steps from the origin), we can do that too by:

\begin{minted}[xleftmargin=3\parindent, linenos=true, fontsize=\small]{python}
grid4 = nimplex.simplex_internal_grid_py(4,10)
grid4[:10]
\end{minted}

\begin{minted}[xleftmargin=3\parindent, fontsize=\small, bgcolor=subtlegray]{output}
[[1, 1, 1, 7],
 [1, 1, 2, 6],
 [1, 1, 3, 5],
 [1, 1, 4, 4],
 [1, 1, 5, 3],
 [1, 1, 6, 2],
 [1, 1, 7, 1],
 [1, 2, 1, 6],
 [1, 2, 2, 5],
 [1, 2, 3, 4]]
\end{minted}

We can now try to do some plotting. Let's create a fairly dense
\texttt{3}-component fractional grid with
\texttt{48} divisions per dimension and plot it in 2D
using \texttt{plotly}:

\begin{minted}[xleftmargin=3\parindent, linenos=true, fontsize=\small]{python}
import plotly.express as px
import pandas as pd
import plotly.io as pio
pio.renderers.default = 'pdf'
\end{minted}

\begin{minted}[xleftmargin=3\parindent, linenos=true, fontsize=\small]{python}
grid5 = nimplex.simplex_internal_grid_fractional_py(3,48)
grid5df = pd.DataFrame(grid5, columns=['x','y','z'])
\end{minted}

\begin{minted}[xleftmargin=3\parindent, linenos=true, fontsize=\small]{python}
px.scatter_ternary(grid5df, a='x', b='y', c='z')
\end{minted}

\begin{figure}[H]
    \centering
    \includegraphics[width=0.55\textwidth]{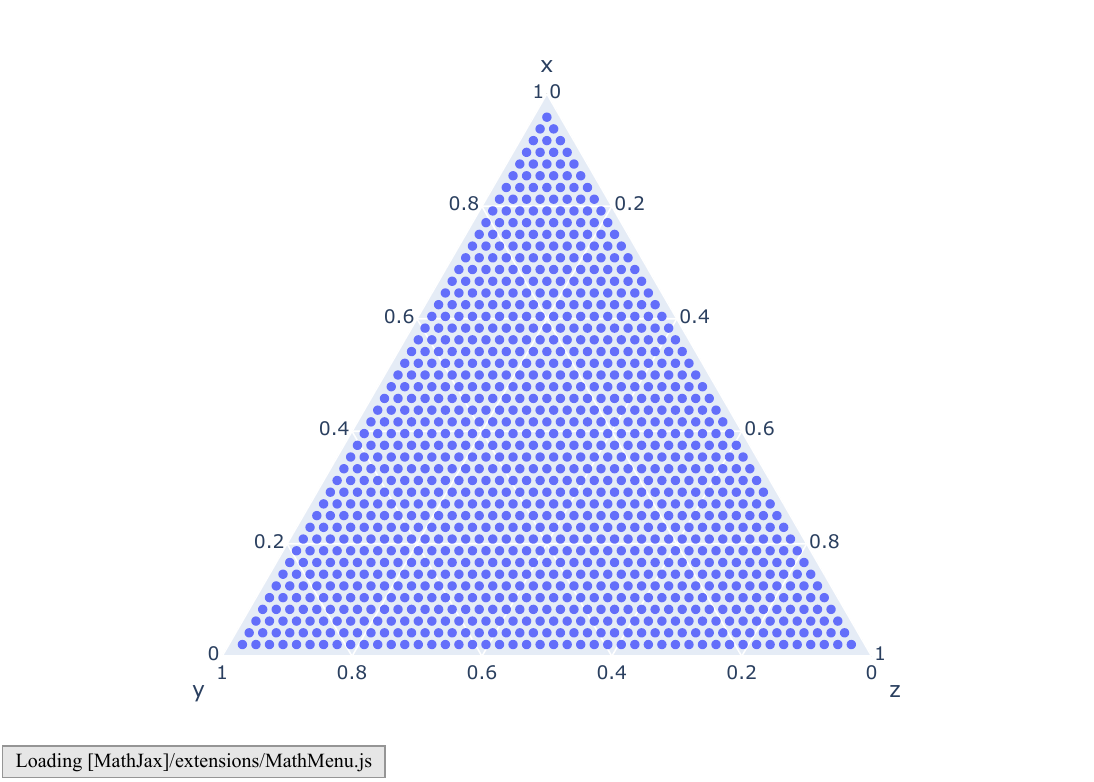}
    \caption{A ternary full fractional simplex grid with $48$ divisions per dimension.}
    \label{nimplextutorial1:fig:ternarygrid}
\end{figure}

\textbf{Neat!}

You can also create a \textbf{uniform sampling} of the simplex space
using \texttt{simplex\_sampling\_mc}. Let us create a
\texttt{1000}-point sample of the simplex in
\texttt{3}-component space:

\begin{minted}[xleftmargin=3\parindent, linenos=true, fontsize=\small]{python}
randomSample1 = nimplex.simplex_sampling_mc_py(3, 2000)
randomSample1[:10]
\end{minted}

\begin{minted}[xleftmargin=3\parindent, fontsize=\small, bgcolor=subtlegray]{output}
[[0.5298803624052214, 0.07086239625884846, 0.3992572413359302],
 [0.0031033677260338954, 0.2388051480096958, 0.7580914842642703],
 [0.18649108765063374, 0.5836686212515504, 0.22984029109781595],
 [0.15016940727387892, 0.2949275122194682, 0.5549030805066528],
 [0.6155094564276237, 0.008656565578592935, 0.3758339779937834],
 [0.5304668735927556, 0.31073677219088075, 0.15879635421636376],
 [0.6207714545708731, 0.32969096943581894, 0.049537575993307915],
 [0.31938546746902047, 0.2709402919340315, 0.4096742405969479],
 [0.5712362080270665, 0.22399413177992233, 0.20476966019301104],
 [0.020101994689313802, 0.019307919101951364, 0.9605900862087349]]
\end{minted}

and plot it in 2D, just like before with the grid:

\begin{minted}[xleftmargin=3\parindent, linenos=true, fontsize=\small]{python}
randomSample1df = pd.DataFrame(randomSample1, columns=['x','y','z'])
px.scatter_ternary(randomSample1df, a='x', b='y', c='z', opacity=0.33)
\end{minted}

\begin{figure}[H]
    \centering
    \includegraphics[width=0.55\textwidth]{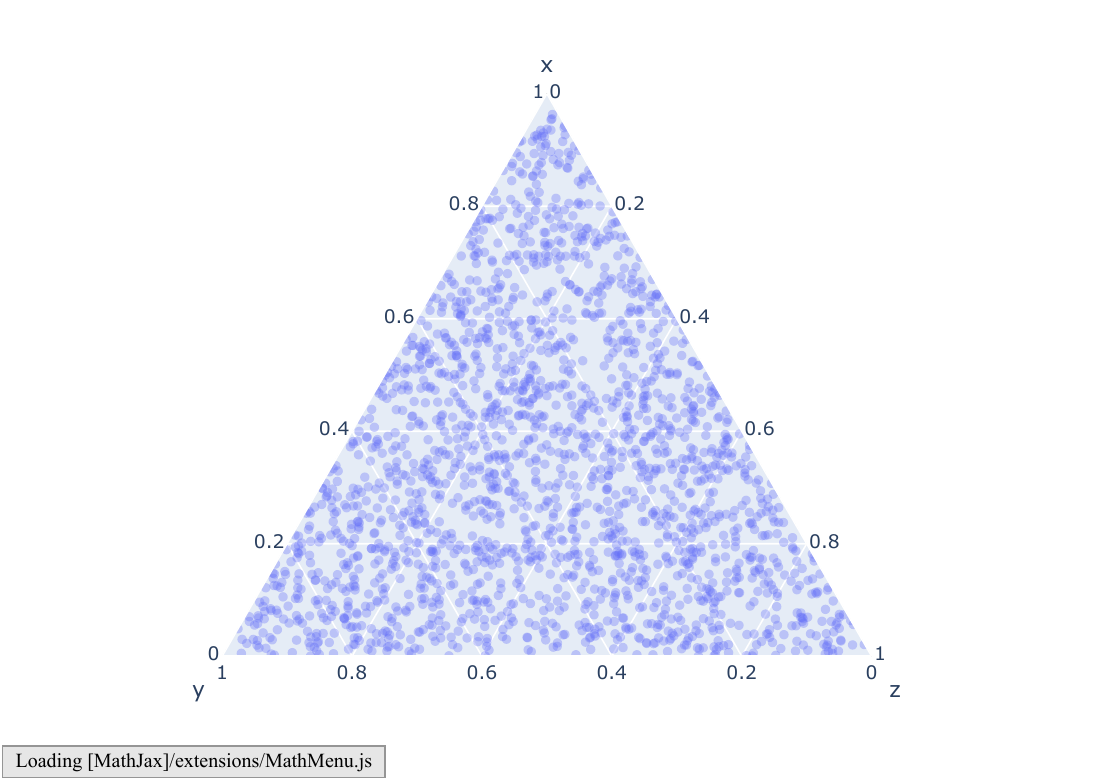}
    \caption{A ternary uniform sampling of a simplex with $1000$ points.}
    \label{nimplextutorial1:fig:ternaryuniformsampling}
\end{figure}

The above works great for ternary (3-component) space, but what if we
want to work in quaternary (4-component) space which
\texttt{plotly} does not support directly? Or we want
to use a different plotting library without function like
\texttt{scatter\_ternary}? We can do so by projecting
the simplex onto the Euclidean space using the
\texttt{utils/plotting.nim} convenience module, which
was compiled as \texttt{plotting.so} and placed in the
\texttt{utils} in the current working directory. Let's
try it out:

\begin{minted}[xleftmargin=3\parindent, linenos=true, fontsize=\small]{python}
from utils import plotting
\end{minted}

\begin{minted}[xleftmargin=3\parindent, linenos=true, fontsize=\small]{python}
grid5_projected = plotting.simplex2cartesian_py(
  nimplex.simplex_internal_grid_fractional_py(3,48))
grid5_projected[:10]
\end{minted}

\begin{minted}[xleftmargin=3\parindent, fontsize=\small, bgcolor=subtlegray]{output}
[[-0.8118984375000001, -0.46875],
 [-0.7758140625000001, -0.46875],
 [-0.7397296875, -0.46875],
 [-0.7036453125000001, -0.46875],
 [-0.6675609375, -0.46875],
 [-0.6314765625, -0.46875],
 [-0.5953921875000001, -0.46875],
 [-0.5593078125000001, -0.46875],
 [-0.5232234375, -0.46875],
 [-0.48713906250000005, -0.46875]]
\end{minted}

As you can immediately notice, all points are now 2D vectors and now we
can plot the same grid as before, but using typical scatter plot
functionality:

\begin{minted}[xleftmargin=3\parindent, linenos=true, fontsize=\small]{python}
grid5_projected_df = pd.DataFrame(grid5_projected, columns=['x','y'])
px.scatter(grid5_projected_df, x='x', y='y', 
  width=600, height=500, template='plotly_white')
\end{minted}

\begin{figure}[H]
    \centering
    \includegraphics[width=0.55\textwidth]{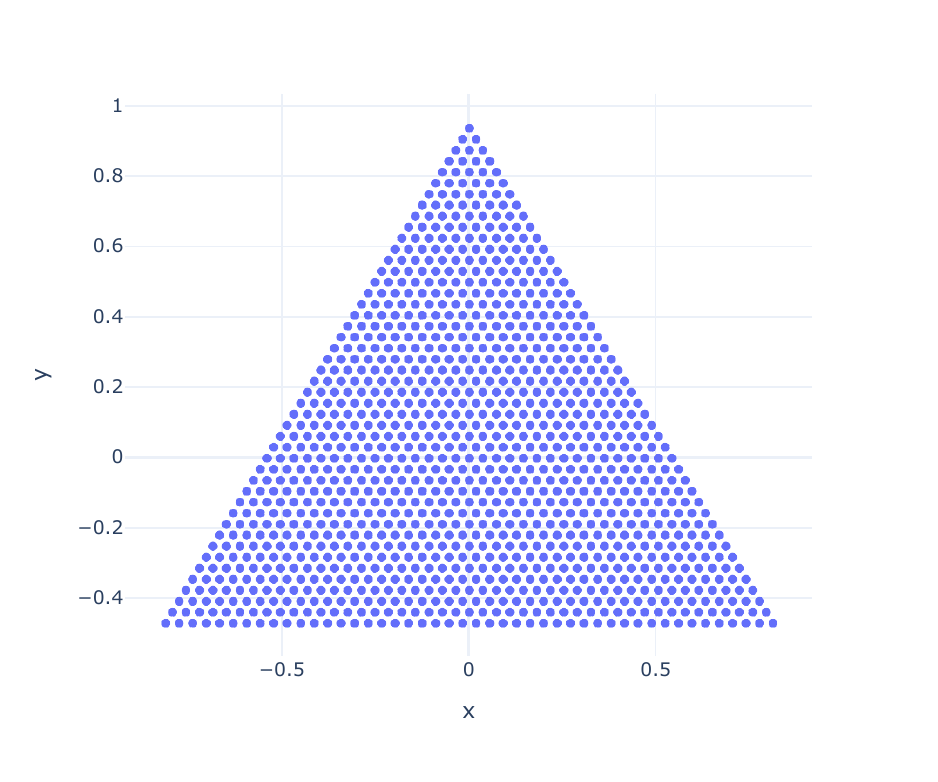}
    \caption{A Cartesian space projection of ternary simplex grid through \texttt{nimplex}'s \texttt{utils.plotting.} \texttt{simplex2cartesian} function enabling plotting of Figure \ref{nimplextutorial1:fig:ternarygrid} in any 2D plotting software.}
    \label{nimplextutorial1:fig:ternarygridprojected}
\end{figure}

and then do the same for 4-component simplex space in 3D cartesian
space:

\begin{minted}[xleftmargin=3\parindent, linenos=true, fontsize=\small]{python}
grid6 = nimplex.simplex_internal_grid_fractional_py(4,12)
grid6_projected = plotting.simplex2cartesian_py(grid6)
grid6_projected[:10]
\end{minted}

\begin{minted}[xleftmargin=3\parindent, fontsize=\small, bgcolor=subtlegray]{output}
[[2.1806839667348754e-18, -8.4887374907083305e-19, 0.66666665],
 [-0.03928370999999999, -0.06804138333333333, 0.5555555333333333],
 [-0.07856742, -0.1360827666666667, 0.4444444166666667],
 [-0.11785112999999998, -0.20412415, 0.3333333],
 [-0.15713484, -0.2721655333333334, 0.22222218333333335],
 [-0.19641855, -0.3402069166666667, 0.11111106666666665],
 [-0.23570226000000002, -0.40824830000000006, -5.000000002919336e-08],
 [-0.27498596999999997, -0.4762896833333333, -0.11111116666666668],
 [-0.31426968, -0.5443310666666666, -0.22222228333333333],
 [-0.03928370999999999, 0.06804138333333333, 0.5555555333333333]]
\end{minted}

\begin{minted}[xleftmargin=3\parindent, linenos=true, fontsize=\small]{python}
grid6_projected_df = pd.DataFrame(grid6_projected, columns=['x','y','z'])
px.scatter_3d(grid6_projected_df, x='x', y='y', z='z', 
              template='plotly_white', width=800, height=700)
\end{minted}

\begin{figure}[H]
    \centering
    \includegraphics[width=0.6\textwidth]{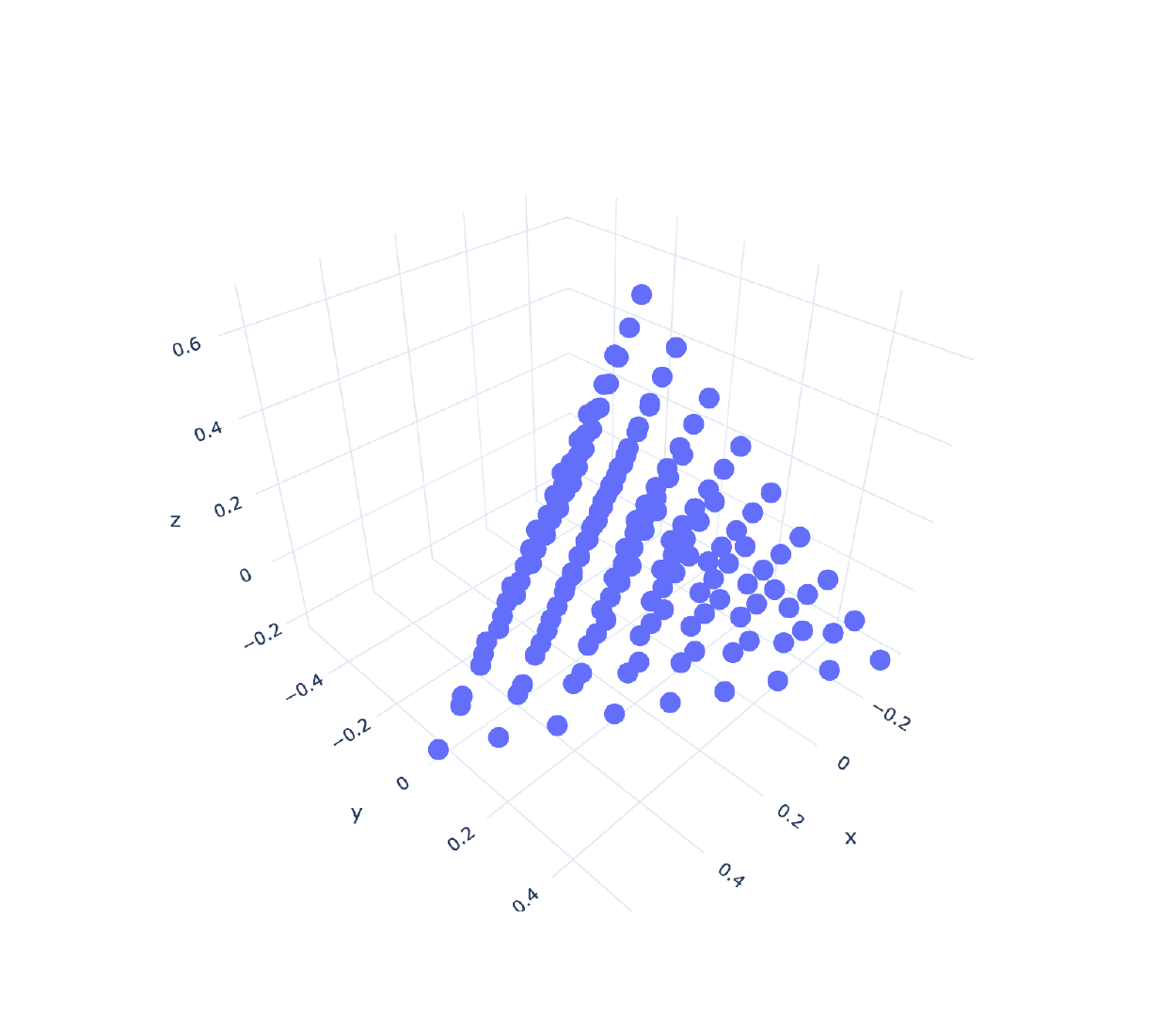}
    \caption{A Cartesian space projection of quaternary simplex grid through \texttt{simplex2cartesian} enabling plotting in any 3D-capable software.}
    \label{nimplextutorial1:fig:quaternarygridprojected}
\end{figure}

or for random sampling:

\begin{minted}[xleftmargin=3\parindent, linenos=true, fontsize=\small]{python}
randomSample2 = plotting.simplex2cartesian_py(
  nimplex.simplex_sampling_mc_py(4, 1000))
randomSample2[:10]
\end{minted}

\begin{minted}[xleftmargin=3\parindent, fontsize=\small, bgcolor=subtlegray]{output}
[[-0.07501012159885143, 0.023590567429365657, -0.23553634907424525],
 [-0.4462490471332081, -0.1756565137829428, -0.2723228511523328],
 [0.07145506465222057, 0.3210603889969648, -0.18732905907855993],
 [-0.18725987909196115, 0.10689040149288015, -0.013098816843650607],
 [0.022355997592579786, 0.010927841672716905, -0.22642456944252606],
 [-0.0432501395787877, -0.17053829898879963, 0.11683087219672761],
 [0.037379243987826286, -0.050016213978314135, -0.031028495845089377],
 [0.01486956568331414, -0.0033756497464057025, 0.9054507810078116],
 [-0.08877119931676818, 0.13681189954017764, 0.6760960009121882],
 [0.38224893038975954, -0.16794340072714056, -0.15924029851917898]]
\end{minted}

\begin{minted}[xleftmargin=3\parindent, linenos=true, fontsize=\small]{python}
randomSample2df = pd.DataFrame(randomSample2, columns=['x','y','z'])
px.scatter_3d(randomSample2df, x='x', y='y', z='z', 
              template='plotly_white', width=800, height=700, opacity=0.2)
\end{minted}

\begin{figure}[H]
    \centering
    \includegraphics[width=0.6\textwidth]{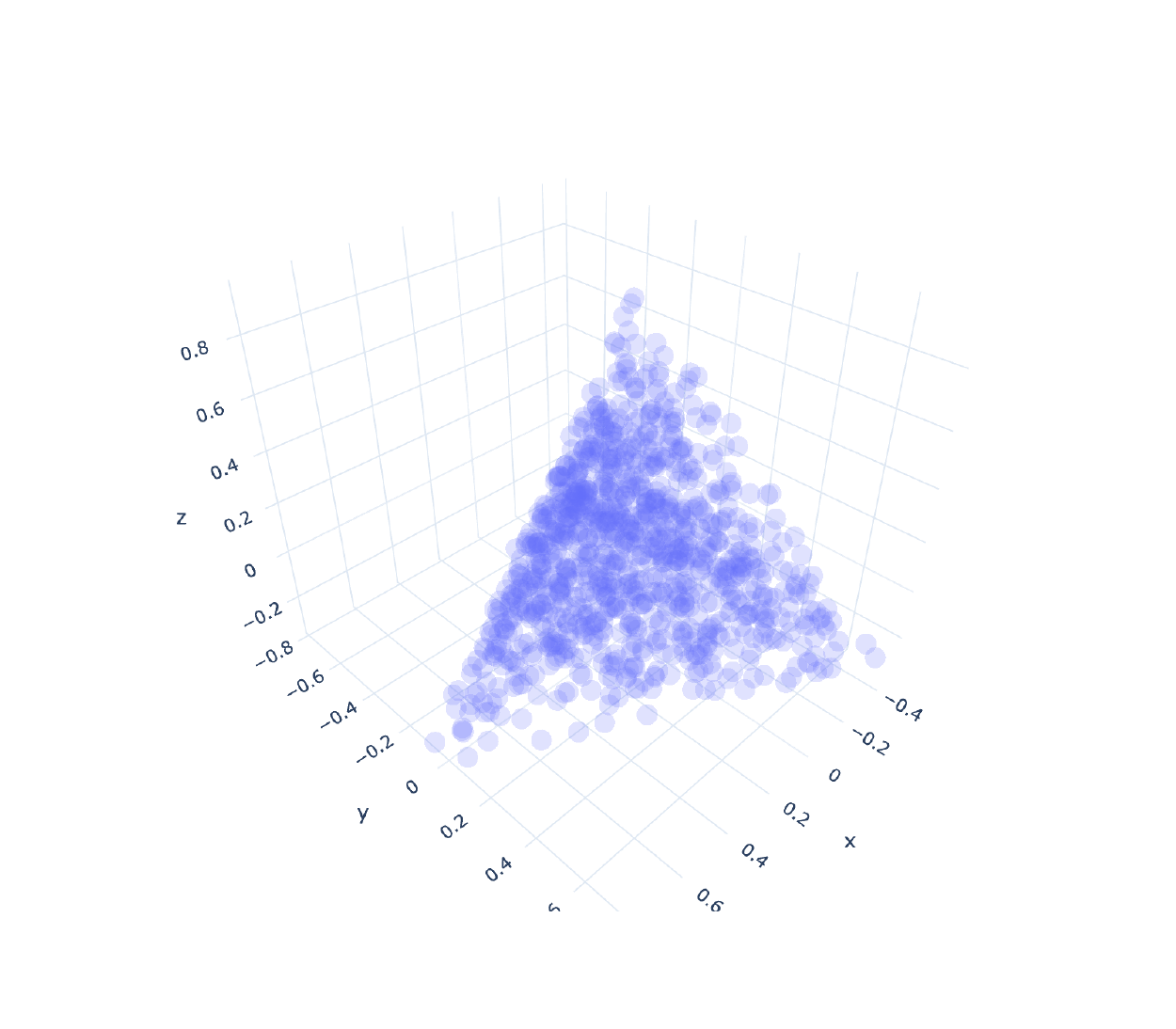}
    \caption{A Cartesian space projection of quaternary simplex uniform sampling through \texttt{simplex2cartesian} enabling plotting in any 3D-capable software.}
    \label{nimplextutorial1:fig:quaternarysamlingprojected}
\end{figure}

Now, we can also attach \emph{some} model to it and see how it looks
like. Let's use Root Mean Squared Atomic Displacement (RMSAD) model for
high-entropy alloys (HEAs) in Ti-Zr-Hf-V-Nb-Ta-Mo-W-Re-Ru space designed
by Tandoc et al.~in
\href{https://doi.org/10.1038/s41524-023-00993-x}{10.1038/s41524-023-00993-x}
and reimplemented for \href{https://ultera.org}{ULTERA Ecosystem} by
nimplex's author. Let's now deploy it for a 4-component space formed by
\texttt{Ti}, \texttt{Zr},
\texttt{Hf}, and \texttt{V}, based on
\texttt{grid6} we generated before.

\begin{minted}[xleftmargin=3\parindent, linenos=true, fontsize=\small]{python}
import pqam_rmsadtandoc2023
\end{minted}

\begin{minted}[xleftmargin=3\parindent, linenos=true, fontsize=\small]{python}
components = ["Ti", "Zr", "Hf", "V"]
rmsadList = []
for point in grid6:
    formula = ' '.join([f'{c}{p}' for c, p in zip(components, point)])
    rmsadList.append(pqam_rmsadtandoc2023.predict(formula))
\end{minted}

\begin{minted}[xleftmargin=3\parindent, linenos=true, fontsize=\small]{python}
px.scatter_3d(grid6_projected_df, x='x', y='y', z='z', color=rmsadList,
              template='plotly_white', width=800, height=700)
\end{minted}

\begin{figure}[H]
    \centering
    \includegraphics[width=0.6\textwidth]{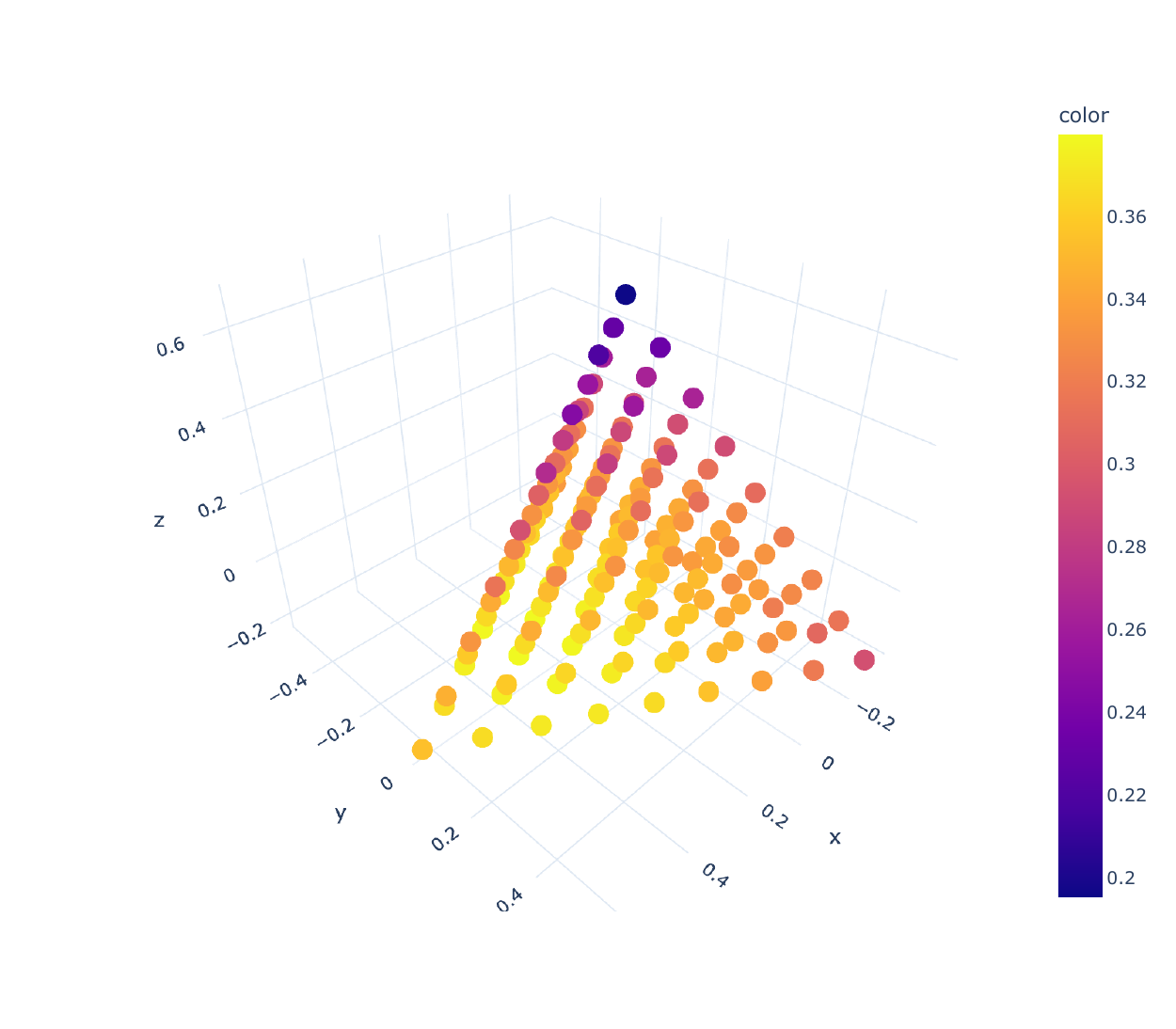}
    \caption{Figure \ref{nimplextutorial1:fig:quaternarygridprojected} colorized by RMSAD predicted by model \cite{Tandoc2023MiningAlloys} deployed over Hf-Ti-V-Zr chemical space.}
    \label{nimplextutorial1:fig:quaternarygridcolored}
\end{figure}

But what if your system is not made of pure components, but rather of
some alloys? For that case you can generate a pair of \textbf{attainable
space} simplex grid and corresponding \textbf{elemental space} positions
serving as inputs to the model. Let's try it out for
\texttt{7}-component elemental space of
\texttt{Ti}, \texttt{Zr},
\texttt{Hf}, \texttt{W},
\texttt{Nb}, \texttt{Ta}, and
\texttt{Mo} formed by \texttt{4}
alloys: - Ti50 Zr50 - Hf95 Zr5 - Mo33 Nb33 Ta33 - Mo10 Nb10 W80

which we can represent as points in the elemental space:

\begin{minted}[xleftmargin=3\parindent, linenos=true, fontsize=\small]{python}
elementalSpaceComponents = 
  ["Ti", "Zr", "Hf", "W", "Nb", "Ta", "Mo"]
attainableSpaceComponents = 
  ["Ti50 Zr50", "Hf95 Zr5", "Mo33 Nb33 Ta33", "Mo10 Nb10 W80"]
attainableSpaceComponentPositions = 
  [[50, 50, 0, 0, 0, 0, 0], [0, 5, 95, 0, 0, 0, 0], 
  [0, 0, 0, 33, 33, 33, 0], [0, 0, 0, 10, 10, 0, 80]]
\end{minted}

and then generate the pair of grids:

\begin{minted}[xleftmargin=3\parindent, linenos=true, fontsize=\small, breaklines]{python}
gridAtt, gridEl = nimplex.embeddedpair_simplex_grid_fractional_py(attainableSpaceComponentPositions, 12)
\end{minted}

\begin{minted}[xleftmargin=3\parindent, linenos=true, fontsize=\small]{python}
gridAtt[:3]
\end{minted}

\begin{minted}[xleftmargin=3\parindent, fontsize=\small, bgcolor=subtlegray]{output}
[[0.0, 0.0, 0.0, 1.0],
 [0.0, 0.0, 0.08333333333333333, 0.9166666666666666],
 [0.0, 0.0, 0.16666666666666666, 0.8333333333333334]]
\end{minted}

\begin{minted}[xleftmargin=3\parindent, linenos=true, fontsize=\small]{python}
gridEl[:3]
\end{minted}

\begin{minted}[xleftmargin=3\parindent, fontsize=\small, bgcolor=subtlegray]{output}
[[0.0, 0.0, 0.0, 0.1, 0.1, 0.0, 0.8],
 [0.0,
  0.0,
  0.0,
  0.11944444444444445,
  0.11944444444444445,
  0.027777777777777776,
  0.7333333333333334],
 [0.0,
  0.0,
  0.0,
  0.1388888888888889,
  0.1388888888888889,
  0.05555555555555555,
  0.6666666666666667]]
\end{minted}

Then, we use the \textbf{elemental} space grid to run the model:

\begin{minted}[xleftmargin=3\parindent, linenos=true, fontsize=\small]{python}
rmsadList = []
for point in gridEl:
    formula = ' '.join([f'{c}{p}' for c, p in zip(elementalSpaceComponents, point)])
    rmsadList.append(pqam_rmsadtandoc2023.predict(formula))
\end{minted}

And \textbf{attainable} space grid to plot the results after projecting
them to the Euclidean space:

\begin{minted}[xleftmargin=3\parindent, linenos=true, fontsize=\small]{python}
gridAtt_projected_df = pd.DataFrame(
  plotting.simplex2cartesian_py(gridAtt), columns=['x','y','z'])
# Add text labels at the corners of the simplex
px.scatter_3d(gridAtt_projected_df, x='x', y='y', z='z', color=rmsadList,
              template='plotly_white', width=800, height=700, 
              labels={'color':'RMSAD', 'x':'', 'y':'', 'z':''})
\end{minted}

\begin{figure}[H]
    \centering
    \includegraphics[width=0.6\textwidth]{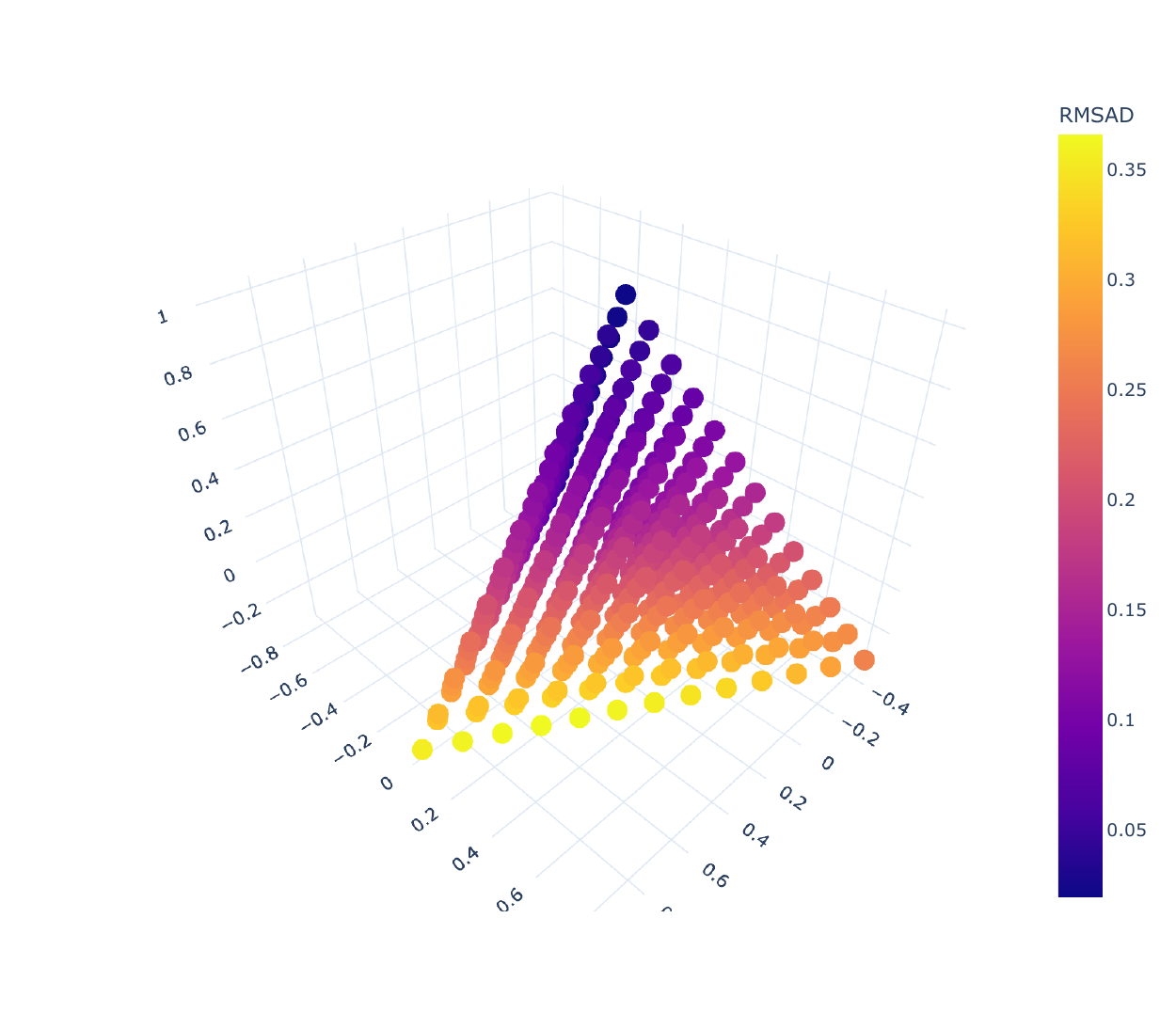}
    \caption{Plot of RMSAD values over compositional tetrahedron (3-simplex) formed by all combinations of Ti50 Zr50, Hf95 Ti5, Mo33 Nb33 Ta33, Mo80 Nb10 W10 discretized at 12 divisions per dimension. The positions in the 7-component elemental space obtained from \texttt{nimplex} \cite{Krajewski2024Nimplex} were used to run RMSAD model by \cite{Tandoc2023MiningAlloys} and projected into Cartesian space for plotting by \texttt{simplex2cartesian} function in \texttt{nimplex}.}
    \label{nimplextutorial1:fig:quaternaryattainablecolored}
\end{figure}

we may also want to label the corners of the simplex with the alloy
names. We can do that by finding the indexes of the corners in the grid
using nimplex

\begin{minted}[xleftmargin=3\parindent, linenos=true, fontsize=\small]{python}
pureComponentIndices = nimplex.pure_component_indexes_py(4, 12)
print(pureComponentIndices)
\end{minted}

\begin{minted}[xleftmargin=3\parindent, fontsize=\small, bgcolor=subtlegray]{output}
[454, 90, 12, 0]
\end{minted}

\begin{minted}[xleftmargin=3\parindent, linenos=true, fontsize=\small]{python}
labels = ['']*len(gridAtt_projected_df)
for comp, idx in zip(attainableSpaceComponents, pureComponentIndices):
    labels[idx] = "<b>"+comp+"</b>"
\end{minted}

\begin{minted}[xleftmargin=3\parindent, linenos=true, fontsize=\small]{python}
gridAtt_projected_df = pd.DataFrame(
  plotting.simplex2cartesian_py(gridAtt), columns=['x','y','z'])
# Add text labels at the corners of the simplex
px.scatter_3d(gridAtt_projected_df, x='x', y='y', z='z', color=rmsadList, text=labels,
              template='plotly_white', width=800, height=700, 
              labels={'color':'RMSAD', 'x':'', 'y':'', 'z':''})
\end{minted}

\begin{figure}[H]
    \centering
    \includegraphics[width=0.6\textwidth]{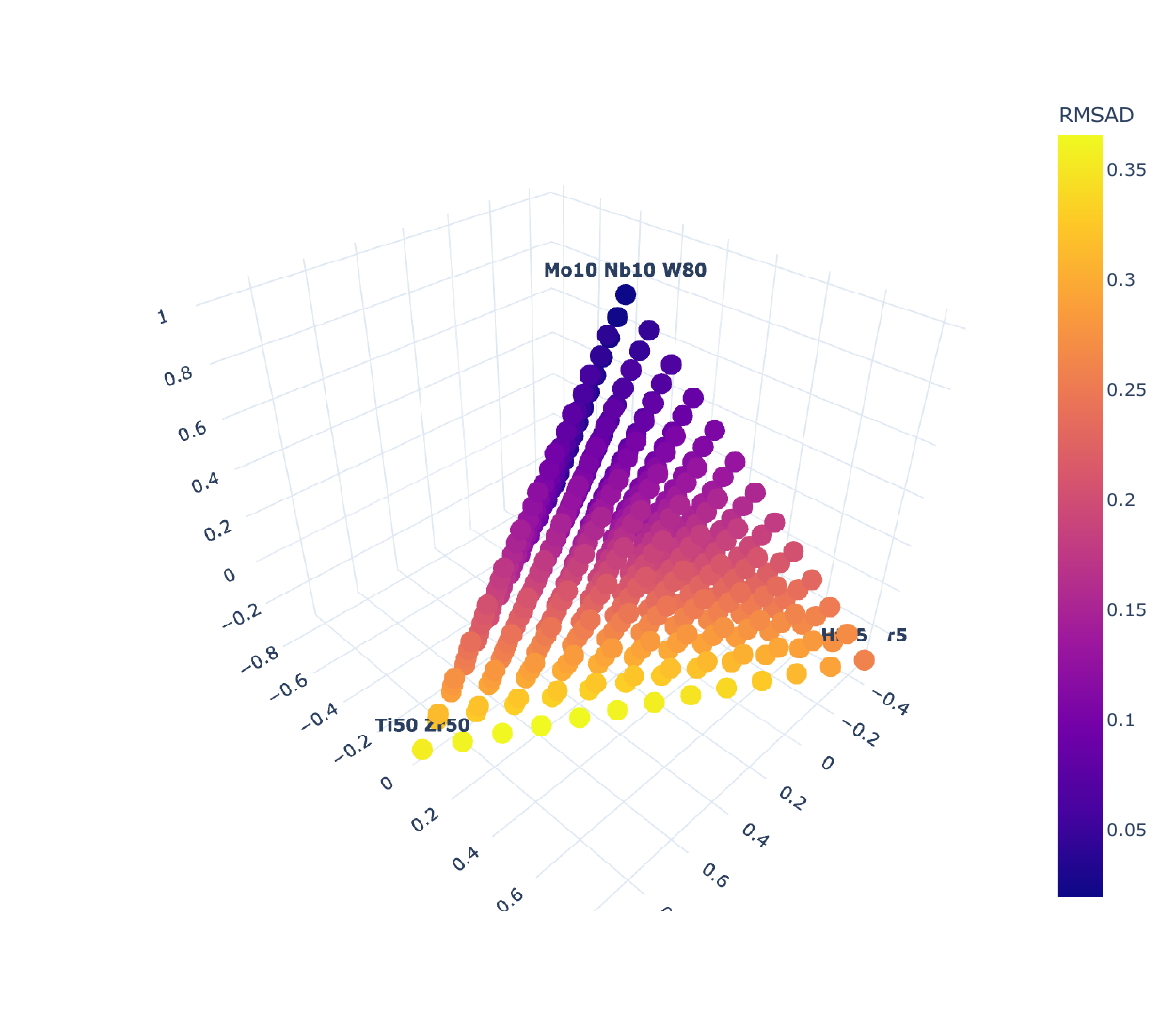}
    \caption{Figure \ref{nimplextutorial1:fig:quaternaryattainablecolored} overlaid with pure-design-component (alloy composition) labels at spatial positions identified using \texttt{nimplex.pure\_component\_indexes}.}
    \label{nimplextutorial1:fig:quaternaryattainablecoloredlabel}
\end{figure}

\section{CLI - Grids and Random in Any Language from Julia to Ada
Samples} 
\label{nimplextutorial1:cli---grids-and-random-samples}

Let's say you don't want to use Python, but rather would like to use the
command line interface (CLI) to generate the grids and use them within,
e.g., \href{https://julialang.org}{Julia Language}. The easiest way to
do it is to output the grid to the common NumPy format, which you can
then then load it in Julia using one of the packages, such as
\href{https://www.juliapackages.com/p/npz}{NPZ.jl}.

The commands are concise, simple, and fully described in both
\href{https://amkrajewski.github.io/nimplex/}{documentation} and CLI
help:

\begin{minted}[xleftmargin=3\parindent, linenos=true, fontsize=\small]{python}
!./nimplex --help
\end{minted}

\begin{minted}[xleftmargin=3\parindent, fontsize=\small, bgcolor=subtlegray]{output}
To run nimplex please either (1) provide no arguments and follow the prompts or 
(2) use "-c" or "--config" to provide the configuration per instructions below:

- Provide the 3-letter configuration for task type:
    1. Grid type or uniform random sampling:
        - F: Full grid (including the simplex boundary)
        - I: Internal grid (only points inside the simplex)
        - R: Random/Monte Carlo uniform sampling over simplex.
        - G: Graph (list of grid nodes and list of their neighbors)
    2. Fractional or Integer positions:
        - F: Fractional grid/graph (points are normalized to fractions of 1)
        - I: Integer grid/graph (points are integers)
    3. Print full result, its shape, or persist in a file:
        - P: Print (presents full result as a table)
        - S: Shape (only the shape / size information)
        - N: Persist to NumPy array file ("nimplex_<configFlags>.npy" or 
             optionally a custom path as an additonal argument)

- Followed by integers of (1) simplex dimension and (2) number of divisions or
  samples depending on the task type. Optionally, custom output file path for 
  NumPy array can be provided as the last argument. E.g.:
    -c FFS [simplex dimension] [number of divisions]
    -c RFP [simplex dimension] [number of samples]
    -c FIN [simplex dimension] [number of divisions] [path/to/outfile.npy]

You can also utilize the following auxiliary flags:
--help       | -h   --> Show help.
--benchmark  | -b   --> Run benchmark for all tasks (9-dimensional space
                        with 12 divisions per dimension / 1M random samples).
\end{minted}

Based on the above, we can quickly generate a nice
\texttt{9}-component \texttt{I}nternal
simplex grid with \texttt{F}ractional with
\texttt{24} divisions per dimension and output it to
\texttt{N}numpy format file
\texttt{gridForJulia.npy}:

\begin{minted}[xleftmargin=3\parindent, linenos=true, fontsize=\small]{python}
!./nimplex --config IFN 9 24 gridForJulia.npy
\end{minted}

\begin{minted}[xleftmargin=3\parindent, fontsize=\small, bgcolor=subtlegray]{output}
Running with configuration:@["IFN", "9", "24", "gridForJulia.npy"]
Persisting to NumPy array file:gridForJulia.npy
Shape:[490314, 9]
\end{minted}

And wihinin a couple hundred miliseconds, we should get around 490,000
points of the grid neatly stored in the file of 36MB (on 64-bit system),
which we can now load it in Julia using

\begin{minted}[xleftmargin=3\parindent, fontsize=\small, bgcolor=subtlegray]{output}
using NPZ
data = npzread("gridForJulia.npy")
\end{minted}

In some cases, you may not have the ability to read the NumPy format,
because for instance, you are writing code for your favorite satellite
or train using a niche language like \href{https://ada-lang.io}{Ada}.
For situations like that, the backup option of formatter text output is
available. It is not nearly as efficient as the binary NumPy format, so
the grid from the last example would take minutes to generate, but it
should work with \emph{everything}.

Let's start by looking at a toy example. To get a small grid of
\texttt{3}-component \texttt{F}ull
simplex grid with \texttt{F}ractional with
\texttt{5} divisions per dimension, we can do:

\begin{minted}[xleftmargin=3\parindent, linenos=true, fontsize=\small]{python}
!./nimplex --config FFP 3 5
\end{minted}

\begin{minted}[xleftmargin=3\parindent, fontsize=\small, bgcolor=subtlegray]{output}
Running with configuration:@["FFP", "3", "5"]
Full Output:Tensor[system.float] of shape "[21, 3]" on backend "Cpu"
|0        0      1|
|0      0.2    0.8|
|0      0.4    0.6|
|0      0.6    0.4|
|0      0.8    0.2|
|0        1      0|
|0.2      0    0.8|
|0.2    0.2    0.6|
|0.2    0.4    0.4|
|0.2    0.6    0.2|
|0.2    0.8      0|
|0.4      0    0.6|
|0.4    0.2    0.4|
|0.4    0.4    0.2|
|0.4    0.6      0|
|0.6      0    0.4|
|0.6    0.2    0.2|
|0.6    0.4      0|
|0.8      0    0.2|
|0.8    0.2      0|
|1        0      0|
\end{minted}

As you can see, the first two lines of the output carry (1) task
metadata on what was run to create the grid, (2) output metadata
including its type(e.g.,
\texttt{int}/\texttt{float}/\texttt{float64}),
shape, and backend. Then the grid itself is printed out in the form of a
2D array with each row representing a single point and structured as
space-separated values between \texttt{|} characters.

To get a larger grid, we may want to stream the output to a file rather
than \texttt{P}rint it to the screen. We can do that by
redirecting the output to a file using \texttt{>}.
Let's create a \texttt{7}-component
\texttt{F}ull simplex grid with
\texttt{F}ractional with \texttt{12}
divisions per dimension and output it to
\texttt{gridForAda.txt}:

\begin{minted}[xleftmargin=3\parindent, linenos=true, fontsize=\small]{python}
!./nimplex --config FFP 7 12 > gridForAda.txt
\end{minted}

,which you can quickly read in Ada using rather elaborate code to parse
the file in strongly typed manner with some extra error checking. For
the sake of conciseness, it is not shown here, but the example below
shows how to do it in pure Python (using just built-in functions):

\begin{minted}[xleftmargin=3\parindent, linenos=true, fontsize=\small]{python}
with open('gridForAda.txt', 'r') as f:
    gridForAda = f.read()
    lines = gridForAda.replace('|', '').strip().split('\n')
    data = [[float(num) for num in line.split()] for line in lines[2:-1]]
\end{minted}

\begin{minted}[xleftmargin=3\parindent, linenos=true, fontsize=\small]{python}
data[:5]
\end{minted}

\begin{minted}[xleftmargin=3\parindent, fontsize=\small, bgcolor=subtlegray]{output}
[[0.0, 0.0, 0.0, 0.0, 0.0, 0.0, 1.0],
 [0.0, 0.0, 0.0, 0.0, 0.0, 0.0833333, 0.916667],
 [0.0, 0.0, 0.0, 0.0, 0.0, 0.166667, 0.833333],
 [0.0, 0.0, 0.0, 0.0, 0.0, 0.25, 0.75],
 [0.0, 0.0, 0.0, 0.0, 0.0, 0.333333, 0.666667]]
\end{minted}

\textbf{And now, you are ready to go and use nimplex in your favorite
language}

\textbf{In the second tutorial, in Appendix \ref{chap:nimplextutorial2} on 
Additive Manufacturing Path Planning, you will learn how to take advantage of the much more advanced graph construction functionality of nimplex to explore compositional spaces
and plan different traversal paths.}


\chapter{Nimplex Workshop No.2 - Additive Manufacturing Path Planning Made
Effortless} \label{chap:nimplextutorial2}

This is the second tutorial going over basics of use of the \texttt{nimplex} software. It goes over its core advantages in additive manufacturing and can be run on the cloud using pre-compiled virtual environment following the current Quick Start link at \href{https://nimplex.phaseslab.org}{nimplex.phaseslab.org}.

\section{Introduction}

In this tutorial, we will demonstrate how effortless it is to
dramatically speed up the exploration of feasible compositional spaces
in high dimensional spaces through employing
\texttt{nimplex}'s graph representations that abstract
the underlying problem and dimensionality.

We will also design several neat, mathematically optimal (given
some criteria) paths in a 7-component chemical space connecting two
alloys of interest by mixing 4 fixed-composition alloy powders to create
a tetrahedral attainable/design space. The beauty of this approach is
that at no point (except for plotting in 3D for ``human consumption'')
will we explicitly consider the dimensionality or the distance as the
connectivity between the points in the space has been abstracted into
graph adjacency. If you wish to add another alloy to the design process,
you add it to the list, and you are done!

\begin{minted}[xleftmargin=3\parindent, linenos=true, fontsize=\small]{python}
# Import nimplex and some of its plotting utilities.
import nimplex
from utils import plotting
\end{minted}

\begin{minted}[xleftmargin=3\parindent, linenos=true, fontsize=\small]{python}
# Python wrapper for Plotly library
import plotly.express as px
import plotly.io as pio
pio.renderers.default = 'pdf'
import pandas as pd
from pprint import pprint
\end{minted}

\section{Recall Last Example}\label{nimplextutorial2:recall-last-example}

Let's get back to our example (QuickStart) of the pair of
\textbf{attainable space} simplex grid and corresponding
\textbf{elemental space} positions defined for
\texttt{7}-component elemental space of
\texttt{Ti}, \texttt{Zr},
\texttt{Hf}, \texttt{W},
\texttt{Nb}, \texttt{Ta}, and
\texttt{Mo} formed by \texttt{4}
alloys: - Ti50 Zr50 - Hf95 Ti5 - Mo33 Nb33 Ta33 - Mo10 Nb10 W80

which we can represent as points in the elemental space:

\begin{minted}[xleftmargin=3\parindent, linenos=true, fontsize=\small]{python}
elementalSpaceComponents = 
  ["Ti", "Zr", "Hf", "W", "Nb", "Ta", "Mo"]
attainableSpaceComponents = 
  ["Ti50 Zr50", "Hf95 Ti5", "Mo33 Nb33 Ta33", "Mo80 Nb10 W10"]
attainableSpaceComponentPositions = 
  [[50, 50, 0, 0, 0, 0, 0], [5, 0, 95, 0, 0, 0, 0], 
  [0, 0, 0, 33, 33, 33, 0], [0, 0, 0, 10, 10, 0, 80]]
\end{minted}

And create tetrahedral grids with their compositions quantized at
\texttt{12} divisions per dimension.

\begin{minted}[xleftmargin=3\parindent, linenos=true, fontsize=\small]{python}
gridAtt, gridEl =
  nimplex.embeddedpair_simplex_grid_fractional_py(
    attainableSpaceComponentPositions, 12)
\end{minted}

We used the \textbf{elemental}, or chemical, space to run the Root Mean
Square Atomic Displacement (RMSAD) model by Tandoc
(10.1038/s41524-023-00993-x) which acts as a lower-cost proxy for yield
stress and hardness estimations in the absence of direct data:

\begin{minted}[xleftmargin=3\parindent, linenos=true, fontsize=\small]{python}
import pqam_rmsadtandoc2023
rmsadList = []
for point in gridEl:
    formula = ' '.join(
      [f'{c}{p}' for c, p in zip(elementalSpaceComponents, point)])
    rmsadList.append(pqam_rmsadtandoc2023.predict(formula))
\end{minted}

And the \textbf{attainable} space grid to plot the results after
projecting them to the Euclidean space:

\begin{minted}[xleftmargin=3\parindent, linenos=true, fontsize=\small]{python}
# Hover approximate formula for each point
formulas = []
for i, comp in enumerate(gridEl):
    formulas.append(f"({i:>3}) "+"".join(
      [f"{el}{100*v:.1f} " if v>0 else "" for 
        el, v in zip(elementalSpaceComponents, comp)]))

# Generate the projected grid
gridAtt_projected_df = 
  pd.DataFrame(plotting.simplex2cartesian_py(gridAtt), columns=['x','y','z'])

# Attach pure component (alloy) labels to corners
pureComponentIndices = nimplex.pure_component_indexes_py(4, 12)
labels = ['']*len(gridAtt_projected_df)
for comp, idx in zip(attainableSpaceComponents, pureComponentIndices):
    labels[idx] = "<b>"+comp+"</b>"

# Add text labels at the corners of the simplex
px.scatter_3d(
  gridAtt_projected_df, x='x', y='y', z='z', 
  color=rmsadList, text=labels, hover_name=formulas,
  template='plotly_white', width=800, height=700, 
  labels={'color':'RMSAD', 'x':'', 'y':'', 'z':''})
\end{minted}

\begin{figure}[H]
    \centering
    \includegraphics[width=0.8\textwidth]{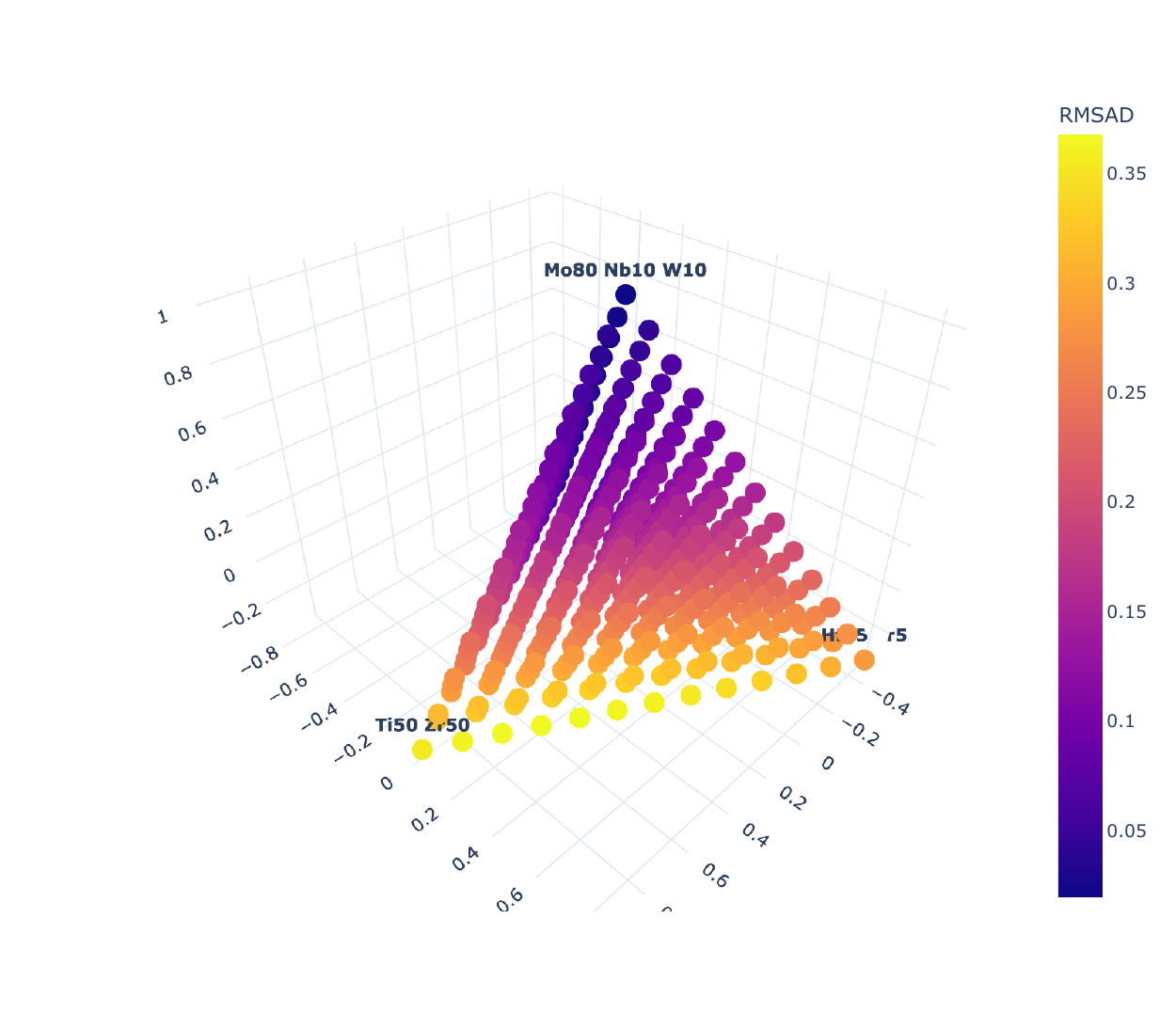}
    \caption{Re-rendering of Figure \ref{nimplextutorial1:fig:quaternaryattainablecoloredlabel} from Appendix \ref{chap:nimplextutorial1}.}
    \label{nimplextutorial2:fig:propertyfield}
\end{figure}

\section{Thermodynamic Equilibria}\label{nimplextutorial2:thermodynamic-equilibria}

\textbf{In this section we will be using
\texttt{pycalphad} which is an amazing free open-source
library for building thermodynamic models and performing calculations on
them, including studying of phase equilibria, or what atomic
arrangements are stable (coexist in equilibrium) at a given temperature
and composition.} You can read more about
\texttt{pycalphad} at
\href{https://pycalphad.org/}{pycalphad.org} and find an great tutorial
at
\href{https://github.com/materialsgenomefoundation/2023-workshop-material/tree/main/pycalphad}{this
2023 workshop materials repository}.

Let's start by loading a database of thermodynamic properties which
defines, among other things, the Gibbs energy of each phase as a
function of temperature and composition. You can find many of such
databases at \href{https://avdwgroup.engin.brown.edu}{TDBDB} maintained
by Axel van de Walle's group at Brown University.

In this example, we will be using a TDB
\texttt{CrHfMoNbTaTiVWZr\_9element\_Feb2023} placed for
you in the \texttt{examples} directory, which is a
9-element database for the elements \texttt{Cr},
\texttt{Hf}, \texttt{Mo},
\texttt{Nb}, \texttt{Ta},
\texttt{Ti}, \texttt{V},
\texttt{W}, and \texttt{Zr}, being
developed by Shuang Lin in our group (Phases Research Lab at PSU). This
wersion is an older \emph{work in progress} one that has not been
published, so while it is perfect for tutorial like this one, please
refrain from using it for any serious work.

\begin{minted}[xleftmargin=3\parindent, linenos=true, fontsize=\small]{python}
from pycalphad import Database
dbf = Database("CrHfMoNbTaTiVWZr_9element_Feb2023.tdb")
phases = list(set(dbf.phases.keys()))
print(elementalSpaceComponents)
print(f'Loaded TDB file with phases considered: {phases}')
\end{minted}

\begin{minted}[xleftmargin=3\parindent, fontsize=\small, bgcolor=subtlegray]{output}
['Ti', 'Zr', 'Hf', 'W', 'Nb', 'Ta', 'Mo']
Loaded TDB file with phases considered: 
['LAVES_C14', 'BCC_A2', 'HCP_A3', 'LAVES_C15', 'LIQUID', 'LAVES_C36', 'FCC_A1']
\end{minted}

As you can see, we will be looking at several different phases here. To
keep things as simple as possible, we can split them in three groups: -
\textbf{liquid} phases: \texttt{LIQUID} (which we
obviously want to avoid in a solid part design) - \textbf{solid
solution} phases: \texttt{FCC\_A1},
\texttt{BCC\_A2}, and \texttt{HCP\_A3}
- \textbf{intermetallic} phases: \texttt{LAVES\_C14},
\texttt{LAVES\_C15}, and
\texttt{LAVES\_C36}

Designing an alloy is a complex procedure but for the sake of this
tutorial, we will be focusing on the most common issue which is the
formation of intermetallic phases which cause embrittlement and reduce
the ductility of the material. \textbf{Thus, we will apply a simple
constraint that the alloy should only contain the solid solution phases
FCC, BCC, and HCP.}

\textbf{Now, knowing what we are looking for (phases at equilibrium), we
can start by writing a short Python script around
\texttt{pycalphad} to calculate the equilibrium phases
for a given chemical elements composition \texttt{elP}.
It is already placed in the \texttt{examples} directory
as \texttt{myPycalphadCallable.py} which defines a
function \texttt{equilibrium\_callable} that takes a
composition and returns the equilibrium phases.}

\textbf{We will arbitrarily pick \texttt{1000K}} as the
temperature for the sake of this tutorial, but you can change it to any
other value you like. Or even make it a list and add phases present at
each temperature to a set to apply our constraint over a range of
temperatures.

Please note that much more information is generated in the process
(e.g., chemical composition of each phase and its fraction) but we are
only interested in the phase presence. If you wish to do so, modifying
the script to, e.g., allow for up to 5\% of intermetallic phases, is a
trivial task. Advanced users may also want to have a look at the
\texttt{scheil\_callable} we do not use in this
tutorial for the sake of runtime, but which can be used to simulate
solidification of the alloy from a liquid state in an additive
manufacturing process.

\begin{minted}[xleftmargin=3\parindent, linenos=true, fontsize=\small]{python}
from myPycalphadCallable import equilibrium_callable
\end{minted}

Let's test it on some composition in our space starting with the first
point!

\begin{minted}[xleftmargin=3\parindent, linenos=true, fontsize=\small]{python}
print(formulas[0])
equilibrium_callable(gridEl[0])
\end{minted}

\begin{minted}[xleftmargin=3\parindent, fontsize=\small, bgcolor=subtlegray]{output}
(  0) W10.0 Nb10.0 Mo80.0 
['BCC_A2']
\end{minted}

You should see \texttt{['BCC\_A2']} in a second or so
if you've run it at the default \texttt{1000K}. Quick
and neat, right? Now, let's pick some compositionally complex alloy that
does not lay around the corner of the attainable space tetrahedron and
presents an actual challenge.

\begin{minted}[xleftmargin=3\parindent, linenos=true, fontsize=\small]{python}
print(formulas[63])
equilibrium_callable(gridEl[63])
\end{minted}

\begin{minted}[xleftmargin=3\parindent, fontsize=\small, bgcolor=subtlegray]{output}
( 63) Ti2.5 Hf47.5 W5.0 Nb5.0 Mo40.0 
['HCP_A3', 'LAVES_C15', 'BCC_A2']
\end{minted}

Now, you should have seen an example of infeasible point composed of
\texttt{['HCP\_A3', 'LAVES\_C15', 'BCC\_A2']}. Let's
deploy this in parallel over all the points in the elemental space
\texttt{gridEl} and see how it looks like! We will use
the \texttt{process\_map} function from the
\texttt{tqdm} library to show a neat progress bar while
the calculations are running in parallel. On the 4-core Codespaces VM
you can expect it to take around 2-3 minutes.

\begin{minted}[xleftmargin=3\parindent, linenos=true, fontsize=\small]{python}
from tqdm import tqdm
from tqdm.contrib.concurrent import process_map
\end{minted}

\begin{minted}[xleftmargin=3\parindent, linenos=true, fontsize=\small]{python}
gridPhases = process_map(equilibrium_callable, gridEl)
\end{minted}

\begin{minted}[xleftmargin=3\parindent, fontsize=\small, bgcolor=subtlegray]{output}
  0%|          | 0/455 [00:00<?, ?it/s]
\end{minted}

Let's see how some of the data looks like.

\begin{minted}[xleftmargin=3\parindent, linenos=true, fontsize=\small]{python}
gridPhases[120:130]
\end{minted}

\begin{minted}[xleftmargin=3\parindent, fontsize=\small, bgcolor=subtlegray]{output}
[['HCP_A3', 'BCC_A2', 'LAVES_C15'],
 ['HCP_A3', 'BCC_A2'],
 ['HCP_A3', 'BCC_A2'],
 [],
 ['LAVES_C15', 'BCC_A2'],
 ['LAVES_C15', 'BCC_A2', 'HCP_A3'],
 ['HCP_A3', 'LAVES_C15', 'BCC_A2'],
 ['LAVES_C15', 'HCP_A3', 'BCC_A2'],
 ['HCP_A3', 'BCC_A2', 'LAVES_C15'],
 ['HCP_A3', 'BCC_A2', 'LAVES_C15']]
\end{minted}

Now, let's turn that list of phases into a list of feasibility based on
the constraint we defined earlier. Note that in some cases, the
\texttt{pycalphad} library may return an empty list of
phases, which we will treat as infeasible.

\begin{minted}[breaklines, xleftmargin=3\parindent, linenos=true, fontsize=\small]{python}
gridFeasible = 
  [len(set(p) & set(['LAVES_C15', 'LAVES_C36', 'LAVES_C14', 'LIQUID']))==0 and p!=[] for p in gridPhases]
gridFeasible[120:130]
\end{minted}

\begin{minted}[xleftmargin=3\parindent, fontsize=\small, bgcolor=subtlegray]{output}
[False, True, True, False, False, False, False, False, False, False]
\end{minted}

Finally, let's plot the result in 3D using the
\texttt{plotly} library and our spacial-transformed
attainable space grid we obtained with
\texttt{plotting.simplex2cartesian\_py(gridAtt)}
earlier.

\textbf{Once you run the cell below, you should be seeing an interactive
3D plot with 455 split roughly 50/50 between feasible and infeasible
points. You can rotate the plot, zoom in and out, and hover over the
points to see their composition and feasibility. You can also click on
the legend to hide/show the points based on their feasibility.}

\begin{minted}[xleftmargin=3\parindent, linenos=true, fontsize=\small]{python}
fig = px.scatter_3d(
  gridAtt_projected_df, x='x', y='y', z='z', color=gridFeasible, 
  text=labels, hover_name=formulas, template='plotly_white', 
  width=800, height=700, opacity=0.333, 
  color_discrete_sequence=['green', 'red'], 
  labels={'color':'Solid Solution Phases', 'x':'', 'y':'', 'z':''})
fig.update_scenes({'camera': {'eye': {'x': -2.3, 'y': 0.2, 'z': 0.2}}})
\end{minted}

\begin{figure}[H]
    \centering
    \includegraphics[width=0.8\textwidth]{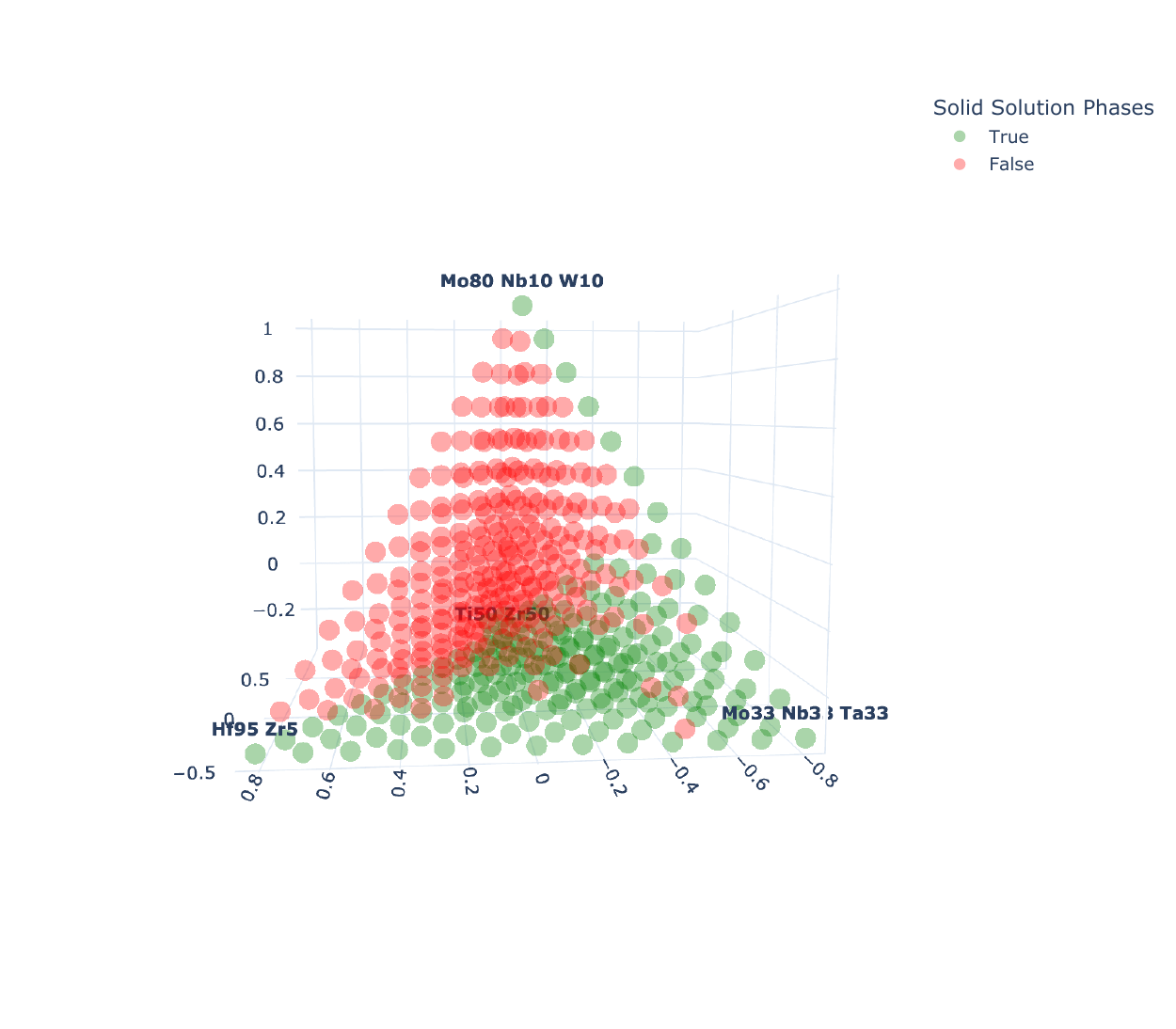}
    \caption{Feasibility map constrained by limiting phases present at equilibrium at 1000K to single or many solid solution phases demonstrating roughly half of the system to be infeasible.}
    \label{nimplextutorial2:fig:fullfeasibility}
\end{figure}

\section{Graph Walk aka Infeasibility Gliding (Exciting
Stuff!)}\label{nimplextutorial2:graph-walk-aka-infeasibility-gliding-exciting-stuff}

\textbf{While the last two sections highlighted the elegant abstraction
of grid generation and space transformation using
\texttt{nimplex}, they did not really show the true
power of the library or the novel combinatorics-based algorithm we
developed in the associated paper. That is the simplex graph
construction in arbitrary dimensions and the graph traversal it
enables.}

\textbf{Let's start by generating the graph for our problem. We can use
the function called
\texttt{nimplex.embeddedpair\_simplex\_graph\_fractional\_py}, which also first generates the two grids we already used in
the previous sections.}

\begin{minted}[xleftmargin=3\parindent, linenos=true, fontsize=\small]{python}
_, _, graphN = 
  nimplex.embeddedpair_simplex_graph_fractional_py(
    attainableSpaceComponentPositions, 12)
\end{minted}

As you can see, we immediately obtain list of adjecent nodes
(compostions) for each node in the graph, with few neighbors for the
corners and many for the interior points. This is the power of the
abstraction we are talking about!

\begin{minted}[xleftmargin=3\parindent, linenos=true, fontsize=\small]{python}
graphN[:5]
\end{minted}

\begin{minted}[xleftmargin=3\parindent, fontsize=\small, bgcolor=subtlegray]{output}
[[1, 13, 91],
 [0, 2, 13, 14, 91, 92],
 [1, 3, 14, 15, 92, 93],
 [2, 4, 15, 16, 93, 94],
 [3, 5, 16, 17, 94, 95]]
\end{minted}

\begin{minted}[xleftmargin=3\parindent, linenos=true, fontsize=\small]{python}
graphN[200:205]
\end{minted}

\begin{minted}[xleftmargin=3\parindent, fontsize=\small, bgcolor=subtlegray]{output}
[[134, 126, 125, 192, 191, 199, 201, 207, 208, 255, 262, 263],
 [135, 127, 126, 193, 192, 200, 202, 208, 209, 256, 263, 264],
 [136, 128, 127, 194, 193, 201, 203, 209, 210, 257, 264, 265],
 [137, 129, 128, 195, 194, 202, 204, 210, 211, 258, 265, 266],
 [138, 130, 129, 196, 195, 203, 205, 211, 212, 259, 266, 267]]
\end{minted}

We can now use this graph to do a lot of things. Let's start by
considering that \textbf{the infeasible region, for thermodynamic
reasons beyond the scope of this tutorial, are generally difficult to
predict but are generally bound by continous surfaces called phase
boundaries.}

\textbf{Thus, if we iteratively traverse the graph expanding only the
feasible nodes, while noting infeasible nodes, we can glide along the
phase boundaries and explore the feasible space without ever wasting our
resources on calculations inside the insides of the infeasible regions.}

Let's do this starting from the
\texttt{W10.0 Nb10.0 Mo80.0} and terminating when we
reach the \texttt{Hf95.0 Ti5.0} point.

\textbf{Bonus Exercises:} - Please note that you do not need to specify
the termination point, as the algorithm will stop when it completeley
explores the feasible space. In this example, we are just showing you
how to specify it and it has no effect on the result. If you want to
play a bit, you can set \texttt{endNode} to
\texttt{12} to see how path to
\texttt{W33.3 Nb33.3 Ta33.3} gets explored in just 87
nodes. - If you remove the \texttt{endNode} termination
(change \texttt{break} to
\texttt{pass}), you can add the endNode (or any other
node you believe is feasible) to the initial
\texttt{queue} list and see how the algorithm explores
the feasible space from multiple starting points much faster thanks to
better parallelization.

\begin{minted}[xleftmargin=3\parindent, linenos=true, fontsize=\small]{python}
startingNode = 0
endNode = 90

print(f"Starting node: {formulas[startingNode]}")
print(f"Ending node: {formulas[endNode]}")
\end{minted}

\begin{minted}[xleftmargin=3\parindent, fontsize=\small]{output}
Starting node: (  0) W10.0 Nb10.0 Mo80.0 
Ending node: ( 90) Ti5.0 Hf95.0 
\end{minted}

\begin{minted}[xleftmargin=3\parindent, linenos=true, fontsize=\small]{python}
gridFeasible = [None]*len(graphN)
queue = [startingNode]
explored = set()
calcCount = 0
\end{minted}

\textbf{Now, we will be using a \texttt{queue} of nodes
to keep track of the nodes we need to visit and a set of visited nodes
to avoid revisiting them. This simple procedure is a type of
\texttt{depth-first search} algorithm that is
guaranteed to find the shortest path between two points in a graph if it
exists, while exploring the feasible space in the process in an unbiased
way. In a more elaborate problem, you would likely want to implement a
priority queue to explore the space in a more efficient way, but for
this tutorial, this is more than enough and allows for better, more
direct comparions with the typical complete exploration approach.} On
the 4-core Codespaces VM you can expect it to take around 1-2 minutes.
In a relatively simple problem like this one, the difference between the
two approaches is not that big, coming partially from overhead of
limited Python parallelization capabilities, but in a more complex
problem, the difference can be dramatic.

\begin{minted}[breaklines, xleftmargin=3\parindent, linenos=true, fontsize=\small]{python}
while len(queue)>0:
    print(f"Queue: {queue}")
    # Assign feasibilities to the current queue
    elPositions = [gridEl[i] for i in queue]
    if len(queue)>3:
        phases = process_map(
          equilibrium_callable, elPositions, max_workers=4)
    else:
        phases = [equilibrium_callable(elP) for elP in elPositions]
    feasibilities = [len(set(p) & set(['LAVES_C15', 'LAVES_C36', 'LAVES_C14', 'LIQUID']))==0 and p!=[] for p in phases]

    calcCount += len(feasibilities)
    explored = explored.union(queue)

    # Create next queue based on neighbors of feasible points
    nextQueue = set()
    for f, i in zip(feasibilities, queue):
        gridFeasible[i] = f
        # Only if feasible
        if f:
            for n in graphN[i]:
                if n not in explored:
                    nextQueue.add(n)

    # Early termination criteria if we just evaluated the target
    if endNode in queue:
        break

    print(f"Calculations done: {calcCount}")
    queue = list(nextQueue)
\end{minted}

\begin{minted}[xleftmargin=3\parindent, fontsize=\small, bgcolor=subtlegray]{output}
Queue: [0]
Calculations done: 1
Queue: [1, 91, 13]
Calculations done: 4
Queue: [2, 92, 14]
...
Calculations done: 87
Queue: [129, 132, 261, 138, 139, 266, 267, 268, 140, 12, 24, 35, 42, 298, 44, 
45, 306, 179, 52, 53, 311, 312, 313, 189, 198, 203, 206, 211, 212, 213, 342, 
346, 347, 348, 349, 102, 113, 244, 374, 375, 376, 377, 123, 253]
  0%|          | 0/44 [00:00<?, ?it/s]
Calculations done: 131
Queue: [273, 146, 147, 274, 401, 402, 403, 404, 54, 310, 317, 62, 61, 318, 319, 
218, 219, 353, 354, 355, 373, 380, 381, 382, 383]
  0%|          | 0/25 [00:00<?, ?it/s]
Calculations done: 156
...
Queue: [396, 398, 287, 289, 417, 419, 164, 167, 434, 444, 450, 453, 454, 332, 
334, 83, 87, 232, 234, 368, 370]
  0%|          | 0/21 [00:00<?, ?it/s]
Calculations done: 277
Queue: [168, 89, 166, 86]
  0%|          | 0/4 [00:00<?, ?it/s]
Calculations done: 281
Queue: [88, 90]
\end{minted}

\textbf{You should now see that only 281 phase equilibria calculations
were performed to get all the feasible points! That's only a bit more
than half of what we had to do in the previous section. Let's plot the
path we found in the 3D space.} If you rotate the figure, you can
clearly see the path gliding along the infeasible space boundary.

\begin{minted}[breaklines, xleftmargin=3\parindent, linenos=true, fontsize=\small]{python}
fig = px.scatter_3d(
  gridAtt_projected_df, x='x', y='y', z='z', color=gridFeasible, 
  text=labels, hover_name=formulas, template='plotly_white', 
  width=800, height=700, opacity=0.333, 
  color_discrete_sequence=['green', 'red'], 
  labels={'color':'Solid Solution Phases', 'x':'', 'y':'', 'z':''})
fig.update_scenes({'camera': {'eye': {'x': -2.3, 'y': 0.2, 'z': 0.2}}})
\end{minted}

\begin{figure}[H]
    \centering
    \includegraphics[width=0.8\textwidth]{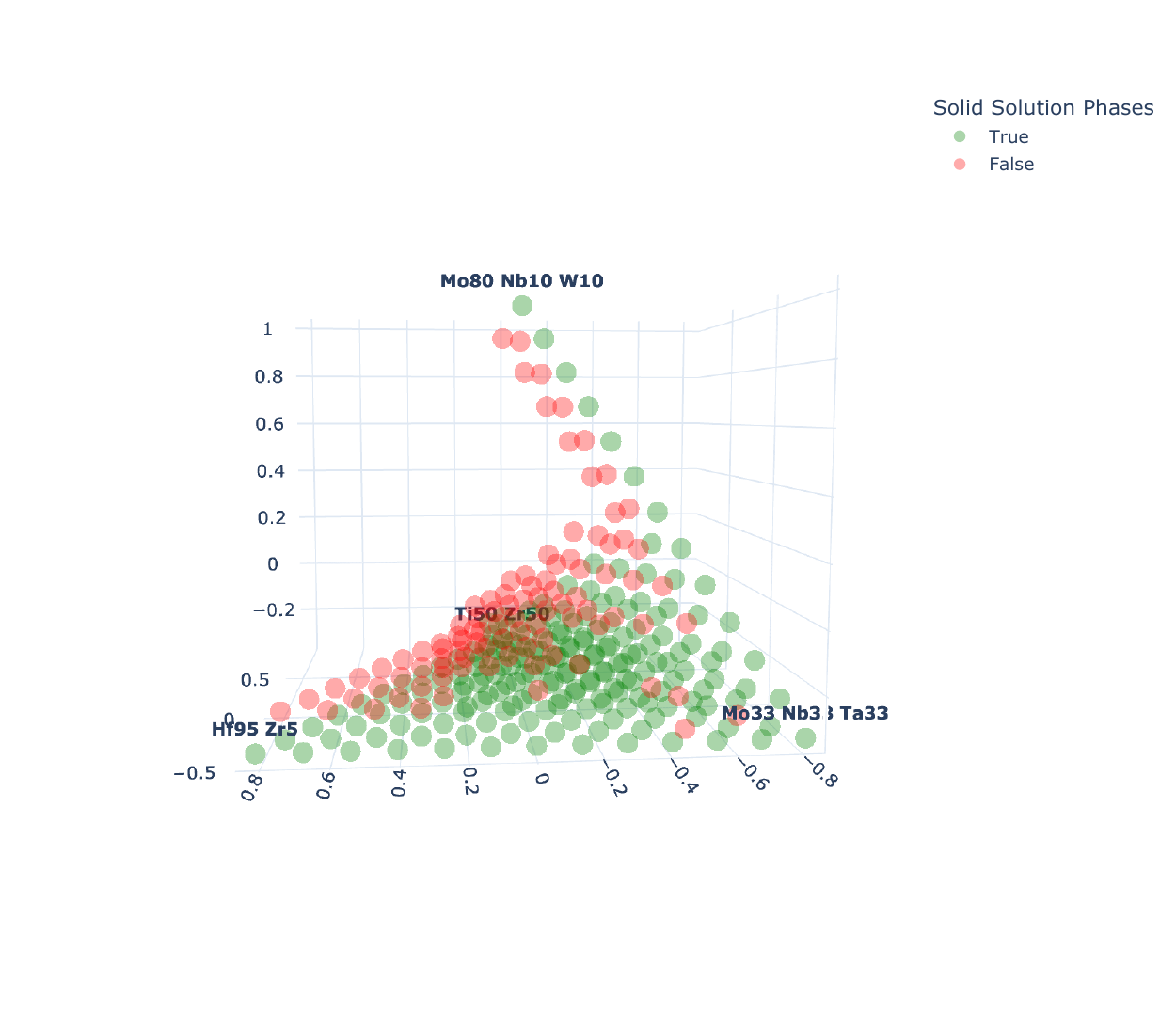}
    \caption{Feasibility map created by gliding over the infeasible region boundary, efficiently calculating all feasible points at the minimum number of evaluations.}
    \label{nimplextutorial2:fig:infeasibilityglide}
\end{figure}

\section{Path Finding}\label{nimplextutorial2:path-finding}

\textbf{Now, let's take the appeal of graph representation a few steps
forward! Since they are very general and widely applicable, there are
countless off-the-shelf high-performance libraries implementing
mathematically optimal path finding algorithms on graphs one can deploy
on challenging problems.}

In this example, for the sake of simplicity, we will use pure-Python
\texttt{pathfinding} library which implements some of
the most popular path finding algorithms without high performance (we do
not need here) but also without any dependencies. In real applications,
you would likely want to use a more advanced library like
\texttt{networkx} which is more efficient and have more
features, or write one yourself. Please note that
\texttt{nimplex}'s CLI can generate massive graphs in
native NumPy format you may find useful for such applications (e.g., to
use Julia or C++ libraries).

\begin{minted}[xleftmargin=3\parindent, linenos=true]{python}
from pathfinding.core.graph import Graph
from pathfinding.finder.dijkstra import DijkstraFinder
\end{minted}

Let's now turn our graph definition into a format that
\texttt{pathfinding} library can understand, assigning
equal cost to all edges. \textbf{Note, that we only need to consider the
feasible points leading to other feasible point}

\begin{minted}[breaklines, xleftmargin=3\parindent, linenos=true, fontsize=\small]{python}
edges = []
for i, nList in enumerate(graphN):
    if gridFeasible[i]:
        for n in nList:
            if gridFeasible[n]:
                edges.append([i, n, 1])
print(edges[:5])
\end{minted}

\begin{minted}[xleftmargin=3\parindent, fontsize=\small, bgcolor=subtlegray]{output}
[[0, 1, 1], [1, 0, 1], [1, 2, 1], [2, 1, 1], [2, 3, 1]]
\end{minted}

Let's initialize the \texttt{Graph} object.

\begin{minted}[breaklines, xleftmargin=3\parindent, linenos=true, fontsize=\small]{python}
pathfindingGraph = Graph(edges=edges, bi_directional=False)
\end{minted}

Now, let's create the \texttt{Finder} object that will
perform the path finding. Here we use the
\texttt{DijkstraFinder} which is a very popular
algorithm for path finding in graphs (see
\href{https://en.wikipedia.org/wiki/Dijkstra's_algorithm}{it's Wikipedia
page} for more details). It \textbf{finds the shortest possible path
between two points in the graph}, which is exactly what we need here. In
the future tutorials, we will use some more advanced algorithms, like
A*, which accomplishes the same task but in the mathematcally minimal
number of evaluations.

\begin{minted}[breaklines, xleftmargin=3\parindent, linenos=true, fontsize=\small]{python}
finder = DijkstraFinder()
\end{minted}

For hundreds of points, this shouldn't take more than tenth of a second
or so.

\begin{minted}[breaklines, xleftmargin=3\parindent, linenos=true, fontsize=\small]{python}
path, runs = finder.find_path(
    pathfindingGraph.node(0), 
    pathfindingGraph.node(90), 
    pathfindingGraph)
\end{minted}

Let's now turn the path into a list of nodes on the path.

\begin{minted}[breaklines, xleftmargin=3\parindent, linenos=true, fontsize=\small]{python}
shortestPath = [p.node_id for p in path]
\end{minted}

And plot corresponding compositions. As you can see, we immediately get
a neat list of compositions that we can use to design our additive
manufacturing path.

\begin{minted}[breaklines, xleftmargin=3\parindent, linenos=true, fontsize=\small]{python}
for step, i in enumerate(shortestPath):
    print(f"{step+1:>2}: {formulas[i]}")
\end{minted}

\begin{minted}[xleftmargin=3\parindent, fontsize=\small, bgcolor=subtlegray]{output}
 1: (  0) W10.0 Nb10.0 Mo80.0 
 2: (  1) W11.9 Nb11.9 Ta2.8 Mo73.3 
 3: (  2) W13.9 Nb13.9 Ta5.6 Mo66.7 
 4: (  3) W15.8 Nb15.8 Ta8.3 Mo60.0 
 5: (  4) W17.8 Nb17.8 Ta11.1 Mo53.3 
 6: (  5) W19.7 Nb19.7 Ta13.9 Mo46.7 
 7: (  6) W21.7 Nb21.7 Ta16.7 Mo40.0 
 8: (  7) W23.6 Nb23.6 Ta19.4 Mo33.3 
 9: ( 98) Ti4.2 Zr4.2 W22.8 Nb22.8 Ta19.4 Mo26.7 
10: (110) Ti4.6 Zr4.2 Hf7.9 W21.9 Nb21.9 Ta19.4 Mo20.0 
11: (121) Ti5.0 Zr4.2 Hf15.8 W21.1 Nb21.1 Ta19.4 Mo13.3 
12: ( 43) Ti1.2 Hf23.8 W21.1 Nb21.1 Ta19.4 Mo13.3 
13: ( 53) Ti1.7 Hf31.7 W20.3 Nb20.3 Ta19.4 Mo6.7 
14: ( 61) Ti2.1 Hf39.6 W17.5 Nb17.5 Ta16.7 Mo6.7 
15: ( 68) Ti2.5 Hf47.5 W14.7 Nb14.7 Ta13.9 Mo6.7 
16: ( 74) Ti2.9 Hf55.4 W11.9 Nb11.9 Ta11.1 Mo6.7 
17: ( 80) Ti3.3 Hf63.3 W11.1 Nb11.1 Ta11.1 
18: ( 84) Ti3.8 Hf71.2 W8.3 Nb8.3 Ta8.3 
19: ( 87) Ti4.2 Hf79.2 W5.6 Nb5.6 Ta5.6 
20: ( 89) Ti4.6 Hf87.1 W2.8 Nb2.8 Ta2.8 
21: ( 90) Ti5.0 Hf95.0 
\end{minted}

\textbf{Even better, since we know the positions in the attainable
design space, we can get a list of exact quantized instructions (e.g.,
powder flow rates) for the additive manufacturing apparatus to follow at
each step.}

\textbf{This is a very powerful abstraction, eliminating possible
mistakes and miscommunications between design and manufacturing teams.
Furthermore, it also allows for easy manual (or mechanized) per-layer
mixing of powders to create compositional paths in single-feed additive
manufacturing systems.}

\begin{minted}[breaklines, xleftmargin=3\parindent, linenos=true, fontsize=\small]{python}
integerGrid = nimplex.simplex_grid_py(4, 12)
for step, i in enumerate(shortestPath):
    print(f"{step+1:>2}: {integerGrid[i]}")
\end{minted}

\begin{minted}[xleftmargin=3\parindent, fontsize=\small, bgcolor=subtlegray]{output}
 1: [0, 0, 0, 12]
 2: [0, 0, 1, 11]
 3: [0, 0, 2, 10]
 4: [0, 0, 3, 9]
 5: [0, 0, 4, 8]
 6: [0, 0, 5, 7]
 7: [0, 0, 6, 6]
 8: [0, 0, 7, 5]
 9: [1, 0, 7, 4]
10: [1, 1, 7, 3]
11: [1, 2, 7, 2]
12: [0, 3, 7, 2]
13: [0, 4, 7, 1]
14: [0, 5, 6, 1]
15: [0, 6, 5, 1]
16: [0, 7, 4, 1]
17: [0, 8, 4, 0]
18: [0, 9, 3, 0]
19: [0, 10, 2, 0]
20: [0, 11, 1, 0]
21: [0, 12, 0, 0]
\end{minted}

We can also plot the path in 3D like we did before! As you can see, it
\textbf{glides through the inside of the tetrahedron in a shortest path}
avoiding an obstacle extending far on the face.

\begin{minted}[breaklines, xleftmargin=3\parindent, linenos=true, fontsize=\small]{python}
gridFeasibleMarked = ['path' if i in shortestPath else f for i, f in enumerate(gridFeasible)]
fig = px.scatter_3d(
  gridAtt_projected_df, x='x', y='y', z='z', color=gridFeasibleMarked, 
  text=labels, hover_name=formulas, template='plotly_white', width=800, 
  height=700, opacity=0.333, color_discrete_sequence=['blue', 'green', 'red'],
  labels={'color':'Solid Solution Phases', 'x':'', 'y':'', 'z':''})
fig.update_scenes({'camera': {'eye': {'x': -1.8, 'y': 1.2, 'z': 1.5}}})
\end{minted}

\begin{figure}[H]
    \centering
    \includegraphics[width=0.8\textwidth]{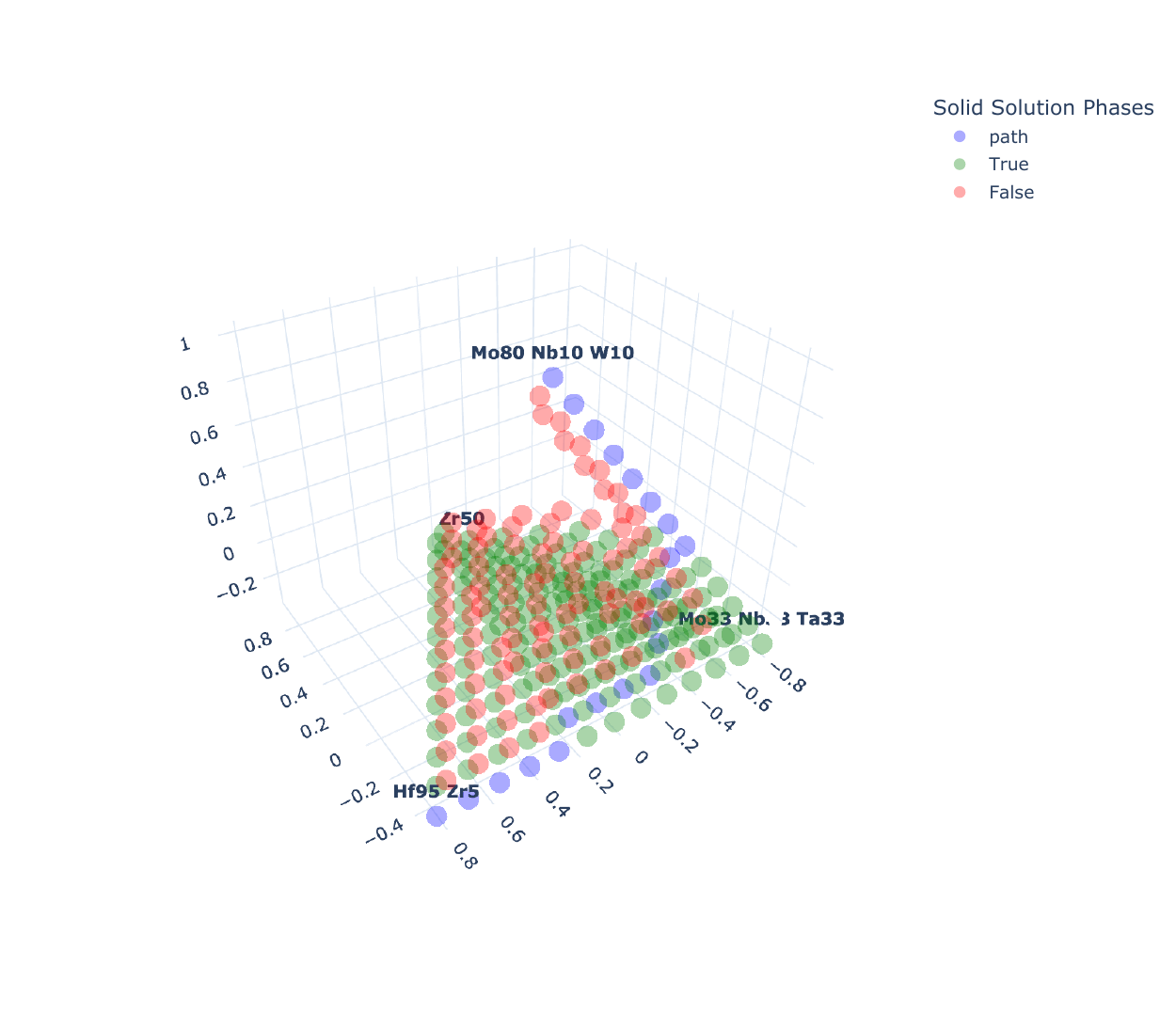}
    \caption{An optimal shortest path between W10 Nb10 Mo80 and Ti5 Hf95 connecting them in 21 steps.}
    \label{nimplextutorial2:fig:shortest}
\end{figure}

\section{Path Planning considering the Property
Gradient}\label{nimplextutorial2:path-planning-considering-the-property-gradient}

\textbf{Let's go one step further to demonstrate the beaty of the graph
representations in the context of efortless path planning in
high-dimensional spaces by including the gradient into consideration in
just a line or two of code. We will now stretch the space}

\begin{minted}[breaklines, xleftmargin=3\parindent, linenos=true, fontsize=\small]{python}
edges = []
gradientsList = []
penaltyFactor = 100
for i, nList in enumerate(graphN):
    if gridFeasible[i]:
        for n in nList:
            if gridFeasible[n]:
                rmsadGradient = abs(rmsadList[i]-rmsadList[n])
                gradientsList.append(rmsadGradient)
                edges.append([i, n, 1+round(rmsadGradient*penaltyFactor, 3)])
print(edges[:5])
print(f"{len(gradientsList)} gradients calculated with absolute value range between {min(gradientsList)} to {max(gradientsList)}")
print(f"Penalty at {penaltyFactor}x gradient *magnitude* per path step assigned.")
\end{minted}

\begin{minted}[xleftmargin=3\parindent]{output}
[[0, 1, 1.2770000000000001], [1, 0, 1.2770000000000001], [1, 2, 1.232], 
[2, 1, 1.232], [2, 3, 1.208]]
1374 gradients calculated with absolute value range between 
8.898859123501746e-06 to 0.0473823732310002
Penalty at 100x gradient *magnitude* per path step assigned.
\end{minted}

\textbf{As you can see, we just increased the predicted cost of moving
between nodes by a small factor ranging from near 0 to 2.5 depending on
the RMSAD gradient. In another words, we are prioritizing paths that
change RMSAD the least over the entire path.} This can be done, e.g., to
minimize the inter-layer stresses in the component caused by yield
stress mismatch between the layers that may cause delamination. Similar
considerations can also be made for other properties, like thermal
expansion coefficient, missmatch of which causes internal stress or
wrapping.

\textbf{Since the RMSAD is generally smoothly changing, we can expect
that this will not impact the result in terms of number of nodes. Just
finding a better path of equivalent length. However, if you try to
square the gradient by changing penalty to
\texttt{(rmsadGradient*penaltyFactor)**2}, you will see
how viciously the path will avoid any RMSAD changes through zig-zag
patterns.}

\begin{minted}[breaklines, xleftmargin=3\parindent, linenos=true, fontsize=\small]{python}
pathfindingGraph = Graph(edges=edges, bi_directional=False)
finder = DijkstraFinder()
path, runs = finder.find_path(
    pathfindingGraph.node(0), 
    pathfindingGraph.node(90), 
    pathfindingGraph)
lowGradientPath = [p.node_id for p in path]
\end{minted}

\begin{minted}[breaklines, xleftmargin=3\parindent, linenos=true, fontsize=\small]{python}
for step, i in enumerate(lowGradientPath):
    print(f"{step+1:>2}: {formulas[i]}")
\end{minted}

\begin{minted}[xleftmargin=3\parindent, fontsize=\small, bgcolor=subtlegray]{output}
 1: (  0) W10.0 Nb10.0 Mo80.0 
 2: (  1) W11.9 Nb11.9 Ta2.8 Mo73.3 
 3: (  2) W13.9 Nb13.9 Ta5.6 Mo66.7 
 4: (  3) W15.8 Nb15.8 Ta8.3 Mo60.0 
 5: (  4) W17.8 Nb17.8 Ta11.1 Mo53.3 
 6: (  5) W19.7 Nb19.7 Ta13.9 Mo46.7 
 7: (  6) W21.7 Nb21.7 Ta16.7 Mo40.0 
 8: ( 97) Ti4.2 Zr4.2 W20.8 Nb20.8 Ta16.7 Mo33.3 
 9: (175) Ti8.3 Zr8.3 W20.0 Nb20.0 Ta16.7 Mo26.7 
10: (186) Ti8.8 Zr8.3 Hf7.9 W19.2 Nb19.2 Ta16.7 Mo20.0 
11: (196) Ti9.2 Zr8.3 Hf15.8 W18.3 Nb18.3 Ta16.7 Mo13.3 
12: (205) Ti9.6 Zr8.3 Hf23.8 W17.5 Nb17.5 Ta16.7 Mo6.7 
13: (139) Ti5.8 Zr4.2 Hf31.7 W17.5 Nb17.5 Ta16.7 Mo6.7 
14: ( 61) Ti2.1 Hf39.6 W17.5 Nb17.5 Ta16.7 Mo6.7 
15: ( 68) Ti2.5 Hf47.5 W14.7 Nb14.7 Ta13.9 Mo6.7 
16: ( 74) Ti2.9 Hf55.4 W11.9 Nb11.9 Ta11.1 Mo6.7 
17: ( 80) Ti3.3 Hf63.3 W11.1 Nb11.1 Ta11.1 
18: ( 84) Ti3.8 Hf71.2 W8.3 Nb8.3 Ta8.3 
19: ( 87) Ti4.2 Hf79.2 W5.6 Nb5.6 Ta5.6 
20: ( 89) Ti4.6 Hf87.1 W2.8 Nb2.8 Ta2.8 
21: ( 90) Ti5.0 Hf95.0 
\end{minted}

As one can see below, the path still is optimal (21 steps) but it now
passes much closer to the center of the tetrahedron, avoiding the lower
RMSAD region around \texttt{Mo33.3 Nb33.3 Ta33.3} from
which it would have to climb a bit more rapidly to reach the
\texttt{Hf95.0 Ti5.0} point.

\begin{minted}[breaklines, xleftmargin=3\parindent, linenos=true, fontsize=\small]{python}
gridFeasibleMarked = ['path' if i in lowGradientPath else f for i, f in enumerate(gridFeasible)]
fig = px.scatter_3d(
  gridAtt_projected_df, x='x', y='y', z='z', color=gridFeasibleMarked, 
  text=labels, hover_name=formulas, template='plotly_white', width=800, 
  height=700, opacity=0.333, color_discrete_sequence=['blue', 'green', 'red'],
  labels={'color':'Solid Solution Phases', 'x':'', 'y':'', 'z':''})
fig.update_scenes({'camera': {'eye': {'x': -1.8, 'y': 1.2, 'z': 1.5}}})
\end{minted}

\begin{figure}[H]
    \centering
    \includegraphics[width=0.8\textwidth]{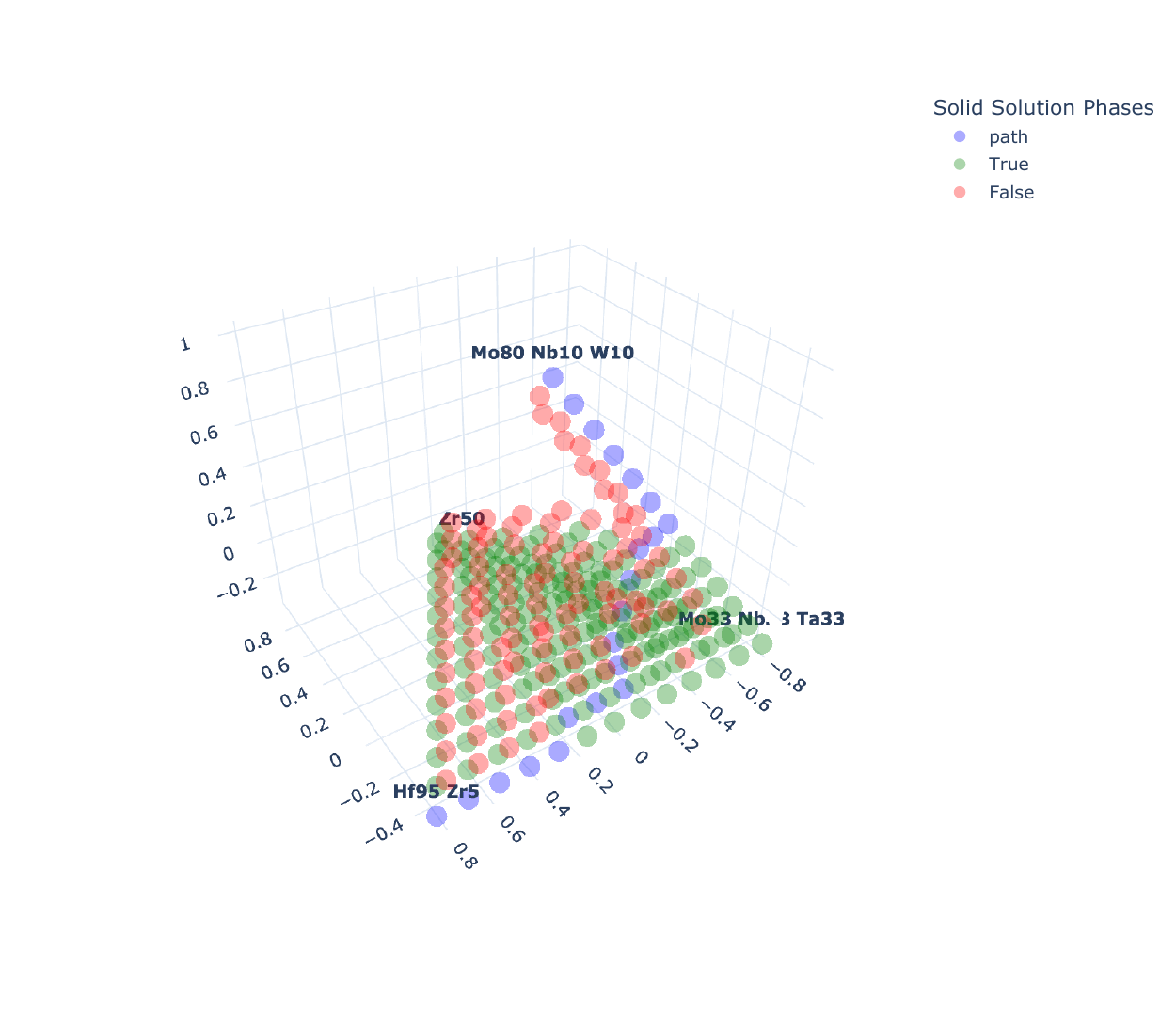}
    \caption{An biased selection from a set of equally optimal shortest paths between W10 Nb10 Mo80 and Ti5 Hf95 connecting them in 21 steps, penalized by the magnitude of gradient of RMSAD over the path. Relative to one in Figure \ref{nimplextutorial2:fig:shortest}, it passes closer to the center of the compositional space to avoid low-RMSAD region around MoNbTa and possibly pass through iso-value surface. It can also be seen as the shortest path in a stretched space.}
    \label{nimplextutorial2:fig:lowgradient}
\end{figure}

\section{Property Maximization Along the
Path}\label{nimplextutorial2:property-maximization-along-the-path}

\textbf{Finally, let's demonstrate how easy it is to maximize a property
along the path by exploiting the unidirectionaliy of the graph edge
definitions in \texttt{nimplex}, which allow us to
efortlessly enocde the property maximization goal as directional reward
or penalty applied to edges.} This is as simple as applying the property
gradient directly rather than its magnitude.

\begin{minted}[breaklines, xleftmargin=3\parindent, linenos=true, fontsize=\small]{python}
edges = []
gradientsList = []
penaltyFactor = 20
for i, nList in enumerate(graphN):
    if gridFeasible[i]:
        for n in nList:
            if gridFeasible[n]:
                rmsadGradient = rmsadList[i]-rmsadList[n]
                gradientsList.append(rmsadGradient)
                edges.append([i, n, 1+round((rmsadGradient*penaltyFactor), 3)])
print(edges[:15])
print(f"{len(gradientsList)} gradients calculated with values up to {max(gradientsList)}")
print(f"Penalty at -{penaltyFactor}x gradient *value* per path step assigned (higher RMSAD preferred)")
\end{minted}

\begin{minted}[xleftmargin=3\parindent, fontsize=\small, bgcolor=subtlegray]{output}
[[0, 1, 0.945], [1, 0, 1.055], [1, 2, 0.954], [2, 1, 1.046], [2, 3, 0.958], 
[3, 2, 1.042], [3, 4, 0.961], [4, 3, 1.039], [4, 5, 0.962], [5, 4, 1.038], 
[5, 6, 0.963], [6, 5, 1.037], [6, 7, 0.963], [6, 97, 0.577], [7, 6, 1.037]]
1374 gradients calculated with values up to 0.0473823732310002
Penalty at -20x gradient *value* per path step assigned (higher RMSAD preferred)
\end{minted}

\begin{minted}[xleftmargin=3\parindent, linenos=true]{python}
pathfindingGraph = Graph(edges=edges, bi_directional=False)
finder = DijkstraFinder()
path, runs = finder.find_path(
    pathfindingGraph.node(0), 
    pathfindingGraph.node(90), 
    pathfindingGraph)
pathList = [p.node_id for p in path]
\end{minted}

\begin{minted}[breaklines, xleftmargin=3\parindent, linenos=true, fontsize=\small]{python}
for step, i in enumerate(pathList):
    print(f"{step+1:>2}: {formulas[i]}")
\end{minted}

\begin{minted}[xleftmargin=3\parindent, fontsize=\small, bgcolor=subtlegray]{output}
 1: (  0) W10.0 Nb10.0 Mo80.0 
 2: (  1) W11.9 Nb11.9 Ta2.8 Mo73.3 
 3: (  2) W13.9 Nb13.9 Ta5.6 Mo66.7 
 4: (  3) W15.8 Nb15.8 Ta8.3 Mo60.0 
 5: (  4) W17.8 Nb17.8 Ta11.1 Mo53.3 
 6: (  5) W19.7 Nb19.7 Ta13.9 Mo46.7 
 7: (  6) W21.7 Nb21.7 Ta16.7 Mo40.0 
 8: ( 97) Ti4.2 Zr4.2 W20.8 Nb20.8 Ta16.7 Mo33.3 
 9: (175) Ti8.3 Zr8.3 W20.0 Nb20.0 Ta16.7 Mo26.7 
10: (186) Ti8.8 Zr8.3 Hf7.9 W19.2 Nb19.2 Ta16.7 Mo20.0 
11: (196) Ti9.2 Zr8.3 Hf15.8 W18.3 Nb18.3 Ta16.7 Mo13.3 
12: (204) Ti9.6 Zr8.3 Hf23.8 W15.6 Nb15.6 Ta13.9 Mo13.3 
13: (212) Ti10.0 Zr8.3 Hf31.7 W14.7 Nb14.7 Ta13.9 Mo6.7 
14: (219) Ti10.4 Zr8.3 Hf39.6 W13.9 Nb13.9 Ta13.9 
15: (153) Ti6.7 Zr4.2 Hf47.5 W13.9 Nb13.9 Ta13.9 
16: (158) Ti7.1 Zr4.2 Hf55.4 W11.1 Nb11.1 Ta11.1 
17: (162) Ti7.5 Zr4.2 Hf63.3 W8.3 Nb8.3 Ta8.3 
18: (165) Ti7.9 Zr4.2 Hf71.2 W5.6 Nb5.6 Ta5.6 
19: (167) Ti8.3 Zr4.2 Hf79.2 W2.8 Nb2.8 Ta2.8 
20: (168) Ti8.8 Zr4.2 Hf87.1 
21: ( 90) Ti5.0 Hf95.0 
\end{minted}

\textbf{As you can see, we again complete the path in 21 steps but
following a different path that maximizes the RMSAD along the way thanks
to the small bias we introduced.}

\begin{minted}[breaklines, xleftmargin=3\parindent, linenos=true, fontsize=\small]{python}
gridFeasibleMarked = ['path' if i in pathList else f for i, f in enumerate(gridFeasible)]
fig = px.scatter_3d(
  gridAtt_projected_df, x='x', y='y', z='z', color=gridFeasibleMarked, 
  text=labels, hover_name=formulas, template='plotly_white', width=800, 
  height=700, opacity=0.333, color_discrete_sequence=['blue', 'green', 'red'], 
  labels={'color':'Solid Solution Phases', 'x':'', 'y':'', 'z':''})
fig.update_scenes({'camera': {'eye': {'x': -1.8, 'y': 1.2, 'z': 1.5}}})
\end{minted}

\begin{figure}[H]
    \centering
    \includegraphics[width=0.8\textwidth]{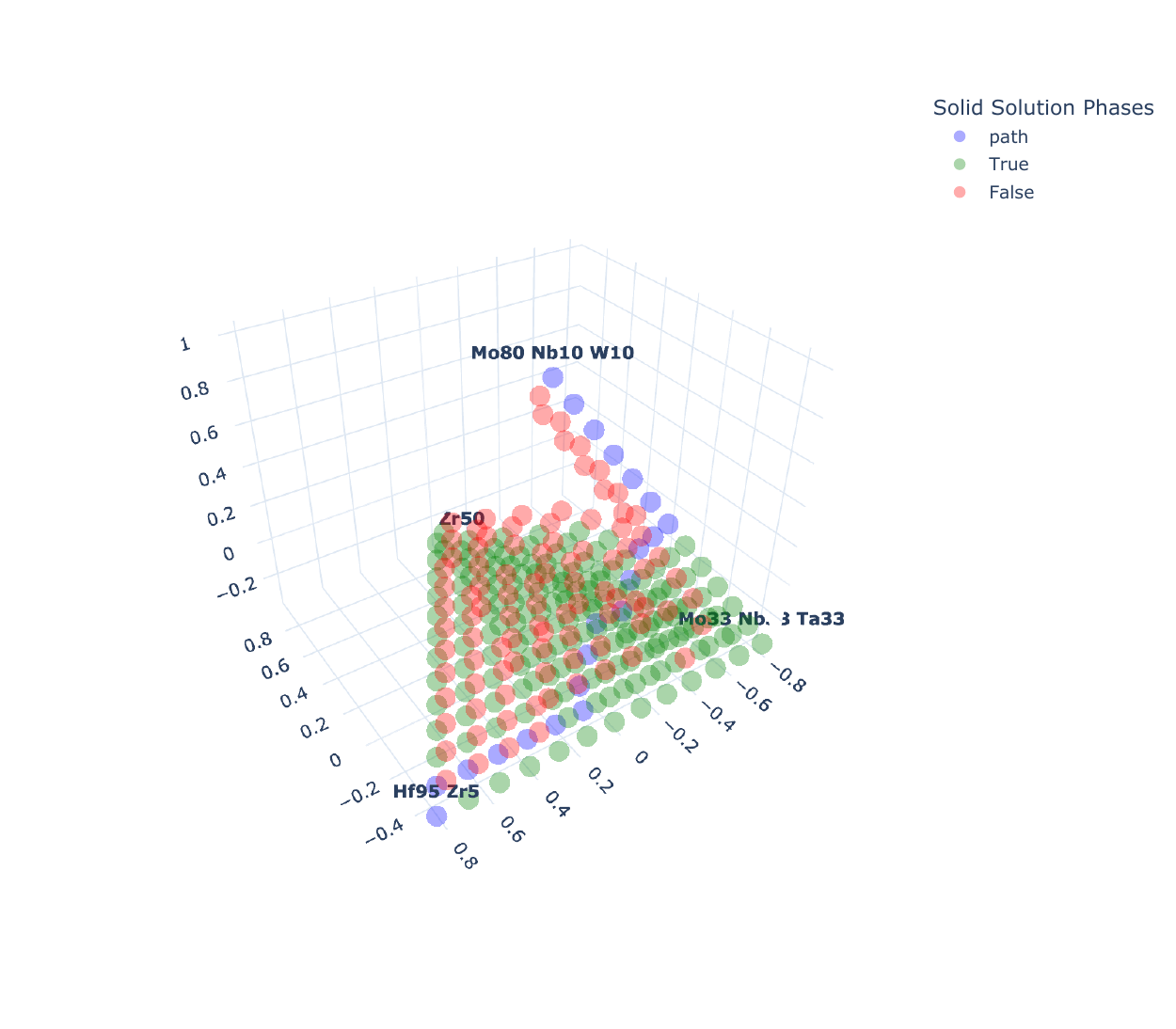}
    \caption{An biased selection from a set of equally optimal shortest paths between W10 Nb10 Mo80 and Ti5 Hf95 connecting them in 21 steps, penalized for negative the value of gradient of RMSAD over the path. It can also be seen as the shortest path in a bidirectionally stretched space.}
    \label{nimplextutorial2:fig:highvalue}
\end{figure}

\textbf{And, if we care more about going through high RMSAD regions than
we do about the number of steps taken, we can always simply increase the
bias to move the path into the high RMSAD regions even more at the cost
of the number of steps taken.}

\begin{minted}[breaklines, xleftmargin=3\parindent, linenos=true, fontsize=\small]{python}
edges = []
gradientsList = []
penaltyFactor = 200
for i, nList in enumerate(graphN):
    if gridFeasible[i]:
        for n in nList:
            if gridFeasible[n]:
                rmsadGradient = rmsadList[i]-rmsadList[n]
                gradientsList.append(rmsadGradient)
                edges.append([i, n, 1+round((rmsadGradient*penaltyFactor), 3)])
print(edges[:15])
print(f"{len(gradientsList)} gradients calculated with values up to {max(gradientsList)}")
print(f"Penalty at -{penaltyFactor}x gradient *value* per path step assigned (higher RMSAD preferred)")

pathfindingGraph = Graph(edges=edges, bi_directional=False)
finder = DijkstraFinder()
path, runs = finder.find_path(
    pathfindingGraph.node(0), 
    pathfindingGraph.node(90), 
    pathfindingGraph)
pathList = [p.node_id for p in path]
\end{minted}

\begin{minted}[xleftmargin=3\parindent, fontsize=\small, bgcolor=subtlegray]{output}
[[0, 1, 0.44599999999999995], [1, 0, 1.554], [1, 2, 0.536], [2, 1, 1.464], 
[2, 3, 0.5840000000000001], [3, 2, 1.416], [3, 4, 0.609], [4, 3, 1.391], 
[4, 5, 0.621], [5, 4, 1.379], [5, 6, 0.625], [6, 5, 1.375], [6, 7, 0.626], 
[6, 97, -3.2300000000000004], [7, 6, 1.374]]
1374 gradients calculated with values up to 0.0473823732310002
Penalty at -200x gradient *value* per path step assigned (higher RMSAD preferred)
\end{minted}

\begin{minted}[breaklines, xleftmargin=3\parindent, linenos=true, fontsize=\small]{python}
for step, i in enumerate(pathList):
    print(f"{step+1:>2}: {formulas[i]}")
\end{minted}

\begin{minted}[xleftmargin=3\parindent, fontsize=\small, bgcolor=subtlegray]{output}
1: (  0) W10.0 Nb10.0 Mo80.0 
 2: (  1) W11.9 Nb11.9 Ta2.8 Mo73.3 
 3: (  2) W13.9 Nb13.9 Ta5.6 Mo66.7 
 4: (  3) W15.8 Nb15.8 Ta8.3 Mo60.0 
 5: (  4) W17.8 Nb17.8 Ta11.1 Mo53.3 
 6: (  5) W19.7 Nb19.7 Ta13.9 Mo46.7 
 7: (  6) W21.7 Nb21.7 Ta16.7 Mo40.0 
 8: ( 97) Ti4.2 Zr4.2 W20.8 Nb20.8 Ta16.7 Mo33.3 
 9: (175) Ti8.3 Zr8.3 W20.0 Nb20.0 Ta16.7 Mo26.7 
10: (186) Ti8.8 Zr8.3 Hf7.9 W19.2 Nb19.2 Ta16.7 Mo20.0 
11: (196) Ti9.2 Zr8.3 Hf15.8 W18.3 Nb18.3 Ta16.7 Mo13.3 
12: (205) Ti9.6 Zr8.3 Hf23.8 W17.5 Nb17.5 Ta16.7 Mo6.7 
13: (213) Ti10.0 Zr8.3 Hf31.7 W16.7 Nb16.7 Ta16.7 
14: (219) Ti10.4 Zr8.3 Hf39.6 W13.9 Nb13.9 Ta13.9 
15: (279) Ti14.6 Zr12.5 Hf39.6 W11.1 Nb11.1 Ta11.1 
16: (328) Ti18.8 Zr16.7 Hf39.6 W8.3 Nb8.3 Ta8.3 
17: (367) Ti22.9 Zr20.8 Hf39.6 W5.6 Nb5.6 Ta5.6 
18: (369) Ti23.3 Zr20.8 Hf47.5 W2.8 Nb2.8 Ta2.8 
19: (333) Ti19.6 Zr16.7 Hf55.4 W2.8 Nb2.8 Ta2.8 
20: (288) Ti15.8 Zr12.5 Hf63.3 W2.8 Nb2.8 Ta2.8 
21: (233) Ti12.1 Zr8.3 Hf71.2 W2.8 Nb2.8 Ta2.8 
22: (167) Ti8.3 Zr4.2 Hf79.2 W2.8 Nb2.8 Ta2.8 
23: (168) Ti8.8 Zr4.2 Hf87.1 
24: ( 90) Ti5.0 Hf95.0 
\end{minted}

\begin{minted}[breaklines, xleftmargin=3\parindent, linenos=true, fontsize=\small]{python}
gridFeasibleMarked = ['path' if i in pathList else f for i, f in enumerate(gridFeasible)]
fig = px.scatter_3d(
  gridAtt_projected_df, x='x', y='y', z='z', color=gridFeasibleMarked, 
  text=labels, hover_name=formulas, template='plotly_white', width=800, 
  height=700, opacity=0.333, color_discrete_sequence=['blue', 'green', 'red'],
  labels={'color':'Solid Solution Phases', 'x':'', 'y':'', 'z':''})
fig.update_scenes({'camera': {'eye': {'x': -1.8, 'y': 1.2, 'z': 1.5}}})
\end{minted}

\begin{figure}[H]
    \centering
    \includegraphics[width=0.8\textwidth]{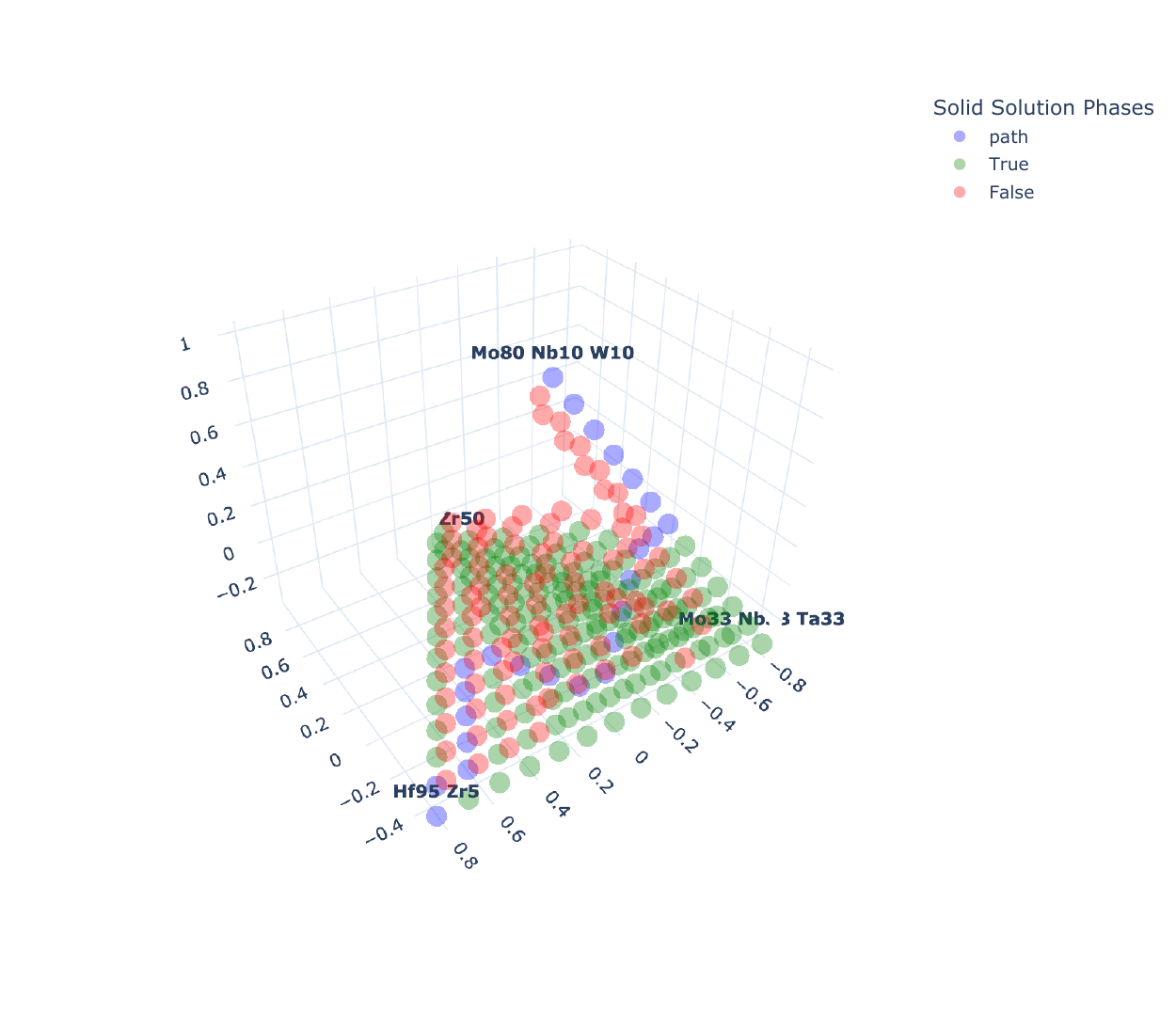}
    \caption{A paths between W10 Nb10 Mo80 and Ti5 Hf95, connecting them in 24 steps, very strongly biased towards regions of high RMSAD.}
    \label{nimplextutorial2:fig:veryhighvalue}
\end{figure}

\section{Final Remarks} \label{nimplextutorial2:final-remarks}

\textbf{And that's it! We hope you enjoyed this tutorial and that you
see the potential of \texttt{nimplex} in your work. If
you have any questions, feel free to ask them in the
\texttt{nimplex} repository under the Issues tab. You
are also welcome to shoot an email to the author at
\href{mailto:adam@phaseslab.org}{\nolinkurl{adam@phaseslab.org}}. We
will be excited to hear about your applications.}


\chapter{MatSE580 Guest Lecture 1 - Quick Guide to Manipulating Materials With \texttt{pymatgen}, Setting up \texttt{MongoDB}, and Getting Started with \texttt{pySIPFENN}} 
\label{chap:pysipfenntutorial1} 

\hypertarget{introduction}{%
\section{Introduction}\label{pysipfenntutorial:introduction}}

In this guest lecture, we will cover: 

\begin{enumerate}
    \item \protect\hyperlink{Manipulating-and-analyzing-materials}{Manipulating and analyzing materials} - using \href{https://github.com/materialsproject/pymatgen}{pymatgen}

    \item \protect\hyperlink{Setting-up-MongoDB}{Setting up a small NoSQL database on the cloud to synchronize decentralized processing} - using \href{https://www.mongodb.com/atlas}{MongoDB Atlas} Free Tier 

    \item \protect\hyperlink{pymongo}{Interacting with the database} - using \href{https://github.com/mongodb/mongo-python-driver}{pymongo} library

    \item \protect\hyperlink{pysipfenn-install}{Installing machine learning (ML) tools} to predict stability of materials - using \href{https://pysipfenn.readthedocs.io/en/stable/}{pySIPFENN}
\end{enumerate}

Before you begin, you will need to set up a few essential development
tools.

While not required, it is recommended first to set up a virtual
environment using venv or Conda. This ensures that one of the required
versions of Python (3.9+) is used and there are no dependency conflicts.
It often comes preinstalled, like in GitHub Codespaces and some Linux
distributions. You can quickly check that by running.

\begin{minted}[xleftmargin=3\parindent, fontsize=\small, bgcolor=subtlegray]{output}
conda --version
\end{minted}

And if it is not installed, you can follow the
(\href{https://docs.conda.io/en/latest/miniconda.html}{miniconda
instructions} ) for a quick clean setup.

Once you have Conda installed on your system, you can create a new
environment with:

\begin{minted}[xleftmargin=3\parindent, fontsize=\small, bgcolor=subtlegray]{output}
conda create -n 580demo python=3.10 jupyter numpy scipy
conda init
\end{minted}

Restart your terminal, and activate the environment with:

\begin{minted}[xleftmargin=3\parindent, fontsize=\small, bgcolor=subtlegray]{output}
conda activate 580demo
\end{minted}

At this point, you should be able to run
\texttt{jupyter notebook} and open this notebook in
your browser with it or select the kernel
\texttt{580demo} in VS Code (top-right corner) or other
IDEs.

First, we will import some libraries that ship with Python so that we
don't need to worry about getting them, and are used in this notebook:

\begin{minted}[xleftmargin=3\parindent, linenos=true, fontsize=\small]{python}
from pprint import pprint            # pretty printing
from collections import defaultdict  # convenience in the example
import os                            # file handling
from datetime import datetime        # time handling
from zoneinfo import ZoneInfo        # time handling
\end{minted}

Now, we need to use \texttt{pip} package manager to
install the rest of the libraries we will use. If you are using Conda,
you could also use \texttt{conda install} instead, but
it is more elaborate for non-Anaconda-default packages.

We start with \texttt{pymatgen}, used in the next part
of this notebook. To install it, simply remove the
\texttt{\#} in the following line and run it, or open a
terminal and run \texttt{pip install pymatgen}.

\begin{minted}[xleftmargin=3\parindent, linenos=true, fontsize=\small]{python}
!pip install pymatgen
\end{minted}

And then install \texttt{pymongo} used in the 2nd part:

\begin{minted}[xleftmargin=3\parindent, linenos=true, fontsize=\small]{python}
!pip install pymongo
\end{minted}

Now, you should be ready to go!

\hypertarget{manipulating-and-analyzing-materials}{%
\section{Manipulating and analyzing
materials}\label{pysipfenntutorial:manipulating-and-analyzing-materials}}

To start working with atomic structures, often referred to as atomic
configurations or simply materials, we must be able to represent and
manipulate them. One of the most powerful and mature tools to do so is
\href{https://github.com/materialsproject/pymatgen}{pymatgen}, which we
just installed. The critical component of pymatgen is its library of
representations of fundamental materials objects, such as
\texttt{Structure} and
\texttt{Molecule}, contained in the
\texttt{pymatgen.core} module. Let's import it and
create a simple cubic structure of Al just as we did in the DFTTK
tutorial last week:

\hypertarget{basics}{%
\subsection{Basics}\label{pysipfenntutorial:basics}}

\begin{minted}[xleftmargin=3\parindent, linenos=true, fontsize=\small]{python}
from pymatgen.core import Structure

s = Structure(
        lattice=[[4.0384, 0, 0], [0, 4.0384, 0], [0, 0, 4.0384]],
        species=['Al', 'Al', 'Al', 'Al'],
        coords=[[0.0, 0.0, 0.0], [0, 0.5, 0.5], [0.5, 0.0, 0.5], [0.5, 0.5, 0.0]]
    )
\end{minted}

Now, \texttt{s} holds our initialized structure, and we
can apply print on it to see what it looks like:

\begin{minted}[xleftmargin=3\parindent, linenos=true, fontsize=\small]{python}
print(s)
\end{minted}

\begin{minted}[xleftmargin=3\parindent, fontsize=\small, bgcolor=subtlegray]{output}
Full Formula (Al4)
Reduced Formula: Al
abc   :   4.038400   4.038400   4.038400
angles:  90.000000  90.000000  90.000000
pbc   :       True       True       True
Sites (4)
  #  SP      a    b    c
---  ----  ---  ---  ---
  0  Al    0    0    0
  1  Al    0    0.5  0.5
  2  Al    0.5  0    0.5
  3  Al    0.5  0.5  0
\end{minted}

\textbf{Initialized} is a critical word here because the
\texttt{Structure} object is not just a collection of
``numbers''. It holds a lot of information we can access using the
\texttt{Structure} object's attributes and methods. For
example, the density of the material is immediately available:

\begin{minted}[xleftmargin=3\parindent, linenos=true, fontsize=\small]{python}
s.density
\end{minted}

\begin{minted}[xleftmargin=3\parindent, fontsize=\small, bgcolor=subtlegray]{output}
2.721120664587368
\end{minted}

We can also ``mutate'' the object with a few intuitive methods like
\texttt{apply\_strain}:

\begin{minted}[xleftmargin=3\parindent, linenos=true, fontsize=\small]{python}
s.apply_strain(0.1)
\end{minted}

\begin{minted}[xleftmargin=3\parindent, fontsize=\small, bgcolor=subtlegray]{output}
Structure Summary
Lattice
    abc : 4.442240000000001 4.442240000000001 4.442240000000001
 angles : 90.0 90.0 90.0
 volume : 87.66092623767148
      A : 4.442240000000001 0.0 0.0
      B : 0.0 4.442240000000001 0.0
      C : 0.0 0.0 4.442240000000001
    pbc : True True True
PeriodicSite: Al (0.0, 0.0, 0.0) [0.0, 0.0, 0.0]
PeriodicSite: Al (0.0, 2.221, 2.221) [0.0, 0.5, 0.5]
PeriodicSite: Al (2.221, 0.0, 2.221) [0.5, 0.0, 0.5]
PeriodicSite: Al (2.221, 2.221, 0.0) [0.5, 0.5, 0.0]
\end{minted}

Importantly, as you can see, \texttt{s} has been
printed out when we ran the command, as if the
\texttt{s.apply\_strain} returned a modified
\texttt{Structure} object. This is true! However, by
default, pymatgen will also strain the original object, as you can see
looking at the \texttt{s} density:

\begin{minted}[xleftmargin=3\parindent, linenos=true, fontsize=\small]{python}
s.density
\end{minted}

\begin{minted}[xleftmargin=3\parindent, fontsize=\small, bgcolor=subtlegray]{output}
2.0444182303436262
\end{minted}

This is a very convenient feature, but it can be dangerous if you are
not careful and, for instance, try to generate 10 structures with
increasing strains:

\begin{minted}[xleftmargin=3\parindent, linenos=true, fontsize=\small]{python}
strainedList = [s.apply_strain(0.1 * i) for i in range(1, 11)]
for strained in strainedList[:2]:
    print(strained)
\end{minted}

\begin{minted}[xleftmargin=3\parindent, fontsize=\small, bgcolor=subtlegray]{output}
Full Formula (Al4)
Reduced Formula: Al
abc   : 297.826681 297.826681 297.826681
angles:  90.000000  90.000000  90.000000
pbc   :       True       True       True
Sites (4)
  #  SP      a    b    c
---  ----  ---  ---  ---
  0  Al    0    0    0
  1  Al    0    0.5  0.5
  2  Al    0.5  0    0.5
  3  Al    0.5  0.5  0
Full Formula (Al4)
Reduced Formula: Al
abc   : 297.826681 297.826681 297.826681
angles:  90.000000  90.000000  90.000000
pbc   :       True       True       True
Sites (4)
  #  SP      a    b    c
---  ----  ---  ---  ---
  0  Al    0    0    0
  1  Al    0    0.5  0.5
  2  Al    0.5  0    0.5
  3  Al    0.5  0.5  0
\end{minted}

We will now end up with a single object with 67 times the original
volume (1.1 * 1.2 * \ldots{} * 2.0) repeated 10 times. To avoid this, we
can get (or regenerate) original \texttt{s} and use the
\texttt{copy} method to create a new object each time:

\begin{minted}[xleftmargin=3\parindent, linenos=true, fontsize=\small]{python}
from copy import copy

s = Structure(
        lattice=[[4.0384, 0, 0], [0, 4.0384, 0], [0, 0, 4.0384]],
        species=['Al', 'Al', 'Al', 'Al'],
        coords=[[0.0, 0.0, 0.0], [0, 0.5, 0.5], [0.5, 0.0, 0.5], [0.5, 0.5, 0.0]]
    )
\end{minted}

\begin{minted}[xleftmargin=3\parindent, linenos=true, fontsize=\small]{python}
strainedList = [copy(s).apply_strain(0.1 * i) for i in range(0, 11)]
for strained in strainedList[:2]:
    print(strained)
\end{minted}

\begin{minted}[xleftmargin=3\parindent, fontsize=\small, bgcolor=subtlegray]{output}
Full Formula (Al4)
Reduced Formula: Al
abc   :   4.038400   4.038400   4.038400
angles:  90.000000  90.000000  90.000000
pbc   :       True       True       True
Sites (4)
  #  SP      a    b    c
---  ----  ---  ---  ---
  0  Al    0    0    0
  1  Al    0    0.5  0.5
  2  Al    0.5  0    0.5
  3  Al    0.5  0.5  0
Full Formula (Al4)
Reduced Formula: Al
abc   :   4.442240   4.442240   4.442240
angles:  90.000000  90.000000  90.000000
pbc   :       True       True       True
Sites (4)
  #  SP      a    b    c
---  ----  ---  ---  ---
  0  Al    0    0    0
  1  Al    0    0.5  0.5
  2  Al    0.5  0    0.5
  3  Al    0.5  0.5  0
\end{minted}

And now everything works as expected! We can also easily do some
modifications to the structure, like replacing one of the atoms with
another

\begin{minted}[xleftmargin=3\parindent, linenos=true, fontsize=\small]{python}
s.replace(0, "Au")
print(s)
\end{minted}

\begin{minted}[xleftmargin=3\parindent, fontsize=\small, bgcolor=subtlegray]{output}
Full Formula (Al3 Au1)
Reduced Formula: Al3Au
abc   :   4.038400   4.038400   4.038400
angles:  90.000000  90.000000  90.000000
pbc   :       True       True       True
Sites (4)
  #  SP      a    b    c
---  ----  ---  ---  ---
  0  Au    0    0    0
  1  Al    0    0.5  0.5
  2  Al    0.5  0    0.5
  3  Al    0.5  0.5  0
\end{minted}

or all of the atoms of a given element at once

\begin{minted}[xleftmargin=3\parindent, linenos=true, fontsize=\small]{python}
s.replace_species({"Al": "Ni"})
\end{minted}

\begin{minted}[xleftmargin=3\parindent, fontsize=\small, bgcolor=subtlegray]{output}
Structure Summary
Lattice
    abc : 4.0384 4.0384 4.0384
 angles : 90.0 90.0 90.0
 volume : 65.860951343104
      A : 4.0384 0.0 0.0
      B : 0.0 4.0384 0.0
      C : 0.0 0.0 4.0384
    pbc : True True True
PeriodicSite: Au (0.0, 0.0, 0.0) [0.0, 0.0, 0.0]
PeriodicSite: Ni (0.0, 4.038, 4.038) [0.0, 0.5, 0.5]
PeriodicSite: Ni (4.038, 0.0, 4.038) [0.5, 0.0, 0.5]
PeriodicSite: Ni (4.038, 4.038, 0.0) [0.5, 0.5, 0.0]
\end{minted}

Lastly, with \texttt{Structure} objects, we also have
access to lower-order primitives, such as
\texttt{Composition}

\begin{minted}[xleftmargin=3\parindent, linenos=true, fontsize=\small]{python}
c = s.composition
c
\end{minted}

\begin{minted}[xleftmargin=3\parindent, fontsize=\small, bgcolor=subtlegray]{output}
Composition('Au1 Ni3')
\end{minted}

which may look like a simple string but is actually a powerful object
that can be used to do things like calculate the fraction of each
element in the structure:

\begin{minted}[xleftmargin=3\parindent, linenos=true, fontsize=\small]{python}
c.fractional_composition
\end{minted}

\begin{minted}[xleftmargin=3\parindent, fontsize=\small, bgcolor=subtlegray]{output}
Composition('Au0.25 Ni0.75')
\end{minted}

including the weight fractions (I wrote this part of pymatgen :)):

\begin{minted}[xleftmargin=3\parindent, linenos=true, fontsize=\small]{python}
c.to_weight_dict
\end{minted}

\begin{minted}[xleftmargin=3\parindent, fontsize=\small, bgcolor=subtlegray]{output}
{'Au': 0.5279943035775228, 'Ni': 0.47200569642247725}
\end{minted}

\hypertarget{symmetry-analysis}{%
\subsection{Symmetry Analysis}\label{pysipfenntutorial:symmetry-analysis}}

With some basics of the way, let's look at some more advanced features
of pymatgen that come from the integration with 3rd party libraries like
\href{https://spglib.readthedocs.io/en/latest/index.html}{spglib}, which
is a high-performance library for symmetry analysis (1) written in C,
(2) wrapped in Python by the authors, and finally (3) wrapped in
pymatgen for convenience.

Such an approach introduces a lot of performance bottlenecks (4-20x
slower and 50x RAM needs compared to my interface written in
\href{https://nim-lang.org}{Nim}), but allows us to get started with
things like symmetry analysis in with just one line of code where
\texttt{SpacegroupAnalyzer} puts
\texttt{s} in a new context:

\begin{minted}[xleftmargin=3\parindent, linenos=true, fontsize=\small]{python}
from pymatgen.symmetry.analyzer import SpacegroupAnalyzer
spgA = SpacegroupAnalyzer(s)
\end{minted}

Now many useful methods are available to us, allowing quickly getting
\texttt{crystal\_system},
\texttt{space\_group\_symbol}, and
\texttt{point\_group\_symbol}:

\begin{minted}[xleftmargin=3\parindent, linenos=true, fontsize=\small]{python}
spgA.get_crystal_system()
\end{minted}

\begin{minted}[xleftmargin=3\parindent, fontsize=\small, bgcolor=subtlegray]{output}
'cubic'
\end{minted}

\begin{minted}[xleftmargin=3\parindent, linenos=true, fontsize=\small]{python}
spgA.get_space_group_symbol()
\end{minted}

\begin{minted}[xleftmargin=3\parindent, fontsize=\small, bgcolor=subtlegray]{output}
'Pm-3m'
\end{minted}

\begin{minted}[xleftmargin=3\parindent, linenos=true, fontsize=\small]{python}
spgA.get_point_group_symbol()
\end{minted}

\begin{minted}[xleftmargin=3\parindent, fontsize=\small, bgcolor=subtlegray]{output}
'm-3m'
\end{minted}

We can also do some more advanced operations involving symmetry. For
example, as some may have noticed, the \texttt{s}
structure we created is primitive, but if we fix its symmetry, we can
describe it with just 1 face-centered atom instead of 3, as they are
symmetrically equivalent. We can do this with the
\texttt{get\_symmetrized\_structure}:

\begin{minted}[xleftmargin=3\parindent, linenos=true, fontsize=\small]{python}
symmetrized = spgA.get_symmetrized_structure()
symmetrized
\end{minted}

\begin{minted}[xleftmargin=3\parindent, fontsize=\small, bgcolor=subtlegray]{output}
SymmetrizedStructure
Full Formula (Ni3 Au1)
Reduced Formula: Ni3Au
Spacegroup: Pm-3m (221)
abc   :   4.038400   4.038400   4.038400
angles:  90.000000  90.000000  90.000000
Sites (4)
  #  SP      a    b    c  Wyckoff
---  ----  ---  ---  ---  ---------
  0  Au      0  0    0    1a
  1  Ni      0  0.5  0.5  3c
\end{minted}

Which we can then use to get the primitive or conventional structure
back. Here, they happen to be the same, but that is often not the case.

\begin{minted}[xleftmargin=3\parindent, linenos=true, fontsize=\small]{python}
symmetrized.to_primitive()
\end{minted}

\begin{minted}[xleftmargin=3\parindent, fontsize=\small, bgcolor=subtlegray]{output}
Structure Summary
Lattice
    abc : 4.0384 4.0384 4.0384
 angles : 90.0 90.0 90.0
 volume : 65.860951343104
      A : 4.0384 0.0 2.472806816838336e-16
      B : -2.472806816838336e-16 4.0384 2.472806816838336e-16
      C : 0.0 0.0 4.0384
    pbc : True True True
PeriodicSite: Ni (-1.236e-16, 2.019, 2.019) [0.0, 0.5, 0.5]
PeriodicSite: Ni (2.019, 0.0, 2.019) [0.5, 0.0, 0.5]
PeriodicSite: Ni (2.019, 2.019, 2.473e-16) [0.5, 0.5, 0.0]
PeriodicSite: Au (0.0, 0.0, 0.0) [0.0, 0.0, 0.0]
\end{minted}

\begin{minted}[xleftmargin=3\parindent, linenos=true, fontsize=\small]{python}
symmetrized.to_conventional()
\end{minted}

\begin{minted}[xleftmargin=3\parindent, fontsize=\small, bgcolor=subtlegray]{output}
Structure Summary
Lattice
    abc : 4.0384 4.0384 4.0384
 angles : 90.0 90.0 90.0
 volume : 65.860951343104
      A : 4.0384 0.0 2.472806816838336e-16
      B : -2.472806816838336e-16 4.0384 2.472806816838336e-16
      C : 0.0 0.0 4.0384
    pbc : True True True
PeriodicSite: Ni (-1.236e-16, 2.019, 2.019) [0.0, 0.5, 0.5]
PeriodicSite: Ni (2.019, 0.0, 2.019) [0.5, 0.0, 0.5]
PeriodicSite: Ni (2.019, 2.019, 2.473e-16) [0.5, 0.5, 0.0]
PeriodicSite: Au (0.0, 0.0, 0.0) [0.0, 0.0, 0.0]
\end{minted}

\hypertarget{more-complex-structures}{%
\subsection{More Complex Structures}\label{pysipfenntutorial:more-complex-structures}}

Armed with all the basics, let's look at some more complex structures
and start to modify them! For that purpose, we will take a topologically
close-packed (TCP) phase from the Cr-Fe-Ni system called Sigma, which is
both difficult to predict and critical to the performance of Ni-based
superalloys.

The structure is available here under
\texttt{assets/0-Cr8Fe18Ni4.POSCAR}, in plain-text
looking like

\begin{minted}[xleftmargin=3\parindent, fontsize=\small, bgcolor=subtlegray]{output}
Cr8 Fe18 Ni4
1.0
8.547048 0.000000 0.000000
0.000000 8.547048 0.000000
0.000000 0.000000 4.477714
Cr Fe Ni
8 18 4
direct
0.737702 0.063709 0.000000 Cr
0.262298 0.936291 0.000000 Cr
...
0.899910 0.100090 0.500000 Ni
\end{minted}

,or when visualized in Figure \ref{pysipfenntutorial:simgaexample} below:

\begin{figure}[H]
    \centering
    \includegraphics[width=0.6\textwidth]{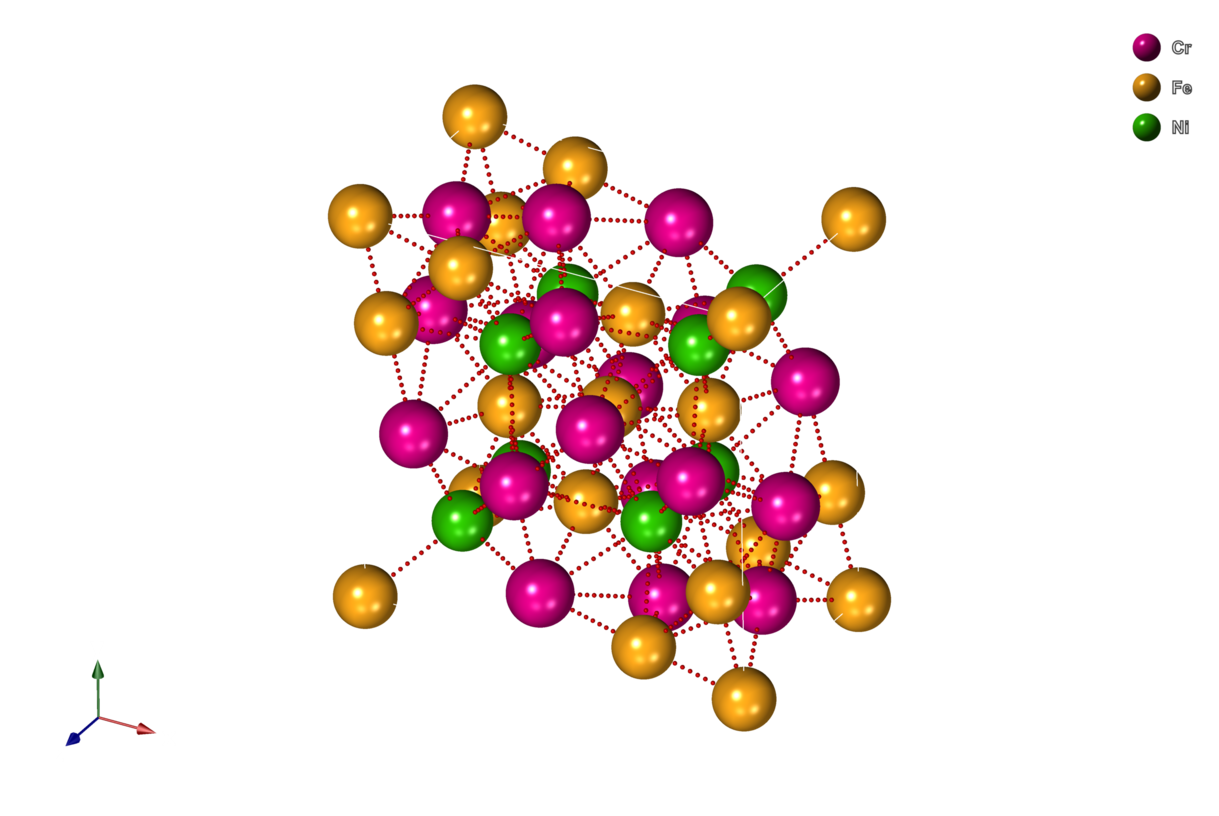}
    \caption{Rendering of $Cr_{12}Fe_{10}Ni_8$ endmember occupancy of the $\sigma$-phase.}
    \label{pysipfenntutorial:simgaexample}
\end{figure}

Now, we can quickly load it into pymatgen with either (1)
\texttt{Structure.from\_file} or (2)
\texttt{pymatgen.io.vasp} module using
\texttt{Poscar} class, with the latter being more
reliable in some cases. Since it is an example of Sigma TCP phase
occupation, we will call it \texttt{baseStructure}.

\begin{minted}[xleftmargin=3\parindent, linenos=true, fontsize=\small]{python}
baseStructure = Structure.from_file("assets/0-Cr8Fe18Ni4.POSCAR")
baseStructure
\end{minted}

\begin{minted}[xleftmargin=3\parindent, fontsize=\small, bgcolor=subtlegray]{output}
Structure Summary
Lattice
    abc : 8.547048 8.547048 4.477714
 angles : 90.0 90.0 90.0
 volume : 327.10609528461225
      A : 8.547048 0.0 0.0
      B : 0.0 8.547048 0.0
      C : 0.0 0.0 4.477714
    pbc : True True True
PeriodicSite: Cr (6.305, 0.5445, 0.0) [0.7377, 0.06371, 0.0]
PeriodicSite: Cr (2.242, 8.003, 0.0) [0.2623, 0.9363, 0.0]
PeriodicSite: Cr (3.729, 2.032, 2.239) [0.4363, 0.2377, 0.5]
PeriodicSite: Cr (6.515, 4.818, 2.239) [0.7623, 0.5637, 0.5]
PeriodicSite: Cr (4.818, 6.515, 2.239) [0.5637, 0.7623, 0.5]
PeriodicSite: Cr (2.032, 3.729, 2.239) [0.2377, 0.4363, 0.5]
PeriodicSite: Cr (0.5445, 6.305, 0.0) [0.06371, 0.7377, 0.0]
PeriodicSite: Cr (8.003, 2.242, 0.0) [0.9363, 0.2623, 0.0]
PeriodicSite: Fe (0.0, 0.0, 0.0) [0.0, 0.0, 0.0]
PeriodicSite: Fe (4.274, 4.274, 2.239) [0.5, 0.5, 0.5]
PeriodicSite: Fe (3.958, 1.107, 0.0) [0.463, 0.1295, 0.0]
PeriodicSite: Fe (4.59, 7.44, 0.0) [0.537, 0.8705, 0.0]
PeriodicSite: Fe (3.167, 8.231, 2.239) [0.3705, 0.963, 0.5]
PeriodicSite: Fe (0.316, 5.38, 2.239) [0.03697, 0.6295, 0.5]
PeriodicSite: Fe (5.38, 0.316, 2.239) [0.6295, 0.03697, 0.5]
PeriodicSite: Fe (8.231, 3.167, 2.239) [0.963, 0.3705, 0.5]
PeriodicSite: Fe (1.107, 3.958, 0.0) [0.1295, 0.463, 0.0]
PeriodicSite: Fe (7.44, 4.59, 0.0) [0.8705, 0.537, 0.0]
PeriodicSite: Fe (1.562, 1.562, 1.127) [0.1827, 0.1827, 0.2517]
PeriodicSite: Fe (6.985, 6.985, 3.351) [0.8173, 0.8173, 0.7483]
PeriodicSite: Fe (6.985, 6.985, 1.127) [0.8173, 0.8173, 0.2517]
PeriodicSite: Fe (2.712, 5.835, 3.366) [0.3173, 0.6827, 0.7517]
PeriodicSite: Fe (2.712, 5.835, 1.112) [0.3173, 0.6827, 0.2483]
PeriodicSite: Fe (1.562, 1.562, 3.351) [0.1827, 0.1827, 0.7483]
PeriodicSite: Fe (5.835, 2.712, 1.112) [0.6827, 0.3173, 0.2483]
PeriodicSite: Fe (5.835, 2.712, 3.366) [0.6827, 0.3173, 0.7517]
PeriodicSite: Ni (3.418, 3.418, 0.0) [0.3999, 0.3999, 0.0]
PeriodicSite: Ni (5.129, 5.129, 0.0) [0.6001, 0.6001, 0.0]
PeriodicSite: Ni (0.8555, 7.692, 2.239) [0.1001, 0.8999, 0.5]
PeriodicSite: Ni (7.692, 0.8555, 2.239) [0.8999, 0.1001, 0.5]
\end{minted}

Now, we can quickly investigate the symmetry with tools we just learned:

\begin{minted}[xleftmargin=3\parindent, linenos=true, fontsize=\small]{python}
spgA = SpacegroupAnalyzer(baseStructure)
spgA.get_symmetrized_structure()
\end{minted}

\begin{minted}[xleftmargin=3\parindent, fontsize=\small, bgcolor=subtlegray]{output}
SymmetrizedStructure
Full Formula (Cr8 Fe18 Ni4)
Reduced Formula: Cr4Fe9Ni2
Spacegroup: P4_2/mnm (136)
abc   :   8.547048   8.547048   4.477714
angles:  90.000000  90.000000  90.000000
Sites (30)
  #  SP           a         b         c  Wyckoff
---  ----  --------  --------  --------  ---------
  0  Cr    0.737702  0.063709  0         8i
  1  Fe    0         0         0         2a
  2  Fe    0.463029  0.129472  0         8i
  3  Fe    0.182718  0.182718  0.251726  8j
  4  Ni    0.39991   0.39991   0         4f
\end{minted}

We can quickly see that our atomic configuration has \textbf{5}
chemically unique sites of different multiplicities occupied by the
\textbf{3} elements of interest. However, performing the analysis like
that can quickly lead to problems if, for instance, we introduce even a
tiny disorder in the structure, like a substitutional defect.

\begin{minted}[xleftmargin=3\parindent, linenos=true, fontsize=\small]{python}
sDilute = copy(baseStructure)
sDilute.replace(0, "Fe")
spgA = SpacegroupAnalyzer(sDilute)
spgA.get_symmetrized_structure()
\end{minted}

\begin{minted}[xleftmargin=3\parindent, fontsize=\small, bgcolor=subtlegray]{output}
SymmetrizedStructure
Full Formula (Cr7 Fe19 Ni4)
Reduced Formula: Cr7Fe19Ni4
Spacegroup: Pm (6)
abc   :   8.547048   8.547048   4.477714
angles:  90.000000  90.000000  90.000000
Sites (30)
  #  SP           a         b         c  Wyckoff
---  ----  --------  --------  --------  ---------
  0  Fe    0.737702  0.063709  0         1a
  1  Cr    0.262298  0.936291  0         1a
  2  Cr    0.436291  0.237702  0.5       1b
  3  Cr    0.762298  0.563709  0.5       1b
  4  Cr    0.563709  0.762298  0.5       1b
  5  Cr    0.237702  0.436291  0.5       1b
  6  Cr    0.063709  0.737702  0         1a
  7  Cr    0.936291  0.262298  0         1a
  8  Fe    0         0         0         1a
  9  Fe    0.5       0.5       0.5       1b
 10  Fe    0.463029  0.129472  0         1a
 11  Fe    0.536971  0.870528  0         1a
 12  Fe    0.370528  0.963029  0.5       1b
 13  Fe    0.036971  0.629472  0.5       1b
 14  Fe    0.629472  0.036971  0.5       1b
 15  Fe    0.963029  0.370528  0.5       1b
 16  Fe    0.129472  0.463029  0         1a
 17  Fe    0.870528  0.536971  0         1a
 18  Fe    0.182718  0.182718  0.251726  2c
 19  Fe    0.817282  0.817282  0.748274  2c
 20  Fe    0.317282  0.682718  0.751726  2c
 21  Fe    0.682718  0.317282  0.248274  2c
 22  Ni    0.39991   0.39991   0         1a
 23  Ni    0.60009   0.60009   0         1a
 24  Ni    0.10009   0.89991   0.5       1b
 25  Ni    0.89991   0.10009   0.5       1b
\end{minted}

Without any change to the other 29 atoms, there are 25 unique sites
rather than 5. Thus, if one wants to see what are the symmetry-enforced
unique sites, determining underlying sublattices, in the structure, one
needs anonymize the atoms first.

\begin{minted}[xleftmargin=3\parindent, linenos=true, fontsize=\small]{python}
for el in set(baseStructure.species):
    baseStructure.replace_species({el: 'dummy'})
print(baseStructure)
\end{minted}

\begin{minted}[xleftmargin=3\parindent, fontsize=\small, bgcolor=subtlegray]{output}
Full Formula (Dummy30)
Reduced Formula: Dummy
abc   :   8.547048   8.547048   4.477714
angles:  90.000000  90.000000  90.000000
pbc   :       True       True       True
Sites (30)
  #  SP              a         b         c
---  -------  --------  --------  --------
  0  Dummy0+  0.737702  0.063709  0
  1  Dummy0+  0.262298  0.936291  0
  2  Dummy0+  0.436291  0.237702  0.5
  3  Dummy0+  0.762298  0.563709  0.5
  4  Dummy0+  0.563709  0.762298  0.5
  5  Dummy0+  0.237702  0.436291  0.5
  6  Dummy0+  0.063709  0.737702  0
  7  Dummy0+  0.936291  0.262298  0
  8  Dummy0+  0         0         0
  9  Dummy0+  0.5       0.5       0.5
 10  Dummy0+  0.463029  0.129472  0
 11  Dummy0+  0.536971  0.870528  0
 12  Dummy0+  0.370528  0.963029  0.5
 13  Dummy0+  0.036971  0.629472  0.5
 14  Dummy0+  0.629472  0.036971  0.5
 15  Dummy0+  0.963029  0.370528  0.5
 16  Dummy0+  0.129472  0.463029  0
 17  Dummy0+  0.870528  0.536971  0
 18  Dummy0+  0.182718  0.182718  0.251726
 19  Dummy0+  0.817282  0.817282  0.748274
 20  Dummy0+  0.817282  0.817282  0.251726
 21  Dummy0+  0.317282  0.682718  0.751726
 22  Dummy0+  0.317282  0.682718  0.248274
 23  Dummy0+  0.182718  0.182718  0.748274
 24  Dummy0+  0.682718  0.317282  0.248274
 25  Dummy0+  0.682718  0.317282  0.751726
 26  Dummy0+  0.39991   0.39991   0
 27  Dummy0+  0.60009   0.60009   0
 28  Dummy0+  0.10009   0.89991   0.5
 29  Dummy0+  0.89991   0.10009   0.5
\end{minted}

Which we then pass to the \texttt{SpacegroupAnalyzer}
to get the symmetry information as before:

\begin{minted}[xleftmargin=3\parindent, linenos=true, fontsize=\small]{python}
spgA = SpacegroupAnalyzer(baseStructure)
spgA.get_symmetrized_structure()
\end{minted}

\begin{minted}[xleftmargin=3\parindent, fontsize=\small, bgcolor=subtlegray]{output}
SymmetrizedStructure
Full Formula (Dummy30)
Reduced Formula: Dummy
Spacegroup: P4_2/mnm (136)
abc   :   8.547048   8.547048   4.477714
angles:  90.000000  90.000000  90.000000
Sites (30)
  #  SP              a         b         c  Wyckoff
---  -------  --------  --------  --------  ---------
  0  Dummy0+  0.737702  0.063709  0         8i
  1  Dummy0+  0         0         0         2a
  2  Dummy0+  0.463029  0.129472  0         8i
  3  Dummy0+  0.182718  0.182718  0.251726  8j
  4  Dummy0+  0.39991   0.39991   0         4f
\end{minted}

Or we can turn into a useful dict for generating all possible
occupancies of the structure.

\begin{minted}[xleftmargin=3\parindent, linenos=true, fontsize=\small]{python}
spgA = SpacegroupAnalyzer(baseStructure)
uniqueDict = defaultdict(list)
for site, unique in enumerate(spgA.get_symmetry_dataset()['equivalent_atoms']):
    uniqueDict[unique] += [site]
pprint(uniqueDict)
\end{minted}

\begin{minted}[xleftmargin=3\parindent, fontsize=\small, bgcolor=subtlegray]{output}
defaultdict(<class 'list'>,
            {0: [0, 1, 2, 3, 4, 5, 6, 7],
             8: [8, 9],
             10: [10, 11, 12, 13, 14, 15, 16, 17],
             18: [18, 19, 20, 21, 22, 23, 24, 25],
             26: [26, 27, 28, 29]})
\end{minted}

\begin{minted}[xleftmargin=3\parindent, linenos=true, fontsize=\small]{python}
from itertools import product
allPermutations = list(product(['Fe', 'Cr', 'Ni'], repeat=5))
print(
    f'Obtained {len(allPermutations)} permutations of the sublattice occupancy\n'
    'E.g.:  {allPermutations[32]}')
\end{minted}

\begin{minted}[xleftmargin=3\parindent, fontsize=\small, bgcolor=subtlegray]{output}
Obtained 243 permutations of the sublattice occupancy
E.g.:  ('Fe', 'Cr', 'Fe', 'Cr', 'Ni')
\end{minted}

We can now generate them iteratively, as done below:

\begin{minted}[xleftmargin=3\parindent, linenos=true, fontsize=\small]{python}
structList = []
for permutation in allPermutations:
    tempStructure = baseStructure.copy()
    for unique, el in zip(uniqueDict, permutation):
        for site in uniqueDict[unique]:
            tempStructure.replace(site, el)
    structList.append(tempStructure)
print(structList[25])
\end{minted}

\begin{minted}[xleftmargin=3\parindent, fontsize=\small, bgcolor=subtlegray]{output}
Full Formula (Cr4 Fe10 Ni16)
Reduced Formula: Cr2Fe5Ni8
abc   :   8.547048   8.547048   4.477714
angles:  90.000000  90.000000  90.000000
pbc   :       True       True       True
Sites (30)
  #  SP           a         b         c
---  ----  --------  --------  --------
  0  Fe    0.737702  0.063709  0
  1  Fe    0.262298  0.936291  0
  2  Fe    0.436291  0.237702  0.5
  3  Fe    0.762298  0.563709  0.5
  4  Fe    0.563709  0.762298  0.5
  5  Fe    0.237702  0.436291  0.5
  6  Fe    0.063709  0.737702  0
  7  Fe    0.936291  0.262298  0
  8  Fe    0         0         0
  9  Fe    0.5       0.5       0.5
 10  Ni    0.463029  0.129472  0
 11  Ni    0.536971  0.870528  0
 12  Ni    0.370528  0.963029  0.5
 13  Ni    0.036971  0.629472  0.5
 14  Ni    0.629472  0.036971  0.5
 15  Ni    0.963029  0.370528  0.5
 16  Ni    0.129472  0.463029  0
 17  Ni    0.870528  0.536971  0
 18  Ni    0.182718  0.182718  0.251726
 19  Ni    0.817282  0.817282  0.748274
 20  Ni    0.817282  0.817282  0.251726
 21  Ni    0.317282  0.682718  0.751726
 22  Ni    0.317282  0.682718  0.248274
 23  Ni    0.182718  0.182718  0.748274
 24  Ni    0.682718  0.317282  0.248274
 25  Ni    0.682718  0.317282  0.751726
 26  Cr    0.39991   0.39991   0
 27  Cr    0.60009   0.60009   0
 28  Cr    0.10009   0.89991   0.5
 29  Cr    0.89991   0.10009   0.5
\end{minted}

\hypertarget{persisting-on-disk}{%
\subsection{Persisting on Disk}\label{pysipfenntutorial:persisting-on-disk}}

The easiest way to persist a structure on disk is to use the
\texttt{to} method of the
\texttt{Structure} object, which will write the
structure in a variety of formats, including
\texttt{POSCAR} and \texttt{CIF}:

\begin{minted}[xleftmargin=3\parindent, linenos=true, fontsize=\small]{python}
os.mkdir('POSCARs')
os.mkdir('CIFs')
for struct, permutation in zip(structList, allPermutations):
    struct.to(filename='POSCARs/' + "".join(permutation) + '.POSCAR')
    struct.to(filename='CIFs/' + "".join(permutation) + '.cif')
\end{minted}

And now we are ready to use them in a variety of other tools like DFTTK
covered last week or
\href{https://pysipfenn.readthedocs.io/en/stable/}{pySIPFENN} covered
during the next lecture!

\hypertarget{setting-up-mongodb}{%
\section{Setting up MongoDB}\label{pysipfenntutorial:setting-up-mongodb}}

With the ability to manipulate structures locally, one will quickly run
into two major problems:

\begin{itemize}
\item
  \textbf{How to pass them between personal laptop, HPC clusters, and
  lab workstations?}
\item
  \textbf{How do I share them with others later?}
\end{itemize}

One of the easiest ways to do so is to use a cloud-based database, which
will allow us to synchronize our work regardless of what machine we use
and then share it with others in a highly secure way or publicly, as
needed. In this lecture, we will use
\href{https://www.mongodb.com/atlas}{MongoDB Atlas} to set up a small
NoSQL database on the cloud. For our needs and most of the other
personal needs of researchers, the Free Tier will be more than enough,
but if you need more, you can always upgrade to a paid plan for a few
dollars a month if you need to store tens of thousands of structures.

\emph{\textbf{Note for Online Students: At this point, we will pause the
Jupiter Notebook and switch to the MongoDB Atlas website to set up the
database.} The process is fairly straightforward but feel free to stop
by during office hours for help}

Now, we should have the following: - A database called
\texttt{matse580} with a collection called
\texttt{structures} - User with read/write access named
\texttt{student} - API key for the user to access the
database (looks like \texttt{2fnc92niu2bnc9o240dc}) -
Resulting connection string to the database (looks like
\texttt{mongodb+srv://student:2fnc92niu2bnc9o240dc@<cluster\_name>/matse580})
and we can move to populating it with data!

\hypertarget{pymongo}{%
\section{Connecting Pymongo}\label{pysipfenntutorial:pymongo}}

The \texttt{pymongo} is a Python library that allows us
to interact with MongoDB databases in a very intuitive way. Let's start
by importing its \texttt{MongoClient} class and
creating a connection to our database:

\begin{minted}[xleftmargin=3\parindent, linenos=true, fontsize=\small]{python}
from pymongo import MongoClient
uri = 'mongodb+srv://amk7137:
  kASMuF5au1069Go8@cluster0.3wlhaan.mongodb.net/?retryWrites=true&w=majority'
client = MongoClient(uri)
\end{minted}

We can see what databases are available:

\begin{minted}[xleftmargin=3\parindent, linenos=true, fontsize=\small]{python}
client.list_database_names()
\end{minted}

Lets now go back to MongoDB Atlas and create a new database called
\texttt{matse580} and a collection called
\texttt{structures} in it, and hopefully see that they
are /available:

\begin{minted}[xleftmargin=3\parindent, linenos=true, fontsize=\small]{python}
client.list_database_names()
\end{minted}

\begin{minted}[xleftmargin=3\parindent, fontsize=\small, bgcolor=subtlegray]{output}
['matse580', 'admin', 'local']
\end{minted}

To go one level deeper and see what collections are available in the
\texttt{matse580} database we just created, we can use
the \texttt{list\_collection\_names} method:

\begin{minted}[xleftmargin=3\parindent, linenos=true, fontsize=\small]{python}
database = client['matse580']
database.list_collection_names()
\end{minted}

\begin{minted}[xleftmargin=3\parindent, fontsize=\small, bgcolor=subtlegray]{output}
['structures']
\end{minted}

And then read the entries in it!

\begin{minted}[xleftmargin=3\parindent, linenos=true, fontsize=\small]{python}
collection = database['structures']
\end{minted}

\begin{minted}[xleftmargin=3\parindent, linenos=true, fontsize=\small]{python}
for entry in collection.find():
    print(entry)
\end{minted}

But that's not very useful, because we didn't put anything in it yet.

\hypertarget{inserting-data}{%
\section{Inserting Data}\label{pysipfenntutorial:inserting-data}}

We start by constructing our idea of how a structure should be
represented in the database. For that purpose, we will use a dictionary
representation of the structure. This process is very flexible as NoSQL
databases like MongoDB do not require a strict schema and can be
modified on the fly and post-processed later. For our purposes, we will
use the following schema:

\begin{minted}[xleftmargin=3\parindent, linenos=true, fontsize=\small]{python}
def struct2entry(s: Structure):
    # convert to pymatgen Structure dictionary default
    strcutreDict = {'structure': s.as_dict()} 
    # convert to pymatgen Composition dictionary default
    compositionDict = {'composition': s.composition.as_dict()} 
    # merge the two dictionaries
    entry = {**strcutreDict, **compositionDict} 
    # add some extra information
    entry.update({'density': s.density,
                  'volume': s.volume,
                  'reducedFormula': s.composition.reduced_formula,
                  'weightFractions': s.composition.to_weight_dict
                  }) 
    # and a full POSCAR for easy ingestion into VASP
    entry.update({'POSCAR': s.to(fmt='poscar')})
    return entry
\end{minted}

\begin{minted}[xleftmargin=3\parindent, linenos=true, fontsize=\small]{python}
pprint(struct2entry(structList[25]))
\end{minted}

\begin{minted}[xleftmargin=3\parindent, fontsize=\small, bgcolor=subtlegray]{output}
{'POSCAR': 'Cr4 Fe10 Ni16\n'
           ...,
 'composition': {'Cr': 4.0, 'Fe': 10.0, 'Ni': 16.0},
 'density': 8.658038607159655,
 'reducedFormula': 'Cr2Fe5Ni8',
 'structure': {'@class': 'Structure',
               '@module': 'pymatgen.core.structure',
               'charge': 0,
               'lattice': {'a': 8.547048,
                           'alpha': 90.0,
                           'b': 8.547048,
                           'beta': 90.0,
                           'c': 4.477714,
                           'gamma': 90.0,
                           'matrix': [[8.547048, 0.0, 0.0],
                                      [0.0, 8.547048, 0.0],
                                      [0.0, 0.0, 4.477714]],
                           'pbc': (True, True, True),
                           'volume': 327.10609528461225},
               'properties': {},
               'sites': [{'abc': [0.737702, 0.063709, 0.0],
                          'label': 'Fe',
                          'properties': {},
                          'species': [{'element': 'Fe', 'occu': 1}],
                          'xyz': [6.305174403696, 0.544523881032, 0.0]},
                         ...
                         ]},
 'volume': 327.10609528461225,
 'weightFractions': {'Cr': 0.12194716383563854,
                     'Fe': 0.3274351039982438,
                     'Ni': 0.5506177321661175}}
\end{minted}

Looks great! Now we can add some metadata to it, like who created it,
when, and what was the permutation label used to generate it earlier; to
then insert it into the database using the
\texttt{insert\_one} method, which is not the fastest,
but the most flexible way to do so:

\begin{minted}[xleftmargin=3\parindent, linenos=true, fontsize=\small]{python}
for struct, permutation in zip(structList, allPermutations):
    entry = struct2entry(struct)
    entry.update({'permutation': "".join(permutation),
                  'autor': 'Happy Student',
                  'creationDate': datetime.now(ZoneInfo('America/New_York'))
                })
    collection.insert_one(entry)
\end{minted}

We can now quickly check if they are present by counting the number of
entries in the collection:

\begin{minted}[xleftmargin=3\parindent, linenos=true, fontsize=\small]{python}
collection.count_documents({})
\end{minted}

\begin{minted}[xleftmargin=3\parindent, fontsize=\small, bgcolor=subtlegray]{output}
243
\end{minted}

If something went wrong halfway, you can start over by deleting all
entries in the collection (be careful with this one!):

\begin{minted}[xleftmargin=3\parindent, linenos=true, fontsize=\small]{python}
# Uncomment to run
#collection.delete_many({})
#collection.count_documents({})
\end{minted}

\hypertarget{updating-data}{%
\subsection{Updating Data}\label{pysipfenntutorial:updating-data}}

This will be reiterated in the next lecture, but in principle updating
the data is easy. For example, we can add a new field to the document,
like \texttt{averageElectronegativity} by iterating
over all entries present in the collection and calculating it:

\begin{minted}[xleftmargin=3\parindent, linenos=true, fontsize=\small]{python}
for entry in collection.find():
    id = entry['_id']
    s = Structure.from_dict(entry['structure'])
    collection.update_one(
        {'_id': id}, 
        {'$set': {'averageElectronegativity': s.composition.average_electroneg}})
\end{minted}

Or, to remove a field, like \texttt{volume}, which
happens to be the same for all structures, we can do it in a similar
way:

\begin{minted}[xleftmargin=3\parindent, linenos=true, fontsize=\small]{python}
for entry in collection.find():
    id = entry['_id']
    collection.update_one({'_id': id}, {'$unset': {'volume': ''}})
\end{minted}

Since we apply it in the same way on all entries, we can do it in a
single line of code using the \texttt{update\_many}
method and an empty filter \texttt{\{\}} querying all
entries:

\begin{minted}[xleftmargin=3\parindent, linenos=true, fontsize=\small]{python}
collection.update_many({}, {'$unset': {'volume': ''}})
\end{minted}

\begin{minted}[xleftmargin=3\parindent, fontsize=\small, bgcolor=subtlegray]{output}
<pymongo.results.UpdateResult at 0x294323340>
\end{minted}

\hypertarget{querying-data}{%
\subsection{Querying Data}\label{pysipfenntutorial:querying-data}}

Now that we have some data in the database, we can start querying it.
MongoDB has state-of-the-art query language that allows us to do very
complex queries and do them with extreme performance. You can find more
information about it
\href{https://www.mongodb.com/docs/manual/reference/method/db.collection.find/\#db.collection.find}{in
this documentation} but for our purposes, we will stick to the basics
like finding all Cr-containing structures.

To find all entries in the collection, we can use the
\texttt{find} method with a dictionary of query
parameters. We can use many different methods, but the simplest would be
to look for a composition dictionary with over-0 or non-empty values for
Cr:

\begin{minted}[xleftmargin=3\parindent, linenos=true, fontsize=\small]{python}
for entry in collection.find({'weightFractions.Cr': {'$gt': 0}}):
    print(entry['reducedFormula'])
\end{minted}

\begin{minted}[xleftmargin=3\parindent, fontsize=\small, bgcolor=subtlegray]{output}
Cr2Fe13
Cr4Fe11
Cr2Fe3
Cr4Fe9Ni2
Cr2Fe9Ni4
Cr4Fe11
...
\end{minted}

\begin{minted}[xleftmargin=3\parindent, linenos=true, fontsize=\small]{python}
for entry in collection.find({'weightFractions.Cr': {'$ne': None}}):
    print(entry['reducedFormula'])
\end{minted}

\begin{minted}[xleftmargin=3\parindent, fontsize=\small, bgcolor=subtlegray]{output}
Cr2Fe13
Cr4Fe11
Cr2Fe3
Cr4Fe9Ni2
Cr2Fe9Ni4
Cr4Fe11
...
\end{minted}

Or to get a specific permutation, we can use
\texttt{find\_one} method, which will return the first
entry matching the query:

\begin{minted}[xleftmargin=3\parindent, linenos=true, fontsize=\small]{python}
originalStruct25 = collection.find_one({'permutation': 'FeFeNiNiCr'})
originalStruct25['reducedFormula']
\end{minted}

\begin{minted}[xleftmargin=3\parindent, fontsize=\small, bgcolor=subtlegray]{output}
'Cr2Fe5Ni8'
\end{minted}

\hypertarget{pysipfenn-install}{%
\section{pySIPFENN Install}\label{pysipfenntutorial:pysipfenn-install}}

The last quick thing we will do today is to install pySIPFENN, which is
a Python framework which, among other things, allows us to quickly
predict stability of materials using machine learning. It can be
installed using \texttt{pip} just like pymatgen:

\begin{minted}[xleftmargin=3\parindent, linenos=true, fontsize=\small]{python}
#!pip install pysipfenn
\end{minted}

The reason we are installing it here is that the employed models are
fairly large and may take a while to download, unless you use cloud
virtual machine like GitHub Codespaces. Thus, we will start it now so
that it is ready for next week's lecture. Process is automated and you
just need to initialize an empty \texttt{Calculator}
object:

\begin{minted}[xleftmargin=3\parindent, linenos=true, fontsize=\small]{python}
from pysipfenn import Calculator
c = Calculator()
\end{minted}

\begin{minted}[xleftmargin=3\parindent, fontsize=\small, bgcolor=subtlegray, breaklines]{output}
*********  Initializing pySIPFENN Calculator  **********
Loading model definitions from: ...
Found 4 network definitions in models.json
  SIPFENN_Krajewski2020 Standard Materials Model
  SIPFENN_Krajewski2020 Novel Materials Model
  SIPFENN_Krajewski2020 Light Model
  SIPFENN_Krajewski2022 KS2022 Novel Materials Model
Loading all available models (autoLoad=True)
Loading models:

100%|||||||||||||||||||||||||||||||| 4/4 [00:14<00:00,  3.63s/it]

*********  pySIPFENN Successfully Initialized  **********
\end{minted}

And then, order it to download the models:

\begin{minted}[xleftmargin=3\parindent, linenos=true, fontsize=\small]{python}
c.downloadModels()
\end{minted}

\begin{minted}[xleftmargin=3\parindent, fontsize=\small, bgcolor=subtlegray]{output}
Fetching all networks!
All networks available!
  SIPFENN_Krajewski2020 Standard Materials Model
  SIPFENN_Krajewski2020 Novel Materials Model
  SIPFENN_Krajewski2020 Light Model
  SIPFENN_Krajewski2022 KS2022 Novel Materials Model
\end{minted}

It should take 1-30 minutes depending on your internet connection, but
once it is done they will be available until the package is uninstalled.
Also, you can run this command as many times as you want, and it will
only download the models that are not yet present on your system.


\chapter{MatSE580 Guest Lecture 2 - Running ML Models in \texttt{pySIPFENN} and Guiding Limited DFT Calculations Based on \texttt{KS2022} Embedding Cluster Exploration} 
\label{chap:pysipfenntutorial2} 

\hypertarget{introduction}{%
\section{Introduction}\label{pysipfenntutorial2:introduction}}

In this guest lecture, we will cover:
\begin{enumerate}
    \item \protect\hyperlink{verify-the-connection-to-the-database}{\textbf{Interacting with the database we set up in Lecture 1}} and \protect\hyperlink{plotting-with-mongodb-charts}{\textbf{visualizing the results}} - using \href{https://github.com/mongodb/mongo-python-driver}{pymongo} library and \href{https://www.mongodb.com/docs/charts/}{MongoDB Charts} service
    
    \item \protect\hyperlink{pysipfenn}{\textbf{Using machine learning (ML) tools to predict the stability of materials}} - using \href{https://pysipfenn.readthedocs.io/en/stable/}{pySIPFENN} 
    
    \item \protect\hyperlink{featurization}{\textbf{Using ML featurization and dimensionality reduction to embed materials in feature space}} - using \href{https://pysipfenn.readthedocs.io/en/stable/}{pySIPFENN} with \href{https://www.mongodb.com/docs/charts/}{MongoDB Charts} visualization 
    
    \item \protect\hyperlink{transfer-learning-on-small-dft-dataset}{\textbf{Using featurization to guide DFT and improve ML models}}
\end{enumerate}

\textbf{This notebook assumes that you already followed the instructions
in Lecture 1} and that you:

\begin{enumerate}
    \item Have a conda environment called \texttt{580demo} (or other) with all the packages installed, including: \texttt{pymatgen}, \texttt{pymongo}, and \texttt{pysipfenn}.

    \item Have a MongoDB database called \texttt{matse580} with collection \texttt{structures} to which you have access:
    \begin{itemize}
      \item username (e.g.~\texttt{student})
      \item API key / password string
        (e.g.~\texttt{sk39mIM2f35Iwc})
      \item Whitelisted your IP address or \texttt{0.0.0.0/0}
        (entire internet)
      \item Know the connection string (URI) to the database)
    \end{itemize}

    \item You populated the database with all Sigma phase end members (see Lecture 1 Section \ref{pysipfenntutorial:inserting-data} on Inserting Data)

    \item After you installed \texttt{pysipfenn}, you have downloaded all the \href{https://zenodo.org/records/7373089}{pre-trained models} by calling \texttt{downloadModels()} and it finished successfully. If not, run this one liner:
    \begin{minted}[xleftmargin=3\parindent, fontsize=\small, bgcolor=subtlegray, breaklines]{output}
    python -c "import pysipfenn; c = pysipfenn.Calculator(); c.downloadModels(); c.loadModels();"
    \end{minted}
\end{enumerate}

If all of the above are true, you are ready to go!

\begin{minted}[xleftmargin=3\parindent, linenos=true, fontsize=\small]{python}
from pprint import pprint            # pretty printing
from collections import defaultdict  # convenience in the example
import os                            # file handling
from datetime import datetime        # time handling
from zoneinfo import ZoneInfo        # time handling
from pymatgen.core import Structure  # pymatgen
import numpy as np                   # numpy for data manipulation
import plotly.express as px          # plotly for plotting
from importlib import resources      # for accessing the data files
\end{minted}

\label{pysipfenntutorial2:verify-the-connection-to-the-database}

As shown in the Lecture 1, pymongo Python library allows us to interact
with MongoDB databases in a very intuitive way. Let's start by importing
its \texttt{MongoClient} class and creating a
connection to our database we have set up.

\begin{minted}[xleftmargin=3\parindent, linenos=true, fontsize=\small, breaklines]{python}
from pymongo import MongoClient
uri = 'mongodb+srv://amk7137:kASMuF5au1069Go8@cluster0.3wlhaan.mongodb.net'
client = MongoClient(uri)
\end{minted}

And see what databases are available. If you followed the instructions
in Lecture 1 (Appendix \ref{chap:pysipfenntutorial1}), you should see \texttt{matse580} database
in the list.

\begin{minted}[xleftmargin=3\parindent, linenos=true, fontsize=\small]{python}
client.list_database_names()
\end{minted}

\begin{minted}[xleftmargin=3\parindent, fontsize=\small, bgcolor=subtlegray]{output}
['matse580', 'admin', 'local']
\end{minted}

Now connect to \texttt{matse580.structures} collection

\begin{minted}[xleftmargin=3\parindent, linenos=true, fontsize=\small]{python}
collection = client['matse580']['structures']
\end{minted}

And verify that the Sigma phase structures we created are present. You can look up one of them.

\begin{minted}[xleftmargin=3\parindent, linenos=true, fontsize=\small]{python}
print(f'Found: {collection.count_documents({})} structures\n')
pprint(collection.find_one({}, skip=100))
\end{minted}

\begin{minted}[xleftmargin=3\parindent, fontsize=\small, bgcolor=subtlegray]{output}
Found: 243 structures

{'DFT_dH': 0.103503167,
 'POSCAR': 'Cr12 Fe10 Ni8\n'
           '1.0\n'
           '   8.5470480000000002    0.0000000000000000    0.0000000000000000\n'
           '   0.0000000000000000    8.5470480000000002    0.0000000000000000\n'
           '   0.0000000000000000    0.0000000000000000    4.4777139999999997\n'
           'Cr Fe Ni Fe Cr\n'
           '8 2 8 8 4\n'
           'direct\n'
           '   0.7377020000000000    0.0637090000000000    0.0000000000000000 '
           'Cr\n'
           '   0.2622980000000000    0.9362910000000000    0.0000000000000000 '
           ...
           ,
 'SIPFENN_Krajewski2020_NN20': 0.1178784966468811,
 'SIPFENN_Krajewski2020_NN24': 0.11623784899711609,
 'SIPFENN_Krajewski2020_NN9': 0.15312525629997253,
 'TSNE_2D_X': 22.739927291870117,
 'TSNE_2D_Y': 63.28989028930664,
 '_id': ObjectId('6531516032703fc61d08d457'),
 'autor': 'Happy Student',
 'averageElectronegativity': 1.7833333333333334,
 'composition': {'Cr': 12.0, 'Fe': 10.0, 'Ni': 8.0},
 'creationDate': datetime.datetime(2023, 10, 19, 15, 55, 12, 965000),
 'density': 8.38605032076435,
 'permutation': 'CrFeNiFeCr',
 'reducedFormula': 'Cr6Fe5Ni4',
 'structure': {'@class': 'Structure',
               '@module': 'pymatgen.core.structure',
               'charge': 0,
               'lattice': {'a': 8.547048,
                           'alpha': 90.0,
                           'b': 8.547048,
                           'beta': 90.0,
                           'c': 4.477714,
                           'gamma': 90.0,
                           'matrix': [[8.547048, 0.0, 0.0],
                                      [0.0, 8.547048, 0.0],
                                      [0.0, 0.0, 4.477714]],
                           'pbc': [True, True, True],
                           'volume': 327.10609528461225},
               'properties': {},
               'sites': [{'abc': [0.737702, 0.063709, 0.0],
                          'label': 'Cr',
                          'properties': {},
                          'species': [{'element': 'Cr', 'occu': 1}],
                          'xyz': [6.305174403696, 0.544523881032, 0.0]},
                         ...
                         ]},
 'weightFractions': {'Cr': 0.37770698200139663,
                     'Fe': 0.33805494402253244,
                     'Ni': 0.2842380739760709}}
\end{minted}

\label{pysipfenntutorial2:plotting-with-mongodb-charts}

MongoDB Charts is an associated service that allows us to quickly
visualize the data in the database online and share it with others,
while keeping the source data secure and private.

\emph{\textbf{Note for Online Students: At this point, we will pause the
Jupiter Notebook and switch to the MongoDB Atlas website to set up the
database or skip until next week, depending on the available time.} The
process is fairly straightforward, but feel free to stop by office hours
for help}

You should end up with some neat charts like the Figure \ref{pysipfenntutorial2:MongoDBChartExample} below.

\begin{figure}[H]
    \centering
    \includegraphics[width=0.7\textwidth]{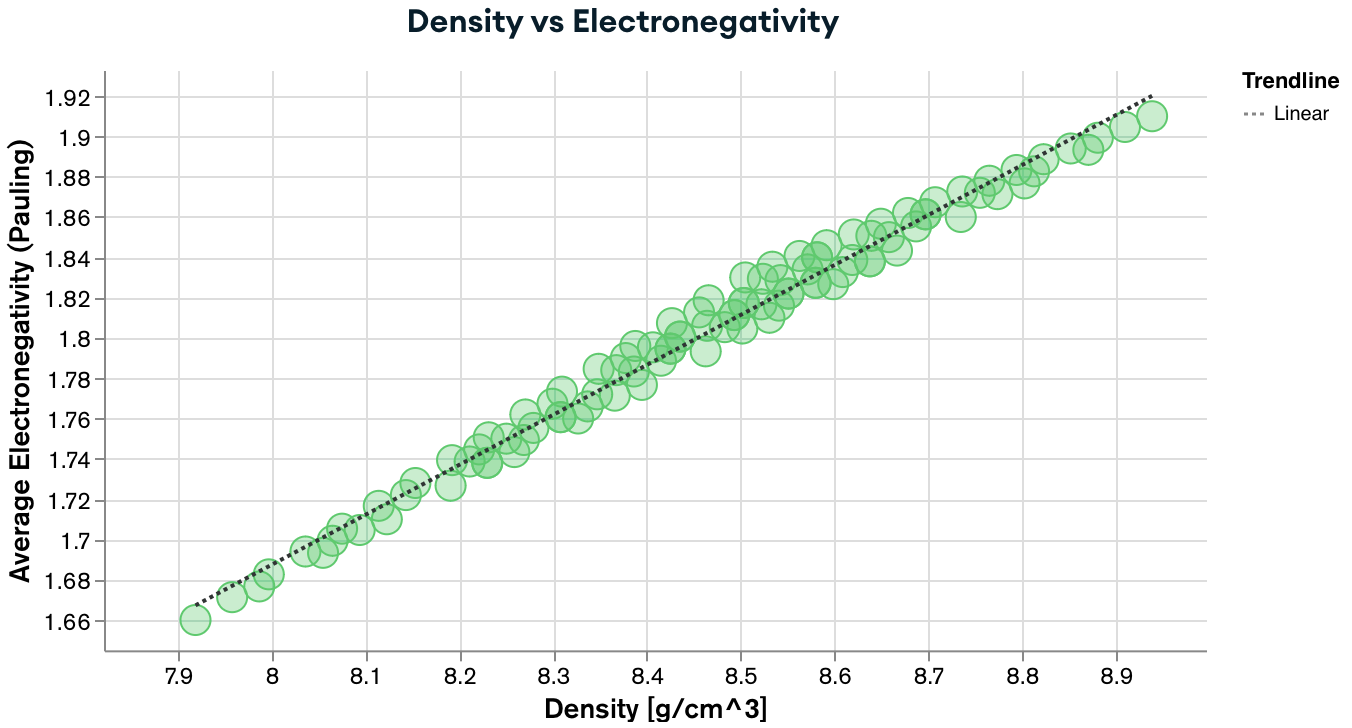}
    \caption{An example of MongoDB Chart showing average Pauling electronegativity vs density of specific intermetallic compounds in a ternary chemical space.}
    \label{pysipfenntutorial2:MongoDBChartExample}
\end{figure}

If you are interested in seeing a couple more examples, you can visit
the dashboard of \href{https://ultera.org}{ULTERA Database} for examples covering high
entropy alloys.

\hypertarget{pysipfenn}{%
\section{pySIPFENN}\label{pysipfenntutorial2:pysipfenn}}

We will now complete a brief walkthrough covering core functionalities
of the \textbf{pySIPFENN} or
\textbf{py}(\textbf{S}tructure-\textbf{I}nformed \textbf{P}rediction of
\textbf{F}ormation \textbf{E}nergy using \textbf{N}eural
\textbf{N}etworks) package; available through the PyPI repository. For a
full up-to-date documentation, please refer to the
\href{https://pysipfenn.org}{pySIPFENN documentation page} or
\href{https://git.pysipfenn.org}{pySIPFENN GitHub repository}. You can
also find news about our projects using SIPFENN at our
\href{https://phaseslab.org}{Phases Research Lab} group website.

On the conceptual level, pySIPFENN is a framework composed of:

\begin{itemize}
\item
  Featurizers / descriptor calculators allowing the user to interpret
  atomic structures (hence \textbf{S}tructure-\textbf{I}nformed) and
  represent them with numbers in a way suitable for machine learning
  (ML) \textbf{P}rediction of properties. A few we created are shipped
  to public; these are Ward2017 (general) and KS2022 (general or
  optimized to different material types) calcualting Ward2017 and KS2022
  feature vectors, respectively. Thanks to how modular pySIPFENN is, you
  can generally just ``steal'' them as standalone modules and use them
  in your own projects.
\item
  It can handle any properties the user wants to predict based purely on
  the model training, but the key fundamental property of interest has
  been \textbf{F}ormation \textbf{E}nergy of materials, and that is what
  is shipped by default with the package.
\item
  It can use any \href{https://onnx.ai}{Open Neural Network Exchange
  (ONNX)} trained on the supported feature vectors (Ward2017 and KS2022
  included). The models shipped by default are \textbf{N}eural
  \textbf{N}etworks, hence the inclusion in the name, but neither
  pySIPFENN nor ONNX is limited to NNs. You can export, for instance,
  complete \texttt{scikit-learn} pipelines (as done
  \href{https://github.com/amkrajewski/cGAN_demo/blob/master/heagan/notebooks/train_surrogates.ipynb}{here
  in heaGAN package}) and use them in pySIPFENN.
\end{itemize}

Figure \ref{pysipfenntutorial2:nncolorized} below shows how they fit together conceptually.

\begin{figure}[H]
    \centering
    \includegraphics[width=0.7\textwidth]{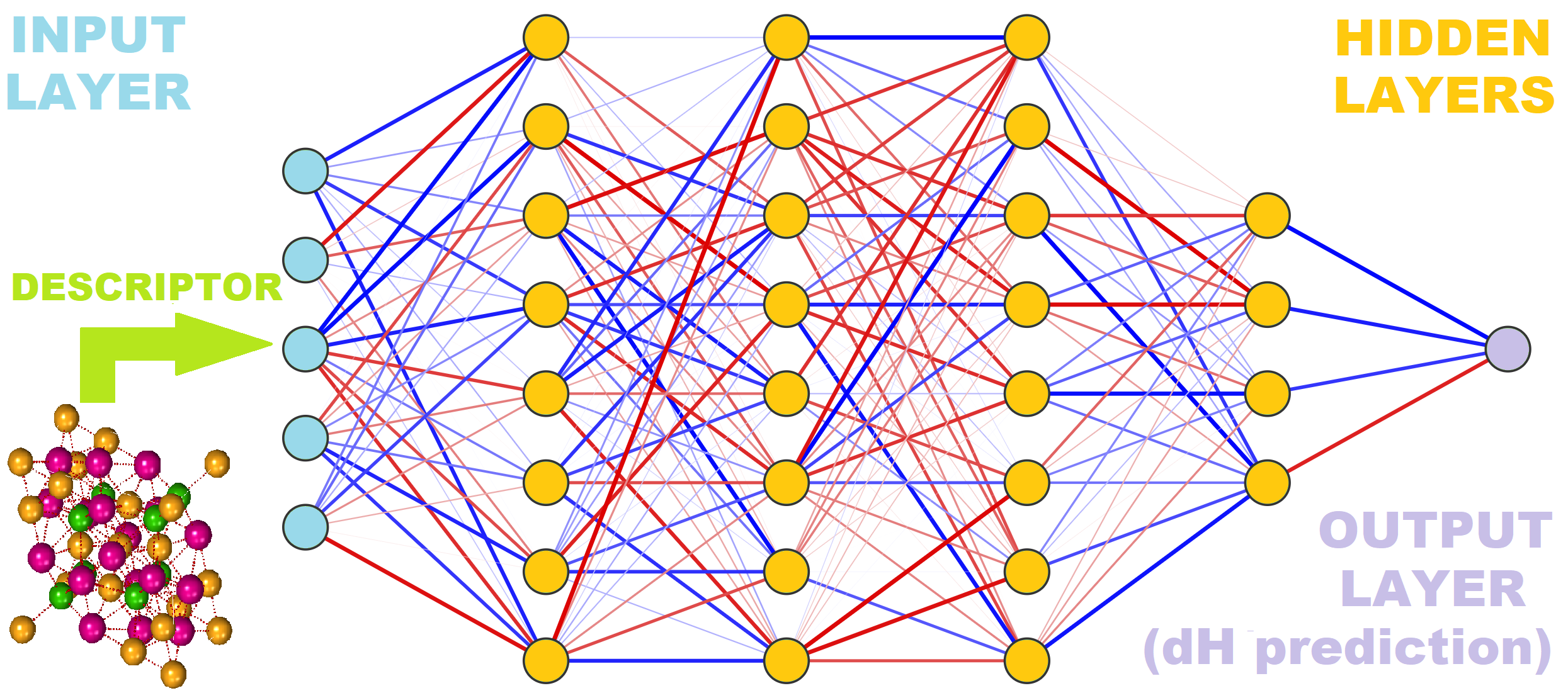}
    \caption{A basic schematic of a ML model pipeline with depictions of featurization step, fully connected feed-forward neural network, and output interpretation.}
    \label{pysipfenntutorial2:nncolorized}
\end{figure}

\hypertarget{getting-started}{%
\subsection{Getting Started}\label{pysipfenntutorial2:getting-started}}

To utilize pySIPFENN for straightforward calculations, \textbf{only the
Calculator class is needed}, which acts as an
\textbf{\emph{environment}} for all package components. Under the hood,
it will do many things for you, including both fetching and
identification of available NN models. Afterwards, it will expose a very
high-level API for you to use.

\begin{minted}[xleftmargin=3\parindent, linenos=true, fontsize=\small]{python}
from pysipfenn import Calculator     # The only thing needed for calculations
\end{minted}

Now initialize the Calculator. When run, this should display all models
detected (e.g.~v SIPFENN\_Krajewski2020 Standard Materials Model) and
those not detected, but declared in the
\texttt{modelsSIPFENN/models.json} file. If some
networks are not detected (prepended with \emph{x}), this may mean the
download (you were to do in Lecture 1) was not completed successfully.
You can try to download them again by calling
\texttt{c.downloadModels()}, which will only download
the missing ones.

\begin{minted}[xleftmargin=3\parindent, linenos=true, fontsize=\small]{python}
c = Calculator()
\end{minted}

\begin{minted}[xleftmargin=3\parindent, fontsize=\small, bgcolor=subtlegray]{output}
*********  Initializing pySIPFENN Calculator  **********
Loading model definitions from: /Users/adam/opt/anaconda3/envs/580demo/lib/
  python3.10/site-packages/pysipfenn/modelsSIPFENN/models.json
Found 4 network definitions in models.json
v SIPFENN_Krajewski2020 Standard Materials Model
v SIPFENN_Krajewski2020 Novel Materials Model
v SIPFENN_Krajewski2020 Light Model
v SIPFENN_Krajewski2022 KS2022 Novel Materials Model
Loading all available models (autoLoad=True)
Loading models:

100%|||||||||||||||||||||||||||||||| 4/4 [00:13<00:00,  3.27s/it]

*********  pySIPFENN Successfully Initialized  **********
\end{minted}

The simplest and most common usage of pySIPFENN is to deploy it on a
directory/folder containing atomic structure files such as POSCAR or
CIF. To of so, one simply specifies its location and which descriptor /
feature vector should be used. The latter determines which ML models
will be run, as they require a list of specific and ordered features as
input.

\begin{minted}[xleftmargin=3\parindent, fontsize=\small, bgcolor=subtlegray]{output}
c.runFromDirectory(directory='myInputFiles', descriptor='KS2022')
\end{minted}

Furthermore, while the exact model can be specified by the user, by
default, all applicable models are run, as the run itself is 1-3 orders
of magnitude faster than descriptor calculation. Following the link
printed during \texttt{Calculator} initialization
reveals which models will be run.

In this demonstration, a set of test files shipped under
\texttt{assets/examplePOSCARS}. Let's run them with
Ward2017 featurizer.

\begin{minted}[xleftmargin=3\parindent, linenos=true, fontsize=\small]{python}
c.runFromDirectory(directory='assets/examplePOSCARS',
                   descriptor='Ward2017');
\end{minted}

\begin{minted}[xleftmargin=3\parindent, fontsize=\small, bgcolor=subtlegray]{output}
Importing structures...

100%|||||||||||||||||||||||||||||||| 6/6 [00:00<00:00, 33.33it/s]

Models that will be run: [
  'SIPFENN_Krajewski2020_NN9', 
  'SIPFENN_Krajewski2020_NN20', 
  'SIPFENN_Krajewski2020_NN24'
]
Calculating descriptors...

100%|||||||||||||||||||||||||||||||| 6/6 [00:04<00:00,  1.35it/s]

Done!
Making predictions...
Prediction rate: 22.0 pred/s
Obtained 6 predictions from:  SIPFENN_Krajewski2020_NN9
Prediction rate: 23.2 pred/s
Obtained 6 predictions from:  SIPFENN_Krajewski2020_NN20
Prediction rate: 132.7 pred/s
Obtained 6 predictions from:  SIPFENN_Krajewski2020_NN24
Done!
\end{minted}

Now, all results are obtained and stored within the \textbf{c}
Calculator object inside a few exposed conveniently named variables
\emph{predictions} and \emph{inputFiles}. Also, the descriptor data is
retained in \emph{descriptorData} if needed. Let's look up all 6
entries. Note that the unit of prediction will depend on the model used;
in this case, it is eV/atom.

\begin{minted}[xleftmargin=3\parindent, linenos=true, fontsize=\small]{python}
pprint(c.inputFiles)
pprint(c.predictions)
\end{minted}

\begin{minted}[xleftmargin=3\parindent, fontsize=\small, bgcolor=subtlegray]{output}
['12-Gd4Cr4O12.POSCAR',
 '13-Fe16Ni14.POSCAR',
 '14-Fe24Ni6.POSCAR',
 '15-Ta4Tl4O12.POSCAR',
 '16-Fe18Ni12.POSCAR',
 '17-Pr4Ga4O12.POSCAR']
[[-3.154766321182251, -3.214848756790161, -3.187128782272339],
 [-0.013867354951798916, 0.04655897989869118, 0.053411152213811874],
 [0.02639671415090561, 0.05997598543763161, 0.06677809357643127],
 [-2.467507839202881, -2.4308743476867676, -2.391871690750122],
 [0.01810809224843979, 0.06462040543556213, 0.10881152749061584],
 [-2.7106518745422363, -2.6583476066589355, -2.727781057357788]]
\end{minted}

For user convenience, a few methods are provided for extracting the
results. E.g., if pySIPFENN has been run from structure files, the
\texttt{get\_resultDictsWithNames()} method is
available to conveniently pass results forward in the code.

\begin{minted}[xleftmargin=3\parindent, linenos=true, fontsize=\small]{python}
c.get_resultDictsWithNames()
\end{minted}

\begin{minted}[xleftmargin=3\parindent, fontsize=\small, bgcolor=subtlegray]{output}
[{'name': '12-Gd4Cr4O12.POSCAR',
  'SIPFENN_Krajewski2020_NN9': -3.154766321182251,
  'SIPFENN_Krajewski2020_NN20': -3.214848756790161,
  'SIPFENN_Krajewski2020_NN24': -3.187128782272339},
 {'name': '13-Fe16Ni14.POSCAR',
  'SIPFENN_Krajewski2020_NN9': -0.013867354951798916,
  'SIPFENN_Krajewski2020_NN20': 0.04655897989869118,
  'SIPFENN_Krajewski2020_NN24': 0.053411152213811874},
 {'name': '14-Fe24Ni6.POSCAR',
  'SIPFENN_Krajewski2020_NN9': 0.02639671415090561,
  'SIPFENN_Krajewski2020_NN20': 0.05997598543763161,
  'SIPFENN_Krajewski2020_NN24': 0.06677809357643127},
 {'name': '15-Ta4Tl4O12.POSCAR',
  'SIPFENN_Krajewski2020_NN9': -2.467507839202881,
  'SIPFENN_Krajewski2020_NN20': -2.4308743476867676,
  'SIPFENN_Krajewski2020_NN24': -2.391871690750122},
 {'name': '16-Fe18Ni12.POSCAR',
  'SIPFENN_Krajewski2020_NN9': 0.01810809224843979,
  'SIPFENN_Krajewski2020_NN20': 0.06462040543556213,
  'SIPFENN_Krajewski2020_NN24': 0.10881152749061584},
 {'name': '17-Pr4Ga4O12.POSCAR',
  'SIPFENN_Krajewski2020_NN9': -2.7106518745422363,
  'SIPFENN_Krajewski2020_NN20': -2.6583476066589355,
  'SIPFENN_Krajewski2020_NN24': -2.727781057357788}]
\end{minted}

Alternatively, if results are to be preserved in a spreadsheet, they can
be exported into a CSV.

\begin{minted}[xleftmargin=3\parindent, linenos=true, fontsize=\small]{python}
c.writeResultsToCSV('myFirstResults_pySIPFENN.csv')
\end{minted}

\hypertarget{predicting-all-sigma-endmembers-from-lecture-1}{%
\subsection{Predicting all Sigma Endmembers from Lecture
1}\label{pysipfenntutorial2:predicting-all-sigma-endmembers-from-lecture-1}}

Now, armed with the power of pySIPFENN, we can quickly get the formation
energies of all Sigma phase endmembers we defined in Lecture 1. We start
by getting all the structures from the database:

\begin{minted}[xleftmargin=3\parindent, linenos=true, fontsize=\small]{python}
structList, idList = [], []
for entry in collection.find({}):
    idList.append(entry['_id'])
    structList.append(Structure.from_dict(entry['structure']))
print(f'Fetched {len(structList)} structures')
\end{minted}

\begin{minted}[xleftmargin=3\parindent, fontsize=\small, bgcolor=subtlegray]{output}
Fetched 243 structures
\end{minted}

Now, we will use \texttt{runModels} function, which is
one layer of abstraction lower than
\texttt{runFromDirectory} as it skips file processing
and directly takes the structure objects. We will set
\texttt{mode='parallel'} to run in parallel, which is
much faster than sequential execution on multi-core machines. Each
thread on a modern CPU should be able to process \textasciitilde1
structure per second, so this should take about a minute.

We will also use \texttt{get\_resultDicts} to get the
results in a convenient format.

\begin{minted}[xleftmargin=3\parindent, linenos=true, fontsize=\small]{python}
c.runModels(structList=structList, descriptor='Ward2017', mode='parallel', max_workers=4)
results = c.get_resultDicts()
\end{minted}

\begin{minted}[xleftmargin=3\parindent, fontsize=\small, bgcolor=subtlegray]{output}
Models that will be run: [
  'SIPFENN_Krajewski2020_NN9', 
  'SIPFENN_Krajewski2020_NN20', 
  'SIPFENN_Krajewski2020_NN24'
]
Calculating descriptors...

Done!
Making predictions...
Prediction rate: 2442.2 pred/s
Obtained 243 predictions from:  SIPFENN_Krajewski2020_NN9
Prediction rate: 2438.8 pred/s
Obtained 243 predictions from:  SIPFENN_Krajewski2020_NN20
Prediction rate: 13656.1 pred/s
Obtained 243 predictions from:  SIPFENN_Krajewski2020_NN24
\end{minted}

\begin{minted}[xleftmargin=3\parindent, linenos=true, fontsize=\small]{python}
pprint(results[0])
\end{minted}

\begin{minted}[xleftmargin=3\parindent, fontsize=\small, bgcolor=subtlegray]{output}
{'SIPFENN_Krajewski2020_NN20': 0.07977379858493805,
 'SIPFENN_Krajewski2020_NN24': 0.03619053587317467,
 'SIPFENN_Krajewski2020_NN9': 0.07845475524663925}
\end{minted}

and now we can easily upload them back to the database, as we learned in
Lecture 1

\begin{minted}[xleftmargin=3\parindent, linenos=true, fontsize=\small]{python}
for id, result in zip(idList, results):
    collection.update_one({'_id': id}, {'$set': result})
\end{minted}

and now they are accessible to anyone with access!

\begin{minted}[xleftmargin=3\parindent, linenos=true, fontsize=\small]{python}
collection.find_one({}, skip=100)['SIPFENN_Krajewski2020_NN9']
\end{minted}

\begin{minted}[xleftmargin=3\parindent, fontsize=\small, bgcolor=subtlegray]{output}
0.15312525629997253
\end{minted}

\hypertarget{featurization}{%
\section{Featurization}\label{pysipfenntutorial2:featurization}}

We've just made some predictions using pySIPFENN and created a dataset
to share with the world. Still, like with most ML materials tools, the
real power comes from featurization, which is often obfuscated from the
user. Fortunately, pySIPFENN is very transparent and allows us to easily
access all \texttt{Ward2017} features for all
structures we just predicted under the
\texttt{descriptorData} variable of the
\texttt{Calculator} object.

\begin{minted}[xleftmargin=3\parindent, linenos=true, fontsize=\small]{python}
print(f'Number of features: {len(c.descriptorData[25])}\nFeature values (first 30):')
pprint(c.descriptorData[25][:30])
print('...\n')
\end{minted}

\begin{minted}[xleftmargin=3\parindent, fontsize=\small, bgcolor=subtlegray]{output}
Number of features: 271
Feature values (first 30):
array([12.66417893,  0.56971497, 11.79388959, 13.794477  ,  0.01616931,
        0.96905691,  1.03318127,  0.05207893,  0.01485447,  0.0272743 ,
        0.0661337 ,  0.03462427,  0.20295946,  0.079708  ,  0.03877594,
        0.57907191,  1.58050955,  0.36763378,  1.15612262,  2.83469893,
        1.6785763 ,  4.74152866,  1.10290133,  3.46836787,  8.50409678,
        5.0357289 ,  2.49699209,  0.65083092,  1.7542278 ,  4.93760804])
...
\end{minted}

With this data available for all 243 endmembers, we have an embedding of
all these structures in so-called feature space. However, it is so
highly dimensional that it is impossible to visualize. Fortunately, we
can use dimensionality reduction techniques to reduce the number of
dimensions to 2 or 3 and visualize the results for human eyes.

We will use TSNE (t-distributed stochastic neighbor embedding) to reduce
the dimensionality to 2 using the \texttt{sklearn}
library. It is not a part of pySIPFENN dependency tree, so you may need
to install it with \texttt{pip} below (after
uncommenting the line).

\begin{minted}[xleftmargin=3\parindent, linenos=true, fontsize=\small]{python}
#!pip install scikit-learn
\end{minted}

\begin{minted}[xleftmargin=3\parindent, linenos=true, fontsize=\small]{python}
from sklearn.manifold import TSNE              # neighborhood  dimensionality reduction
from sklearn.preprocessing import MinMaxScaler # scaling
\end{minted}

We start by copying the \texttt{c.descriptorData},
normalizing it across feature values to minima and maxima using
\texttt{MinMaxScaler} (similar to fixed Max scaler
inside pySIPFENN NNs), and setting up the \texttt{TSNE}
object.

\begin{minted}[xleftmargin=3\parindent, linenos=true, fontsize=\small]{python}
scaler = MinMaxScaler()
descriptorData = scaler.fit_transform(np.array(c.descriptorData))
\end{minted}

We will use \texttt{perplexity=4}, as we don't expect
large clusters, \texttt{n\_iter=1000} to speed up the
calculation, the \texttt{pca} initialization to give a
good starting point, and use
\href{https://docs.scipy.org/doc/scipy/reference/generated/scipy.spatial.distance.correlation.html\#scipy.spatial.distance.correlation}{\texttt{scipy.spatial.distance.correlation}}
as the neighbor distance metric between points.

You are certainly encouraged to play with these parameters to see how
they affect the results!

\begin{minted}[xleftmargin=3\parindent, linenos=true, fontsize=\small]{python}
tsne = TSNE(n_components=2, perplexity=4, init='pca', 
  n_iter=2000, metric='correlation', angle=0.2, n_jobs=-1)
\end{minted}

Now, let's embed the data in 2D space and look at the first 5 results as
a sanity check.

\begin{minted}[xleftmargin=3\parindent, linenos=true, fontsize=\small]{python}
embeddedDescriptorData = tsne.fit_transform(descriptorData)
pprint(embeddedDescriptorData[:5])
\end{minted}

\begin{minted}[xleftmargin=3\parindent, fontsize=\small, bgcolor=subtlegray]{output}
array([[ 41.368214, -60.807182],
       [-22.270252,  72.0951  ],
       [ 45.874958, -56.52266 ],
       [-25.202972,  67.837654],
       [-31.33803 ,  58.68771 ]], dtype=float32)
\end{minted}

Note that the embedding is stochastic and chaotic, so you will get
different results each time you run it. However, you can easily fix it
by setting the \texttt{random\_state} parameter to a
fixed value of \texttt{580}. Note: this value will
persist across the notebook now.

\begin{minted}[xleftmargin=3\parindent, linenos=true, fontsize=\small]{python}
tsne.random_state = 580
\end{minted}

\begin{minted}[xleftmargin=3\parindent, linenos=true, fontsize=\small]{python}
embeddedDescriptorData = tsne.fit_transform(descriptorData)
pprint(embeddedDescriptorData[:5])
\end{minted}

\begin{minted}[xleftmargin=3\parindent, fontsize=\small, bgcolor=subtlegray]{output}
array([[ 41.368214, -60.807182],
       [-22.270252,  72.0951  ],
       [ 45.874958, -56.52266 ],
       [-25.202972,  67.837654],
       [-31.33803 ,  58.68771 ]], dtype=float32)
\end{minted}

Finally, we can plot the results using \texttt{plotly}
library. We will colorize the points by the formation energy coming from
the first model in \texttt{c.predictions} variable. We
will also use \texttt{structList} to get chemical
formulas when hovering over the points.

\begin{minted}[xleftmargin=3\parindent, linenos=true, fontsize=\small]{python}
fig = px.scatter(x=np.transpose(embeddedDescriptorData)[0],
                 y=np.transpose(embeddedDescriptorData)[1],
                 hover_name=[s.formula for s in structList],
                 color=[round(p[0], 3) for p in c.predictions],
                 color_discrete_sequence=px.colors.qualitative.Dark24,
                 template='plotly_white',
                 labels={'x': f'{len(descriptorData[0])}D->2D TSNE1',
                         'y': f'{len(descriptorData[0])}D->2D TSNE2',
                         'color': f'Formation Energy (eV/atom)'},
                 height=400,
                 width=800
                 )
fig.show()
\end{minted}

\begin{figure}
    \centering
    \includegraphics[width=0.9\textwidth]{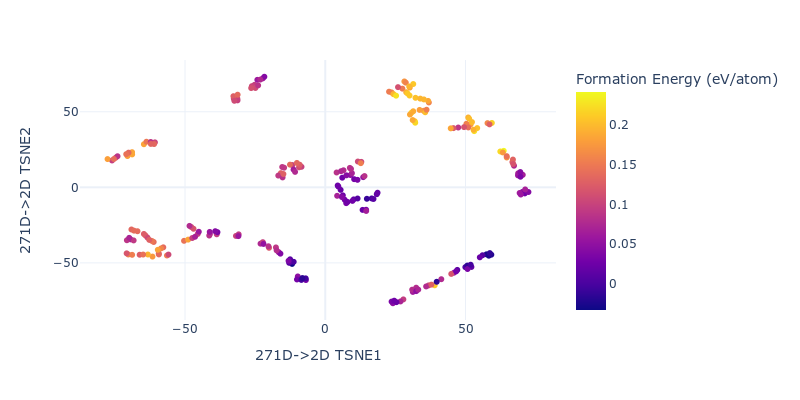}
    \caption{tSNE Embedding of $\sigma$-phase endmembers color-coded based on their predicted formation energies.}
    \label{pysipfenntutorial2:tsnefig1}
\end{figure}

An immediate result one can see in Figure \ref{pysipfenntutorial2:tsnefig1} is that similar structures in the feature space have similar energies in the prediction space, which is a
good sign that the method is working as expected.

Before moving further, let's upload the results to the database so we
can visualize them later in MongoDB Charts if we want to.

\begin{minted}[xleftmargin=3\parindent, linenos=true, fontsize=\small]{python}
for id, embedding in zip(idList, embeddedDescriptorData.tolist()):
    collection.update_one(
        {'_id': id}, 
        {'$set': {'TSNE_2D_X': embedding[0], 'TSNE_2D_Y': embedding[1]}})
\end{minted}

Next, let's check how the data looks like. For clarity, we can ask
\texttt{pymongo} to skip part of the data entry in a
process called
\href{https://docs.mongodb.com/manual/tutorial/project-fields-from-query-results/}{projection}
by setting it to \texttt{0} in a dictionary of fields
directly after the query. In this case, we are skipping the
\texttt{structure} and \texttt{POSCAR}
fields.

\begin{minted}[xleftmargin=3\parindent, linenos=true, fontsize=\small]{python}
collection.find_one({}, {'structure': 0, 'POSCAR': 0}, skip=100)
\end{minted}

\begin{minted}[xleftmargin=3\parindent, fontsize=\small, bgcolor=subtlegray]{output}
{'_id': ObjectId('6531516032703fc61d08d457'),
 'composition': {'Cr': 12.0, 'Fe': 10.0, 'Ni': 8.0},
 'density': 8.38605032076435,
 'reducedFormula': 'Cr6Fe5Ni4',
 'weightFractions': {'Cr': 0.37770698200139663,
  'Fe': 0.33805494402253244,
  'Ni': 0.2842380739760709},
 'permutation': 'CrFeNiFeCr',
 'autor': 'Happy Student',
 'creationDate': datetime.datetime(2023, 10, 19, 15, 55, 12, 965000),
 'averageElectronegativity': 1.7833333333333334,
 'SIPFENN_Krajewski2020_NN20': 0.1178784966468811,
 'SIPFENN_Krajewski2020_NN24': 0.11623784899711609,
 'SIPFENN_Krajewski2020_NN9': 0.15312525629997253,
 'TSNE_2D_X': 22.739927291870117,
 'TSNE_2D_Y': 63.28989028930664,
 'DFT_dH': 0.103503167}
\end{minted}

\hypertarget{transfer-learning-on-small-dft-dataset}{%
\section{Transfer Learning on Small DFT
Dataset}\label{pysipfenntutorial2:transfer-learning-on-small-dft-dataset}}

Now, let's look at an example of how we can use this data to guide DFT
calculations!

There are a few possible ways, but we will focus on transfer learning,
which is a technique of using a model trained on a large dataset
(e.g.~all OQMD, like \texttt{NN20} we used, or
Materials Project) and fine-tuning it on a smaller dataset (e.g.~10 DFT
calculations of Sigma phase endmembers).

\begin{minted}[xleftmargin=3\parindent, linenos=true, fontsize=\small]{python}
dftN = 12
\end{minted}

\hypertarget{selecting-a-subset-of-data}{%
\subsection{Selecting a Subset of the
Data}\label{pysipfenntutorial2:selecting-a-subset-of-data}}

\textbf{\emph{The most critical step to getting good performance here is
the selection of a good small subset we will train on.}}

This can be done in many ways, but the baseline is a random selection
like:

\begin{minted}[xleftmargin=3\parindent, linenos=true, fontsize=\small]{python}
import random

randomEndmembersList = random.sample(range(len(descriptorData)), dftN)
print(randomEndmembersList)
\end{minted}

\begin{minted}[xleftmargin=3\parindent, fontsize=\small, bgcolor=subtlegray]{output}
[76, 79, 189, 45, 195, 124, 219, 112, 6, 186, 86, 75]
\end{minted}

However, with the feature space embedding, we can \emph{on average} do
better than that! Let's look for some representative structures in the
feature space by identifying cluster centers with
\texttt{KMeans} clustering from, again,
\texttt{sklearn} library, and locating points nearest
to them by computing \texttt{pairwise\_distances}
matrix between all points and cluster centers.

\begin{minted}[xleftmargin=3\parindent, linenos=true, fontsize=\small]{python}
from sklearn.cluster import KMeans                          # clustering method
from sklearn.metrics import pairwise_distances_argmin_min   # distance metric

kmeansClustering = KMeans(n_clusters=dftN, n_init=500, max_iter=1000, random_state=580)
clusterData = kmeansClustering.fit(embeddedDescriptorData)
print(clusterData.cluster_centers_)
\end{minted}

\begin{minted}[xleftmargin=3\parindent, fontsize=\small, bgcolor=subtlegray]{output}
[[-64.198555   -38.431873  ]
 [-12.584402    12.041971  ]
 [ 51.93679     41.060352  ]
 [ 52.304935   -50.383396  ]
 [-14.359932   -48.381523  ]
 [-27.307589    65.01916   ]
 [-68.85164     23.871618  ]
 [ 67.888756     9.928222  ]
 [ 10.148277     0.46234655]
 [ 30.642172    57.845394  ]
 [-42.25075    -30.75432   ]
 [ 31.958817   -69.17891   ]]
\end{minted}

\begin{figure}
    \centering
    \includegraphics[width=0.9\textwidth]{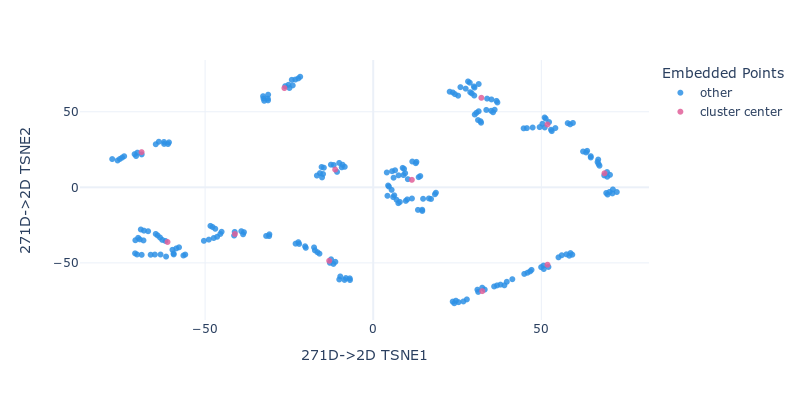}
    \caption{tSNE Embedding of $\sigma$-phase endmembers after clustering, with predefined number of cluster-centers selected as representative examples to be later passed to "virtual" DFT calculations and model improvements.}
    \label{pysipfenntutorial2:tsnefig2}
\end{figure}

\begin{minted}[xleftmargin=3\parindent, linenos=true, fontsize=\small]{python}
clusterCenters, _ = pairwise_distances_argmin_min(
  clusterData.cluster_centers_, embeddedDescriptorData)
print(clusterCenters)
\end{minted}

\begin{minted}[xleftmargin=3\parindent, fontsize=\small, bgcolor=subtlegray]{output}
[ 89 219  14 216 239 108  13 124  63 156  25 182]
\end{minted}

which we can now plot on top of the TSNE embedding we made earlier, as depicted in Figure \ref{pysipfenntutorial2:tsnefig2}.

\begin{minted}[xleftmargin=3\parindent, linenos=true, fontsize=\small]{python}
fig = px.scatter(x=np.transpose(embeddedDescriptorData)[0],
                 y=np.transpose(embeddedDescriptorData)[1],
                 hover_name=[s.formula for s in structList],
                 color=['cluster center' if i in clusterCenters else 'other' 
                        for i in range(len(embeddedDescriptorData))],
                 opacity=0.85,
                 color_discrete_sequence=px.colors.qualitative.Dark24,
                 template='plotly_white',
                 labels={'x': f'{len(descriptorData[0])}D->2D TSNE1',
                         'y': f'{len(descriptorData[0])}D->2D TSNE2',
                         'color': f'Embedded Points'},
                 height=400,
                 width=800
                 )
fig.show()
\end{minted}

\hypertarget{our-virtual-hpc-run}{%
\subsection{Our "Virtual" HPC Run}\label{pysipfenntutorial2:our-virtual-hpc-run}}

Before we go forward, in order to prove that a selection of points for
DFT was good, we need to calculate all of them, even though only a small
subset will be used for training.

To do so, we will now take results pre-calculated with DFTTK (in
\texttt{assets/sigma.csv}) and insert them into the
database; pretending they were calculated on a HPC cluster. The result
will be functionally the same.

Note we will be matching them by permutation (with the same order of
elements we used earlier) and not by the \texttt{id} as
we did earlier, since the order of results is not guaranteed to be the
same.

\begin{minted}[xleftmargin=3\parindent, linenos=true, fontsize=\small]{python}
with open('assets/sigma.csv', 'r') as sigmaData:
    for l in sigmaData.readlines()[1:]:
        lSplit = l.split(',')
        permutation = "".join(lSplit[0:5])
        DFT_dH = float(lSplit[8])
        collection.update_one(
            {'permutation': permutation}, 
            {'$set': {'DFT_dH': DFT_dH}})
\end{minted}

\begin{minted}[xleftmargin=3\parindent, linenos=true, fontsize=\small]{python}
collection.find_one({}, {'structure': 0, 'POSCAR': 0}, skip=100)
\end{minted}

\begin{minted}[xleftmargin=3\parindent, fontsize=\small, bgcolor=subtlegray]{output}
{'_id': ObjectId('6531516032703fc61d08d457'),
 'composition': {'Cr': 12.0, 'Fe': 10.0, 'Ni': 8.0},
 'density': 8.38605032076435,
 'reducedFormula': 'Cr6Fe5Ni4',
 'weightFractions': {'Cr': 0.37770698200139663,
  'Fe': 0.33805494402253244,
  'Ni': 0.2842380739760709},
 'permutation': 'CrFeNiFeCr',
 'autor': 'Happy Student',
 'creationDate': datetime.datetime(2023, 10, 19, 15, 55, 12, 965000),
 'averageElectronegativity': 1.7833333333333334,
 'SIPFENN_Krajewski2020_NN20': 0.1178784966468811,
 'SIPFENN_Krajewski2020_NN24': 0.11623784899711609,
 'SIPFENN_Krajewski2020_NN9': 0.15312525629997253,
 'TSNE_2D_X': 22.739927291870117,
 'TSNE_2D_Y': 63.28989028930664,
 'DFT_dH': 0.103503167}
\end{minted}

\hypertarget{fine-tuning-the-models}{%
\subsection{Fine-Tuning the Models}\label{pysipfenntutorial2:fine-tuning-the-models}}

Now, we are ready to fine-tune the models! In the (near) future release
of pySIPFENN, this will be done using a high-level API (including
hyperparameter tuning); thus, if you are following this lecture in the
future, you may want to check out the documentation for the latest
version. For now, we will do it manually by calling many
\texttt{torch} and \texttt{onnx}
functions, which won't be explained in this lecture, but I am happy to
discuss them during office hours.

If you are interested in learning more about what we are doing here, you
can have a look at
\href{https://www.sciencedirect.com/science/article/pii/S0927025622000593?via\%3Dihub\#sec3}{Section
3.5 in the SIPFENN paper}.

\begin{minted}[xleftmargin=3\parindent, linenos=true, fontsize=\small]{python}
import torch
import onnx
import onnx2torch
import random
\end{minted}

Pull DFT data from the database and convert it to a
\texttt{torch} tensor. The
\texttt{\{'DFT\_dH': 1\}} after the query is a
projection, which means we only want to get the
\texttt{DFT\_dH} field from the database. This is a
good practice to reduce the amount of data transferred over the network.

\begin{minted}[xleftmargin=3\parindent, linenos=true, fontsize=\small]{python}
labelTensor = torch.from_numpy(np.array(
  [[collection.find_one({'_id': id}, {'DFT_dH': 1})['DFT_dH']] 
  for id in idList])).float()
print(labelTensor[:10])
\end{minted}

\begin{minted}[xleftmargin=3\parindent, fontsize=\small, bgcolor=subtlegray]{output}
tensor([[0.0805],
        [0.0778],
        [0.0802],
        [0.0935],
        [0.0809],
        [0.1118],
        [0.0656],
        [0.0678],
        [0.0702],
        [0.0867]])
\end{minted}

Convert the \texttt{numpy} array to
\texttt{torch} tensor.

\begin{minted}[xleftmargin=3\parindent, linenos=true, fontsize=\small]{python}
ddTensor = torch.from_numpy(descriptorData).float()
\end{minted}

Load the underlying model, in this case \texttt{NN24}
as it is the lightest on memory and fasterst to tune, and get it ready.
Alternatively, you could use \texttt{loadedModels} from
\texttt{c} by copying them with
\texttt{copy.deepcopy}, but that would be slower and
use more memory than necessary.

\begin{minted}[xleftmargin=3\parindent, linenos=true, fontsize=\small]{python}
with resources.files('pysipfenn').joinpath(
  'modelsSIPFENN/SIPFENN_Krajewski2020_NN24.onnx') as nn24model:
    model = onnx2torch.convert(onnx.load(nn24model))
    model.eval()
\end{minted}

Set the optimizer and MAE loss function.

\begin{minted}[xleftmargin=3\parindent, linenos=true, fontsize=\small]{python}
optimizer = torch.optim.Adam(model.parameters(), lr=0.00005)
loss = torch.nn.L1Loss()
\end{minted}

And before we go further, let's see the current performance of the model
on the DFT data we just pulled.

\begin{minted}[xleftmargin=3\parindent, linenos=true, fontsize=\small]{python}
dataOut = model(ddTensor, None)
print(dataOut[:10])
\end{minted}

\begin{minted}[xleftmargin=3\parindent, fontsize=\small, bgcolor=subtlegray]{output}
tensor([[ 0.1671],
        [ 0.3394],
        [ 0.1794],
        [ 0.5271],
        [ 0.4225],
        [ 0.4708],
        [ 0.2241],
        [ 0.4732],
        [-0.2594],
        [ 0.4692]], grad_fn=<SliceBackward0>)
\end{minted}

\begin{minted}[xleftmargin=3\parindent, linenos=true, fontsize=\small]{python}
loss(dataOut, labelTensor)
\end{minted}

\begin{minted}[xleftmargin=3\parindent, fontsize=\small, bgcolor=subtlegray]{output}
tensor(0.2319, grad_fn=<MeanBackward0>)
\end{minted}

\hypertarget{random-selection}{%
\subsection{Random Selection}\label{pysipfenntutorial2:random-selection}}

Randomly select the subset of data to train on.

\begin{minted}[xleftmargin=3\parindent, linenos=true, fontsize=\small]{python}
transferIndexes = 
  random.sample(range(len(descriptorData)), dftN)
validationIndexes = 
  list(set(range(len(descriptorData))).difference(transferIndexes))
print(transferIndexes)
\end{minted}

\begin{minted}[xleftmargin=3\parindent, fontsize=\small, bgcolor=subtlegray]{output}
[76, 190, 11, 0, 84, 218, 87, 219, 79, 52, 164, 26]
\end{minted}

Cherry-pick the data creating new tensors

\begin{minted}[xleftmargin=3\parindent, linenos=true, fontsize=\small]{python}
transferData = 
  torch.index_select(ddTensor, 0, torch.LongTensor(transferIndexes)).float()
transferLabels = 
  torch.index_select(labelTensor, 0, torch.LongTensor(transferIndexes)).float()

validationData = 
  torch.index_select(ddTensor, 0, torch.LongTensor(validationIndexes)).float()
validationLabels = 
  torch.index_select(labelTensor, 0, torch.LongTensor(validationIndexes)).float()
\end{minted}

\textbf{\emph{And finally, train the model}}

\begin{minted}[xleftmargin=3\parindent, linenos=true, fontsize=\small, breaklines]{python}
model.eval()
transferLosses = [float(loss(model(transferData, None), transferLabels))]
validationLosses = [float(loss(model(validationData, None), validationLabels))]

model.train()
print('Initial Mean Absolute Errors (MAE):')
print(
    f'Train: {round(1000 * float(transferLosses[0]), 1):>5}  |  Val: {round(1000 * float(validationLosses[0]), 1):>5}   [meV/atom]')
print('Starting Training...')
for i in range(250):
    optimizer.zero_grad()
    transfer_pred = model(transferData, None)
    transfer_loss = loss(transfer_pred, transferLabels)
    transferLosses.append(float(transfer_loss))
    transfer_loss.backward()
    optimizer.step()
    validationLosses.append(float(loss(model(validationData, None), validationLabels)))
    if i % 10 == 0:
        print(
            f'Epoch {i:>3}:  Train: {round(1000 * float(transferLosses[-1]), 1):>5}  |  Val: {round(1000 * float(validationLosses[-1]), 1):>5}   [meV/atom]')
print('Training Complete!')
model.eval()
transferLosses.append(float(loss(model(transferData, None), transferLabels)))
validationLosses.append(float(loss(model(validationData, None), validationLabels)))
print('Final Evaluation Mean Absolute Error (MAE):')
print(f'Train: {round(1000 * float(transferLosses[-1]), 1):>5}  |  Val: {round(1000 * float(validationLosses[-1]), 1):>5}')
\end{minted}

\begin{minted}[xleftmargin=3\parindent, fontsize=\small, bgcolor=subtlegray]{output}
Initial Mean Absolute Errors (MAE):
Train: 263.6  |  Val: 230.2   [meV/atom]
Starting Training...
Epoch   0:  Train: 266.0  |  Val: 202.5   [meV/atom]
Epoch  10:  Train:  86.8  |  Val: 124.6   [meV/atom]
Epoch  20:  Train:  53.9  |  Val: 100.7   [meV/atom]
Epoch  30:  Train:  37.8  |  Val:  69.9   [meV/atom]
Epoch  40:  Train:  33.1  |  Val:  63.9   [meV/atom]
Epoch  50:  Train:  24.9  |  Val:  56.0   [meV/atom]
Epoch  60:  Train:  20.0  |  Val:  56.6   [meV/atom]
Epoch  70:  Train:  25.9  |  Val:  53.5   [meV/atom]
Epoch  80:  Train:  19.7  |  Val:  50.4   [meV/atom]
Epoch  90:  Train:  23.4  |  Val:  45.9   [meV/atom]
Epoch 100:  Train:  19.4  |  Val:  48.1   [meV/atom]
Epoch 110:  Train:  35.5  |  Val:  44.9   [meV/atom]
Epoch 120:  Train:  21.7  |  Val:  45.4   [meV/atom]
Epoch 130:  Train:  13.4  |  Val:  42.8   [meV/atom]
Epoch 140:  Train:  20.2  |  Val:  38.7   [meV/atom]
Epoch 150:  Train:  13.2  |  Val:  44.2   [meV/atom]
Epoch 160:  Train:  17.0  |  Val:  39.5   [meV/atom]
Epoch 170:  Train:  14.5  |  Val:  38.7   [meV/atom]
Epoch 180:  Train:  21.5  |  Val:  38.9   [meV/atom]
Epoch 190:  Train:  15.2  |  Val:  38.8   [meV/atom]
Epoch 200:  Train:  12.5  |  Val:  38.1   [meV/atom]
Epoch 210:  Train:  12.6  |  Val:  38.7   [meV/atom]
Epoch 220:  Train:  12.9  |  Val:  39.5   [meV/atom]
Epoch 230:  Train:  19.4  |  Val:  36.4   [meV/atom]
Epoch 240:  Train:  15.2  |  Val:  38.2   [meV/atom]
Training Complete!
Final Evaluation Mean Absolute Error (MAE):
Train:   6.8  |  Val:  30.1
\end{minted}

\hypertarget{feature-space-informed-selection}{%
\subsection{Feature-Space-Informed
Selection}\label{pysipfenntutorial2:feature-space-informed-selection}}

Now, lets do the same, but using the subset of data we selected based on
the feature space embedding and assigned to
\texttt{clusterCenters} variable.

Start by reloading feature data from pySIPFENN.

\begin{minted}[xleftmargin=3\parindent, linenos=true, fontsize=\small, breaklines]{python}
with resources.files('pysipfenn').joinpath('modelsSIPFENN/SIPFENN_Krajewski2020_NN24.onnx') as nn24model:
    model = onnx2torch.convert(onnx.load(nn24model))
    model.eval()

optimizer = torch.optim.Adam(model.parameters(), lr=0.00005)
loss = torch.nn.L1Loss()
\end{minted}

Select the subset with \texttt{clusterCenters} and
convert it to \texttt{torch} tensor.

\begin{minted}[xleftmargin=3\parindent, linenos=true, fontsize=\small]{python}
transferIndexes = clusterCenters
validationIndexes = list(set(range(len(descriptorData))).difference(clusterCenters))
print(transferIndexes)
\end{minted}

\begin{minted}[xleftmargin=3\parindent, fontsize=\small, bgcolor=subtlegray]{output}
[ 89 219  14 216 239 108  13 124  63 156  25 182]
\end{minted}

\begin{minted}[xleftmargin=3\parindent, linenos=true, fontsize=\small]{python}
transferData = 
    torch.index_select(ddTensor, 0, torch.LongTensor(transferIndexes)).float()
transferLabels = 
    torch.index_select(labelTensor, 0, torch.LongTensor(transferIndexes)).float()

validationData = 
    torch.index_select(ddTensor, 0, torch.LongTensor(validationIndexes)).float()
validationLabels = 
    torch.index_select(labelTensor, 0, torch.LongTensor(validationIndexes)).float()
\end{minted}

\begin{minted}[xleftmargin=3\parindent, linenos=true, fontsize=\small, breaklines]{python}
model.eval()
transferLosses = [float(loss(model(transferData, None), transferLabels))]
validationLosses = [float(loss(model(validationData, None), validationLabels))]

model.train()
print('Initial Mean Absolute Errors (MAE):')
print(f'Train: {round(1000 * float(transferLosses[0]), 1):>5}  |  Val: {round(1000 * float(validationLosses[0]), 1):>5}   [meV/atom]')
print('Starting Training...')
for i in range(250):
    optimizer.zero_grad()
    transfer_pred = model(transferData, None)
    transfer_loss = loss(transfer_pred, transferLabels)
    transferLosses.append(float(transfer_loss))
    transfer_loss.backward()
    optimizer.step()
    validationLosses.append(float(loss(model(validationData, None), validationLabels)))
    if i % 10 == 0:
        print(f'Epoch {i:>3}:  Train: {round(1000 * float(transferLosses[-1]), 1):>5}  |  Val: {round(1000 * float(validationLosses[-1]), 1):>5}')
print('Training Complete!')
model.eval()
transferLosses.append(float(loss(model(transferData, None), transferLabels)))
validationLosses.append(float(loss(model(validationData, None), validationLabels)))
print('Final Evaluation Mean Absolute Error (MAE):')
print(f'Train: {round(1000 * float(transferLosses[-1]), 1):>5}  |  Val: {round(1000 * float(validationLosses[-1]), 1):>5}')
\end{minted}

\begin{minted}[xleftmargin=3\parindent, fontsize=\small, bgcolor=subtlegray]{output}
Initial Mean Absolute Errors (MAE):
Train: 266.1  |  Val: 230.1   [meV/atom]
Starting Training...
Epoch   0:  Train: 297.3  |  Val: 202.5
Epoch  10:  Train: 121.2  |  Val: 118.8
Epoch  20:  Train:  20.9  |  Val:  76.5
Epoch  30:  Train:  35.6  |  Val:  56.4
Epoch  40:  Train:  40.7  |  Val:  50.5
Epoch  50:  Train:  33.6  |  Val:  48.2
Epoch  60:  Train:  31.6  |  Val:  45.6
Epoch  70:  Train:  25.6  |  Val:  42.3
Epoch  80:  Train:  29.8  |  Val:  43.7
Epoch  90:  Train:  20.5  |  Val:  41.7
Epoch 100:  Train:  23.3  |  Val:  41.8
Epoch 110:  Train:  13.6  |  Val:  41.6
Epoch 120:  Train:  18.4  |  Val:  35.7
Epoch 130:  Train:  20.6  |  Val:  38.5
Epoch 140:  Train:  15.7  |  Val:  34.5
Epoch 150:  Train:  17.5  |  Val:  32.5
Epoch 160:  Train:  20.6  |  Val:  33.3
Epoch 170:  Train:  17.3  |  Val:  36.0
Epoch 180:  Train:  12.2  |  Val:  30.6
Epoch 190:  Train:  17.3  |  Val:  31.9
Epoch 200:  Train:  17.4  |  Val:  29.5
Epoch 210:  Train:  16.2  |  Val:  30.0
Epoch 220:  Train:  11.7  |  Val:  30.5
Epoch 230:  Train:   8.4  |  Val:  28.3
Epoch 240:  Train:  13.9  |  Val:  32.8
Training Complete!
Final Evaluation Mean Absolute Error (MAE):
Train:   9.5  |  Val:  26.3
\end{minted}

Now, \textbf{on average you should see a result around 25meV/atom}
depending on the run, which should be around 20-30\% reduction in MAE,
which is a significant improvement! However, the real power of this
approach lays in its consistency, as if you repeat this process many
times, the random selection may sometimes happen to be better, but it
will ocassionally be two or three times higher.

\hypertarget{conclusions-and-further-resources}{%
\section{Conclusions and Further
Resources}\label{pysipfenntutorial2:conclusions-and-further-resources}}

If you were able to complete this notebook, you should now have a good
basic understanding of how manipulte atomic configurations in Python,
send it back and forth to MongoDB, and use pySIPFENN to (1) predict
formation energy, (2) featurize structures, and (3) tune ML models.

Here are some additional resources you may find useful if you want to
learn more:

\begin{itemize}
\item
  \href{https://www.mongodb.com/try/download/compass}{MongoDB Compass
  GUI Application} which will allow you to interact with the database in
  a very user-friendly way, including basic visualization of the data,
  testing your queries, analyzing the structure of the database, and
  more.
\item
  \href{https://docs.mongodb.com/manual/tutorial/query-documents/}{MongoDB
  Query Language (MQL) documentation}
\item
  \href{https://pysipfenn.org}{pySIPFENN documentation} and
  \href{https://git.pysipfenn.org}{pySIPFENN GitHub repository}
\item
  \href{https://www.sciencedirect.com/science/article/pii/S0927025622000593?via\%3Dihub\#sec3}{Results
  section of SIPFENN Paper} for a discussion of how and why we do
  transfer learning. Including reasoning behing some hyperparameters we
  used here.
\end{itemize}


\chapter{\texttt{nimCSO} Basic Tutorial on Selecting Elements for High Entropy Alloy Modeling} \label{chap:nimcsotutorial}

The purpose of this guide is to demonstrate some common use cases of
\texttt{nimCSO} and go in a bit more into the details
of how it could be used, but it is not by any means extensive. If
something is not covered but you would like to see it here, please do
not hesitate to open an issue on GitHub and let use know!

\hypertarget{dataset-config-and-compilation}{%
\section{Dataset, Config, and
Compilation}\label{nimcsotutorial:dataset-config-and-compilation}}

To get started, let's first recap what we need to do to get
\texttt{nimCSO} up and running.

\textbf{1.} Install \texttt{nim} and dependencies, but \textbf{that's already
done for you if you are in the Codespace}. You can see what was run to
get the environment set up in the
\href{../.devcontainer/Dockerfile}{\texttt{Dockerfile}}.

\textbf{2.} Create the dataset. For now, let's just use the default one
(based on ULTERA Database) that comes with the package. Relative to this
notebook, the dataset is located at
\texttt{../dataList.txt}. Let's have a look at the
first few lines of the file to see what it looks like.

\begin{minted}[xleftmargin=3\parindent, linenos=true, fontsize=\small]{shell}
!head -n 8 ../dataList.txt
\end{minted}

\begin{minted}[xleftmargin=3\parindent, fontsize=\small, bgcolor=subtlegray]{output}
Al,Co,Cr,Cu,Fe,Ni
Nb,Ta,Ti
Co,Cr,Ni
Al,Co,Cr,Fe,Mn,Ni
Al,Co,Fe,Mo,Ni
Hf,Nb,Ti,V
Co,Cr,Fe,Nb,Ni
Al,Co,Cr,Cu,Fe,Ni
\end{minted}

\textbf{3.} Now, we need to create task
\texttt{config.yaml} file that will describe what we
are doing and point to our data file. That was already done for you in
the \href{config.yaml}{\texttt{config.yaml}} file, but
you are more than welcome to play and modify it.

\textbf{4.} Finally, we can run the \texttt{nimCSO}
package to get the results. To do so, we will use one long command you
can see below. Let's break it down:

\begin{itemize}
\item
  \texttt{nim} is the official Nim language compiler.
\item
  \texttt{c} instructs \texttt{nim}
  compiler to use \texttt{C} compiler to optimize and
  compile intermediate code. You can also use
  \texttt{cpp} to use \texttt{C++}
  compiler or \texttt{objc} to use
  \texttt{Objective-C} compiler. If you want, you can
  also compile directly with LLVM using
  \href{https://github.com/arnetheduck/nlvm}{\texttt{nlvm}},
  but it isn't pre-installed for you here.
\item
  \texttt{-f} is a flag to force the compiler to
  compile everything, even if the code didn't change. We want this
  because \texttt{config.yaml}, which tells
  \texttt{nimCSO} how to write itself, is not tracked
  by the compiler, but is critical to the compilation process (see two
  point below).
\item
  \texttt{-d:release} is a flag that tells the compiler
  to optimize the code for release. You can also use
  \texttt{-d:debug} to compile the code with better
  debugging support, but it will be slower and it will not prevent bugs
  from happening. There is also \texttt{-d:danger} that
  will disable all runtime checks and run a bit faster, but you no
  longer get memory safety guarantees.
\item
  \texttt{-d:configPath=config.yaml} is a flag pointing
  to \textbf{\texttt{config.yaml} that is read and
  tells \texttt{nimCSO} (not the compiler!) how to
  write itself \emph{before} the compilation starts.} That's the magic
  metaprogramming sauce enabling us to write functions which
  \texttt{C}/\texttt{C++} compiler can
  then turn into single deterministically allocated and exectuted
  machine code through
  \href{https://en.wikipedia.org/wiki/Inline_expansion}{inlining}.
\item
  \texttt{out:nimcso} is just telling the compiler to
  output the compiled binary right here and name it
  \texttt{nimcso}. You can name it whatever you want,
  but it's a good idea to name it something that makes sense.
\item
  \texttt{../src/nimcso} is pointing to the source code
  of \texttt{nimCSO} package to compile, relative to
  this notebook.
\end{itemize}

Let's run the command and see what happens! Shouldn't take more than a
few seconds.

\begin{minted}[xleftmargin=3\parindent, linenos=true, fontsize=\small]{shell}
nim c -f -d:release -d:configPath=config.yaml --out:nimcso ../src/nimcso 
\end{minted}

\begin{minted}[xleftmargin=3\parindent, fontsize=\small, bgcolor=subtlegray]{output}
...
config.yaml
CC: ../../../opt/conda/nim/lib/system/exceptions.nim
CC: ../../../opt/conda/nim/lib/std/private/digitsutils.nim
CC: ../../../opt/conda/nim/lib/std/assertions.nim
...
CC: nimcso/bitArrayAutoconfigured.nim
CC: nimcso.nim
Hint: orc; threads: on; opt: speed; options: -d:release
87026 lines; 7.635s; 257.383MiB peakmem; proj: /workspaces/nimCSO/src/nimcso; 
out: /workspaces/nimCSO/examples/nimcso[SuccessX]
\end{minted}

Now, let's run \texttt{nimCSO} and see what happens!

\begin{minted}[xleftmargin=3\parindent, linenos=true, fontsize=\small]{shell}
./nimcso
\end{minted}

\begin{figure}[H]
  \centering
  \includegraphics[width=0.97\textwidth]{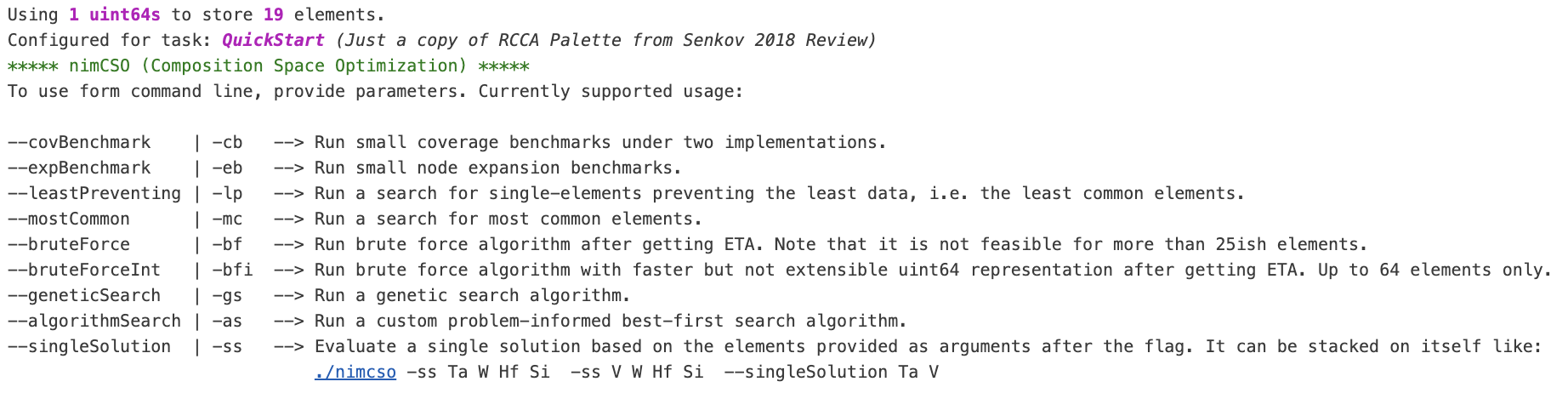}
\end{figure}

You should have seen a neat \texttt{help} message that
tells you how to use \texttt{nimCSO}. Let's start with
a ``coverage'' benchmark to see how fast can we check how many
datapoints will be removed from the dataset if we remove the first 5
elements of \texttt{elementOrder}.

\begin{minted}[xleftmargin=3\parindent, linenos=true, fontsize=\small]{shell}
./nimcso -cb
\end{minted}

\begin{figure}[H]
  \centering
  \includegraphics[width=0.97\textwidth]{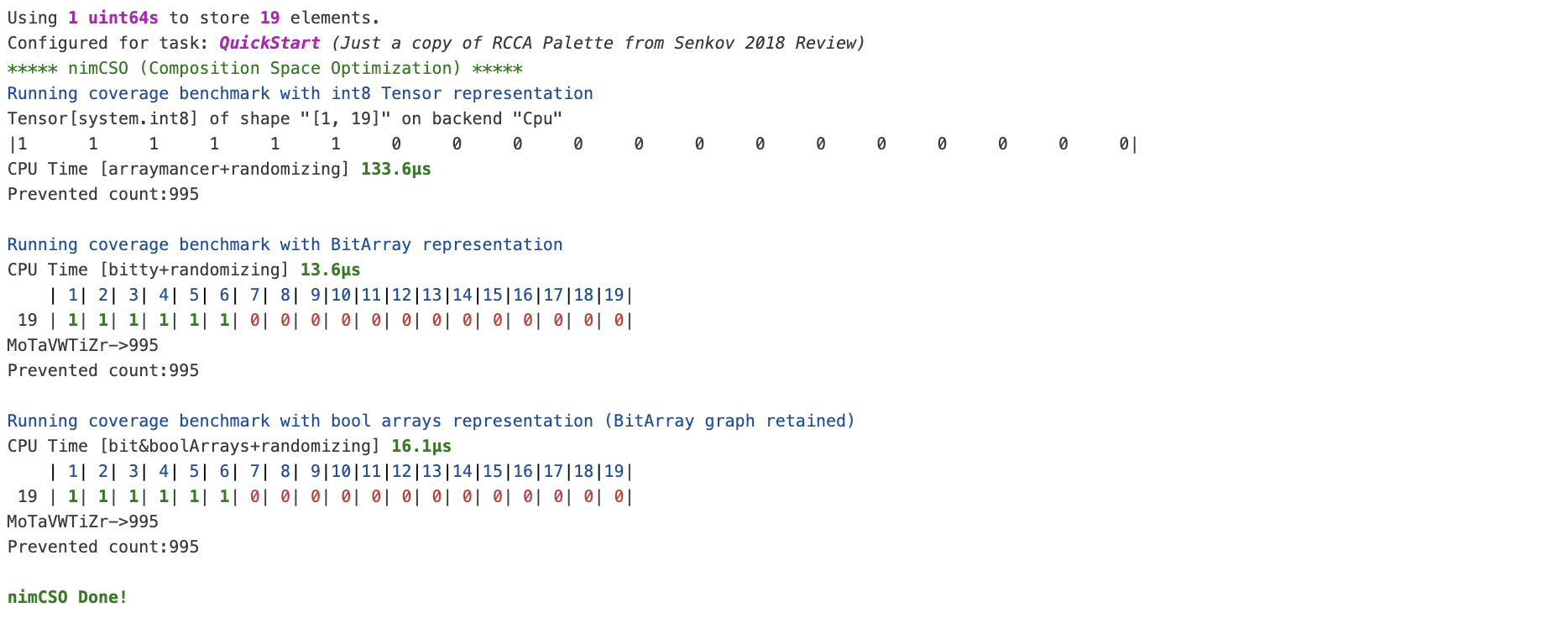}
\end{figure}

\hypertarget{key-routines-and-brute-forcing}{
\section{Key Routines and Brute Forcing}\label{nimcsotutorial:key-routines-and-brute-forcing}}

And if you were able to run that, you are all set to start using
\texttt{nimCSO}!

Let's try the simplest routine \texttt{mostCommon} or
\emph{What are the most common elements in the dataset?}

\begin{minted}[xleftmargin=3\parindent, linenos=true, fontsize=\small]{shell}
./nimcso --mostCommon
\end{minted}

\begin{figure}[H]
  \centering
  \includegraphics[width=0.97\textwidth]{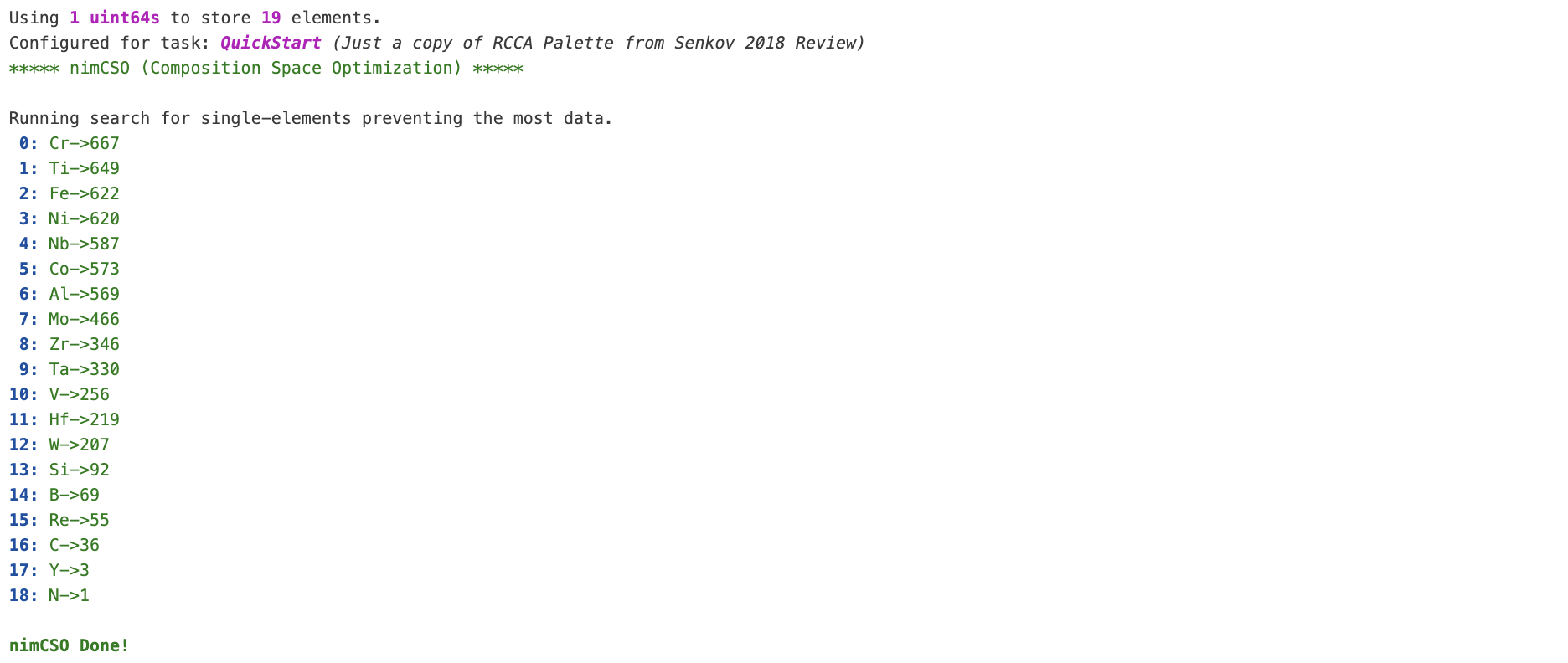}
\end{figure}

If you didn't modify anything, you should now see that elements like
\texttt{N}, \texttt{Y},
\texttt{C}, and \texttt{Re}, are not
very common in the dataset, while \texttt{Cr},
\texttt{Ti}, \texttt{Fe}, and
\texttt{Ni} are very common. When it comes to them, its
pretty obvious that removing the first group will be the first choice,
while the latter will be the last, if we want to keep the dataset as
extensive as possible.

The critical question here is, \emph{which of the intermediate elements
like \texttt{Hf}, \texttt{V},
\texttt{Ta}, or \texttt{Zr} should we
remove first?}

With a dataset spanning 19 elements, the solution space is around 0.5M,
so we can actually just brute force it in seconds :)

\begin{minted}[xleftmargin=3\parindent, linenos=true, fontsize=\small]{shell}
./nimcso -bfi
\end{minted}

\begin{figure}[H]
  \centering
  \includegraphics[width=0.97\textwidth]{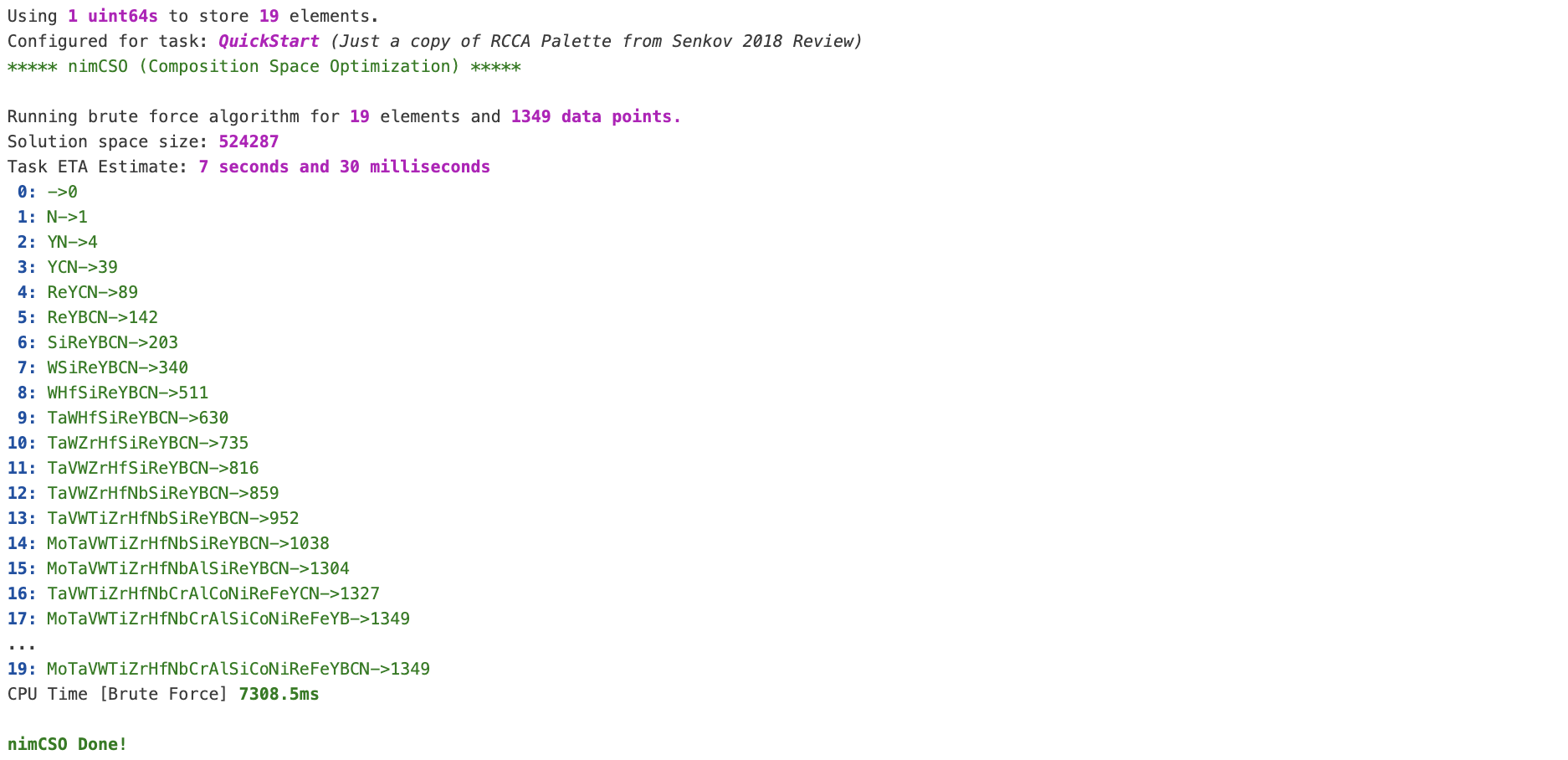}
\end{figure}

Let's look at the result! As expected, \texttt{N},
\texttt{Y}, \texttt{C}, and
\texttt{Re} are removed first (0-4) and then the trend
follows for a bit to \texttt{Hf} \textbf{The first
break is \texttt{V}, you can notice that it's better to
remove either or both \texttt{Ta} or
\texttt{Zr} first, despite the fact that they are
nearly 50\% more common than \texttt{V}} That's
because they often coocur with \texttt{Re} and
\texttt{Hf}, which are not common.

We can test exactly how much more data we will have if we remove
\texttt{Ta} insead of \texttt{V} by
using the \texttt{--singleSolution} /
\texttt{-ss} routine.

\begin{minted}[xleftmargin=3\parindent, linenos=true, fontsize=\small]{shell}
./nimcso -ss Ta W Hf Si Re Y B C N -ss V W Hf Si Re Y B C N
\end{minted}

\begin{figure}[H]
  \centering
  \includegraphics[width=0.97\textwidth]{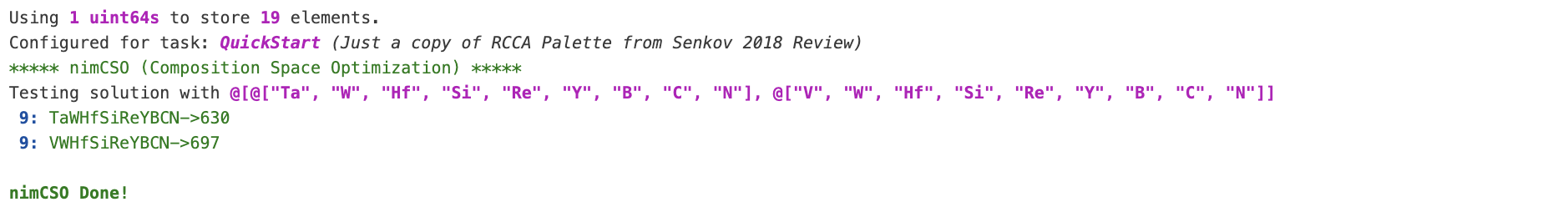}
\end{figure}

Wow! Looking at the \texttt{--mostCommon} output from
earlier, we can see that \textbf{\texttt{Ta} is present
in 74 more datapoints than \texttt{V}, but after
removing \texttt{WHfSiReYBCN}, picking
\texttt{V} as one of 10 elements to model will result
in 67 \emph{more} datapoints.} Relative to a dataset without
interdependencies, that's a 141 datapoint difference!

And another case that breaks from the ordering is
\texttt{Mo}, which is better to keep than much more
common \texttt{Nb}, and after
\texttt{Nb} is removed, even better thank keeping the
\texttt{Ti}, which is the second most common element in
the dataset!

Similarly to what we did with \texttt{V}
vs.~\texttt{Ta}, we can test how much more data we will
have if we remove \texttt{Nb} instead of
\texttt{Mo} by using the
\texttt{--singleSolution} /
\texttt{-ss} routine.

\begin{minted}[xleftmargin=3\parindent, linenos=true, fontsize=\small, breaklines]{shell}
./nimcso -ss Ta V W Zr Hf Nb Si Re Y B C N -ss Ta V W Zr Hf Mo Si Re Y B C N -ss Ta V W Zr Hf Ti Si Re Y B C N
\end{minted}

\begin{figure}[H]
  \centering
  \includegraphics[width=0.97\textwidth]{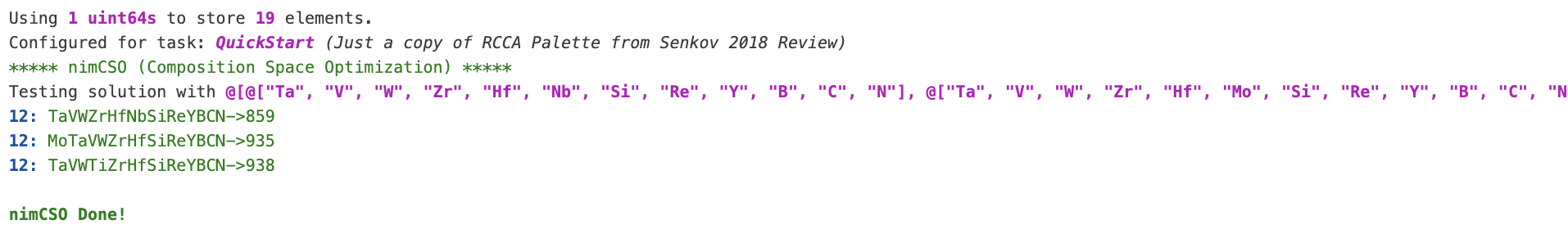}
\end{figure}

We can see that \textbf{\texttt{Nb} is present in 121
more datapoints than \texttt{Mo}, but after removing
\texttt{TaVWZrHfSiReYBCN}, picking
\texttt{Mo} as one of 7 elements to model will result
in 76 \emph{more} datapoints.} Relative to a dataset without
interdependencies, that's a 197 datapoint difference, which is even more
than the \texttt{Ta} vs.~\texttt{V}
case! Additionally, we can see that \texttt{Ti} is only
3 datapoints better than \texttt{Mo}, despite being
present in 183 more datapoints than \texttt{Mo}.

\hypertarget{algorithm-search}{%
\section{Algorithm Search}\label{nimcsotutorial:algorithm-search}}

The
\texttt{--bruteForceInt}/\texttt{-bfi}
routine we used to find the solutions worked great for our 19-element
dataset and took only a few seconds on the low-performance Codespace
machine, but in many cases dimensionality of the problem will be too
high to brute force it.

Let's now try to use the
\texttt{--algorithmSearch}/\texttt{-as}
routine, which takes advantage of some assumptions known to be valid or
likely to be valid (see manuscript), to limit the search space and find
the solution in a reasonable time. Let's try it now!

\begin{minted}[xleftmargin=3\parindent, linenos=true, fontsize=\small]{shell}
./nimcso -as
\end{minted}

\begin{figure}[H]
  \centering
  \includegraphics[width=0.97\textwidth]{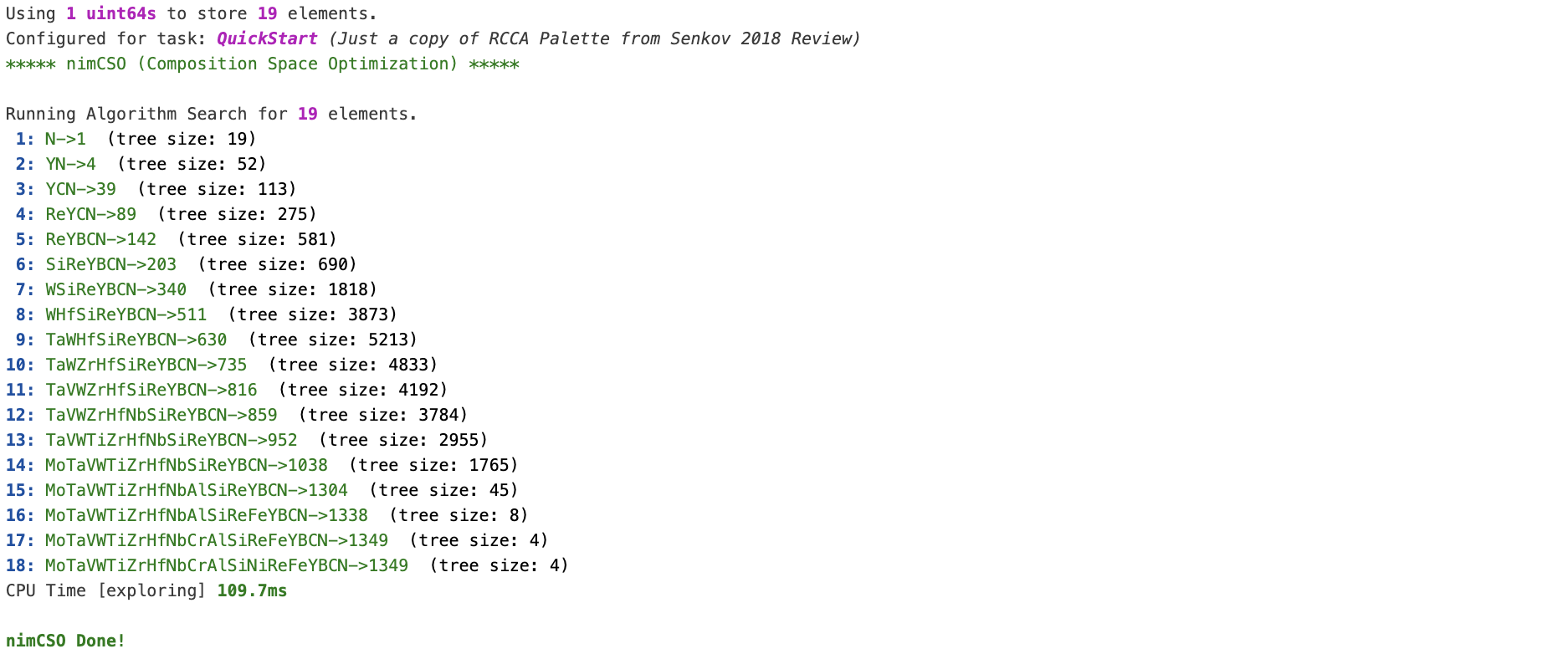}
\end{figure}

As you can see, \textbf{the algorithm reproduced the same results as the
brute force search around 100 times faster}, except for third-to-last
step because dataset had points with at least 3 elements breaking its
backtracking assumptions.

\hypertarget{genetic-search}{%
\section{Genetic Search}\label{nimcsotutorial:genetic-search}}

For cases where the dimensionality of the problem is too high to either
brute-force or use the algorithm search, we can still use the
\texttt{--geneticSearch}/\texttt{-gs}
routine to find the solution in a reasonable time. Let's try it now!

Please note that the results are stochastic, so you might get different
results than ones shown below if you run the command again.

\begin{minted}[xleftmargin=3\parindent, linenos=true, fontsize=\small]{shell}
./nimcso -gs
\end{minted}

\begin{figure}[H]
  \centering
  \includegraphics[width=0.97\textwidth]{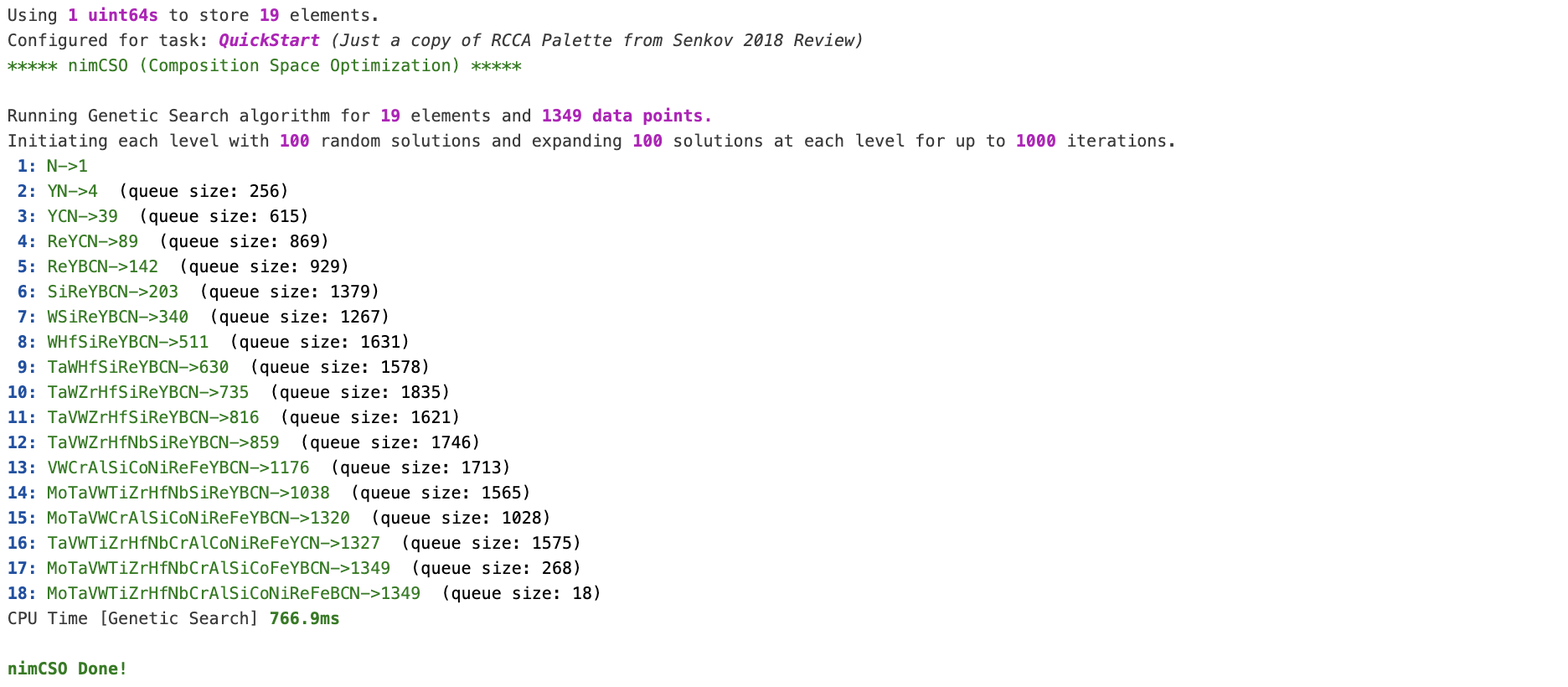}
\end{figure}

\label{nimcsotutorial:summary}

To summarize, now, you should be able to apply \texttt{nimCSO} to
your own dataset and get some valuable insights on how to model it!

If you are working in a Codespace, you can just do everything right in
this notebook by simply modifying the
\texttt{config.yaml} file and running the commands you
just learned about. The Codespace will be persisted until you explicitly
delete it, so you can come back to it later and continue your work by
clicking on the link in the \textbf{Open in Codespaces} badge in the README
of the repository and resuming your work.


\end{appendices}

\printbibliography[
heading=bibintoc,
title={Combined Bibliography}
]

@inbook{Toher2018,
   author = {Cormac Toher and Corey Oses and David Hicks and Eric Gossett and Frisco Rose and Pinku Nath and Demet Usanmaz and Denise C. Ford and Eric Perim and Camilo E. Calderon and Jose J. Plata and Yoav Lederer and Michal Jahnátek and Wahyu Setyawan and Shidong Wang and Junkai Xue and Kevin Rasch and Roman V. Chepulskii and Richard H. Taylor and Geena Gomez and Harvey Shi and Andrew R. Supka and Rabih Al Rahal Al Orabi and Priya Gopal and Frank T. Cerasoli and Laalitha Liyanage and Haihang Wang and Ilaria Siloi and Luis A. Agapito and Chandramouli Nyshadham and Gus L. W Hart and Jesús Carrete and Fleur Legrain and Natalio Mingo and Eva Zurek and Olexandr Isayev and Alexander Tropsha and Stefano Sanvito and Robert M. Hanson and Ichiro Takeuchi and Michael J. Mehl and Aleksey N. Kolmogorov and Kesong Yang and Pino D’Amico and Arrigo Calzolari and Marcio Costa and Riccardo De Gennaro and Marco Buongiorno Nardelli and Marco Fornari and Ohad Levy and Stefano Curtarolo},
   city = {Cham},
   doi = {10.1007/978-3-319-42913-7_63-1},
   journal = {Handbook of Materials Modeling},
   pages = {1-28},
   publisher = {Springer International Publishing},
   title = {The AFLOW Fleet for Materials Discovery},
   url = {http://link.springer.com/10.1007/978-3-319-42913-7_63-1},
   year = {2018},
}

@article{Merchant2023,
   author = {Amil Merchant and Simon Batzner and Samuel S Schoenholz and Muratahan Aykol and Gowoon Cheon and Ekin Dogus Cubuk},
   doi = {10.1038/s41586-023-06735-9},
   journal = {80 | Nature |},
   title = {Scaling deep learning for materials discovery},
   volume = {624},
   url = {https://doi.org/10.1038/s41586-023-06735-9},
   year = {2023},
}

@article{Mehl2016,
   author = {Michael J. Mehl and David Hicks and Cormac Toher and Ohad Levy and Robert M. Hanson and Gus Hart and Stefano Curtarolo},
   month = {7},
   title = {The AFLOW Library of Crystallographic Prototypes},
   url = {http://arxiv.org/abs/1607.02532},
   year = {2016},
}

@article{Schmidt2023,
   author = {Jonathan Schmidt and Noah Hoffmann and Hai-Chen Wang and Pedro Borlido and Pedro J M A Carriço and Tiago F T Cerqueira and Silvana Botti and Miguel A L Marques and J Schmidt and N Hoffmann and H.-C Wang and M A L Marques and P Borlido and P J M A Carriço and T F T Cerqueira and S Botti},
   doi = {10.1002/ADMA.202210788},
   issn = {1521-4095},
   issue = {22},
   journal = {Advanced Materials},
   month = {6},
   pages = {2210788},
   pmid = {36949007},
   publisher = {John Wiley & Sons, Ltd},
   title = {Machine-Learning-Assisted Determination of the Global Zero-Temperature Phase Diagram of Materials},
   volume = {35},
   url = {https://onlinelibrary.wiley.com/doi/full/10.1002/adma.202210788 https://onlinelibrary.wiley.com/doi/abs/10.1002/adma.202210788 https://onlinelibrary.wiley.com/doi/10.1002/adma.202210788},
   year = {2023},
}

@article{Wang2021,
   author = {Hai-Chen Wang and Silvana Botti and Miguel A. L. Marques},
   doi = {10.1038/s41524-020-00481-6},
   issn = {2057-3960},
   issue = {1},
   journal = {npj Computational Materials},
   month = {1},
   pages = {12},
   title = {Predicting stable crystalline compounds using chemical similarity},
   volume = {7},
   url = {https://www.nature.com/articles/s41524-020-00481-6},
   year = {2021},
}

@article{Zunger2019,
   author = {Alex Zunger},
   doi = {10.1038/d41586-019-00676-y},
   issn = {14764687},
   issue = {7745},
   journal = {Nature 2021 566:7745},
   month = {2},
   pages = {447-449},
   pmid = {30814720},
   publisher = {Nature Publishing Group},
   title = {Beware of plausible predictions of fantasy materials},
   volume = {566},
   url = {https://www.nature.com/articles/d41586-019-00676-y},
   year = {2019},
}

@article{LeBail2005,
   author = {Armel Le Bail},
   doi = {10.1107/S0021889805002384},
   issn = {00218898},
   issue = {2},
   journal = {Journal of Applied Crystallography},
   month = {4},
   pages = {389-393},
   publisher = {International Union of Crystallography},
   title = {Inorganic structure prediction with GRINSP},
   volume = {38},
   url = {www.iza-structure.org/databases/},
   year = {2005},
}

@article{Hawthorne1994,
   author = {F. C. Hawthorne},
   doi = {10.1107/S0108768193014466},
   issn = {0108-7681},
   issue = {5},
   journal = {Acta Crystallographica Section B Structural Science},
   month = {10},
   pages = {481-510},
   title = {Structural aspects of oxide and oxysalt crystals},
   volume = {50},
   year = {1994},
}

@article{Olson2023,
   author = {G.B. Olson and Z.K. Liu},
   doi = {10.1016/j.calphad.2023.102590},
   issn = {03645916},
   journal = {Calphad},
   month = {9},
   pages = {102590},
   publisher = {Pergamon},
   title = {Genomic materials design: CALculation of PHAse Dynamics},
   volume = {82},
   url = {https://linkinghub.elsevier.com/retrieve/pii/S0364591623000627},
   year = {2023},
}

@article{Liu2022,
   author = {Zi-Kui Liu and Yi Wang and Shun-Li Shang},
   doi = {10.1007/s11669-022-00942-z},
   issn = {1547-7037},
   journal = {Journal of Phase Equilibria and Diffusion},
   month = {2},
   title = {Zentropy Theory for Positive and Negative Thermal Expansion},
   url = {https://link.springer.com/10.1007/s11669-022-00942-z},
   year = {2022},
}

@article{Dasgupta2022,
   author = {Dasgupta, Anish and He, Haoran and Gong, Rushi and Shang, Shun-Li and Zimmerer, Eric K. and Meyer, Randall J. and Liu, Zi-Kui and Janik, Michael J. and Rioux, Robert M.},
   doi = {10.1038/s41557-021-00855-3},
   issn = {1755-4330},
   journal = {Nature Chemistry},
   month = {2},
   pages = {1--7},
   publisher = {Nature Publishing Group},
   title = {{Atomic control of active-site ensembles in ordered alloys to enhance hydrogenation selectivity}},
   url = {https://www.nature.com/articles/s41557-021-00855-3},
   year = {2022}
}

@article{LeBail1988,
   author = {A. Le Bail and H. Duroy and J.L. Fourquet},
   doi = {10.1016/0025-5408(88)90019-0},
   issn = {00255408},
   issue = {3},
   journal = {Materials Research Bulletin},
   month = {3},
   pages = {447-452},
   title = {Ab-initio structure determination of LiSbWO6 by X-ray powder diffraction},
   volume = {23},
   url = {https://linkinghub.elsevier.com/retrieve/pii/0025540888900190},
   year = {1988},
}

@article{Kohn1965,
   author = {W. Kohn and L. J. Sham},
   doi = {10.1103/PhysRev.140.A1133},
   issn = {0031-899X},
   issue = {4A},
   journal = {Physical Review},
   month = {11},
   pages = {A1133-A1138},
   title = {Self-Consistent Equations Including Exchange and Correlation Effects},
   volume = {140},
   year = {1965},
}

@article{Ferrari2013,
   author = {Andrea C. Ferrari and Denis M. Basko},
   doi = {10.1038/nnano.2013.46},
   issn = {1748-3387},
   issue = {4},
   journal = {Nature Nanotechnology},
   month = {4},
   pages = {235-246},
   title = {Raman spectroscopy as a versatile tool for studying the properties of graphene},
   volume = {8},
   url = {https://www.nature.com/articles/nnano.2013.46},
   year = {2013},
}

@article{Kothakonda2023,
   author = {Manish Kothakonda and Aaron D. Kaplan and Eric B. Isaacs and Christopher J. Bartel and James W. Furness and Jinliang Ning and Chris Wolverton and John P. Perdew and Jianwei Sun},
   doi = {10.1021/ACSMATERIALSAU.2C00059/SUPPL_FILE/MG2C00059_SI_002.PDF},
   issn = {26942461},
   issue = {2},
   journal = {ACS Materials Au},
   month = {3},
   pages = {102-111},
   publisher = {American Chemical Society},
   title = {Testing the r2SCAN Density Functional for the Thermodynamic Stability of Solids with and without a van der Waals Correction},
   volume = {3},
   url = {https://doi.org/10.1021/acsmaterialsau.2c00059},
   year = {2023},
}

@article{Curtarolo2012,
   author = {Stefano Curtarolo and Wahyu Setyawan and Gus L.W. Hart and Michal Jahnatek and Roman V. Chepulskii and Richard H. Taylor and Shidong Wang and Junkai Xue and Kesong Yang and Ohad Levy and Michael J. Mehl and Harold T. Stokes and Denis O. Demchenko and Dane Morgan},
   doi = {10.1016/j.commatsci.2012.02.005},
   issn = {09270256},
   journal = {Computational Materials Science},
   month = {6},
   pages = {218-226},
   title = {AFLOW: An automatic framework for high-throughput materials discovery},
   volume = {58},
   url = {https://linkinghub.elsevier.com/retrieve/pii/S0927025612000717},
   year = {2012},
}

@article{Pickard2011,
   author = {Chris J Pickard and R J Needs},
   doi = {10.1088/0953-8984/23/5/053201},
   issn = {0953-8984},
   issue = {5},
   journal = {Journal of Physics: Condensed Matter},
   month = {2},
   pages = {053201},
   title = {Ab initio random structure searching},
   volume = {23},
   url = {https://iopscience.iop.org/article/10.1088/0953-8984/23/5/053201},
   year = {2011},
}

@article{Glass2006,
   author = {Colin W. Glass and Artem R. Oganov and Nikolaus Hansen},
   doi = {10.1016/j.cpc.2006.07.020},
   issn = {00104655},
   issue = {11-12},
   journal = {Computer Physics Communications},
   month = {12},
   pages = {713-720},
   title = {USPEX—Evolutionary crystal structure prediction},
   volume = {175},
   url = {https://linkinghub.elsevier.com/retrieve/pii/S0010465506002931},
   year = {2006},
}

@article{Ferreira2023,
   author = {Pedro P. Ferreira and Lewis J. Conway and Alessio Cucciari and Simone Di Cataldo and Federico Giannessi and Eva Kogler and Luiz T. F. Eleno and Chris J. Pickard and Christoph Heil and Lilia Boeri},
   doi = {10.1038/s41467-023-41005-2},
   issn = {2041-1723},
   issue = {1},
   journal = {Nature Communications},
   month = {9},
   pages = {5367},
   title = {Search for ambient superconductivity in the Lu-N-H system},
   volume = {14},
   url = {https://www.nature.com/articles/s41467-023-41005-2},
   year = {2023},
}

@article{Shumailov2023,
   author = {Ilia Shumailov and Zakhar Shumaylov and Yiren Zhao and Yarin Gal and Nicolas Papernot and Ross Anderson},
   doi = {10.48550/arXiv.2305.17493},
   month = {5},
   title = {The Curse of Recursion: Training on Generated Data Makes Models Forget},
   url = {http://arxiv.org/abs/2305.17493},
   year = {2023},
}

@article{Alemohammad2023,
   author = {Sina Alemohammad and Josue Casco-Rodriguez and Lorenzo Luzi and Ahmed Imtiaz Humayun and Hossein Babaei and Daniel LeJeune and Ali Siahkoohi and Richard G. Baraniuk},
   doi = {10.48550/arXiv.2307.01850},
   month = {7},
   title = {Self-Consuming Generative Models Go MAD},
   url = {http://arxiv.org/abs/2307.01850},
   year = {2023},
}

@article{Ye2022,
   author = {Weike Ye and Xiangyun Lei and Muratahan Aykol and Joseph H. Montoya},
   doi = {10.1038/s41597-022-01438-8},
   issn = {2052-4463},
   issue = {1},
   journal = {Scientific Data},
   month = {6},
   pages = {302},
   title = {Novel inorganic crystal structures predicted using autonomous simulation agents},
   volume = {9},
   url = {https://www.nature.com/articles/s41597-022-01438-8},
   year = {2022},
}

@article{Yue2024,
   author = {Weiqi Yue and Pawan K. Tripathi and Gabriel Ponon and Zhuldyz Ualikhankyzy and Donald W. Brown and Bjorn Clausen and Maria Strantza and Darren C. Pagan and Matthew A. Willard and Frank Ernst and Erman Ayday and Vipin Chaudhary and Roger H. French},
   doi = {10.1007/s40192-023-00328-0},
   issn = {2193-9764},
   issue = {1},
   journal = {Integrating Materials and Manufacturing Innovation},
   month = {3},
   pages = {36-52},
   title = {Phase Identification in Synchrotron X-ray Diffraction Patterns of Ti–6Al–4V Using Computer Vision and Deep Learning},
   volume = {13},
   url = {https://link.springer.com/10.1007/s40192-023-00328-0},
   year = {2024},
}

@article{Evans2021,
   author = {Matthew L. Evans and Casper W. Andersen and Shyam Dwaraknath and Markus Scheidgen and Ádám Fekete and Donald Winston},
   doi = {10.21105/JOSS.03458},
   issn = {2475-9066},
   issue = {65},
   journal = {Journal of Open Source Software},
   month = {9},
   pages = {3458},
   publisher = {The Open Journal},
   title = {`optimade-python-tools`: a Python library for serving and consuming materials data via OPTIMADE APIs},
   volume = {6},
   url = {https://joss.theoj.org/papers/10.21105/joss.03458},
   year = {2021},
}

@article{Yeh2004,
   author = {Jien Wei Yeh and Swe Kai Chen and Su Jien Lin and Jon Yiew Gan and Tsung Shune Chin and Tao Tsung Shun and Chun Huei Tsau and Shou Yi Chang},
   doi = {10.1002/adem.200300567},
   issn = {14381656},
   issue = {5},
   journal = {Advanced Engineering Materials},
   pages = {299-303},
   publisher = {Wiley-VCH Verlag},
   title = {Nanostructured high-entropy alloys with multiple principal elements: Novel alloy design concepts and outcomes},
   volume = {6},
   year = {2004},
}

@article{Cantor2004,
   author = {B. Cantor and I. T.H. Chang and P. Knight and A. J.B. Vincent},
   doi = {10.1016/j.msea.2003.10.257},
   issn = {09215093},
   issue = {1-2 SPEC. ISS.},
   journal = {Materials Science and Engineering A},
   month = {7},
   pages = {213-218},
   title = {Microstructural development in equiatomic multicomponent alloys},
   volume = {375-377},
   year = {2004},
}

@article{Debnath2021,
   author = {Arindam Debnath and Adam M. Krajewski and Hui Sun and Shuang Lin and Marcia Ahn and Wenjie Li and Shanshank Priya and Jogender Singh and Shunli Shang and Allison M. Beese and Zi-Kui Liu and Wesley F. Reinhart},
   doi = {10.20517/jmi.2021.05},
   volume = {1},
   issue = {3},
   year = {2021},
   month = {9},
   journal = {Journal of Materials Informatics}
}

@article{Li2024,
   author = {Wenjie Li and Lavanya Raman and Arindam Debnath and Marcia Ahn and Shuang Lin and Adam M Krajewski and Shunli Shang and Shashank Priya and Wesley F Reinhart and Zi-Kui Liu and Allison M Beese},
   doi = {10.2139/ssrn.4689687},
   month = {1},
   title = {Design and Validation of Refractory Alloys Using Machine Learning, CALPHAD, and Experiments},
   url = {https://ssrn.com/abstract=4689687},
   year = {2024},
}

@article{Senkov2018,
   author = {Oleg N. Senkov and Daniel B. Miracle and Kevin J. Chaput and Jean-Philippe Couzinie},
   doi = {10.1557/jmr.2018.153},
   issn = {0884-2914},
   issue = {19},
   journal = {Journal of Materials Research},
   month = {10},
   pages = {3092-3128},
   publisher = {Cambridge University Press},
   title = {Development and exploration of refractory high entropy alloys—A review},
   volume = {33},
   url = {https://www.cambridge.org/core/product/identifier/S088429141800153X/type/journal_article},
   year = {2018},
}

@book{Goldberg1989, 
author = {Goldberg, David E.}, 
title = {Genetic Algorithms in Search, Optimization and Machine Learning}, 
year = {1989}, 
isbn = {0201157675}, 
doi = {10.5860/choice.27-0936},
publisher = {Addison-Wesley Longman Publishing Co., Inc.}, 
address = {USA}, 
edition = {1st} 
}

@book{Knuth, 
author = {Knuth, Donald E.}, 
title = {The Art of Computer Programming, Volume 4, Fascicle 1: Bitwise Tricks \& Techniques; Binary Decision Diagrams}, 
year = {2009}, 
isbn = {0321580508}, 
publisher = {Addison-Wesley Professional}, 
edition = {12th} 
}

@article{Evans2024,
   author = {Matthew L. Evans and Johan Bergsma and Andrius Merkys and Casper W. Andersen and Oskar B. Andersson and Daniel Beltrán and Evgeny Blokhin and Tara M. Boland and Rubén Castañeda Balderas and Kamal Choudhary and Alberto Díaz Díaz and Rodrigo Domínguez García and Hagen Eckert and Kristjan Eimre and María Elena Fuentes Montero and Adam M. Krajewski and Jens Jørgen Mortensen and José Manuel Nápoles Duarte and Jacob Pietryga and Ji Qi and Felipe de Jesús Trejo Carrillo and Antanas Vaitkus and Jusong Yu and Adam Zettel and Pedro Baptista de Castro and Johan Carlsson and Tiago F. T. Cerqueira and Simon Divilov and Hamidreza Hajiyani and Felix Hanke and Kevin Jose and Corey Oses and Janosh Riebesell and Jonathan Schmidt and Donald Winston and Christen Xie and Xiaoyu Yang and Sara Bonella and Silvana Botti and Stefano Curtarolo and Claudia Draxl and Luis Edmundo Fuentes Cobas and Adam Hospital and Zi-Kui Liu and Miguel A. L. Marques and Nicola Marzari and Andrew J. Morris and Shyue Ping Ong and Modesto Orozco and Kristin A. Persson and Kristian S. Thygesen and Chris Wolverton and Markus Scheidgen and Cormac Toher and Gareth J. Conduit and Giovanni Pizzi and Saulius Gražulis and Gian-Marco Rignanese and Rickard Armiento},
   doi = {10.48550/arXiv.2402.00572},
   month = {2},
   title = {Developments and applications of the OPTIMADE API for materials discovery, design, and data exchange},
   url = {http://arxiv.org/abs/2402.00572},
   year = {2024},
}

@misc{Arraymancer,
   author = {Mamy Ratsimbazafy},
   title = {Arraymancer v0.7.24: A fast, ergonomic and portable tensor library in Nim with a deep learning focus for CPU, GPU and embedded devices via OpenMP, Cuda and OpenCL backends},
   url = {https://github.com/mratsim/Arraymancer},
}

@article{lecun2015deep,
  title={Deep learning},
  author={LeCun, Yann and Bengio, Yoshua and Hinton, Geoffrey},
  journal={nature},
  volume={521},
  number={7553},
  pages={436--444},
  year={2015},
  publisher={Nature Publishing Group}
}

@article{vapnik1999overview,
  title={An overview of statistical learning theory},
  author={Vapnik, Vladimir N},
  journal={IEEE transactions on neural networks},
  volume={10},
  number={5},
  pages={988--999},
  year={1999},
  publisher={IEEE}
}

@book{goodfellow2016deep,
  title={Deep learning},
  author={Goodfellow, Ian and Bengio, Yoshua and Courville, Aaron},
  year={2016},
  publisher={MIT press}
}

@inproceedings{lecun1995comparison,
  title={Comparison of learning algorithms for handwritten digit recognition},
  author={LeCun, Yann and Jackel, LD and Bottou, Leon and Brunot, A and Cortes, Corinna and Denker, JS and Drucker, Harris and Guyon, I and Muller, UA and Sackinger, Eduard and others},
  booktitle={International conference on artificial neural networks},
  volume={60},
  pages={53--60},
  year={1995},
  organization={Perth, Australia}
}

@article{lecun1998gradient,
  title={Gradient-based learning applied to document recognition},
  author={LeCun, Yann and Bottou, L{\'e}on and Bengio, Yoshua and Haffner, Patrick},
  journal={Proceedings of the IEEE},
  volume={86},
  number={11},
  pages={2278--2324},
  year={1998},
  publisher={Ieee}
}

@article{kingma2014adam,
  title={Adam: A method for stochastic optimization},
  author={Kingma, Diederik P and Ba, Jimmy},
  journal={arXiv preprint arXiv:1412.6980},
  year={2014}
}

@inproceedings{krizhevsky2012imagenet,
  title={Imagenet classification with deep convolutional neural networks},
  author={Krizhevsky, Alex and Sutskever, Ilya and Hinton, Geoffrey E},
  booktitle={Advances in neural information processing systems},
  pages={1097--1105},
  year={2012}
}

@article{goodfellow2013multi,
  title={Multi-digit number recognition from street view imagery using deep convolutional neural networks},
  author={Goodfellow, Ian J and Bulatov, Yaroslav and Ibarz, Julian and Arnoud, Sacha and Shet, Vinay},
  journal={arXiv preprint arXiv:1312.6082},
  year={2013}
}

@article{dahl2011context,
  title={Context-dependent pre-trained deep neural networks for large-vocabulary speech recognition},
  author={Dahl, George E and Yu, Dong and Deng, Li and Acero, Alex},
  journal={IEEE Transactions on audio, speech, and language processing},
  volume={20},
  number={1},
  pages={30--42},
  year={2011},
  publisher={IEEE}
}

@book{hastie2009elements,
  title={The elements of statistical learning: data mining, inference, and prediction},
  author={Hastie, Trevor and Tibshirani, Robert and Friedman, Jerome},
  year={2009},
  publisher={Springer Science \& Business Media}
}

@article{srivastava2014dropout,
  title={Dropout: a simple way to prevent neural networks from overfitting},
  author={Srivastava, Nitish and Hinton, Geoffrey and Krizhevsky, Alex and Sutskever, Ilya and Salakhutdinov, Ruslan},
  journal={The journal of machine learning research},
  volume={15},
  number={1},
  pages={1929--1958},
  year={2014},
  publisher={JMLR. org}
}

@techreport{rycroft2009voro++,
  title={Voro++: A three-dimensional Voronoi cell library in C++},
  author={Rycroft, Chris},
  year={2009},
  institution={Lawrence Berkeley National Lab.(LBNL), Berkeley, CA (United States)}
}

@article{silver2016mastering,
  title={Mastering the game of Go with deep neural networks and tree search},
  author={Silver, David and Huang, Aja and Maddison, Chris J and Guez, Arthur and Sifre, Laurent and Van Den Driessche, George and Schrittwieser, Julian and Antonoglou, Ioannis and Panneershelvam, Veda and Lanctot, Marc and others},
  journal={nature},
  volume={529},
  number={7587},
  pages={484},
  year={2016},
  publisher={Nature Publishing Group}
}

@article{devlin2018bert,
  title={Bert: Pre-training of deep bidirectional transformers for language understanding},
  author={Devlin, Jacob and Chang, Ming-Wei and Lee, Kenton and Toutanova, Kristina},
  journal={arXiv preprint arXiv:1810.04805},
  year={2018}
}

@article{carleo2017solving,
  title={Solving the quantum many-body problem with artificial neural networks},
  author={Carleo, Giuseppe and Troyer, Matthias},
  journal={Science},
  volume={355},
  number={6325},
  pages={602--606},
  year={2017},
  publisher={American Association for the Advancement of Science}
}

@book{vapnik2013nature,
  title={The nature of statistical learning theory},
  author={Vapnik, Vladimir},
  year={2013},
  publisher={Springer science \& business media}
}

@book{everitt2002cambridge,
  title={The Cambridge dictionary of statistics},
  author={Everitt, Brian and Skrondal, Anders},
  volume={106},
  year={2002},
  publisher={Cambridge University Press Cambridge}
}

@inproceedings{tikhonov1963solution,
  title={On the solution of ill-posed problems and the method of regularization},
  author={Tikhonov, Andrei Nikolaevich},
  booktitle={Doklady Akademii Nauk},
  volume={151},
  number={3},
  pages={501--504},
  year={1963},
  organization={Russian Academy of Sciences}
}

@article{hoerl1970ridge,
  title={Ridge regression: Biased estimation for nonorthogonal problems},
  author={Hoerl, Arthur E and Kennard, Robert W},
  journal={Technometrics},
  volume={12},
  number={1},
  pages={55--67},
  year={1970},
  publisher={Taylor \& Francis Group}
}

@article{rosenblatt1958perceptron,
  title={The perceptron: a probabilistic model for information storage and organization in the brain.},
  author={Rosenblatt, Frank},
  journal={Psychological review},
  volume={65},
  number={6},
  pages={386},
  year={1958},
  publisher={American Psychological Association}
}

@inproceedings{lecun1990handwritten,
  title={Handwritten digit recognition with a back-propagation network},
  author={LeCun, Yann and Boser, Bernhard E and Denker, John S and Henderson, Donnie and Howard, Richard E and Hubbard, Wayne E and Jackel, Lawrence D},
  booktitle={Advances in neural information processing systems},
  pages={396--404},
  year={1990}
}

@inproceedings{tan2018survey,
  title={A survey on deep transfer learning},
  author={Tan, Chuanqi and Sun, Fuchun and Kong, Tao and Zhang, Wenchang and Yang, Chao and Liu, Chunfang},
  booktitle={International conference on artificial neural networks},
  pages={270--279},
  year={2018},
  organization={Springer}
}

@inproceedings{cirecsan2012transfer,
  title={Transfer learning for Latin and Chinese characters with deep neural networks},
  author={Cire{\c{s}}an, Dan C and Meier, Ueli and Schmidhuber, J{\"u}rgen},
  booktitle={The 2012 International Joint Conference on Neural Networks (IJCNN)},
  pages={1--6},
  year={2012},
  organization={IEEE}
}

@article{chang2017unsupervised,
  title={Unsupervised transfer learning via multi-scale convolutional sparse coding for biomedical applications},
  author={Chang, Hang and Han, Ju and Zhong, Cheng and Snijders, Antoine M and Mao, Jian-Hua},
  journal={IEEE transactions on pattern analysis and machine intelligence},
  volume={40},
  number={5},
  pages={1182--1194},
  year={2017},
  publisher={IEEE}
}

@article{george2018deep,
  title={Deep Learning for real-time gravitational wave detection and parameter estimation: Results with Advanced LIGO data},
  author={George, Daniel and Huerta, EA},
  journal={Physics Letters B},
  volume={778},
  pages={64--70},
  year={2018},
  publisher={Elsevier}
}

@article{gotmare2018closer,
  title={A closer look at deep learning heuristics: Learning rate restarts, warmup and distillation},
  author={Gotmare, Akhilesh and Keskar, Nitish Shirish and Xiong, Caiming and Socher, Richard},
  journal={arXiv preprint arXiv:1810.13243},
  year={2018}
}

@article{Curtarolo2013AFLOW:Discovery,
    title = {{AFLOW: An automatic framework for high-throughput materials discovery}},
    year = {2012},
    journal = {Computational Materials Science},
    author = {Curtarolo, Stefano and Setyawan, Wahyu and Hart, Gus L.W. and Jahnatek, Michal and Chepulskii, Roman V. and Taylor, Richard H. and Wang, Shidong and Xue, Junkai and Yang, Kesong and Levy, Ohad and Mehl, Michael J. and Stokes, Harold T. and Demchenko, Denis O. and Morgan, Dane},
    month = {6},
    pages = {218--226},
    volume = {58},
    url = {https://linkinghub.elsevier.com/retrieve/pii/S0927025612000717},
    doi = {10.1016/j.commatsci.2012.02.005},
    issn = {09270256}
}

@article{Krajewski2024Nimplex,
    title = {{Efficient Generation of Grids and Traversal Graphs in Compositional Spaces towards Exploration and Path Planning Exemplified in Materials}},
    year = {2024},
    author = {Krajewski, Adam M. and Beese, Allison M. and Reinhart, Wesley F. and Liu, Zi-Kui},
    month = {2},
    url = {http://arxiv.org/abs/2402.03528},
    doi = {10.48550/arXiv.2402.03528},
    arxivId = {2402.03528}
}

@article{Ward2017,
    title = {{Including crystal structure attributes in machine learning models of formation energies via Voronoi tessellations}},
    year = {2017},
    journal = {Physical Review B},
    author = {Ward, Logan and Liu, Ruoqian and Krishna, Amar and Hegde, Vinay I. and Agrawal, Ankit and Choudhary, Alok and Wolverton, Chris},
    number = {2},
    month = {7},
    pages = {024104},
    volume = {96},
    publisher = {American Physical Society},
    url = {http://link.aps.org/doi/10.1103/PhysRevB.96.024104},
    doi = {10.1103/PhysRevB.96.024104},
    issn = {2469-9950}
}

@misc{Krajewski2020SIPFENNModels,
    title = {{Neural Networks for Structure-Informed Prediction of Formation Energy (employed in SIPFENN)}},
    year = {2020},
    author = {Krajewski, Adam and Siegel, Jonathan and Xu, Jinhao and Liu, Zi-Kui},
    month = {8},
    publisher = {Zenodo},
    url = {https://doi.org/10.5281/zenodo.4006802},
    doi = {10.5281/zenodo.4006802}
}

@misc{Krajewski20232023YouTube,
    title = {{2023 03 02 MGF workshop on pySIPFENN by Adam Krajewski - YouTube}},
    year = {2023},
    author = {Krajewski, Adam},
    url = {https://www.youtube.com/watch?v=OHgkRuE0UQM}
}

@article{Saleh202030Challenges,
    title = {{30 Years of functionally graded materials: An overview of manufacturing methods, Applications and Future Challenges}},
    year = {2020},
    journal = {Composites Part B: Engineering},
    author = {Saleh, Bassiouny and Jiang, Jinghua and Fathi, Reham and Al-hababi, Tareq and Xu, Qiong and Wang, Lisha and Song, Dan and Ma, Aibin},
    month = {11},
    pages = {108376},
    volume = {201},
    publisher = {Elsevier},
    doi = {10.1016/J.COMPOSITESB.2020.108376},
    issn = {1359-8368}
}

@article{CharlesW.BauschlicherJr.1995AFunctionals,
    title = {{A comparison of the accuracy of different functionals}},
    year = {1995},
    journal = {Chemical Physics Letters},
    author = {{Charles W. Bauschlicher Jr.}},
    month = {11},
    pages = {40--44},
    volume = {246}
}

@article{Schmidt2022AFunctionals,
    title = {{A dataset of 175k stable and metastable materials calculated with the PBEsol and SCAN functionals}},
    year = {2022},
    journal = {Scientific Data 2022 9:1},
    author = {Schmidt, Jonathan and Wang, Hai Chen and Cerqueira, Tiago F.T. and Botti, Silvana and Marques, Miguel A.L.},
    number = {1},
    month = {3},
    pages = {1--8},
    volume = {9},
    publisher = {Nature Publishing Group},
    url = {https://www.nature.com/articles/s41597-022-01177-w},
    doi = {10.1038/s41597-022-01177-w},
    issn = {2052-4463},
    pmid = {35236866}
}

@article{Ester1996ANoise,
    title = {{A Density-Based Algorithm for Discovering Clusters in Large Spatial Databases with Noise}},
    year = {1996},
    author = {Ester, Martin and Kriegel, Hans-Peter and Sander, Jiirg and Xu, Xiaowei},
    url = {www.aaai.org}
}

@article{Long2019AProperties,
    title = {{A fine-grained NbMoTaWVCr refractory high-entropy alloy with ultra-high strength: Microstructural evolution and mechanical properties}},
    year = {2019},
    journal = {Journal of Alloys and Compounds},
    author = {Long, Yan and Liang, Xiaobiao and Su, Kai and Peng, Haiyan and Li, Xiaozhen},
    month = {4},
    pages = {607--617},
    volume = {780},
    publisher = {Elsevier},
    doi = {10.1016/J.JALLCOM.2018.11.318},
    issn = {0925-8388}
}

@article{Hart1968APaths,
    title = {{A Formal Basis for the Heuristic Determination of Minimum Cost Paths}},
    year = {1968},
    journal = {IEEE Transactions on Systems Science and Cybernetics},
    author = {Hart, Peter and Nilsson, Nils and Raphael, Bertram},
    number = {2},
    pages = {100--107},
    volume = {4},
    url = {http://ieeexplore.ieee.org/document/4082128/},
    doi = {10.1109/TSSC.1968.300136},
    issn = {0536-1567}
}

@article{Ward2016AMaterials,
    title = {{A general-purpose machine learning framework for predicting properties of inorganic materials}},
    year = {2016},
    journal = {npj Computational Materials},
    author = {Ward, Logan and Agrawal, Ankit and Choudhary, Alok and Wolverton, Christopher},
    month = {8},
    volume = {2},
    publisher = {Nature Publishing Group},
    doi = {10.1038/npjcompumats.2016.28},
    issn = {20573960}
}

@article{Yang2022AHardness,
    title = {{A machine learning-based alloy design system to facilitate the rational design of high entropy alloys with enhanced hardness}},
    year = {2022},
    journal = {Acta Materialia},
    author = {Yang, Chen and Ren, Chang and Jia, Yuefei and Wang, Gang and Li, Minjie and Lu, Wencong},
    month = {1},
    pages = {117431},
    volume = {222},
    publisher = {Pergamon},
    url = {https://linkinghub.elsevier.com/retrieve/pii/S1359645421008107},
    doi = {10.1016/j.actamat.2021.117431},
    issn = {13596454}
}

@article{Wu2023ATechnology,
    title = {{A new strategy of Al0.1CoCrFeNi high entropy alloys and Inconel 625 alloys joining technology}},
    year = {2023},
    journal = {Journal of Materials Processing Technology},
    author = {Wu, Baolei and Yu, Weiyuan and Zhu, Wenqi and Li, Yang and Hu, Zhen},
    month = {12},
    pages = {118251},
    publisher = {Elsevier},
    url = {https://linkinghub.elsevier.com/retrieve/pii/S0924013623003965},
    doi = {10.1016/J.JMATPROTEC.2023.118251},
    issn = {0924-0136}
}

@techreport{Dijkstra1959AGraphs,
    title = {{A Note on Two Problems in Connexion with Graphs}},
    year = {1959},
    booktitle = {Numerische Mathematik {\~{}}},
    author = {Dijkstra, E W},
    pages = {269--271},
    doi = {10.1007/FBF01386390}
}

@article{Hattrick-Simpers2018AMaterials,
    title = {{A simple constrained machine learning model for predicting high-pressure-hydrogen-compressor materials}},
    year = {2018},
    journal = {Cite this: Mol. Syst. Des. Eng},
    author = {Hattrick-Simpers, Jason R and Choudhary, Kamal and Corgnale, Claudio},
    pages = {509},
    volume = {3},
    doi = {10.1039/c8me00005k}
}

@article{Allen2022AAlloys,
    title = {{A Subspace-Inclusive Sampling Method for the Computational Design of Compositionally Graded Alloys}},
    year = {2022},
    author = {Allen, Marshall and Kirk, Tanner and Malak, Richard and Arroyave, Raymundo},
    url = {http://asmedigitalcollection.asme.org/mechanicaldesign/article-pdf/144/4/041704/6847101/md_144_4_041704.pdf},
    doi = {10.1115/1.4040816}
}

@article{Hargather2022ANi,
    title = {{A Systematic First-Principles Study of Computational Parameters Affecting Self-diffusion Coefficients in FCC Ag, Cu, and Ni}},
    year = {2022},
    journal = {Journal of Phase Equilibria and Diffusion},
    author = {Hargather, Chelsey Z. and O’Connell, John M.},
    month = {9},
    url = {https://link.springer.com/10.1007/s11669-022-00991-4},
    doi = {10.1007/s11669-022-00991-4},
    issn = {1547-7037}
}

@article{Ben-David2010ADomains,
    title = {{A theory of learning from different domains}},
    year = {2010},
    journal = {Mach Learn},
    author = {Ben-David, Shai and Blitzer, John and Crammer, Koby and Kulesza, Alex and Pereira, Fernando and Wortman Vaughan, Jennifer and Ben-David David Cheriton, S R and Blitzer, J and Crammer, K and Kulesza, A and Pereira, F and Vaughan, JW},
    pages = {151--175},
    volume = {79},
    doi = {10.1007/s10994-009-5152-4}
}

@article{Choi2019ADesign,
    title = {{A thermodynamic description of the Co-Cr-Fe-Ni-V system for high-entropy alloy design}},
    year = {2019},
    journal = {Calphad},
    author = {Choi, Won-Mi and Jo, Yong Hee and Kim, Dong Geun and Sohn, Seok Su and Lee, Sunghak and Lee, Byeong-Joo},
    month = {9},
    pages = {101624},
    volume = {66},
    url = {https://linkinghub.elsevier.com/retrieve/pii/S0364591618302165},
    doi = {10.1016/j.calphad.2019.05.001},
    issn = {03645916}
}

@article{Kresse1993AbMetals,
    title = {{Ab initio molecular dynamics for liquid metals}},
    year = {1993},
    journal = {Physical Review B},
    author = {Kresse, G. and Hafner, J.},
    number = {1},
    month = {1},
    pages = {558--561},
    volume = {47},
    url = {https://link.aps.org/doi/10.1103/PhysRevB.47.558},
    doi = {10.1103/PhysRevB.47.558},
    issn = {0163-1829}
}

@article{Yan2021AcceleratedLearning,
    title = {{Accelerated discovery of single-phase refractory high entropy alloys assisted by machine learning}},
    year = {2021},
    journal = {Computational Materials Science},
    author = {Yan, Yonggang and Lu, Dan and Wang, Kun},
    month = {11},
    pages = {110723},
    volume = {199},
    publisher = {Elsevier},
    doi = {10.1016/J.COMMATSCI.2021.110723},
    issn = {0927-0256}
}

@article{Reichardt2021AdvancesMaterials,
    title = {{Advances in additive manufacturing of metal-based functionally graded materials}},
    year = {2021},
    journal = {International Materials Reviews},
    author = {Reichardt, Ashley and Shapiro, Andrew A. and Otis, Richard and Dillon, R. Peter and Borgonia, John Paul and McEnerney, Bryan W. and Hosemann, Peter and Beese, Allison M.},
    number = {1},
    month = {1},
    pages = {1--29},
    volume = {66},
    doi = {10.1080/09506608.2019.1709354},
    issn = {0950-6608}
}

@article{Pizzi2016AiiDA:Science,
    title = {{AiiDA: automated interactive infrastructure and database for computational science}},
    year = {2016},
    journal = {Computational Materials Science},
    author = {Pizzi, Giovanni and Cepellotti, Andrea and Sabatini, Riccardo and Marzari, Nicola and Kozinsky, Boris},
    month = {1},
    pages = {218--230},
    volume = {111},
    publisher = {Elsevier},
    url = {https://linkinghub.elsevier.com/retrieve/pii/S0927025615005820},
    doi = {10.1016/j.commatsci.2015.09.013},
    issn = {09270256},
    arxivId = {1504.01163}
}

@misc{Sandrock2017Alchemyst/ternplothttps://github.com/alchemyst/ternplot,
    title = {{alchemyst/ternplot (https://github.com/alchemyst/ternplot)}},
    year = {2017},
    author = {Sandrock, Carl},
    publisher = {GitHub}
}

@article{Williams1964AlgorithmHeapsort,
    title = {{Algorithm 232 - Heapsort}},
    year = {1964},
    journal = {Communications of the ACM},
    author = {Williams, J. W. J.},
    editor = {Forsythe, G. E.},
    number = {6},
    month = {6},
    pages = {347--349},
    volume = {7},
    url = {https://dl.acm.org/doi/10.1145/512274.512284},
    doi = {10.1145/512274.512284},
    issn = {0001-0782}
}

@article{Chasalow1995AlgorithmPoints,
    title = {{Algorithm AS 299: Generation of Simplex Lattice Points}},
    year = {1995},
    journal = {Applied Statistics},
    author = {Chasalow, Scott D. and Brand, Richard J.},
    number = {4},
    pages = {534},
    volume = {44},
    publisher = {JSTOR},
    doi = {10.2307/2986144},
    issn = {00359254}
}

@article{Cowley1950AnAlloys,
    title = {{An Approximate Theory of Order in Alloys}},
    year = {1950},
    journal = {Physical Review},
    author = {Cowley, J. M.},
    number = {5},
    month = {3},
    pages = {669},
    volume = {77},
    publisher = {American Physical Society},
    url = {https://journals.aps.org/pr/abstract/10.1103/PhysRev.77.669},
    doi = {10.1103/PhysRev.77.669},
    issn = {0031899X}
}

@article{SheikhAnAlloys,
    title = {{An Automated Fully-Computational Framework to Construct Printability Maps for Additively Manufactured Metal Alloys}},
    author = {Sheikh, Sofia and Ranaiefar, Meelad and Honarmandi, Pejman and Vela, Brent and Morcos, Peter and Shoukr, David and Karaman, Ibrahim and Elwany, Alaa and Arr{\'{o}}yave, Raymundo},
    arxivId = {2304.04113v1}
}

@article{Lu2023AnCells,
    title = {{An extension to Voro++ for multithreaded computation of Voronoi cells}},
    year = {2023},
    journal = {Computer Physics Communications},
    author = {Lu, Jiayin and Lazar, Emanuel A. and Rycroft, Chris H.},
    month = {10},
    pages = {108832},
    volume = {291},
    url = {https://linkinghub.elsevier.com/retrieve/pii/S0010465523001777},
    doi = {10.1016/j.cpc.2023.108832},
    issn = {00104655}
}

@article{Otis2017AnSystems,
    title = {{An improved sampling strategy for global energy minimization of multi-component systems}},
    year = {2017},
    journal = {Computational Materials Science},
    author = {Otis, Richard and Emelianenko, Maria and Liu, Zi-Kui},
    month = {4},
    pages = {282--291},
    volume = {130},
    url = {https://linkinghub.elsevier.com/retrieve/pii/S0927025617300332},
    doi = {10.1016/j.commatsci.2017.01.019},
    issn = {09270256}
}

@article{Zha2024ApplyingEnergy,
    title = {{Applying enhanced active learning to predict formation energy}},
    year = {2024},
    journal = {Computational Materials Science},
    author = {Zha, Yang and Liu, Wei and Fan, Jiayi and Jiang, Lu and Li, Ying and Lu, Xiao Gang},
    month = {2},
    pages = {112825},
    volume = {235},
    publisher = {Elsevier},
    doi = {10.1016/J.COMMATSCI.2024.112825},
    issn = {0927-0256}
}

@article{Harris2020ArrayNumPy,
    title = {{Array programming with NumPy}},
    year = {2020},
    journal = {Nature},
    author = {Harris, Charles R. and Millman, K. Jarrod and van der Walt, Stéfan J. and Gommers, Ralf and Virtanen, Pauli and Cournapeau, David and Wieser, Eric and Taylor, Julian and Berg, Sebastian and Smith, Nathaniel J. and Kern, Robert and Picus, Matti and Hoyer, Stephan and van Kerkwijk, Marten H. and Brett, Matthew and Haldane, Allan and del R{\'{i}}o, Jaime Fernández and Wiebe, Mark and Peterson, Pearu and G{\'{e}}rard-Marchant, Pierre and Sheppard, Kevin and Reddy, Tyler and Weckesser, Warren and Abbasi, Hameer and Gohlke, Christoph and Oliphant, Travis E.},
    number = {7825},
    month = {9},
    pages = {357--362},
    volume = {585},
    url = {https://www.nature.com/articles/s41586-020-2649-2},
    doi = {10.1038/s41586-020-2649-2},
    issn = {0028-0836}
}

@misc{ASMInternational,
    title = {{ASM Alloy Phase Diagram Database™ - ASM International}},
    url = {https://www.asminternational.org/materials-resources/online-databases/asm-alloy-phase-diagram-database/}
}

@article{Wang2009AtomicAlloy,
    title = {{Atomic packing efficiency and phase transition in a high entropy alloy}},
    year = {2009},
    journal = {Journal of Alloys and Compounds},
    author = {Wang, F. J. and Zhang, Y. and Chen, G. L.},
    number = {1-2},
    month = {6},
    pages = {321--324},
    volume = {478},
    publisher = {Elsevier},
    doi = {10.1016/J.JALLCOM.2008.11.059},
    issn = {0925-8388}
}

@article{Choudhary2021AtomisticPredictions,
    title = {{Atomistic Line Graph Neural Network for improved materials property predictions}},
    year = {2021},
    journal = {npj Computational Materials},
    author = {Choudhary, Kamal and DeCost, Brian},
    number = {1},
    month = {11},
    pages = {185},
    volume = {7},
    url = {https://www.nature.com/articles/s41524-021-00650-1},
    doi = {10.1038/s41524-021-00650-1},
    issn = {2057-3960}
}

@article{Hu2021Atomtransmachine:Learning,
    title = {{Atomtransmachine: An atomic feature representation model for machine learning}},
    year = {2021},
    journal = {Computational Materials Science},
    author = {Hu, Mengxian and Yuan, Jianmei and Sun, Tao and Huang, Meng and Liang, Qingyun},
    month = {12},
    pages = {110841},
    volume = {200},
    publisher = {Elsevier},
    doi = {10.1016/J.COMMATSCI.2021.110841},
    issn = {0927-0256}
}

@article{Jung2019BayesianSteels,
    title = {{Bayesian approach in predicting mechanical properties of materials: Application to dual phase steels}},
    year = {2019},
    journal = {Materials Science and Engineering A},
    author = {Jung, Jaimyun and Yoon, Jae Ik and Park, Hyung Keun and Kim, Jin You and Kim, Hyoung Seop},
    month = {1},
    pages = {382--390},
    volume = {743},
    publisher = {Elsevier Ltd},
    doi = {10.1016/j.msea.2018.11.106},
    issn = {09215093}
}

@article{Zunger2019BewareMaterials,
    title = {{Beware of plausible predictions of fantasy materials}},
    year = {2019},
    journal = {Nature 2021 566:7745},
    author = {Zunger, Alex},
    number = {7745},
    month = {2},
    pages = {447--449},
    volume = {566},
    publisher = {Nature Publishing Group},
    url = {https://www.nature.com/articles/d41586-019-00676-y},
    doi = {10.1038/d41586-019-00676-y},
    issn = {14764687},
    pmid = {30814720}
}

@article{Agren2023CALPHADAnniversary,
    title = {{CALPHAD and the materials genome A 10 year anniversary}},
    year = {2023},
    journal = {Calphad},
    author = {{\AA}gren, John},
    month = {3},
    pages = {102532},
    volume = {80},
    publisher = {Pergamon},
    doi = {10.1016/J.CALPHAD.2023.102532},
    issn = {0364-5916}
}

@misc{CHADWICKArpa-e.energy.gov,
    title = {{CHADWICK | arpa-e.energy.gov}},
    url = {https://arpa-e.energy.gov/technologies/programs/chadwick}
}

@article{Deng2023CHGNetModelling,
    title = {{CHGNet as a pretrained universal neural network potential for charge-informed atomistic modelling}},
    year = {2023},
    journal = {Nature Machine Intelligence |},
    author = {Deng, Bowen and Zhong, Peichen and Jun, Kyujung and Riebesell, Janosh and Han, Kevin and Bartel, Christopher J and Ceder, Gerbrand},
    pages = {1031--1041},
    volume = {5},
    url = {https://doi.org/10.1038/s42256-023-00716-3},
    doi = {10.1038/s42256-023-00716-3}
}

@book{Nijenhuis1978CombinatorialCalculators,
    title = {{Combinatorial Algorithms For Computers and Calculators}},
    year = {1978},
    author = {Nijenhuis, Albert and Wlif, Herbert S.},
    edition = {2},
    pages = {46--51},
    publisher = {Academic Press},
    url = {https://linkinghub.elsevier.com/retrieve/pii/C20130112433},
    isbn = {9780125192606},
    doi = {10.1016/C2013-0-11243-3}
}

@article{GambaroCombinedAlloys,
    title = {{Combined Experimental and Calphad Investigation of Alcocrfeni-X (X=Mo,Ta,W) High-Entropy Alloys}},
    author = {Gambaro, Sofia and Fenocchio, Lorenzo and Valenza, Fabrizio and Riani, Paola and Cacciamani, Gabriele},
    url = {https://papers.ssrn.com/abstract=4625326},
    doi = {10.2139/SSRN.4625326}
}

@article{Jain2013Commentary:Innovation,
    title = {{Commentary: The Materials Project: A materials genome approach to accelerating materials innovation}},
    year = {2013},
    journal = {APL Materials},
    author = {Jain, Anubhav and Ong, Shyue Ping and Hautier, Geoffroy and Chen, Wei and Richards, William Davidson and Dacek, Stephen and Cholia, Shreyas and Gunter, Dan and Skinner, David and Ceder, Gerbrand and Persson, Kristin A.},
    number = {1},
    month = {7},
    pages = {011002},
    volume = {1},
    url = {http://aip.scitation.org/doi/10.1063/1.4812323},
    doi = {10.1063/1.4812323},
    issn = {2166-532X}
}

@article{Debnath2023ComparingAlloys,
    title = {{Comparing forward and inverse design paradigms: A case study on refractory high-entropy alloys}},
    year = {2023},
    journal = {Journal of Materials Research},
    author = {Debnath, Arindam and Raman, Lavanya and Li, Wenjie and Krajewski, Adam M. and Ahn, Marcia and Lin, Shuang and Shang, Shunli and Beese, Allison M. and Liu, Zi-Kui and Reinhart, Wesley F.},
    number = {17},
    month = {9},
    pages = {4107--4117},
    volume = {38},
    url = {https://link.springer.com/10.1557/s43578-023-01122-6},
    doi = {10.1557/s43578-023-01122-6},
    issn = {0884-2914}
}

@article{Alturk2017ComparisonMaterial,
    title = {{Comparison of different hybrid DFT methods on structural, spectroscopic, electronic and NLO parameters for a potential NLO material}},
    year = {2017},
    journal = {Computational and Theoretical Chemistry},
    author = {Alturk, Sumeyye and Avcı, Davut and Tamer, Omer and Atalay, Yusuf},
    month = {1},
    pages = {34--45},
    volume = {1100},
    url = {http://dx.doi.org/10.1016/j.comptc.2016.12.007 https://linkinghub.elsevier.com/retrieve/pii/S2210271X16304935},
    doi = {10.1016/j.comptc.2016.12.007},
    issn = {2210271X}
}

@article{Kirk2021ComputationalMonotonicity,
    title = {{Computational design of compositionally graded alloys for property monotonicity}},
    year = {2021},
    journal = {Journal of Mechanical Design},
    author = {Kirk, Tanner and Malak, Richard and Arroyave, Raymundo},
    number = {3},
    month = {3},
    volume = {143},
    publisher = {American Society of Mechanical Engineers (ASME)},
    url = {https://dx.doi.org/10.1115/1.4048627},
    doi = {10.1115/1.4048627/1087579},
    issn = {10500472}
}

@article{Elder2023ComputationalDown-selection,
    title = {{Computational discovery of ultra-strong, stable, and lightweight refractory multi-principal element alloys. Part I: design principles and rapid down-selection}},
    year = {2023},
    journal = {npj Computational Materials},
    author = {Elder, Kate L. M. and Berry, Joel and Bocklund, Brandon and McCall, Scott K. and Perron, Aurélien and McKeown, Joseph T.},
    number = {1},
    month = {5},
    pages = {84},
    volume = {9},
    publisher = {Nature Publishing Group},
    url = {https://www.nature.com/articles/s41524-023-01030-7},
    doi = {10.1038/s41524-023-01030-7},
    issn = {2057-3960}
}

@article{Elder2023ComputationalValidation,
    title = {{Computational discovery of ultra-strong, stable, and lightweight refractory multi-principal element alloys. Part II: comprehensive ternary design and validation}},
    year = {2023},
    journal = {npj Computational Materials},
    author = {Elder, Kate L. M. and Berry, Joel and Perron, Aurélien and Bocklund, Brandon and Shittu, Jibril and Rietema, Connor J. and Henderson, Hunter B. and McCall, Scott K. and McKeown, Joseph T.},
    number = {1},
    month = {5},
    pages = {88},
    volume = {9},
    publisher = {Nature Publishing Group},
    url = {https://www.nature.com/articles/s41524-023-01031-6},
    doi = {10.1038/s41524-023-01031-6},
    issn = {2057-3960}
}

@article{Ouyang2020ComputationalConductors,
    title = {{Computational Investigation of Halogen-Substituted Na Argyrodites as Solid-State Superionic Conductors}},
    year = {2020},
    journal = {Chemistry of Materials},
    author = {Ouyang, Bin and Wang, Yan and Sun, Yingzhi and Ceder, Gerbrand},
    number = {5},
    pages = {1896--1903},
    volume = {32},
    url = {https://dx.doi.org/10.1021/acs.chemmater.9b04541},
    doi = {10.1021/acs.chemmater.9b04541}
}

@misc{Kaufman1970ComputerMetals,
    title = {{Computer calculation of phase diagrams. With special reference to refractory metals}},
    year = {1970},
    author = {Kaufman, L. and Bernstein, H.},
    url = {http://inis.iaea.org/Search/search.aspx?orig_q=RN:2004171}
}

@article{Hermann2013CondensedMetallic,
    title = {{Condensed astatine: Monatomic and metallic}},
    year = {2013},
    journal = {Physical Review Letters},
    author = {Hermann, Andreas and Hoffmann, Roald and Ashcroft, N. W.},
    number = {11},
    month = {9},
    pages = {116404},
    volume = {111},
    publisher = {American Physical Society},
    url = {https://journals.aps.org/prl/abstract/10.1103/PhysRevLett.111.116404},
    doi = {10.1103/PHYSREVLETT.111.116404/FIGURES/4/MEDIUM},
    issn = {00319007}
}

@article{Mirza2014ConditionalNets,
    title = {{Conditional Generative Adversarial Nets}},
    year = {2014},
    author = {Mirza, Mehdi and Osindero, Simon},
    month = {11},
    url = {http://arxiv.org/abs/1411.1784},
    arxivId = {1411.1784}
}

@article{Chong2021CorrelationAlloys,
    title = {{Correlation analysis of materials properties by machine learning: illustrated with stacking fault energy from first-principles calculations in dilute fcc-based alloys}},
    year = {2021},
    journal = {Journal of Physics: Condensed Matter},
    author = {Chong, Xiaoyu and Shang, Shun Li and Krajewski, Adam M. and Shimanek, John D. and Du, Weihang and Wang, Yi and Feng, Jing and Shin, Dongwon and Beese, Allison M. and Liu, Zi Kui},
    number = {29},
    month = {6},
    pages = {295702},
    volume = {33},
    publisher = {IOP Publishing},
    url = {https://iopscience.iop.org/article/10.1088/1361-648X/ac0195},
    doi = {10.1088/1361-648X/AC0195},
    issn = {0953-8984},
    pmid = {34132202}
}

@misc{Feurer2019Cr-Fe-NiCalculations,
    title = {{Cr-Fe-Ni Sigma Phase Finite Temperature Calculations}},
    year = {2019},
    booktitle = {Citrine Informatics},
    author = {Feurer, Matthew and Bocklund, Brandon and Shang, Shunli and Beese, Alison and Liu, Zi-Kui},
    url = {https://doi.org/10.25920/yjrc-zj59},
    doi = {10.25920/YJRC-ZJ59}
}

@article{Joubert2008CrystalPhase,
    title = {{Crystal chemistry and Calphad modeling of the {$\sigma$} phase}},
    year = {2008},
    journal = {Progress in Materials Science},
    author = {Joubert, J.-M.},
    number = {3},
    month = {3},
    pages = {528--583},
    volume = {53},
    url = {https://linkinghub.elsevier.com/retrieve/pii/S0079642507000242},
    doi = {10.1016/j.pmatsci.2007.04.001},
    issn = {00796425}
}

@article{Grazulis2012CrystallographyCollaboration,
    title = {{Crystallography Open Database (COD): An open-access collection of crystal structures and platform for world-wide collaboration}},
    year = {2012},
    journal = {Nucleic Acids Research},
    author = {Gra{\v{z}}ulis, Saulius and Da{\v{s}}kevi{\v{c}}, Adriana and Merkys, Andrius and Chateigner, Daniel and Lutterotti, Luca and Quir{\'{o}}s, Miguel and Serebryanaya, Nadezhda R. and Moeck, Peter and Downs, Robert T. and Le Bail, Armel},
    number = {D1},
    month = {1},
    pages = {D420-D427},
    volume = {40},
    publisher = {Oxford Academic},
    url = {https://academic.oup.com/nar/article/40/D1/D420/2903497},
    doi = {10.1093/nar/gkr900},
    issn = {03051048},
    pmid = {22070882}
}

@article{Grazulis2009CrystallographyStructures,
    title = {{Crystallography Open Database – an open-access collection of crystal structures}},
    year = {2009},
    journal = {Journal of Applied Crystallography},
    author = {Gra{\v{z}}ulis, Saulius and Chateigner, Daniel and Downs, Robert T. and Yokochi, A. F. T. and Quir{\'{o}}s, Miguel and Lutterotti, Luca and Manakova, Elena and Butkus, Justas and Moeck, Peter and Le Bail, Armel},
    number = {4},
    month = {8},
    pages = {726--729},
    volume = {42},
    publisher = {International Union of Crystallography},
    doi = {10.1107/S0021889809016690},
    issn = {0021-8898}
}

@incollection{Grazulis2019CrystallographyPerspectives,
    title = {{Crystallography Open Database: History, Development, and Perspectives}},
    year = {2019},
    booktitle = {Materials Informatics},
    author = {Gra{\v{z}}ulis, Saulius and Merkys, Andrius and Vaitkus, Antanas and Chateigner, Daniel and Lutterotti, Luca and Moeck, Peter and Quiros, Miguel and Downs, Robert T. and Kaminsky, Werner and Le Bail, Armel},
    month = {10},
    pages = {1--39},
    publisher = {Wiley},
    url = {https://onlinelibrary.wiley.com/doi/10.1002/9783527802265.ch1},
    doi = {10.1002/9783527802265.ch1}
}

@article{Castleton2009DensitySupercells,
    title = {{Density functional theory calculations of defect energies using supercells}},
    year = {2009},
    journal = {Modelling and Simulation in Materials Science and Engineering},
    author = {Castleton, C W M and H{\"{o}}glund, A and Mirbt, S},
    number = {8},
    month = {12},
    pages = {084003},
    volume = {17},
    url = {https://iopscience.iop.org/article/10.1088/0965-0393/17/8/084003},
    doi = {10.1088/0965-0393/17/8/084003},
    issn = {0965-0393}
}

@article{Kohn1996DensityStructure,
    title = {{Density Functional Theory of Electronic Structure}},
    year = {1996},
    journal = {Journal of Physical Chemistry},
    author = {Kohn, W. and Becke, A. D. and Parr, R. G.},
    number = {31},
    month = {8},
    pages = {12974--12980},
    volume = {100},
    publisher = { American Chemical Society },
    url = {https://pubs.acs.org/doi/abs/10.1021/jp960669l},
    doi = {10.1021/JP960669L},
    issn = {00223654}
}

@article{Atta-Fynn2015DensityElements,
    title = {{Density functional theory Studies of condensed Phases of 6d super heavy elements}},
    year = {2015},
    journal = {Solid State Communications},
    author = {Atta-Fynn, Raymond and Ray, Asok K.},
    month = {1},
    pages = {88--94},
    volume = {201},
    url = {https://linkinghub.elsevier.com/retrieve/pii/S0038109814004463},
    doi = {10.1016/j.ssc.2014.10.025},
    issn = {00381098}
}

@article{Li2024DesignExperiments,
    title = {{Design and validation of refractory alloys using machine learning, CALPHAD, and experiments}},
    year = {2024},
    journal = {International Journal of Refractory Metals and Hard Materials},
    author = {Li, Wenjie and Raman, Lavanya and Debnath, Arindam and Ahn, Marcia and Lin, Shuang and Krajewski, Adam M. and Shang, Shunli and Priya, Shashank and Reinhart, Wesley F. and Liu, Zi-Kui and Beese, Allison M.},
    month = {6},
    pages = {106673},
    volume = {121},
    url = {https://linkinghub.elsevier.com/retrieve/pii/S0263436824001215},
    doi = {10.1016/j.ijrmhm.2024.106673},
    issn = {02634368}
}

@article{Yang2023DesignCracking,
    title = {{Design methodology for functionally graded materials: Framework for considering cracking}},
    year = {2023},
    journal = {Additive Manufacturing},
    author = {Yang, Zhening and Sun, Hui and Liu, Zi-Kui and Beese, Allison M},
    pages = {103672},
    volume = {73},
    url = {https://doi.org/10.1016/j.addma.2023.103672},
    doi = {10.1016/j.addma.2023.103672}
}

@article{Bobbio2022DesignCompositions,
    title = {{Design of an additively manufactured functionally graded material of 316 stainless steel and Ti-6Al-4V with Ni-20Cr, Cr, and V intermediate compositions}},
    year = {2022},
    journal = {Additive Manufacturing},
    author = {Bobbio, Lourdes D. and Bocklund, Brandon and Simsek, Emrah and Ott, Ryan T. and Kramer, Matt J. and Liu, Zi Kui and Beese, Allison M.},
    month = {3},
    pages = {102649},
    volume = {51},
    publisher = {Elsevier},
    doi = {10.1016/J.ADDMA.2022.102649},
    issn = {2214-8604}
}

@article{Hofmann2014DevelopingManufacturing,
    title = {{Developing gradient metal alloys through radial deposition additive manufacturing}},
    year = {2014},
    journal = {Scientific Reports},
    author = {Hofmann, Douglas C. and Roberts, Scott and Otis, Richard and Kolodziejska, Joanna and Dillon, R. Peter and Suh, Jong Ook and Shapiro, Andrew A. and Liu, Zi Kui and Borgonia, John Paul},
    month = {6},
    volume = {4},
    publisher = {Nature Publishing Groups},
    doi = {10.1038/SREP05357},
    issn = {20452322}
}

@article{Reichardt2016DevelopmentManufacturing,
    title = {{Development and characterization of Ti-6Al-4V to 304L stainless steel gradient components fabricated with laser deposition additive manufacturing}},
    year = {2016},
    journal = {Materials {\&} Design},
    author = {Reichardt, Ashley and Dillon, R. Peter and Borgonia, John Paul and Shapiro, Andrew A. and McEnerney, Bryan W. and Momose, Tatsuki and Hosemann, Peter},
    month = {8},
    pages = {404--413},
    volume = {104},
    publisher = {Elsevier},
    doi = {10.1016/J.MATDES.2016.05.016},
    issn = {0264-1275}
}

@article{Senkov2016DevelopmentSuperalloy,
    title = {{Development of a Refractory High Entropy Superalloy}},
    year = {2016},
    journal = {Entropy},
    author = {Senkov, Oleg and Isheim, Dieter and Seidman, David and Pilchak, Adam},
    number = {3},
    month = {3},
    pages = {102},
    volume = {18},
    url = {http://www.mdpi.com/1099-4300/18/3/102},
    doi = {10.3390/e18030102},
    issn = {1099-4300}
}

@article{Kim2016DevelopmentRatios,
    title = {{Development of Mo-Ni-Si-B metallic glass with high thermal stability and H versus E ratios}},
    year = {2016},
    author = {Kim, Jinwoo and Kyeong, Joon Seok and Ham, Moon-Ho and Minor, Andrew M and Kim, Hyang and Park, Eun Soo},
    url = {http://dx.doi.org/10.1016/j.matdes.2016.02.090},
    doi = {10.1016/j.matdes.2016.02.090}
}

@article{Evans2024DevelopmentsExchange,
    title = {{Developments and applications of the OPTIMADE API for materials discovery, design, and data exchange}},
    year = {2024},
    journal = {Digital Discovery},
    author = {Evans, Matthew and Bergsma, Johan and Merkys, Andrius and Andersen, Casper and Andersson, Oskar B. and Beltr{\'{a}}n, Daniel and Blokhin, Evgeny and Boland, Tara M. and Casta{\~{n}}eda Balderas, Rubén and Choudhary, Kamal and D{\'{i}}az D{\'{i}}az, Alberto and Dom{\'{i}}nguez Garc{\'{i}}a, Rodrigo and Eckert, Hagen and Eimre, Kristjan and Fuentes-Montero, Maria Elena and Krajewski, Adam M. and Mortensen, Jens Jørgen and N{\'{a}}poles-Duarte, José Manuel and Pietryga, Jacob and Qi, Ji and Trejo Carrillo, Felipe de Jesús and Vaitkus, Antanas and Yu, Jusong and Zettel, Adam and de Castro, Pedro Baptista and Carlsson, Johan Martin and Cerqueira, Tiago F. T. and Divilov, Simon and Hajiyani, Hamidreza and Hanke, Felix and Jose, Kevin and Oses, Corey and Riebesell, Janosh and Schmidt, Jonathan and Winston, Donald and Xie, Christen and Yang, Xiaoyu and Bonella, Sara and Botti, Silvana and Curtarolo, Stefano and Draxl, Claudia and Fuentes-Cobas, Luis E. E and Hospital, Adam and Liu, Zi-Kui and Marques, Miguel A., L. and Marzari, Nicola and Morris, Andrew James and Ong, Shyue Ping and Orozco, Modesto and Persson, Kristin and Thygesen, Kristian Sommer and Wolverton, Chris M and Scheidgen, Markus and Toher, Cormac and Conduit, Gareth and Pizzi, Giovanni and Grazulis, Saulius and Rignanese, Gian-Marco and Armiento, Rickard},
    month = {2},
    url = {http://pubs.rsc.org/en/Content/ArticleLanding/2024/DD/D4DD00039K},
    doi = {10.1039/D4DD00039K},
    issn = {2635-098X}
}

@article{Wang2021DFTTK:Calculations,
    title = {{DFTTK: Density Functional Theory ToolKit for high-throughput lattice dynamics calculations}},
    year = {2021},
    journal = {Calphad},
    author = {Wang, Yi and Liao, Mingqing and Bocklund, Brandon J. and Gao, Peng and Shang, Shun Li and Kim, Hojong and Beese, Allison M. and Chen, Long Qing and Liu, Zi Kui},
    month = {12},
    pages = {102355},
    volume = {75},
    publisher = {Pergamon},
    doi = {10.1016/J.CALPHAD.2021.102355},
    issn = {0364-5916}
}

@article{Lu2017DirectlyRange,
    title = {{Directly cast bulk eutectic and near-eutectic high entropy alloys with balanced strength and ductility in a wide temperature range}},
    year = {2017},
    journal = {Acta Materialia},
    author = {Lu, Yiping and Gao, Xuzhou and Jiang, Li and Chen, Zongning and Wang, Tongmin and Jie, Jinchuan and Kang, Huijun and Zhang, Yubo and Guo, Sheng and Ruan, Haihui and Zhao, Yonghao and Cao, Zhiqiang and Li, Tingju},
    month = {2},
    pages = {143--150},
    volume = {124},
    publisher = {Pergamon},
    doi = {10.1016/J.ACTAMAT.2016.11.016},
    issn = {1359-6454}
}

@techreport{Srivastava2014Dropout:Overfitting,
    title = {{Dropout: A Simple Way to Prevent Neural Networks from Overfitting}},
    year = {2014},
    booktitle = {Journal of Machine Learning Research},
    author = {Srivastava, Nitish and Hinton, Geoffrey and Krizhevsky, Alex and Salakhutdinov, Ruslan},
    pages = {1929--1958},
    volume = {15}
}

@article{Stepanov2019EffectContent,
    title = {{Effect of Al on structure and mechanical properties of Fe-Mn-Cr-Ni-Al non-equiatomic high entropy alloys with high Fe content}},
    year = {2019},
    journal = {Journal of Alloys and Compounds},
    author = {Stepanov, N. D. and Shaysultanov, D. G. and Chernichenko, R. S. and Tikhonovsky, M. A. and Zherebtsov, S. V.},
    month = {1},
    pages = {194--203},
    volume = {770},
    publisher = {Elsevier},
    doi = {10.1016/J.JALLCOM.2018.08.093},
    issn = {0925-8388}
}

@article{vandeWalle2013EfficientStructures,
    title = {{Efficient stochastic generation of special quasirandom structures}},
    year = {2013},
    journal = {Calphad},
    author = {van de Walle, A. and Tiwary, P. and de Jong, M. and Olmsted, D.L. and Asta, M. and Dick, A. and Shin, D. and Wang, Y. and Chen, L.-Q. and Liu, Z.-K.},
    month = {9},
    pages = {13--18},
    volume = {42},
    url = {https://linkinghub.elsevier.com/retrieve/pii/S0364591613000540},
    doi = {10.1016/j.calphad.2013.06.006},
    issn = {03645916}
}

@article{Krajewski2024EfficientStructures,
    title = {{Efficient Structure-Informed Featurization and Property Prediction of Ordered, Dilute, and Random Atomic Structures}},
    year = {2024},
    author = {Krajewski, Adam M. and Siegel, Jonathan W. and Liu, Zi-Kui},
    month = {4},
    url = {http://arxiv.org/abs/2404.02849},
    doi = {10.48550/arXiv.2404.02849},
    arxivId = {2404.02849}
}

@article{Liu2022EightyFuture,
    title = {{Eighty Years of the Finite Element Method: Birth, Evolution, and Future}},
    year = {2022},
    journal = {Archives of Computational Methods in Engineering 2022 29:6},
    author = {Liu, Wing Kam and Li, Shaofan and Park, Harold S.},
    number = {6},
    month = {6},
    pages = {4431--4453},
    volume = {29},
    publisher = {Springer},
    url = {https://link.springer.com/article/10.1007/s11831-022-09740-9},
    isbn = {0123456789},
    doi = {10.1007/S11831-022-09740-9},
    issn = {1886-1784},
    arxivId = {2107.04960}
}

@article{Koufos2013ElectronicFrancium,
    title = {{Electronic structure of francium}},
    year = {2013},
    journal = {International Journal of Quantum Chemistry},
    author = {Koufos, Alexander P. and Papaconstantopoulos, Dimitrios A.},
    number = {17},
    month = {9},
    pages = {2070--2077},
    volume = {113},
    publisher = {John Wiley {\&} Sons, Ltd},
    url = {https://onlinelibrary.wiley.com/doi/10.1002/qua.24466},
    doi = {10.1002/qua.24466},
    issn = {00207608}
}

@article{Jha2018ElemNet:Composition,
    title = {{ElemNet: Deep Learning the Chemistry of Materials From Only Elemental Composition}},
    year = {2018},
    journal = {Scientific Reports},
    author = {Jha, Dipendra and Ward, Logan and Paul, Arindam and Liao, Wei-keng and Choudhary, Alok and Wolverton, Chris and Agrawal, Ankit},
    number = {1},
    month = {12},
    pages = {17593},
    volume = {8},
    publisher = {Nature Publishing Group},
    url = {http://www.nature.com/articles/s41598-018-35934-y},
    doi = {10.1038/s41598-018-35934-y},
    issn = {2045-2322}
}

@article{Bucior2019Energy-based,
    title = {{Energy-based descriptors to rapidly predict hydrogen storage in metal-organic frameworks †}},
    year = {2019},
    journal = {Mol. Syst. Des. Eng},
    author = {Bucior, Benjamin J and Bobbitt, N Scott and Islamoglu, Timur and Goswami, Subhadip and Gopalan, Arun and Yildirim, Taner and Farha, Omar K and Bagheri, Neda and Snurr, Randall Q},
    pages = {162},
    volume = {4},
    doi = {10.1039/c8me00050f},
    issn = {2058-9689}
}

@article{Bocklund2019ESPEICuMg,
    title = {{ESPEI for efficient thermodynamic database development, modification, and uncertainty quantification: application to Cu–Mg}},
    year = {2019},
    journal = {MRS Communications},
    author = {Bocklund, Brandon and Otis, Richard and Egorov, Aleksei and Obaied, Abdulmonem and Roslyakova, Irina and Liu, Zi-Kui},
    number = {2},
    month = {6},
    pages = {618--627},
    volume = {9},
    url = {https://doi.org/10.1557/mrc.2019.59 http://link.springer.com/10.1557/mrc.2019.59},
    doi = {10.1557/mrc.2019.59},
    issn = {2159-6859}
}

@article{Wu2023EstimatingApproach,
    title = {{Estimating the temperature dependent zero-phase-fraction features in ternary phase diagram via Bayesian approach}},
    year = {2023},
    journal = {Scripta Materialia},
    author = {Wu, Biao and Zhang, Haihui and Zhang, Lanting and Wang, Hong},
    month = {10},
    pages = {115615},
    volume = {235},
    publisher = {Pergamon},
    url = {https://linkinghub.elsevier.com/retrieve/pii/S135964622300338X},
    doi = {10.1016/j.scriptamat.2023.115615},
    issn = {13596462}
}

@article{Banik2024EvaluatingMaterials,
    title = {{Evaluating generalized feature importance via performance assessment of machine learning models for predicting elastic properties of materials}},
    year = {2024},
    journal = {Computational Materials Science},
    author = {Banik, Suvo and Balasubramanian, Karthik and Manna, Sukriti and Derrible, Sybil and Sankaranarayananan, Subramanian K.R.S.},
    month = {3},
    pages = {112847},
    volume = {236},
    publisher = {Elsevier},
    url = {https://linkinghub.elsevier.com/retrieve/pii/S0927025624000685},
    doi = {10.1016/j.commatsci.2024.112847},
    issn = {09270256}
}

@article{Borg2020ExpandedAlloys,
    title = {{Expanded dataset of mechanical properties and observed phases of multi-principal element alloys}},
    year = {2020},
    journal = {Scientific Data},
    author = {Borg, Christopher K. H. and Frey, Carolina and Moh, Jasper and Pollock, Tresa M. and Gorsse, Stéphane and Miracle, Daniel B. and Senkov, Oleg N. and Meredig, Bryce and Saal, James E.},
    number = {1},
    month = {12},
    pages = {430},
    volume = {7},
    url = {https://www.nature.com/articles/s41597-020-00768-9},
    doi = {10.1038/s41597-020-00768-9},
    issn = {2052-4463}
}

@article{Zhao2014ExperimentalSystem,
    title = {{Experimental investigation and thermodynamic calculation of the phase equilibria in the Fe–Ni–V system}},
    year = {2014},
    journal = {Calphad},
    author = {Zhao, C.C. and Yang, S.Y. and Lu, Y and Guo, Y.H. and Wang, C.P. and Liu, X.J.},
    month = {9},
    pages = {80--86},
    volume = {46},
    url = {http://dx.doi.org/10.1016/j.calphad.2014.02.004 https://linkinghub.elsevier.com/retrieve/pii/S0364591614000200},
    doi = {10.1016/j.calphad.2014.02.004},
    issn = {03645916}
}

@article{Bocklund2020ExperimentalMaterials,
    title = {{Experimental validation of Scheil–Gulliver simulations for gradient path planning in additively manufactured functionally graded materials}},
    year = {2020},
    journal = {Materialia},
    author = {Bocklund, Brandon and Bobbio, Lourdes D. and Otis, Richard A. and Beese, Allison M. and Liu, Zi-Kui},
    month = {6},
    pages = {100689},
    volume = {11},
    url = {https://linkinghub.elsevier.com/retrieve/pii/S258915292030106X},
    doi = {10.1016/j.mtla.2020.100689},
    issn = {25891529}
}

@article{Trombach2019ExploringTheory,
    title = {{Exploring the chemical nature of super-heavy main-group elements by means of efficient plane-wave density-functional theory}},
    year = {2019},
    journal = {Physical Chemistry Chemical Physics},
    author = {Trombach, Lukas and Ehlert, Sebastian and Grimme, Stefan and Schwerdtfeger, Peter and Mewes, Jan-Michael},
    number = {33},
    month = {8},
    pages = {18048--18058},
    volume = {21},
    publisher = {Royal Society of Chemistry},
    url = {https://xlink.rsc.org/?DOI=C9CP02455G},
    doi = {10.1039/C9CP02455G},
    issn = {1463-9076}
}

@article{Krajewski2022ExtensibleNetworks,
    title = {{Extensible Structure-Informed Prediction of Formation Energy with improved accuracy and usability employing neural networks}},
    year = {2022},
    journal = {Computational Materials Science},
    author = {Krajewski, Adam M. and Siegel, Jonathan W. and Xu, Jinchao and Liu, Zi-Kui},
    month = {6},
    pages = {111254},
    volume = {208},
    publisher = {Elsevier B.V.},
    url = {https://linkinghub.elsevier.com/retrieve/pii/S0927025622000593},
    doi = {10.1016/j.commatsci.2022.111254},
    issn = {09270256}
}

@article{Shen2016FabricationProcess,
    title = {{Fabrication of Fe-FeAl Functionally Graded Material Using the Wire-Arc Additive Manufacturing Process}},
    year = {2016},
    journal = {Metallurgical and Materials Transactions B},
    author = {Shen, Chen and Pan, Zengxi and Cuiuri, Dominic and Roberts, Jon and Li, Huijun},
    number = {1},
    month = {2},
    pages = {763--772},
    volume = {47},
    url = {https://link.springer.com/10.1007/s11663-015-0509-5},
    doi = {10.1007/s11663-015-0509-5},
    issn = {1073-5615}
}

@misc{FactHouse,
    title = {{Fact Sheet: Developing a Bold Vision for Commercial Fusion Energy | OSTP | The White House}},
    url = {https://www.whitehouse.gov/ostp/news-updates/2022/03/15/fact-sheet-developing-a-bold-vision-for-commercial-fusion-energy/}
}

@misc{FAIRFAIR,
    title = {{FAIR Principles - GO FAIR}},
    url = {https://www.go-fair.org/fair-principles/}
}

@article{Atarah2020FirstNihonium,
    title = {{First principle study of the structural and electronic properties of Nihonium}},
    year = {2020},
    journal = {MRS Advances},
    author = {Atarah, Samuel A. and Egblewogbe, Martin N. H. and Hagoss, Gebreyesus G.},
    number = {23-24},
    month = {5},
    pages = {1175--1183},
    volume = {5},
    url = {http://link.springer.com/10.1557/adv.2020.159},
    doi = {10.1557/adv.2020.159},
    issn = {2059-8521}
}

@article{Jiang2004First-principlesStructures,
    title = {{First-principles study of binary bcc alloys using special quasirandom structures}},
    year = {2004},
    journal = {Physical Review B},
    author = {Jiang, Chao and Wolverton, C and Sofo, Jorge and Chen, Long-Qing and Liu, Zi-Kui},
    number = {21},
    month = {6},
    pages = {214202},
    volume = {69},
    url = {https://link.aps.org/doi/10.1103/PhysRevB.69.214202},
    doi = {10.1103/PhysRevB.69.214202},
    issn = {1098-0121}
}

@article{Davariashtiyani2023FormationRepresentation,
    title = {{Formation energy prediction of crystalline compounds using deep convolutional network learning on voxel image representation}},
    year = {2023},
    journal = {Communications Materials},
    author = {Davariashtiyani, Ali and Kadkhodaei, Sara},
    number = {1},
    month = {12},
    pages = {105},
    volume = {4},
    url = {https://www.nature.com/articles/s43246-023-00433-9},
    doi = {10.1038/s43246-023-00433-9},
    issn = {2662-4443}
}

@article{Shang2021FormingJoints,
    title = {{Forming mechanism of equilibrium and non-equilibrium metallurgical phases in dissimilar aluminum/steel (Al–Fe) joints}},
    year = {2021},
    journal = {Scientific Reports},
    author = {Shang, Shun-Li and Sun, Hui and Pan, Bo and Wang, Yi and Krajewski, Adam M. and Banu, Mihaela and Li, Jingjing and Liu, Zi-Kui},
    number = {1},
    month = {12},
    pages = {24251},
    volume = {11},
    url = {https://www.nature.com/articles/s41598-021-03578-0},
    doi = {10.1038/s41598-021-03578-0},
    issn = {2045-2322}
}

@article{Dechter1985GeneralizedA,
    title = {{Generalized best-first search strategies and the optimality of A*}},
    year = {1985},
    journal = {Journal of the ACM},
    author = {Dechter, Rina and Pearl, Judea},
    number = {3},
    month = {7},
    pages = {505--536},
    volume = {32},
    url = {https://dl.acm.org/doi/10.1145/3828.3830},
    doi = {10.1145/3828.3830},
    issn = {0004-5411}
}

@article{Goodfellow2020GenerativeNetworks,
    title = {{Generative adversarial networks}},
    year = {2020},
    journal = {Communications of the ACM},
    author = {Goodfellow, Ian and Pouget-Abadie, Jean and Mirza, Mehdi and Xu, Bing and Warde-Farley, David and Ozair, Sherjil and Courville, Aaron and Bengio, Yoshua},
    number = {11},
    month = {10},
    pages = {139--144},
    volume = {63},
    url = {https://dl.acm.org/doi/10.1145/3422622},
    doi = {10.1145/3422622},
    issn = {0001-0782}
}

@article{Debnath2021GenerativeAlloys,
    title = {{Generative deep learning as a tool for inverse design of high entropy refractory alloys}},
    year = {2021},
    journal = {Journal of Materials Informatics},
    author = {Debnath, Arindam and Krajewski, Adam M. and Sun, Hui and Lin, Shuang and Ahn, Marcia and Li, Wenjie and Priya, Shanshank and Singh, Jogender and Shang, Shunli and Beese, Allison M. and Liu, Zi-Kui and Reinhart, Wesley F.},
    number = {1},
    month = {9},
    pages = {3},
    volume = {1},
    publisher = {OAE Publishing Inc.},
    url = {https://www.oaepublish.com/articles/jmi.2021.05},
    doi = {10.20517/jmi.2021.05}
}

@misc{Ching-YuanBaiJuliaElliott2019GenerativeImages,
    title = {{Generative Dog Images}},
    year = {2019},
    author = {Ching-Yuan Bai Julia Elliott, Wendy Kan},
    publisher = {Kaggle},
    url = {https://kaggle.com/competitions/generative-dog-images}
}

@article{Olson2023GenomicDynamics,
    title = {{Genomic materials design: CALculation of PHAse Dynamics}},
    year = {2023},
    journal = {Calphad},
    author = {Olson, G.B. and Liu, Z.K.},
    month = {9},
    pages = {102590},
    volume = {82},
    publisher = {Pergamon},
    url = {https://linkinghub.elsevier.com/retrieve/pii/S0364591623000627},
    doi = {10.1016/j.calphad.2023.102590},
    issn = {03645916}
}

@article{Zaki2023Glassomics:Intelligence,
    title = {{Glassomics: An omics approach toward understanding glasses through modeling, simulations, and artificial intelligence}},
    year = {2023},
    journal = {MRS Bulletin},
    author = {Zaki, Mohd and Jan, Amreen and Krishnan, N. M. Anoop and Mauro, John C.},
    number = {10},
    month = {10},
    pages = {1026--1039},
    volume = {48},
    publisher = {Springer Nature},
    url = {https://link.springer.com/10.1557/s43577-023-00560-1},
    doi = {10.1557/s43577-023-00560-1},
    issn = {0883-7694}
}

@article{Chen2019GraphCrystals,
    title = {{Graph Networks as a Universal Machine Learning Framework for Molecules and Crystals}},
    year = {2019},
    journal = {Cite This: Chem. Mater},
    author = {Chen, Chi and Ye, Weike and Zuo, Yunxing and Zheng, Chen and Ong, Shyue Ping},
    pages = {3572},
    volume = {31},
    url = {https://pubs.acs.org/sharingguidelines},
    doi = {10.1021/acs.chemmater.9b01294}
}

@book{Torrey2010HandbookLearning,
    title = {{Handbook of Research on Machine Learning Applications and Trends - Chapter 11: Transfer Learning}},
    year = {2010},
    author = {Torrey, Lisa and Shavlik, Jude},
    editor = {Olivas, Emilio Soria and Guerrero, José David Martín and Martinez-Sober, Marcelino and Magdalena-Benedito, Jose Rafael and Serrano L{\'{o}}pez, Antonio José},
    publisher = {IGI Global},
    url = {http://services.igi-global.com/resolvedoi/resolve.aspx?doi=10.4018/978-1-60566-766-9},
    isbn = {9781605667669},
    doi = {10.4018/978-1-60566-766-9}
}

@article{Senkov2019HighAlloys,
    title = {{High temperature strength of refractory complex concentrated alloys}},
    year = {2019},
    journal = {Acta Materialia},
    author = {Senkov, O. N. and Gorsse, S. and Miracle, D. B.},
    month = {8},
    pages = {394--405},
    volume = {175},
    publisher = {Acta Materialia Inc},
    doi = {10.1016/j.actamat.2019.06.032},
    issn = {13596454}
}

@article{Feng2021High-throughputAlloys,
    title = {{High-throughput design of high-performance lightweight high-entropy alloys}},
    year = {2021},
    journal = {Nature Communications},
    author = {Feng, Rui and Zhang, Chuan and Gao, Michael C. and Pei, Zongrui and Zhang, Fan and Chen, Yan and Ma, Dong and An, Ke and Poplawsky, Jonathan D. and Ouyang, Lizhi and Ren, Yang and Hawk, Jeffrey A. and Widom, Michael and Liaw, Peter K.},
    number = {1},
    month = {7},
    pages = {4329},
    volume = {12},
    url = {https://www.nature.com/articles/s41467-021-24523-9},
    doi = {10.1038/s41467-021-24523-9},
    issn = {2041-1723}
}

@article{Legrain2017HowSolids,
    title = {{How Chemical Composition Alone Can Predict Vibrational Free Energies and Entropies of Solids}},
    year = {2017},
    journal = {Chemistry of Materials},
    author = {Legrain, Fleur and Carrete, Jesús and Van Roekeghem, Ambroise and Curtarolo, Stefano and Mingo, Natalio},
    number = {15},
    month = {8},
    pages = {6220--6227},
    volume = {29},
    publisher = {American Chemical Society},
    doi = {10.1021/acs.chemmater.7b00789},
    issn = {15205002},
    arxivId = {1703.02309}
}

@article{Schutt2014HowProperties,
    title = {{How to represent crystal structures for machine learning: Towards fast prediction of electronic properties}},
    year = {2014},
    journal = {Physical Review B},
    author = {Sch{\"{u}}tt, K. T. and Glawe, H. and Brockherde, F. and Sanna, A. and M{\"{u}}ller, K. R. and Gross, E. K. U.},
    number = {20},
    month = {5},
    pages = {205118},
    volume = {89},
    publisher = {American Physical Society},
    url = {https://link.aps.org/doi/10.1103/PhysRevB.89.205118},
    doi = {10.1103/PhysRevB.89.205118},
    issn = {1098-0121},
    arxivId = {1307.1266}
}

@article{Sayler2024HypersonicCongress,
    title = {{Hypersonic Weapons: Background and Issues for Congress}},
    year = {2024},
    author = {Sayler, Kelley M},
    url = {https://crsreports.congress.gov/product/pdf/r/r45811}
}

@article{Ward2017IncludingTessellations,
    title = {{Including crystal structure attributes in machine learning models of formation energies via Voronoi tessellations}},
    year = {2017},
    journal = {Physical Review B},
    author = {Ward, Logan and Liu, Ruoqian and Krishna, Amar and Hegde, Vinay I. and Agrawal, Ankit and Choudhary, Alok and Wolverton, Chris},
    number = {2},
    month = {7},
    pages = {024104},
    volume = {96},
    publisher = {American Physical Society},
    url = {http://link.aps.org/doi/10.1103/PhysRevB.96.024104},
    doi = {10.1103/PhysRevB.96.024104},
    issn = {2469-9950}
}

@article{Chen2022InteractionStudy,
    title = {{Interaction of elements in dilute Mg alloys: a DFT and machine learning study}},
    year = {2022},
    journal = {Journal of Materials Research and Technology},
    author = {Chen, Tao and Yuan, Yuan and Mi, Xiaoxi and Wu, Jiajia and Tang, Aitao and Wang, Jingfeng and Moelans, Nele and Pan, Fusheng},
    month = {11},
    pages = {4512--4525},
    volume = {21},
    publisher = {Elsevier},
    doi = {10.1016/J.JMRT.2022.11.071},
    issn = {2238-7854}
}

@inproceedings{Jha2019IRNet,
    title = {{IRNet}},
    year = {2019},
    booktitle = {Proceedings of the 25th ACM SIGKDD International Conference on Knowledge Discovery {\&} Data Mining},
    author = {Jha, Dipendra and Ward, Logan and Yang, Zijiang and Wolverton, Christopher and Foster, Ian and Liao, Wei-keng and Choudhary, Alok and Agrawal, Ankit},
    month = {7},
    pages = {2385--2393},
    publisher = {ACM},
    url = {https://dl.acm.org/doi/10.1145/3292500.3330703},
    address = {New York, NY, USA},
    isbn = {9781450362016},
    doi = {10.1145/3292500.3330703}
}

@article{Ury2023Kawin:Model,
    title = {{Kawin: An open source Kampmann-Wagner Numerical (KWN) phase precipitation and coarsening model}},
    year = {2023},
    journal = {Acta Materialia},
    author = {Ury, Nicholas and Neuberger, Raymond and Sargent, Noah and Xiong, Wei and Arr{\'{o}}yave, Raymundo and Otis, Richard},
    pages = {118988},
    volume = {255},
    url = {https://doi.org/10.1016/j.actamat.2023.118988},
    doi = {10.1016/j.actamat.2023.118988}
}

@misc{L2Regularization,
    title = {{L2 Regularization}},
    url = {https://www.textbook.ds100.org/ch/16/reg_ridge.html}
}

@article{Tynes2024LinearCheminformatics,
    title = {{Linear Graphlet Models for Accurate and Interpretable Cheminformatics}},
    year = {2024},
    author = {Tynes, Michael and Taylor, Michael G and Janssen, Jan and Burrill, Daniel J and Perez, Danny and Yang, Ping and Lubbers, Nicholas},
    month = {2},
    url = {https://chemrxiv.org/engage/chemrxiv/article-details/65d9282fe9ebbb4db916761e},
    doi = {10.26434/chemrxiv-2024-r81c8}
}

@article{Gryko2000Low-densityGap,
    title = {{Low-density framework form of crystalline silicon with a wide optical band gap}},
    year = {2000},
    journal = {Physical Review B},
    author = {Gryko, Jan and McMillan, Paul F. and Marzke, Robert F. and Ramachandran, Ganesh K. and Patton, Derek and Deb, Sudip K. and Sankey, Otto F.},
    number = {12},
    month = {9},
    pages = {R7707-R7710},
    volume = {62},
    url = {https://link.aps.org/doi/10.1103/PhysRevB.62.R7707},
    doi = {10.1103/PhysRevB.62.R7707},
    issn = {0163-1829}
}

@article{Wen2019MachineProperty,
    title = {{Machine learning assisted design of high entropy alloys with desired property}},
    year = {2019},
    journal = {Acta Materialia},
    author = {Wen, Cheng and Zhang, Yan and Wang, Changxin and Xue, Dezhen and Bai, Yang and Antonov, Stoichko and Dai, Lanhong and Lookman, Turab and Su, Yanjing},
    month = {5},
    pages = {109--117},
    volume = {170},
    publisher = {Acta Materialia Inc},
    url = {https://linkinghub.elsevier.com/retrieve/pii/S1359645419301430},
    doi = {10.1016/j.actamat.2019.03.010},
    issn = {13596454}
}

@article{Catal2023MachineProperties,
    title = {{Machine learning assisted design of novel refractory high entropy alloys with enhanced mechanical properties}},
    year = {2023},
    author = {Catal, A A and Bedir, E and Yilmaz, R and Swider, M A and Lee, C and El-Atwani, O and Maier, H J and Ozdemir, H C and Canadinc, D},
    url = {https://doi.org/10.1016/j.commatsci.2023.112612},
    doi = {10.1016/j.commatsci.2023.112612}
}

@article{Tao2021MachineDiscovery,
    title = {{Machine learning for perovskite materials design and discovery}},
    year = {2021},
    journal = {npj Computational Materials},
    author = {Tao, Qiuling and Xu, Pengcheng and Li, Minjie and Lu, Wencong},
    number = {1},
    month = {1},
    pages = {23},
    volume = {7},
    url = {https://www.nature.com/articles/s41524-021-00495-8},
    doi = {10.1038/s41524-021-00495-8},
    issn = {2057-3960}
}

@article{Pilania1987MachineSuperlattices,
    title = {{Machine learning properties of binary wurtzite superlattices}},
    year = {1987},
    journal = {Journal of Materials Science},
    author = {Pilania, G and Liu, X.-Y},
    volume = {53},
    url = {https://doi.org/10.1007/s10853-018-1987-z},
    doi = {10.1007/s10853-018-1987-z}
}

@article{Jha2023MachineChallenges,
    title = {{Machine learning-assisted materials development and device management in batteries and supercapacitors: performance comparison and challenges}},
    year = {2023},
    journal = {Journal of Materials Chemistry A},
    author = {Jha, Swarn and Yen, Matthew and Salinas, Yazmin Soto and Palmer, Evan and Villafuerte, John and Liang, Hong},
    number = {8},
    month = {2},
    pages = {3904--3936},
    volume = {11},
    publisher = {The Royal Society of Chemistry},
    url = {https://xlink.rsc.org/?DOI=D2TA07148G},
    doi = {10.1039/D2TA07148G},
    issn = {2050-7488}
}

@article{Rao2022MachineDiscovery,
    title = {{Machine learning–enabled high-entropy alloy discovery}},
    year = {2022},
    journal = {Science},
    author = {Rao, Ziyuan and Tung, Po-Yen and Xie, Ruiwen and Wei, Ye and Zhang, Hongbin and Ferrari, Alberto and Klaver, T.P.C. and K{\"{o}}rmann, Fritz and Sukumar, Prithiv Thoudden and Kwiatkowski da Silva, Alisson and Chen, Yao and Li, Zhiming and Ponge, Dirk and Neugebauer, Jörg and Gutfleisch, Oliver and Bauer, Stefan and Raabe, Dierk},
    number = {6615},
    month = {10},
    pages = {78--85},
    volume = {378},
    publisher = {American Association for the Advancement of Science},
    url = {https://www.science.org/doi/10.1126/science.abo4940},
    doi = {10.1126/science.abo4940},
    issn = {0036-8075}
}

@article{Huang2019Machine-learningAlloys,
    title = {{Machine-learning phase prediction of high-entropy alloys}},
    year = {2019},
    journal = {Acta Materialia},
    author = {Huang, Wenjiang and Martin, Pedro and Zhuang, Houlong L.},
    month = {5},
    pages = {225--236},
    volume = {169},
    publisher = {Acta Materialia Inc},
    url = {https://linkinghub.elsevier.com/retrieve/pii/S1359645419301454},
    doi = {10.1016/j.actamat.2019.03.012},
    issn = {13596454}
}

@article{Kim2018Machine-learning-acceleratedCompounds,
    title = {{Machine-learning-accelerated high-throughput materials screening: Discovery of novel quaternary Heusler compounds Machine Learning Accelerated High-Throughput Materials Screening: Discovery of Novel Quaternary Heusler Compounds}},
    year = {2018},
    journal = {Phys. Rev. Materials},
    author = {Kim, Kyoungdoc and Ward, Logan and He, Jiangang and Krishna, Amar and Agrawal, Ankit and Wolverton, C},
    pages = {123801},
    volume = {2},
    doi = {10.1103/PhysRevMaterials.2.123801}
}

@article{Schmidt2023Machine-Learning-AssistedMaterials,
    title = {{Machine-Learning-Assisted Determination of the Global Zero-Temperature Phase Diagram of Materials}},
    year = {2023},
    journal = {Advanced Materials},
    author = {Schmidt, Jonathan and Hoffmann, Noah and Wang, Hai-Chen and Borlido, Pedro and Carri{\c{c}}o, Pedro J M A and Cerqueira, Tiago F T and Botti, Silvana and Marques, Miguel A L and Schmidt, J and Hoffmann, N and Wang, H.-C and Marques, M A L and Borlido, P and Carri{\c{c}}o, P J M A and Cerqueira, T F T and Botti, S},
    number = {22},
    month = {6},
    pages = {2210788},
    volume = {35},
    publisher = {John Wiley {\&} Sons, Ltd},
    url = {https://onlinelibrary.wiley.com/doi/full/10.1002/adma.202210788 https://onlinelibrary.wiley.com/doi/abs/10.1002/adma.202210788 https://onlinelibrary.wiley.com/doi/10.1002/adma.202210788},
    doi = {10.1002/ADMA.202210788},
    issn = {1521-4095},
    pmid = {36949007}
}

@article{Silver2016MasteringSearch,
    title = {{Mastering the game of Go with deep neural networks and tree search}},
    year = {2016},
    journal = {Nature 2016 529:7587},
    author = {Silver, David and Huang, Aja and Maddison, Chris J. and Guez, Arthur and Sifre, Laurent and Van Den Driessche, George and Schrittwieser, Julian and Antonoglou, Ioannis and Panneershelvam, Veda and Lanctot, Marc and Dieleman, Sander and Grewe, Dominik and Nham, John and Kalchbrenner, Nal and Sutskever, Ilya and Lillicrap, Timothy and Leach, Madeleine and Kavukcuoglu, Koray and Graepel, Thore and Hassabis, Demis},
    number = {7587},
    month = {1},
    pages = {484--489},
    volume = {529},
    publisher = {Nature Publishing Group},
    url = {https://www.nature.com/articles/nature16961},
    doi = {10.1038/nature16961},
    issn = {1476-4687},
    pmid = {26819042}
}

@article{Saal2013MaterialsOQMD,
    title = {{Materials Design and Discovery with High-Throughput Density Functional Theory: The Open Quantum Materials Database (OQMD)}},
    year = {2013},
    journal = {JOM},
    author = {Saal, James E. and Kirklin, Scott and Aykol, Muratahan and Meredig, Bryce and Wolverton, C.},
    number = {11},
    month = {11},
    pages = {1501--1509},
    volume = {65},
    publisher = {Springer},
    url = {http://link.springer.com/10.1007/s11837-013-0755-4},
    doi = {10.1007/s11837-013-0755-4},
    issn = {1047-4838}
}

@article{SubcommitteeontheMaterialsGenomeInitiative2021MaterialsPlan,
    title = {{Materials Genome Initiative Strategic Plan}},
    year = {2021},
    author = {Subcommittee on the Materials Genome Initiative, NSTC},
    url = {http://www.whitehouse.gov/ostp.}
}

@article{Vasudevan2019MaterialsPhysics,
    title = {{Materials science in the artificial intelligence age: high-throughput library generation, machine learning, and a pathway from correlations to the underpinning physics}},
    year = {2019},
    journal = {MRS Communications},
    author = {Vasudevan, Rama K and Choudhary, Kamal and Mehta, Apurva and Smith, Ryan and Kusne, Gilad and Tavazza, Francesca and Vlcek, Lukas and Ziatdinov, Maxim and Kalinin, Sergei V and Hattrick-Simpers, Jason},
    number = {3},
    month = {9},
    pages = {821--838},
    volume = {9},
    url = {https://doi.org/10.1557/mrc.2019.95 http://link.springer.com/10.1557/mrc.2019.95},
    doi = {10.1557/mrc.2019.95},
    issn = {2159-6859}
}

@article{Sun2024MaterialsMap:Ag-Al-Cu,
    title = {{MaterialsMap: A CALPHAD-Based Tool to Design Composition Pathways through feasibility map for Desired Dissimilar Materials, demonstrated with RSW Joining of Ag-Al-Cu}},
    year = {2024},
    author = {Sun, Hui and Pan, Bo and Yang, Zhening and Krajewski, Adam M. and Bocklund, Brandon and Shang, Shun-Li and Li, Jingjing and Beese, Allison M. and Liu, Zi-Kui},
    month = {3},
    url = {http://arxiv.org/abs/2403.19084},
    doi = {10.48550/arXiv.2403.19084},
    arxivId = {2403.19084}
}

@article{Ward2018Matminer:Mining,
    title = {{Matminer: An open source toolkit for materials data mining}},
    year = {2018},
    journal = {Computational Materials Science},
    author = {Ward, Logan and Dunn, Alexander and Faghaninia, Alireza and Zimmermann, Nils E.R. and Bajaj, Saurabh and Wang, Qi and Montoya, Joseph and Chen, Jiming and Bystrom, Kyle and Dylla, Maxwell and Chard, Kyle and Asta, Mark and Persson, Kristin A. and Snyder, G. Jeffrey and Foster, Ian and Jain, Anubhav},
    month = {9},
    pages = {60--69},
    volume = {152},
    publisher = {Elsevier B.V.},
    url = {https://linkinghub.elsevier.com/retrieve/pii/S0927025618303252},
    doi = {10.1016/j.commatsci.2018.05.018},
    issn = {09270256}
}

@article{Amigo2019MechanicalApplications,
    title = {{Mechanical Properties and the Microstructure of {$\beta$} Ti-35Nb-10Ta-xFe Alloys Obtained by Powder Metallurgy for Biomedical Applications}},
    year = {2019},
    journal = {Metals},
    author = {Amig{\'{o}}, Angélica and Vicente, Angel and Afonso, Conrado and Amig{\'{o}}, Vicente},
    number = {1},
    month = {1},
    pages = {76},
    volume = {9},
    url = {http://www.mdpi.com/2075-4701/9/1/76},
    doi = {10.3390/met9010076},
    issn = {2075-4701}
}

@article{Li2019MechanicalAlloys,
    title = {{Mechanical properties of high-entropy alloys with emphasis on face-centered cubic alloys}},
    year = {2019},
    journal = {Progress in Materials Science},
    author = {Li, Zezhou and Zhao, Shiteng and Ritchie, Robert O. and Meyers, Marc A.},
    month = {5},
    pages = {296--345},
    volume = {102},
    publisher = {Pergamon},
    doi = {10.1016/J.PMATSCI.2018.12.003},
    issn = {0079-6425}
}

@article{Maresca2020Mechanistic1900K,
    title = {{Mechanistic origin of high strength in refractory BCC high entropy alloys up to 1900K}},
    year = {2020},
    journal = {Acta Materialia},
    author = {Maresca, Francesco and Curtin, William A.},
    month = {1},
    pages = {235--249},
    volume = {182},
    publisher = {Pergamon},
    url = {https://linkinghub.elsevier.com/retrieve/pii/S1359645419306755},
    doi = {10.1016/j.actamat.2019.10.015},
    issn = {13596454}
}

@article{Hong2022MeltingMaterials,
    title = {{Melting temperature prediction using a graph neural network model: From ancient minerals to new materials}},
    year = {2022},
    journal = {Proceedings of the National Academy of Sciences of the United States of America},
    author = {Hong, Qi Jun and Ushakov, Sergey V. and de Walle, Axel van and Navrotsky, Alexandra},
    number = {36},
    month = {9},
    pages = {e2209630119},
    volume = {119},
    publisher = {National Academy of Sciences},
    url = {https://www.pnas.org/doi/abs/10.1073/pnas.2209630119},
    doi = {10.1073/PNAS.2209630119/SUPPL{\_}FILE/PNAS.2209630119.SD02.XLSX},
    issn = {10916490},
    pmid = {36044552}
}

@article{Cantor2004MicrostructuralAlloys,
    title = {{Microstructural development in equiatomic multicomponent alloys}},
    year = {2004},
    journal = {Materials Science and Engineering A},
    author = {Cantor, B. and Chang, I. T.H. and Knight, P. and Vincent, A. J.B.},
    number = {1-2 SPEC. ISS.},
    month = {7},
    pages = {213--218},
    volume = {375-377},
    doi = {10.1016/j.msea.2003.10.257},
    issn = {09215093}
}

@article{Tandoc2023MiningAlloys,
    title = {{Mining of lattice distortion, strength, and intrinsic ductility of refractory high entropy alloys}},
    year = {2023},
    journal = {npj Computational Materials},
    author = {Tandoc, Christopher and Hu, Yong-Jie and Qi, Liang and Liaw, Peter K.},
    number = {1},
    month = {4},
    pages = {53},
    volume = {9},
    url = {https://www.nature.com/articles/s41524-023-00993-x},
    doi = {10.1038/s41524-023-00993-x},
    issn = {2057-3960}
}

@article{Gehringer2023ModelsSimple,
    title = {{Models of configurationally-complex alloys made simple}},
    year = {2023},
    journal = {Computer Physics Communications},
    author = {Gehringer, Dominik and Fri{\'{a}}k, Martin and Holec, David},
    number = {27},
    month = {5},
    pages = {108664},
    volume = {286},
    url = {https://linkinghub.elsevier.com/retrieve/pii/S0010465523000097},
    doi = {10.1016/j.cpc.2023.108664},
    issn = {00104655}
}

@book{Lemieux2009MonteSampling,
    title = {{Monte Carlo and Quasi-Monte Carlo Sampling}},
    year = {2009},
    author = {Lemieux, Christiane},
    series = {Springer Series in Statistics},
    publisher = {Springer New York},
    url = {https://link.springer.com/10.1007/978-0-387-78165-5},
    address = {New York, NY},
    isbn = {978-0-387-78164-8},
    doi = {10.1007/978-0-387-78165-5}
}

@misc{Krajewski2021MPDD:Database,
    title = {{MPDD: The Material-Property-Descriptor Database}},
    year = {2021},
    author = {Krajewski, Adam M. and Siegel, Jonathan W. and Shang, Shun-Li and Wang, Yi and Xu, Jinchao and Liu, Zi-Kui},
    url = {https://phaseslab.com/mpdd}
}

@misc{RatsimbazafyMratsim/Arraymancer:Backends,
    title = {{mratsim/Arraymancer: A fast, ergonomic and portable tensor library in Nim with a deep learning focus for CPU, GPU and embedded devices via OpenMP, Cuda and OpenCL backends}},
    author = {Ratsimbazafy, Mamy},
    url = {https://github.com/mratsim/Arraymancer}
}

@article{Rycroft2007MultiscaleFlow,
    title = {{Multiscale modeling in granular flow}},
    year = {2007},
    author = {Rycroft, Christopher Harley},
    publisher = {Massachusetts Institute of Technology},
    url = {https://dspace.mit.edu/handle/1721.1/41557}
}

@techreport{ChenMXNet:Systems,
    title = {{MXNet: A Flexible and Efficient Machine Learning Library for Heterogeneous Distributed Systems}},
    author = {Chen, Tianqi and Li, Mu and Li, Yutian and Lin, Min and Wang, Naiyan and Wang, Minjie and Cmu, U Washington and Nus, Stanford and Nyu, Tusimple and Xiao, Tianjun and Xu, Bing and Zhang, Chiyuan and Zhang, Zheng and Alberta, Microsoft U}
}

@article{Yeh2004NanostructuredOutcomes,
    title = {{Nanostructured high-entropy alloys with multiple principal elements: Novel alloy design concepts and outcomes}},
    year = {2004},
    journal = {Advanced Engineering Materials},
    author = {Yeh, Jien Wei and Chen, Swe Kai and Lin, Su Jien and Gan, Jon Yiew and Chin, Tsung Shune and Shun, Tao Tsung and Tsau, Chun Huei and Chang, Shou Yi},
    number = {5},
    pages = {299--303},
    volume = {6},
    publisher = {Wiley-VCH Verlag},
    doi = {10.1002/adem.200300567},
    issn = {14381656}
}

@article{Aykol2019NetworkDiscovery,
    title = {{Network analysis of synthesizable materials discovery}},
    year = {2019},
    journal = {Nature Communications},
    author = {Aykol, Muratahan and Hegde, Vinay I and Hung, Linda and Suram, Santosh and Herring, Patrick and Wolverton, Chris and Hummelsh{\o}j, Jens S},
    volume = {10},
    url = {https://doi.org/10.1038/s41467-019-10030-5},
    doi = {10.1038/s41467-019-10030-5}
}

@article{Lyakhov2013NewUSPEX,
    title = {{New developments in evolutionary structure prediction algorithm USPEX}},
    year = {2013},
    journal = {Computer Physics Communications},
    author = {Lyakhov, Andriy O. and Oganov, Artem R. and Stokes, Harold T. and Zhu, Qiang},
    number = {4},
    pages = {1172--1182},
    volume = {184},
    doi = {10.1016/j.cpc.2012.12.009},
    issn = {00104655}
}

@misc{Rumpf2006Nim:github.com/nim-lang/Nim,
    title = {{Nim: Nim is a statically typed compiled systems programming language. It combines successful concepts from mature languages like Python, Ada and Modula. Its design focuses on efficiency, expressiveness, and elegance, in that order of priority (github.com/nim-lang/Nim)}},
    year = {2006},
    author = {Rumpf, Andreas},
    url = {https://github.com/nim-lang/Nim}
}

@article{Krajewski2024NimCSO:Optimization,
    title = {{nimCSO: A Nim package for Compositional Space Optimization}},
    year = {2024},
    author = {Krajewski, Adam M. and Debnath, Arindam and Reinhart, Wesley F. and Beese, Allison M. and Liu, Zi-Kui},
    month = {3},
    url = {https://arxiv.org/abs/2403.02340v1},
    doi = {10.48550/arXiv.2403.02340},
    arxivId = {2403.02340}
}

@article{Ye2022NovelAgents,
    title = {{Novel inorganic crystal structures predicted using autonomous simulation agents}},
    year = {2022},
    journal = {Scientific Data},
    author = {Ye, Weike and Lei, Xiangyun and Aykol, Muratahan and Montoya, Joseph H.},
    number = {1},
    month = {6},
    pages = {302},
    volume = {9},
    url = {https://www.nature.com/articles/s41597-022-01438-8},
    doi = {10.1038/s41597-022-01438-8},
    issn = {2052-4463}
}

@article{Liu2018OceanLearning,
    title = {{Ocean of Data: Integrating First-Principles Calculations and CALPHAD Modeling with Machine Learning}},
    year = {2018},
    journal = {Journal of Phase Equilibria and Diffusion},
    author = {Liu, Zi Kui},
    number = {5},
    month = {10},
    pages = {635--649},
    volume = {39},
    publisher = {Springer New York LLC},
    url = {https://link.springer.com/article/10.1007/s11669-018-0654-z},
    doi = {10.1007/S11669-018-0654-Z/FIGURES/8},
    issn = {15477037}
}

@article{Bhatia2021OnA,
    title = {{On the Benefits of Randomly Adjusting Anytime Weighted A*}},
    year = {2021},
    journal = {Proceedings of the International Symposium on Combinatorial Search},
    author = {Bhatia, Abhinav and Svegliato, Justin and Zilberstein, Shlomo},
    number = {1},
    month = {7},
    pages = {116--120},
    volume = {12},
    publisher = {Association for the Advancement of Artificial Intelligence},
    url = {https://ojs.aaai.org/index.php/SOCS/article/view/18558},
    isbn = {9781713834557},
    doi = {10.1609/SOCS.V12I1.18558},
    issn = {2832-9163}
}

@misc{Kalgin2021Onnx2torch:PyTorch,
    title = {{onnx2torch: Convert ONNX models to PyTorch}},
    year = {2021},
    author = {Kalgin, Igor and Yanchenko, Arseny and Ivanov, Pyoter and Goncharenko, Alexander and {ENOT Developers}},
    publisher = {GitHub},
    url = {https://enot.ai/ https://github.com/ENOT-AutoDL/onnx2torch},
    howpublished = {https://enot.ai/}
}

@misc{Bai2019ONNX:Exchange,
    title = {{ONNX: Open Neural Network Exchange}},
    year = {2019},
    booktitle = {GitHub repository},
    author = {Bai, Junjie and Lu, Fang and Zhang, Ke and {other developers}},
    publisher = {GitHub},
    url = {https://github.com/onnx/onnx},
    howpublished = {https://github.com/onnx/onnx}
}

@article{Andersen2021OPTIMADEData,
    title = {{OPTIMADE, an API for exchanging materials data}},
    year = {2021},
    journal = {Scientific Data},
    author = {Andersen, Casper W. and Armiento, Rickard and Blokhin, Evgeny and Conduit, Gareth J. and Dwaraknath, Shyam and Evans, Matthew L. and Fekete, Ádám and Gopakumar, Abhijith and Gra{\v{z}}ulis, Saulius and Merkys, Andrius and Mohamed, Fawzi and Oses, Corey and Pizzi, Giovanni and Rignanese, Gian-Marco and Scheidgen, Markus and Talirz, Leopold and Toher, Cormac and Winston, Donald and Aversa, Rossella and Choudhary, Kamal and Colinet, Pauline and Curtarolo, Stefano and Di Stefano, Davide and Draxl, Claudia and Er, Suleyman and Esters, Marco and Fornari, Marco and Giantomassi, Matteo and Govoni, Marco and Hautier, Geoffroy and Hegde, Vinay and Horton, Matthew K. and Huck, Patrick and Huhs, Georg and Hummelsh{\o}j, Jens and Kariryaa, Ankit and Kozinsky, Boris and Kumbhar, Snehal and Liu, Mohan and Marzari, Nicola and Morris, Andrew J. and Mostofi, Arash A. and Persson, Kristin A. and Petretto, Guido and Purcell, Thomas and Ricci, Francesco and Rose, Frisco and Scheffler, Matthias and Speckhard, Daniel and Uhrin, Martin and Vaitkus, Antanas and Villars, Pierre and Waroquiers, David and Wolverton, Chris and Wu, Michael and Yang, Xiaoyu},
    number = {1},
    month = {8},
    pages = {217},
    volume = {8},
    url = {https://www.nature.com/articles/s41597-021-00974-z},
    doi = {10.1038/s41597-021-00974-z},
    issn = {2052-4463}
}

@article{Hsieh2012OverviewSteels,
    title = {{Overview of Intermetallic Sigma () Phase Precipitation in Stainless Steels}},
    year = {2012},
    journal = {ISRN Metallurgy},
    author = {Hsieh, Chih-Chun and Wu, Weite},
    month = {3},
    pages = {1--16},
    volume = {2012},
    url = {https://www.hindawi.com/journals/isrn/2012/732471/},
    doi = {10.5402/2012/732471},
    issn = {2090-8717}
}

@article{Gyanchandani2011PhysicalMetals,
    title = {{Physical properties of the 6d-series elements from density functional theory: Close similarity to lighter transition metals}},
    year = {2011},
    journal = {PHYSICAL REVIEW B},
    author = {Gyanchandani, Jyoti and Sikka, S K},
    number = {4},
    pages = {172101--172102},
    volume = {83},
    doi = {10.1103/PhysRevB.83.172101}
}

@misc{POSCARFile,
    title = {{POSCAR file}},
    url = {https://cms.mpi.univie.ac.at/vasp/vasp/POSCAR_file.html}
}

@article{Zhang2019PrecipitationAlloy,
    title = {{Precipitation behavior and its impact on mechanical properties in an aged carbon-containing Al0.3Cu0.5CrFeNi2 high-entropy alloy}},
    year = {2019},
    journal = {Materials Characterization},
    author = {Zhang, Kai and Wen, Hongyuan and Zhao, Bingbing and Dong, Xianping and Zhang, Lanting},
    month = {9},
    pages = {109792},
    volume = {155},
    publisher = {Elsevier},
    url = {https://linkinghub.elsevier.com/retrieve/pii/S1044580319312410},
    doi = {10.1016/j.matchar.2019.109792},
    issn = {10445803}
}

@article{Goodall2020PredictingStoichiometry,
    title = {{Predicting materials properties without crystal structure: deep representation learning from stoichiometry}},
    year = {2020},
    journal = {Nature Communications},
    author = {Goodall, Rhys E. A. and Lee, Alpha A.},
    number = {1},
    month = {12},
    pages = {6280},
    volume = {11},
    publisher = {Nature Research},
    url = {http://www.nature.com/articles/s41467-020-19964-7},
    doi = {10.1038/s41467-020-19964-7},
    issn = {2041-1723}
}

@article{Balachandran2018PredictionsTheory,
    title = {{Predictions of new perovskite compounds by combining machine learning and density functional theory}},
    year = {2018},
    journal = {Physical Review Materials},
    author = {Balachandran, Prasanna V and Emery, Antoine A and Gubernatis, James E and Lookman, Turab and Wolverton, Chris and Zunger, Alex},
    pages = {43802},
    volume = {2},
    doi = {10.1103/PhysRevMaterials.2.043802}
}

@article{Mukherjee2016PrintabilityManufacturing,
    title = {{Printability of alloys for additive manufacturing}},
    year = {2016},
    journal = {Scientific Reports},
    author = {Mukherjee, T. and Zuback, J. S. and De, A. and DebRoy, T.},
    month = {1},
    volume = {6},
    publisher = {Nature Publishing Group},
    doi = {10.1038/SREP19717},
    issn = {20452322},
    pmid = {26796864}
}

@article{Otis2017Pycalphad:Python,
    title = {{pycalphad: CALPHAD-based Computational Thermodynamics in Python}},
    year = {2017},
    journal = {Journal of Open Research Software},
    author = {Otis, Richard and Liu, Zi-Kui},
    pages = {1--11},
    volume = {5},
    doi = {10.5334/jors.140},
    issn = {2049-9647}
}

@article{Ong2013PythonAnalysis,
    title = {{Python Materials Genomics (pymatgen): A robust, open-source python library for materials analysis}},
    year = {2013},
    journal = {Computational Materials Science},
    author = {Ong, Shyue Ping and Richards, William Davidson and Jain, Anubhav and Hautier, Geoffroy and Kocher, Michael and Cholia, Shreyas and Gunter, Dan and Chevrier, Vincent L. and Persson, Kristin A. and Ceder, Gerbrand},
    month = {2},
    pages = {314--319},
    volume = {68},
    url = {https://linkinghub.elsevier.com/retrieve/pii/S0927025612006295},
    doi = {10.1016/j.commatsci.2012.10.028},
    issn = {09270256}
}

@article{Paszke2019PyTorch:Library,
    title = {{PyTorch: An Imperative Style, High-Performance Deep Learning Library}},
    year = {2019},
    author = {Paszke, Adam and Gross, Sam and Massa, Francisco and Lerer, Adam and Bradbury Google, James and Chanan, Gregory and Killeen, Trevor and Lin, Zeming and Gimelshein, Natalia and Antiga, Luca and Desmaison, Alban and Xamla, Andreas Köpf and Yang, Edward and Devito, Zach and Raison Nabla, Martin and Tejani, Alykhan and Chilamkurthy, Sasank and Ai, Qure and Steiner, Benoit and Facebook, Lu Fang and Facebook, Junjie Bai and Chintala, Soumith},
    url = {https://dl.acm.org/doi/10.5555/3454287.3455008},
    doi = {10.5555/3454287.3455008}
}

@article{Yoshihara1975Rare-earthBismuthides,
    title = {{Rare-earth bismuthides}},
    year = {1975},
    journal = {Journal of The Less-Common Metals},
    author = {Yoshihara, K. and Taylor, J. B. and Calvert, L. D. and Despault, J. G.},
    number = {2},
    pages = {329--337},
    volume = {41},
    doi = {10.1016/0022-5088(75)90038-7},
    issn = {00225088}
}

@article{Schmidt2019RecentScience,
    title = {{Recent advances and applications of machine learning in solid-state materials science}},
    year = {2019},
    journal = {npj Computational Materials},
    author = {Schmidt, Jonathan and Marques, Mário R. G. and Botti, Silvana and Marques, Miguel A L},
    number = {1},
    month = {12},
    pages = {83},
    volume = {5},
    url = {https://doi.org/10.1038/s41524-019-0221-0 http://www.nature.com/articles/s41524-019-0221-0},
    doi = {10.1038/s41524-019-0221-0},
    issn = {2057-3960}
}

@article{Shen2022ReflectionsOQMD,
    title = {{Reflections on one million compounds in the open quantum materials database (OQMD)}},
    year = {2022},
    journal = {Journal of Physics: Materials},
    author = {Shen, Jiahong and Griesemer, Sean D and Gopakumar, Abhijith and Baldassarri, Bianca and Saal, James E and Aykol, Muratahan and Hegde, Vinay I and Wolverton, Chris},
    number = {3},
    month = {7},
    pages = {031001},
    volume = {5},
    url = {https://iopscience.iop.org/article/10.1088/2515-7639/ac7ba9},
    doi = {10.1088/2515-7639/ac7ba9},
    issn = {2515-7639}
}

@article{Senkov2010RefractoryAlloys,
    title = {{Refractory high-entropy alloys}},
    year = {2010},
    journal = {Intermetallics},
    author = {Senkov, O.N. and Wilks, G.B. and Miracle, D.B. and Chuang, C.P. and Liaw, P.K.},
    number = {9},
    month = {9},
    pages = {1758--1765},
    volume = {18},
    publisher = {Elsevier},
    url = {https://linkinghub.elsevier.com/retrieve/pii/S0966979510002475},
    doi = {10.1016/j.intermet.2010.05.014},
    issn = {09669795}
}

@incollection{Muller2006RemarksPositions,
    title = {{Remarks on Wyckoff positions}},
    year = {2006},
    booktitle = {International Tables for Crystallography},
    author = {M{\"{u}}ller, Ulrich},
    month = {10},
    pages = {24--26},
    publisher = {International Union of Crystallography},
    url = {https://onlinelibrary.wiley.com/iucr/itc/A1a/ch1o3v0001/},
    address = {Chester, England},
    doi = {10.1107/97809553602060000539}
}

@article{Seko2017RepresentationProperties,
    title = {{Representation of compounds for machine-learning prediction of physical properties}},
    year = {2017},
    journal = {Phys. Rev. B},
    author = {Seko, Atsuto and Hayashi, Hiroyuki and Nakayama, Keita and Takahashi, Akira and Tanaka, Isao},
    number = {144110},
    volume = {95},
    url = {https://journals.aps.org/prb/abstract/10.1103/PhysRevB.95.144110},
    doi = {10.1103/PhysRevB.95.144110}
}

@article{MaizHadjAhmed2017RevisitingFlerovium,
    title = {{Revisiting the ground state phase stability of super-heavy element Flerovium}},
    year = {2017},
    journal = {Cogent Physics},
    author = {Maiz Hadj Ahmed, H. and Zaoui, A. and Ferhat, M.},
    editor = {Ahuja, Rajeev},
    number = {1},
    month = {1},
    pages = {1380454},
    volume = {4},
    publisher = {Informa UK Limited},
    url = {https://www.tandfonline.com/doi/full/10.1080/23311940.2017.1380454},
    doi = {10.1080/23311940.2017.1380454},
    issn = {2331-1940}
}

@article{Merchant2023ScalingDiscovery,
    title = {{Scaling deep learning for materials discovery}},
    year = {2023},
    journal = {80 | Nature |},
    author = {Merchant, Amil and Batzner, Simon and Schoenholz, Samuel S and Aykol, Muratahan and Cheon, Gowoon and Dogus Cubuk, Ekin},
    volume = {624},
    url = {https://doi.org/10.1038/s41586-023-06735-9},
    doi = {10.1038/s41586-023-06735-9}
}

@article{Schutt2018SchNetMaterials,
    title = {{SchNet – A deep learning architecture for molecules and materials}},
    year = {2018},
    journal = {The Journal of Chemical Physics},
    author = {Sch{\"{u}}tt, K. T. and Sauceda, H. E. and Kindermans, P.-J. and Tkatchenko, A. and M{\"{u}}ller, K.-R.},
    number = {24},
    month = {6},
    pages = {241722},
    volume = {148},
    publisher = {American Institute of Physics Inc.},
    url = {http://aip.scitation.org/doi/10.1063/1.5019779},
    doi = {10.1063/1.5019779},
    issn = {0021-9606},
    arxivId = {1712.06113}
}

@article{PedregosaFABIANPEDREGOSA2011Scikit-learn:Python,
    title = {{Scikit-learn: Machine Learning in Python}},
    year = {2011},
    journal = {Journal of Machine Learning Research},
    author = {Pedregosa FABIANPEDREGOSA, Fabian and Michel, Vincent and Grisel OLIVIERGRISEL, Olivier and Blondel, Mathieu and Prettenhofer, Peter and Weiss, Ron and Vanderplas, Jake and Cournapeau, David and Pedregosa, Fabian and Varoquaux, Gaël and Gramfort, Alexandre and Thirion, Bertrand and Grisel, Olivier and Dubourg, Vincent and Passos, Alexandre and Brucher, Matthieu and Perrot and{\'{E}}douardand, Matthieu and Duchesnay, andÉdouard and Duchesnay EDOUARDDUCHESNAY, FRÉdouard},
    number = {85},
    pages = {2825--2830},
    volume = {12},
    url = {http://jmlr.org/papers/v12/pedregosa11a.html},
    issn = {1533-7928}
}

@article{Blackman2018ScrambledGenerators,
    title = {{Scrambled Linear Pseudorandom Number Generators}},
    year = {2018},
    author = {Blackman, David and Vigna, Sebastiano},
    month = {5},
    url = {http://arxiv.org/abs/1805.01407},
    arxivId = {1805.01407}
}

@article{Hu2021ScreeningAlloys,
    title = {{Screening of generalized stacking fault energies, surface energies and intrinsic ductile potency of refractory multicomponent alloys}},
    year = {2021},
    journal = {Acta Materialia},
    author = {Hu, Yong-Jie and Sundar, Aditya and Ogata, Shigenobu and Qi, Liang},
    month = {5},
    pages = {116800},
    volume = {210},
    publisher = {Acta Materialia Inc},
    url = {https://linkinghub.elsevier.com/retrieve/pii/S1359645421001804},
    doi = {10.1016/j.actamat.2021.116800},
    issn = {13596454}
}

@article{Wang2023SearchingExperiments,
    title = {{Searching for medium entropy alloys with desired mechanical property by adaptive design combined with experiments}},
    year = {2023},
    journal = {International Journal of Refractory Metals and Hard Materials},
    author = {Wang, Xuefei and Luo, Hong and Zhao, Qiancheng and Cheng, Hongxu and Pan, Zhimin and Li, Xiaogang},
    month = {8},
    pages = {106246},
    volume = {114},
    publisher = {Elsevier},
    url = {https://linkinghub.elsevier.com/retrieve/pii/S0263436823001464},
    doi = {10.1016/j.ijrmhm.2023.106246},
    issn = {02634368}
}

@article{Gao2016SenaryHfNbTaTiVZr,
    title = {{Senary Refractory High-Entropy Alloy HfNbTaTiVZr}},
    year = {2016},
    journal = {Metallurgical and Materials Transactions A: Physical Metallurgy and Materials Science},
    author = {Gao, M. C. and Zhang, B. and Yang, S. and Guo, S. M.},
    number = {7},
    month = {7},
    pages = {3333--3345},
    volume = {47},
    publisher = {Springer Boston},
    doi = {10.1007/s11661-015-3105-z},
    issn = {10735623}
}

@misc{LinShuangLin212/refractory-elements-database:ZR,
    title = {{ShuangLin212/refractory-elements-database: This is the refractory elements thermodynamic database contains CR HF MO NB TA TI V W ZR}},
    author = {Lin, Shuang},
    url = {https://github.com/ShuangLin212/refractory-elements-database}
}

@article{vandeWalle2017SoftwareData,
    title = {{Software tools for high-throughput CALPHAD from first-principles data}},
    year = {2017},
    journal = {Calphad},
    author = {van de Walle, Axel and Sun, Ruoshi and Hong, Qi Jun and Kadkhodaei, Sara},
    month = {9},
    pages = {70--81},
    volume = {58},
    publisher = {Pergamon},
    doi = {10.1016/J.CALPHAD.2017.05.005},
    issn = {0364-5916}
}

@article{Chandrasekaran2019SolvingLearning,
    title = {{Solving the electronic structure problem with machine learning}},
    year = {2019},
    journal = {npj Computational Materials},
    author = {Chandrasekaran, Anand and Kamal, Deepak and Batra, Rohit and Kim, Chiho and Chen, Lihua and Ramprasad, Rampi},
    number = {22},
    month = {2},
    volume = {5},
    url = {https://doi.org/10.1038/s41524-019-0162-7},
    doi = {10.1038/s41524-019-0162-7}
}

@article{Zunger1990SpecialStructures,
    title = {{Special quasirandom structures}},
    year = {1990},
    journal = {Physical Review Letters},
    author = {Zunger, Alex and Wei, S.-H. and Ferreira, L. G. and Bernard, James E.},
    number = {3},
    month = {7},
    pages = {353--356},
    volume = {65},
    url = {https://link.aps.org/doi/10.1103/PhysRevLett.65.353},
    doi = {10.1103/PhysRevLett.65.353},
    issn = {0031-9007}
}

@article{Togo2018Spglib:Search,
    title = {{Spglib: a software library for crystal symmetry search}},
    year = {2018},
    author = {Togo, Atsushi and Tanaka, Isao},
    month = {8},
    url = {http://arxiv.org/abs/1808.01590},
    arxivId = {1808.01590}
}

@article{HintonStochasticEmbedding,
    title = {{Stochastic Neighbor Embedding}},
    author = {Hinton, Geoffrey and Roweis, Sam}
}

@article{Asmussen2007StochasticAnalysis,
    title = {{Stochastic Simulation: Algorithms and Analysis}},
    year = {2007},
    author = {Asmussen, Søren and Glynn, Peter W.},
    series = {Stochastic Modelling and Applied Probability},
    volume = {57},
    publisher = {Springer New York},
    url = {http://link.springer.com/10.1007/978-0-387-69033-9},
    address = {New York, NY},
    isbn = {978-0-387-30679-7},
    doi = {10.1007/978-0-387-69033-9}
}

@article{Gupta2024Structure-awareDatasets,
    title = {{Structure-aware graph neural network based deep transfer learning framework for enhanced predictive analytics on diverse materials datasets}},
    year = {2024},
    journal = {npj Computational Materials},
    author = {Gupta, Vishu and Choudhary, Kamal and DeCost, Brian and Tavazza, Francesca and Campbell, Carelyn and Liao, Wei-keng and Choudhary, Alok and Agrawal, Ankit},
    number = {1},
    month = {1},
    pages = {1},
    volume = {10},
    publisher = {Nature Publishing Group},
    url = {https://www.nature.com/articles/s41524-023-01185-3},
    doi = {10.1038/s41524-023-01185-3},
    issn = {2057-3960}
}

@article{Kang2021SuperiorProcess,
    title = {{Superior mechanical properties and strengthening mechanisms of lightweight AlxCrNbVMo refractory high-entropy alloys (x = 0, 0.5, 1.0) fabricated by the powder metallurgy process}},
    year = {2021},
    journal = {Journal of Materials Science {\&} Technology},
    author = {Kang, Byungchul and Kong, Taeyeong and Ryu, Ho Jin and Hong, Soon Hyung},
    month = {4},
    pages = {32--41},
    volume = {69},
    publisher = {Elsevier},
    url = {https://linkinghub.elsevier.com/retrieve/pii/S1005030220307179},
    doi = {10.1016/j.jmst.2020.07.012},
    issn = {10050302}
}

@article{Crivello2022SupervisedExample,
    title = {{Supervised deep learning prediction of the formation enthalpy of complex phases using a DFT database: The {$\sigma$}-phase as an example}},
    year = {2022},
    journal = {Computational Materials Science},
    author = {Crivello, Jean-Claude and Joubert, Jean-Marc and Sokolovska, Nataliya},
    month = {1},
    pages = {110864},
    volume = {201},
    url = {https://linkinghub.elsevier.com/retrieve/pii/S0927025621005796},
    doi = {10.1016/j.commatsci.2021.110864},
    issn = {09270256}
}

@article{Yu2012TensileTemperatures,
    title = {{Tensile properties of multiphase Mo–Si–B refractory alloys at elevated temperatures}},
    year = {2012},
    journal = {Materials Science and Engineering: A},
    author = {Yu, J. L. and Li, Z. K. and Zheng, X. and Zhang, J. J. and Liu, H. and Bai, R. and Wang, H.},
    month = {1},
    pages = {392--395},
    volume = {532},
    publisher = {Elsevier},
    doi = {10.1016/J.MSEA.2011.11.001},
    issn = {0921-5093}
}

@article{Kononova2019Text-minedRecipes,
    title = {{Text-mined dataset of inorganic materials synthesis recipes}},
    year = {2019},
    journal = {Scientific Data},
    author = {Kononova, Olga and Huo, Haoyan and He, tanjin and Rong, Ziqin and Botari, tiago and Sun, Wenhao and tshitoyan, Vahe and Ceder, Gerbrand},
    number = {203},
    volume = {6},
    url = {https://doi.org/10.1038/s41597-019-0224-1},
    doi = {10.1038/s41597-019-0224-1}
}

@article{Alberi2019TheRoadmap,
    title = {{The 2019 materials by design roadmap}},
    year = {2019},
    journal = {Journal of Physics D: Applied Physics},
    author = {Alberi, Kirstin and Nardelli, Marco Buongiorno and Zakutayev, Andriy and Mitas, Lubos and Curtarolo, Stefano and Jain, Anubhav and Fornari, Marco and Marzari, Nicola and Takeuchi, Ichiro and Green, Martin L and Kanatzidis, Mercouri and Toney, Mike F and Butenko, Sergiy and Meredig, Bryce and Lany, Stephan and Kattner, Ursula and Davydov, Albert and Toberer, Eric S and Stevanovic, Vladan and Walsh, Aron and Park, Nam-Gyu and Aspuru-Guzik, Alán and Tabor, Daniel P and Nelson, Jenny and Murphy, James and Setlur, Anant and Gregoire, John and Li, Hong and Xiao, Ruijuan and Ludwig, Alfred and Martin, Lane W and Rappe, Andrew M and Wei, Su-Huai and Perkins, John},
    number = {1},
    month = {10},
    pages = {013001},
    volume = {52},
    publisher = {IOP Publishing},
    doi = {10.1088/1361-6463/AAD926},
    issn = {0022-3727}
}

@incollection{Toher2018TheDiscovery,
    title = {{The AFLOW Fleet for Materials Discovery}},
    year = {2018},
    booktitle = {Handbook of Materials Modeling},
    author = {Toher, Cormac and Oses, Corey and Hicks, David and Gossett, Eric and Rose, Frisco and Nath, Pinku and Usanmaz, Demet and Ford, Denise C. and Perim, Eric and Calderon, Camilo E. and Plata, Jose J. and Lederer, Yoav and Jahn{\'{a}}tek, Michal and Setyawan, Wahyu and Wang, Shidong and Xue, Junkai and Rasch, Kevin and Chepulskii, Roman V. and Taylor, Richard H. and Gomez, Geena and Shi, Harvey and Supka, Andrew R. and Al Orabi, Rabih Al Rahal and Gopal, Priya and Cerasoli, Frank T. and Liyanage, Laalitha and Wang, Haihang and Siloi, Ilaria and Agapito, Luis A. and Nyshadham, Chandramouli and Hart, Gus L. W and Carrete, Jesús and Legrain, Fleur and Mingo, Natalio and Zurek, Eva and Isayev, Olexandr and Tropsha, Alexander and Sanvito, Stefano and Hanson, Robert M. and Takeuchi, Ichiro and Mehl, Michael J. and Kolmogorov, Aleksey N. and Yang, Kesong and D’Amico, Pino and Calzolari, Arrigo and Costa, Marcio and Gennaro, Riccardo De and Nardelli, Marco Buongiorno and Fornari, Marco and Levy, Ohad and Curtarolo, Stefano},
    pages = {1--28},
    publisher = {Springer International Publishing},
    url = {http://link.springer.com/10.1007/978-3-319-42913-7_63-1},
    address = {Cham},
    doi = {10.1007/978-3-319-42913-7{\_}63-1}
}

@article{Mehl2016ThePrototypes,
    title = {{The AFLOW Library of Crystallographic Prototypes}},
    year = {2016},
    author = {Mehl, Michael J. and Hicks, David and Toher, Cormac and Levy, Ohad and Hanson, Robert M. and Hart, Gus and Curtarolo, Stefano},
    month = {7},
    url = {http://arxiv.org/abs/1607.02532},
    arxivId = {1607.02532}
}

@article{Rubin1981TheBootstrap,
    title = {{The Bayesian Bootstrap}},
    year = {1981},
    journal = {https://doi.org/10.1214/aos/1176345338},
    author = {Rubin, Donald B.},
    number = {1},
    month = {1},
    pages = {130--134},
    volume = {9},
    publisher = {Institute of Mathematical Statistics},
    url = {https://projecteuclid.org/journals/annals-of-statistics/volume-9/issue-1/The-Bayesian-Bootstrap/10.1214/aos/1176345338.full https://projecteuclid.org/journals/annals-of-statistics/volume-9/issue-1/The-Bayesian-Bootstrap/10.1214/aos/1176345338.short},
    doi = {10.1214/AOS/1176345338},
    issn = {0090-5364}
}

@article{2006TheElements,
    title = {{The Chemistry of the Actinide and Transactinide Elements}},
    year = {2006},
    journal = {The Chemistry of the Actinide and Transactinide Elements},
    publisher = {Springer Netherlands},
    doi = {10.1007/1-4020-3598-5}
}

@article{Hall1991TheCrystallography,
    title = {{The crystallographic information file (CIF): a new standard archive file for crystallography}},
    year = {1991},
    journal = {Acta Crystallographica Section A Foundations of Crystallography},
    author = {Hall, S. R. and Allen, F. H. and Brown, I. D.},
    number = {6},
    month = {11},
    pages = {655--685},
    volume = {47},
    url = {https://onlinelibrary.wiley.com/doi/10.1107/S010876739101067X},
    doi = {10.1107/S010876739101067X},
    issn = {01087673}
}

@article{Choudhary2020TheDesign,
    title = {{The joint automated repository for various integrated simulations (JARVIS) for data-driven materials design}},
    year = {2020},
    journal = {npj Computational Materials},
    author = {Choudhary, Kamal and Garrity, Kevin F. and Reid, Andrew C.E. and DeCost, Brian and Biacchi, Adam J. and Hight Walker, Angela R. and Trautt, Zachary and Hattrick-Simpers, Jason and Kusne, A. Gilad and Centrone, Andrea and Davydov, Albert and Jiang, Jie and Pachter, Ruth and Cheon, Gowoon and Reed, Evan and Agrawal, Ankit and Qian, Xiaofeng and Sharma, Vinit and Zhuang, Houlong and Kalinin, Sergei V. and Sumpter, Bobby G. and Pilania, Ghanshyam and Acar, Pinar and Mandal, Subhasish and Haule, Kristjan and Vanderbilt, David and Rabe, Karin and Tavazza, Francesca},
    number = {1},
    month = {12},
    pages = {1--13},
    volume = {6},
    publisher = {Nature Research},
    url = {https://doi.org/10.1038/s41524-020-00440-1},
    doi = {10.1038/s41524-020-00440-1},
    issn = {20573960}
}

@article{Kirklin2015TheEnergies,
    title = {{The Open Quantum Materials Database (OQMD): assessing the accuracy of DFT formation energies}},
    year = {2015},
    journal = {npj Computational Materials},
    author = {Kirklin, Scott and Saal, James E and Meredig, Bryce and Thompson, Alex and Doak, Jeff W and Aykol, Muratahan and R{\"{u}}hl, Stephan and Wolverton, Chris},
    number = {1},
    month = {12},
    pages = {15010},
    volume = {1},
    url = {www.oqmd.org/download. http://www.nature.com/articles/npjcompumats201510},
    doi = {10.1038/npjcompumats.2015.10},
    issn = {2057-3960}
}

@article{Blokhin2018TheGenome,
    title = {{The PAULING FILE Project and Materials Platform for Data Science: From Big Data Toward Materials Genome}},
    year = {2018},
    journal = {Handbook of Materials Modeling},
    author = {Blokhin, Evgeny and Villars, Pierre},
    pages = {1--26},
    publisher = {Springer, Cham},
    url = {https://link.springer.com/referenceworkentry/10.1007/978-3-319-42913-7_62-1},
    isbn = {978-3-319-42913-7},
    doi = {10.1007/978-3-319-42913-7{\_}62-1}
}

@article{Rosenblatt1957TheAutomaton,
    title = {{The Perceptron—a perceiving and recognizing automaton}},
    year = {1957},
    journal = {Report 85-460-1},
    author = {Rosenblatt, Frank}
}

@article{I.Hegde2020TheMaterials,
    title = {{The phase stability network of all inorganic materials}},
    year = {2020},
    journal = {Scientific Advances},
    author = {I. Hegde, Vinay and Aykol, Muratahan and Kirklin, Scott and Wolverton, Chris},
    volume = {6},
    doi = {10.1126/sciadv.aay5606}
}

@article{vandeWalle2018TheDatabase,
    title = {{The Thermodynamic Database Database}},
    year = {2018},
    journal = {Calphad: Computer Coupling of Phase Diagrams and Thermochemistry},
    author = {van de Walle, Axel and Nataraj, Chiraag and Liu, Zi Kui},
    month = {6},
    pages = {173--178},
    volume = {61},
    publisher = {Elsevier Ltd},
    doi = {10.1016/j.calphad.2018.04.003},
    issn = {03645916}
}

@article{Liu2022TheoryTheorem,
    title = {{Theory of cross phenomena and their coefficients beyond Onsager theorem}},
    year = {2022},
    journal = {Materials Research Letters},
    author = {Liu, Zi-Kui},
    number = {7},
    month = {7},
    pages = {393--439},
    volume = {10},
    publisher = {Taylor and Francis Ltd.},
    url = {https://www.tandfonline.com/doi/full/10.1080/21663831.2022.2054668},
    doi = {10.1080/21663831.2022.2054668},
    issn = {2166-3831}
}

@article{Ostrowska2022ThermodynamicExperiments,
    title = {{Thermodynamic modeling of the Al-Co-Cr-Fe-Ni high entropy alloys supported by key experiments}},
    year = {2022},
    journal = {Journal of Alloys and Compounds},
    author = {Ostrowska, Marlena and Riani, Paola and Bocklund, Brandon and Liu, Zi Kui and Cacciamani, Gabriele},
    month = {3},
    pages = {162722},
    volume = {897},
    publisher = {Elsevier},
    doi = {10.1016/J.JALLCOM.2021.162722},
    issn = {0925-8388}
}

@article{Sun2023ThermodynamicESPEI,
    title = {{Thermodynamic modeling of the Nb-Ni system with uncertainty quantification using PyCalphad and ESPEI}},
    year = {2023},
    journal = {Calphad},
    author = {Sun, Hui and Shang, Shun-Li and Gong, Rushi and Bocklund, Brandon J. and Beese, Allison M. and Liu, Zi-Kui},
    month = {9},
    pages = {102563},
    volume = {82},
    publisher = {Pergamon},
    url = {https://linkinghub.elsevier.com/retrieve/pii/S0364591623000354},
    doi = {10.1016/j.calphad.2023.102563},
    issn = {03645916}
}

@article{Ostrowska2020ThermodynamicW,
    title = {{Thermodynamic modelling of the {$\sigma$} and {$\mu$} phases in several ternary systems containing Co, Cr, Fe, Mo, Re and W}},
    year = {2020},
    journal = {Journal of Alloys and Compounds},
    author = {Ostrowska, Marlena and Cacciamani, Gabriele},
    month = {12},
    pages = {156122},
    volume = {845},
    publisher = {Elsevier},
    doi = {10.1016/J.JALLCOM.2020.156122},
    issn = {0925-8388}
}

@article{Shin2006ThermodynamicStructures,
    title = {{Thermodynamic properties of binary hcp solution phases from special quasirandom structures}},
    year = {2006},
    journal = {Physical Review B},
    author = {Shin, Dongwon and Arr{\'{o}}yave, Raymundo and Liu, Zi-Kui and Van de Walle, Axel},
    number = {2},
    month = {7},
    pages = {024204},
    volume = {74},
    url = {https://link.aps.org/doi/10.1103/PhysRevB.74.024204},
    doi = {10.1103/PhysRevB.74.024204},
    issn = {1098-0121}
}

@article{Im2022ThermodynamicModeling,
    title = {{Thermodynamic properties of the Nd-Bi system via emf measurements, DFT calculations, machine learning, and CALPHAD modeling}},
    year = {2022},
    journal = {Acta Materialia},
    author = {Im, Sanghyeok and Shang, Shun Li and Smith, Nathan D. and Krajewski, Adam M. and Lichtenstein, Timothy and Sun, Hui and Bocklund, Brandon J. and Liu, Zi Kui and Kim, Hojong},
    month = {1},
    pages = {117448},
    volume = {223},
    publisher = {Pergamon},
    doi = {10.1016/J.ACTAMAT.2021.117448},
    issn = {1359-6454}
}

@article{Li2019ThermodynamicLearning,
    title = {{Thermodynamic Stability Landscape of Halide Double Perovskites via High-Throughput Computing and Machine Learning}},
    year = {2019},
    journal = {Advanced Functional Materials},
    author = {Li, Zhenzhu and Xu, Qichen and Sun, Qingde and Hou, Zhufeng and Yin, Wan-Jian},
    number = {9},
    month = {2},
    pages = {1807280},
    volume = {29},
    publisher = {Wiley-VCH Verlag},
    url = {http://doi.wiley.com/10.1002/adfm.201807280},
    doi = {10.1002/adfm.201807280},
    issn = {1616301X}
}

@article{Liu2023ThermodynamicsPerspectives,
    title = {{Thermodynamics and its prediction and CALPHAD modeling: Review, state of the art, and perspectives}},
    year = {2023},
    journal = {Calphad},
    author = {Liu, Zi-Kui},
    month = {9},
    pages = {102580},
    volume = {82},
    url = {https://linkinghub.elsevier.com/retrieve/pii/S0364591623000524},
    doi = {10.1016/j.calphad.2023.102580},
    issn = {03645916}
}

@misc{ULTIMATEArpa-e.energy.gov,
    title = {{ULTIMATE | arpa-e.energy.gov}},
    url = {https://arpa-e.energy.gov/technologies/programs/ultimate}
}

@article{Isayev2017UniversalCrystals,
    title = {{Universal fragment descriptors for predicting properties of inorganic crystals}},
    year = {2017},
    journal = {Nature Communications},
    author = {Isayev, Olexandr and Oses, Corey and Toher, Cormac and Gossett, Eric and Curtarolo, Stefano and Tropsha, Alexander},
    number = {1},
    month = {8},
    pages = {15679},
    volume = {8},
    publisher = {Nature Publishing Group},
    url = {http://www.nature.com/articles/ncomms15679},
    doi = {10.1038/ncomms15679},
    issn = {2041-1723},
    pmid = {28580961},
    arxivId = {1608.04782}
}

@article{Metz2016UnrolledNetworks,
    title = {{Unrolled Generative Adversarial Networks}},
    year = {2016},
    journal = {ICLR conference papers},
    author = {Metz, Luke and Brain, Google and Poole, Ben and Pfau, David and Deepmind, Google and Sohl-Dickstein, Jascha},
    month = {11},
    url = {https://arxiv.org/abs/1611.02163v4},
    isbn = {1611.02163v4},
    arxivId = {1611.02163}
}

@article{Feng2019UsingDefects,
    title = {{Using deep neural network with small dataset to predict material defects}},
    year = {2019},
    journal = {Materials and Design},
    author = {Feng, Shuo and Zhou, Huiyu and Dong, Hongbiao},
    month = {1},
    pages = {300--310},
    volume = {162},
    publisher = {Elsevier Ltd},
    doi = {10.1016/j.matdes.2018.11.060},
    issn = {18734197}
}

@article{Scime2019UsingProcess,
    title = {{Using machine learning to identify in-situ melt pool signatures indicative of flaw formation in a laser powder bed fusion additive manufacturing process}},
    year = {2019},
    journal = {Additive Manufacturing},
    author = {Scime, Luke and Beuth, Jack},
    month = {1},
    pages = {151--165},
    volume = {25},
    publisher = {Elsevier B.V.},
    doi = {10.1016/j.addma.2018.11.010},
    issn = {22148604}
}

@misc{VASPPOSCAR,
    title = {{VASP Wiki - POSCAR}},
    url = {https://www.vasp.at/wiki/index.php/POSCAR}
}

@techreport{Rycroft2009Voro++:C++,
    title = {{Voro++: a three-dimensional Voronoi cell library in C++}},
    year = {2009},
    author = {Rycroft, Chris},
    month = {1},
    url = {http://www.osti.gov/servlets/purl/946741-A8FxbI/},
    institution = {Lawrence Berkeley National Laboratory (LBNL)},
    address = {Berkeley, CA},
    doi = {10.2172/946741}
}

@article{Bohm1996VoronoiLattices,
    title = {{Voronoi polyhedra: A useful tool to determine the symmetry and bravais class of crystal lattices}},
    year = {1996},
    journal = {Crystal Research and Technology},
    author = {Bohm, J. and Bohm, M. and Heimann, R. B.},
    number = {8},
    pages = {1069--1075},
    volume = {31},
    publisher = {Wiley-VCH Verlag},
    doi = {10.1002/CRAT.2170310816},
    issn = {02321300}
}

@article{Arjovsky2017WassersteinNetworks,
    title = {{Wasserstein Generative Adversarial Networks}},
    year = {2017},
    author = {Arjovsky, Martin and Chintala, Soumith and Bottou, Léon}
}

@misc{WelcomePydantic,
    title = {{Welcome to Pydantic - Pydantic}},
    url = {https://docs.pydantic.dev/latest/}
}

@article{Liu2022ZentropyExpansion,
    title = {{Zentropy Theory for Positive and Negative Thermal Expansion}},
    year = {2022},
    journal = {Journal of Phase Equilibria and Diffusion},
    author = {Liu, Zi-Kui and Wang, Yi and Shang, Shun-Li},
    month = {2},
    url = {https://link.springer.com/10.1007/s11669-022-00942-z},
    doi = {10.1007/s11669-022-00942-z},
    issn = {1547-7037}
}

\newpage
\chapter*{Vita}
\thispagestyle{empty}
\vspace{-6pt}

{\small
Adam Krajewski was born in Europe, where he spent his childhood and received pre-college education at a school nationally recognized for its university-level chemistry curriculum. He first came to the United States in 2013 and moved completely in 2015 to join the Materials Science Department at Case Western Reserve University. Within the first two months, Adam began research in Prof. Welsch's group. After just one year, he enrolled in graduate courses and also joined Prof. Willard's group, progressively moving from experiments towards theory, modeling, and simulations. In the Fall of 2017, he enrolled in graduate courses in Artificial Intelligence, starting to specialize in applying AI techniques, including Machine Learning, to his research which became focused on hidden process modeling, materials data processing, and data-driven design of magnetocaloric metallic glasses.

After earning his B.S.E. degree in 2019, Adam moved directly to pursue PhD under world-renowned thermodynamics expert Prof. Zi-Kui Liu at Penn State. He had the pleasure of working on implementing various computational techniques, ranging from atomistic machine learning through materials data curation, to purely theoretical considerations, while having the support of colleagues who are specialists in applied ab-initio modeling, thermodynamic calculations, and materials discovery. Since 2022, he has also extensively collaborated with Lawrence Livermore National Lab, where he spent two summers on-site.

Adam has published several computational tools and scientific publications listed under his ORCID record (\href{https://orcid.org/0000-0002-2266-0099}{0000-0002-2266-0099}) and Google Scholar (id:\href{https://scholar.google.com/citations?user=3tvHo8kAAAAJ}{3tvHo8kAAAAJ}) including 4 first-author publications listed below and 9 co-author publications. Furthermore, eight first-author papers are under preparation.

\begin{itemize}

    \item \textit{Extensible Structure-Informed Prediction of Formation Energy with improved accuracy and usability employing neural networks}, Comp. Mat. Sci., Jun. 2022, \href{https://doi.org/10.1016/j.commatsci.2022.111254}{10.1016/j.commatsci.2022.111254}

    \item \textit{Efficient Structure-Informed Featurization and Property Prediction of Ordered, Dilute, and Random Atomic Structures}, arXiv, Apr. 2024., \href{https://doi.org/10.48550/arXiv.2404.02849}{10.48550/arXiv.2404.02849}

    \item \textit{Efficient Generation of Grids and Traversal Graphs in Compositional Spaces towards Exploration and Path Planning Exemplified in Materials}, arXiv, Feb. 2024., \href{https://doi.org/10.48550/arXiv.2402.03528}{10.48550/arXiv.2402.03528}

    \item \textit{nimCSO: A Nim package for Compositional Space Optimization}, arXiv, Mar. 2024.,\\ \href{https://doi.org/10.48550/arXiv.2403.02340}{10.48550/arXiv.2403.02340}
    
\end{itemize}
}

\end{document}